\documentclass{article}
\usepackage{latexsym}
\usepackage{amsmath, amsfonts, amsthm}
\usepackage{epsfig}
\usepackage{stmaryrd}
\usepackage{multicol}
\usepackage{pdfsync}
\usepackage{dsfont}
\usepackage{comment}
\usepackage{algorithmic}
\usepackage{algorithm}
\usepackage{caption}
\usepackage{psfrag}
\usepackage{subfig}
\usepackage{xcolor}
\usepackage{graphics}
\usepackage{enumitem}
\usepackage{mathtools}
\usepackage{bbm}
\usepackage{mathrsfs} 
\usepackage{color}
\usepackage[title]{appendix}
\usepackage{bm}
\usepackage{epstopdf}
\usepackage{cite}
\usepackage[margin=1.0in]{geometry}
\usepackage{booktabs}
\usepackage{multirow}

\usepackage{enumitem}

\usepackage{tikz-cd}

\newcommand{\norm}[1]{\left\lVert #1 \right\rVert}
\newcommand{\bae}{\begin{equation}\begin{aligned}}
		\newcommand{\eae}{\end{aligned}\end{equation}}
\newcommand{\beq}{\begin{equation}}
	\newcommand{\eeq}{\end{equation}}

\theoremstyle{definition}

\theoremstyle{remark}

\numberwithin{equation}{section}

\def \Rey {\mathrm{Re}}

\newcommand{\DNSsub}{{\text{\textsc{dns}}}}

\newcommand{\olDelta}{{\overline{\Delta}}}

\newcommand{\bx}{{\mathbf{x}}}
\newcommand{\by}{{\mathbf{y}}}
\newcommand{\br}{{\mathbf{r}}}

\newcommand{\ol}[1]{\overline{#1}}
\newcommand{\avg}[1]{\langle #1 \rangle}
\newcommand{\olu}{\overline{u}}

\title{Deep Learning Closure Models for Large-Eddy Simulation of Flows around Bluff Bodies}
\author{Justin Sirignano\footnote{Mathematical Institute, University of Oxford, Justin.Sirignano@maths.ox.ac.uk} \phantom{.} and Jonathan F. MacArt\footnote{Department of Aerospace and Mechanical Engineering, University of Notre Dame, jmacart@nd.edu}}
\date{\today}

\begin{document}
\maketitle

\begin{abstract}
  A deep learning (DL) closure model for large-eddy simulation (LES) is developed and evaluated for incompressible flows around a rectangular cylinder at moderate Reynolds numbers. Near-wall flow simulation remains a central challenge in aerodynamic modeling: RANS predictions of separated flows are often inaccurate, while LES can require prohibitively small near-wall mesh sizes.
  The DL-LES model is trained using adjoint PDE optimization methods to match, as closely as possible, direct numerical simulation (DNS) data. It is then evaluated out-of-sample (i.e., for new aspect ratios and Reynolds numbers not included in the training data) and compared against a standard LES model (the dynamic Smagorinsky model). The DL-LES model outperforms dynamic Smagorinsky and is able to achieve accurate LES predictions on a relatively coarse mesh (downsampled from the DNS grid by a factor of four in each Cartesian direction). We study the accuracy of the DL-LES model for predicting the drag coefficient, mean flow, and Reynolds stress. A crucial challenge is that the LES quantities of interest are the steady-state flow statistics; for example, the time-averaged mean velocity $\bar{u}(x) = \displaystyle \lim_{t \rightarrow \infty} \frac{1}{t} \int_0^t u(s,x) ds$. Calculating the steady-state flow statistics therefore requires simulating the DL-LES equations over a large number of flow times through the domain; it is a non-trivial question whether an unsteady partial differential equation model whose functional form is defined by a deep neural network can remain stable and accurate on $t \in [0, \infty)$. Our results demonstrate that the DL-LES model is accurate and stable over large physical time spans, enabling the estimation of the steady-state statistics for the velocity, fluctuations, and drag coefficient of turbulent flows around bluff bodies relevant to aerodynamic applications.  
\end{abstract}

\section{Introduction}

LES and RANS enable computationally tractable predictions by reducing the spatial and/or temporal resolution of the flow field. This comes at the cost of significant physical approximations: unclosed terms, representing the neglected scales, require modeling, which significantly degrades predictive accuracy for realistic flight geometries and conditions. The NASA 2030 CFD Vision Report~\cite{NASA2030} highlights the inadequacy of RANS predictions of separated flows, while explicitly modeled LES has its own inadequacies. For example, the (dynamic) Smagorinsky subgrid-scale (SGS) models~\cite{Smagorinsky1963,Germano1991,Lilly1992} have been successful for free shear flows, but significant challenges remain for LES of wall-bounded flows~\cite{SpalartCFDinIndustry}, wing geometries~\cite{Bose2018a}, and flow-separation predictions at high angles of attack~\cite{Kaltenbach1995}. Applications of LES to realistic geometries are further limited by prohibitively small near-wall mesh sizes required for accuracy~\cite{SpalartCFDinIndustry}, and full-scale modeling remains a heroic effort even with current supercomputing capabilities~\cite{Moin2021}. Therefore, it is critically important to develop more-accurate closure models for LES to enable coarse-mesh simulations \emph{with comparable predictive accuracy to high-fidelity LES and DNS}. Doing so will enable faster, more-aggressive design cycles and a higher degree of simulation-driven design optimization.

As an initial step towards this goal, this article develops and evaluates DL closure models for LES of incompressible flows around rectangular cylinders. Our approach models the unclosed LES terms using a deep neural network (DNN). The solution to the LES PDE system then becomes a function of the embedded DNN, and the DNN parameters are selected such that the solution is a good approximation to the target data. Training therefore requires optimizing over the PDE system, which is both mathematically and computationally challenging. To address this, we derive and implement adjoint PDEs, which enables efficient evaluation of the gradients with respect to the DNN parameters for large datasets. 

In contrast to our optimization approach, the vast majority of current DL closure methods for RANS and LES continue to rely primarily on \emph{a priori} optimization, which estimates the neural network parameters offline, without solving the governing equations (e.g., \cite{Ling1,Ling2}). In this approach, standard supervised-learning methods are used to  predict the DNS-evaluated Reynolds stress/SGS stress using the \emph{a priori} DNS-evaluated time-averaged/filtered velocities as model inputs. The objective function for \emph{a priori} training  is therefore completely decoupled from the PDE model, which hinders predictive accuracy and even stability in  \emph{a posteriori} predictions (i.e., substituting the closure model into LES or RANS simulations). Thus the \emph{a priori} optimization approach  is sub-optimal for DL closure models for physics: for example, the neural network is trained with DNS variables as inputs but will receive RANS/LES variables during predictive simulations. 

Our adjoint-trained DL-LES approach has been previously implemented for decaying isotropic turbulence \cite{Sirignano2020} and turbulent jet flows \cite{MacArt2021}. These flows, of course, are very different than the flow around a bluff body considered in this paper. Two new problems are considered in this paper. The first is modeling near-wall turbulent flows, which is a central modeling challenge in aerodynamics. The second is modeling the steady-state flow statistics of a turbulent flow. \cite{Sirignano2020} and \cite{MacArt2021} only study the accuracy of DL-LES models for transient flows over a short time span. In many applications (e.g., an airfoil or wing), although the flow itself does not converge (it is unsteady), the flow statistics -- such as the drag coefficient, mean profile, and Reynolds stress averaged over time $t$ -- do converge as $t \rightarrow \infty$. These steady-state flow statistics evaluate the long-run, average performance of an aerodynamic design and therefore their accurate prediction is crucial.

\section{Governing Equations and Numerical Methods}

The incompressible Navier--Stokes equations are
\begin{align}
\frac{\partial u_i}{\partial t} &= - \frac{\partial p}{\partial x_i} - \frac{\partial u_i u_j}{\partial x_j} + \frac{1}{\textrm{Re} } \frac{\partial^2 u_i}{\partial x_j^2}, \phantom{....} x \in \Omega, \notag \\
0 &= \frac{\partial u_j}{\partial x_j}, \phantom{....} x \in \Omega,
\end{align}
where $\Omega$ is the interior of the domain and $u$ satisfies the standard no-slip boundary conditions for the velocity on the rectangular body. The pressure satisfies Neumann boundary conditions on all boundaries, including the rectangular body. The inlet boundary condition is Dirichlet with $u = (1,0,0)$, while the outflow boundary condition is governed by a convective outflow equation. The spanwise direction $x_3$ has periodic boundary conditions, while the transverse direction $x_2$ has boundary conditions $u_2 = 0$ and Neumann conditions for $(u_1, u_3)$.

Define the filtered quantity $\tilde{\phi}(\bx,t) \equiv \int_\Omega G(\bx-\br)\phi(\br,t) d \br$, where $\bx = (x_1, x_2, x_3)$, and  $G(\by)$  is a filter kernel. We use a box filter with unit support $G(\by)=1$ inside cubes with sides of length $\olDelta$ and zero weights otherwise. The filtered velocity $\tilde{u}$ satisfies the LES equation
\begin{align}
  \frac{\partial \tilde{u}_i}{\partial t} &=     - \frac{\partial \tilde{p}}{\partial x_i} - \frac{\partial \tilde{u}_i \tilde{u}_j}{\partial x_j}
    + \frac{1}{\textrm{Re}} \frac{\partial^2 \tilde{u}_i}{\partial x_j^2}
     \textcolor{red}{ - \frac{\partial \tau_{ij}^r}{\partial x_j} + \Gamma_i} , \notag \\
      0 &= \frac{\partial \tilde{u}_k}{\partial x_k} .
\end{align}
The residual subgrid-scale stress $\tau_{ij}^r\equiv\ol{u_iu_j}-\ol{u}_i\ol{u}_j$ is unclosed and cannot be evaluated using the resolved flow state $\tilde{u}$. $\Gamma_i$ are additional unclosed terms produced by integration by parts at the boundary of the bluff body (i.e., the pressure term will produce an extra unclosed term at the boundary). In addition, if a nonuniform filter is used, additional unclosed terms will be produced and will be included in $\Gamma_i$. In our approach, we model the unclosed terms $- \frac{\partial \tau_{ij}^r}{\partial x_j} + \Gamma_i$ using a deep learning model $h(\tilde{u}_x; \theta)$, where $\theta$ are the neural network parameters. This leads to the DL-LES model:
\begin{align}
  \frac{\partial \tilde{u}_i}{\partial t} &=     - \frac{\partial \tilde{p}}{\partial x_i} - \frac{\partial \tilde{u}_i \tilde{u}_j}{\partial x_j}
    + \frac{1}{\textrm{Re}} \frac{\partial^2 \tilde{u}_i}{\partial x_j^2}
     \textcolor{blue}{ - \frac{\partial h_i}{\partial x_j}(\tilde{u}_x; \theta) }, \notag \\
      0 &= \frac{\partial \tilde{u}_k}{\partial x_k}.
      \label{DLLES}
\end{align}
Let $\tilde{u}^{\theta}$ be the solution to (\ref{DLLES}). The solution depends upon the parameters $\theta$ for the embedded neural network $h(\cdot; \theta)$. The parameters $\theta$ must be selected such that solution $\tilde{u}^{\theta}$ matches trusted data (high-fidelity numerical and/or experimental data) as closely as possible. For example, the parameters could be selected to minimize the objective function
\begin{equation}
J(\theta) =  \frac{1}{2} \int_0^T \int_{\Omega} \norm{ \tilde{u}^{\theta}(t,x) - \tilde{u}^{\textrm{DNS}}(t,x) } dx dt, 
\label{Obj1}
\end{equation}
where $\tilde{u}^{\textrm{DNS}}$ is the filtered DNS data. The optimization problem (\ref{Obj1}) requires optimizing over the system of nonlinear PDEs (\ref{DLLES}) which are a function of a high-dimensional set of neural network parameters:
\begin{equation*}
\textrm{Parameters } \theta \longrightarrow \textrm{Neural network } h \longrightarrow \textrm{Solve PDEs (\ref{DLLES}) for } \tilde{u}^{\theta} \longrightarrow J(\theta).
\end{equation*}
This is a computationally challenging optimization problem, for the neural network parameters $\theta$ are high-dimensional. We derive and solve adjoint PDEs to enable computationally efficient optimization of the DL-LES equations (\ref{DLLES}).

\subsection{Optimization of the deep learning closure model} \label{TrainingAlgorithm}

A nonuniform mesh is crucial for the numerical simulation of (\ref{DLLES}). Smaller mesh sizes are required near the walls of the rectangular cylinder while, to reduce computational cost, larger mesh sizes are used farther away in the flow. Before deriving the adjoint equations, we re-write (\ref{DLLES}) on a nonuniform mesh using a coordinate transformation:

\begin{align}
  \frac{\partial \tilde{u}_i}{\partial t} &=     - c_i(x_i) \frac{\partial \tilde{p}}{\partial x_i} - c_j(x_j) \frac{\partial \tilde{u}_i \tilde{u}_j}{\partial x_j}
    +  \frac{1}{\textrm{Re}} c_j( x_j) \frac{\partial}{\partial x_j} [ c_j( x_j) \frac{\partial \tilde{u}_i}{\partial x_j} ]
      - c_j(x_j) \frac{\partial h_i}{\partial x_j}(c^{\top} \tilde{u}_x; \theta) , \notag \\
      0 &= c_k(x_k) \frac{\partial \tilde{u}_k}{\partial x_k},
      \label{DLLESnonuniform}
\end{align}
where the nonuniform mesh is $z_j = g_j(x_j)$ and $c_j = ( \frac{\partial g_j}{\partial x}(x_j) )^{-1}$. In the spanwise direction $x_3$, $g_3(x) = x$ (i.e., a uniform mesh is used). The training algorithm, at a high-level, optimizes over (\ref{DLLESnonuniform}) to match the filtered DNS velocity on a large set of short time intervals ($10^{-1}$ seconds). That is, we train a closure model to learn the correct short-term evolution of the Navier--Stokes equations.  Parameters are updated using a gradient descent algorithm (RMSprop). The gradients are efficiently evaluated using the adjoint PDE for (\ref{DLLESnonuniform}). In summary, the training algorithm is:
\begin{itemize}
    \item We select time intervals $[t_n, t_n + \tau]$ for $n =1, \ldots, N$.
    \item Initialize the DL-LES model at time $t_n$ with the filtered DNS solution for each sample $n = 1, \ldots, N$.
    \item Parallel simulation of the DL-LES equation for $n =1, \ldots, N$.
    \item Construct the objective function
    \begin{align}
    J(\theta) &= \sum_{n=1}^N J^n(\theta), \notag \\
    J_n(\theta) &= \int_{\Omega} \norm{\tilde{u}^{\theta}(t_n+ \tau,x) - \tilde{u}^{\textrm{DNS}}(t_n + \tau,x) } dx,
    \end{align}
    where $\tilde{u}^{\theta}$ is the solution to (\ref{DLLESnonuniform}) with parameters $\theta$.
    \item Parallel solution of the $N$ adjoint PDEs to evaluate the gradient of the objective function $\nabla_{\theta} J(\theta)$.
    \item Update the parameters $\theta$ using the RMSprop algorithm.
\end{itemize}

We can efficiently evaluate the gradient of the objective function with respect to the high-dimensional neural network parameters $\theta$ by solving the following adjoint PDE on $[t_n, t_n + \tau]$:

\begin{align}
- \frac{\partial \hat u_i}{\partial t} &= -\frac{\partial}{\partial x_i}[ c_i \hat p] + \tilde u_j \frac{\partial }{\partial x_j} [ c_j \hat u_i ] + \tilde u_j \frac{\partial }{\partial x_i} [c_i \hat u_j ] +  \frac{1}{\textrm{Re}} \frac{\partial}{\partial x_j} \big{[} c_j \frac{\partial}{\partial x_j} [ \hat u_i c_j ] \big{]} \notag \\
&+ \frac{\partial}{\partial x_m} \big{[} c_m \hat{u}_k c_j \frac{\partial^2 h_k}{\partial x_j \partial z_{mi}}(c^{\top} \tilde u_x; \theta) \big{]}, \notag \\
0 &= \frac{\partial}{\partial x_k} [ c_k \hat u_k],
\label{DLadjoint}
\end{align}
where $z_{mi} = \frac{\partial \tilde u_i}{\partial x_m}$ and the final condition is $\hat u_i(t_n + \tau, x) = \nabla_{u_i} \norm{\tilde{u}(t_n+ \tau,x) - \tilde{u}^{\textrm{DNS}}(t_n + \tau,x) }$. $\hat u = 0$ on the boundary of the rectangular cylinder and at the inlet. $\hat u_2 = 0$ at the transverse boundaries. To simplify computational implementation for the training algorithm, a constant Dirichlet boundary condition is imposed at the exit, which leads to $\hat u = 0$ at the outflow boundary. (The convective outflow boundary condition is however used for the \emph{a posteriori simulations} to evaluate the DL-LES model.) Assuming $\frac{\partial c_j}{\partial x_j} = 0$ at $x_1 = 0,L$ and $x_2 = -L, L$, $\hat u_2$ and $\hat u_3$ satisfy Neumann boundary conditions at $x_2 = -L, L$.  $\hat p$ satisfies Neumann boundary conditions. We assume that the solutions (and their derivatives) to (\ref{DLadjoint}) and (\ref{DLLESnonuniform}) are periodic in $x_3$. Then, by multiplying equation (\ref{DLLESnonuniform}) by the adjoint variables $\hat u_i$ and using integration by parts, we can derive that the gradient $\nabla_{\theta} J(\theta)$ satisfies the formula:
\begin{align}
\nabla_{\theta} J_n(\theta) = \sum_{i,j=1}^3 \int_0^T \int_{\Omega} \hat u_i c_j \frac{\partial^2 h_i}{\partial x_j \partial \theta}(c^{\top} \tilde{u}_x ; \theta) dx dt. 
\end{align}
Therefore, the gradient $\nabla_{\theta} J(\theta)$ can be evaluated using the solution to the adjoint equation (\ref{DLadjoint}). This allows fast, computationally efficient training of the DL-LES model since the adjoint equation has the same number of equations (3 momentum and 1 continuity equation) as the original Navier--Stokes equation, \emph{no matter the dimension of the neural network parameters $\theta$}. 

The training algorithm optimizes the DL-LES partial differential equation to match the evolution of the filtered DNS velocity over short time intervals. Our ultimate goal of course is a long-run simulation of the DL-LES model, over many flow times through the domain, to estimate time-averaged statistics for the drag coefficient, mean velocity (i.e., $\bar{u}(x) = \displaystyle \lim_{t \rightarrow \infty} \int_0^t u(s,x) ds$), and Reynolds stress. In our \emph{a posteriori} simulations to evaluate the model in Section \ref{Evaluation}, we test whether learning the short-term evolution can produce accurate long-run simulations for the time-averaged statistics.

\subsection{Multi-GPU Accelerated PDE Optimization} \label{Computational}

The training algorithm in Section \ref{TrainingAlgorithm} requires the solution of large numbers of LES equations and adjoint equations for each optimization iteration. Furthermore, the DL-LES model requires evaluating a neural network at each mesh point and every time step. We parallelize these computations using GPUs. Each GPU can simultaneously solve multiple DL-LES forward and adjoint equations by representing a PDE solution as a tensor $4 \times N_1 \times N_2 \times N_3 \times M$, where $M$ is the number of independent PDE systems, and $(N_1, N_2, N_3)$ is the number of mesh points for $(x_1, x_2, x_3)$. The overall training is furthermore distributed across multiple GPUs. Model parameters are updated using synchronized distributed (RMSprop) gradient descent, where the communication between GPUs is implemented using the Message Passing Interface (MPI).

\section{Numerical Results and Analysis} \label{Evaluation}

\subsection{Numerical Experiments}

\begin{figure}
  \centering
  \includegraphics[width=0.40\textwidth]{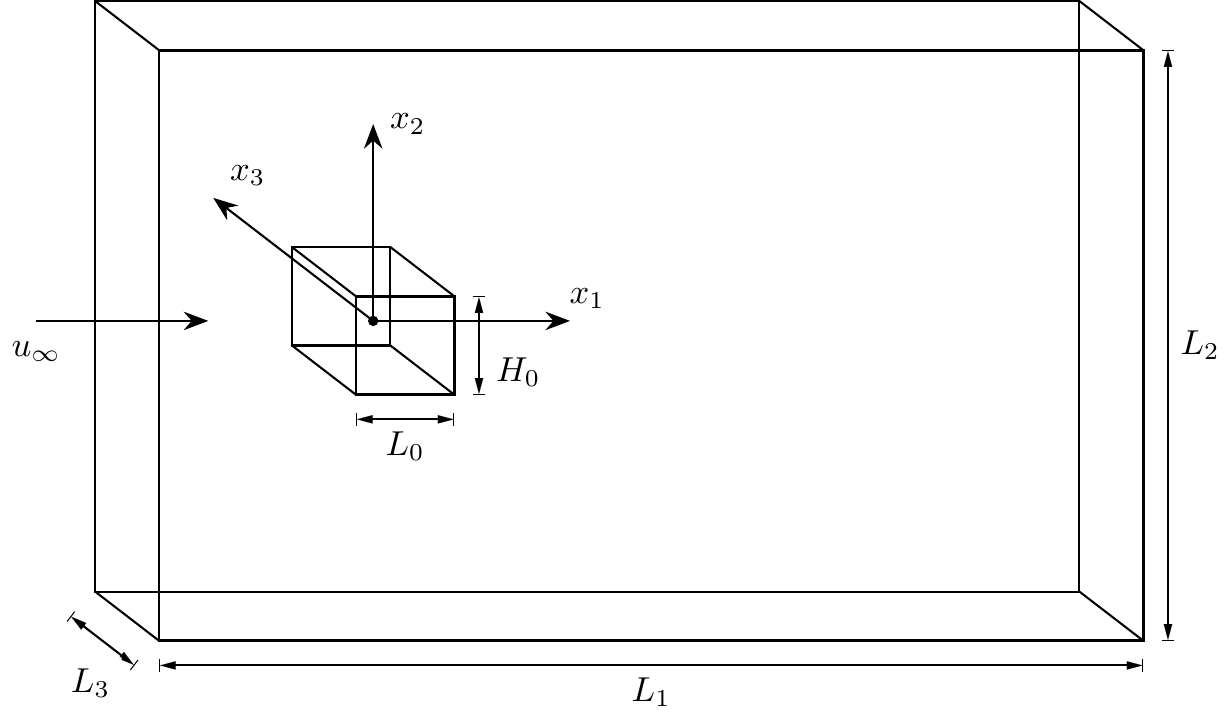}
  \includegraphics[width=0.40\textwidth]{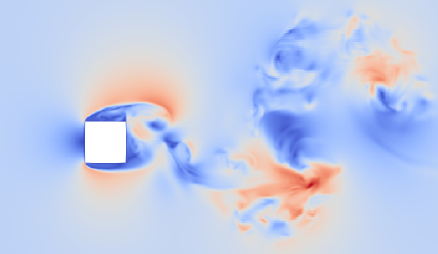}
  \caption{Schematic of the wake problem geometry.}
  \label{fig:dim}
\end{figure}

Data for model training and evaluation is obtained using DNS. Four configurations are simulated for different blockage aspect ratios and bulk Reynolds numbers $\Rey_0=\rho u_\infty H_0/\mu$, where $\rho=1$ is the constant density, $u_\infty=1$ is the uniform freestream velocity, $H_0=1$ is the blockage height (kept constant for all the cases), and $\mu$ is the dynamic viscosity. The blockage length $L_0$ is varied to generate distinct geometries. Mesh sizes, dimensions, aspect ratios $\mathrm{AR}=L_0/H_0$, and bulk Reynolds numbers are listed in Table~\ref{tab:config}.
A schematic of the computational domain is shown in Figure~\ref{fig:dim}. 
\begin{table}
\centering
\caption{DNS configuration parameters. All units are dimensionless.}
\label{tab:config}
\begin{tabular}{ c c c c c c c c }
  \toprule
  Case & Identifier   & $N_1$, $N_2$, $N_3$ & $L_1$, $L_2$, $L_3$ & $H_0$, $L_0$ & AR & $\Rey_{0}$ & $t_f-t_s$ \\
  \midrule
  (i)   & AR1 $\Rey_0=1000$ & 512, 256, 64   & 20, 12, 4 & 1, 1 & 1 & 1000 & 427 \\
  (ii)  & AR2 $\Rey_0=1000$ & 512, 256, 64   & 20, 12, 4 & 1, 2 & 2 & 1000 & 77.0 \\
  (iii) & AR2 $\Rey_0=2000$ & 1024, 512, 128 & 20, 12, 4 & 1, 2 & 2 & 2000 & 96.5 \\
  (iv)  & AR4 $\Rey_0=1000$ & 512, 256, 64   & 20, 12, 4 & 1, 4 & 4 & 1000 & 194 \\
  \bottomrule
\end{tabular}
\end{table}

In the DNS simulations, the dimensionless Navier--Stokes equations are solved in the incompressible limit using second-order centered differences on a staggered mesh~\cite{Harlow1965}. Time integration is by a fractional-step method~\cite{Kim1985} and a linearized trapezoid method in an alternating-direction implicit (ADI) framework~\cite{Desjardins2008,MacArt2016}. The mesh is nonuniform in $x_1$-$x_2$ planes, with local refinement within the blockage boundary layers,  and uniform in the $x_3$ direction.

The time step $\Delta t_\DNSsub$ was fixed, with steady-state CFL numbers of approximately 0.4 for all cases. At the inflow boundary ($-x_1$), Dirichlet conditions prescribe the uniform freestream velocity profile. The streamwise ($\pm x_2$) boundaries are slip walls, and a convective outflow is imposed at the $+x_1$ boundary. The  cross-stream ($x_3$) direction is periodic.

Instantaneous DNS fields $u_i(\bx,t)$ were postprocessed to obtain filtered fields $\olu_i(\bx,t)$. This used a box filter kernel with local coarsening ratios $\olDelta/\Delta_\DNSsub=4$ (cases i, ii, iv) or 8 (case iii), where $\olDelta$ is the local LES grid spacing and $\Delta_\DNSsub$ is the local DNS grid spacing. These filtered fields were then downsampled by the same filter-to-DNS grid ratios to obtain coarse-grained fields representative of LES solutions.

\emph{A posteriori} LES calculations use grids coarsened by the same constant factors as the filtered, downsampled DNS fields, which is consistent with the common practice of implicitly filtered LES. Therefore, the LES calculations use 64 times ($\olDelta/\Delta_\DNSsub=4$) or  or 512 times ($\olDelta/\Delta_\DNSsub=8$) fewer mesh points than the corresponding DNS calculations. The DL-LES simulation uses the same numerical methods as in \cite{Sirignano2020} and \cite{MacArt2021}. The benchmark LES simulations, such as dynamic Smagorinsky, use the same numerical method as in the DNS simulation. 

Statistics were obtained by averaging in the cross-stream ($x_3$) direction and time,
\begin{equation}
  \avg{\phi}(x_1,x_2) = \frac{1}{L_3(t_f-t_s)}\int_{-L_3/2}^{L_3/2}\int_{t_s}^{t_f} \phi(x_1,x_2,x_3,t)\, dt dx_3,
\end{equation}
where $t_s$ and $t_f$ are the start and end times for averaging, respectively. The DNS calculations were advanced to steady state before recording statistics; the minimum time to steady state was $t_s=41.5$ dimensionless units (AR2 $\Rey_0=2000$), and the maximum to steady state was $t_s=393$ units (AR1 $\Rey_0=1000$). The minimum DNS statistics recording window was $t_f-t_s=77$ time units (AR2 $\Rey_0=1000$), and the maximum was $t_f-t_s=427$ time units (AR1 $\Rey_0=1000$). The statistics recording windows for all cases are listed in Table~\ref{tab:config}. Spectra for the DNS-evaluated lift coefficients show at minimum two decades of wavenumbers below the shedding frequency for each case, indicating that the long-time vortex dynamics are sufficiently well represented in the statistics. LES calculations were restarted from filtered, downsampled DNS fields and were allowed to advance for several thousand time units before recording statistics.

\subsection{Model Training and Evaluation}

Two DL-LES models are trained, one for each of cases (i) and (iii). These models will  be denoted ``DL-AR1" and ``DL-AR2," respectively. As benchmarks, the DL-LES models are compared against the dynamic Smagorinsky model and ``no model" LES. In ``no model" LES, the unclosed terms are set to zero.

Models are trained for a single configuration and tested in \emph{a posteriori} LES on all configurations. The three out-of-sample configurations for each trained model have new physical characteristics (either AR or $\Rey$) for which the model has not been trained. For example, one model is trained on data from case (iii) and then tested on cases (i), (ii), (iii), and (iv). Cases (i), (ii), and (iv) would then be completely out-of-sample. Case (ii) would be ``quasi-out-of-sample" in that the model has been trained on data from (ii) but only on short time intervals (see Section \ref{TrainingAlgorithm}) and now it is simulated for (ii) over many flow times.

\subsection{Drag Coefficient}

The drag coefficients for the LES models and test cases are reported in Table~\ref{DragCoeff}.
The deep learning models consistently perform as well or better than the dynamic Smagorinsky model compared to the filtered DNS data. ``No model" LES consistently underperforms the learned models. For several configurations, the DL-LES models are substantially more accurate than the benchmark models. The outperformance of DL-LES for predicting the drag coefficient is a direct consequence of its improved accuracy for modeling the mean velocity and Reynolds stress. Section \ref{ScatterPlots} evaluates the DL-LES models' accuracy for these statistics.

\begin{table}
\centering
\caption{Drag coefficients $c_d$ evaluated from filtered DNS and LES with different models.}
\label{DragCoeff}
\begin{tabular}{ c c | c | c c c c }
  \toprule
  & & & \multicolumn{4}{c}{Model} \\
  Case & Identifier   & f-DNS & NM   & DS   & DL-AR1 & DL-AR2  \\
  \midrule
  (i)   & AR1 $\Rey=1000$ & 2.45  & 1.88 & 2.15 & 2.43   & 2.51    \\
  (ii)  & AR2 $\Rey=1000$ & 1.94  & 1.61 & 1.65 & 1.90   & 1.95    \\
  (iii) & AR2 $\Rey=2000$ & 2.09  & 1.62 & 1.68 & 1.84   & 1.96    \\
  (iv)  & AR4 $\Rey=1000$ & 1.61  & 1.53 & 1.54 & 1.54   & 1.61   \\

  \bottomrule
\end{tabular}
\end{table}

\subsection{Steady-State Statistics} \label{ScatterPlots}

\begin{figure}
\centering
\includegraphics[width=0.32\textwidth]{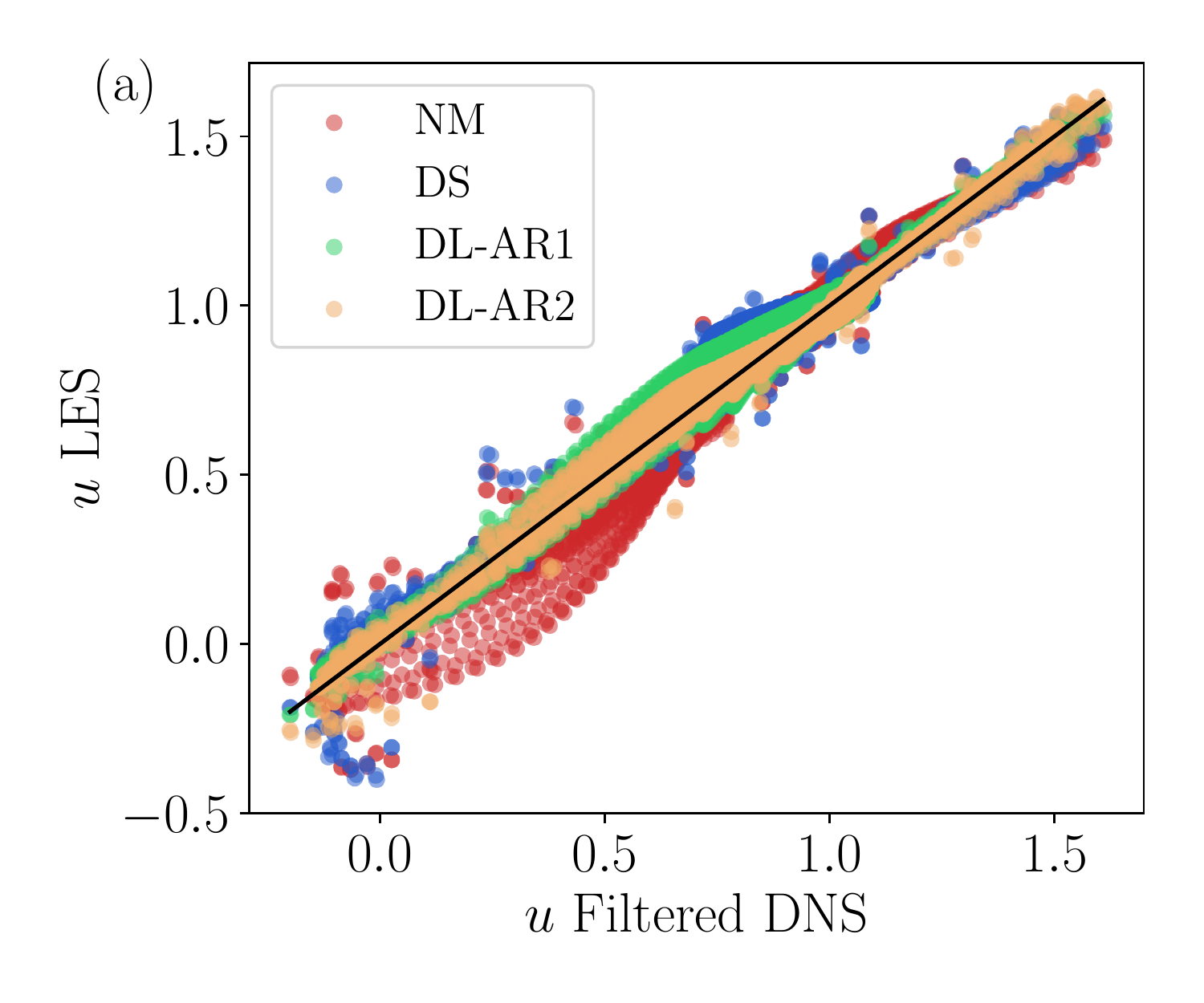}
\includegraphics[width=0.32\textwidth]{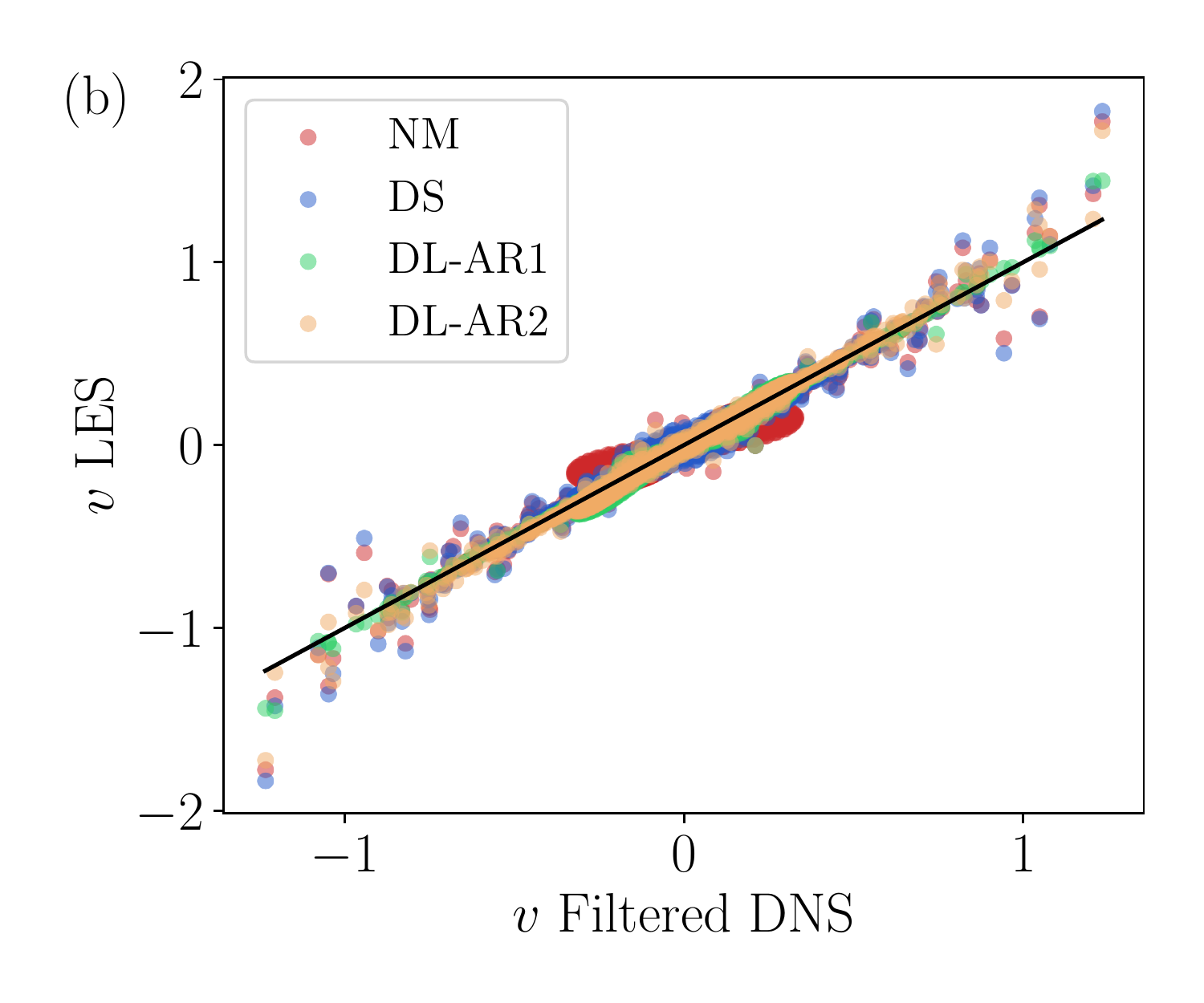}
\includegraphics[width=0.32\textwidth]{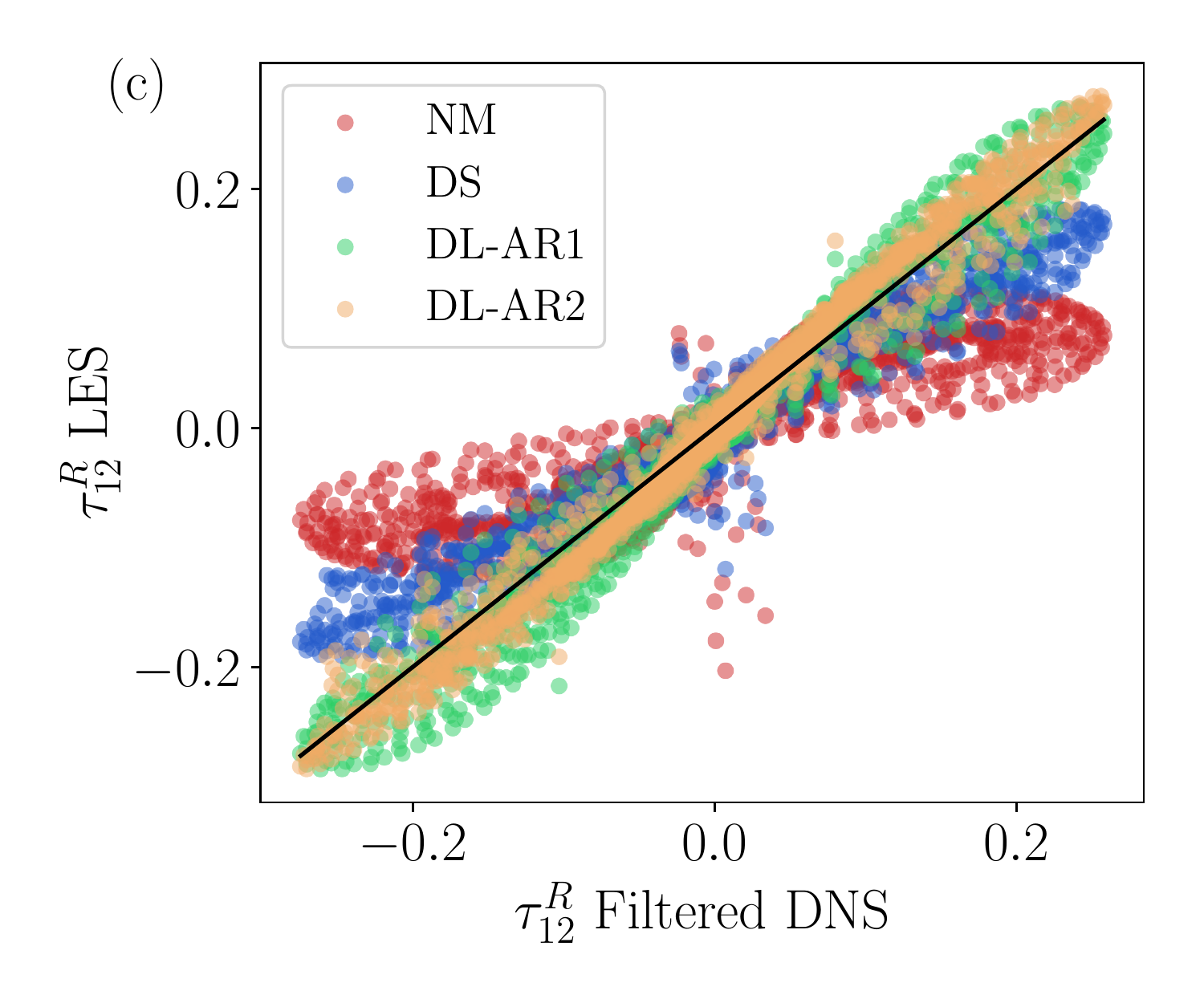}
\includegraphics[width=0.32\textwidth]{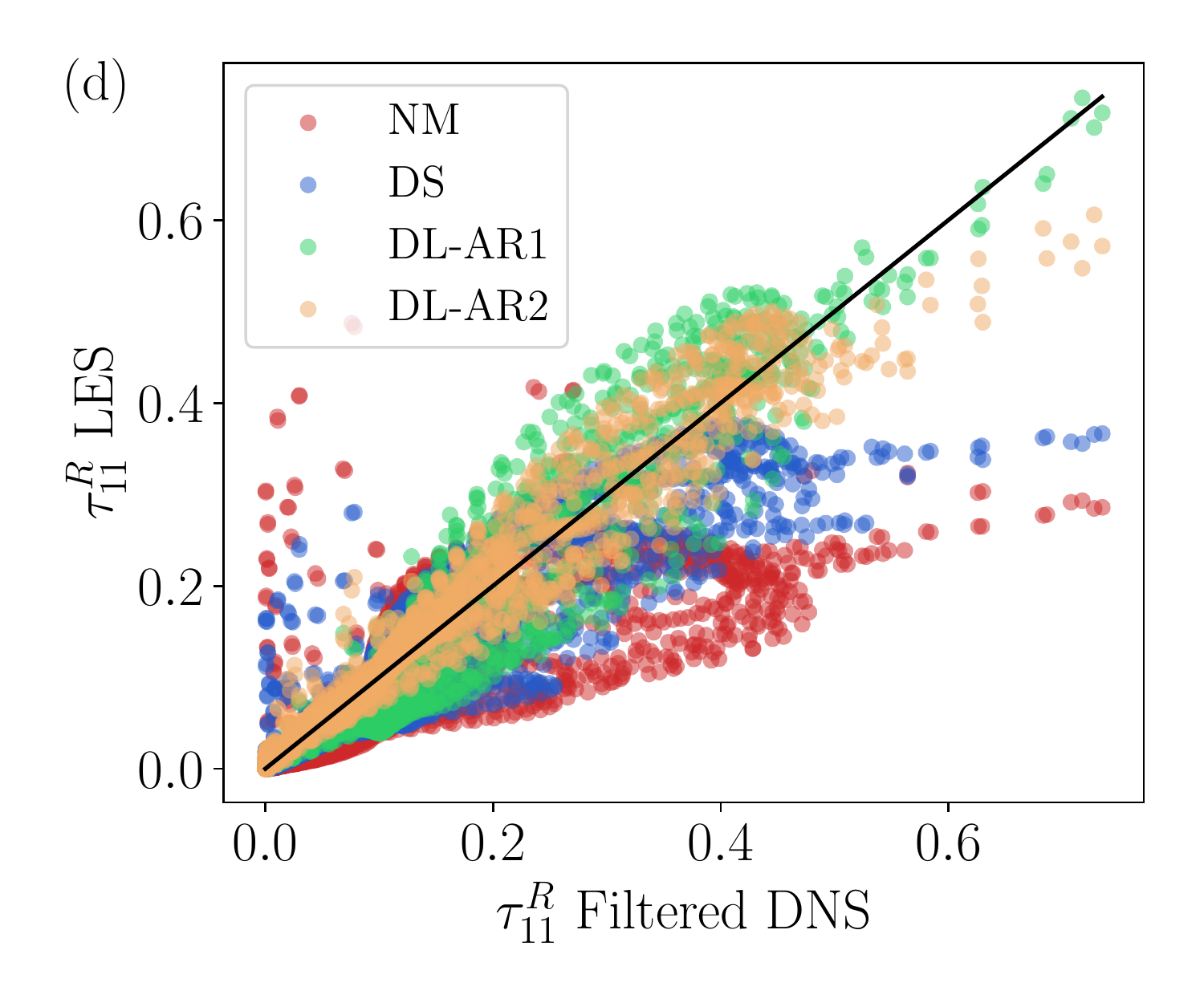}
\includegraphics[width=0.32\textwidth]{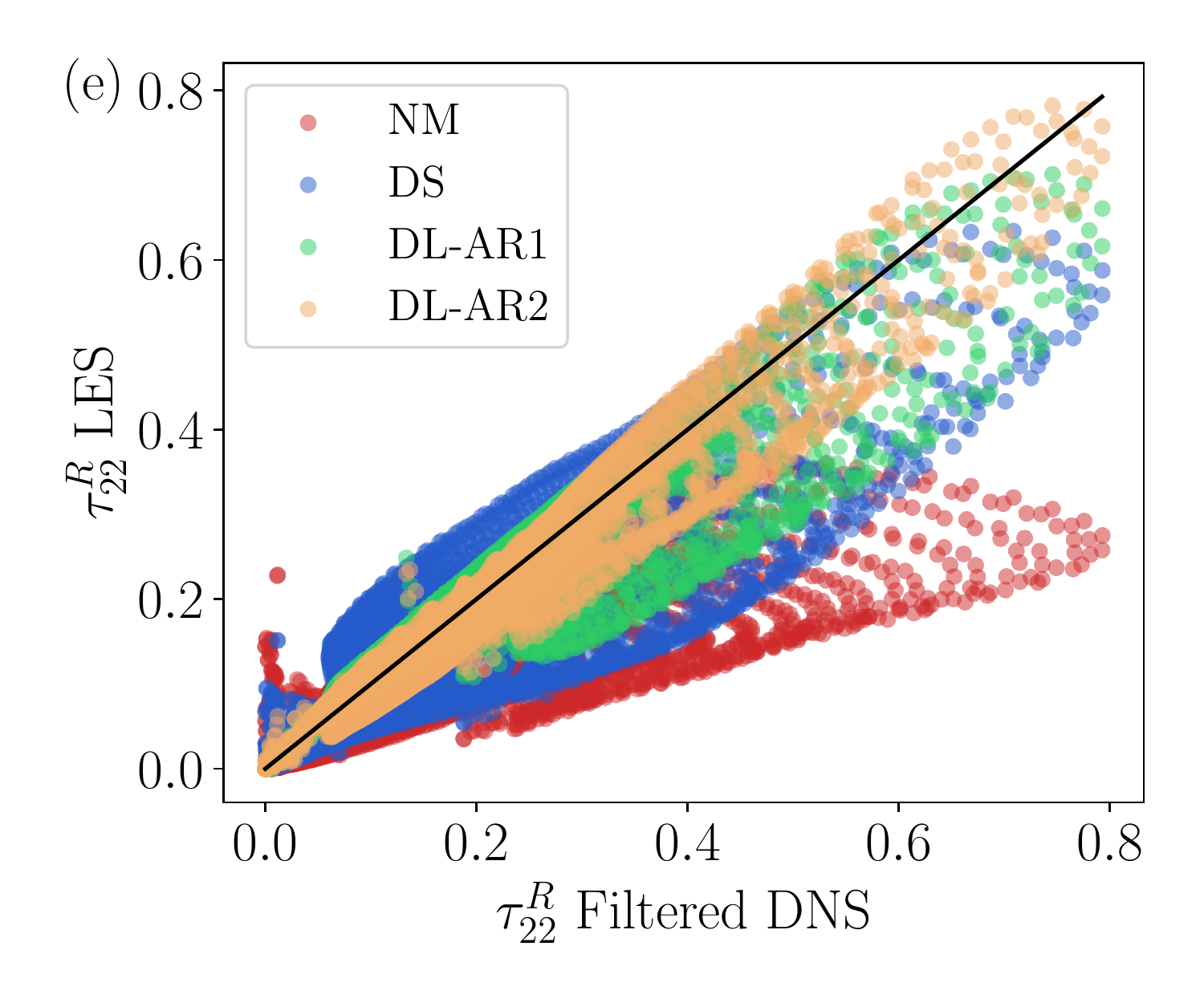}
\includegraphics[width=0.32\textwidth]{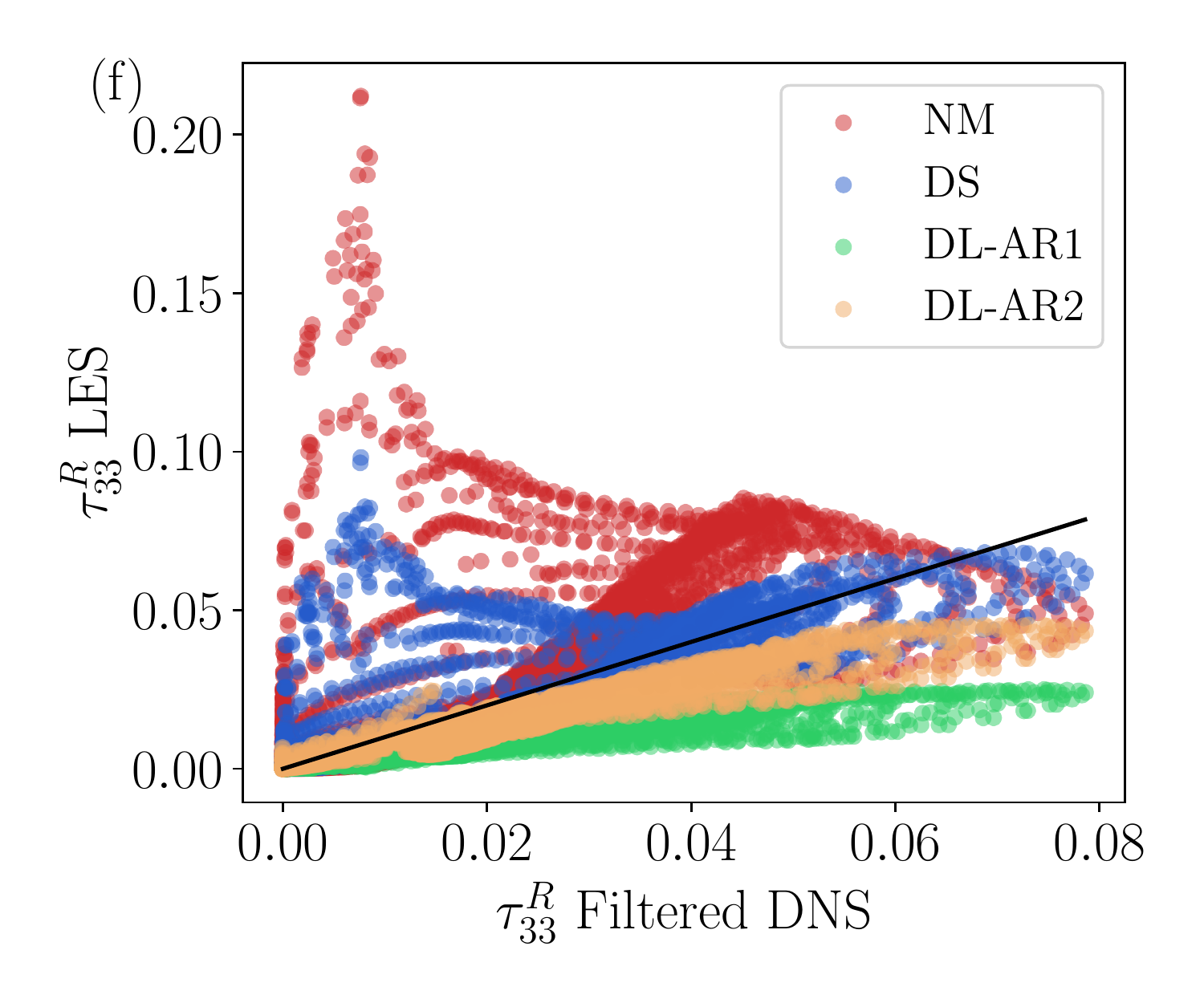}
\caption{AR1 $\Rey_0=1000$ (i): comparison of \emph{a posteriori} LES fields with \emph{a priori} filtered DNS fields. (a) $u$, (b) $v$, (c) $\tau^R_{12}$, (d) $\tau^R_{11}$, (e) $\tau^R_{22}$, and (f) $\tau^R_{33}$. The black lines are 1:1. Shown for no-model LES (NM), dynamic Smagorinsky (DS), and DL models trained for AR1 and AR2.}
\label{ScatterPlotAR1}
\end{figure}

\begin{figure}
\centering
\includegraphics[width=0.32\textwidth]{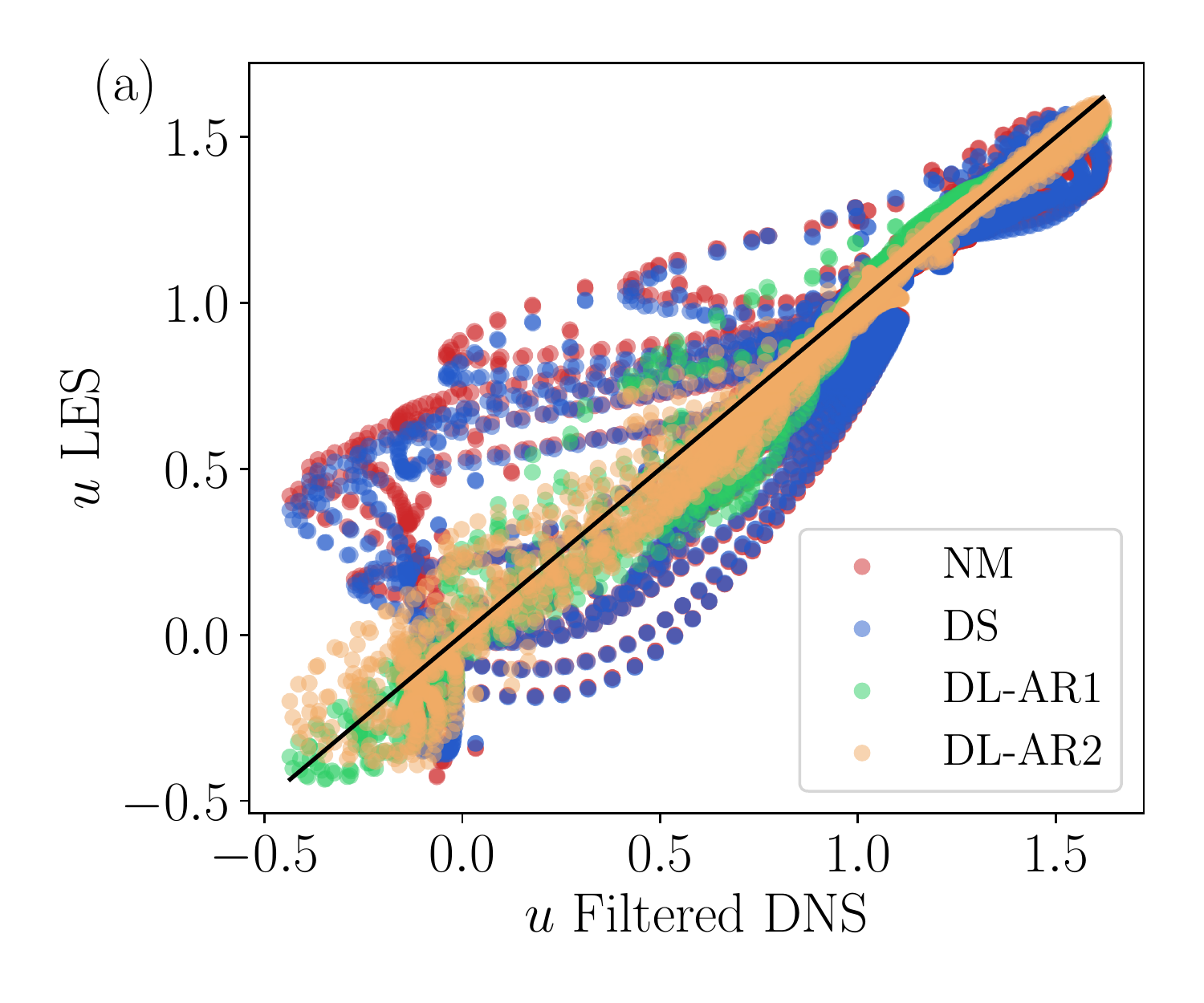}
\includegraphics[width=0.32\textwidth]{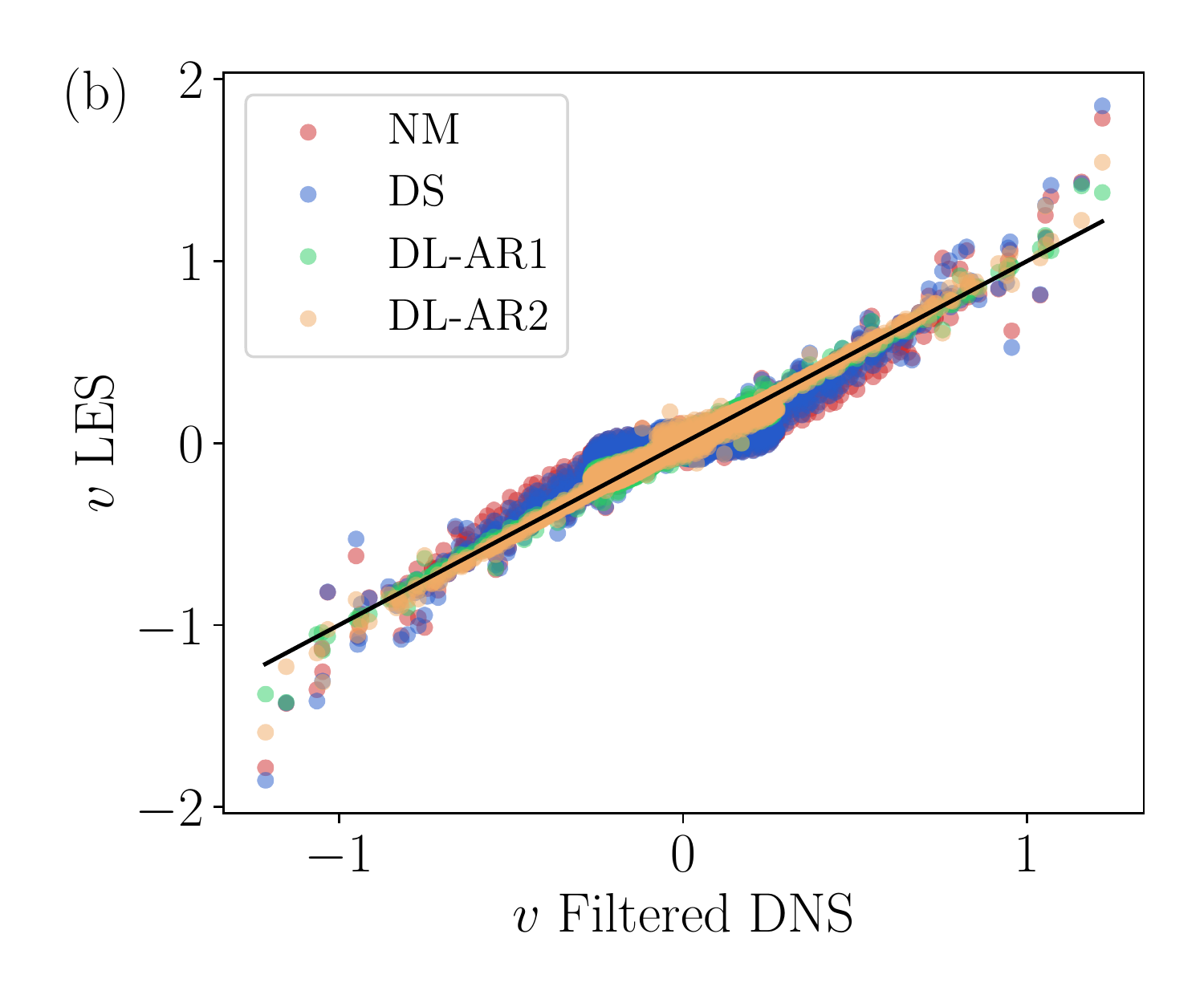}
\includegraphics[width=0.32\textwidth]{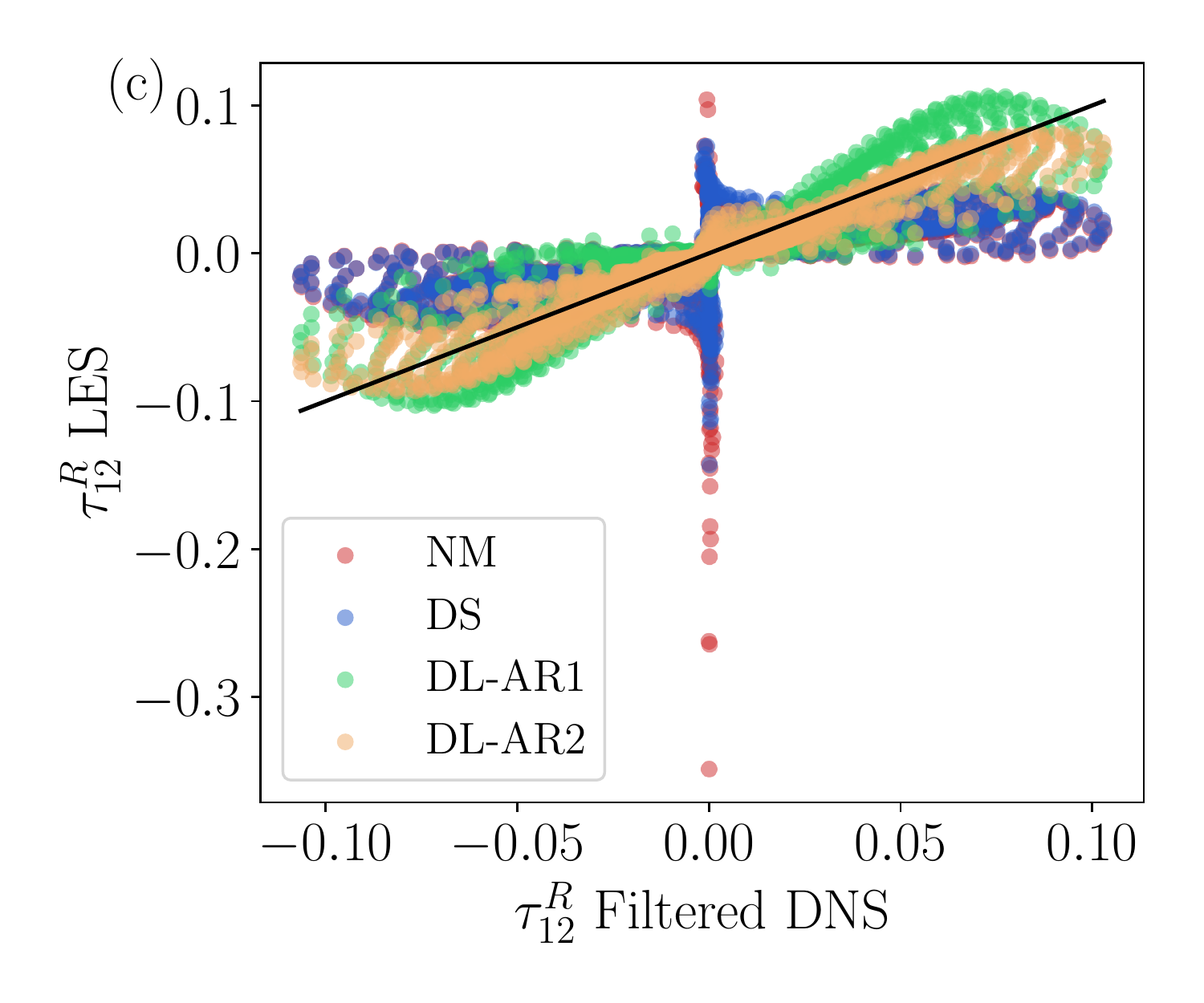}
\includegraphics[width=0.32\textwidth]{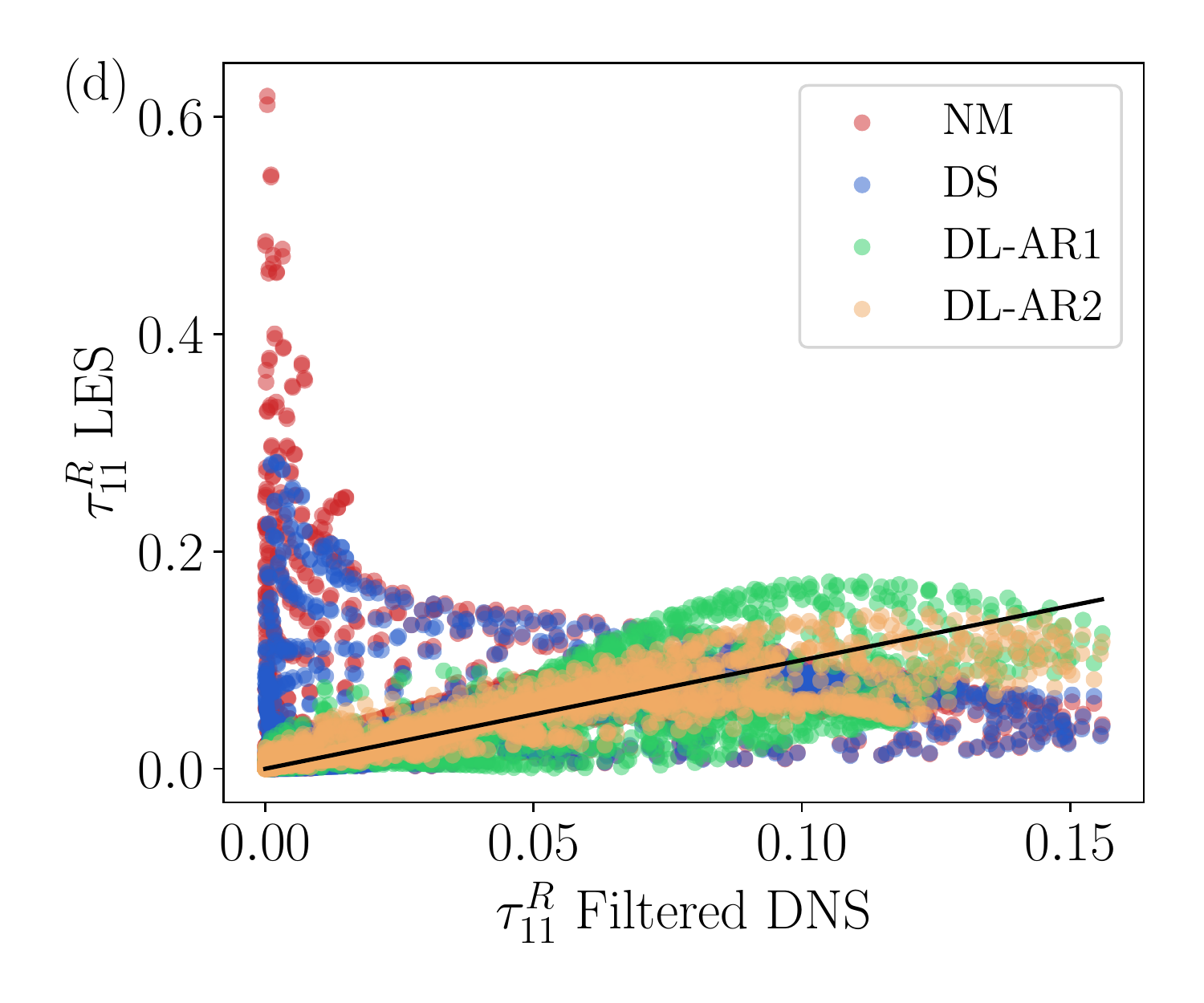}
\includegraphics[width=0.32\textwidth]{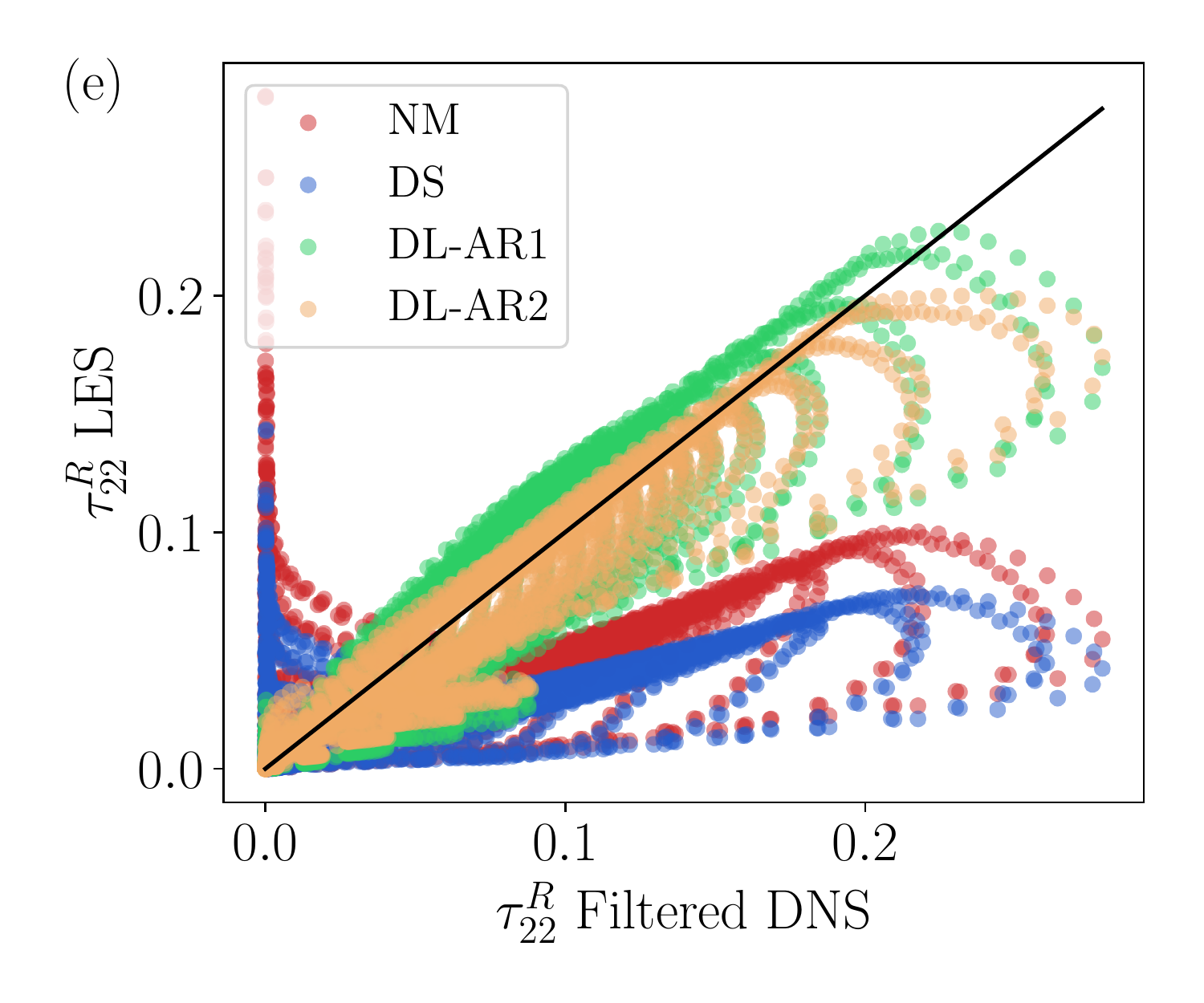}
\includegraphics[width=0.32\textwidth]{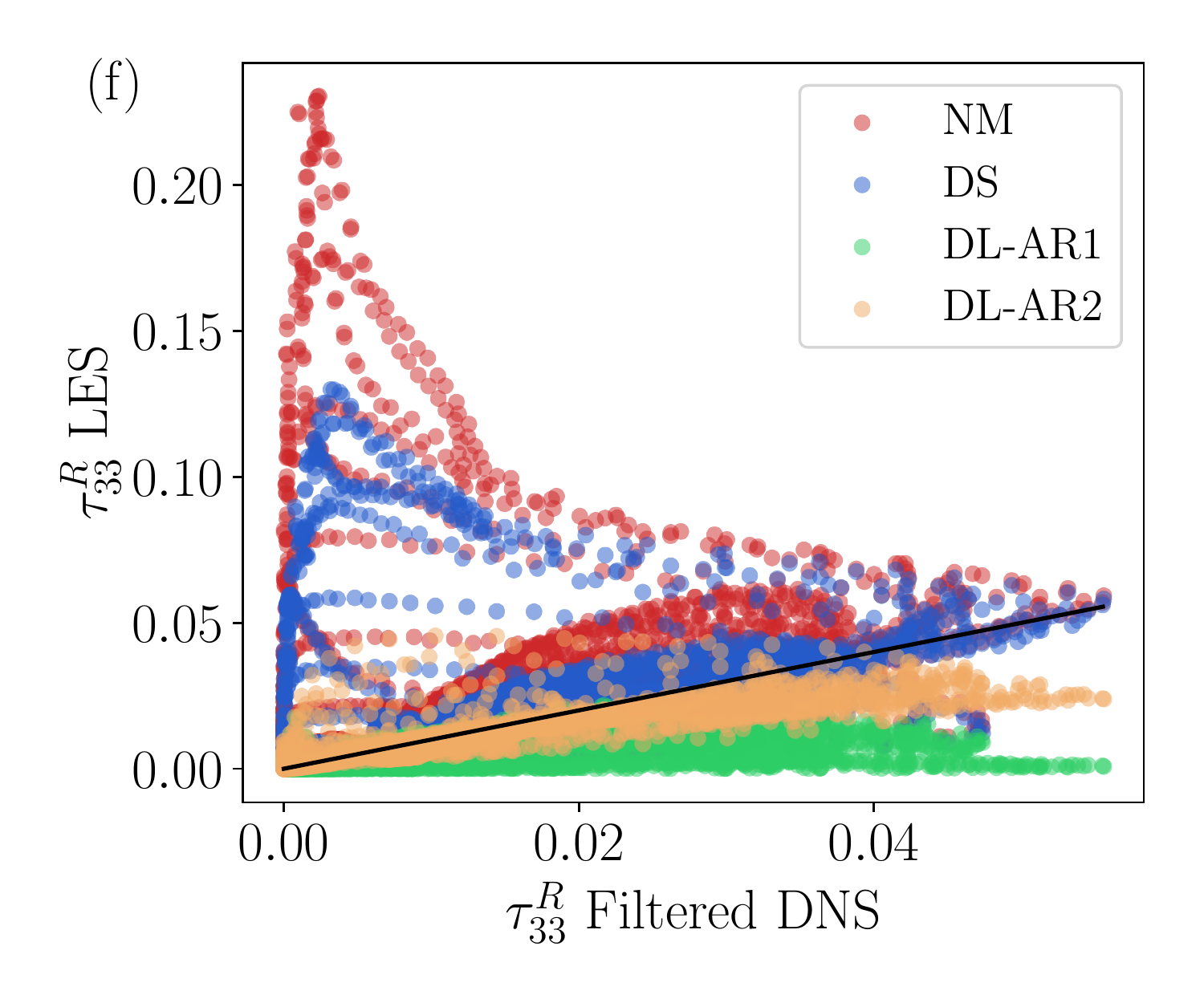}
\caption{AR4 $\Rey_0=1000$ (iv): comparison of \emph{a posteriori} LES fields with \emph{a priori} filtered DNS fields. (a) $u$, (b) $v$, (c) $\tau^R_{12}$, (d) $\tau^R_{11}$, (e) $\tau^R_{22}$, and (f) $\tau^R_{33}$. The black lines are 1:1. Shown for no-model LES (NM), dynamic Smagorinsky (DS), and DL models trained for AR1 and AR2.}
\label{ScatterPlotAR4}
\end{figure}

The averaged filtered velocity $\avg{\olu_i}$ and the resolved Reynolds stress tensor $\tau^R_{ij}\equiv \avg{\olu_i\olu_j} - \avg{\olu_i}\avg{\olu_j}$ are computed from the filtered DNS data and \emph{a posteriori} LES calculations. Figures \ref{ScatterPlotAR1} and \ref{ScatterPlotAR4} plot the the LES-predicted fields versus the filtered DNS fields, which are taken as the ``truth'' values, for cases (i) and (iv), respectively. The scatter plots provide a compact representation of the accuracy of the LES models across the simulation domain. The 1:1 line represents predictions with $100\%$ accuracy relative to the DNS.  The DL-LES models consistently outperform the benchmark LES models, and in several cases, the DL-LES models dramatically improve the predictive accuracy.

\begin{figure}
\centering
\includegraphics[width=0.245\textwidth,trim={0 0 80 0},clip]{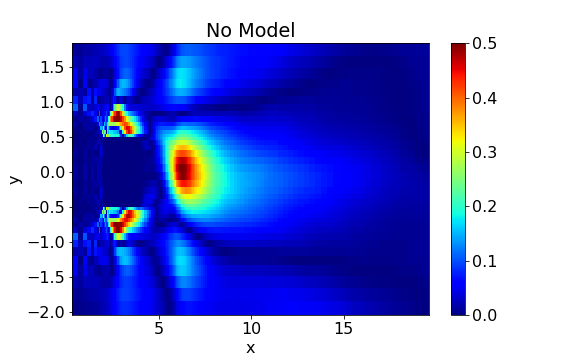}
\includegraphics[width=0.245\textwidth,trim={0 0 80 0},clip]{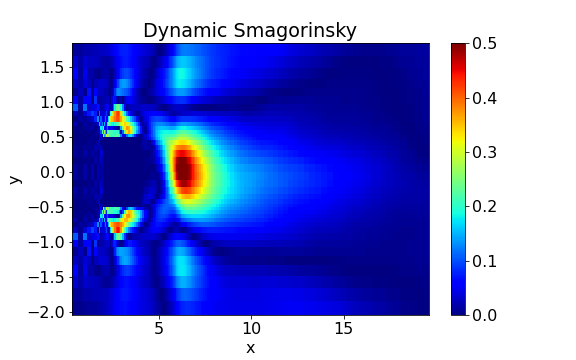}
\includegraphics[width=0.245\textwidth,trim={0 0 80 0},clip]{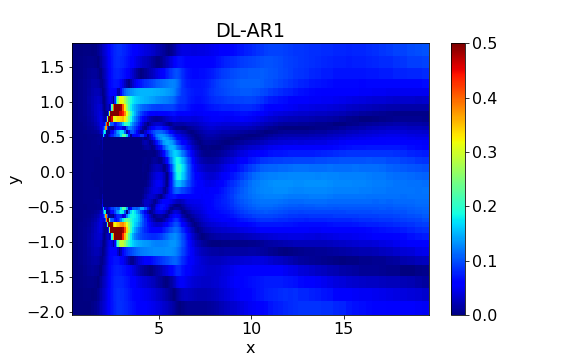}
\includegraphics[width=0.245\textwidth,trim={0 0 80 0},clip]{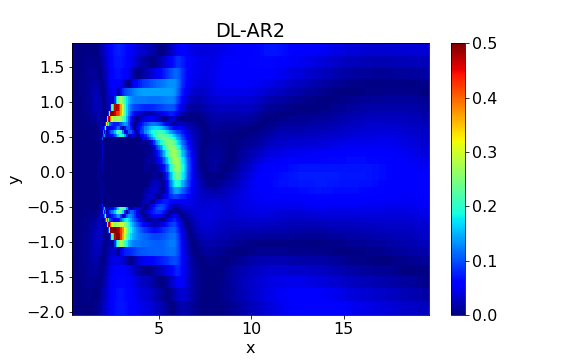}
\caption{Error for  $\avg{\olu_1}$ for the AR2 $\Rey_0=1000$ configuration.}
\label{ContourPlot1}
\end{figure}

\begin{figure}
\centering
\includegraphics[width=0.245\textwidth,trim={0 0 80 0},clip]{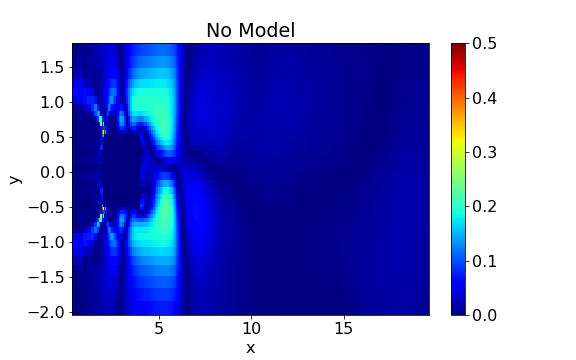}
\includegraphics[width=0.245\textwidth,trim={0 0 80 0},clip]{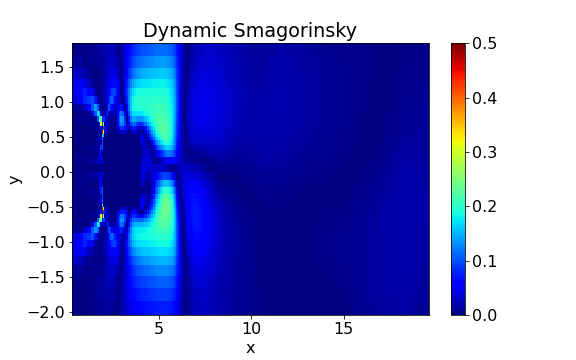}
\includegraphics[width=0.245\textwidth,trim={0 0 80 0},clip]{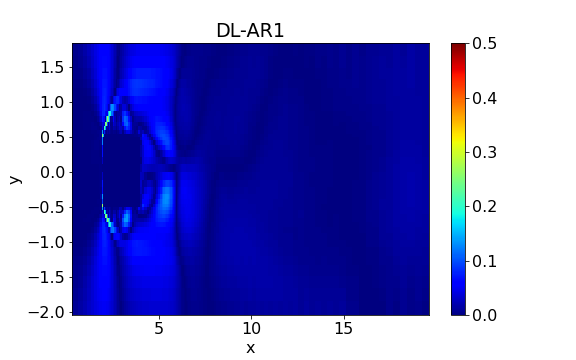}
\includegraphics[width=0.245\textwidth,trim={0 0 80 0},clip]{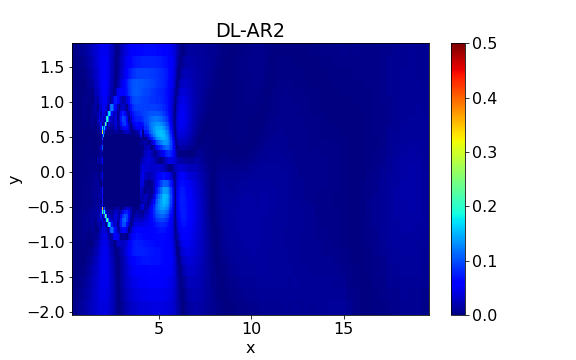}
\caption{Error for $\avg{\olu_2}$ for the AR2 $\Rey_0=1000$ configuration.}
\label{ContourPlot2}
\end{figure}

\begin{figure}
\centering
\includegraphics[width=0.245\textwidth,trim={0 0 80 0},clip]{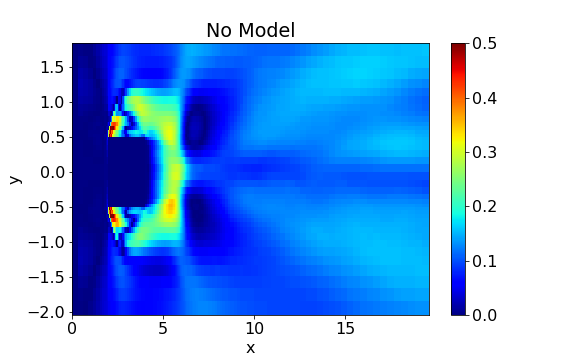}
\includegraphics[width=0.245\textwidth,trim={0 0 80 0},clip]{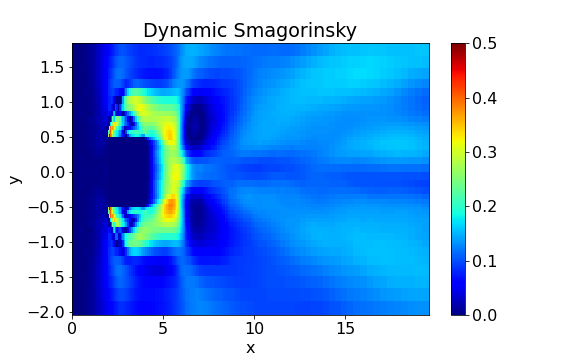}
\includegraphics[width=0.245\textwidth,trim={0 0 80 0},clip]{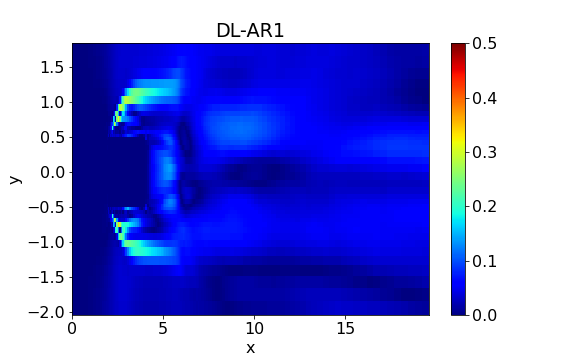}
\includegraphics[width=0.245\textwidth,trim={0 0 80 0},clip]{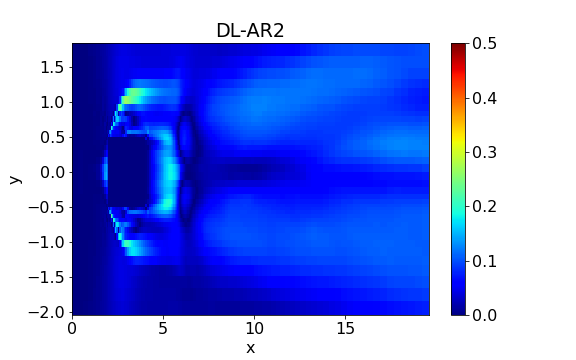}
\caption{Error for $\sqrt{\tau^R_{11}}$ for the AR2 $\Rey_0=1000$ configuration.}
\label{ContourPlot3}
\end{figure}

\begin{figure}
\centering
\includegraphics[width=0.245\textwidth,trim={0 0 80 0},clip]{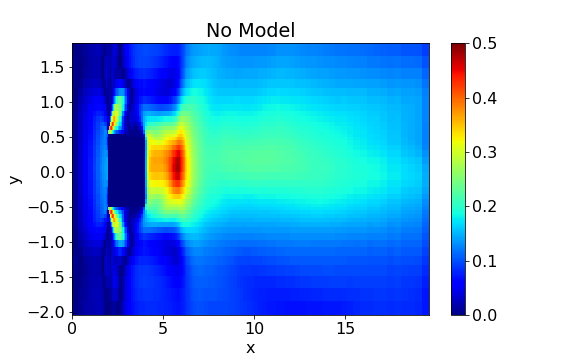}
\includegraphics[width=0.245\textwidth,trim={0 0 80 0},clip]{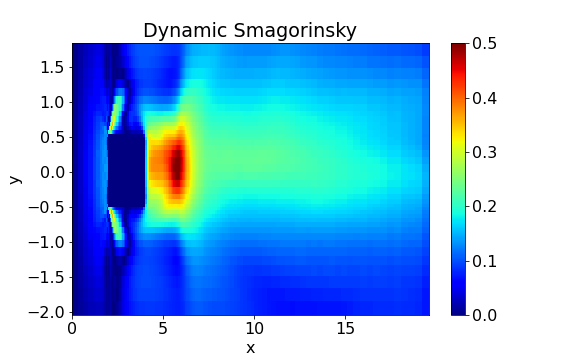}
\includegraphics[width=0.245\textwidth,trim={0 0 80 0},clip]{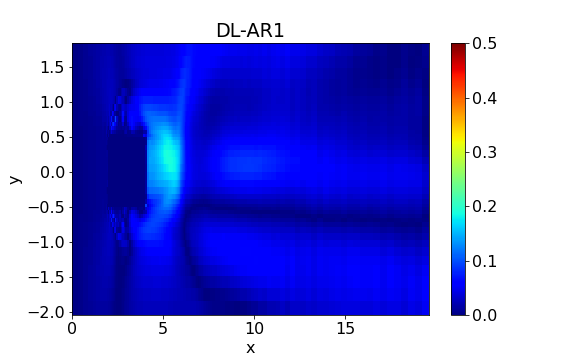}
\includegraphics[width=0.245\textwidth,trim={0 0 80 0},clip]{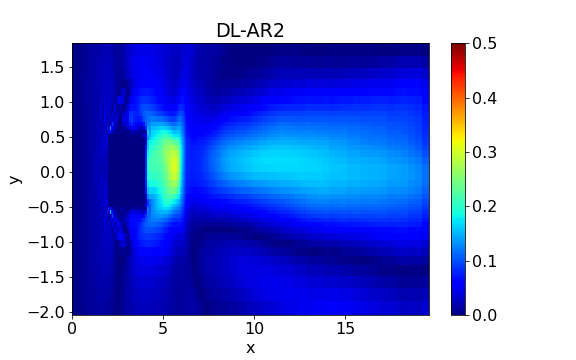}
\caption{Error for $\sqrt{\tau^R_{22}}$ for the AR2 $\Rey_0=1000$ configuration.}
\label{ContourPlot4}
\end{figure}

\begin{figure}
\centering
\includegraphics[width=0.245\textwidth,trim={0 0 80 0},clip]{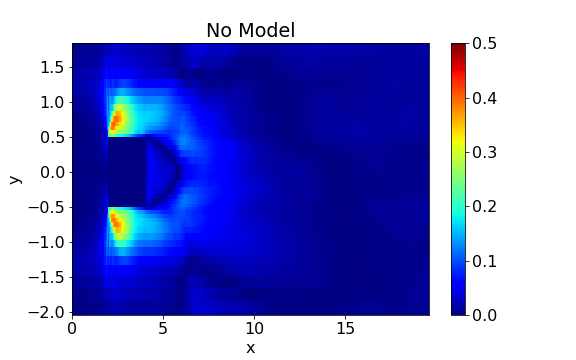}
\includegraphics[width=0.245\textwidth,trim={0 0 80 0},clip]{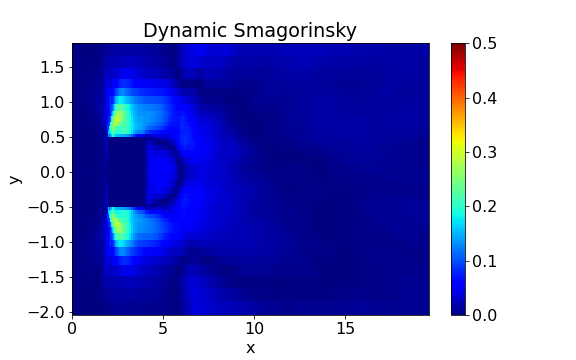}
\includegraphics[width=0.245\textwidth,trim={0 0 80 0},clip]{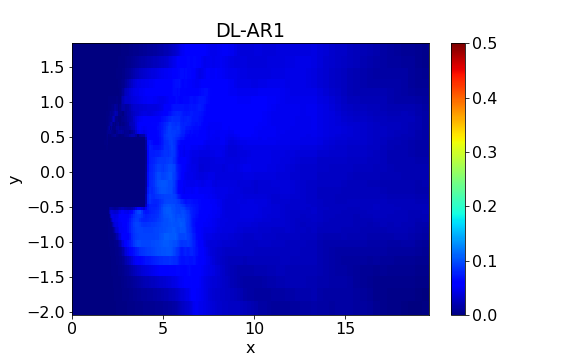}
\includegraphics[width=0.245\textwidth,trim={0 0 80 0},clip]{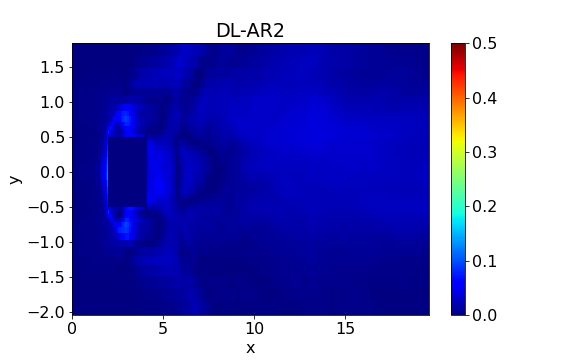}
\caption{Error for $\sqrt{ \tau^R_{33}}$ for the AR2 $\Rey_0=1000$ configuration.}
\label{ContourPlot5}
\end{figure}

\begin{figure}
\centering
\includegraphics[width=0.245\textwidth,trim={0 0 80 0},clip]{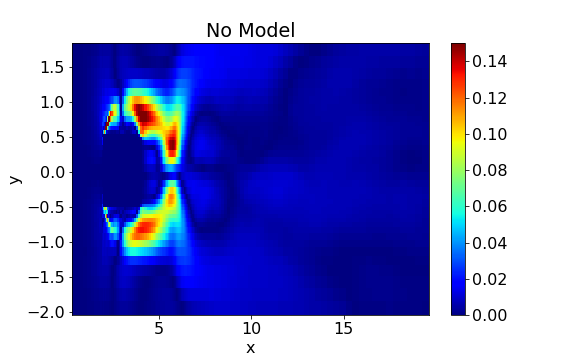}
\includegraphics[width=0.245\textwidth,trim={0 0 80 0},clip]{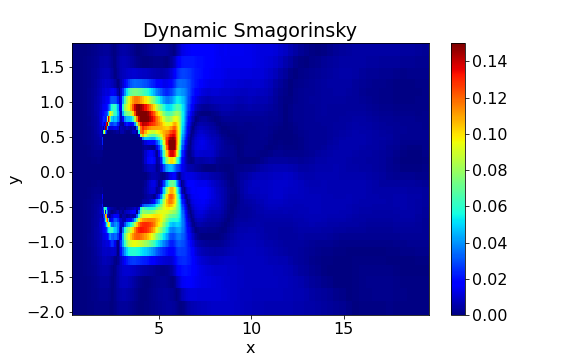}
\includegraphics[width=0.245\textwidth,trim={0 0 80 0},clip]{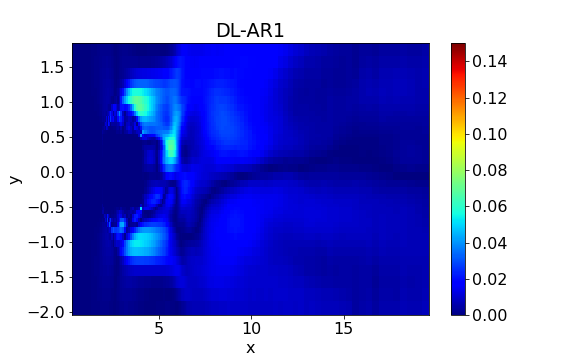}
\includegraphics[width=0.245\textwidth,trim={0 0 80 0},clip]{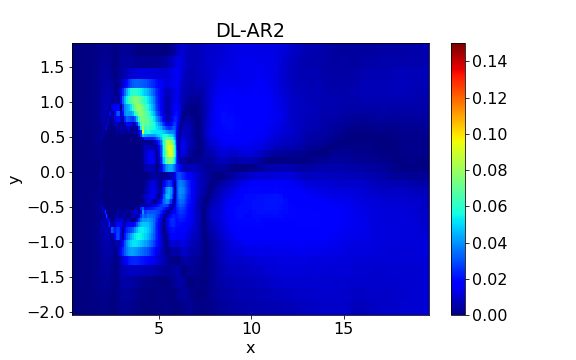}
\caption{Error for $\tau^R_{12}$ for the AR2 $\Rey_0=1000$ configuration.}
\label{ContourPlot6}
\end{figure}

Contour plots  visualize the LES models' accuracy across the domain. At each point $(x_1,x_2)$, the LES $\avg{\olu_i}$ and $\tau^R_{ij}$ are compared against the ``exact'' field evaluated from the filtered DNS data. The $\ell_1$ error is calculated and displayed in Figures \ref{ContourPlot1}, \ref{ContourPlot2}, \ref{ContourPlot3}, \ref{ContourPlot4}, \ref{ContourPlot5}, and \ref{ContourPlot6} for the AR2 $\Rey_0=1000$ case (ii). The DL-LES models consistently outperform the benchmark LES models. We can observe that the DL-LES model especially improves the accuracy near the walls of the rectangular cylinder, which directly leads to a more accurate estimate for the drag coefficient.   

\subsection{Additional Analysis} \label{AdditionalAnalysis}

A complete set of contour plots for all configurations is included in the Appendix; see Section \ref{ContourPlotsAppendix}. We also compare the mean profile of the velocity and resolved Reynolds stress (as a function of the transverse $x_2$-dimension) of the models versus the filtered DNS for all of the configurations at a set of fixed $x_1$ locations;  see Section \ref{FixedPositionsAppendix}. These additional figures provide a complete description of the model performance for the different test cases. A remaining challenge that should be addressed in future research is to increase the symmetry of the DL-LES simulations; in several instances, it can be observed that the mean velocity profiles and/or resolved Reynolds stress profiles are not symmetric.

\section{Conclusion}

LES calculations for wall-bounded flows can be computationally expensive due to the prohibitively small mesh sizes needed to accurately resolve the boundary-layer turbulence. A deep learning closure model for LES was developed and demonstrated, in numerical studies for turbulent flows around a rectangular cylinder, to be accurate even on a coarse LES mesh. The deep learning LES models outperformed the dynamic Smagorinsky model, yielding accurate predictions for the drag coefficient, mean profile, and resolved Reynolds stress. The DL-LES models were trained using adjoint PDE optimization methods, which optimize over the entire PDE solution to select the neural network parameters for the closure model. Our results indicate that deep learning closure models have the potential to improve the accuracy of LES simulations in aerodynamics.

\section{Appendix: Comparison of Models}

\subsection{Contour Plots} \label{ContourPlotsAppendix}

\subsubsection{AR2-Re2,000 Contour Plots}

\begin{figure}[H]
\centering
\includegraphics[width=4cm]{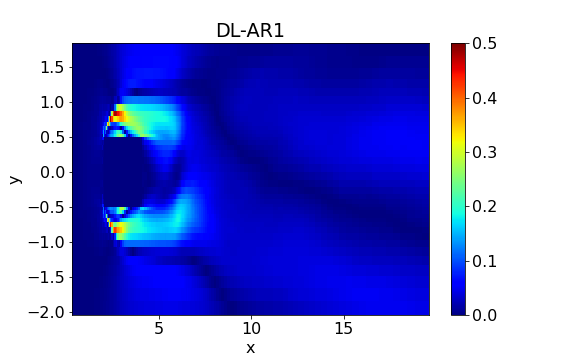}
\includegraphics[width=4cm]{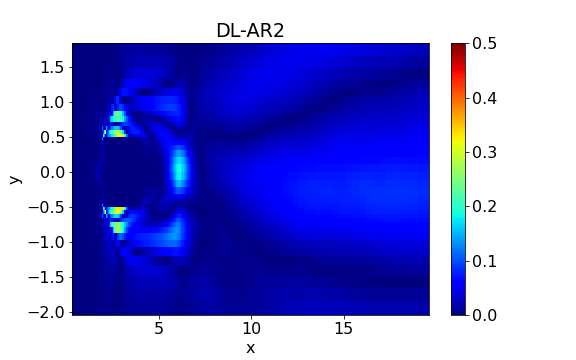}
\includegraphics[width=4cm]{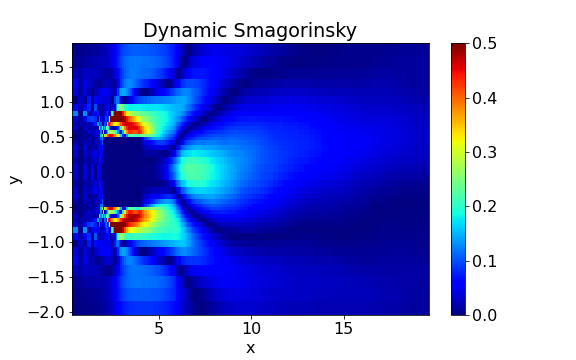}
\includegraphics[width=4cm]{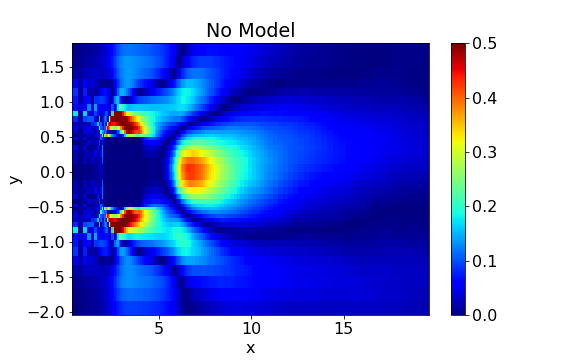}
\label{f1}
\caption{Error for mean profile of $u_1$ for AR2-Re2,000 configuration.}
\end{figure}
\begin{figure}[H]
\centering
\includegraphics[width=4cm]{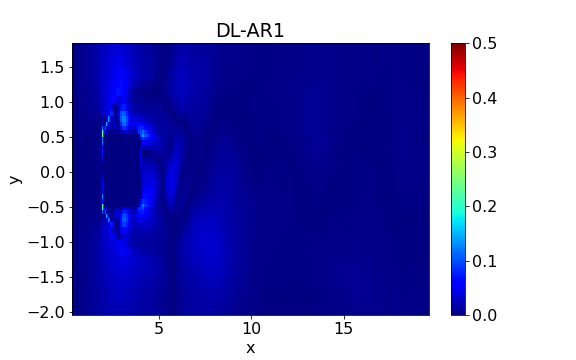}
\includegraphics[width=4cm]{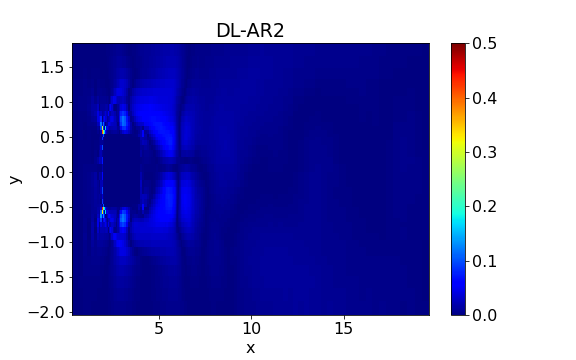}
\includegraphics[width=4cm]{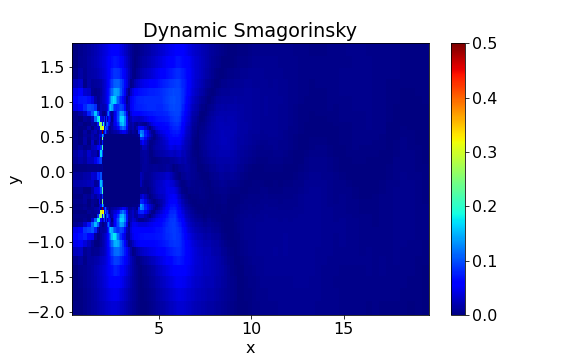}
\includegraphics[width=4cm]{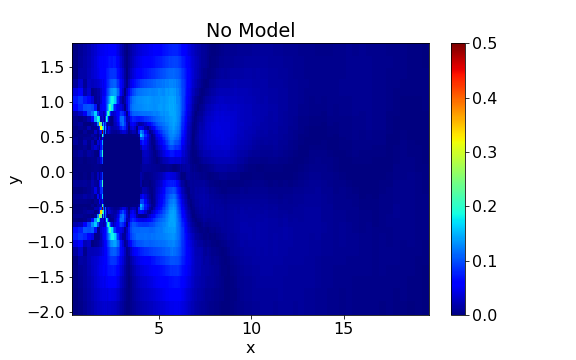}
\label{f1}
\caption{Error for mean profile of $u_2$ for AR2-Re2000 configuration.}
\end{figure}
\begin{figure}[H]
\centering
\includegraphics[width=4cm]{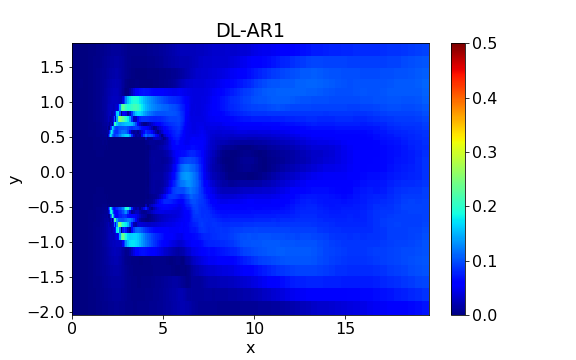}
\includegraphics[width=4cm]{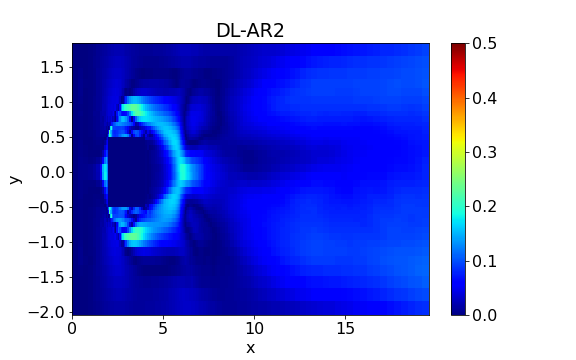}
\includegraphics[width=4cm]{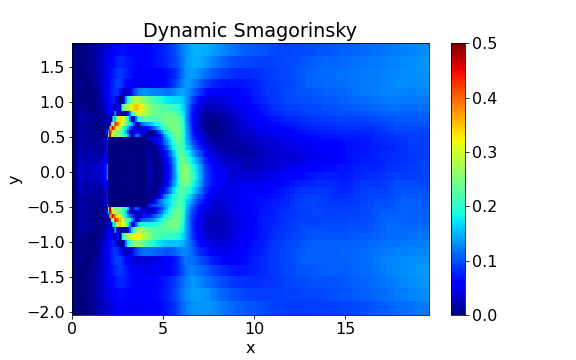}
\includegraphics[width=4cm]{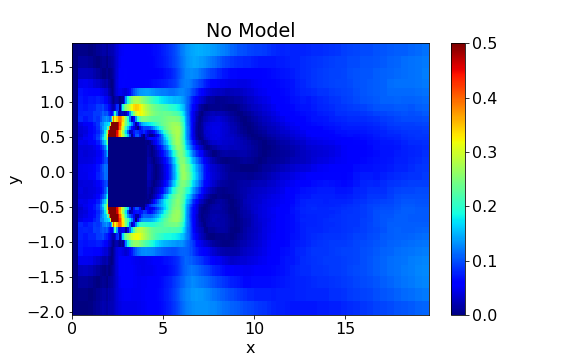}
\label{f1}
\caption{Error for RMS of $u_1$ for AR2-Re2000 configuration.}
\end{figure}
\begin{figure}[H]
\centering
\includegraphics[width=4cm]{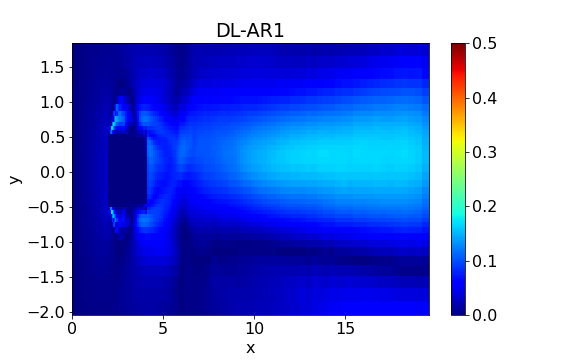}
\includegraphics[width=4cm]{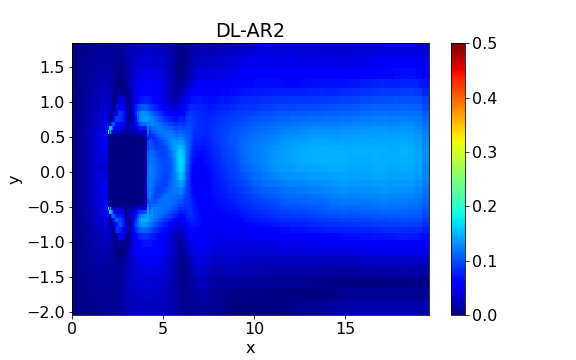}
\includegraphics[width=4cm]{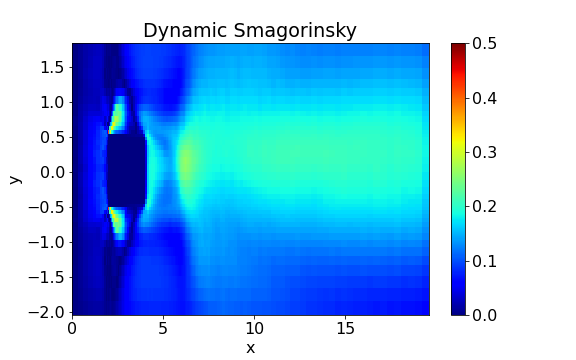}
\includegraphics[width=4cm]{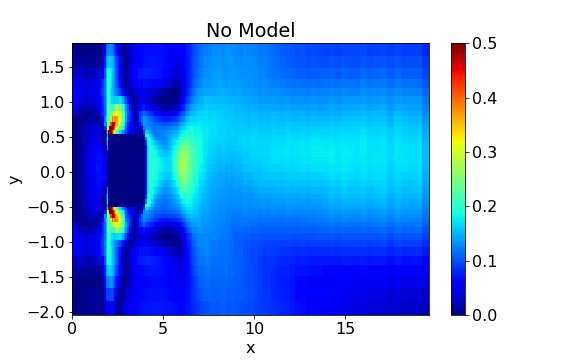}
\label{f1}
\caption{Error for RMS of $u_2$ for AR2-Re2000 configuration.}
\end{figure}

\begin{figure}[H]
\centering
\includegraphics[width=4cm]{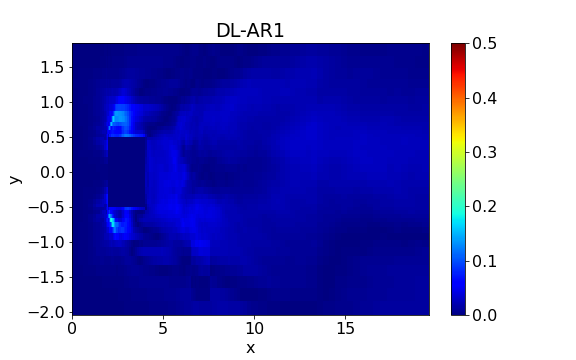}
\includegraphics[width=4cm]{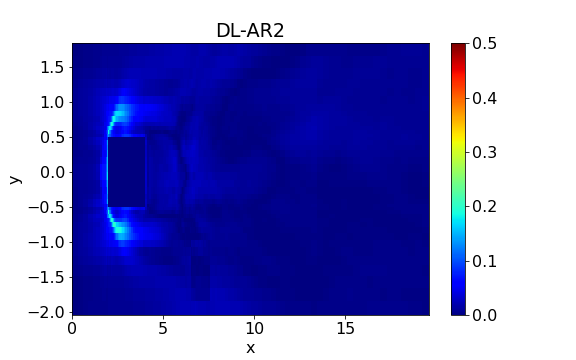}
\includegraphics[width=4cm]{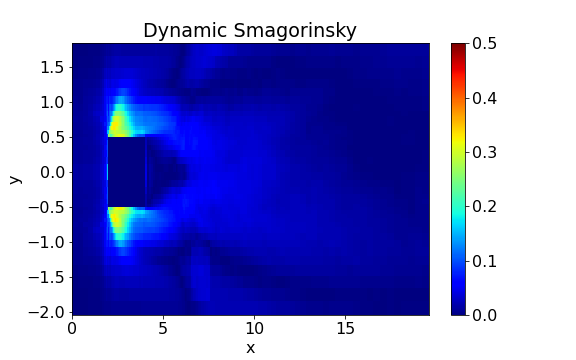}
\includegraphics[width=4cm]{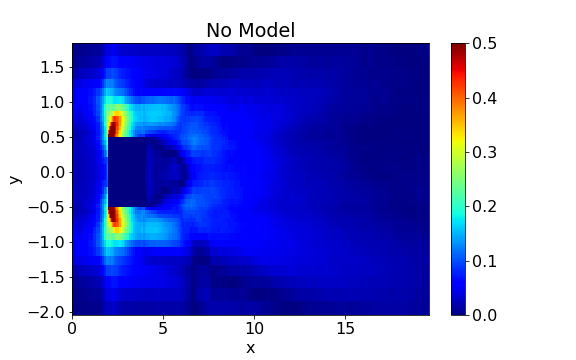}
\label{f1}
\caption{Error for RMS of $u_3$ for AR2-Re2000 configuration.}
\end{figure}

\begin{figure}[H]
\centering
\includegraphics[width=4cm]{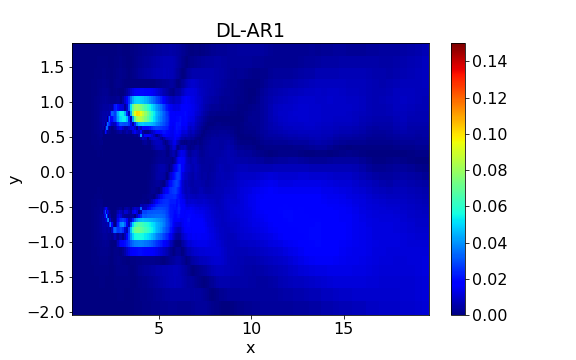}
\includegraphics[width=4cm]{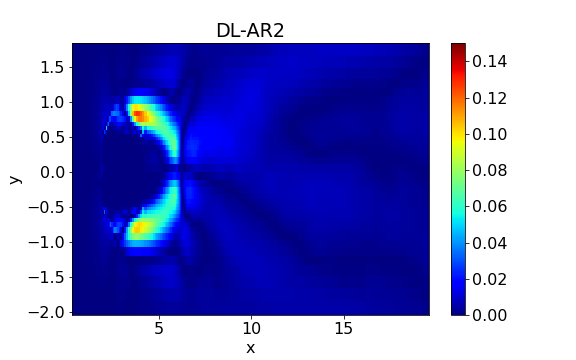}
\includegraphics[width=4cm]{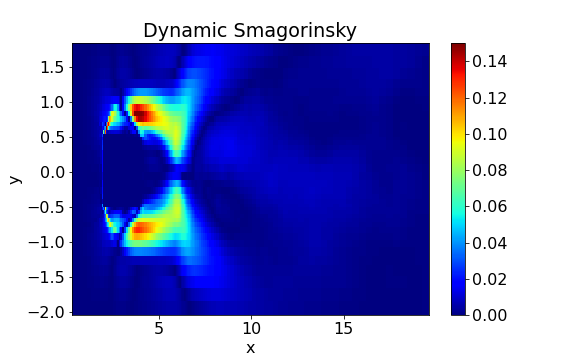}
\includegraphics[width=4cm]{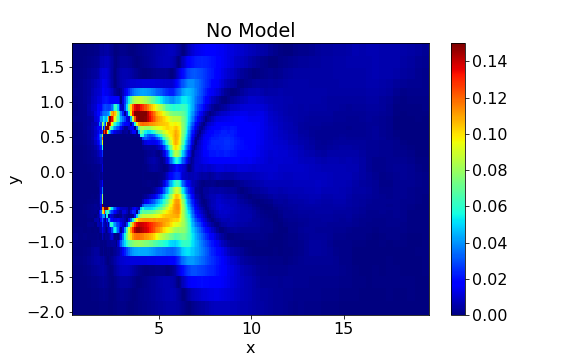}
\label{f1}
\caption{Error for $\tau_{12}$ for AR2-Re2000 configuration.}
\end{figure}

\subsubsection{AR1 Contour Plots}

\begin{figure}[H]
\centering
\includegraphics[width=4cm]{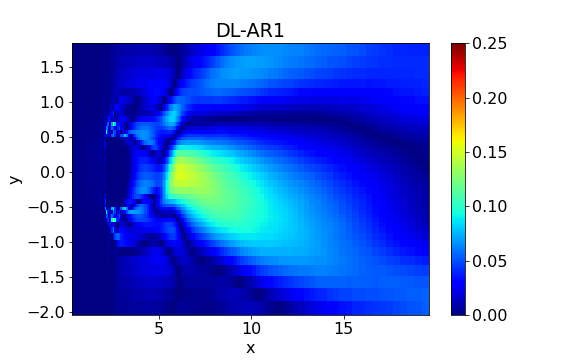}
\includegraphics[width=4cm]{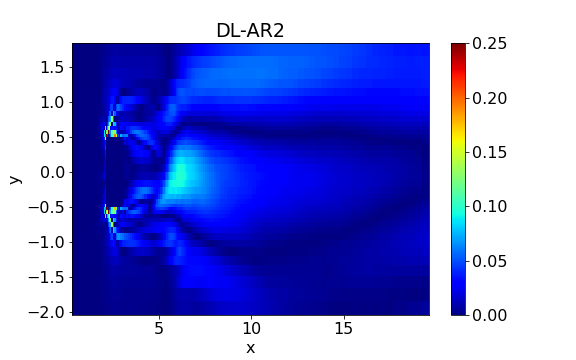}
\includegraphics[width=4cm]{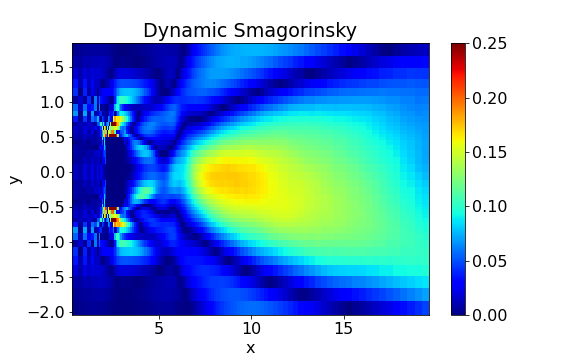}
\includegraphics[width=4cm]{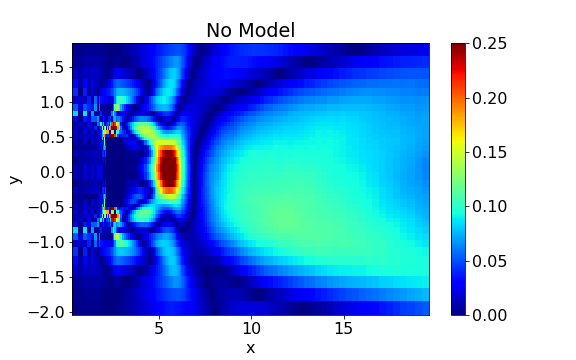}
\label{f1}
\caption{Error for mean profile of $u_1$ for AR1 configuration.}
\end{figure}
\begin{figure}[H]
\centering
\includegraphics[width=4cm]{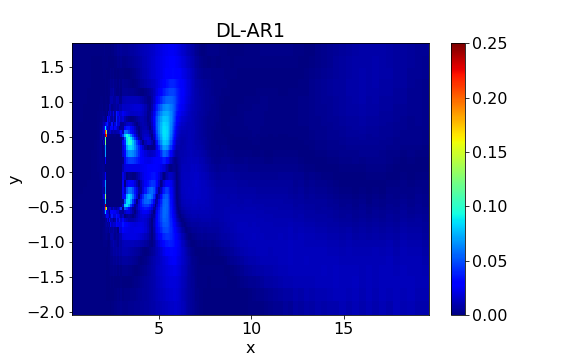}
\includegraphics[width=4cm]{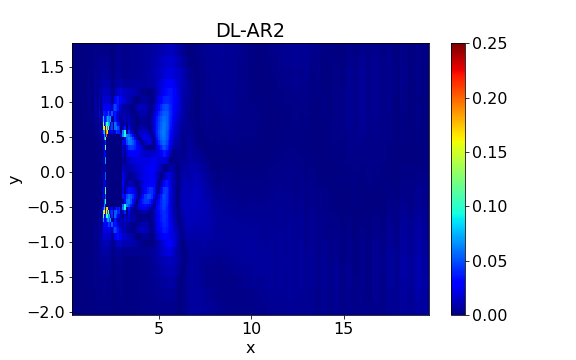}
\includegraphics[width=4cm]{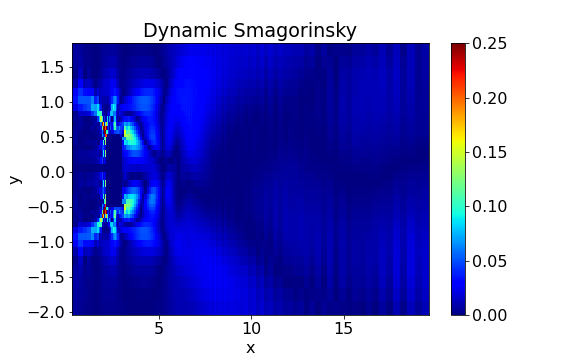}
\includegraphics[width=4cm]{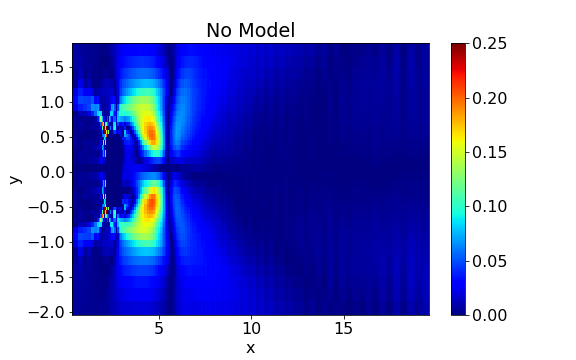}
\label{f1}
\caption{Error for mean profile of $u_2$ for AR1 configuration.}
\end{figure}
\begin{figure}[H]
\centering
\includegraphics[width=4cm]{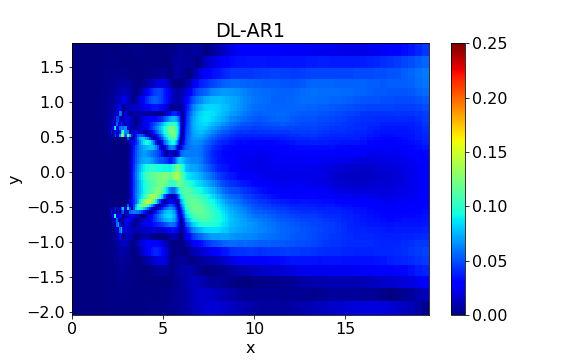}
\includegraphics[width=4cm]{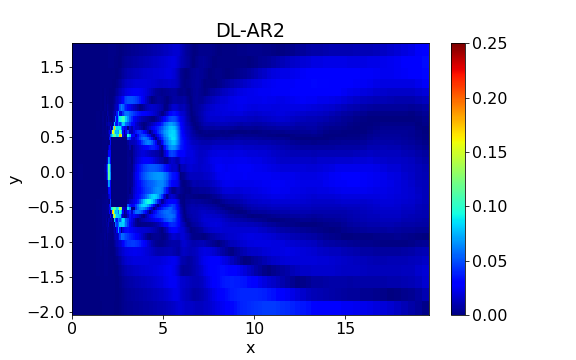}
\includegraphics[width=4cm]{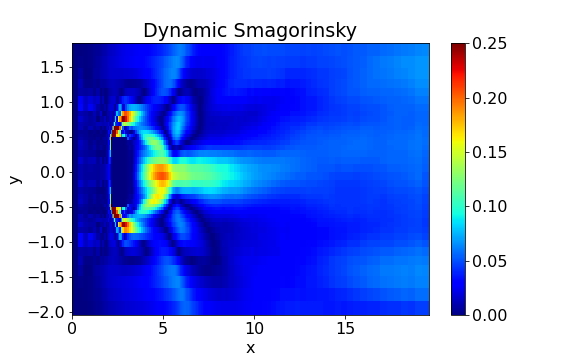}
\includegraphics[width=4cm]{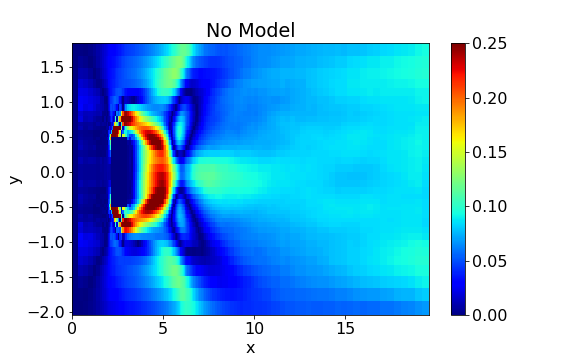}
\label{f1}
\caption{Error for RMS of $u_1$ for AR1 configuration.}
\end{figure}
\begin{figure}[H]
\centering
\includegraphics[width=4cm]{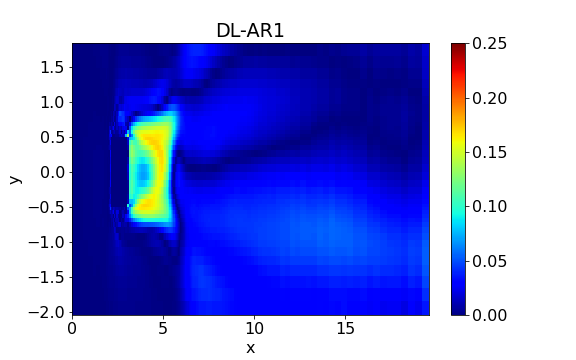}
\includegraphics[width=4cm]{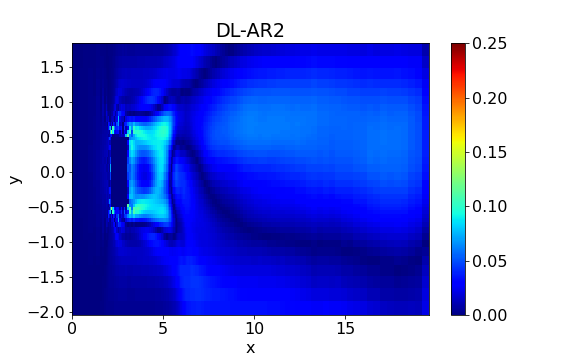}
\includegraphics[width=4cm]{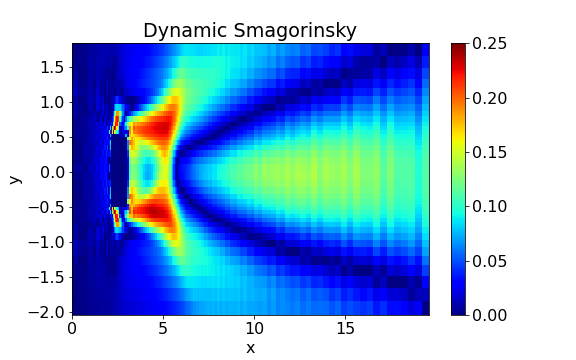}
\includegraphics[width=4cm]{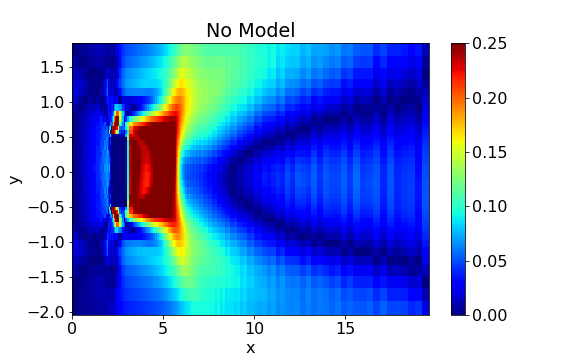}
\label{f1}
\caption{Error for RMS of $u_2$ for AR1 configuration.}
\end{figure}

\begin{figure}[H]
\centering
\includegraphics[width=4cm]{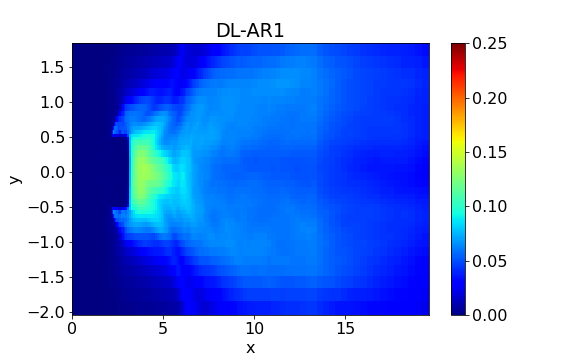}
\includegraphics[width=4cm]{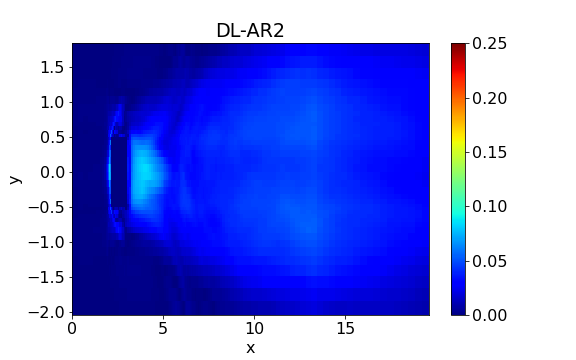}
\includegraphics[width=4cm]{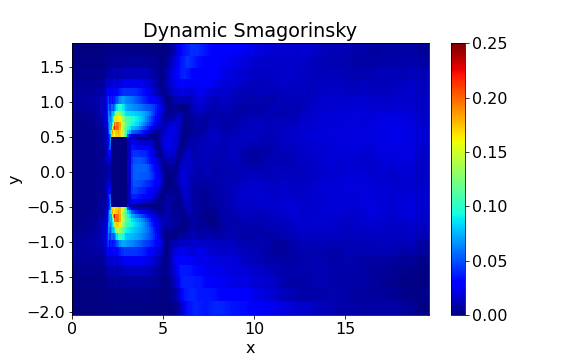}
\includegraphics[width=4cm]{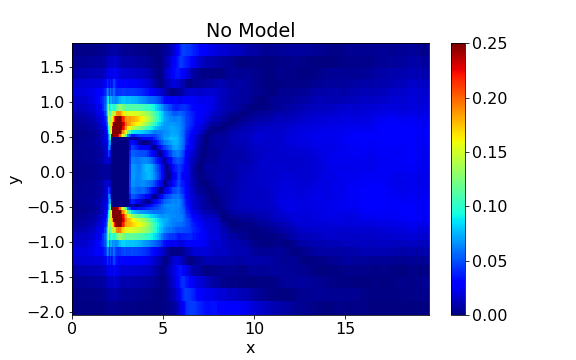}
\label{f1}
\caption{Error for RMS of $u_3$ for AR1 configuration.}
\end{figure}

\begin{figure}[H]
\centering
\includegraphics[width=4cm]{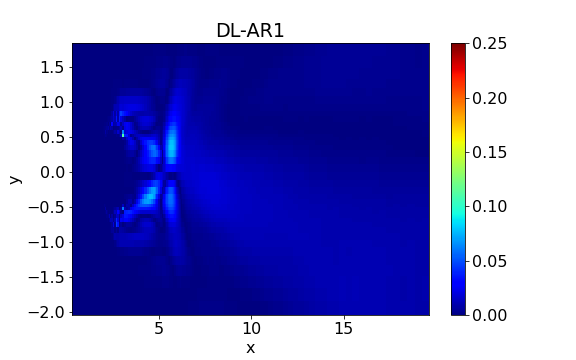}
\includegraphics[width=4cm]{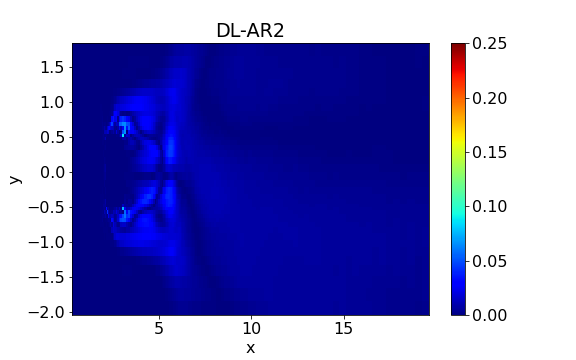}
\includegraphics[width=4cm]{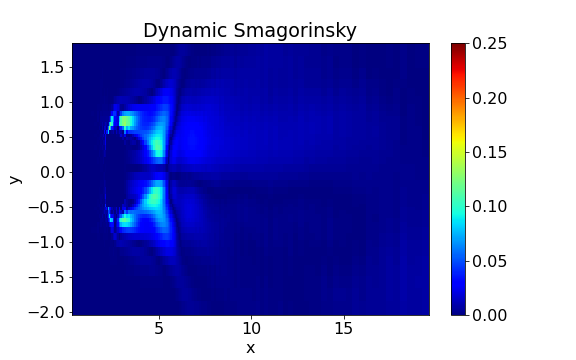}
\includegraphics[width=4cm]{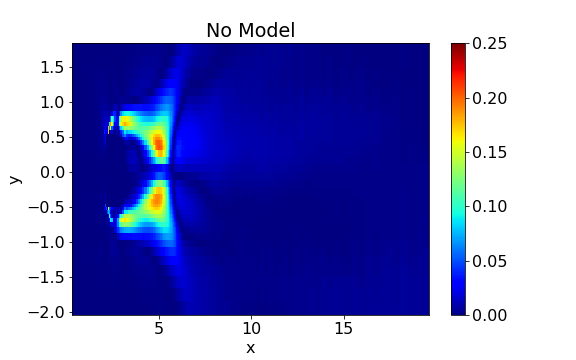}
\label{f1}
\caption{Error for $\tau_{12}$ for AR1 configuration.}
\end{figure}

\subsubsection{AR4 Contour Plots}

\begin{figure}[H]
\centering
\includegraphics[width=4cm]{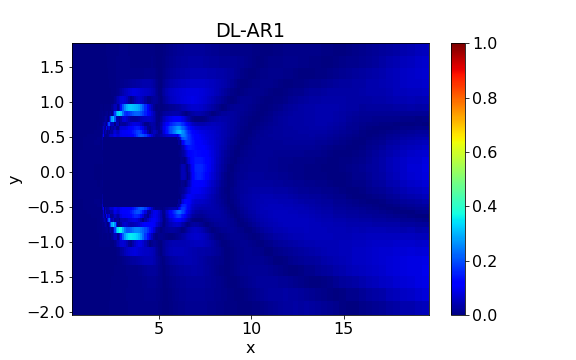}
\includegraphics[width=4cm]{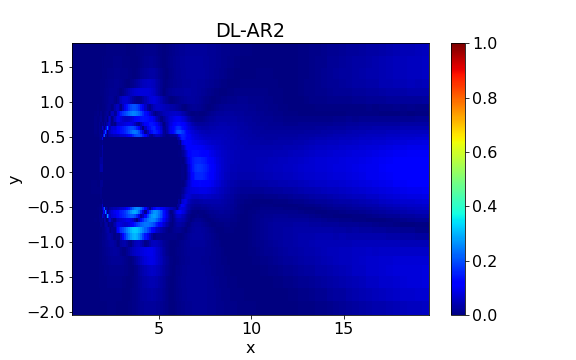}
\includegraphics[width=4cm]{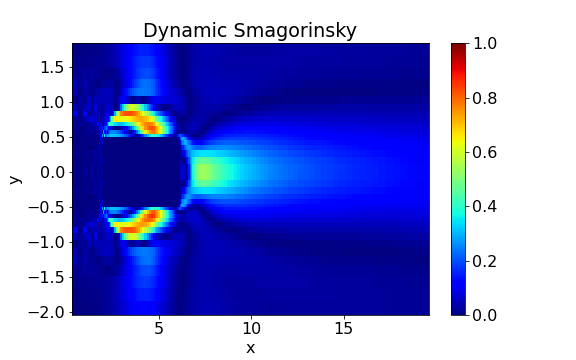}
\includegraphics[width=4cm]{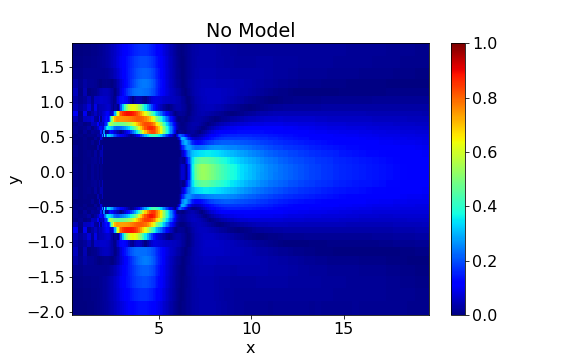}
\label{f1}
\caption{Error for mean profile of $u_1$ for AR4 configuration.}
\end{figure}
\begin{figure}[H]
\centering
\includegraphics[width=4cm]{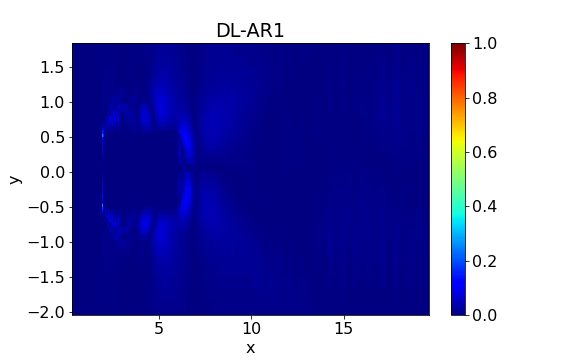}
\includegraphics[width=4cm]{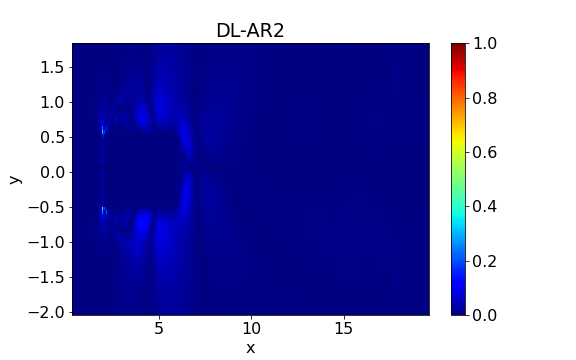}
\includegraphics[width=4cm]{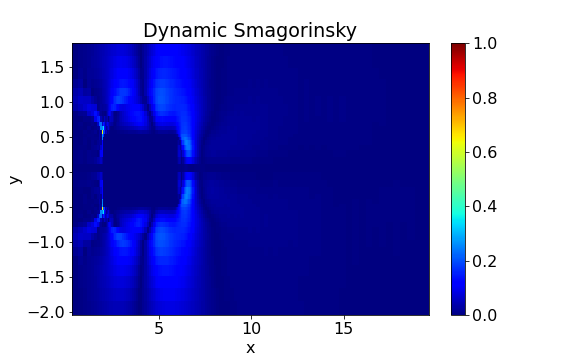}
\includegraphics[width=4cm]{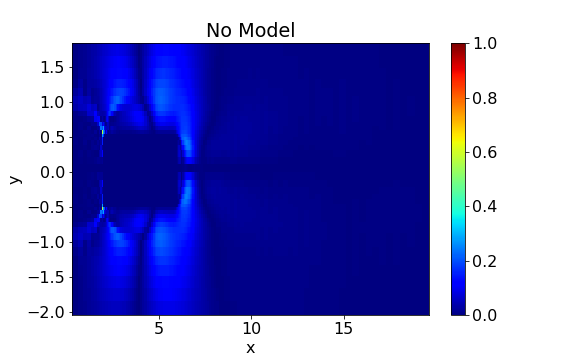}
\label{f1}
\caption{Error for mean profile of $u_2$ for AR4 configuration.}
\end{figure}
\begin{figure}[H]
\centering
\includegraphics[width=4cm]{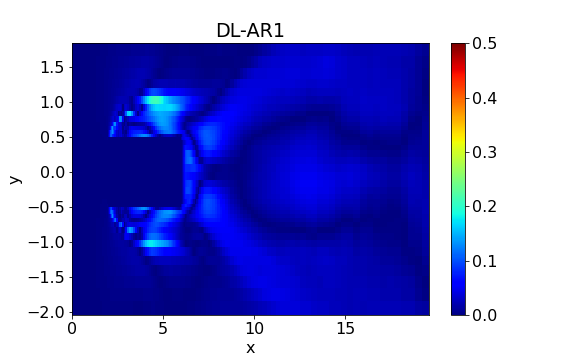}
\includegraphics[width=4cm]{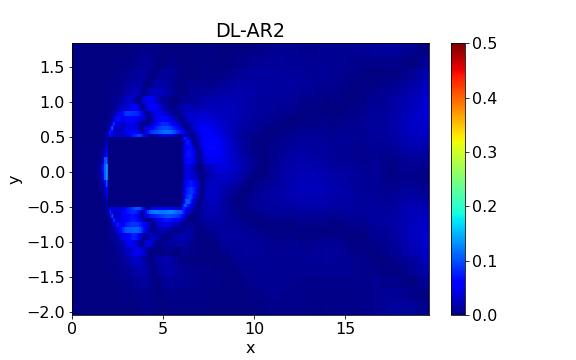}
\includegraphics[width=4cm]{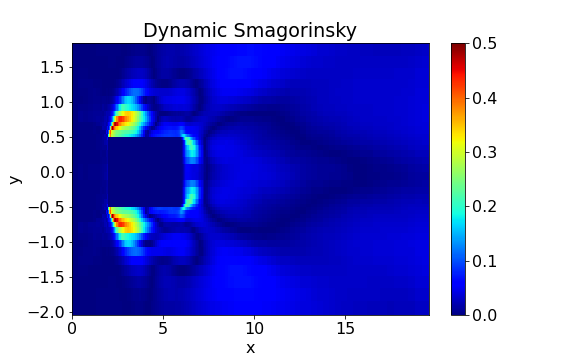}
\includegraphics[width=4cm]{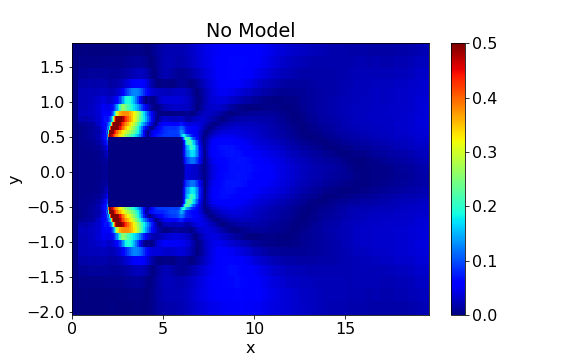}
\label{f1}
\caption{Error for RMS of $u_1$ for AR4 configuration.}
\end{figure}
\begin{figure}[H]
\centering
\includegraphics[width=4cm]{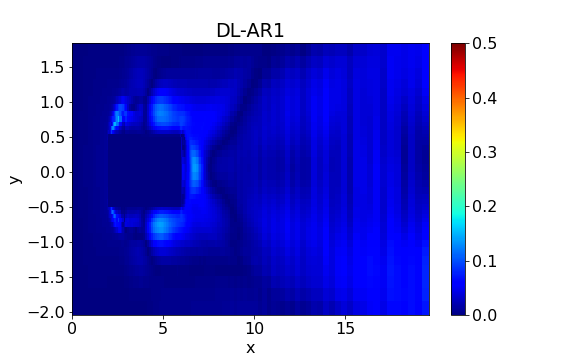}
\includegraphics[width=4cm]{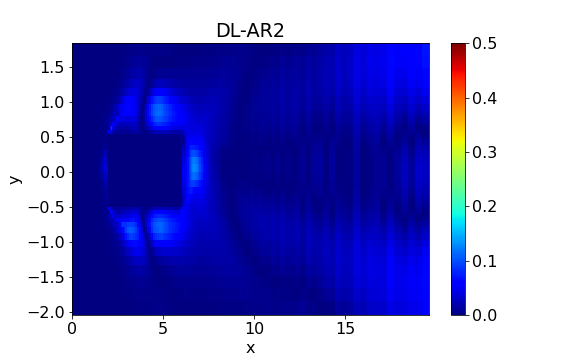}
\includegraphics[width=4cm]{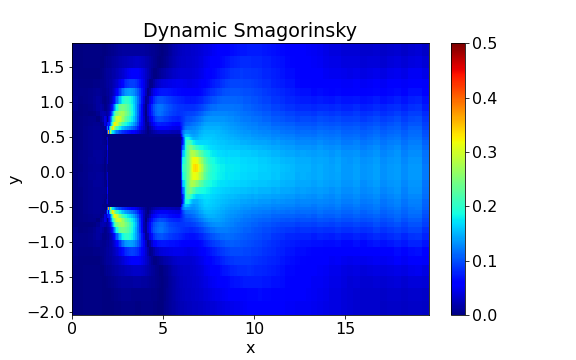}
\includegraphics[width=4cm]{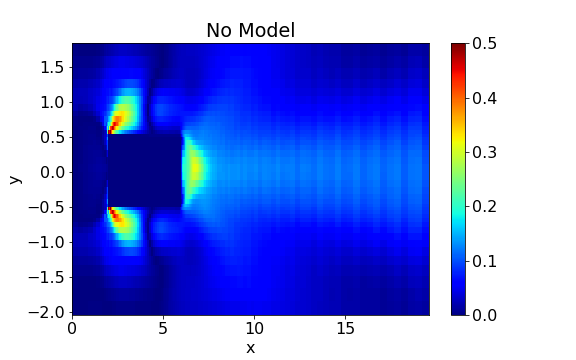}
\label{f1}
\caption{Error for RMS of $u_2$ for AR4 configuration.}
\end{figure}

\begin{figure}[H]
\centering
\includegraphics[width=4cm]{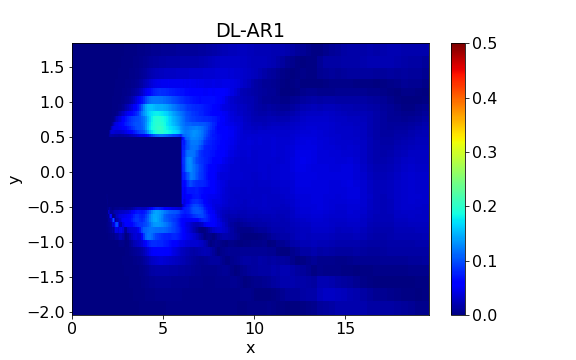}
\includegraphics[width=4cm]{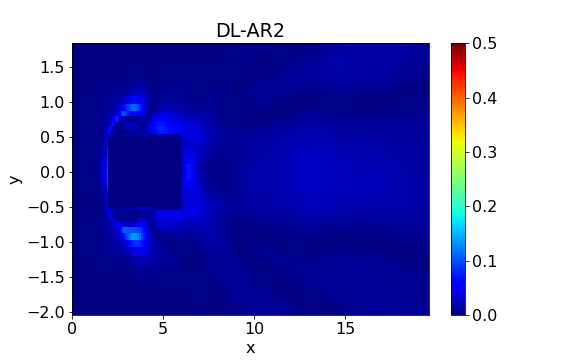}
\includegraphics[width=4cm]{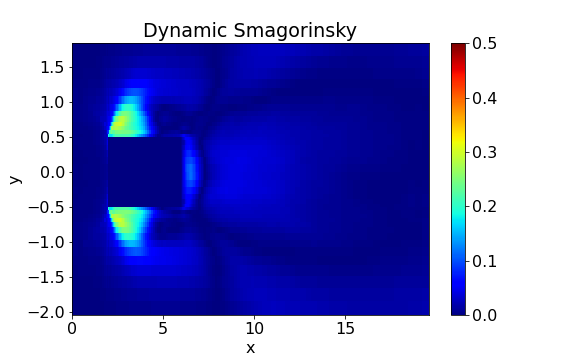}
\includegraphics[width=4cm]{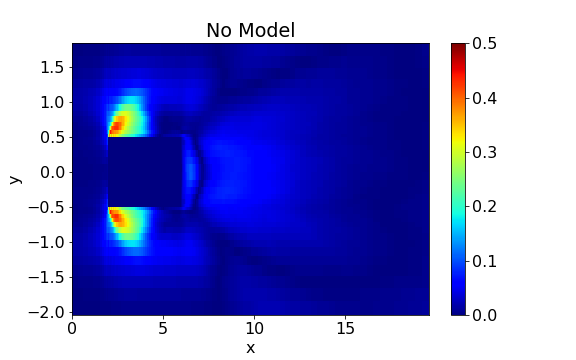}
\label{f1}
\caption{Error for RMS of $u_3$ for AR4 configuration.}
\end{figure}

\begin{figure}[H]
\centering
\includegraphics[width=4cm]{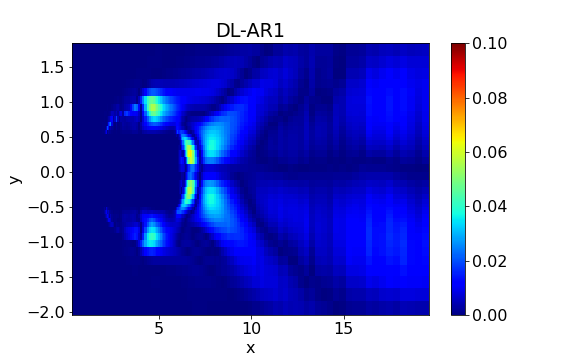}
\includegraphics[width=4cm]{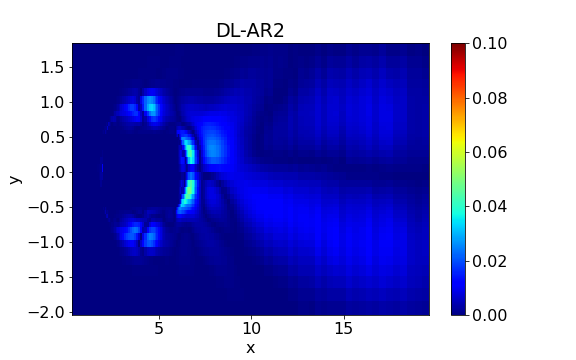}
\includegraphics[width=4cm]{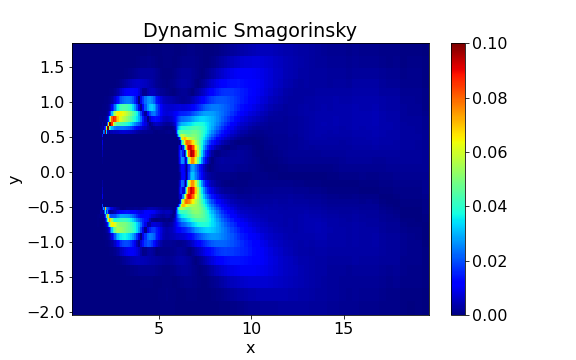}
\includegraphics[width=4cm]{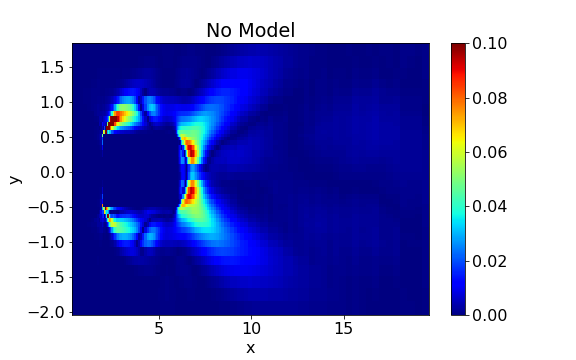}
\label{f1}
\caption{Error for $\tau_{12}$ for AR4 configuration.}
\end{figure}

\subsection{Comparison at Fixed $x_1$ positions} \label{FixedPositionsAppendix}

\subsubsection{AR1}

\begin{figure}[H]
\centering
\includegraphics[width=5cm]{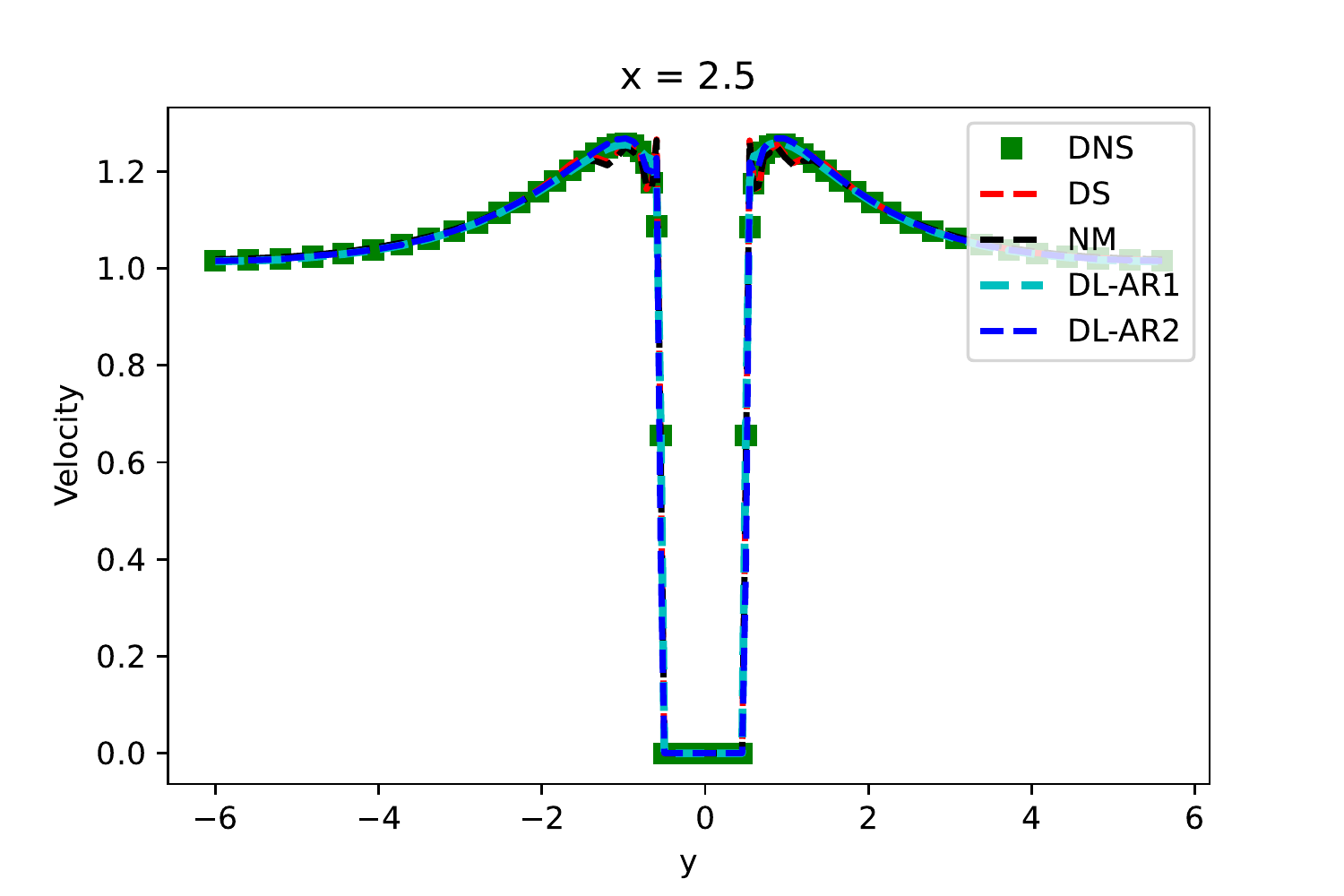}
\includegraphics[width=5cm]{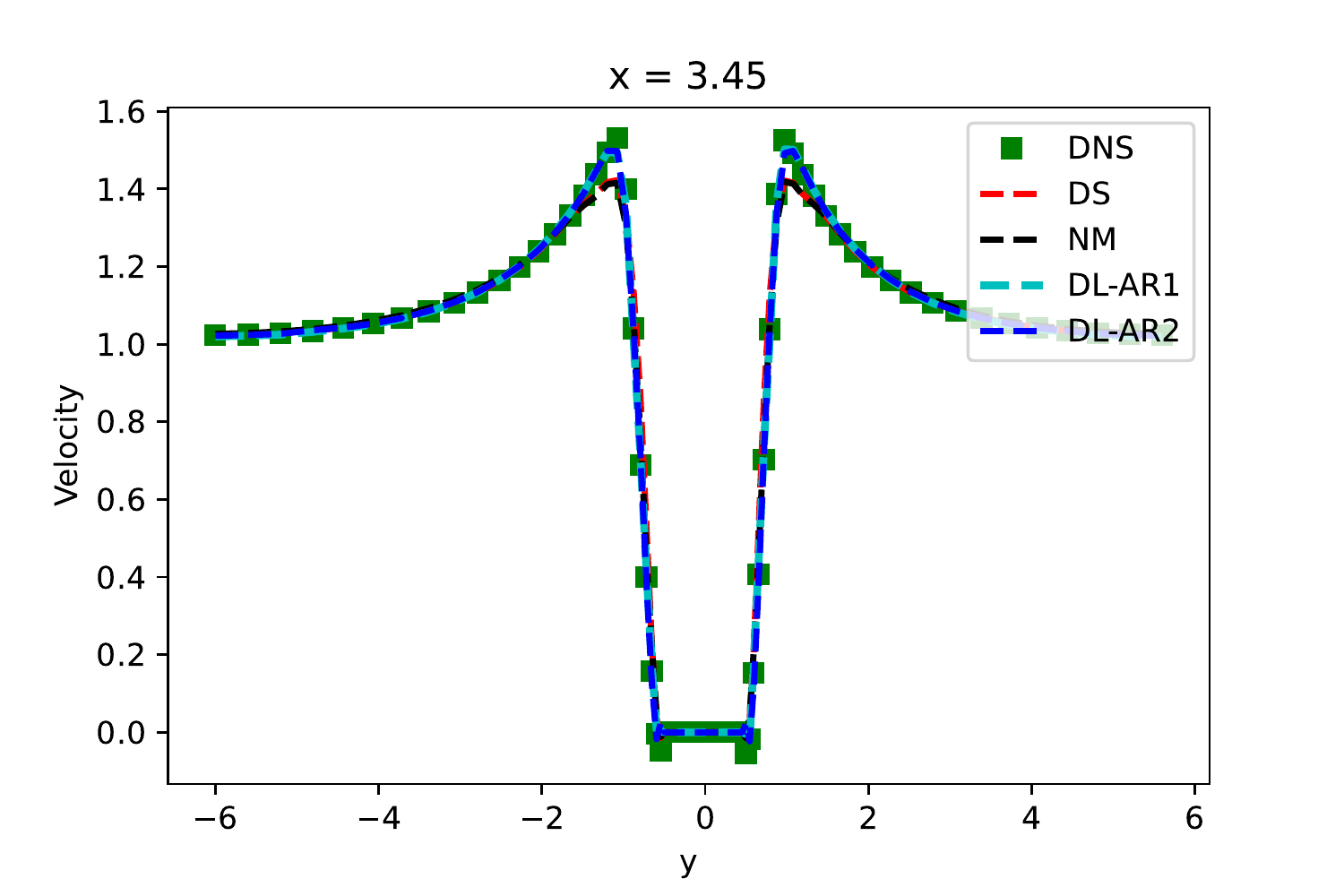}
\includegraphics[width=5cm]{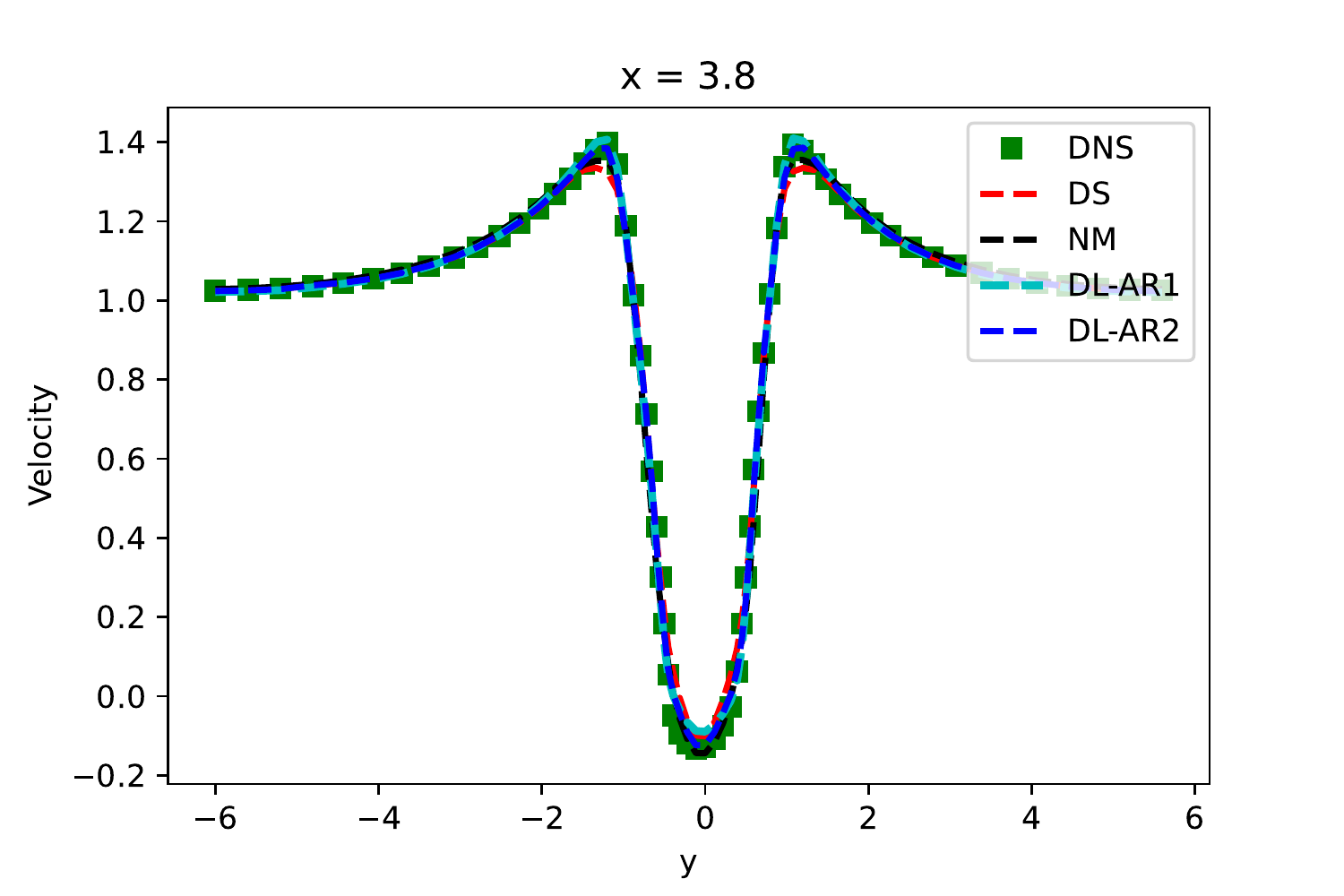}
\includegraphics[width=5cm]{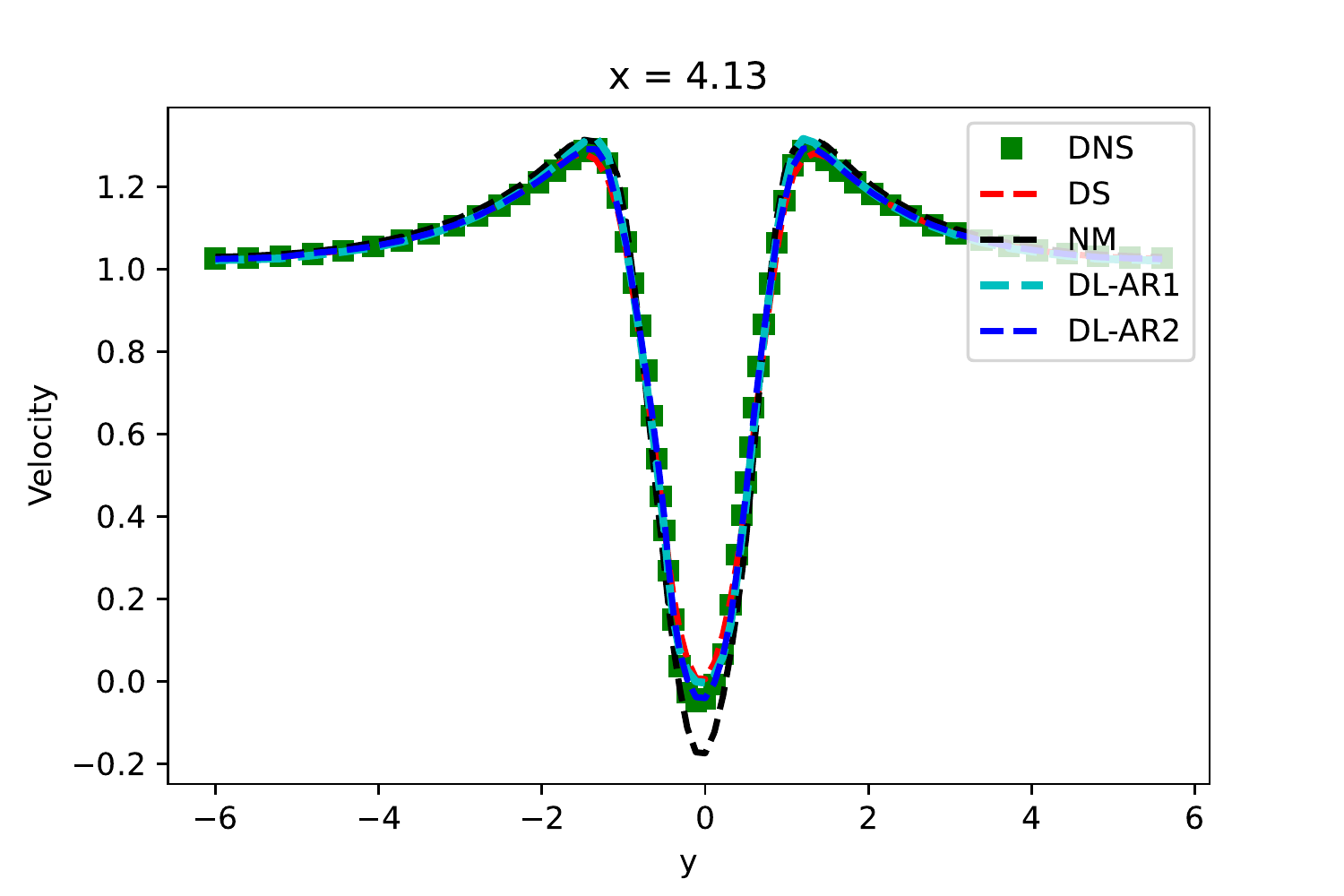}
\includegraphics[width=5cm]{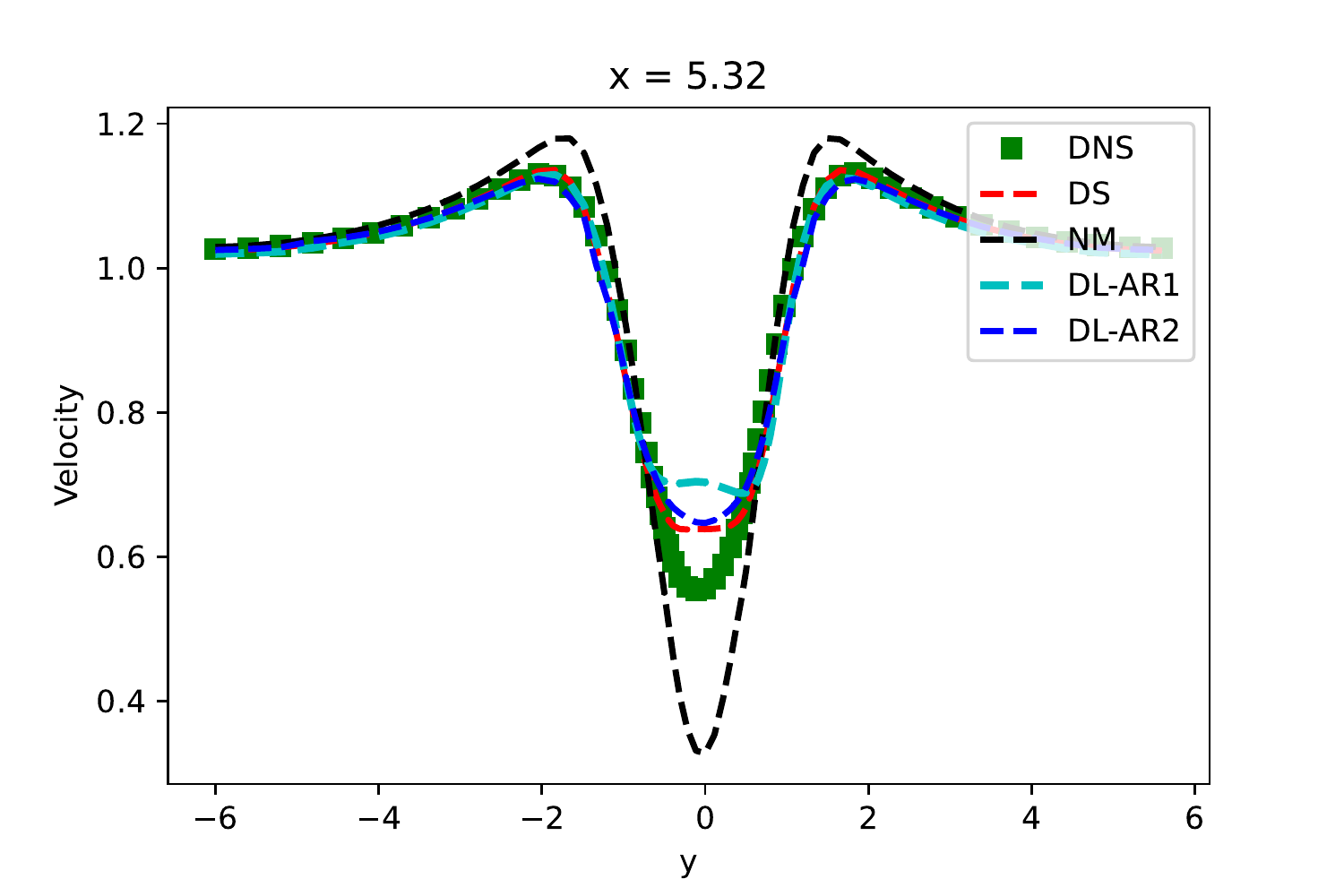}
\includegraphics[width=5cm]{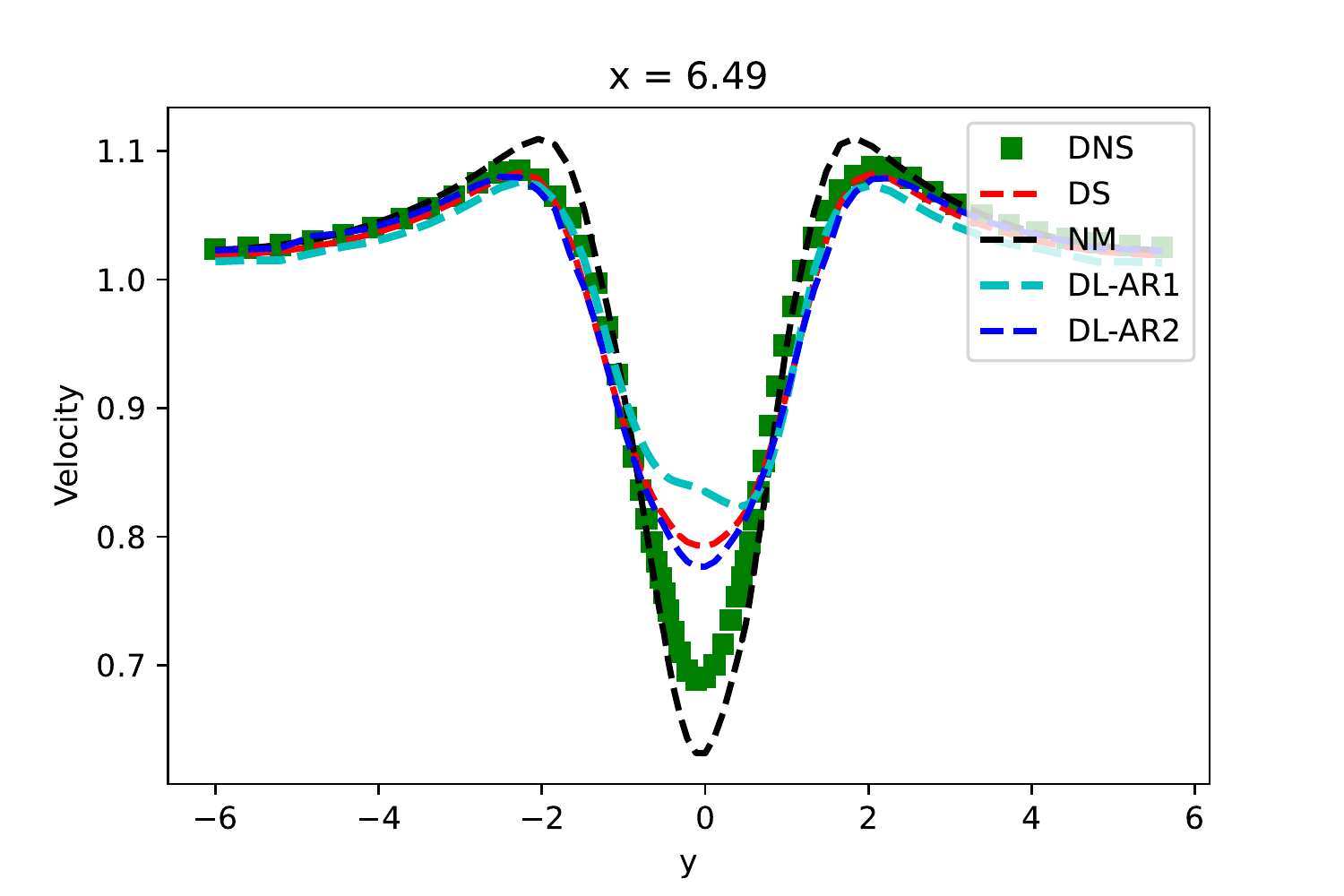}
\includegraphics[width=5cm]{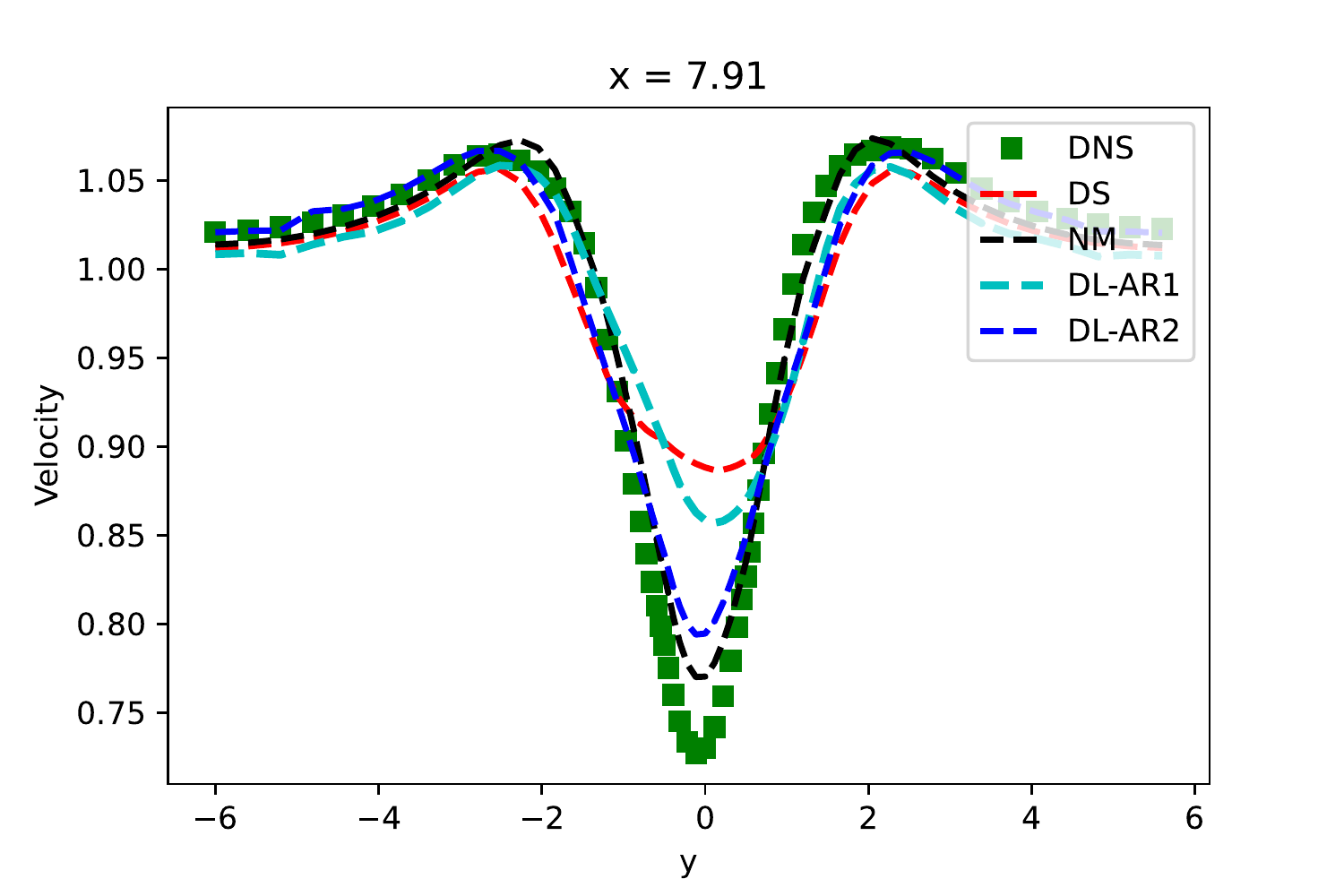}
\includegraphics[width=5cm]{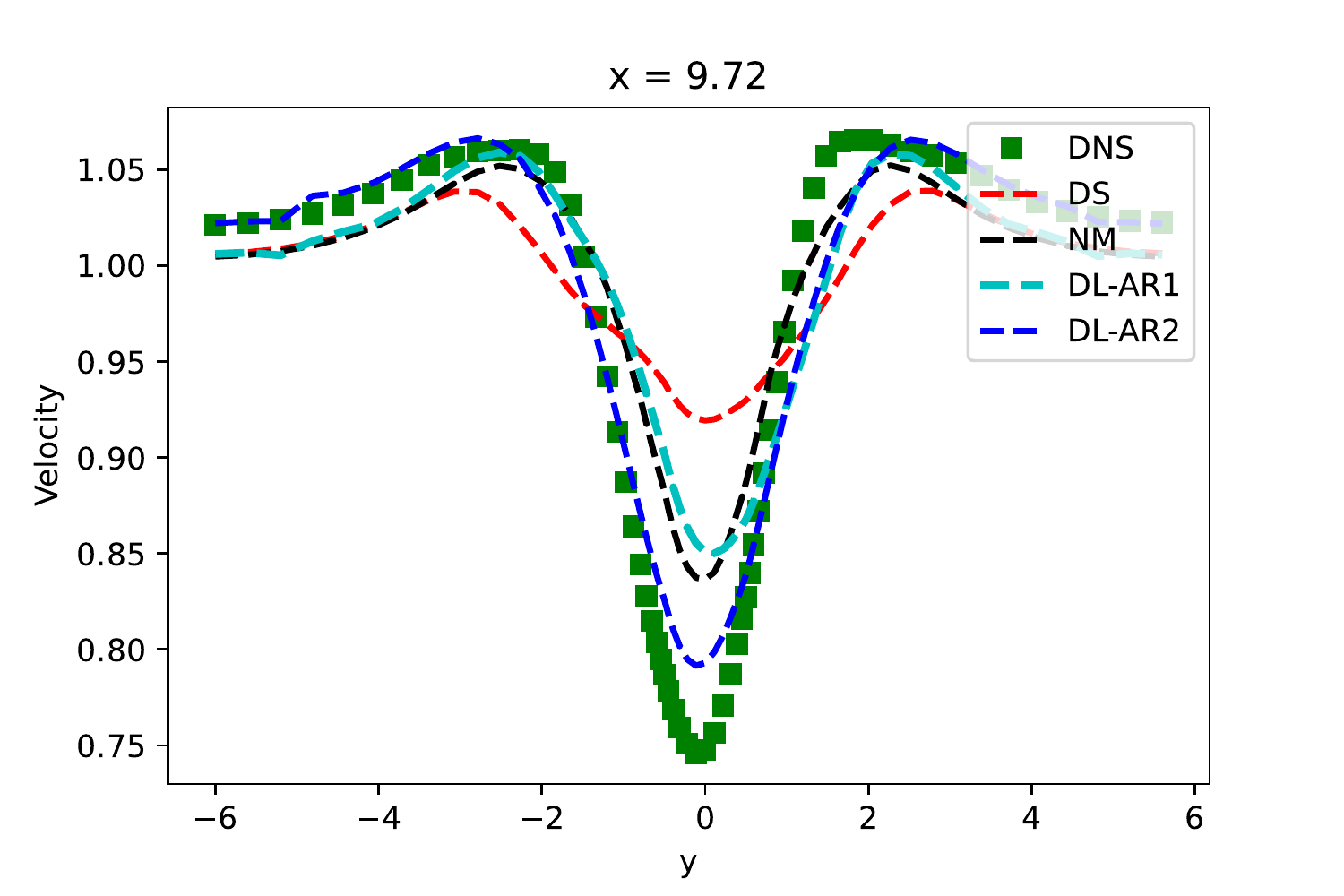}
\includegraphics[width=5cm]{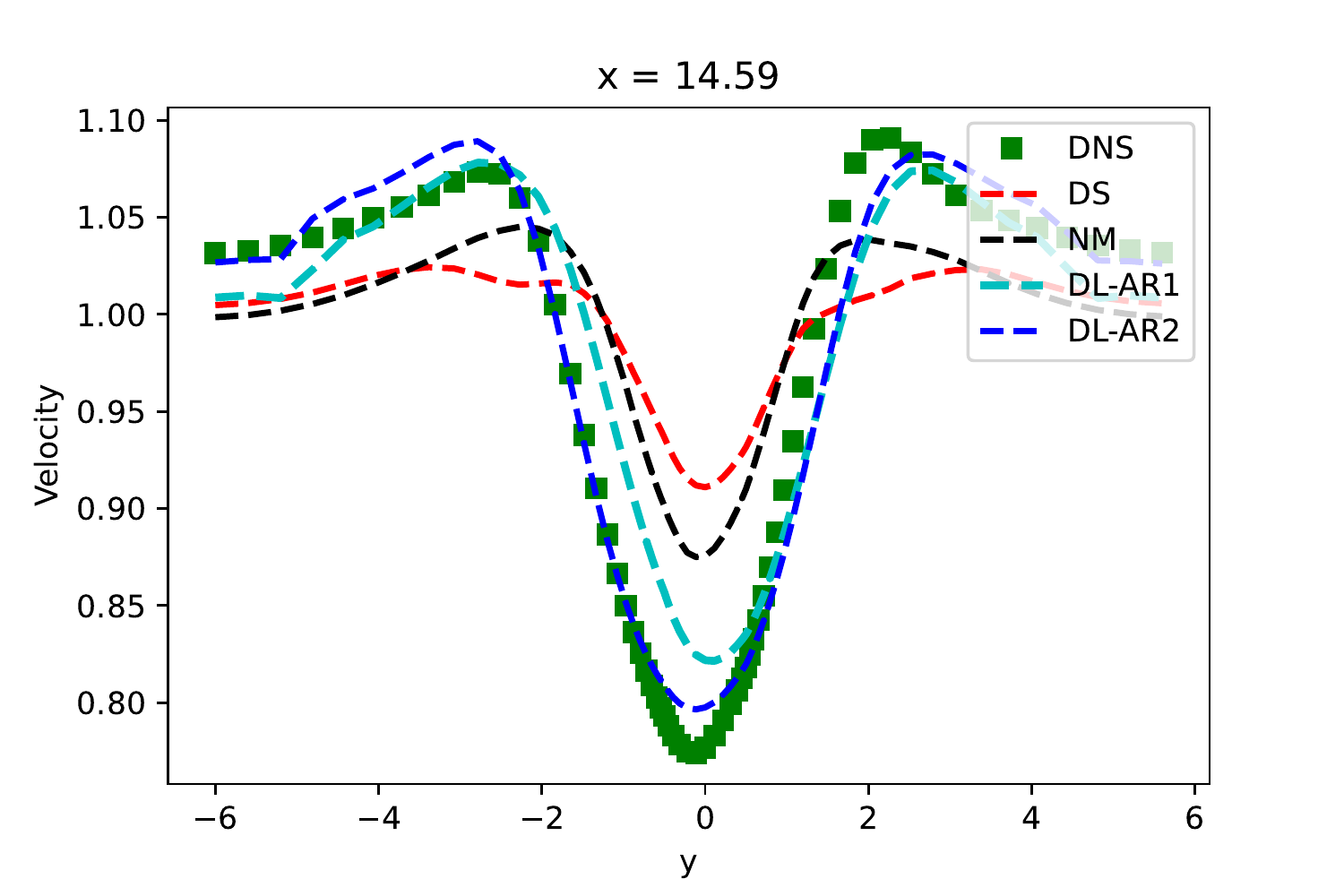}
\label{f1}
\caption{Mean profile for $u_1$ for AR1 configuration.}
\end{figure}

\begin{figure}[H]
\centering
\includegraphics[width=5cm]{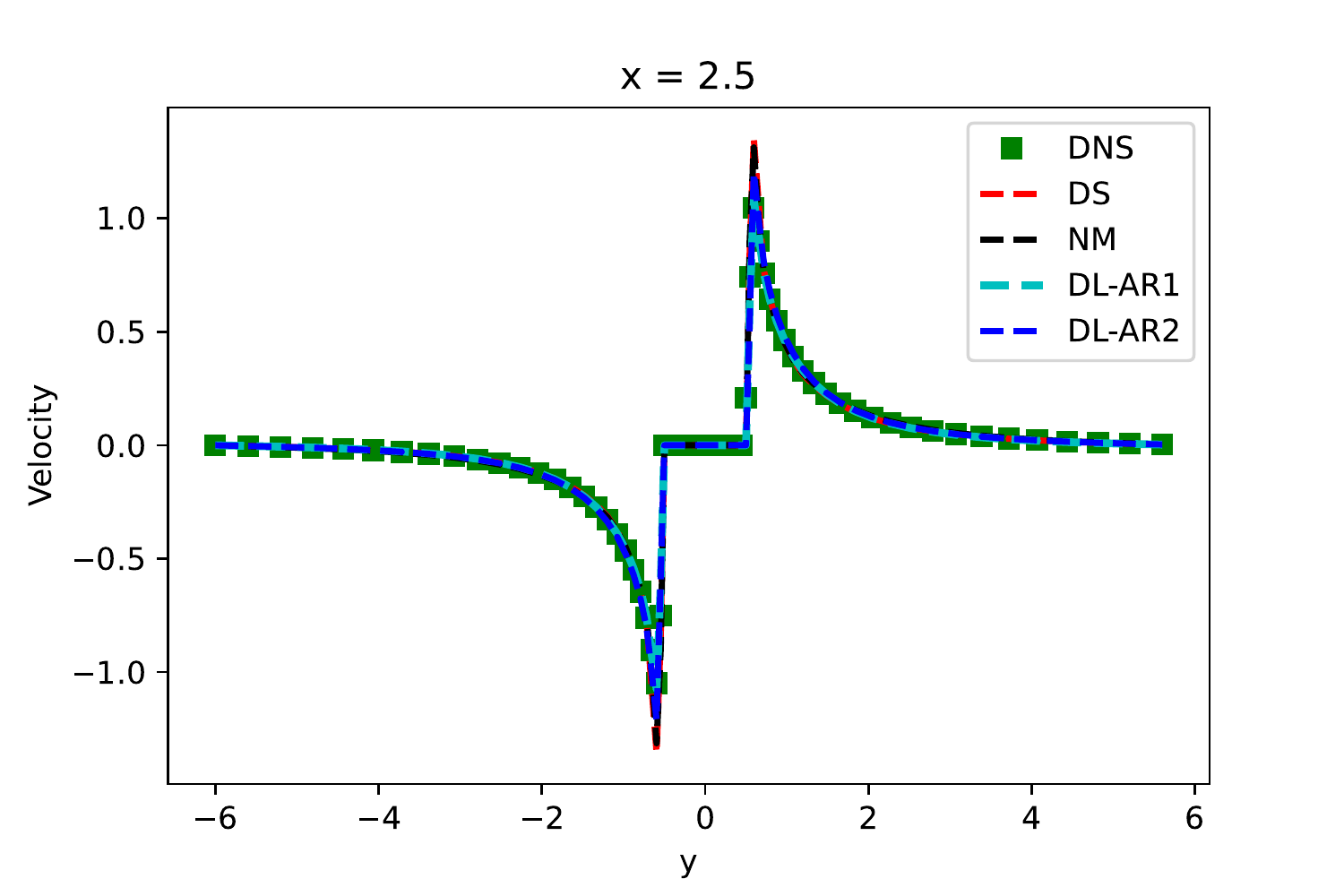}
\includegraphics[width=5cm]{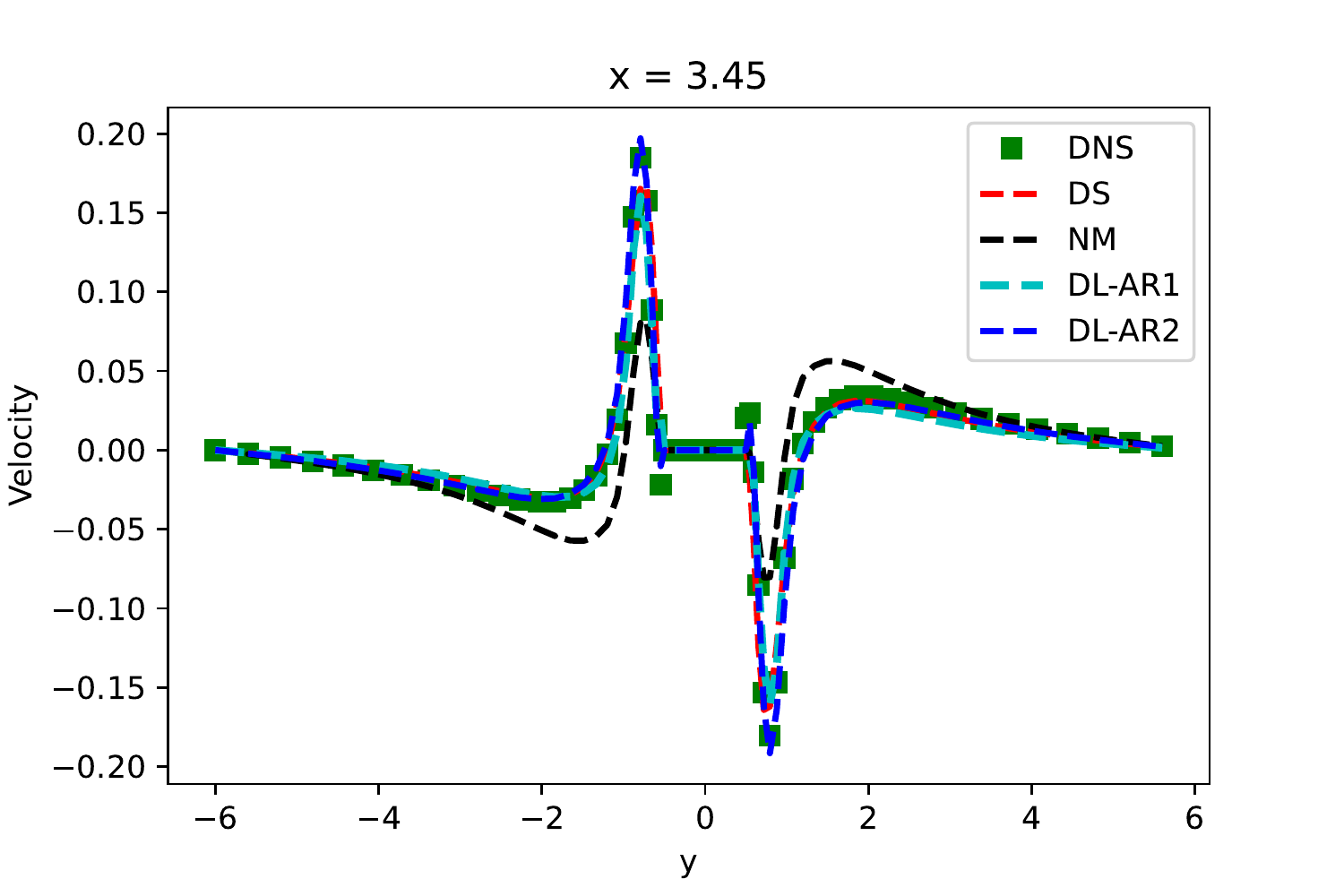}
\includegraphics[width=5cm]{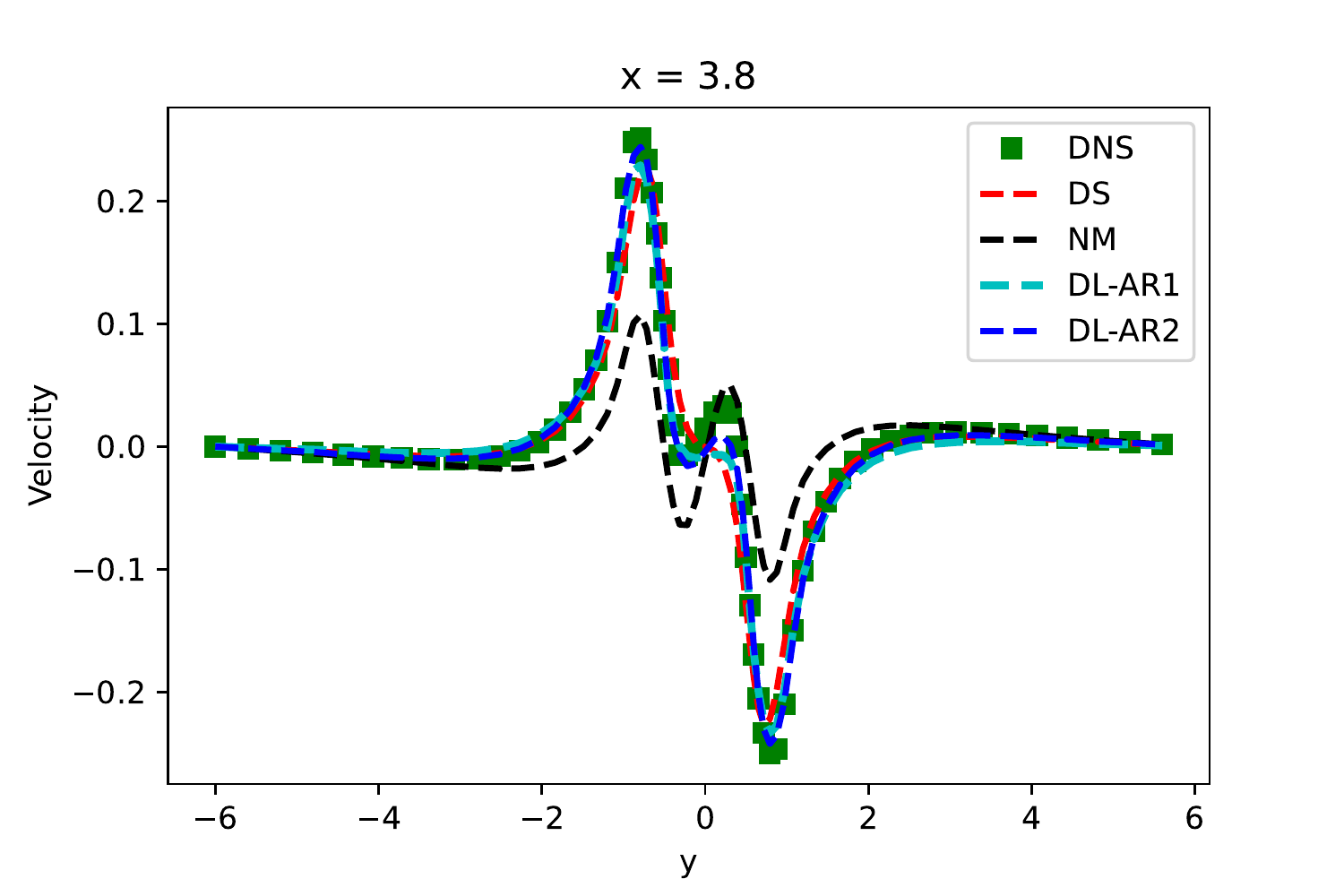}
\includegraphics[width=5cm]{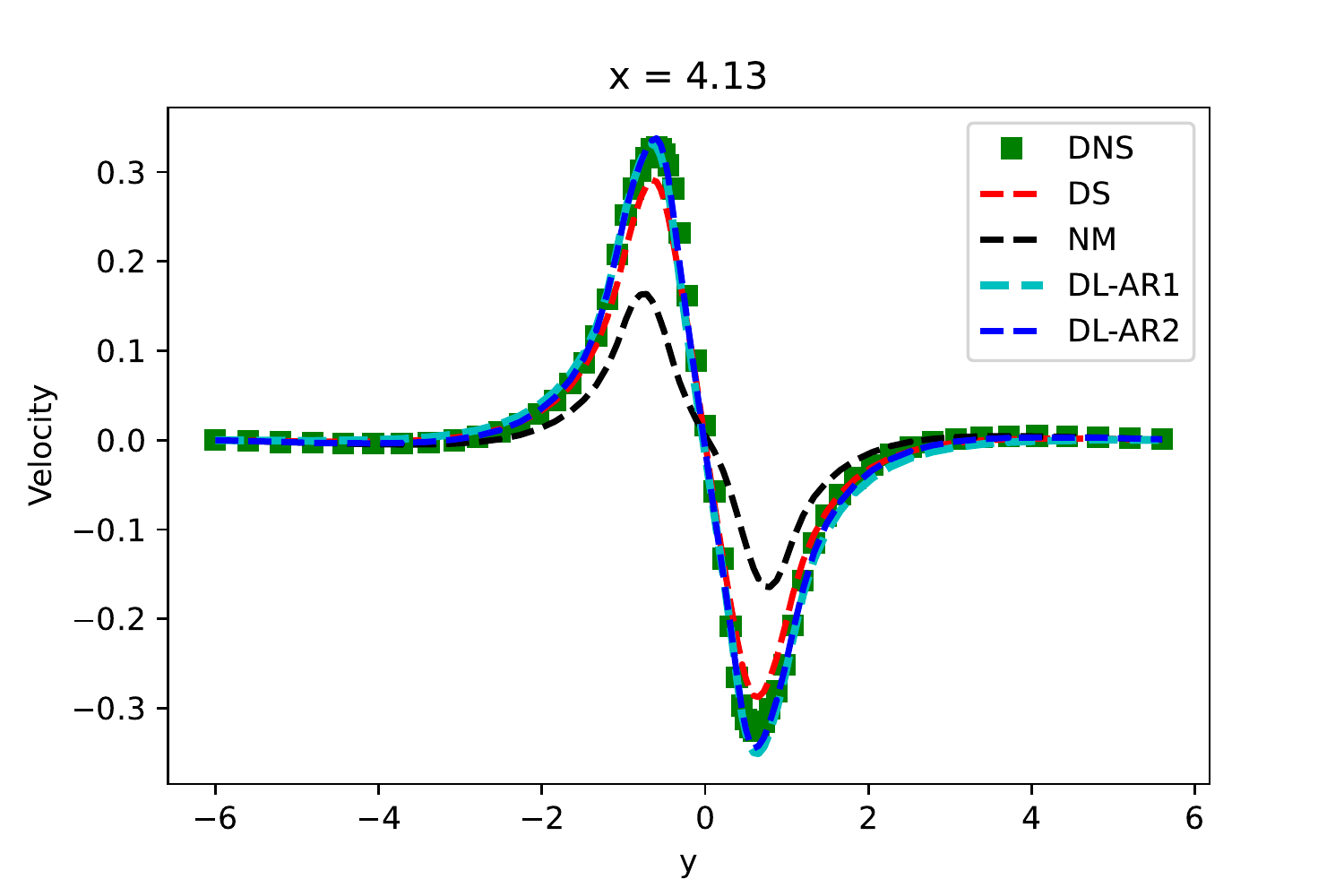}
\includegraphics[width=5cm]{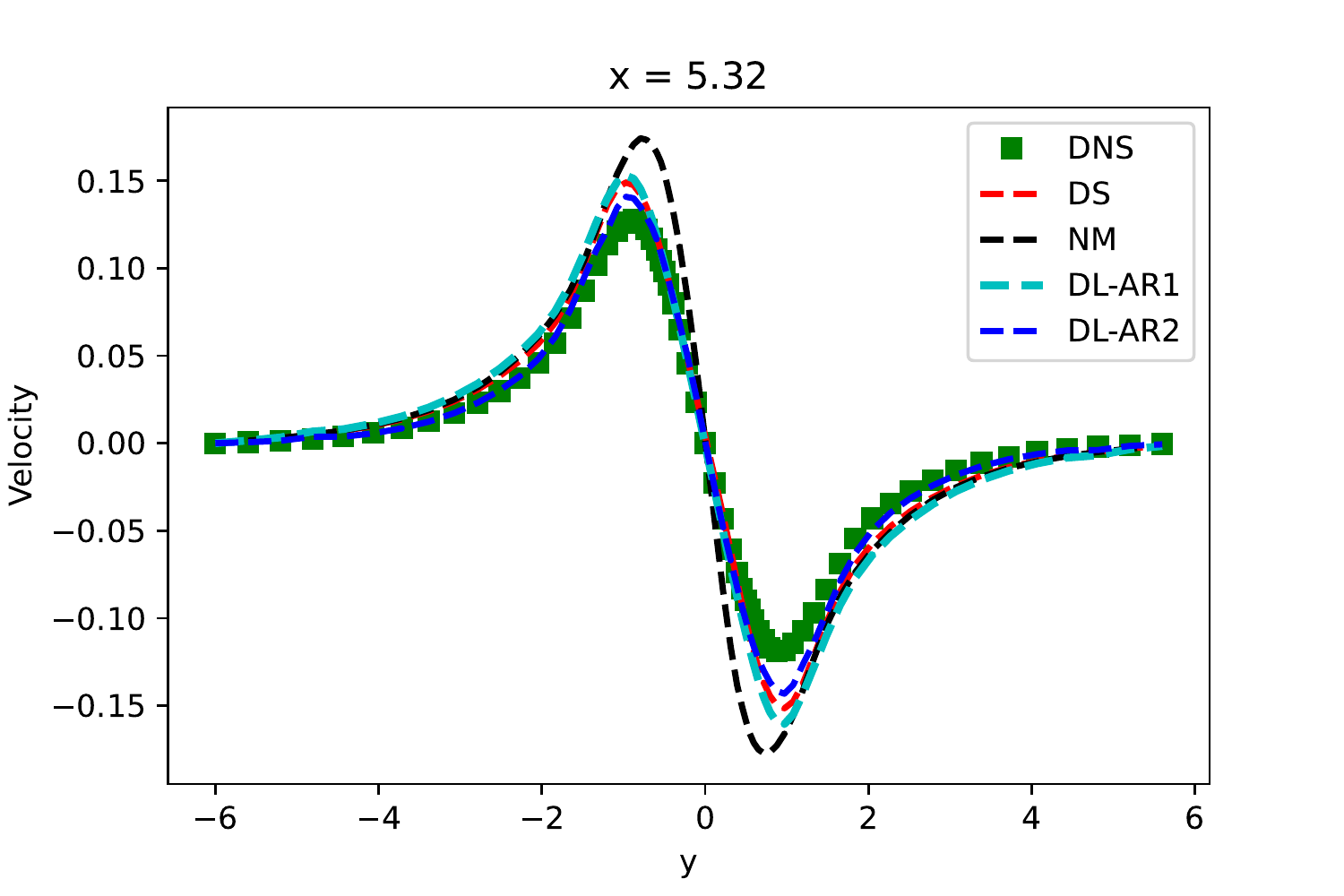}
\includegraphics[width=5cm]{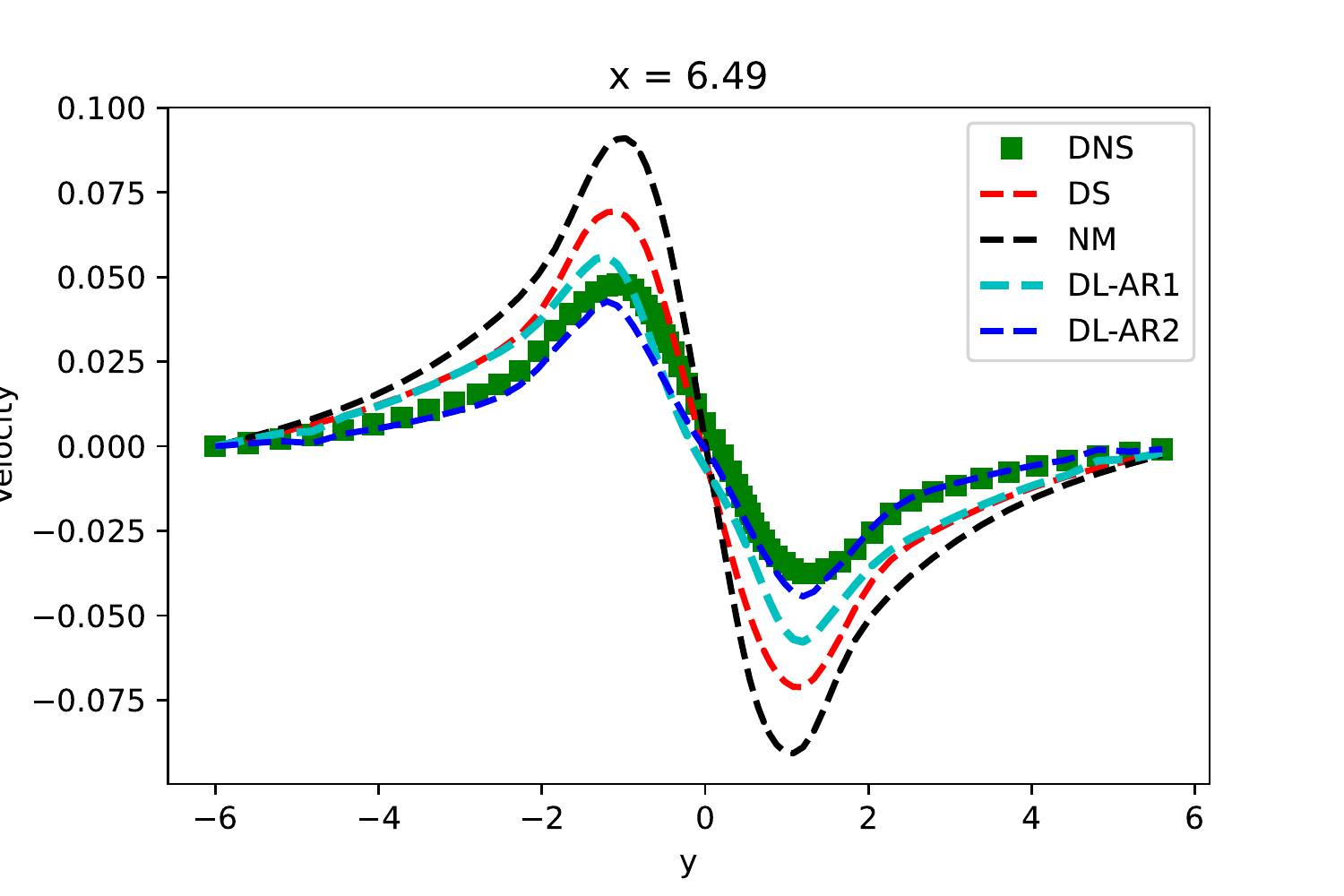}
\includegraphics[width=5cm]{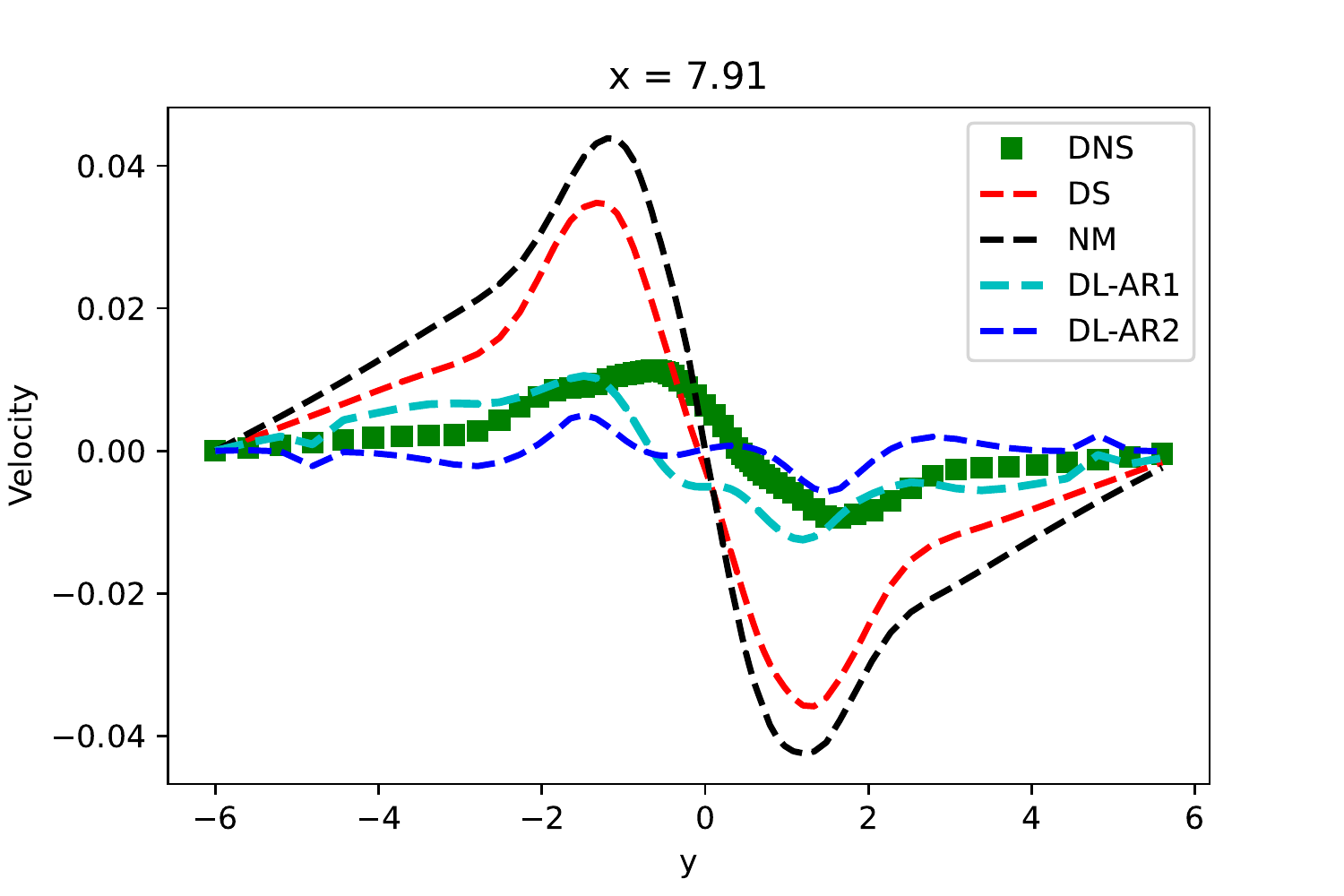}
\includegraphics[width=5cm]{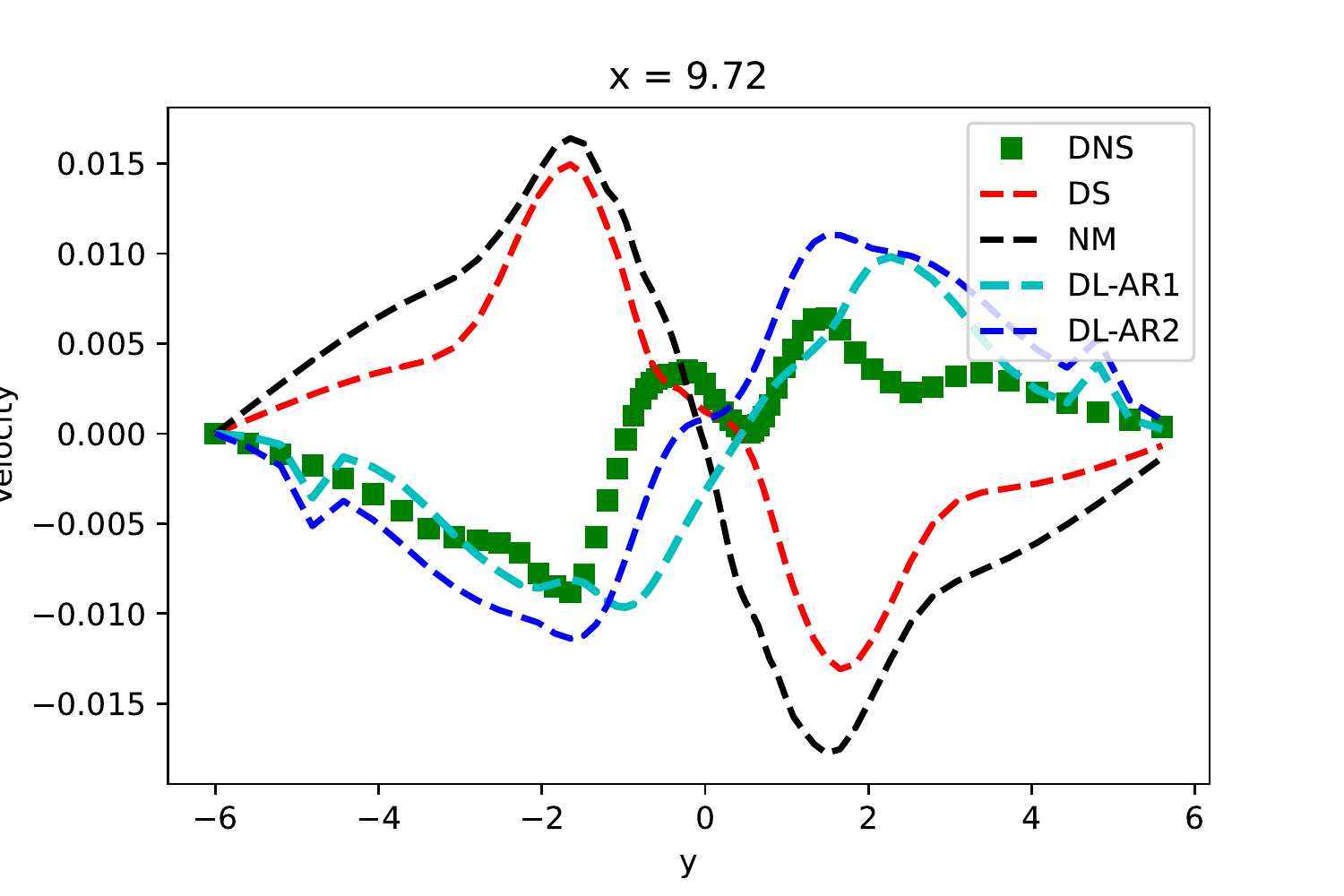}
\includegraphics[width=5cm]{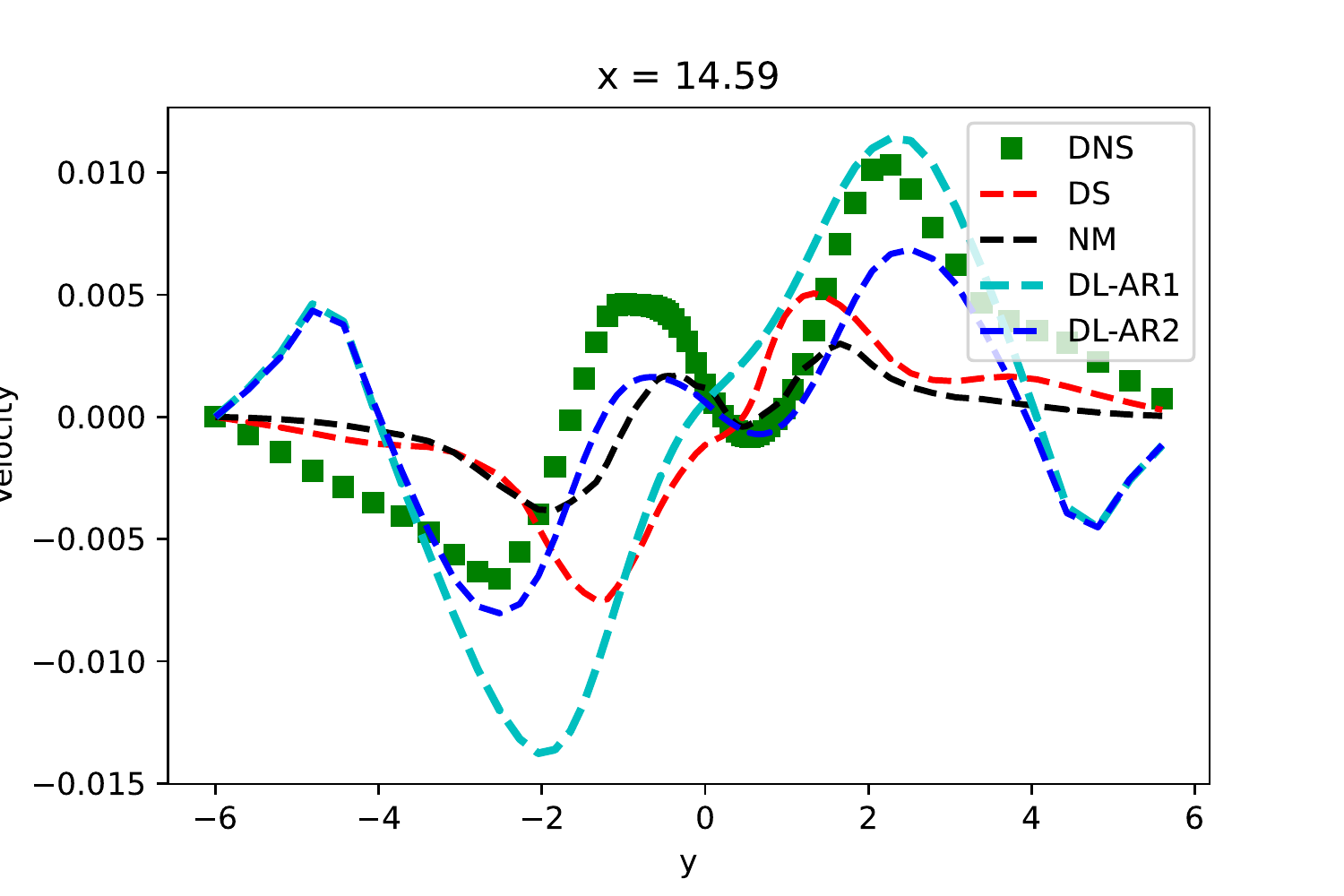}
\label{f1}
\caption{Mean profile for $u_2$ for AR1 configuration.}
\end{figure}

\begin{figure}[H]
\centering
\includegraphics[width=5cm]{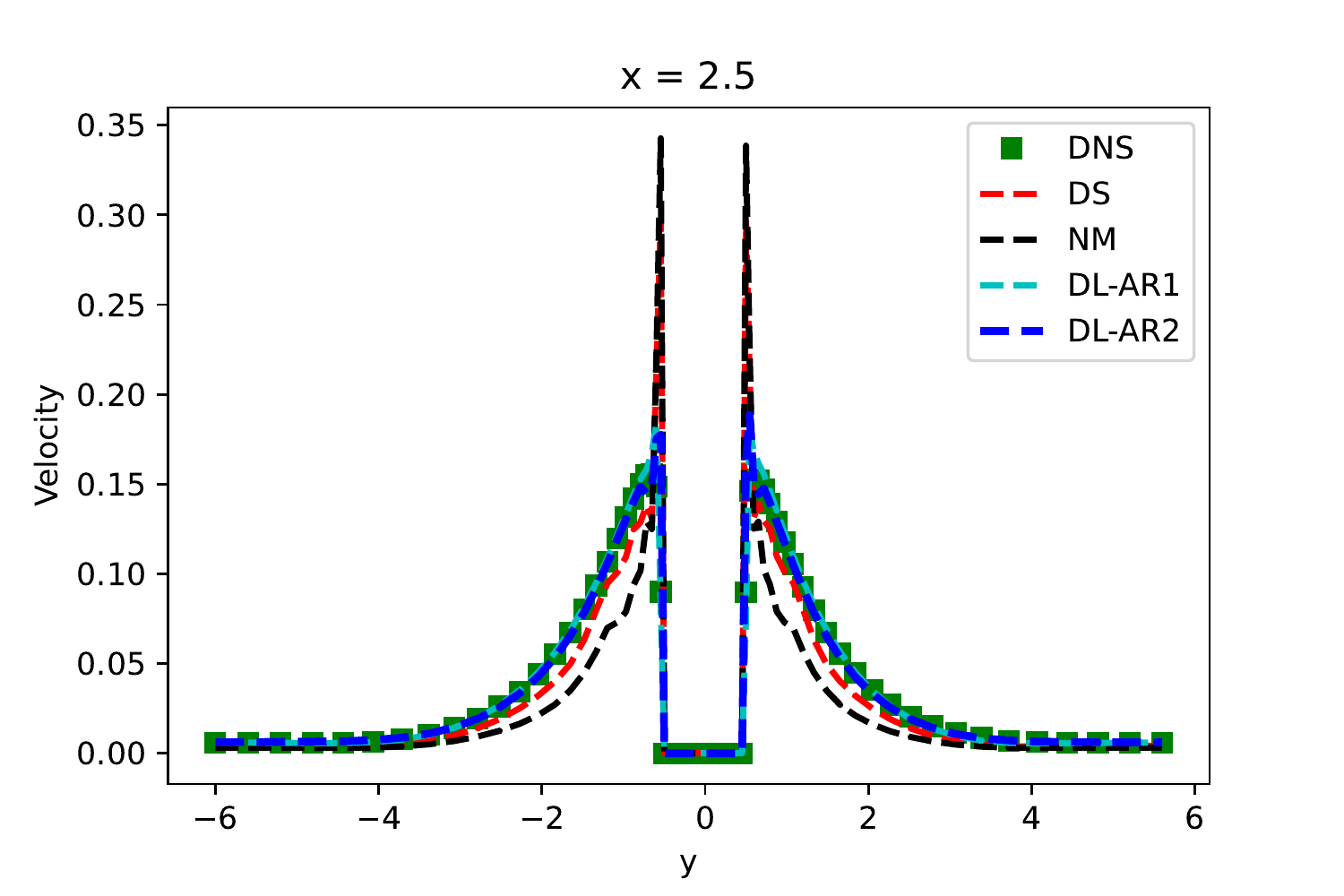}
\includegraphics[width=5cm]{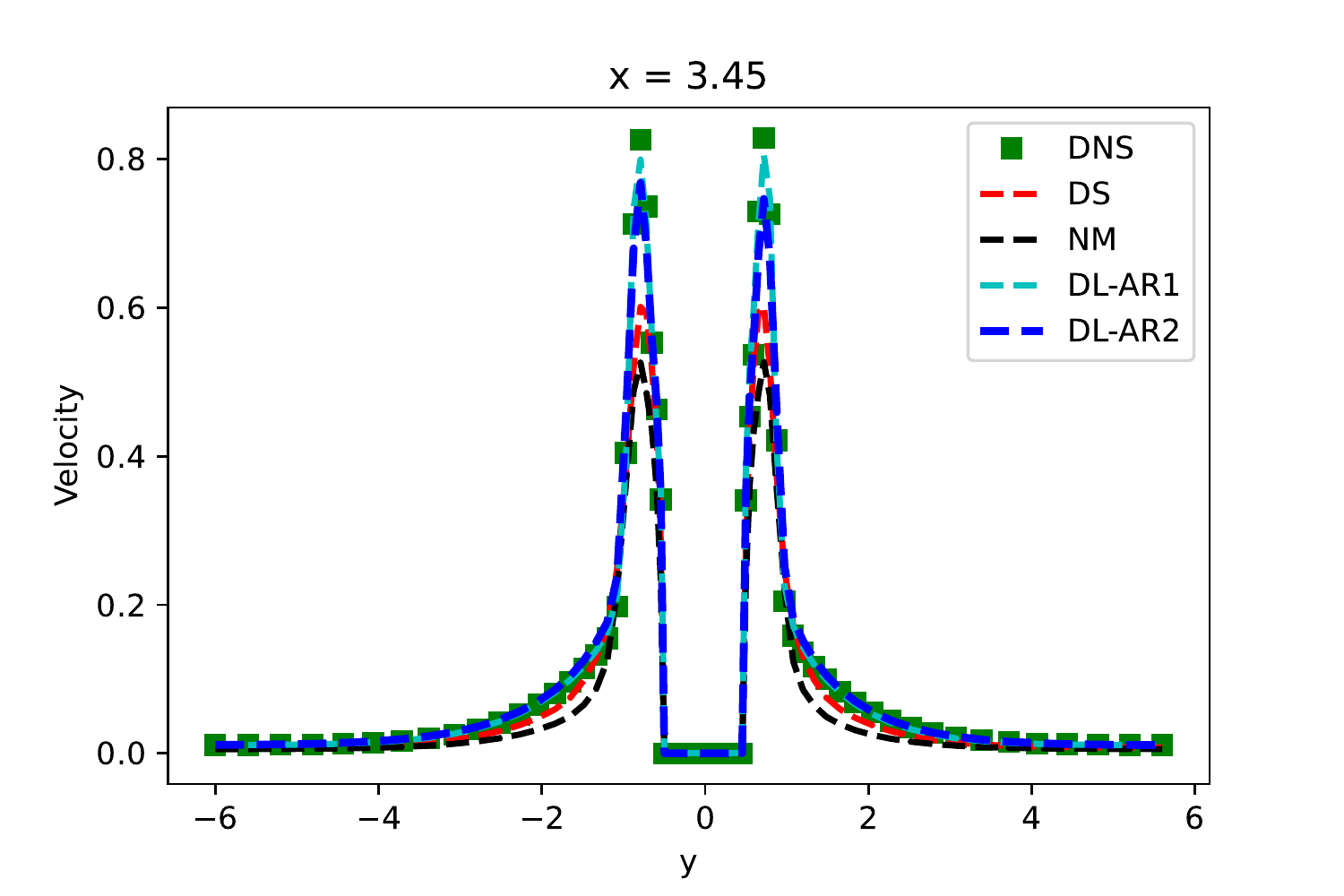}
\includegraphics[width=5cm]{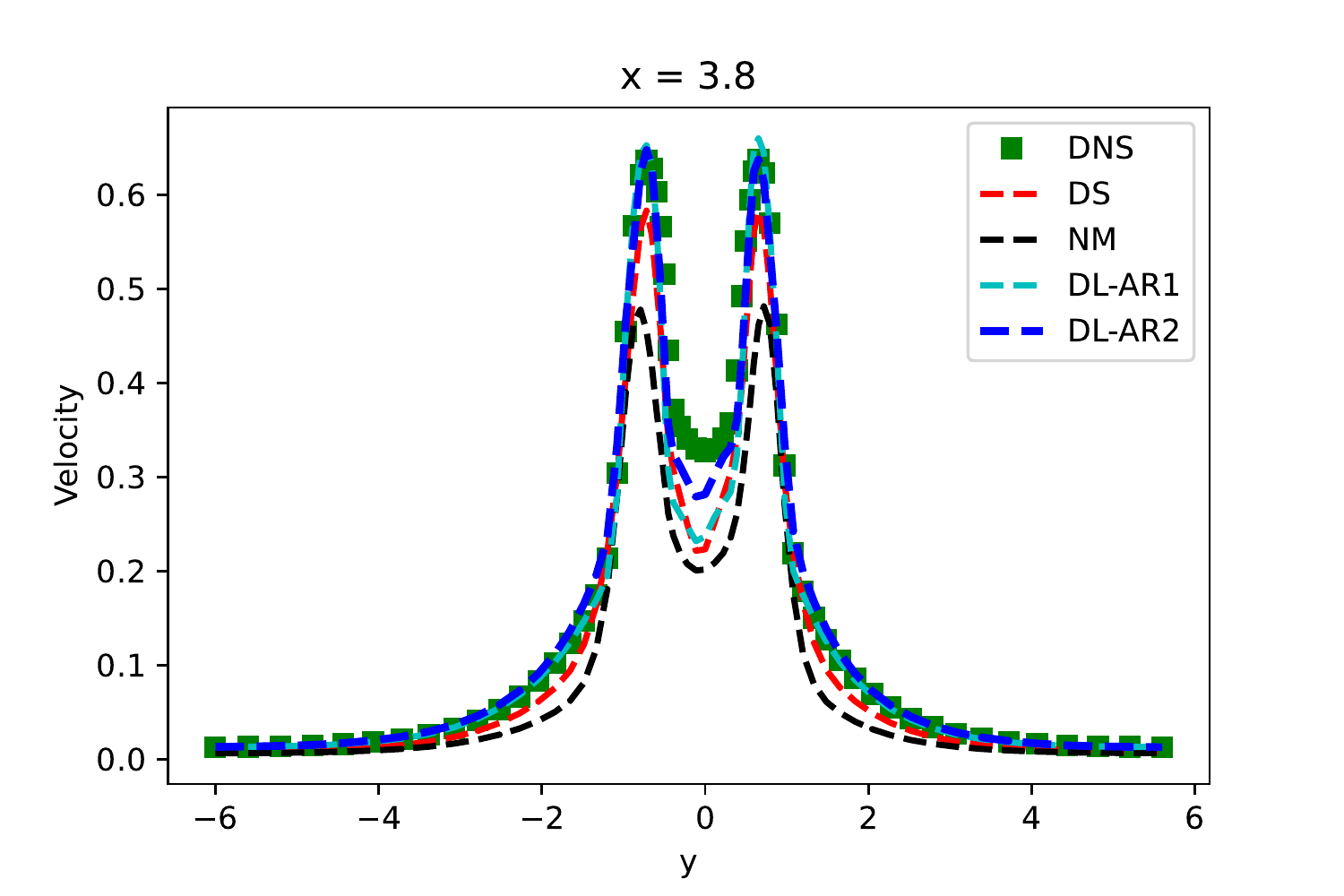}
\includegraphics[width=5cm]{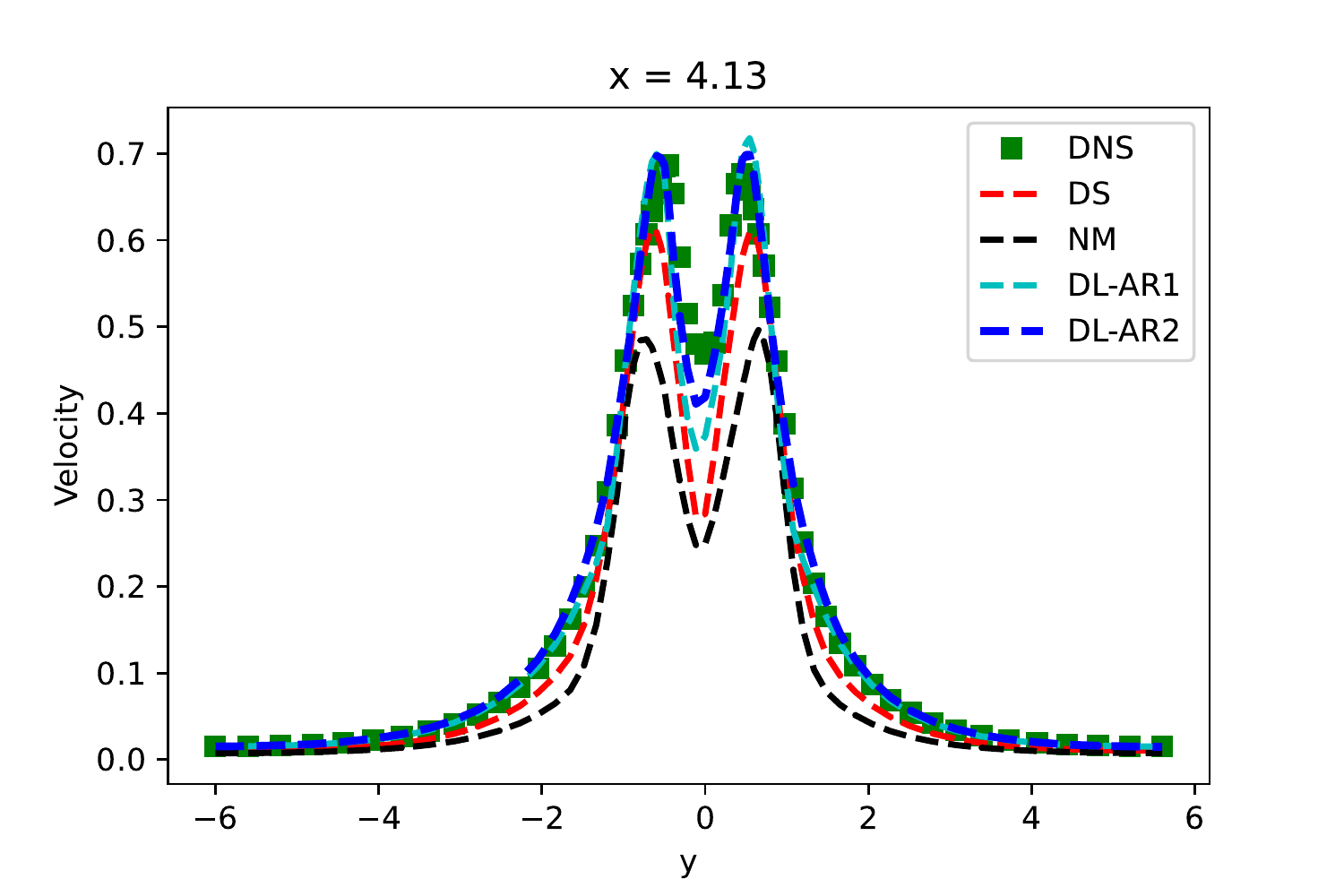}
\includegraphics[width=5cm]{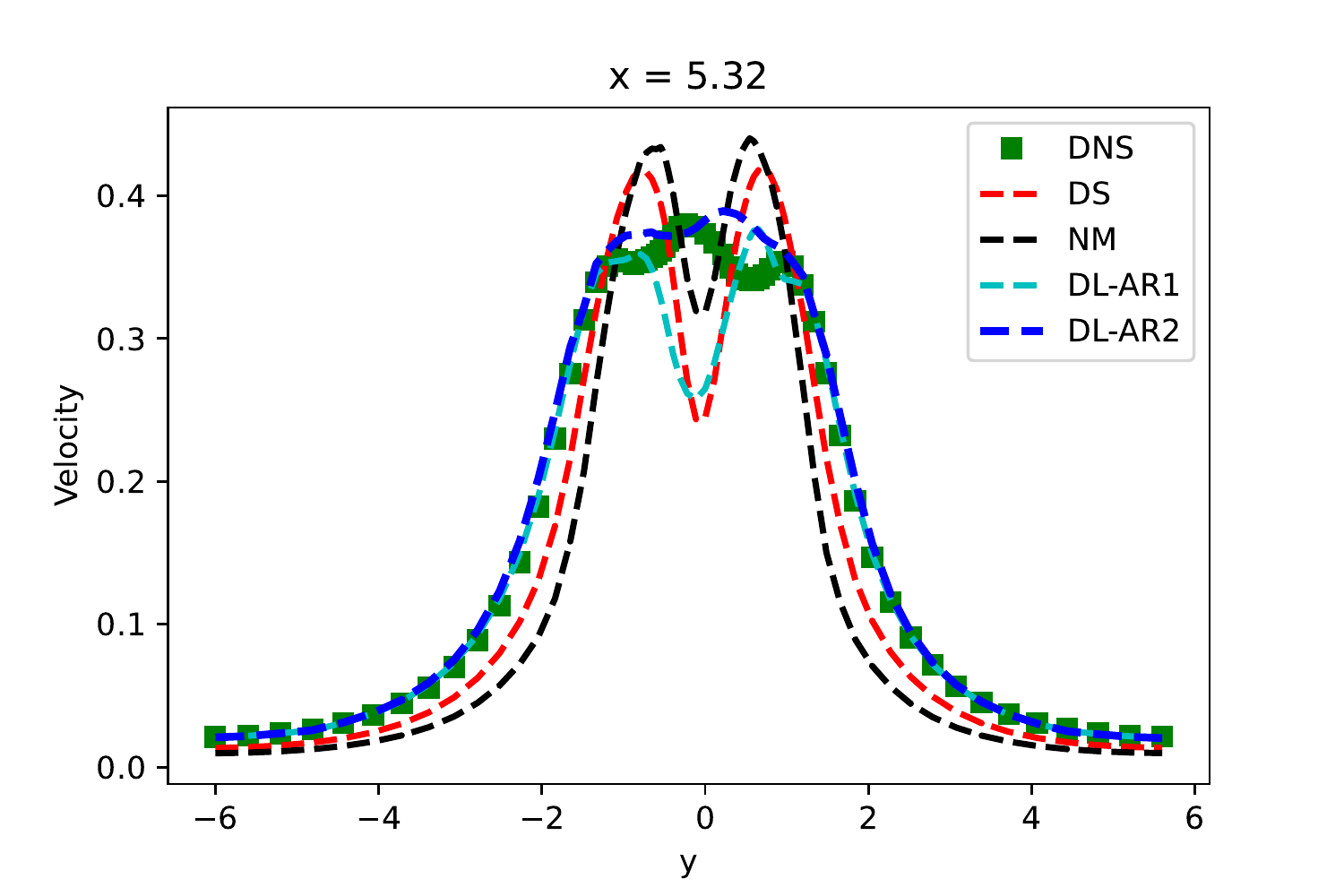}
\includegraphics[width=5cm]{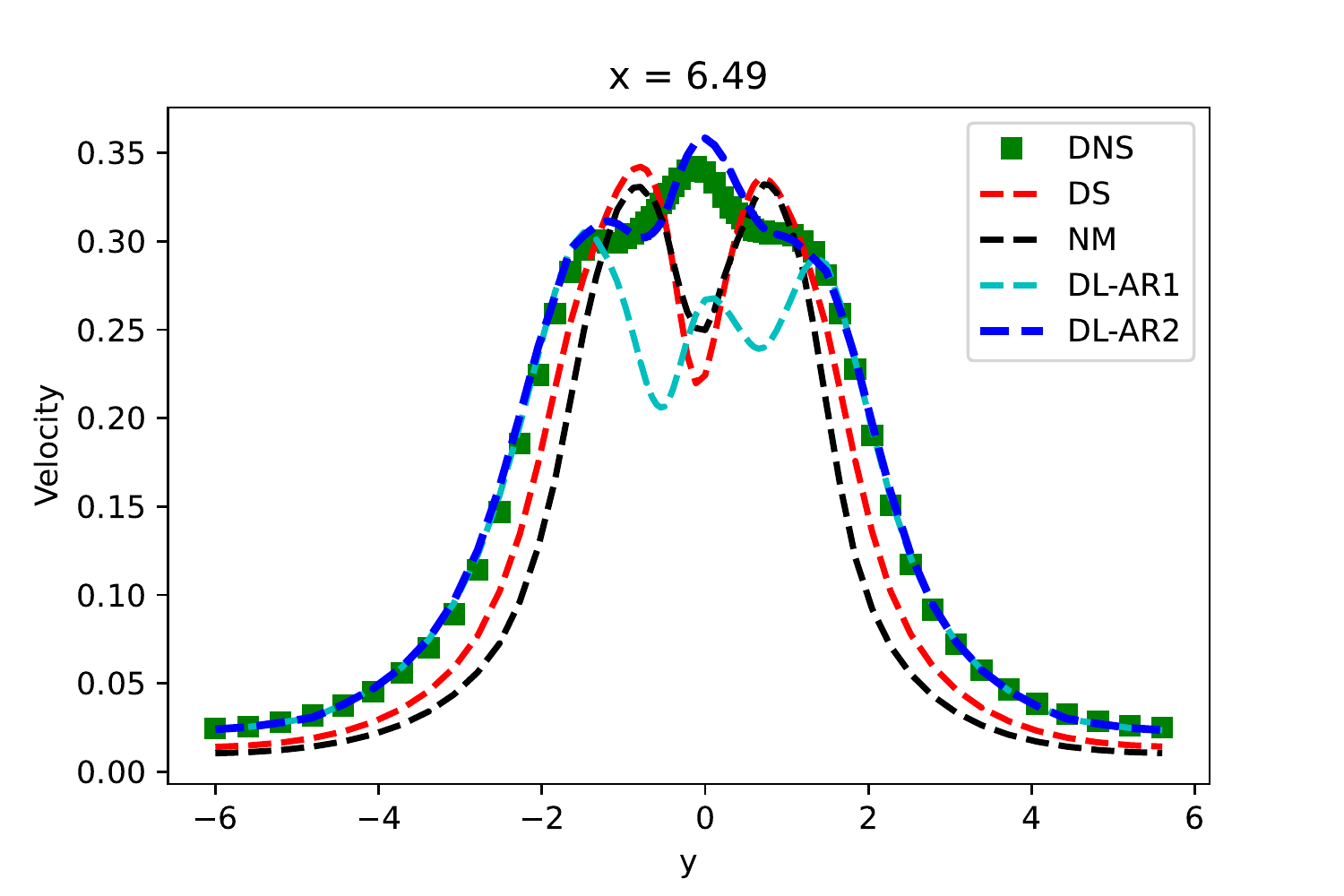}
\includegraphics[width=5cm]{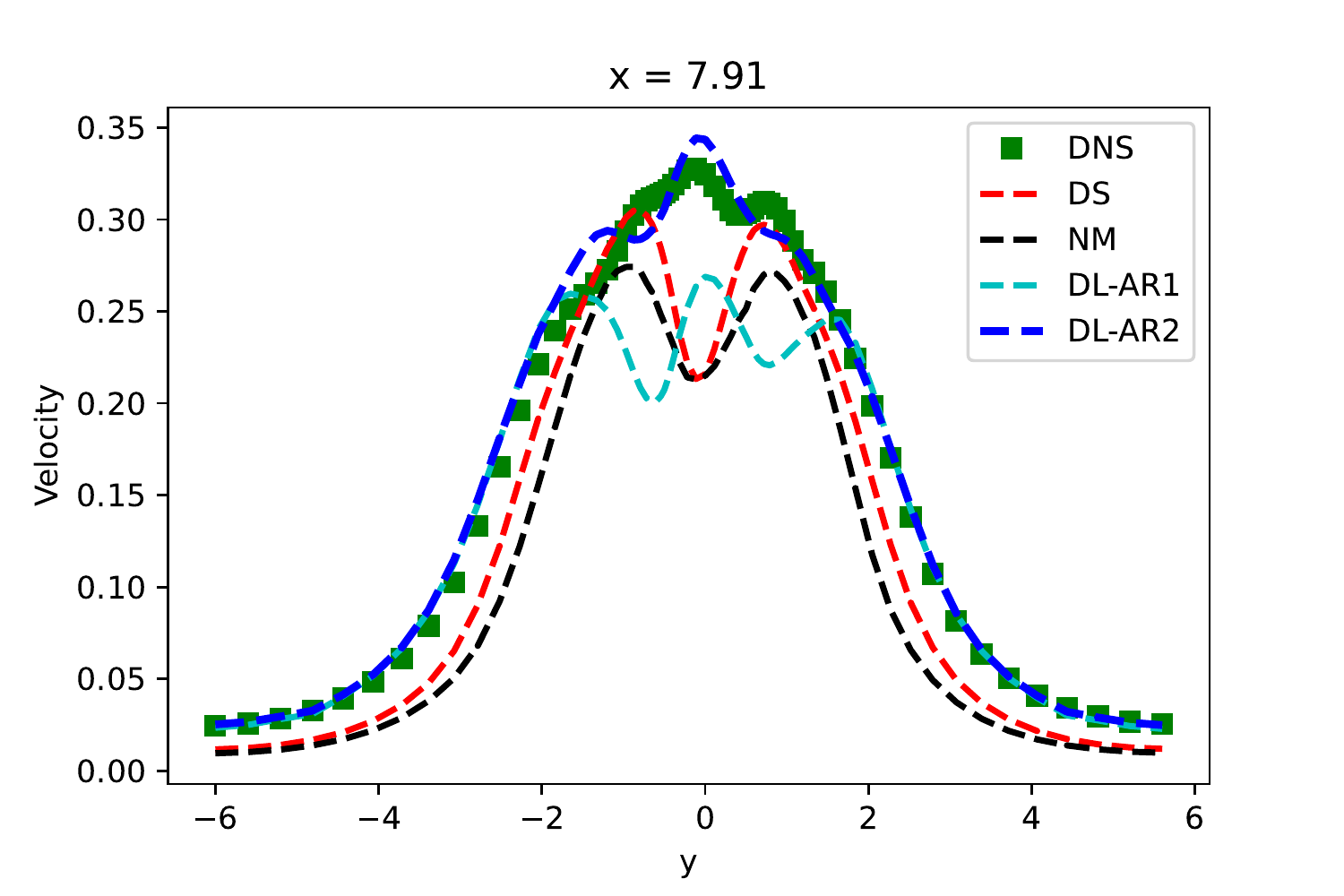}
\includegraphics[width=5cm]{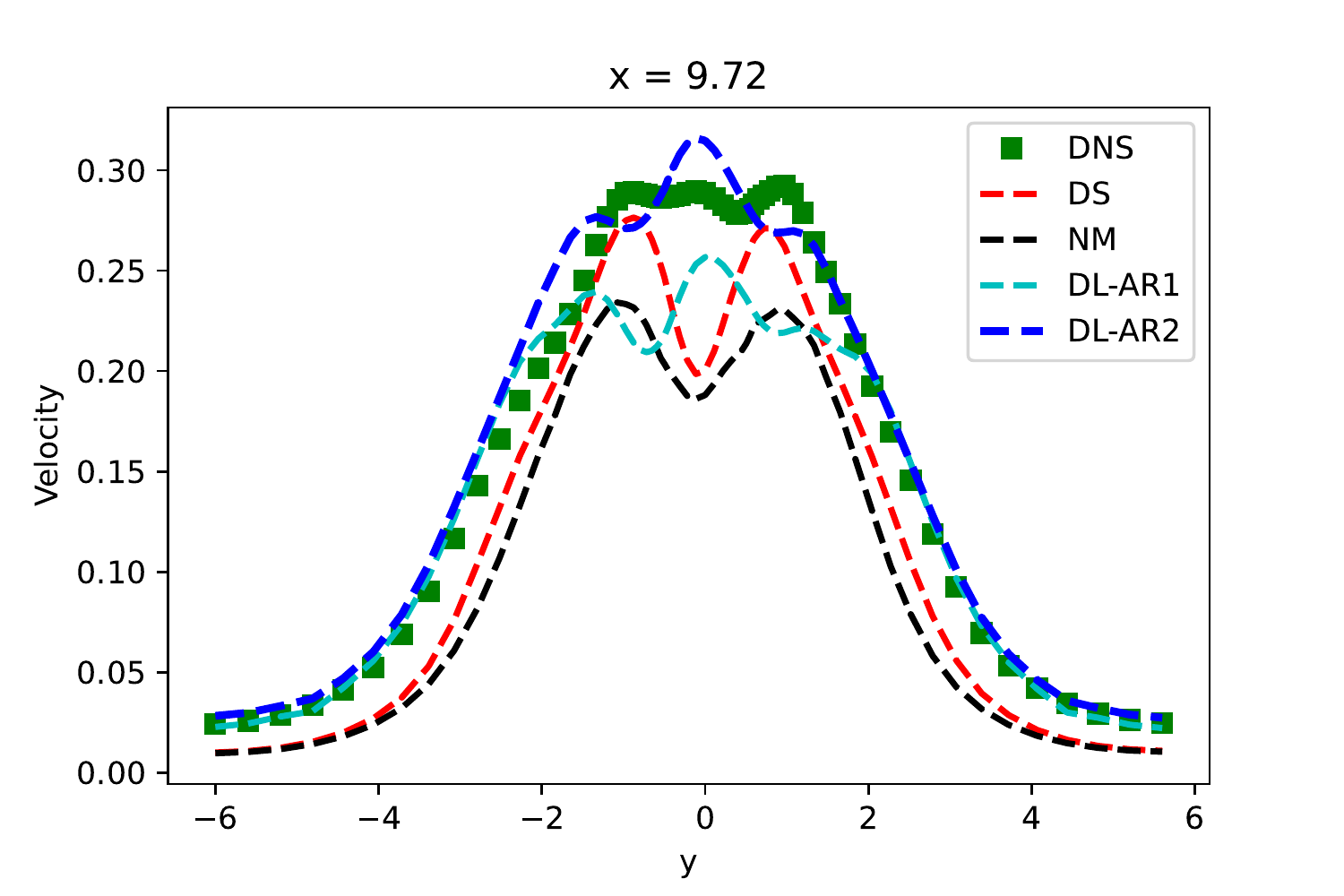}
\includegraphics[width=5cm]{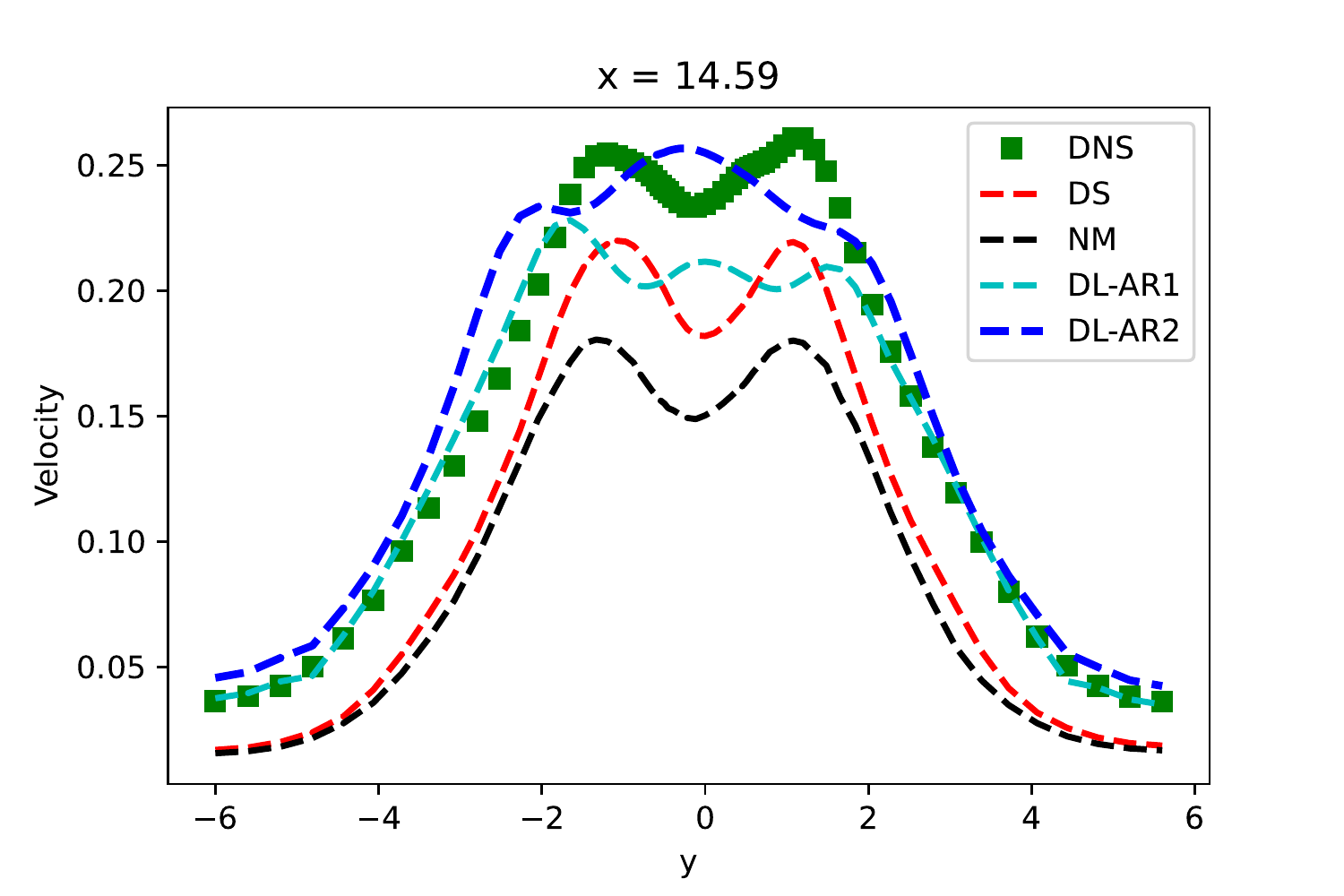}
\label{f1}
\caption{RMS profile for $u_1$ for AR1 configuration.}
\end{figure}

\begin{figure}[H]
\centering
\includegraphics[width=5cm]{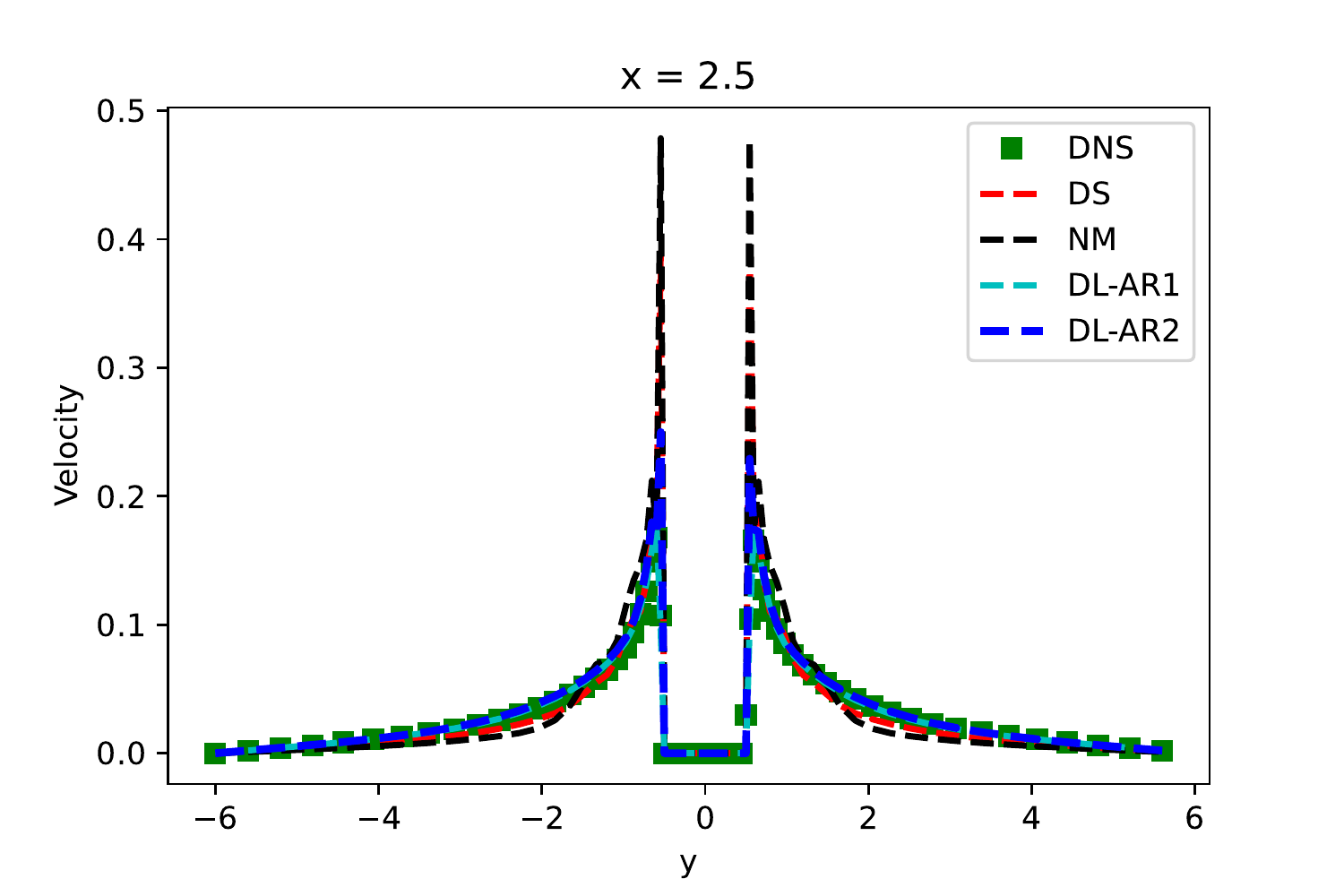}
\includegraphics[width=5cm]{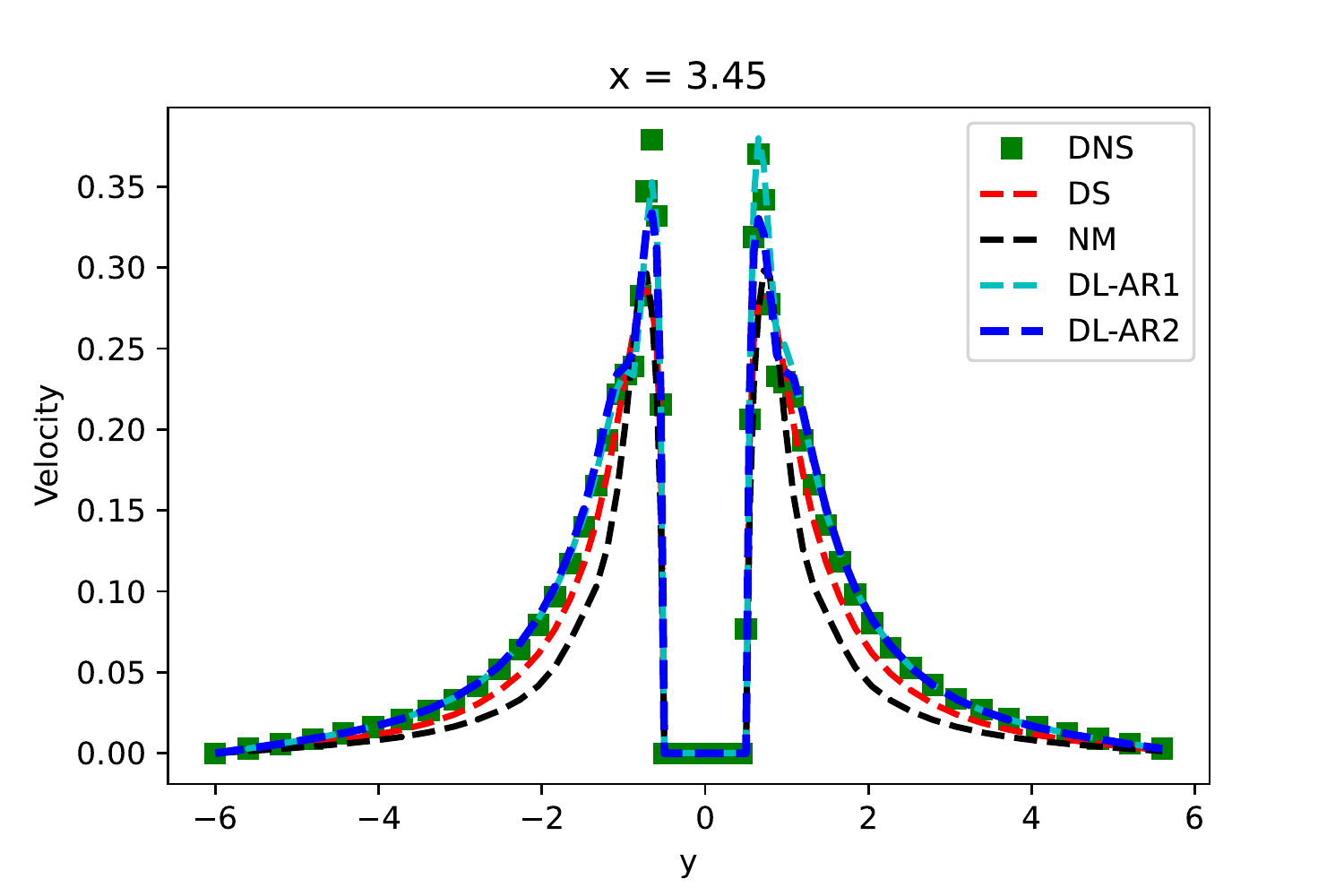}
\includegraphics[width=5cm]{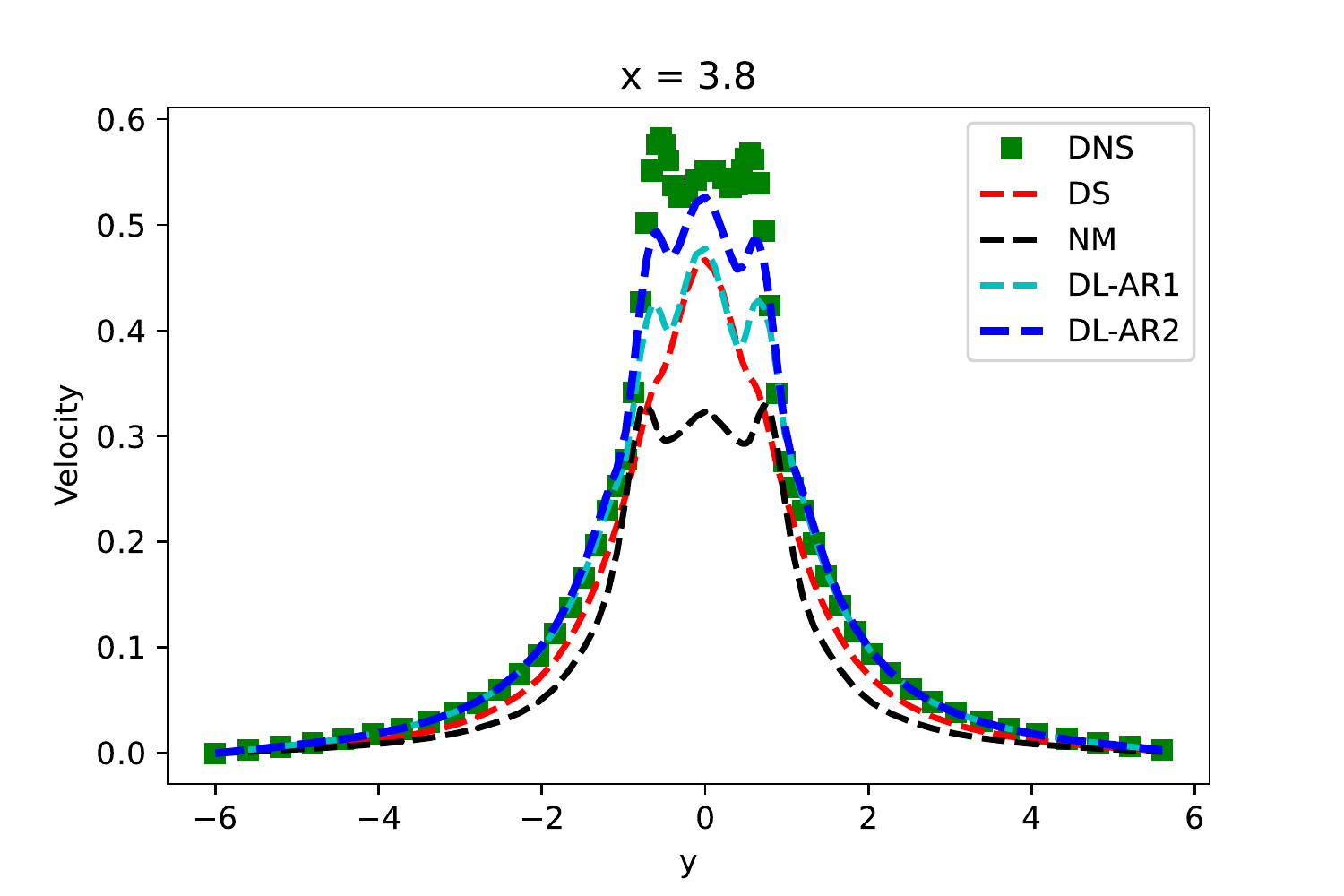}
\includegraphics[width=5cm]{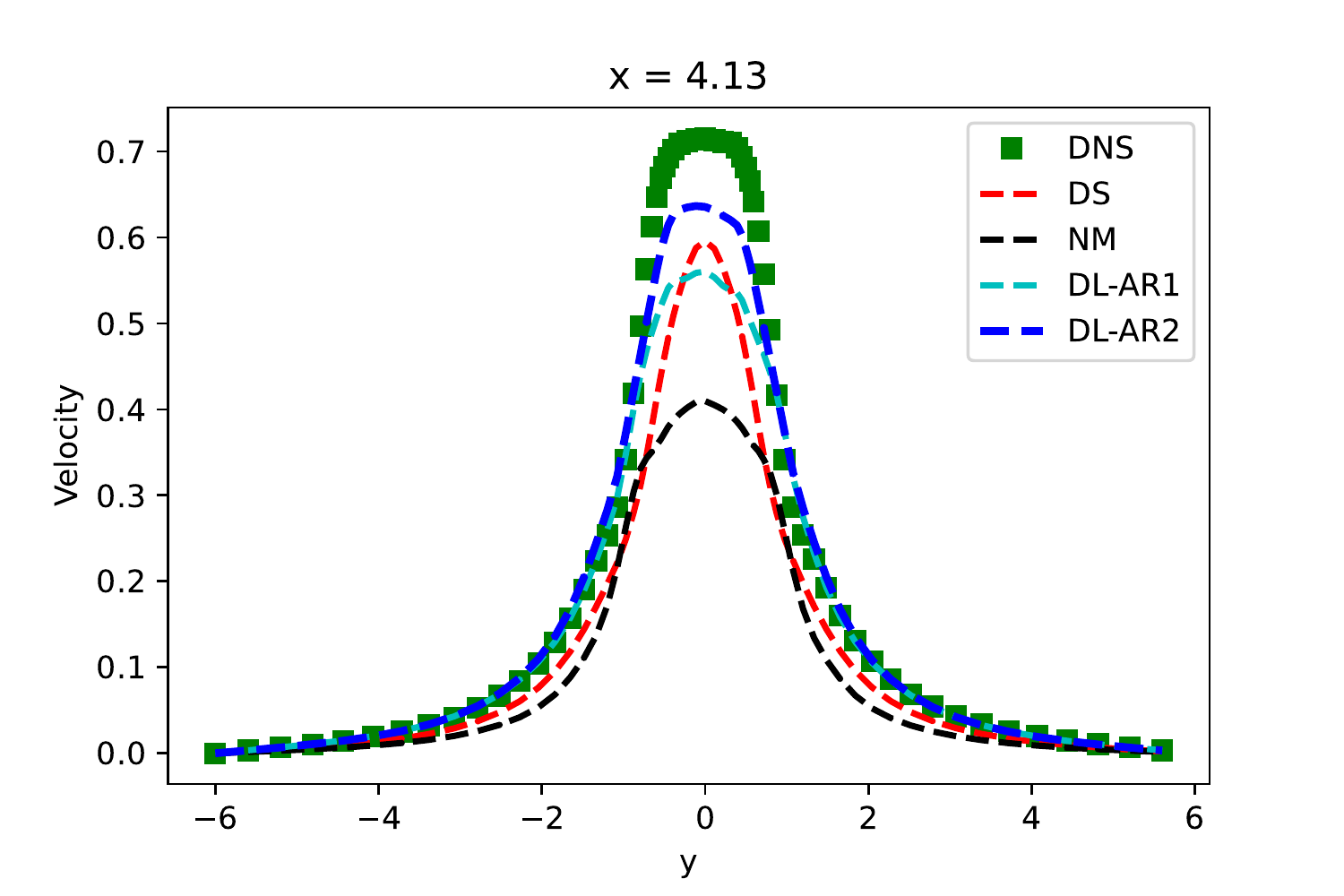}
\includegraphics[width=5cm]{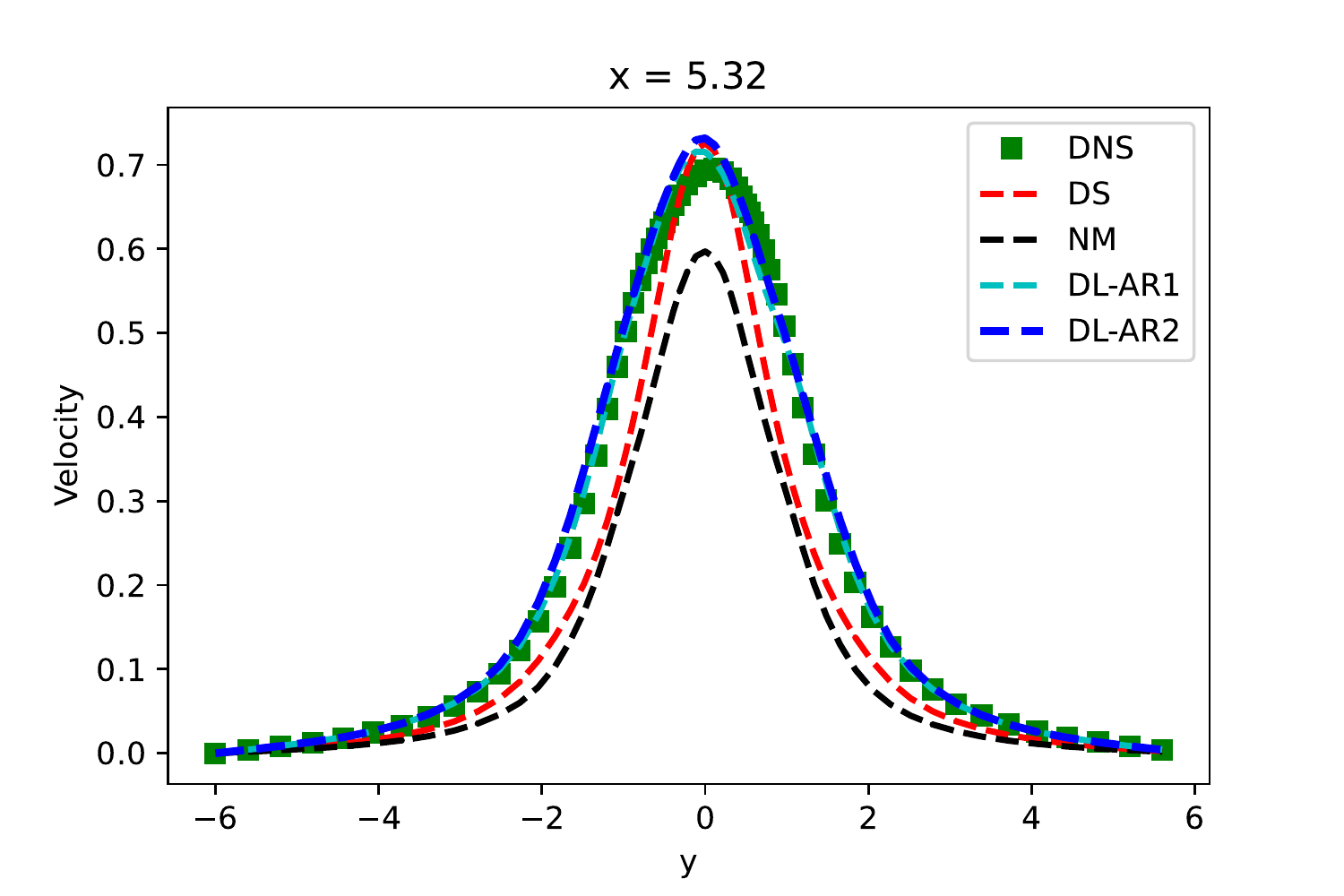}
\includegraphics[width=5cm]{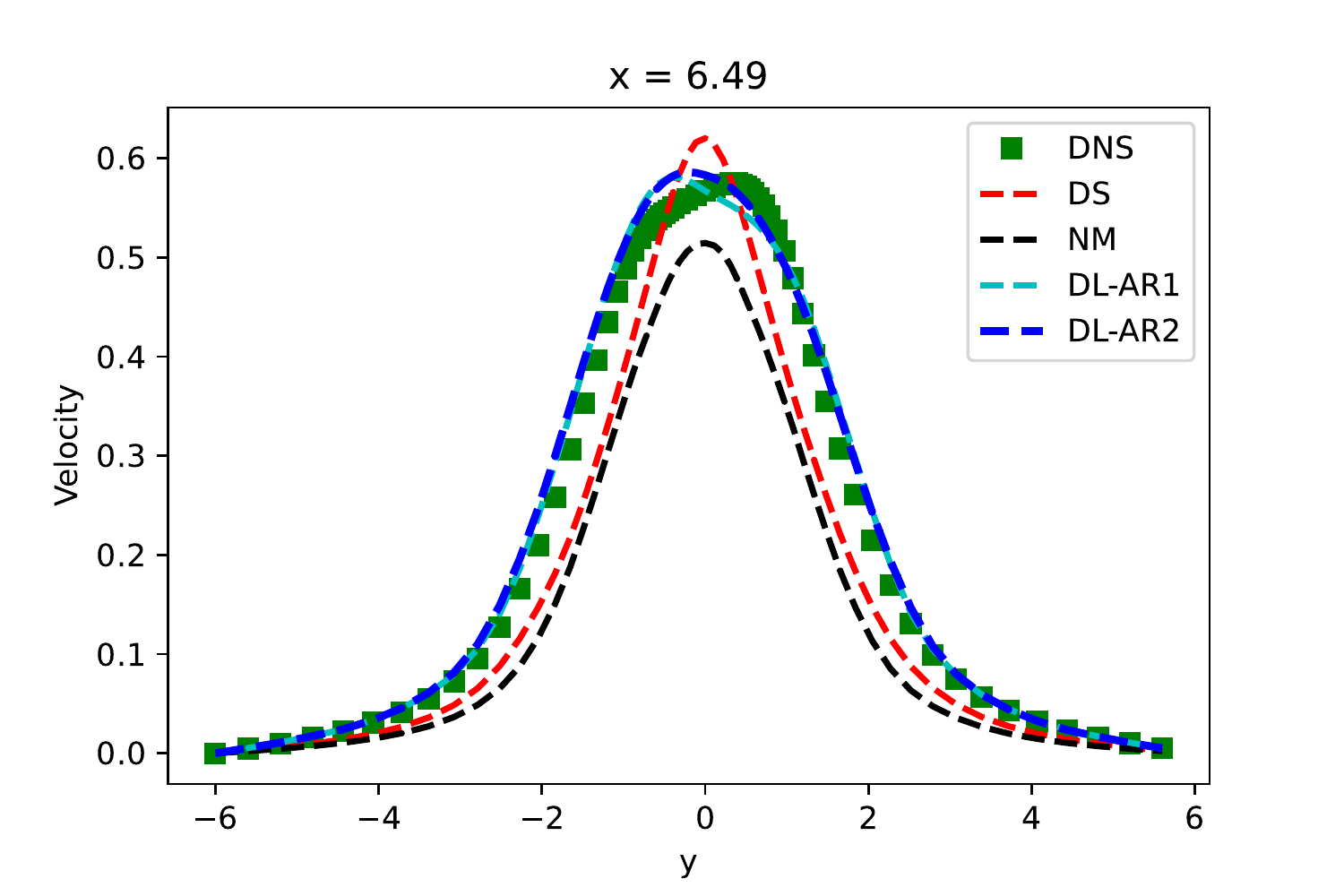}
\includegraphics[width=5cm]{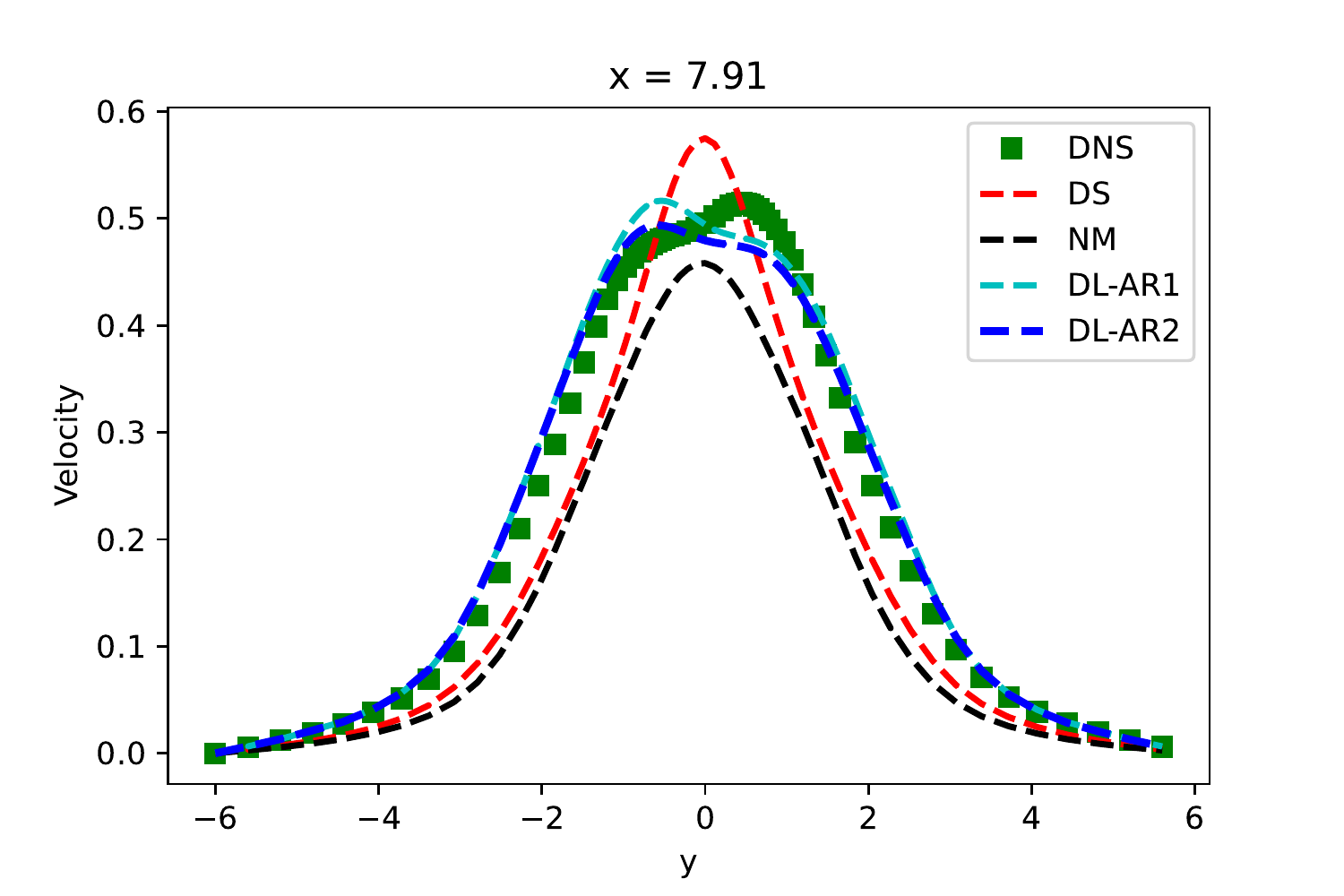}
\includegraphics[width=5cm]{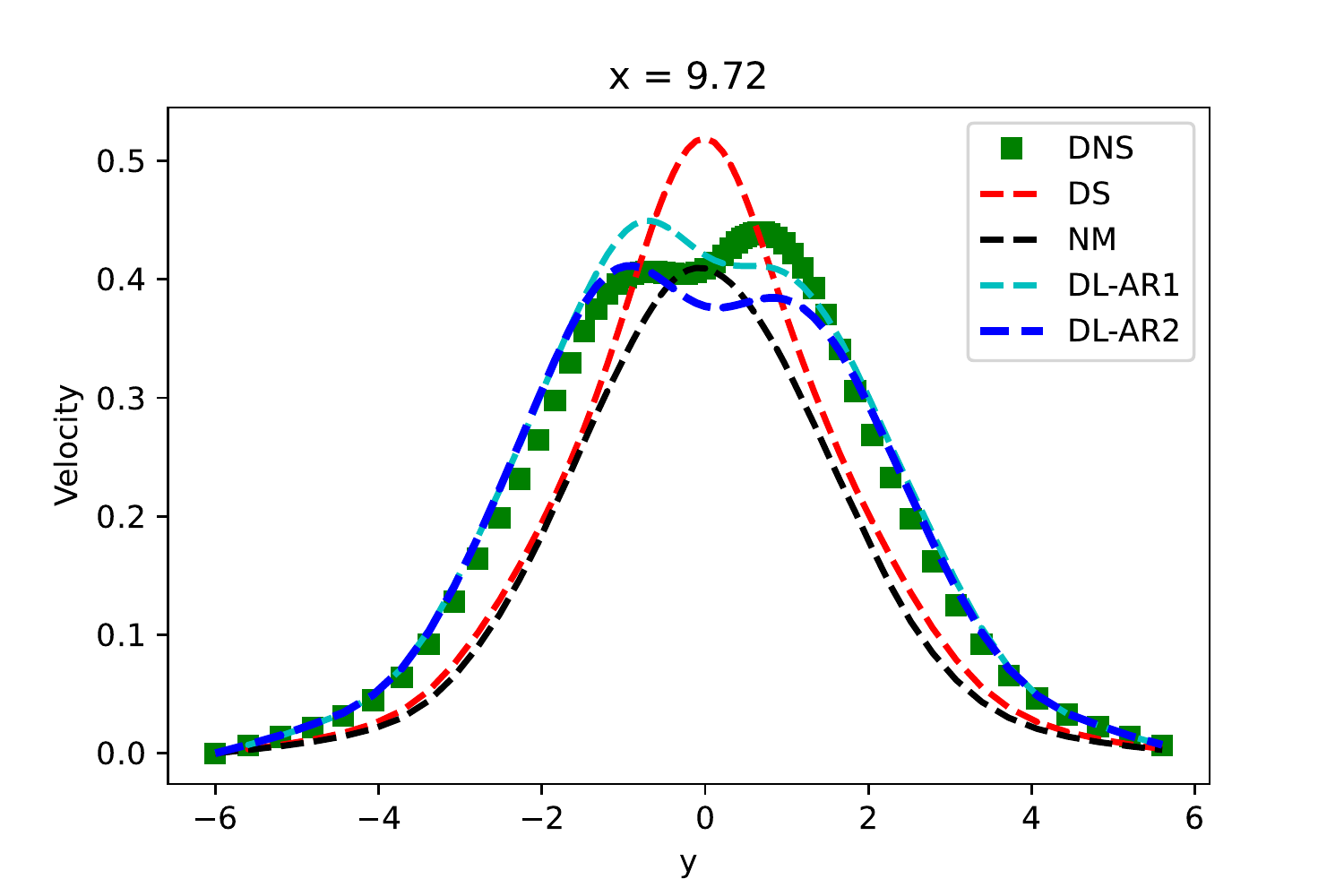}
\includegraphics[width=5cm]{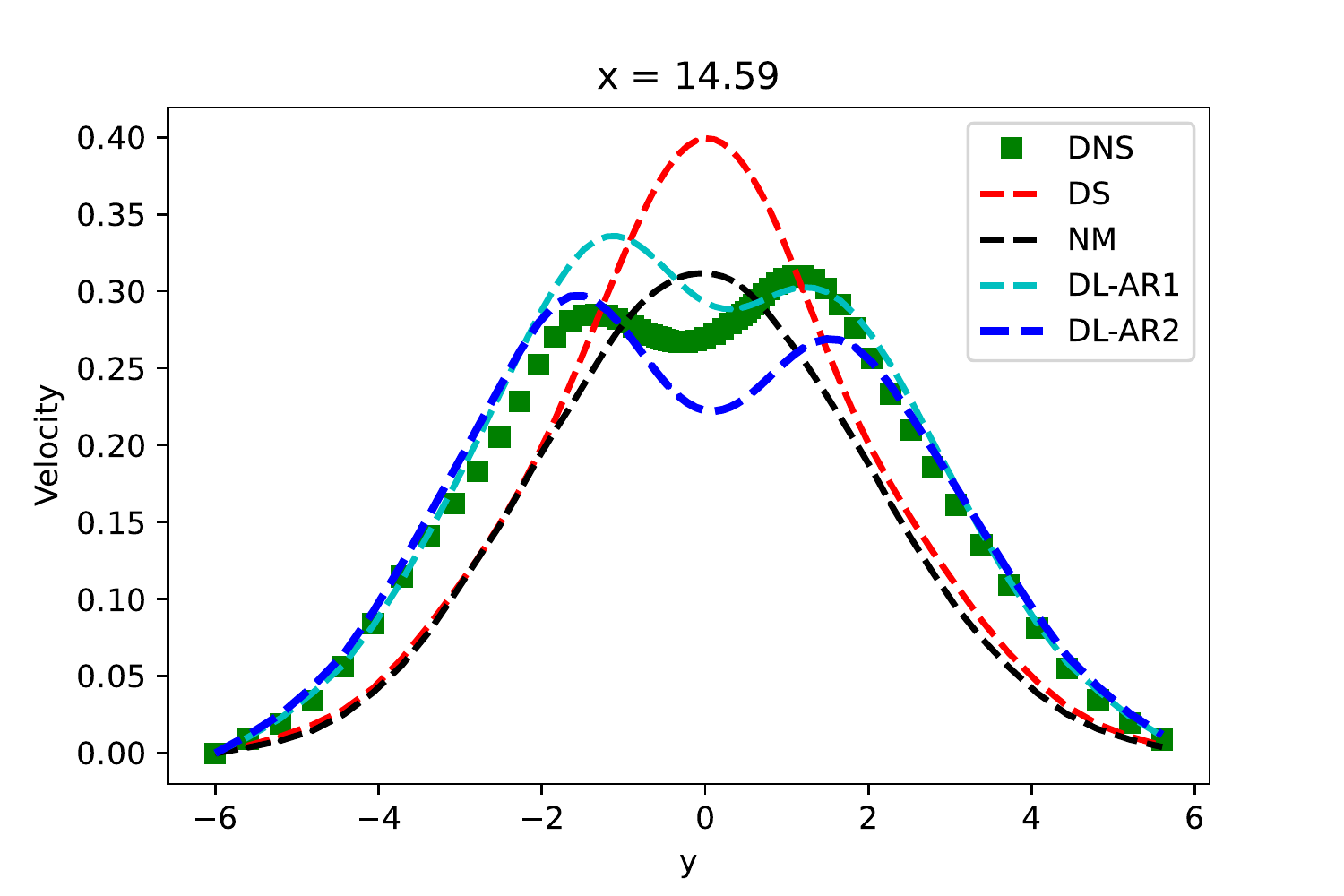}
\label{f1}
\caption{RMS profile for $u_2$ for AR1 configuration.}
\end{figure}

\begin{figure}[H]
\centering
\includegraphics[width=5cm]{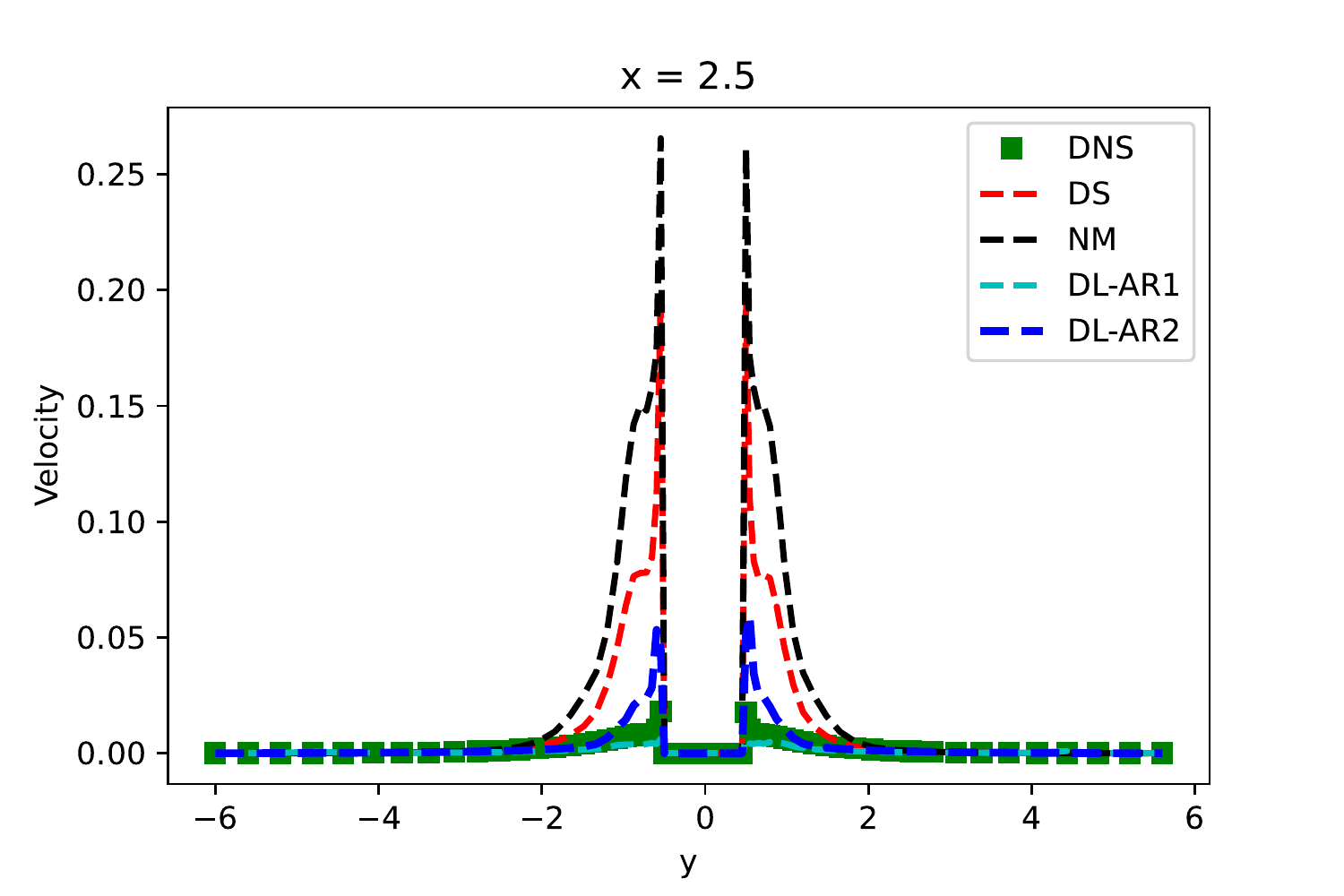}
\includegraphics[width=5cm]{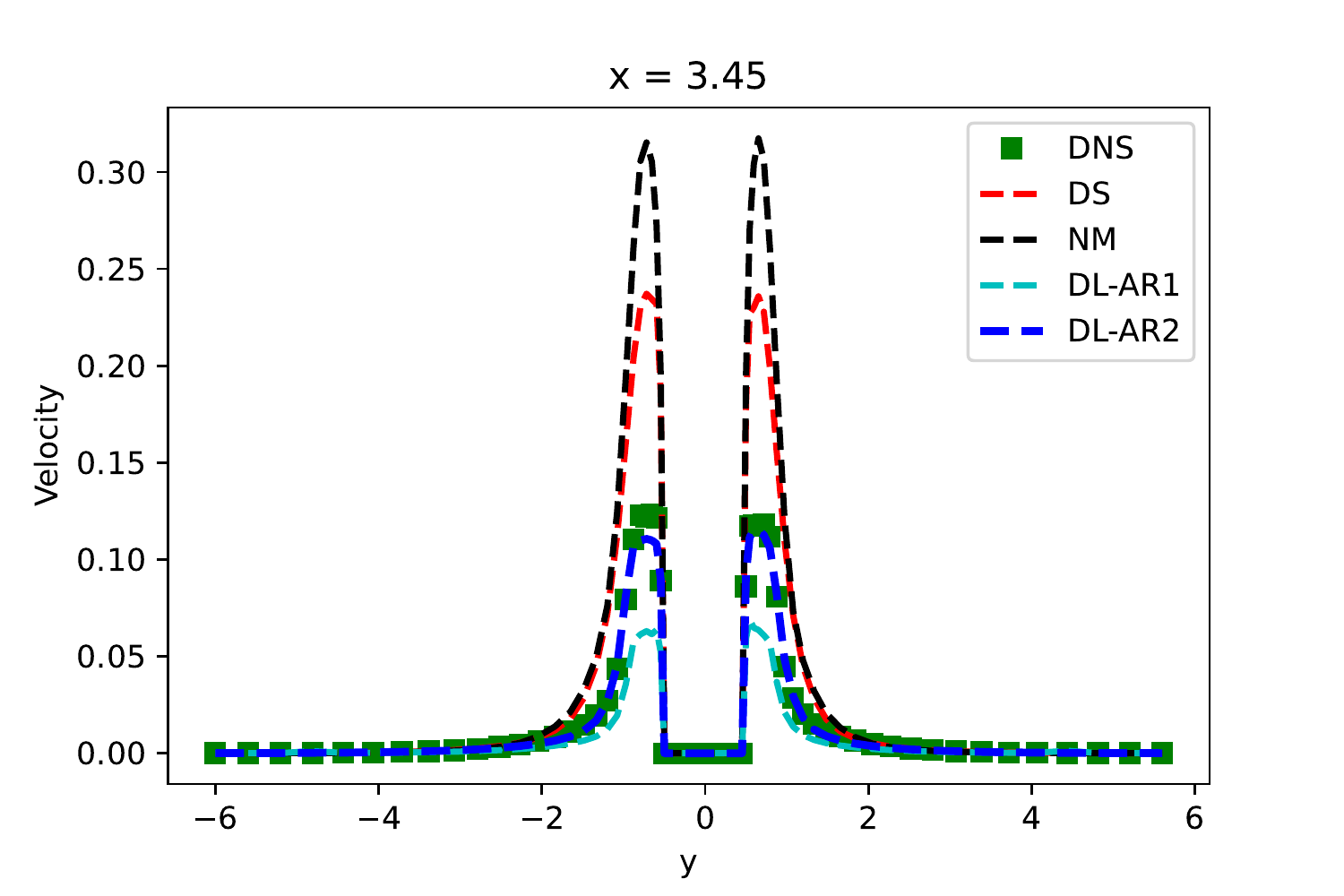}
\includegraphics[width=5cm]{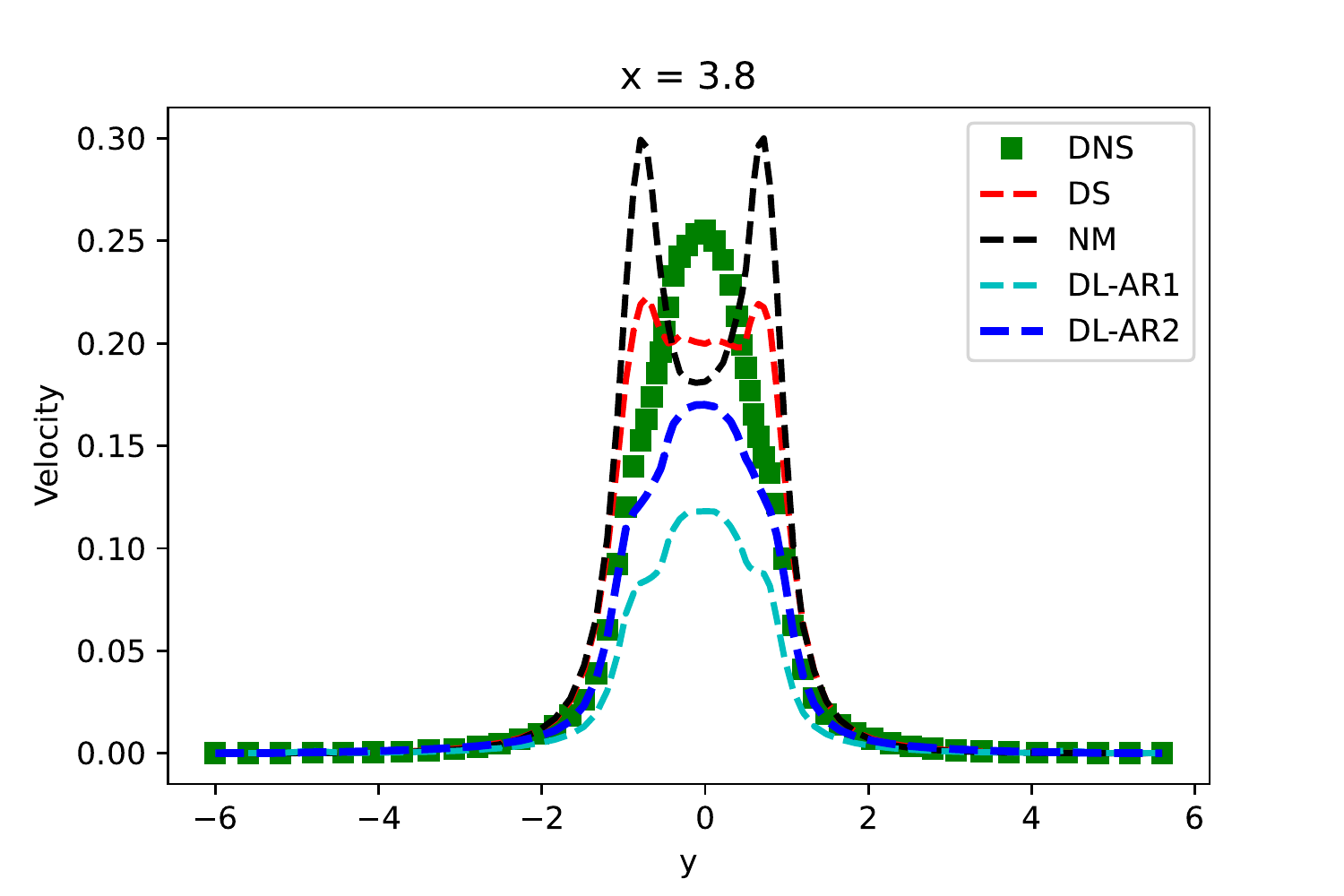}
\includegraphics[width=5cm]{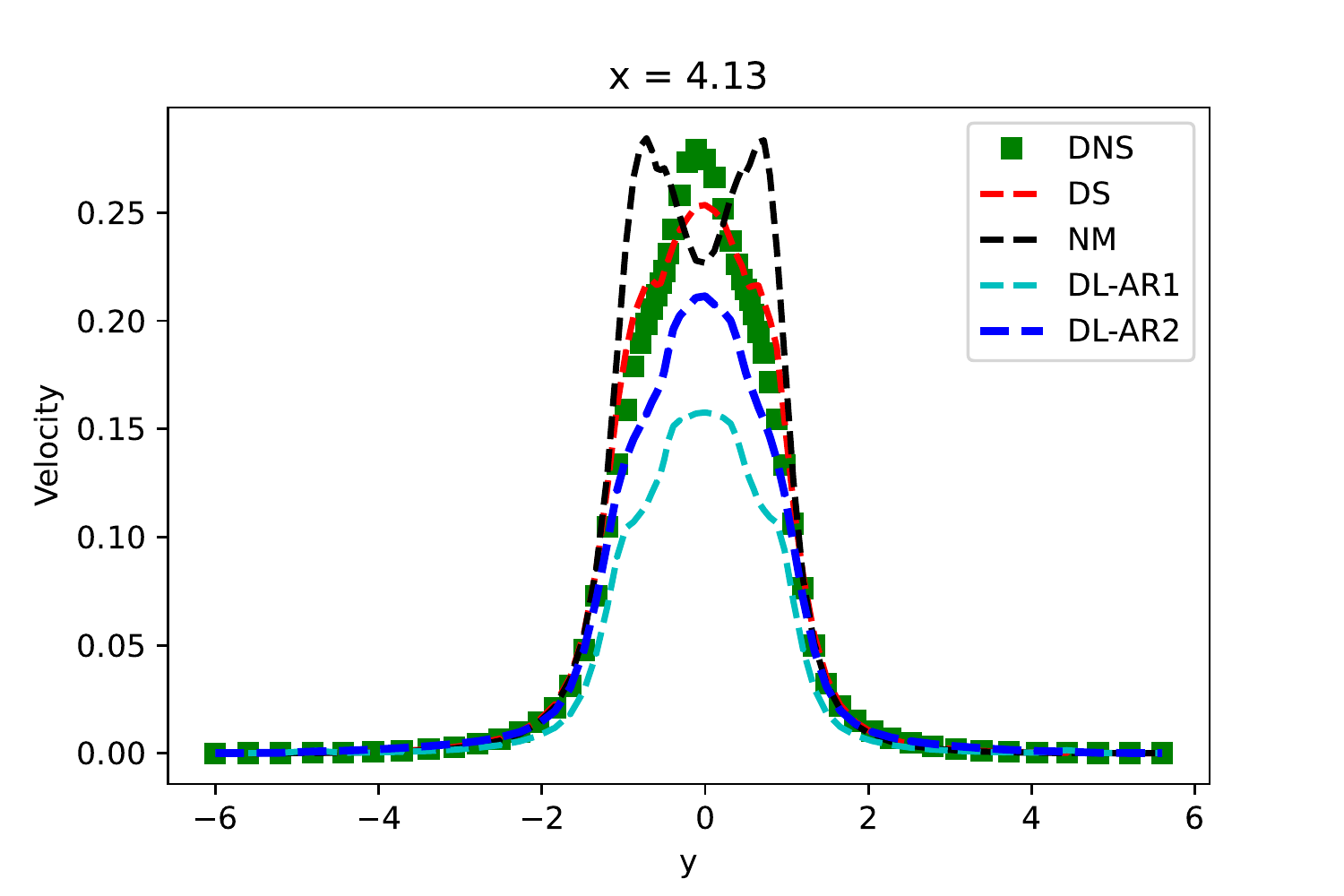}
\includegraphics[width=5cm]{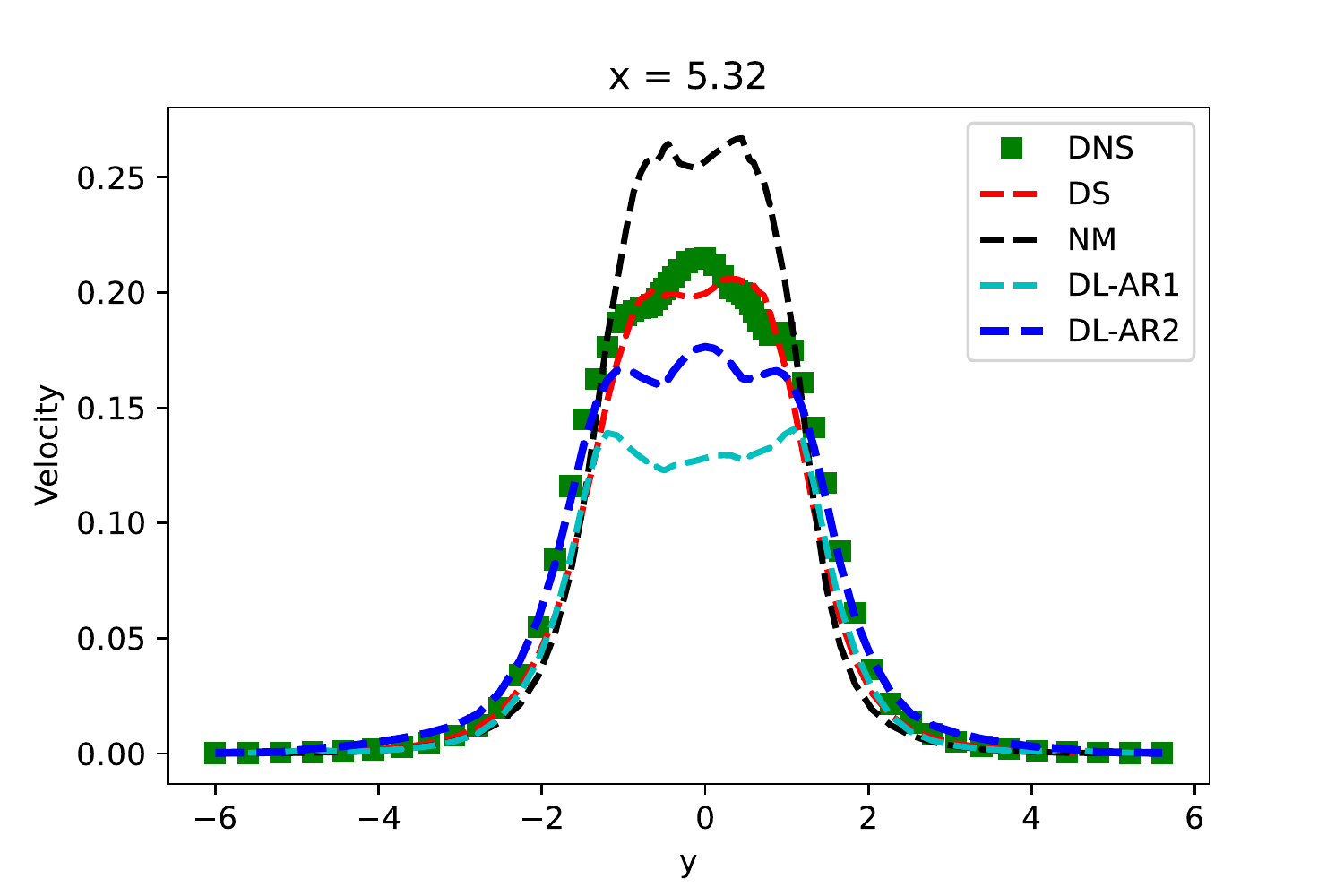}
\includegraphics[width=5cm]{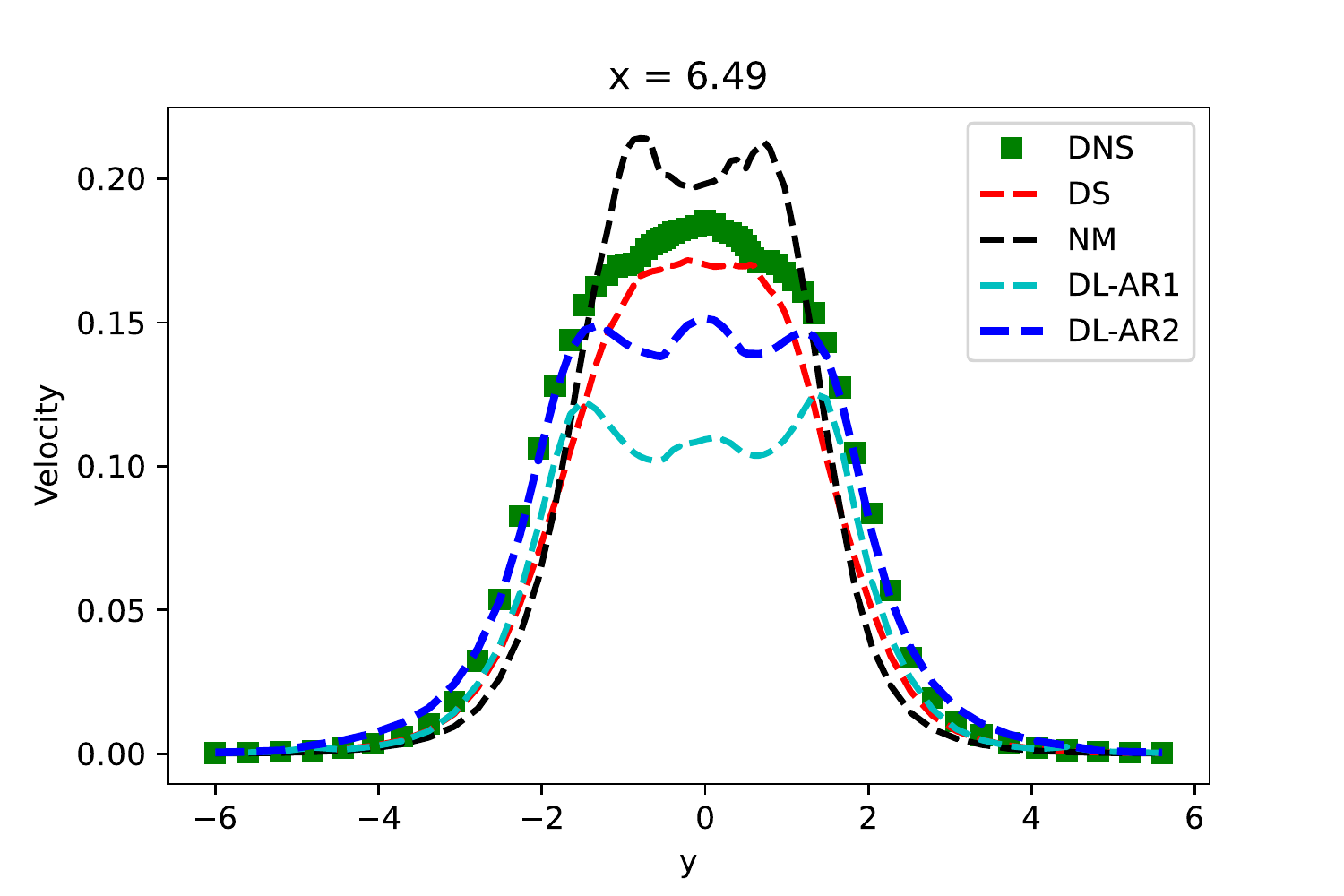}
\includegraphics[width=5cm]{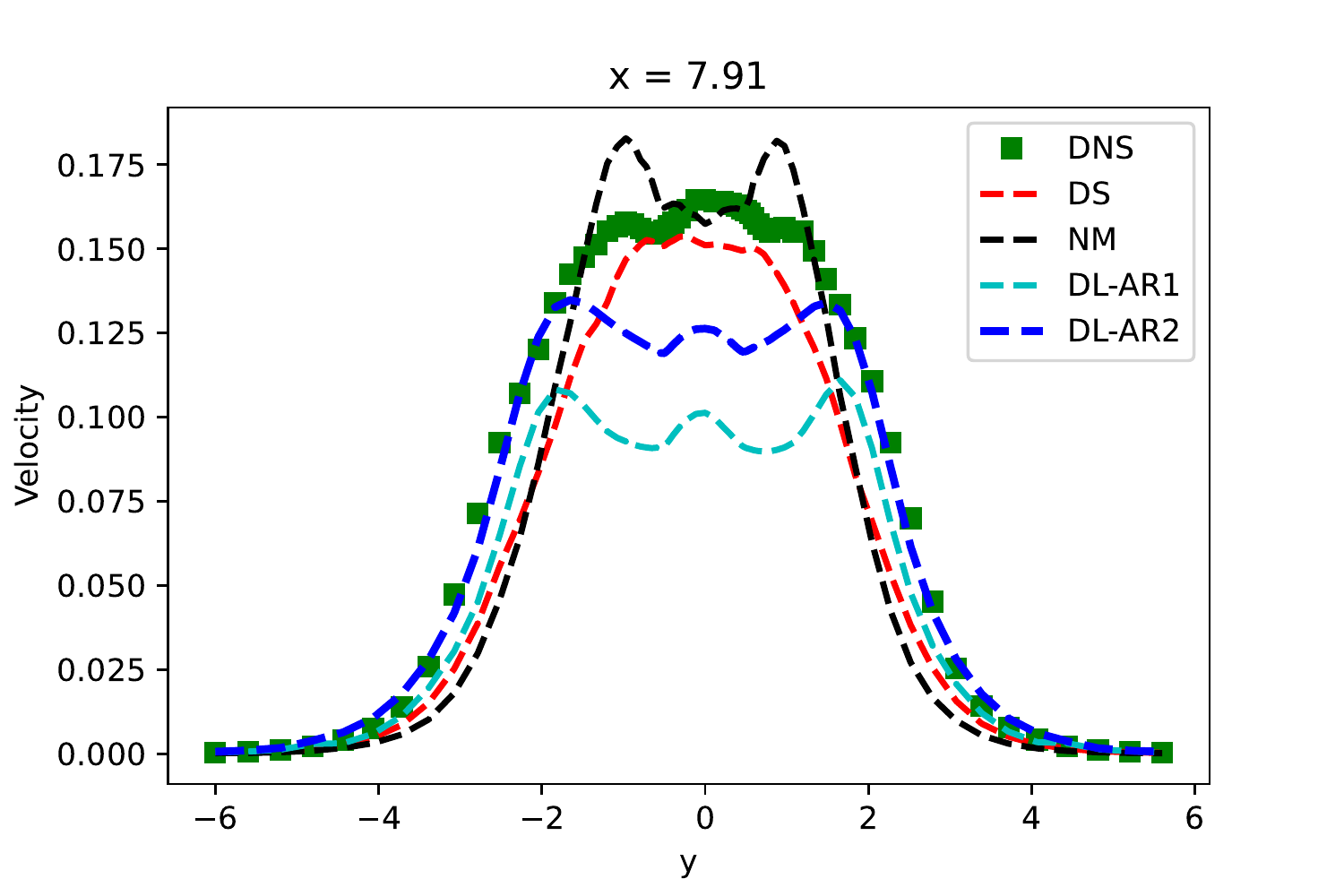}
\includegraphics[width=5cm]{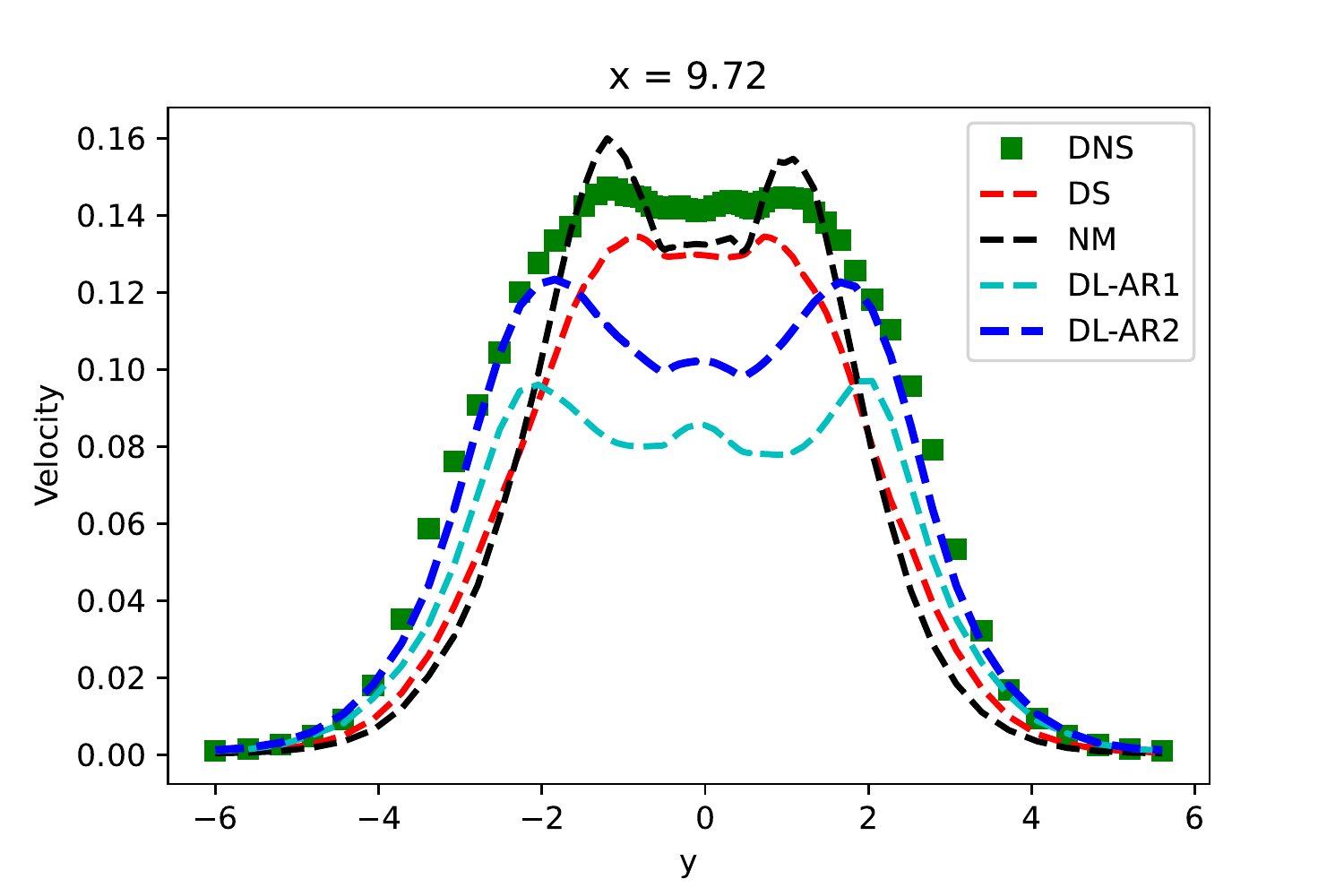}
\includegraphics[width=5cm]{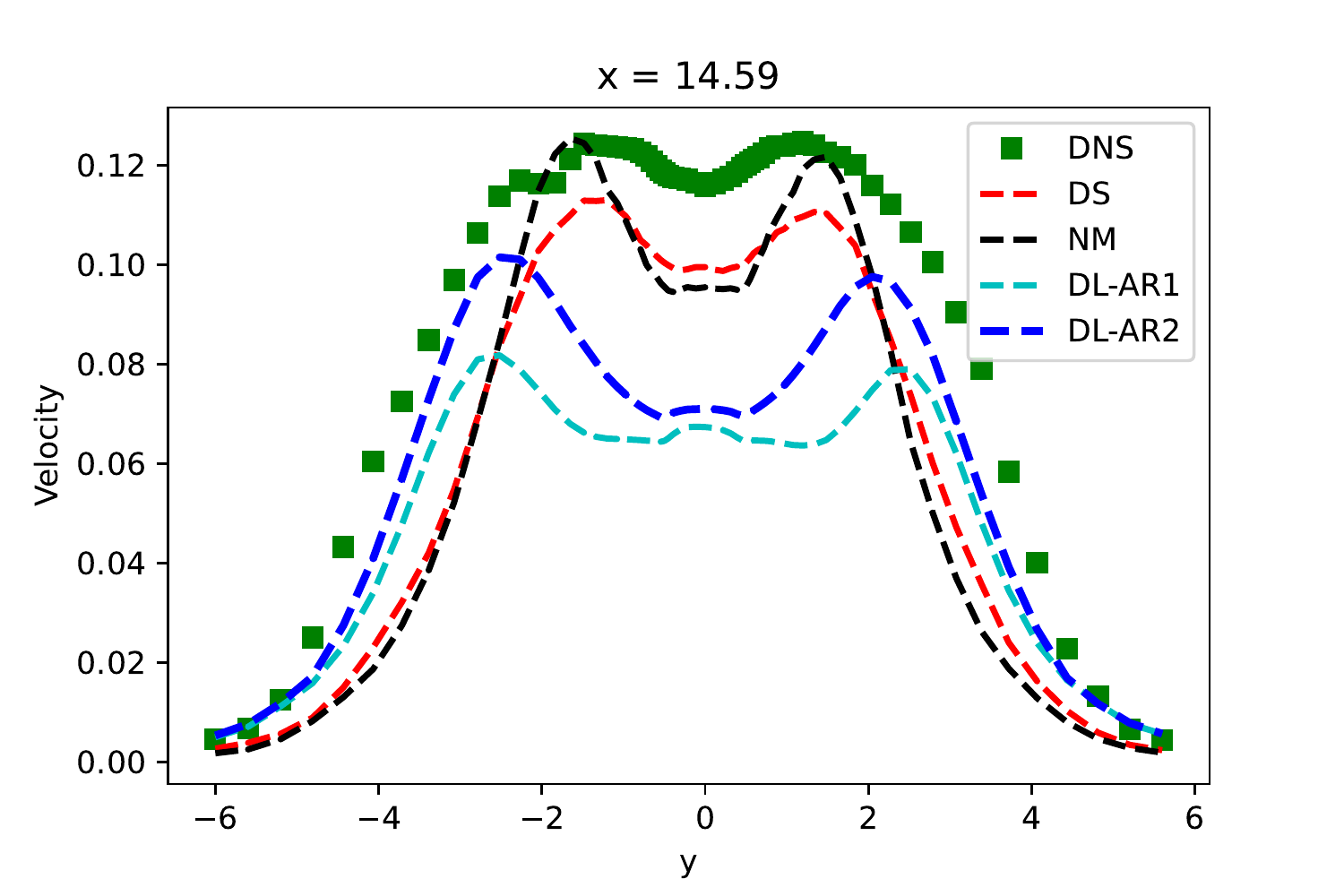}
\label{f1}
\caption{RMS profile for $u_3$ for AR1 configuration.}
\end{figure}

\begin{figure}[H]
\centering
\includegraphics[width=5cm]{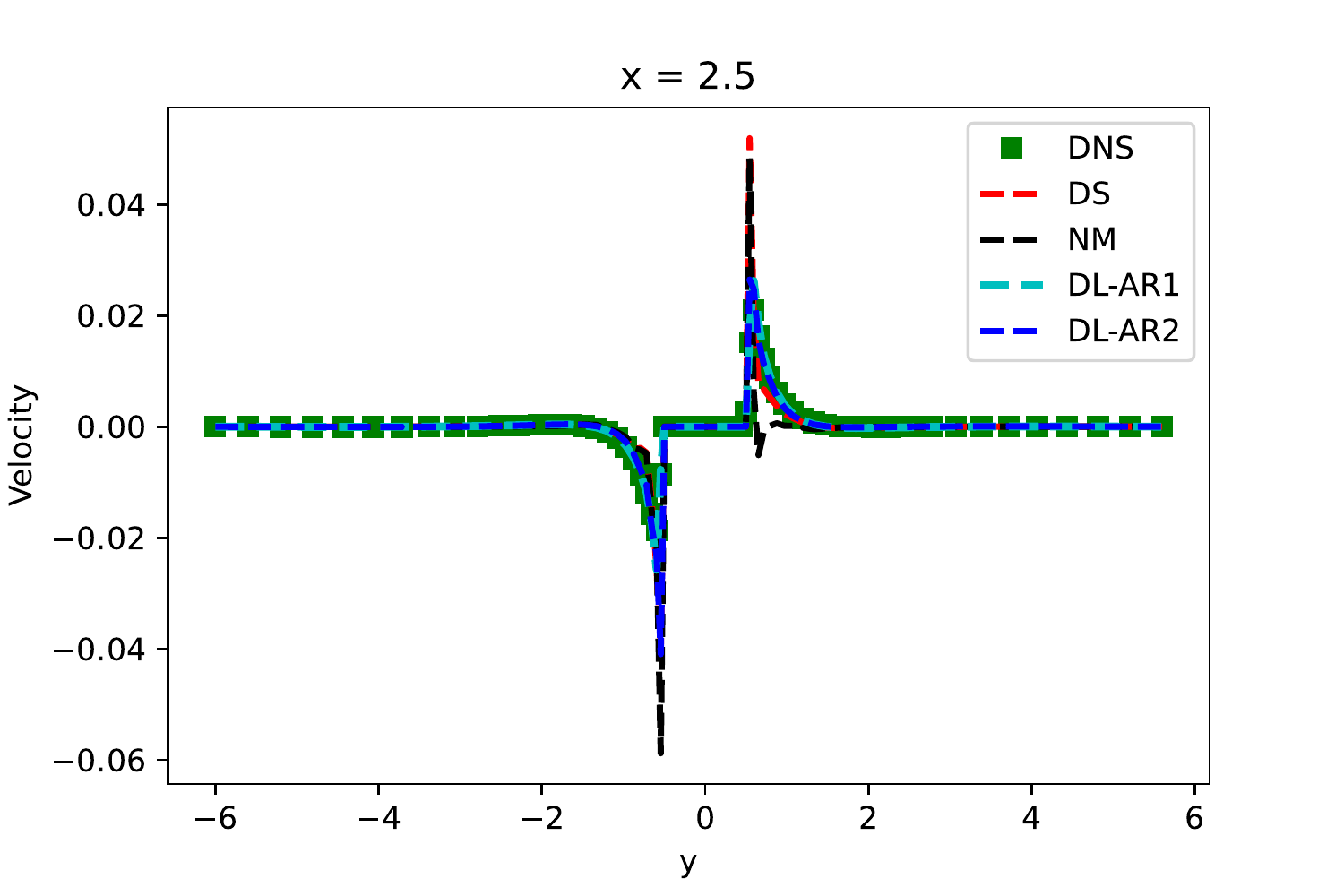}
\includegraphics[width=5cm]{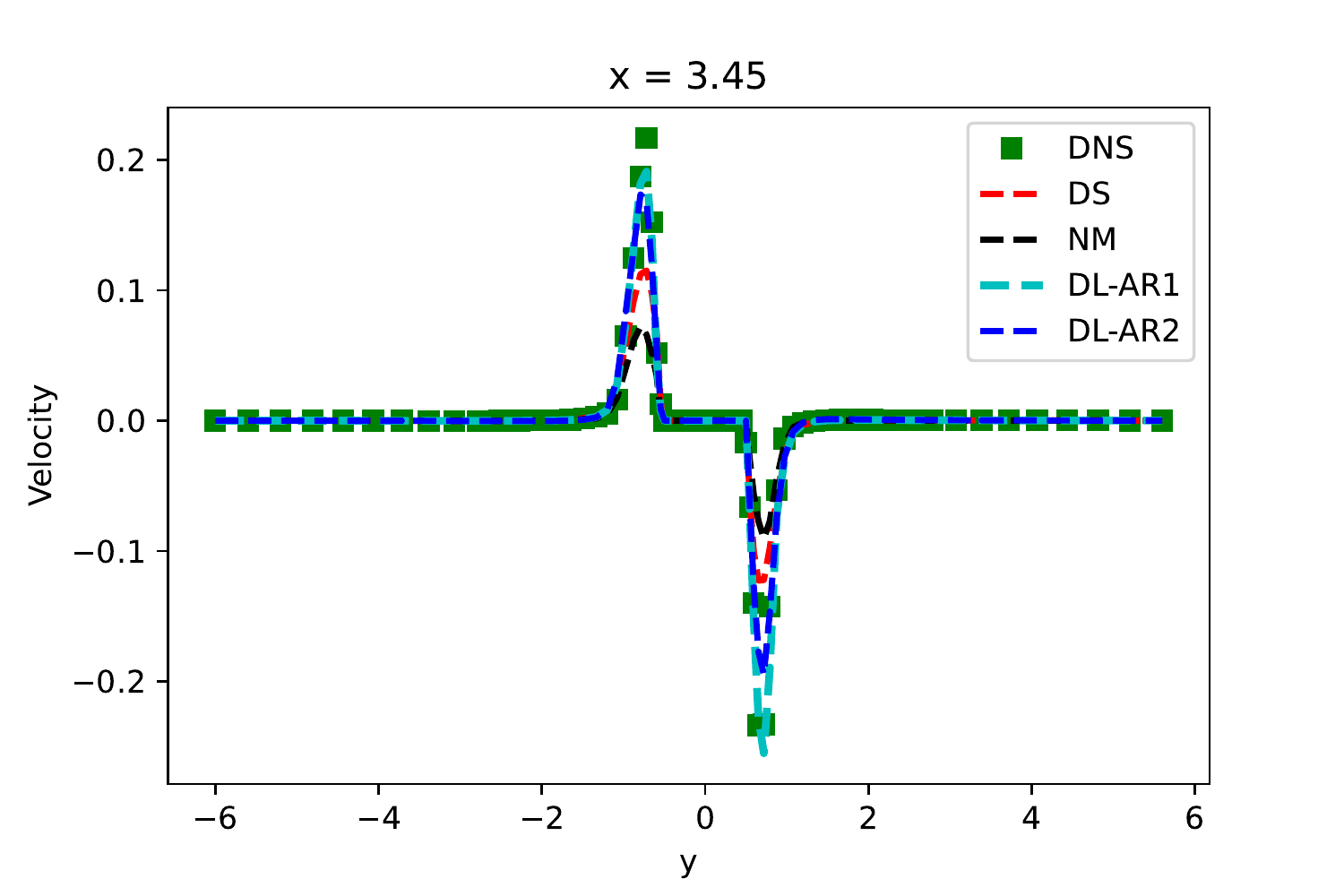}
\includegraphics[width=5cm]{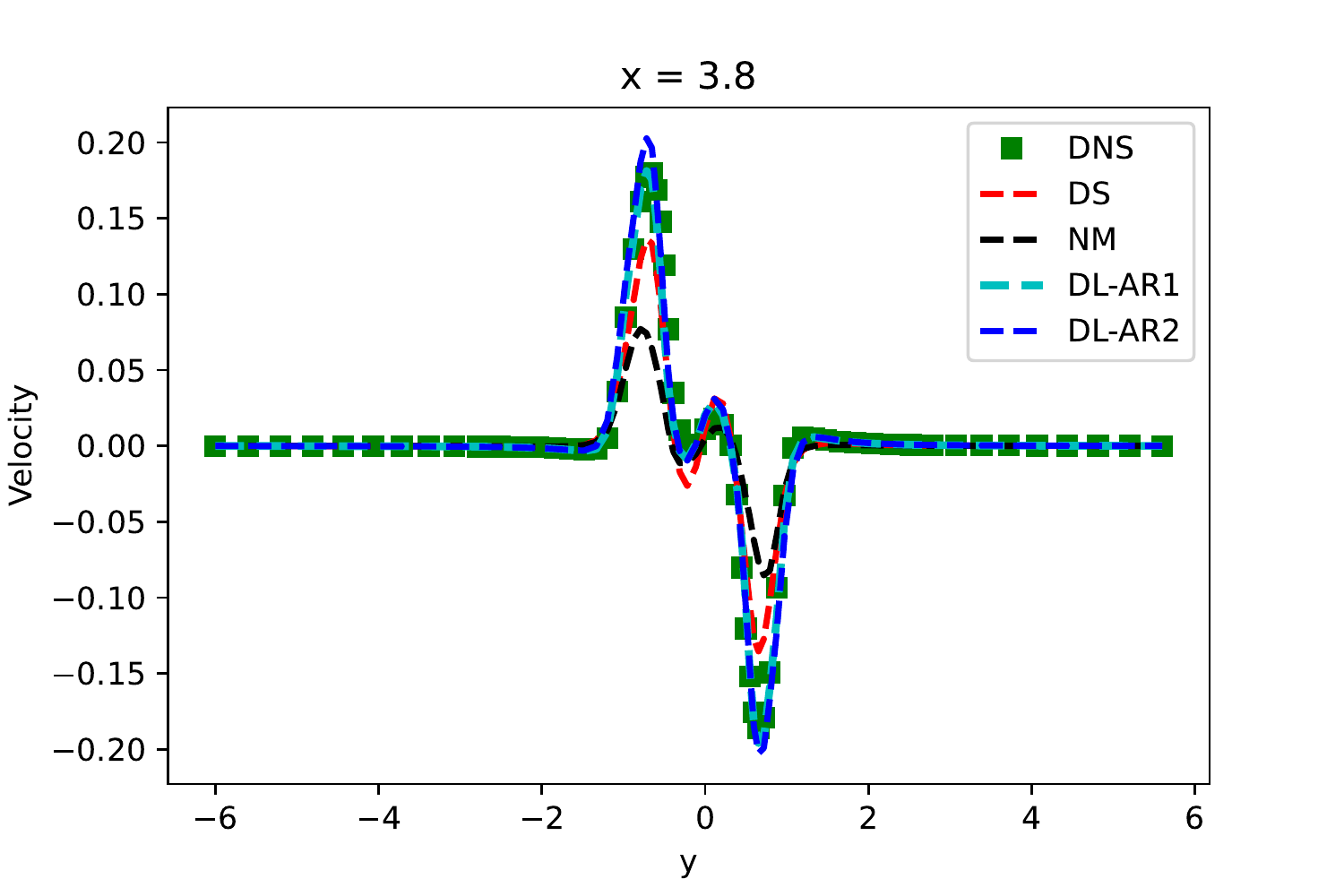}
\includegraphics[width=5cm]{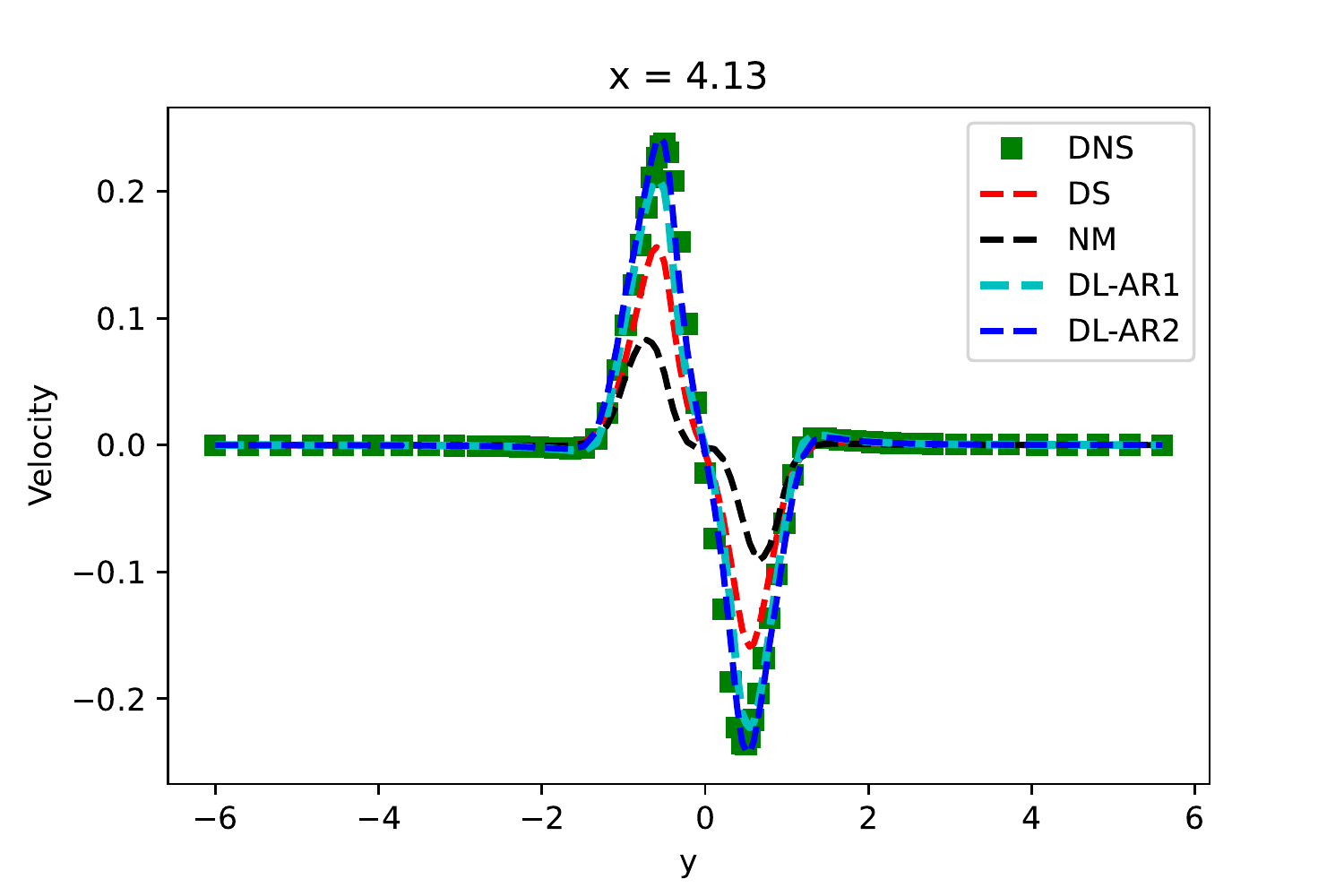}
\includegraphics[width=5cm]{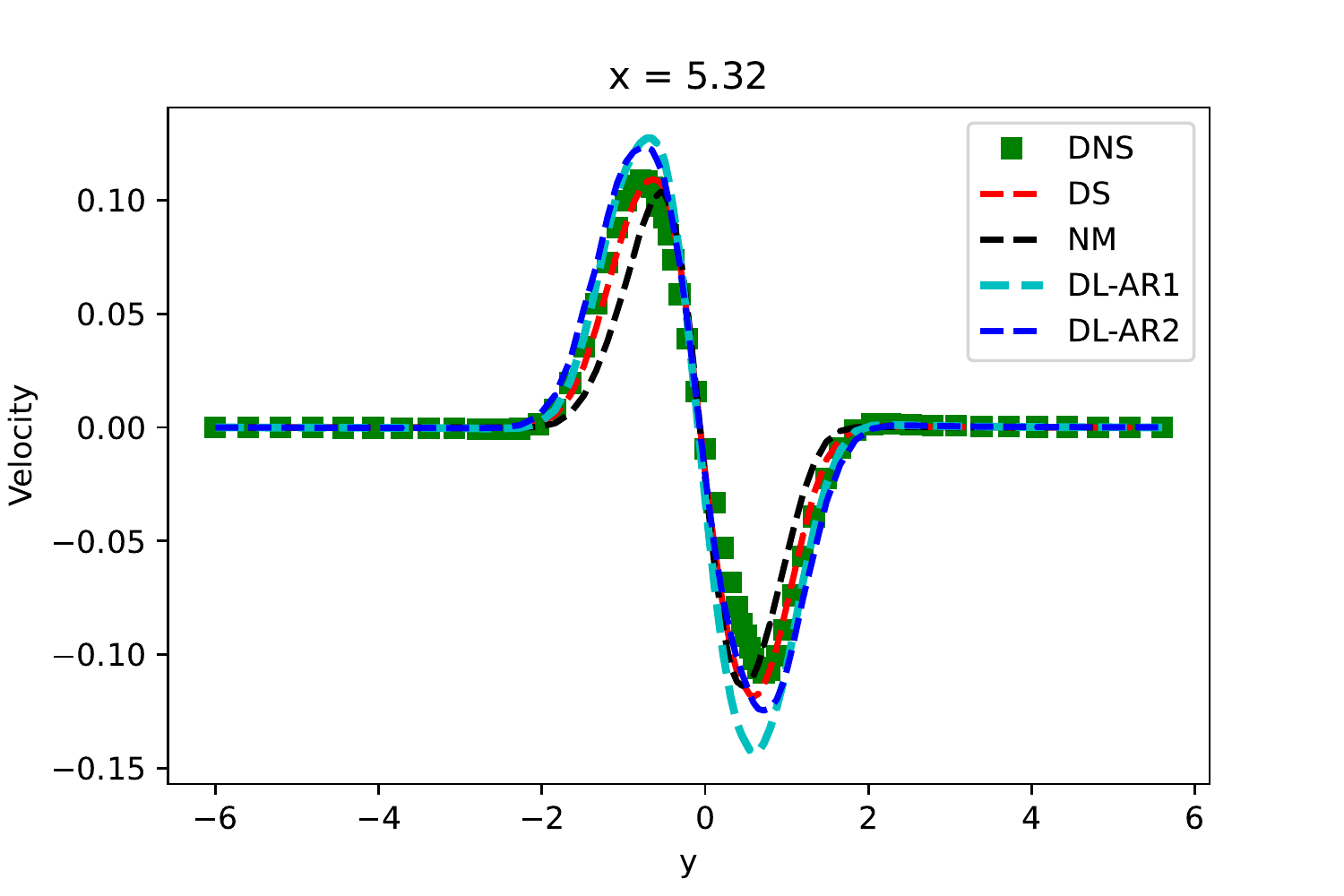}
\includegraphics[width=5cm]{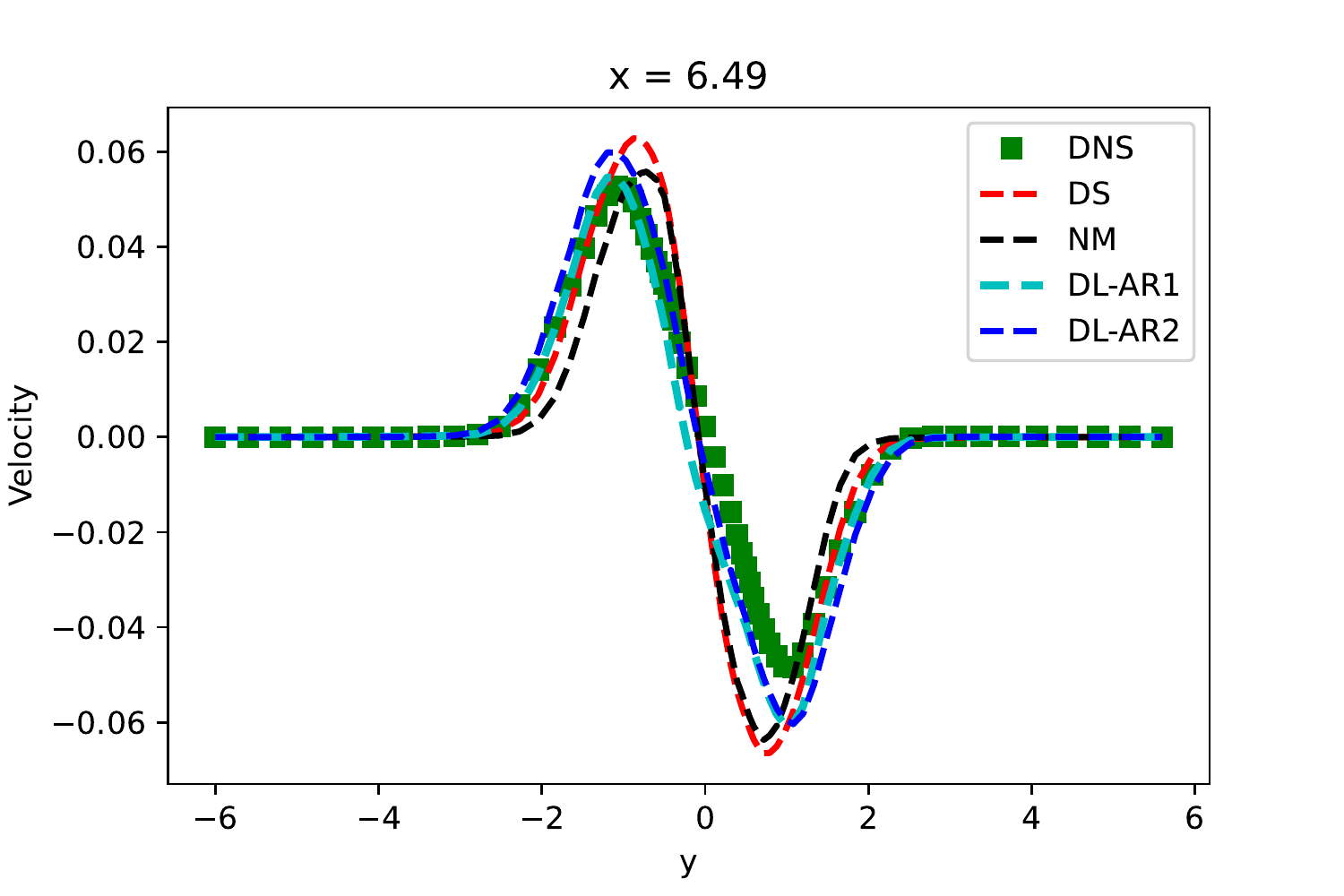}
\includegraphics[width=5cm]{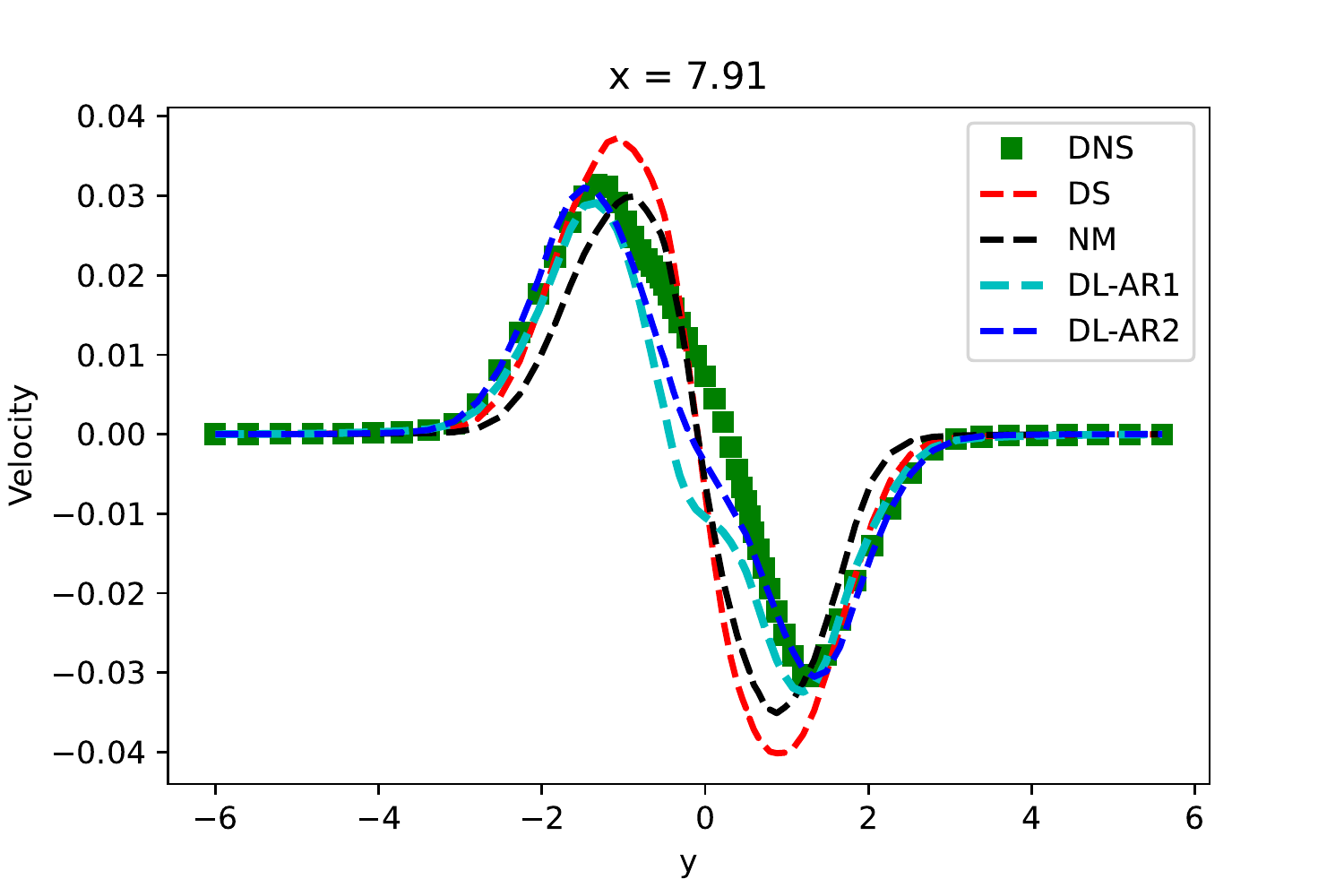}
\includegraphics[width=5cm]{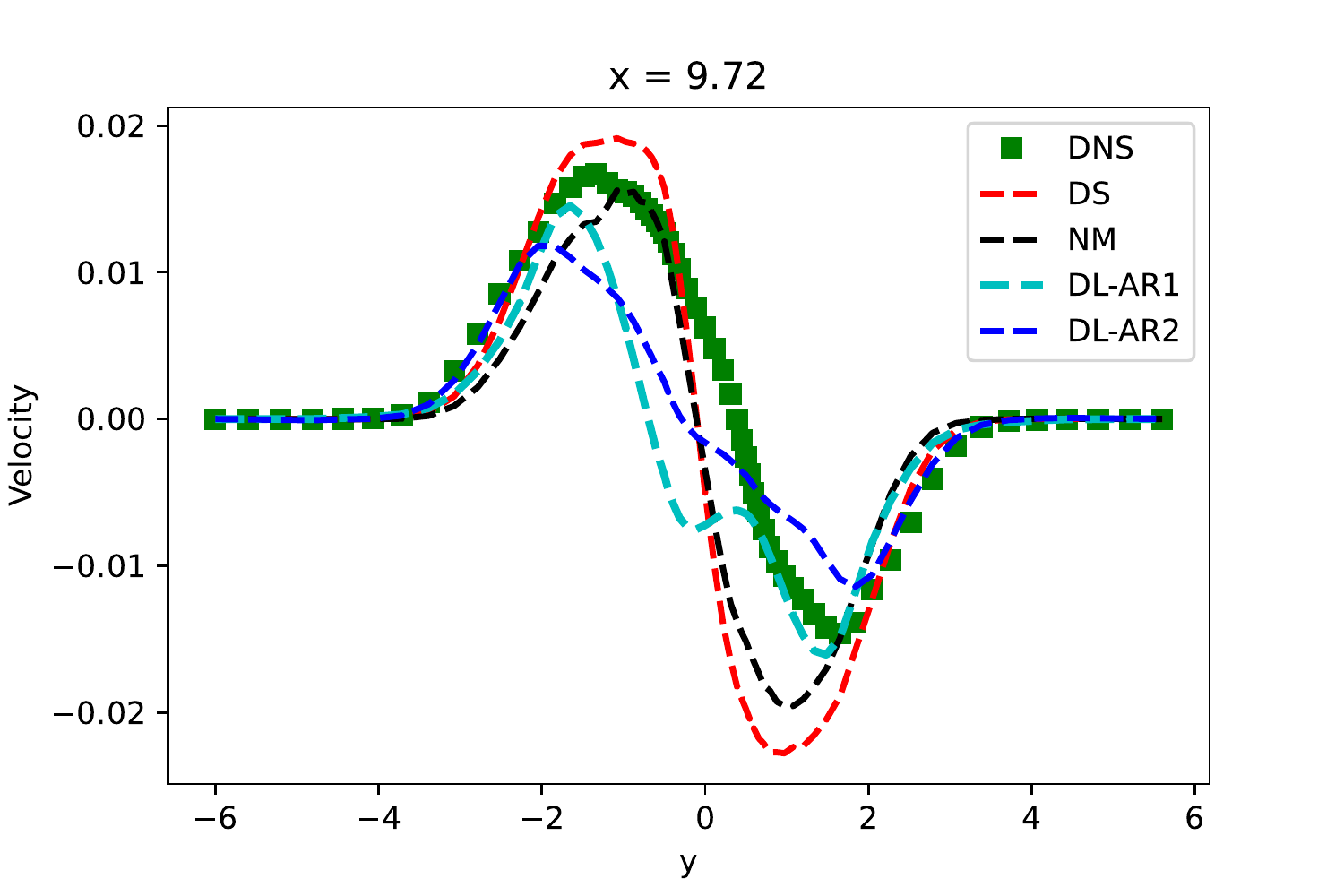}
\includegraphics[width=5cm]{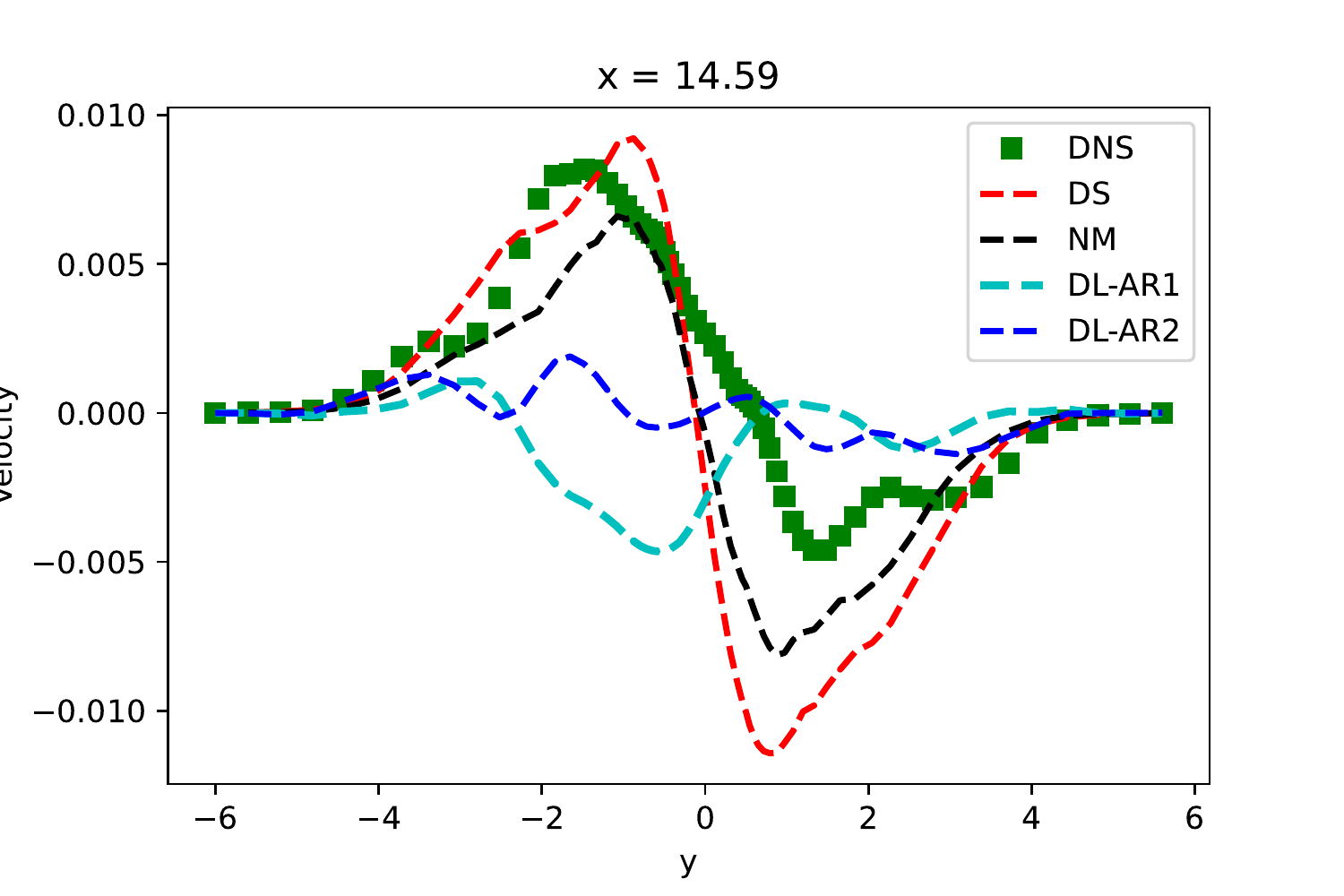}
\label{f1}
\caption{$\tau_{12}$ for AR1 configuration.}
\end{figure}

\subsubsection{AR2 $+$ Re$1,000$}

\begin{figure}[H]
\centering
\includegraphics[width=5cm]{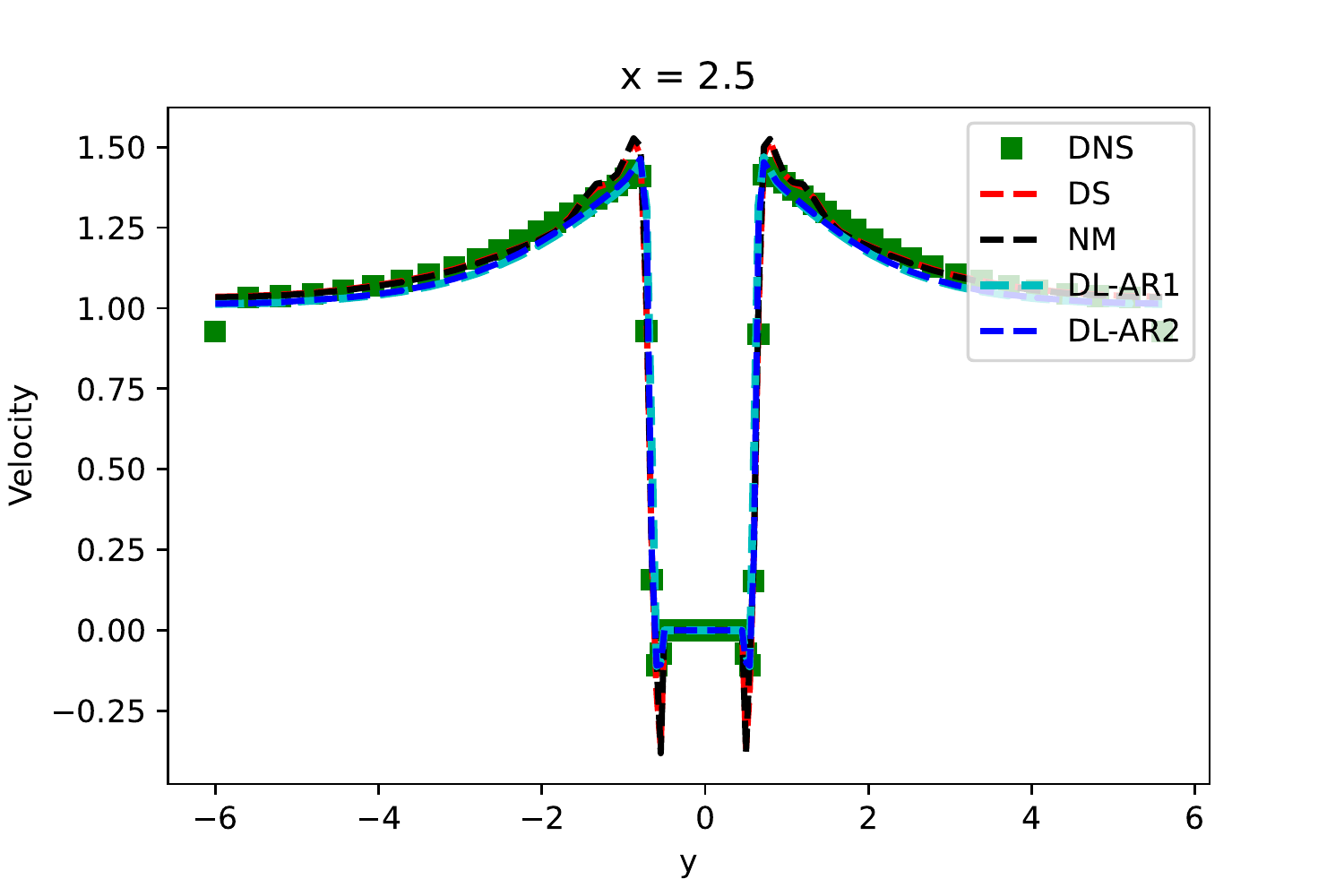}
\includegraphics[width=5cm]{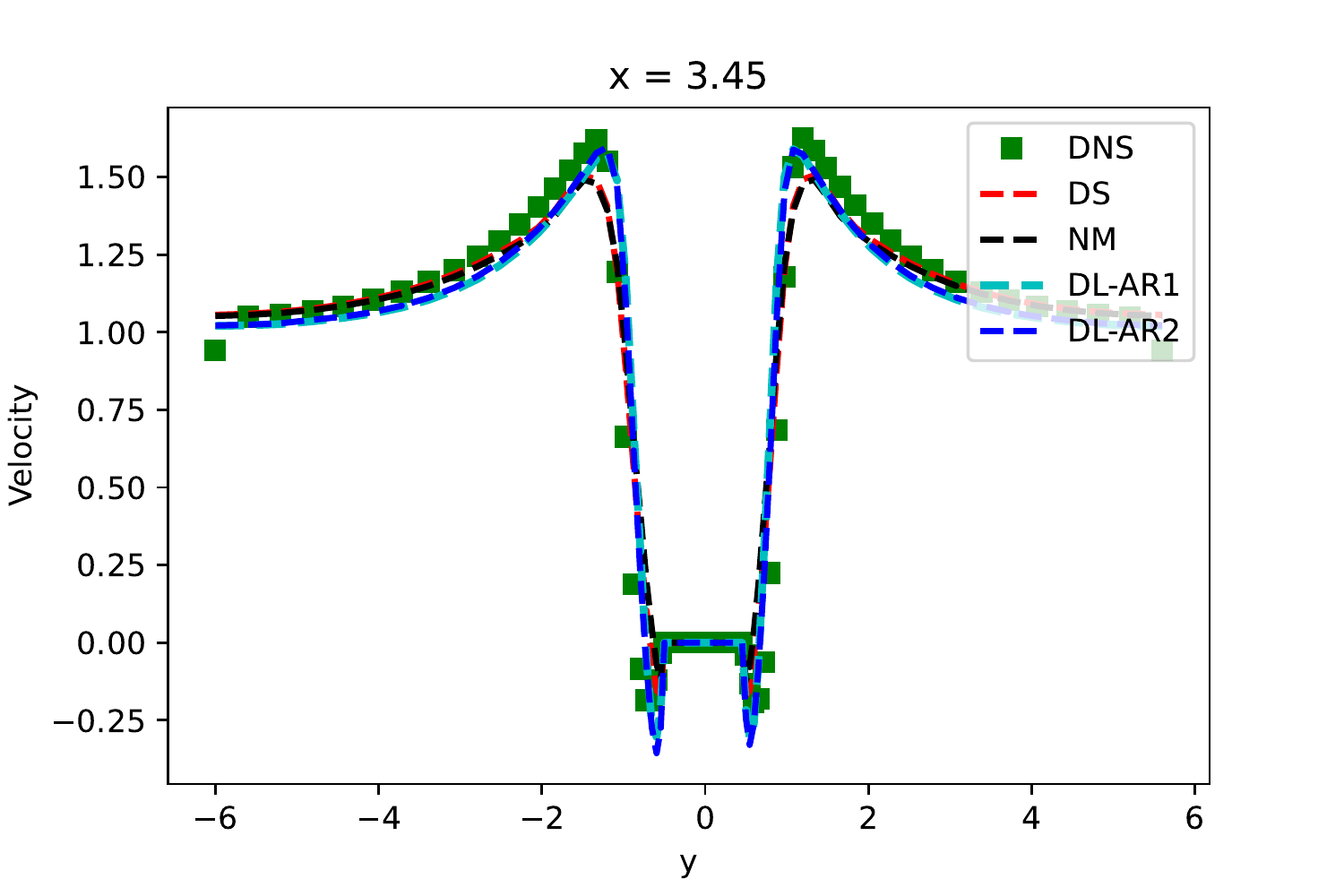}
\includegraphics[width=5cm]{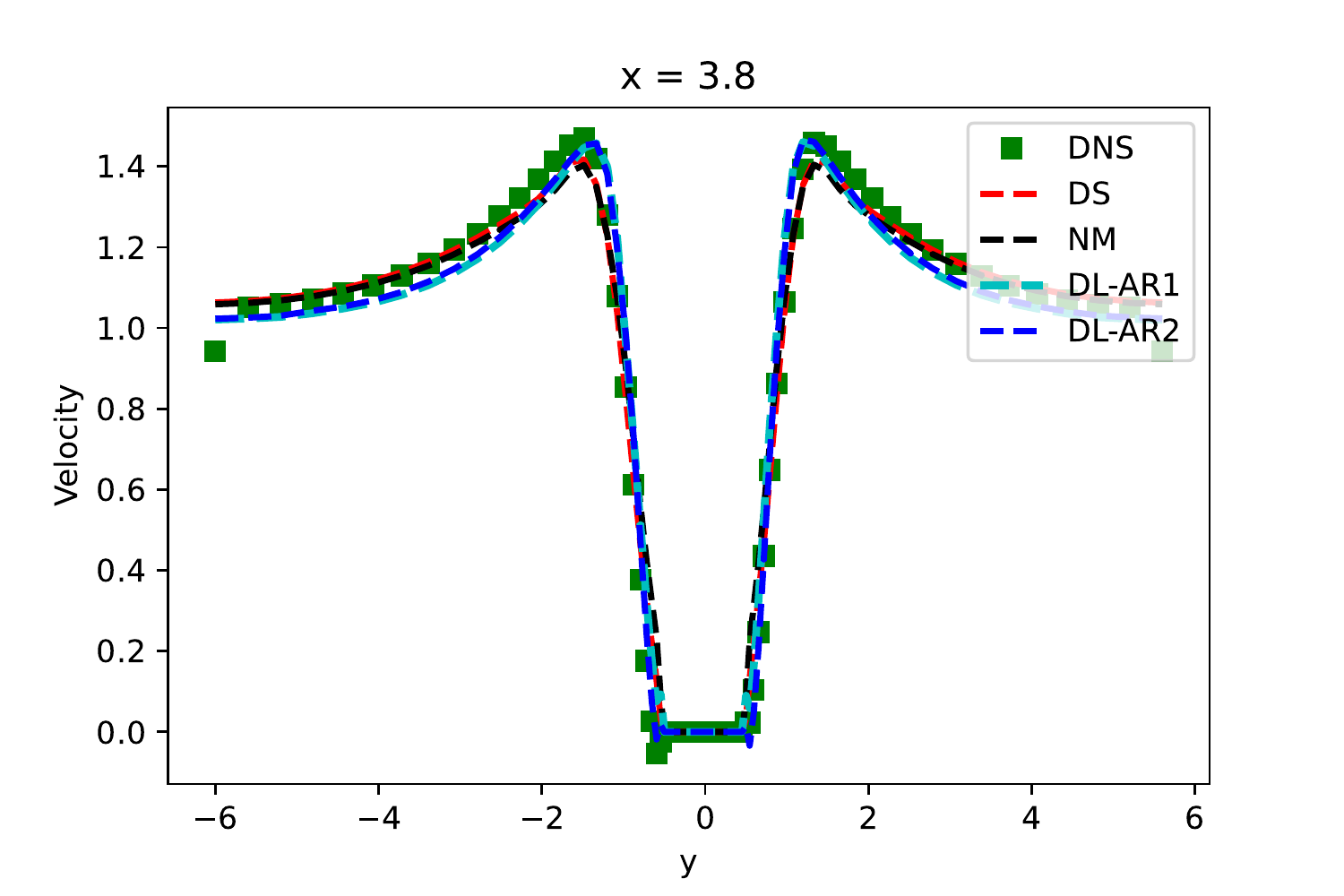}
\includegraphics[width=5cm]{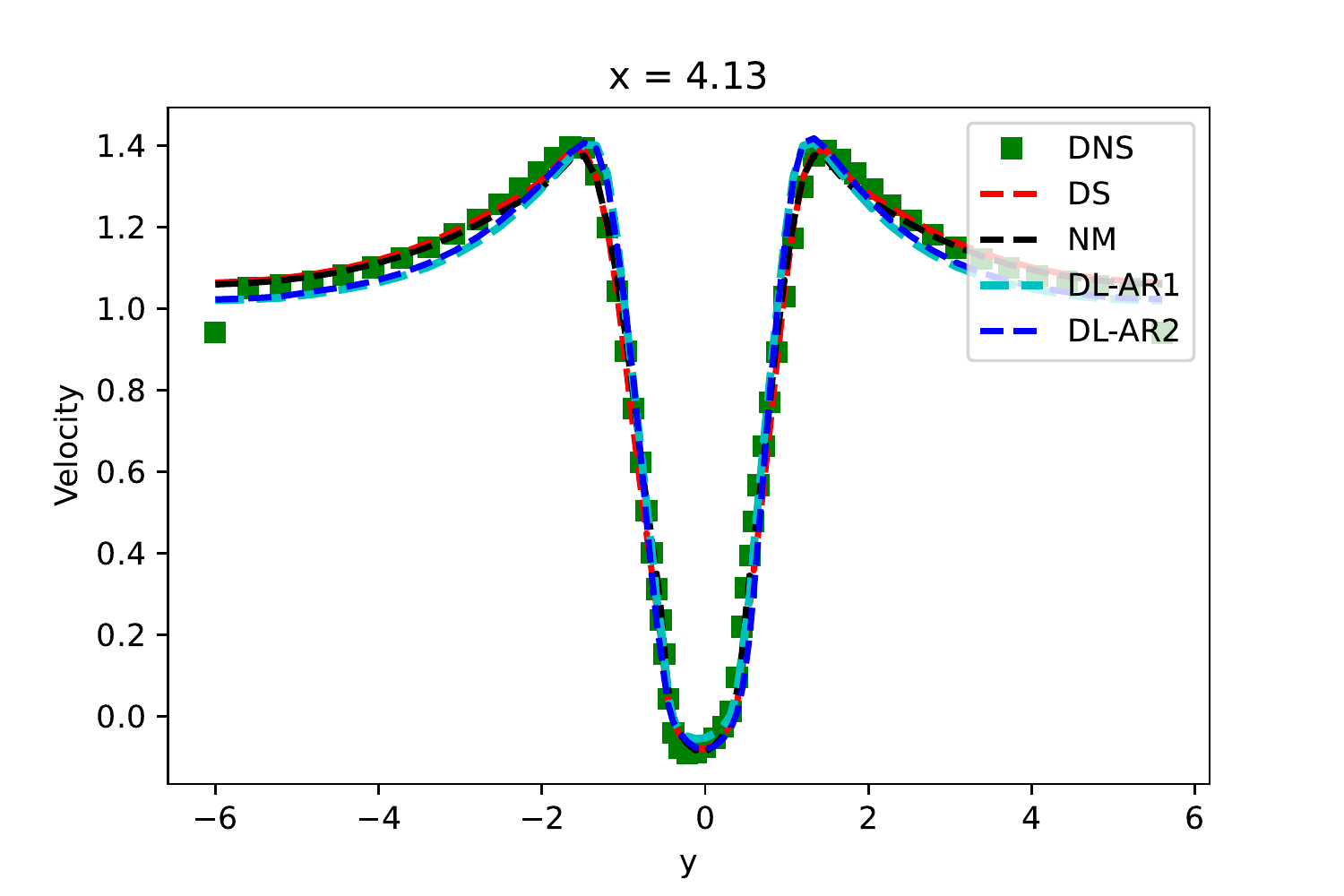}
\includegraphics[width=5cm]{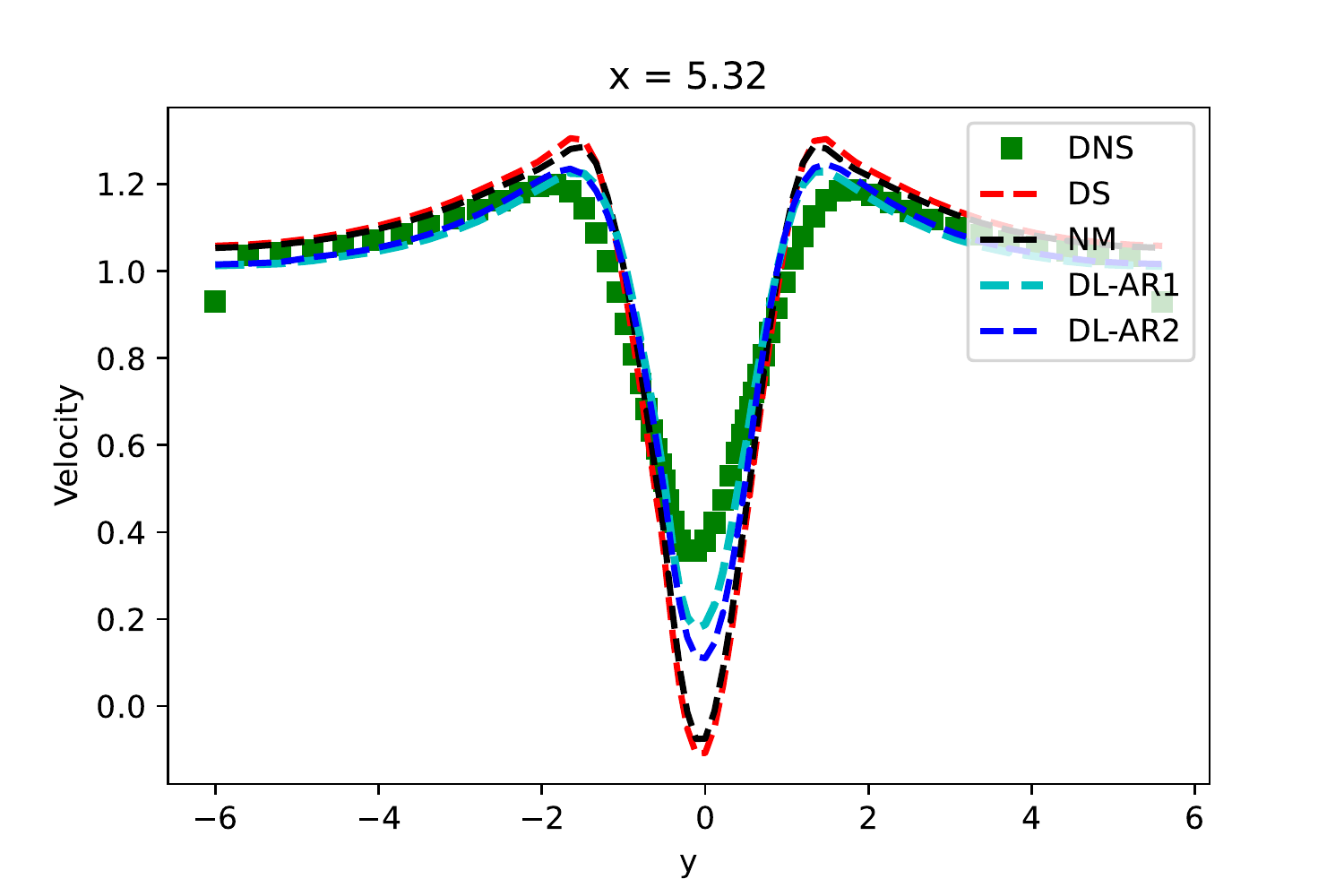}
\includegraphics[width=5cm]{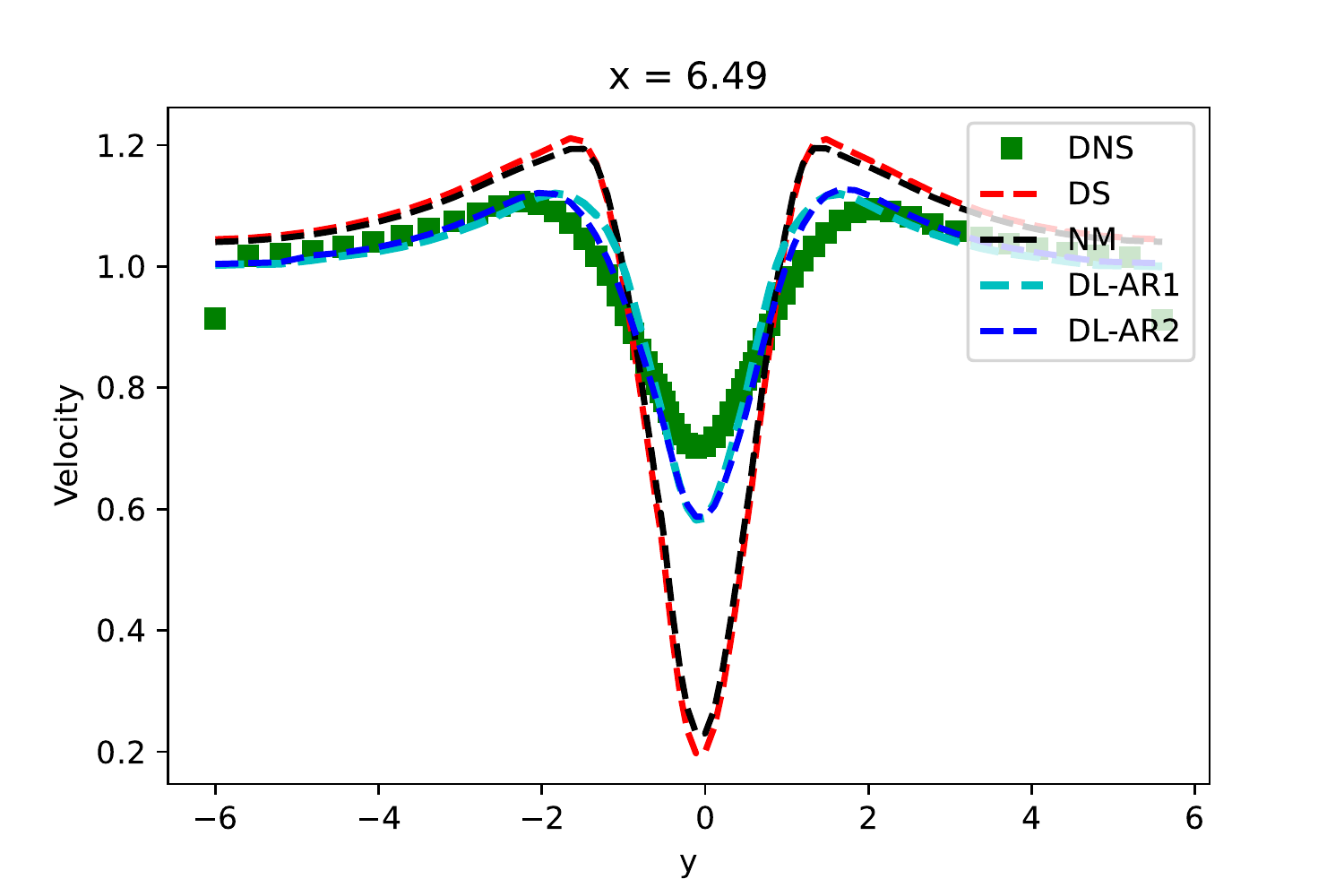}
\includegraphics[width=5cm]{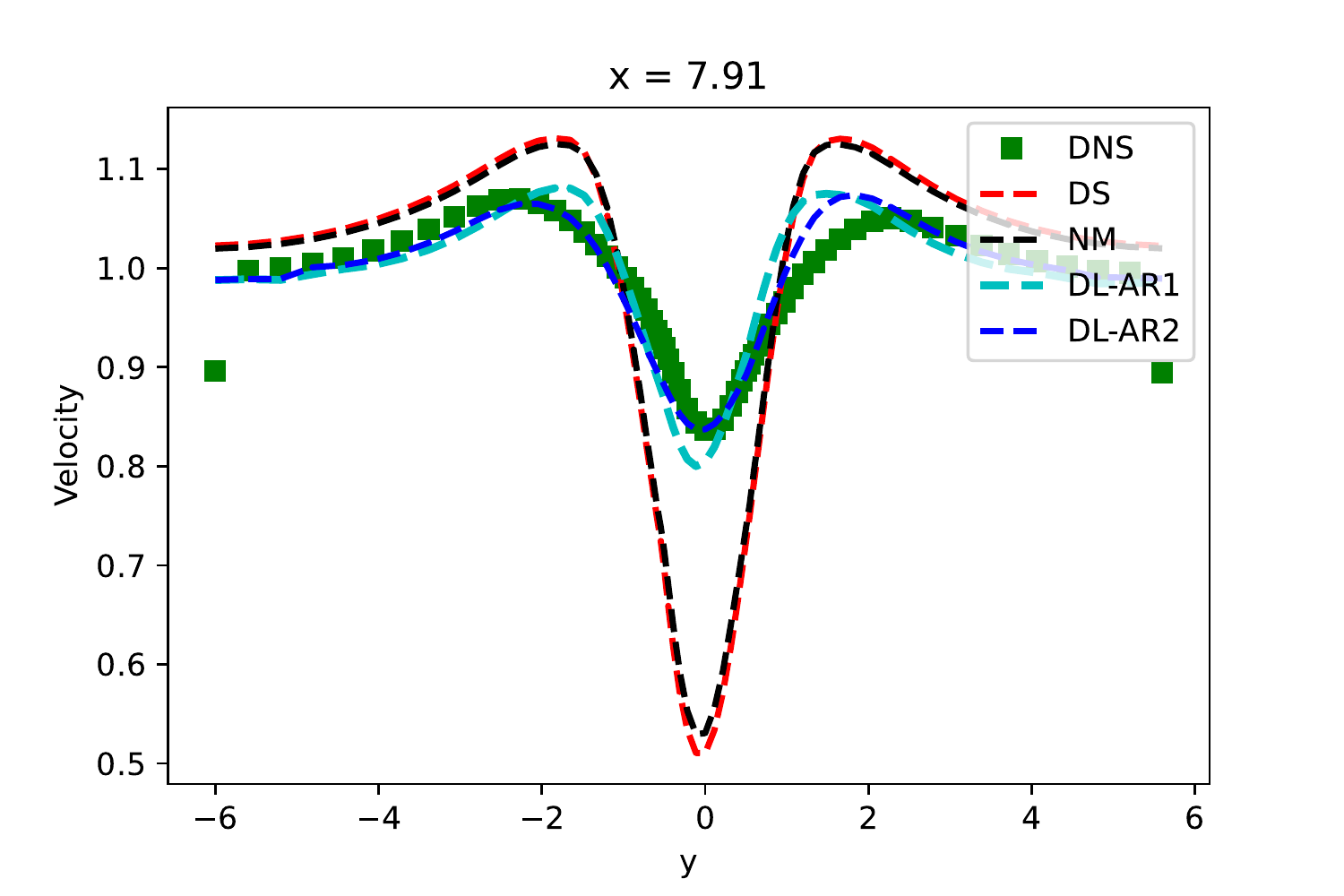}
\includegraphics[width=5cm]{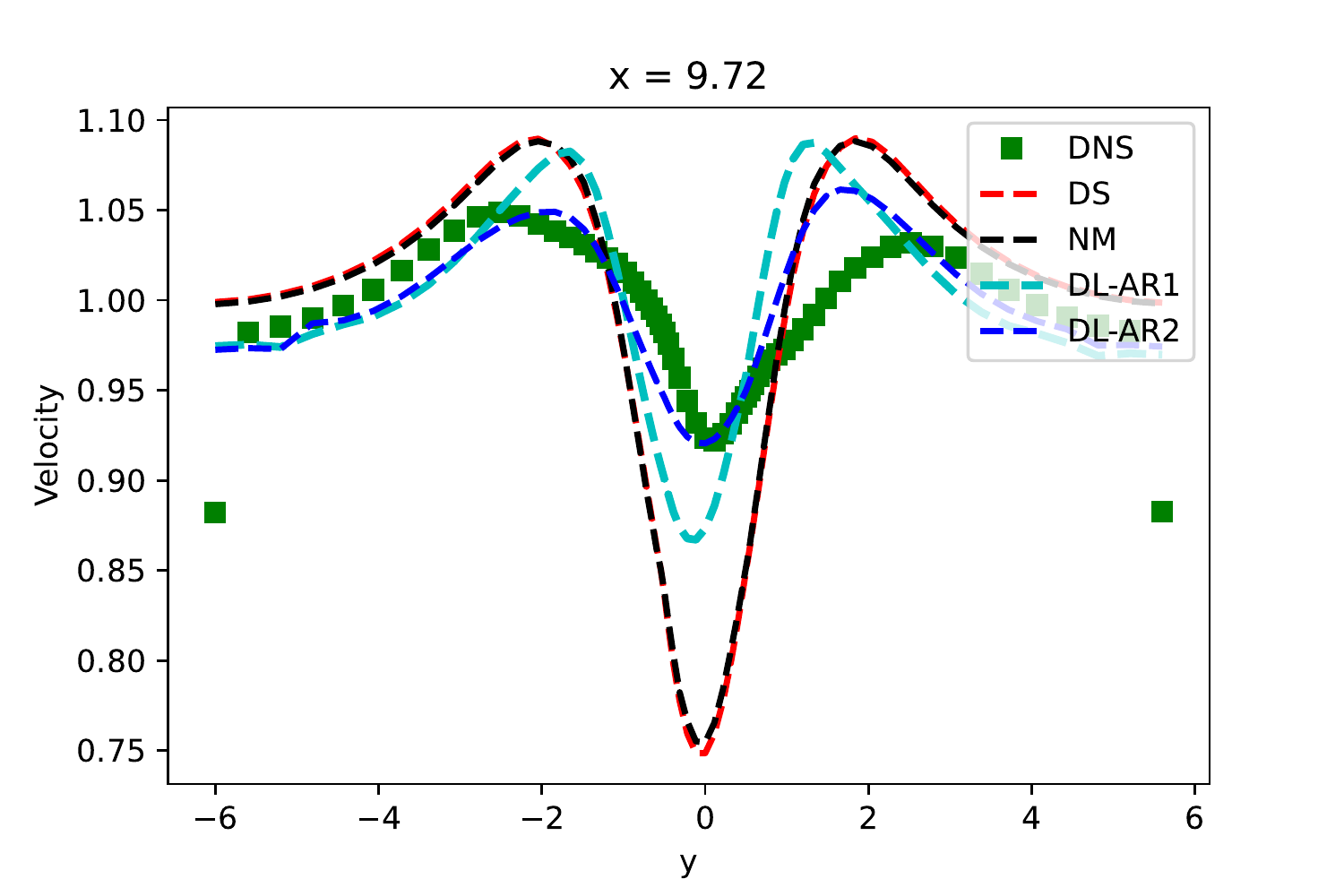}
\includegraphics[width=5cm]{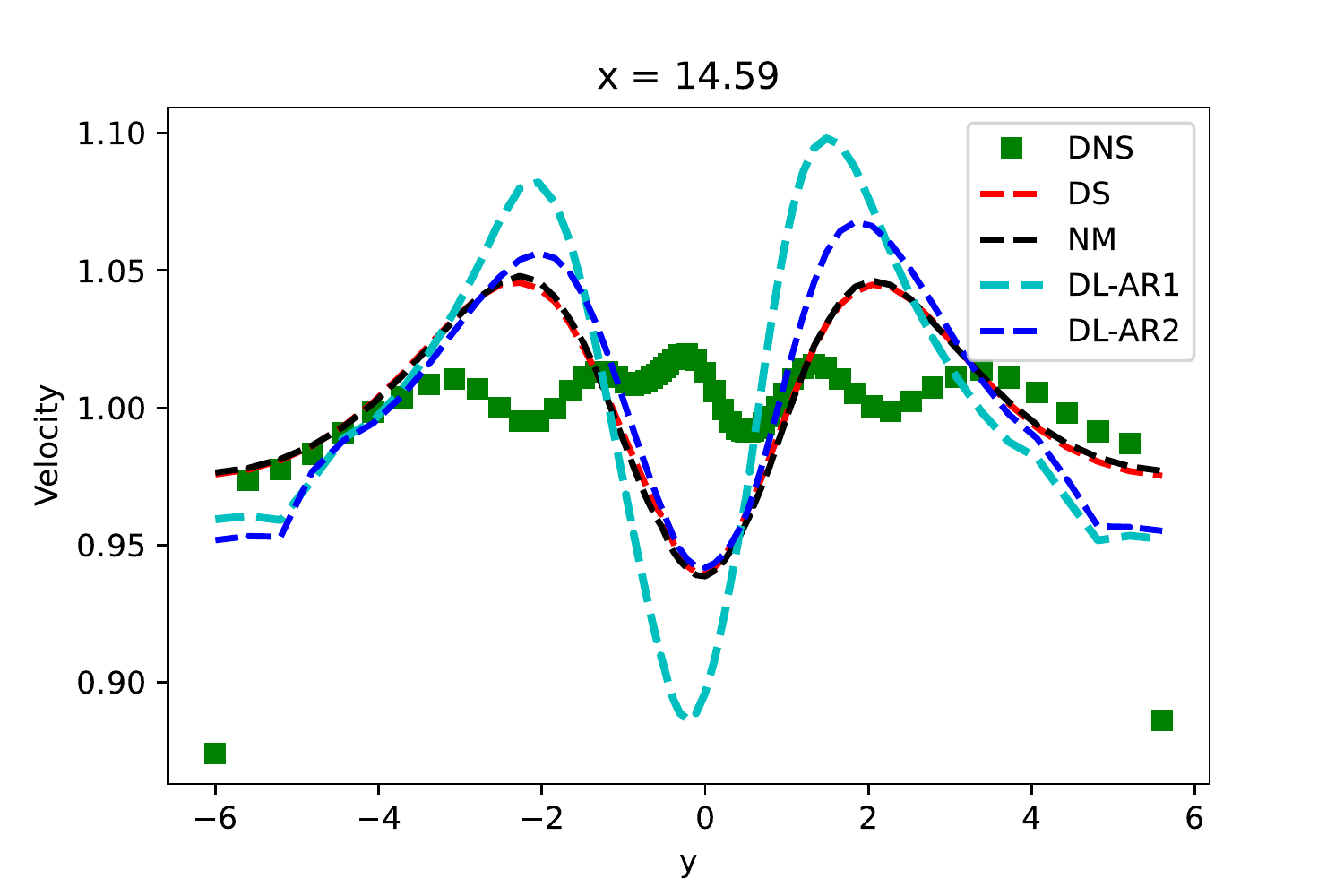}
\label{f1}
\caption{Mean profile for $u_1$ for AR2-Re$1,000$ configuration.}
\end{figure}

\begin{figure}[H]
\centering
\includegraphics[width=5cm]{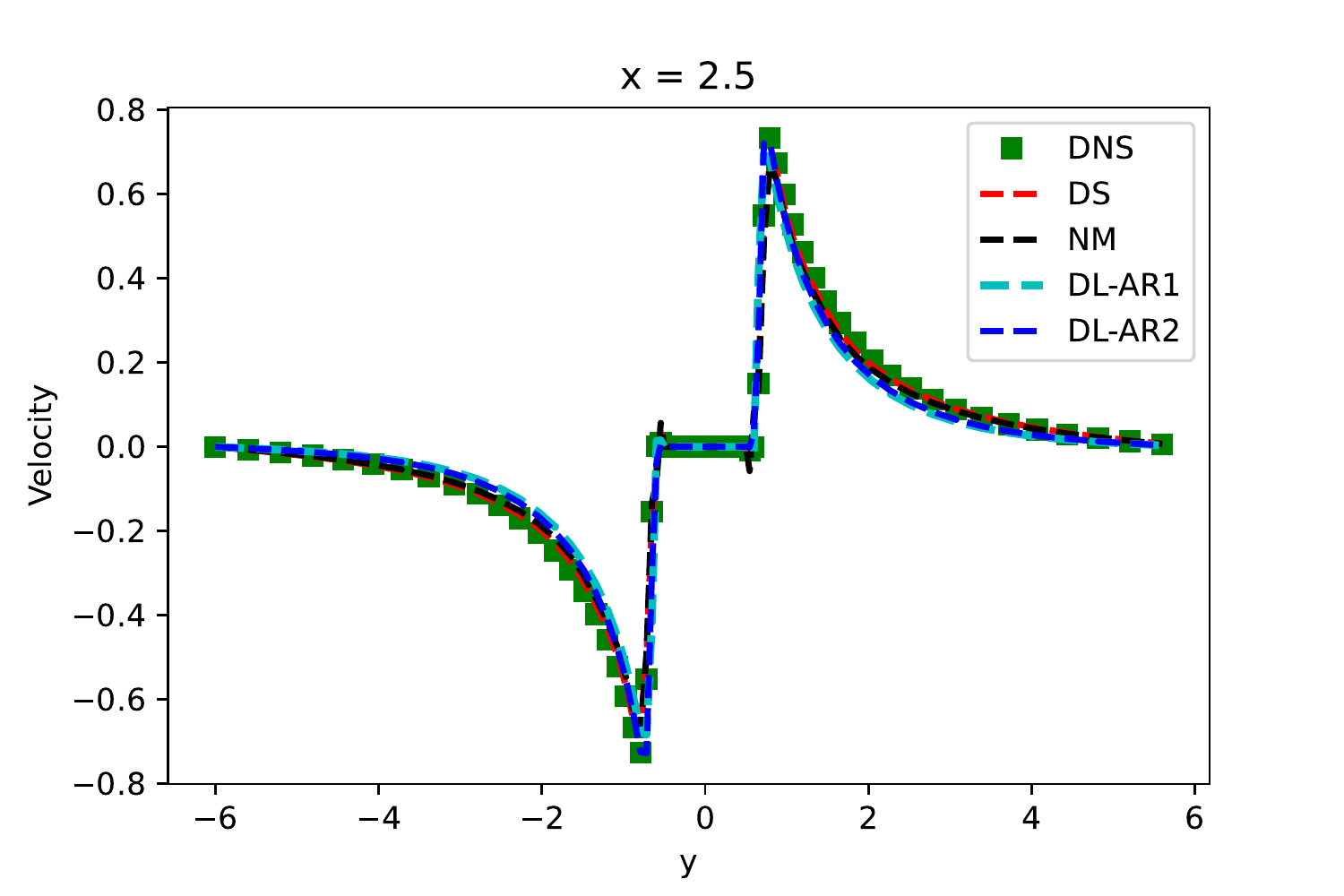}
\includegraphics[width=5cm]{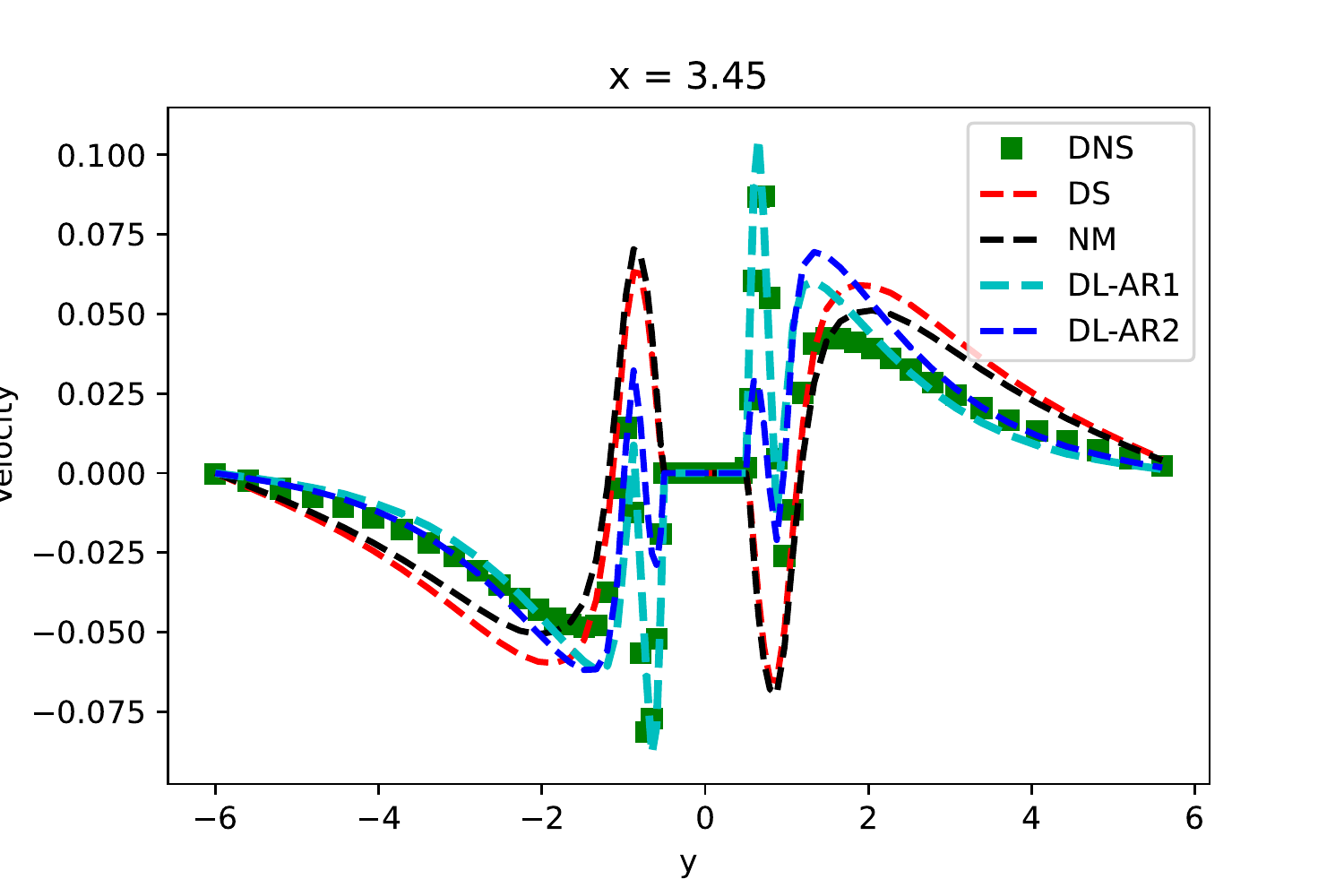}
\includegraphics[width=5cm]{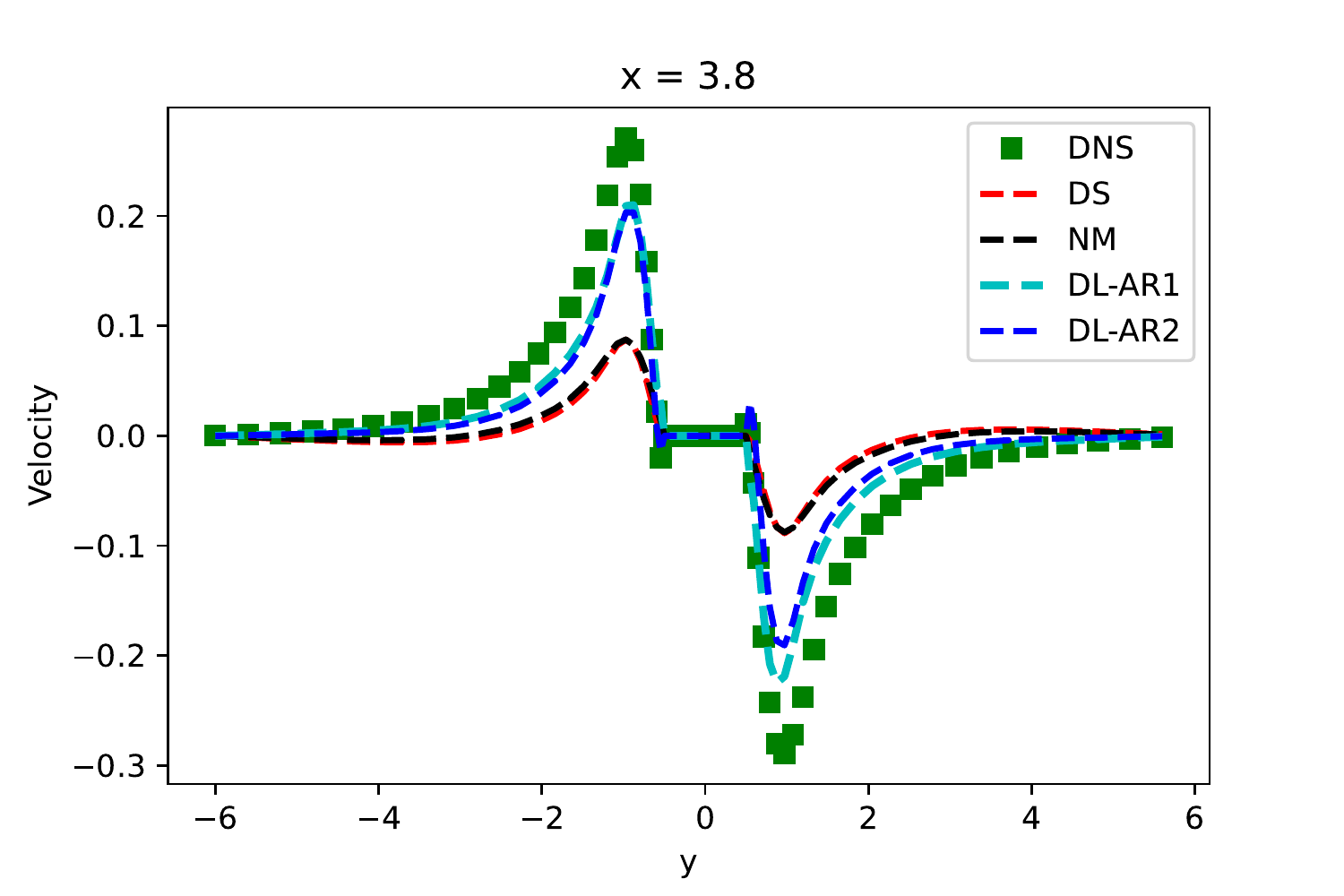}
\includegraphics[width=5cm]{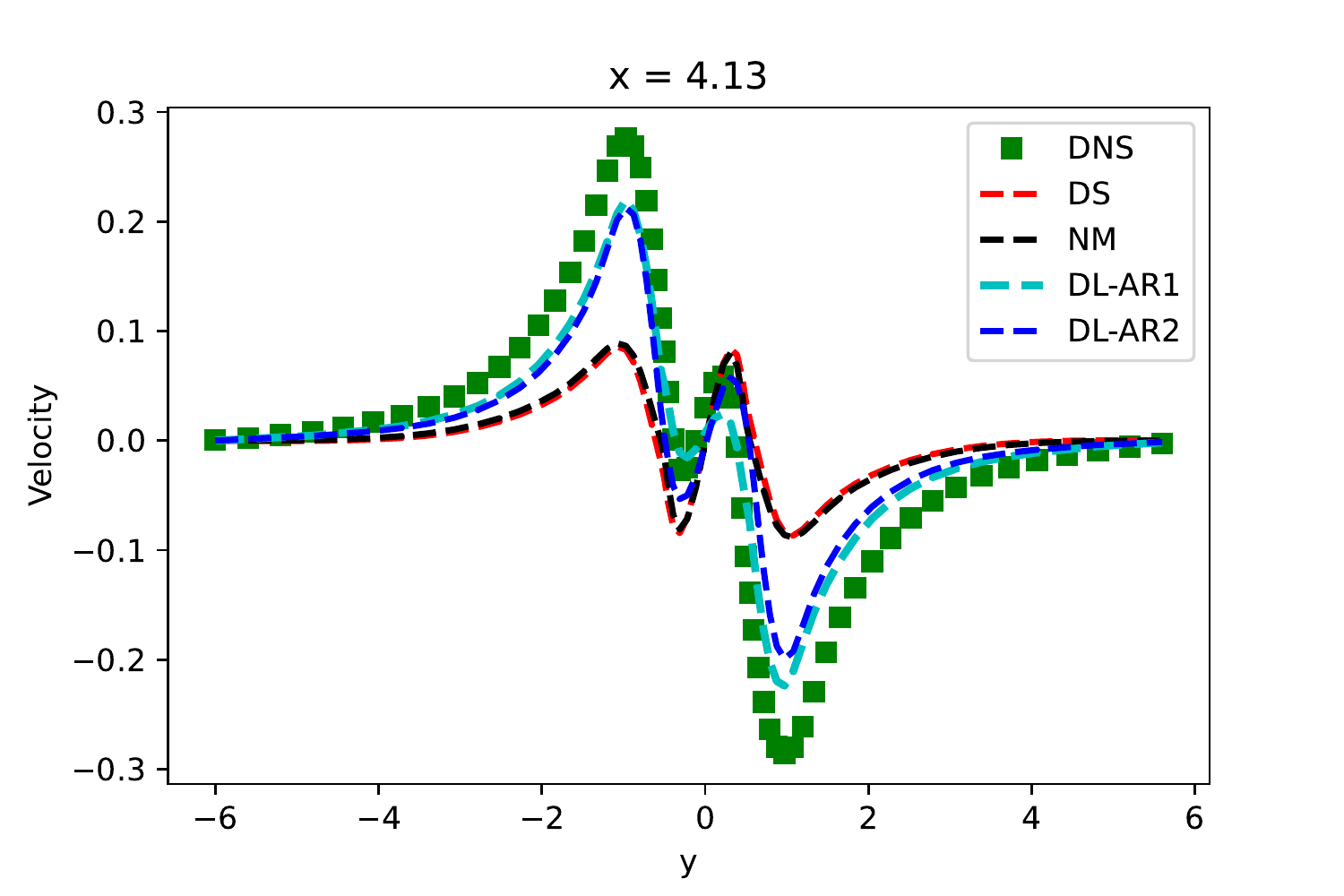}
\includegraphics[width=5cm]{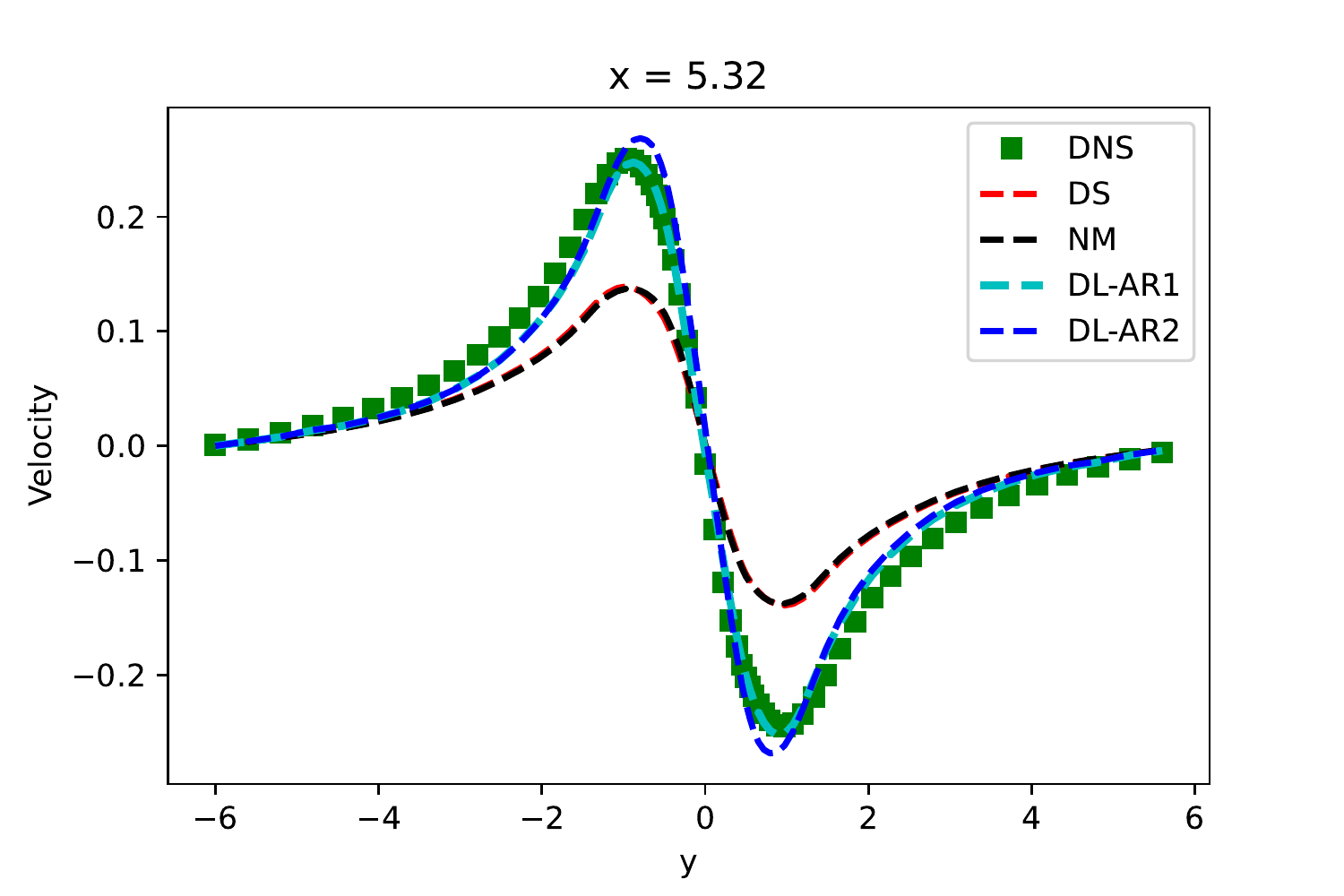}
\includegraphics[width=5cm]{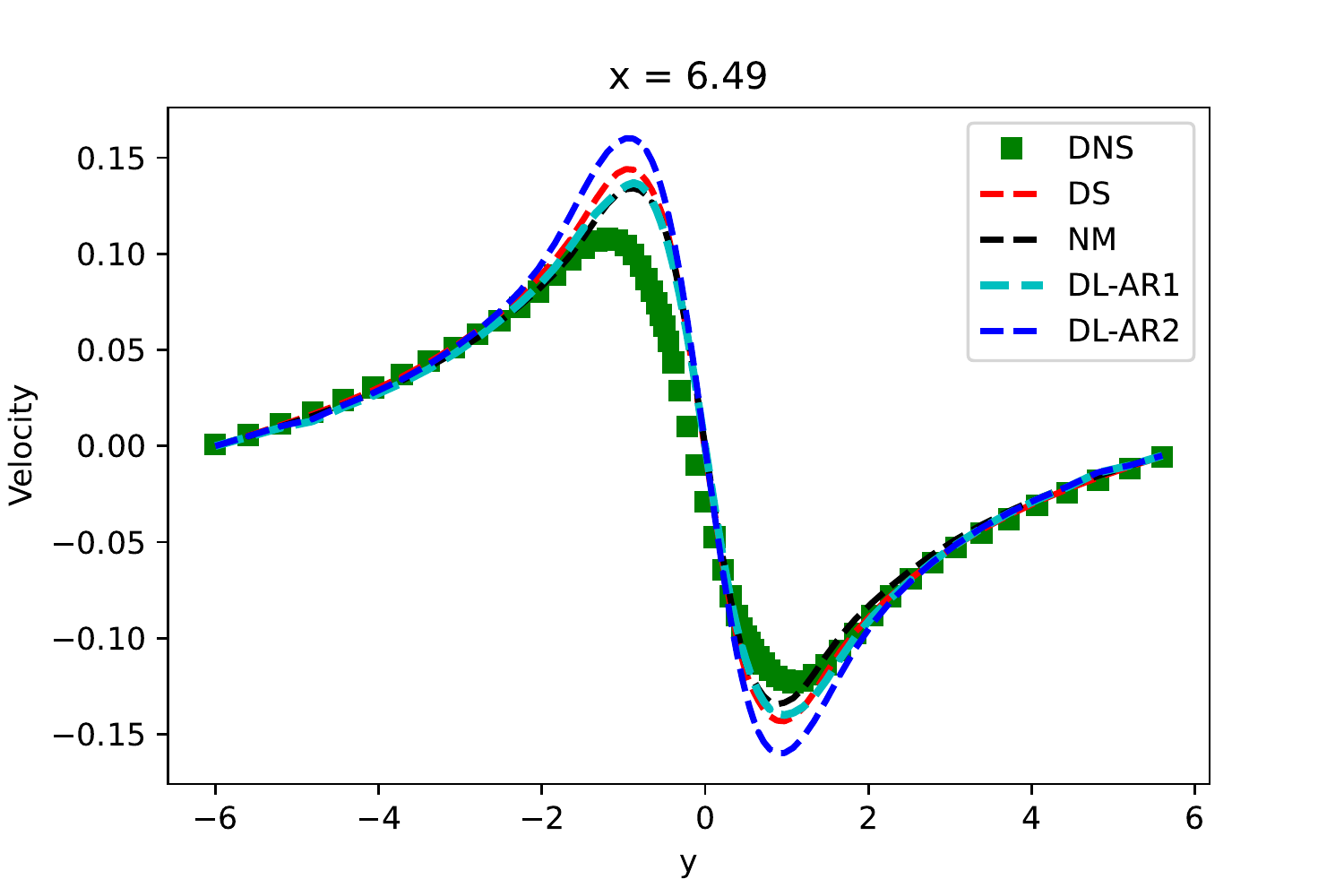}
\includegraphics[width=5cm]{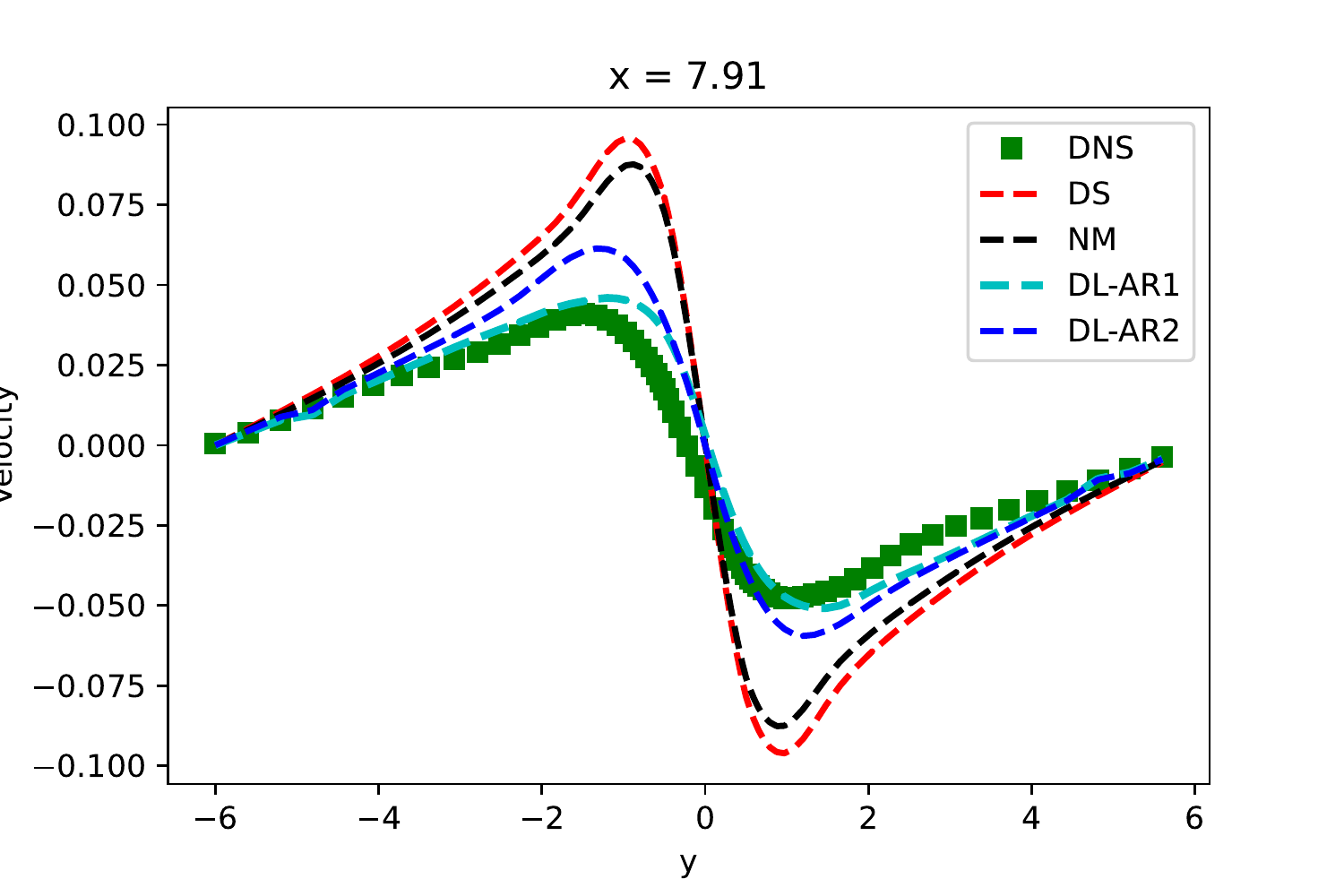}
\includegraphics[width=5cm]{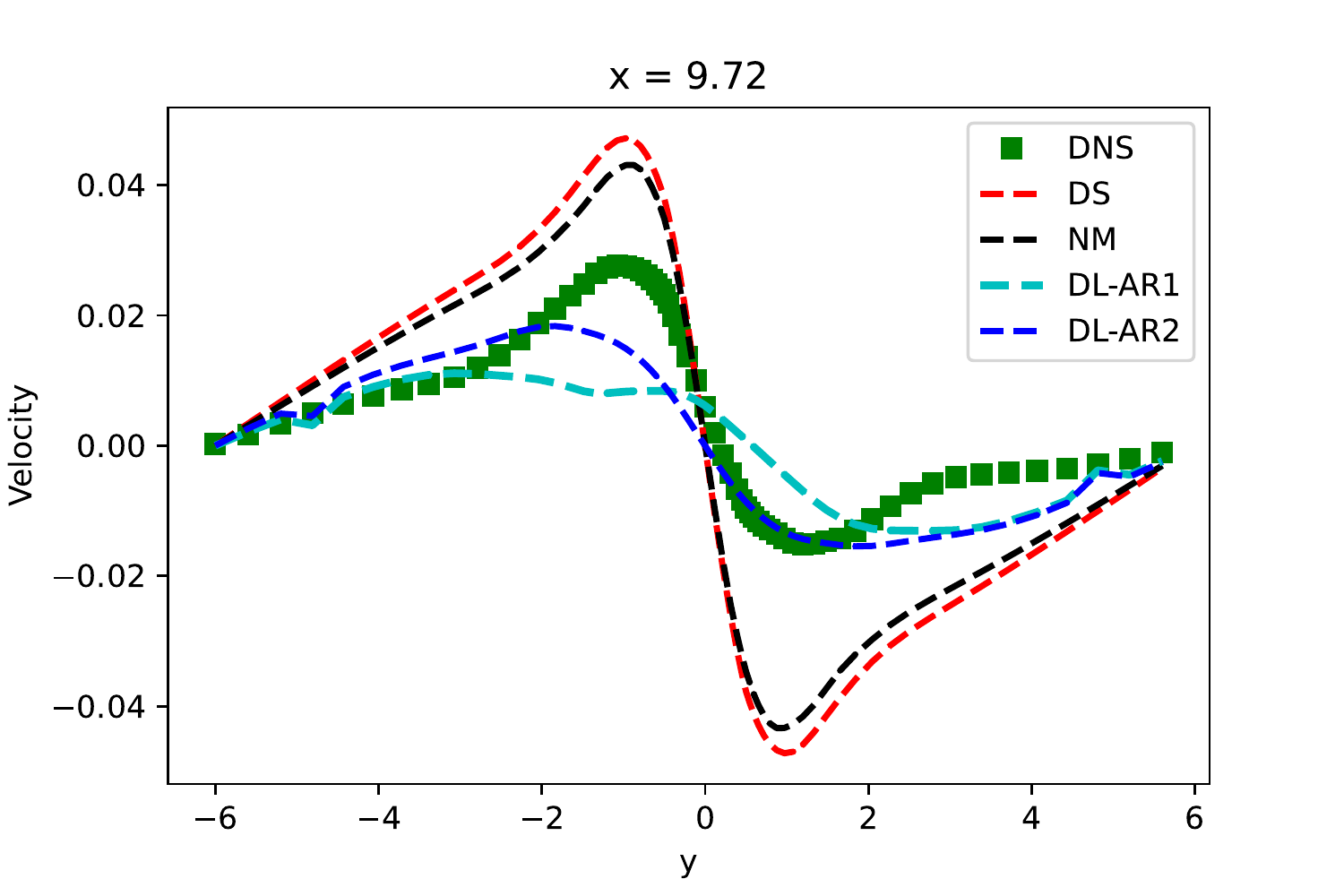}
\includegraphics[width=5cm]{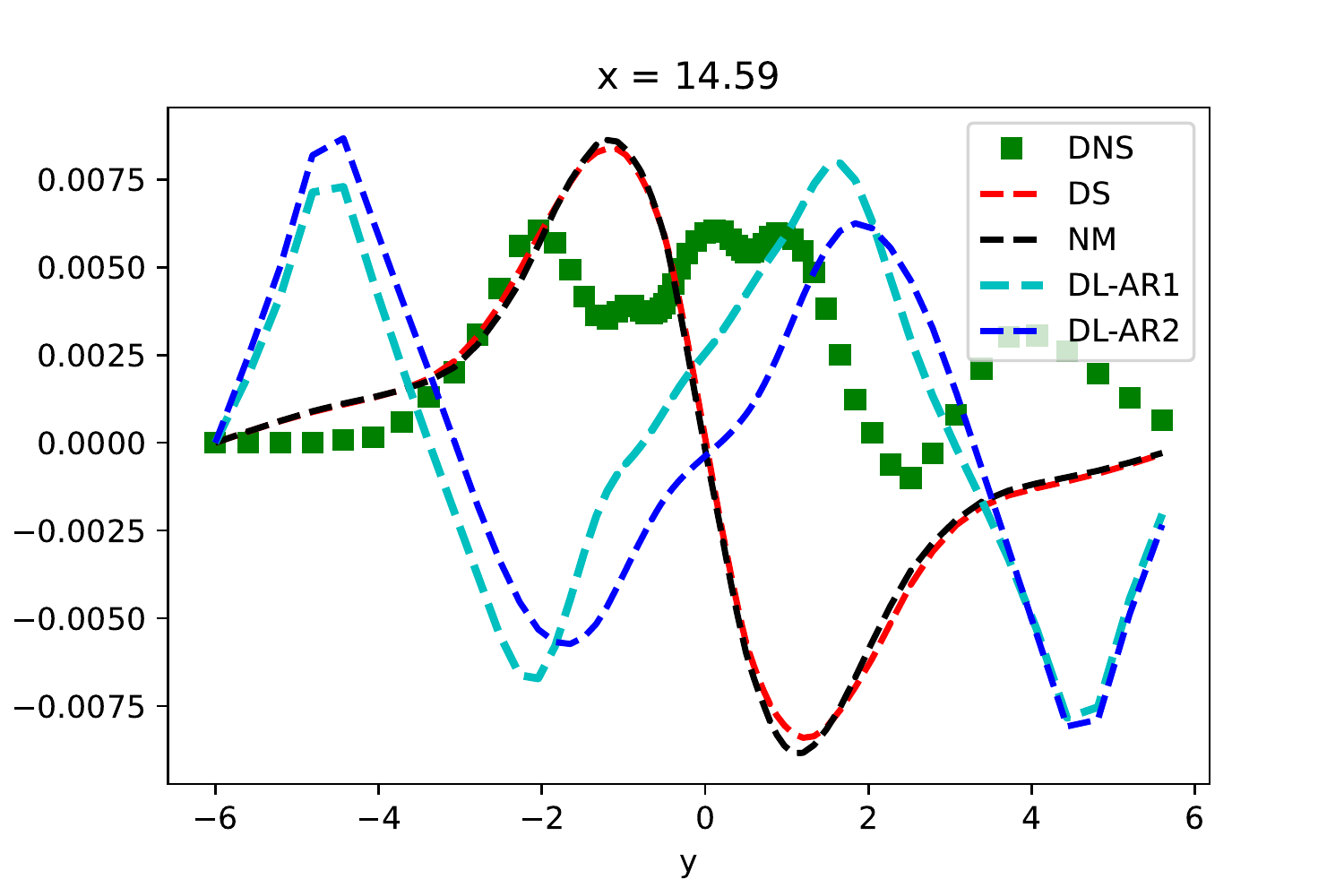}
\label{f1}
\caption{Mean profile for $u_2$ for AR2-Re$1,000$ configuration.}
\end{figure}

\begin{figure}[H]
\centering
\includegraphics[width=5cm]{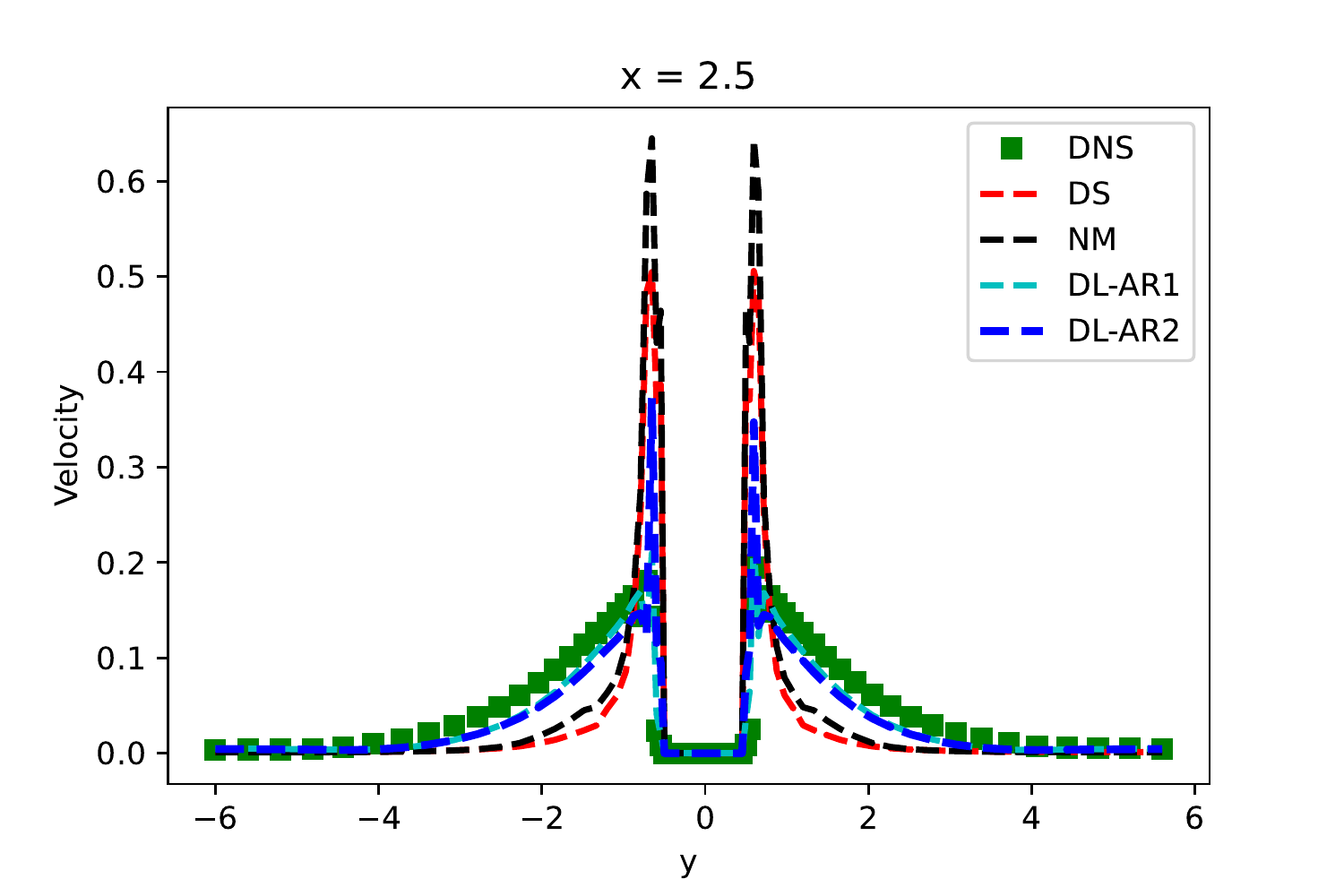}
\includegraphics[width=5cm]{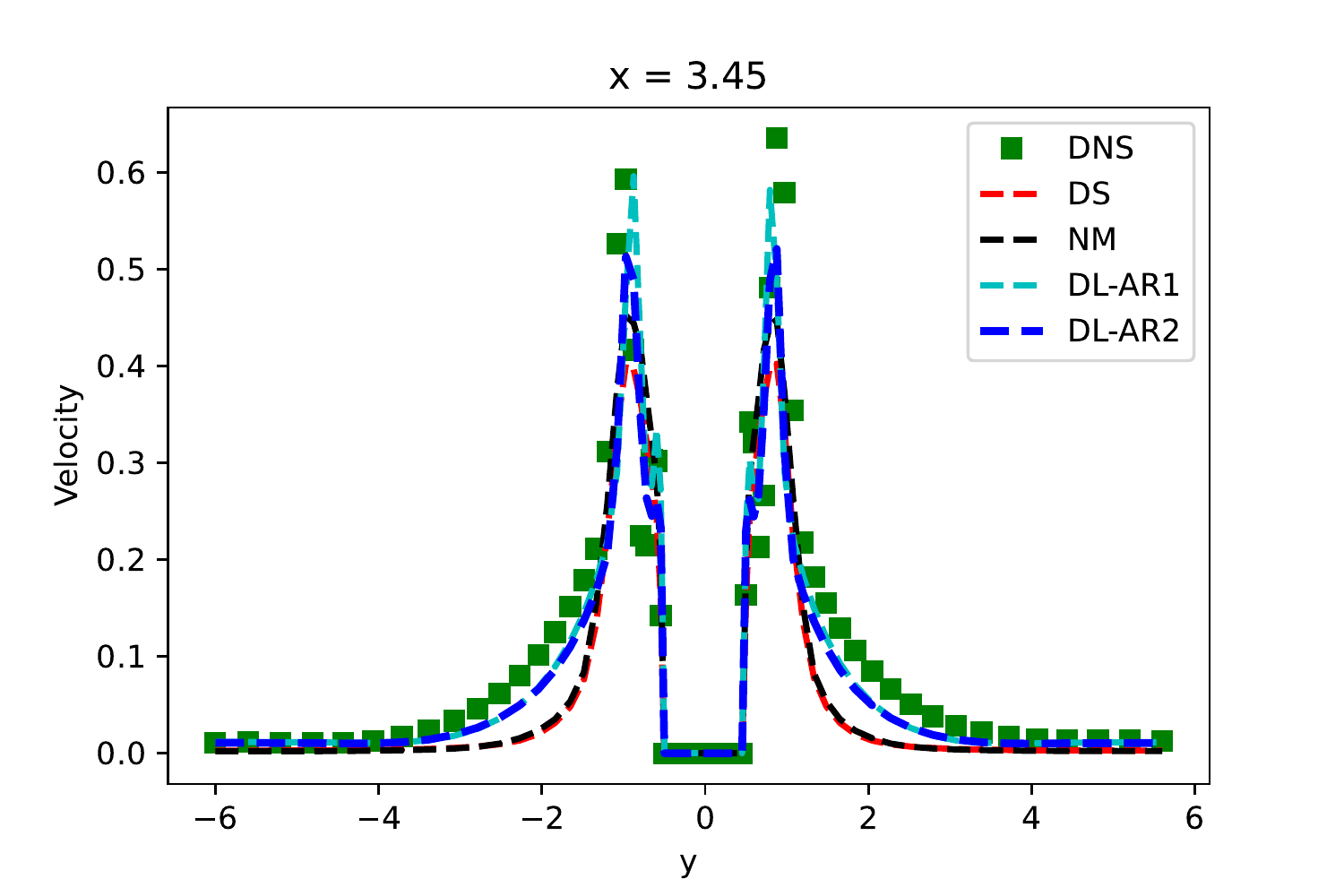}
\includegraphics[width=5cm]{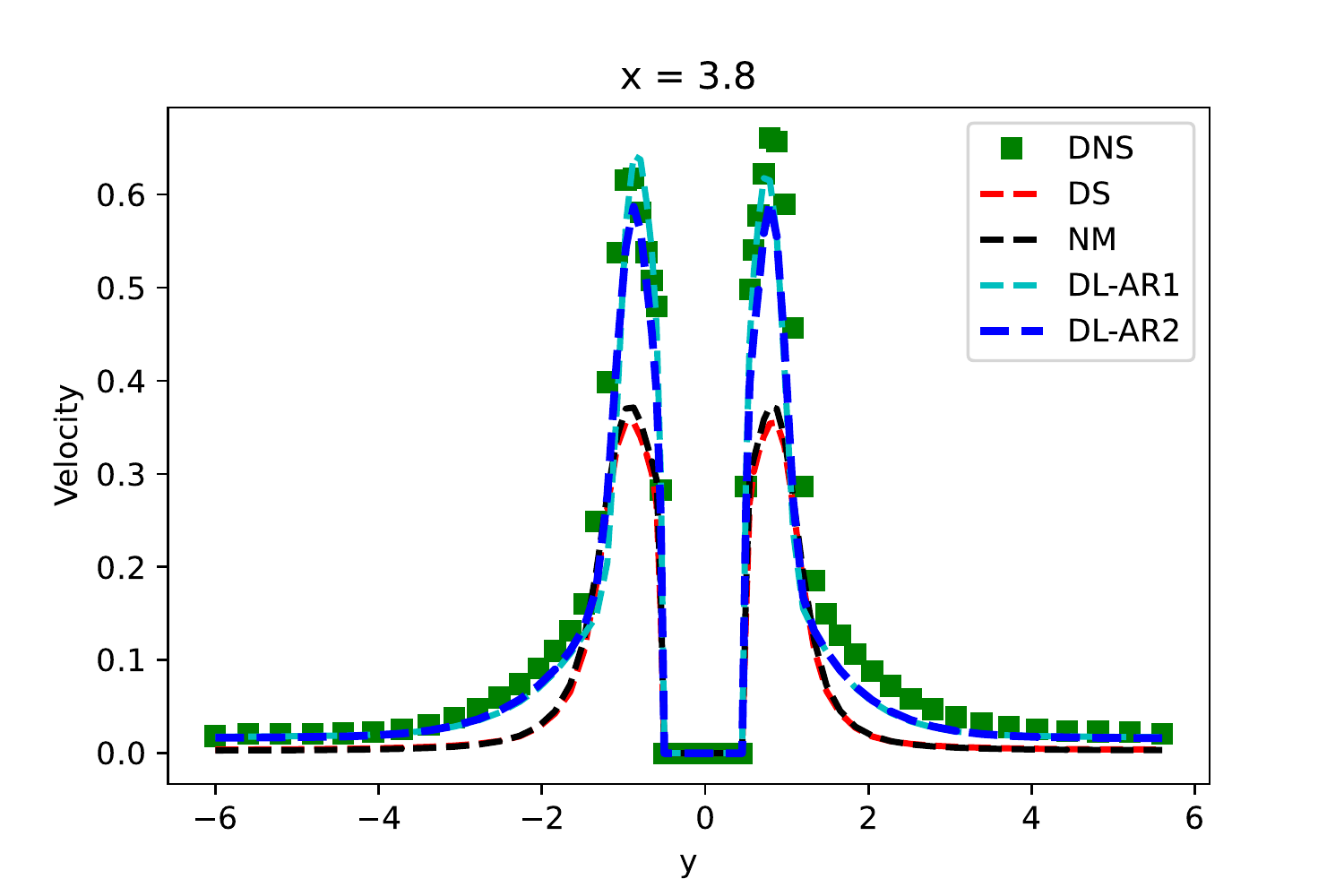}
\includegraphics[width=5cm]{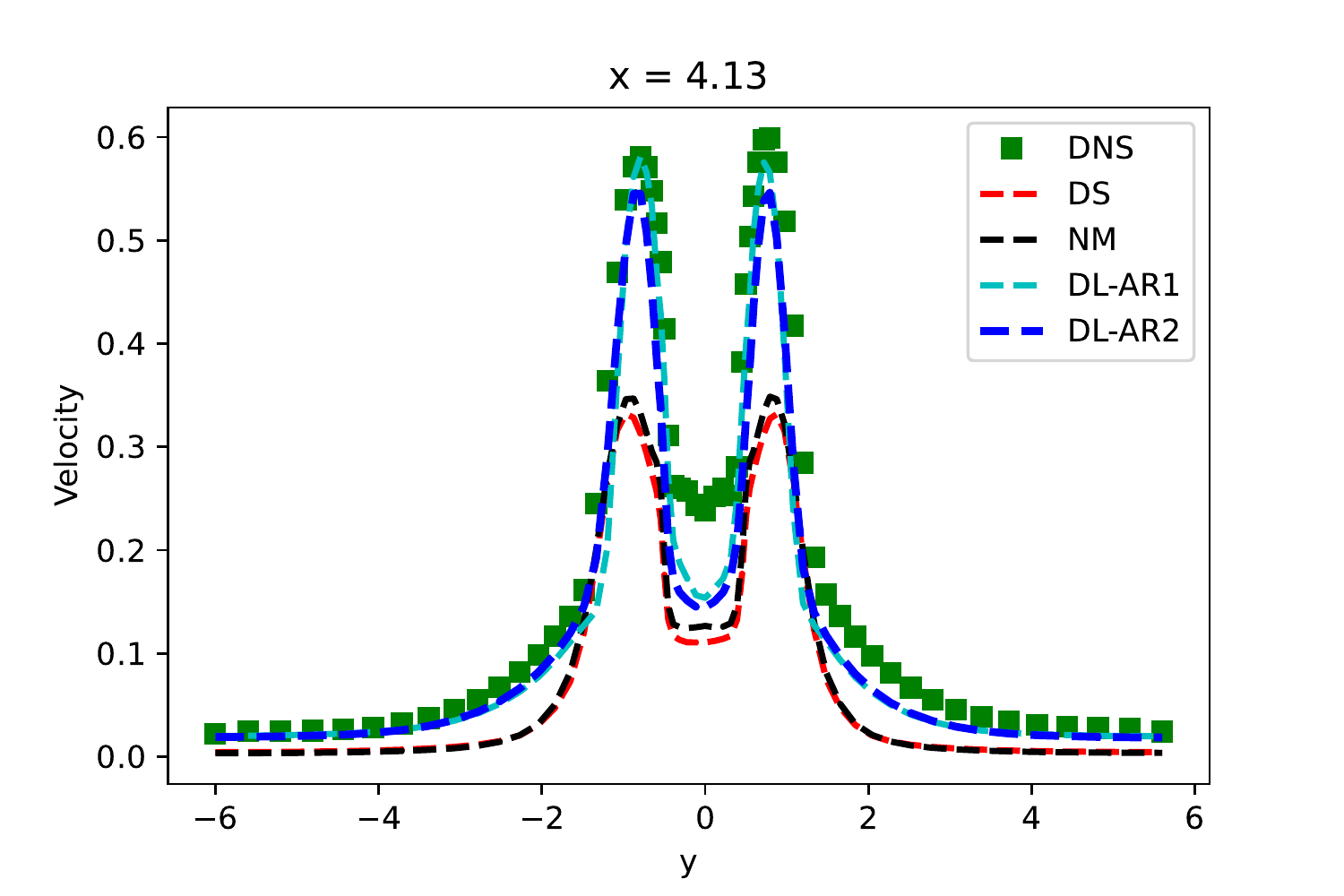}
\includegraphics[width=5cm]{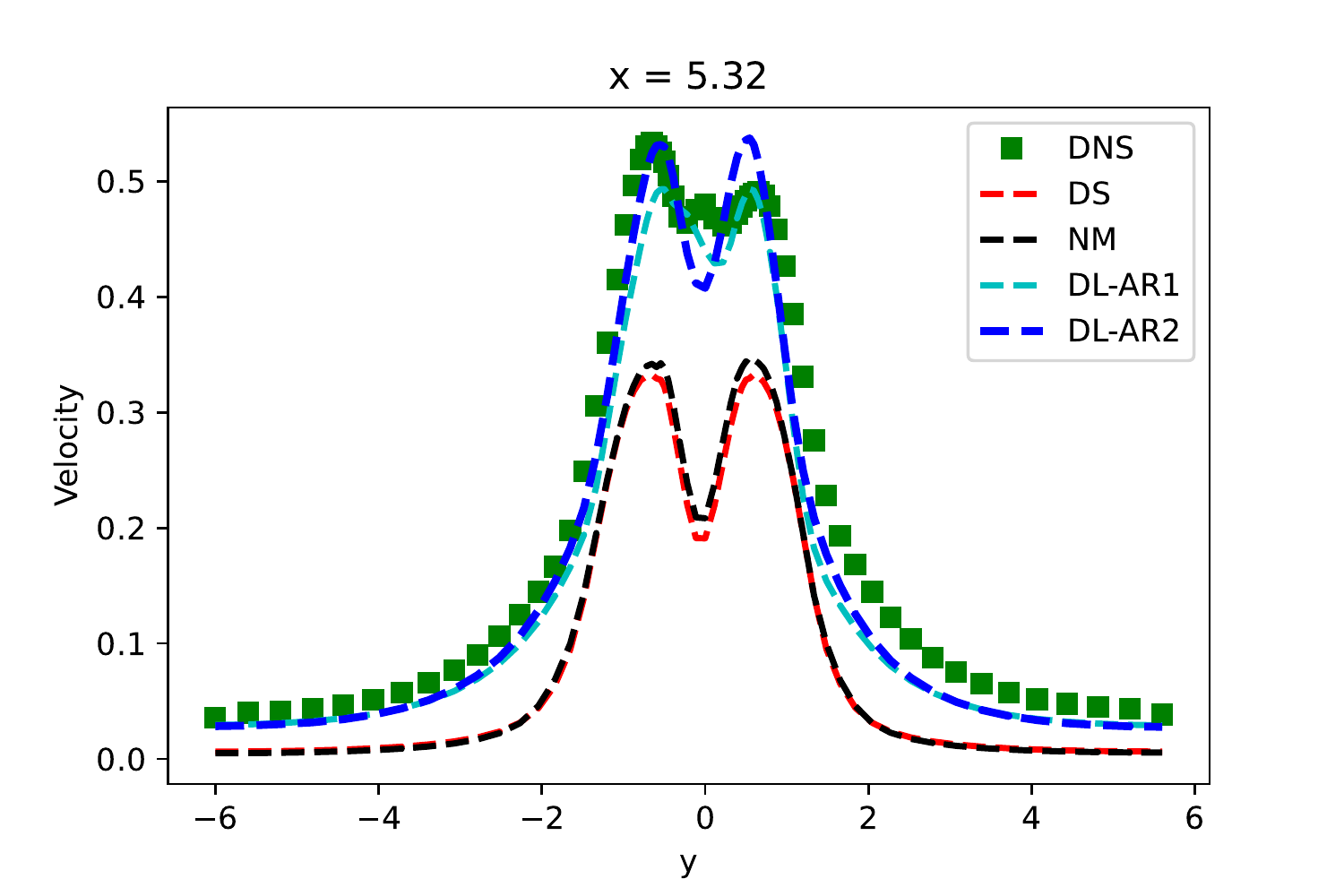}
\includegraphics[width=5cm]{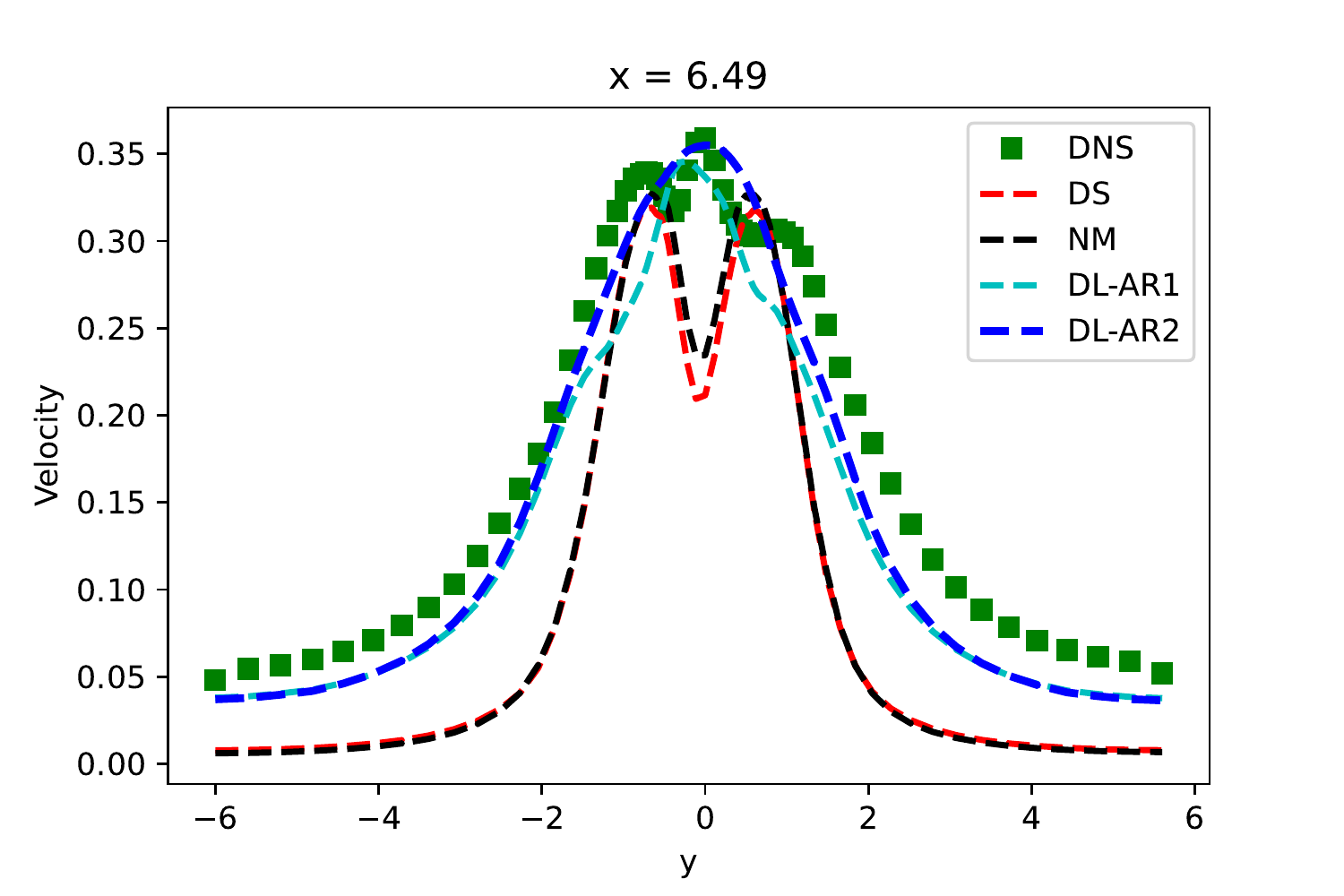}
\includegraphics[width=5cm]{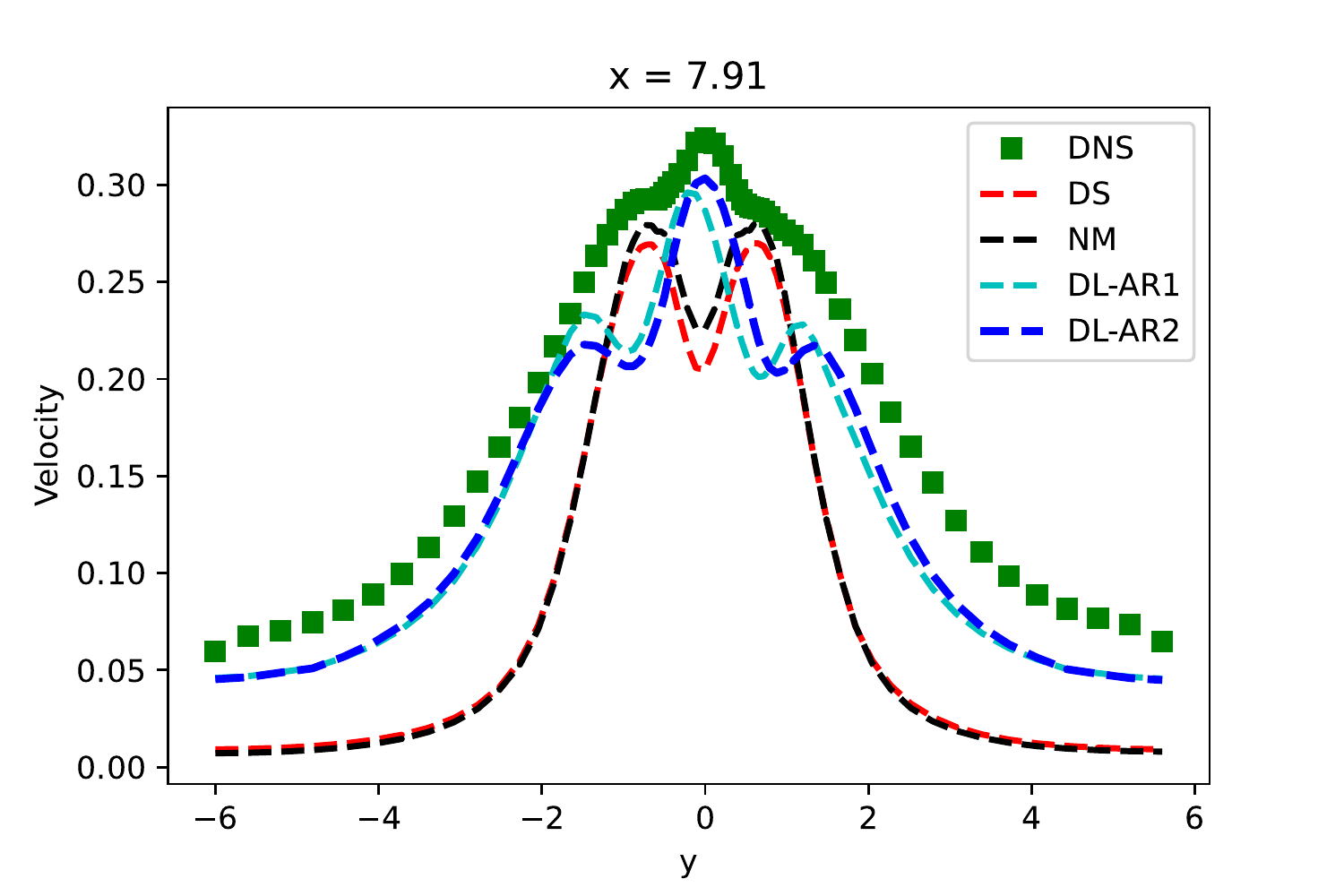}
\includegraphics[width=5cm]{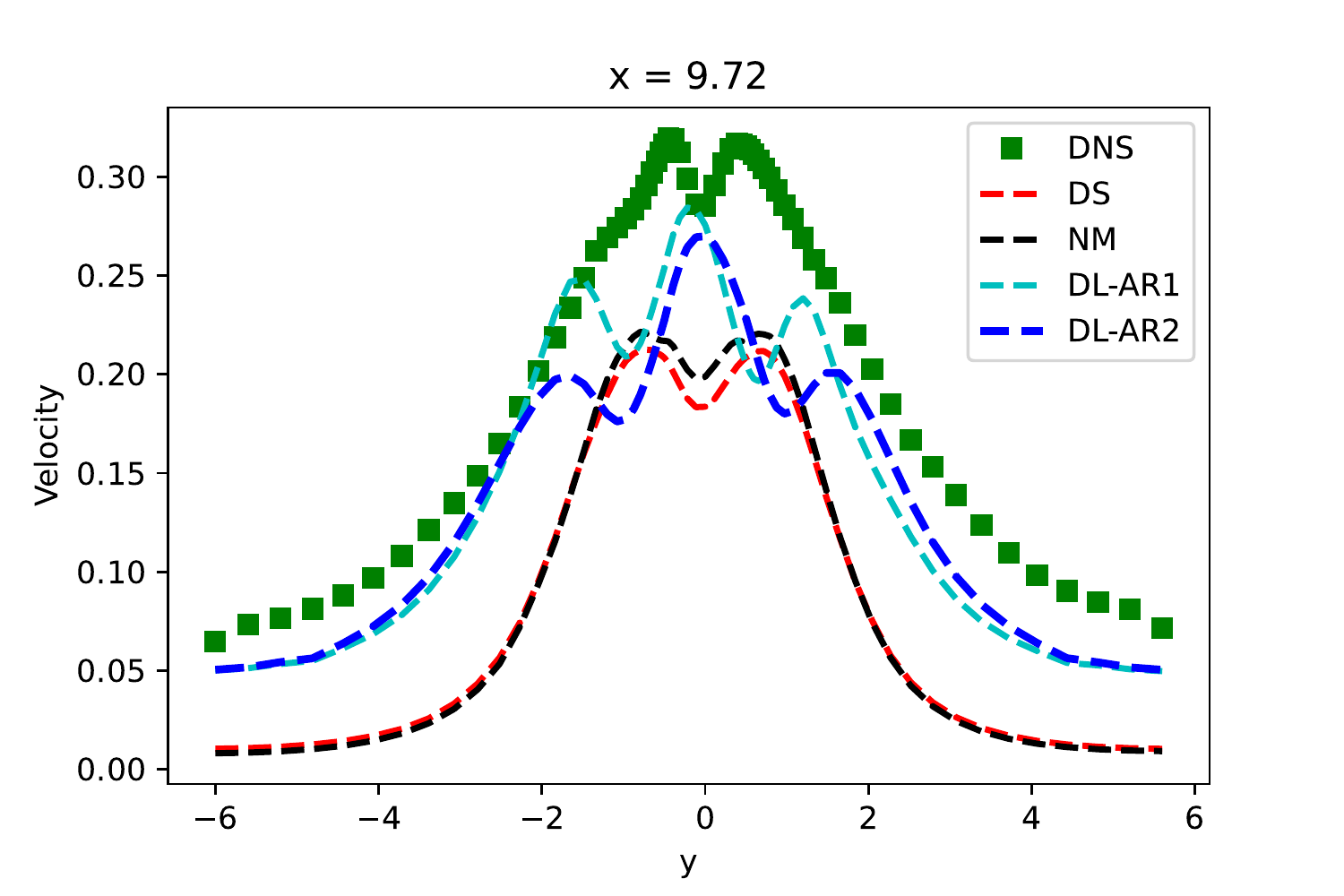}
\includegraphics[width=5cm]{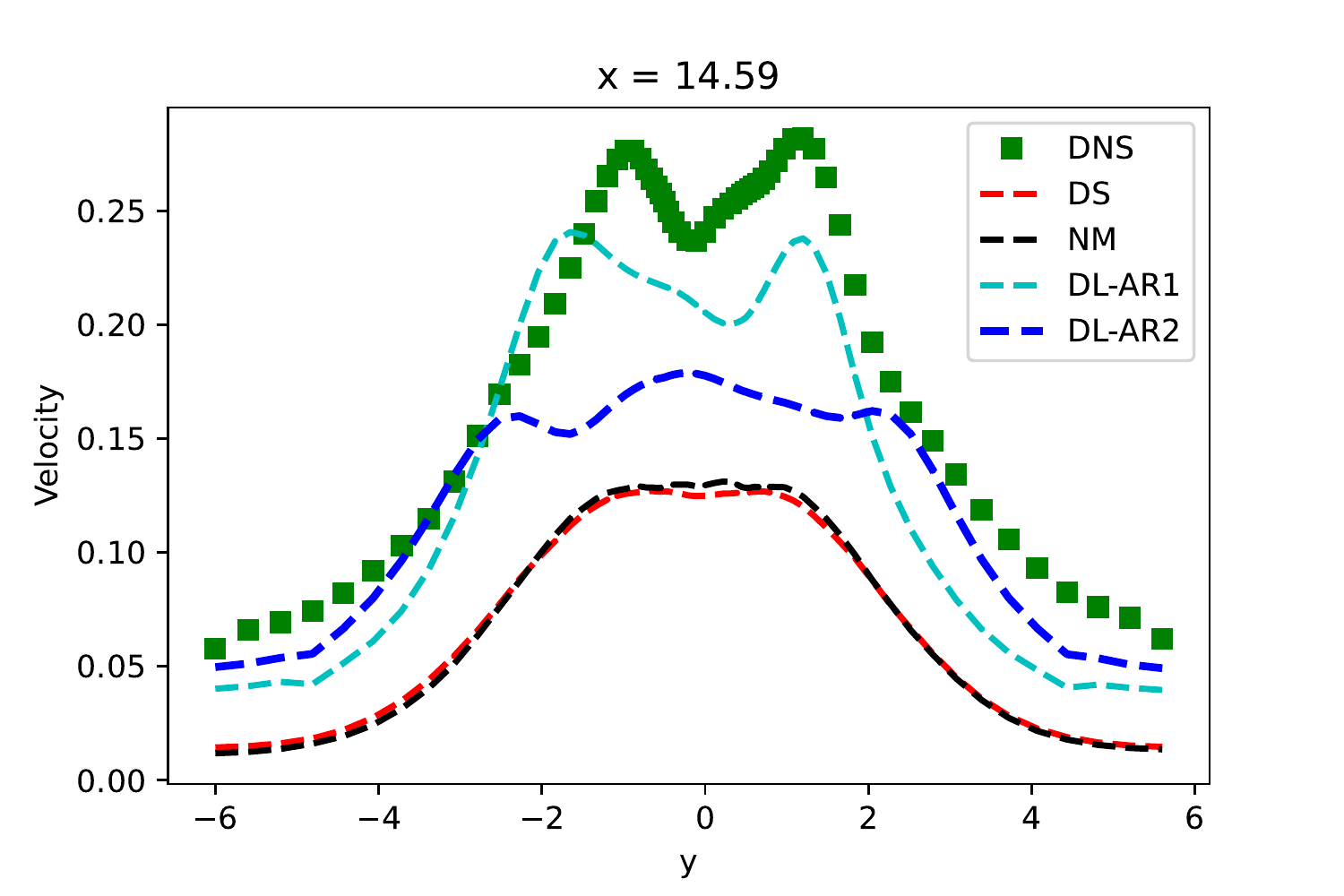}
\label{f1}
\caption{RMS profile for $u_1$ for AR2-Re$1,000$ configuration.}
\end{figure}

\begin{figure}[H]
\centering
\includegraphics[width=5cm]{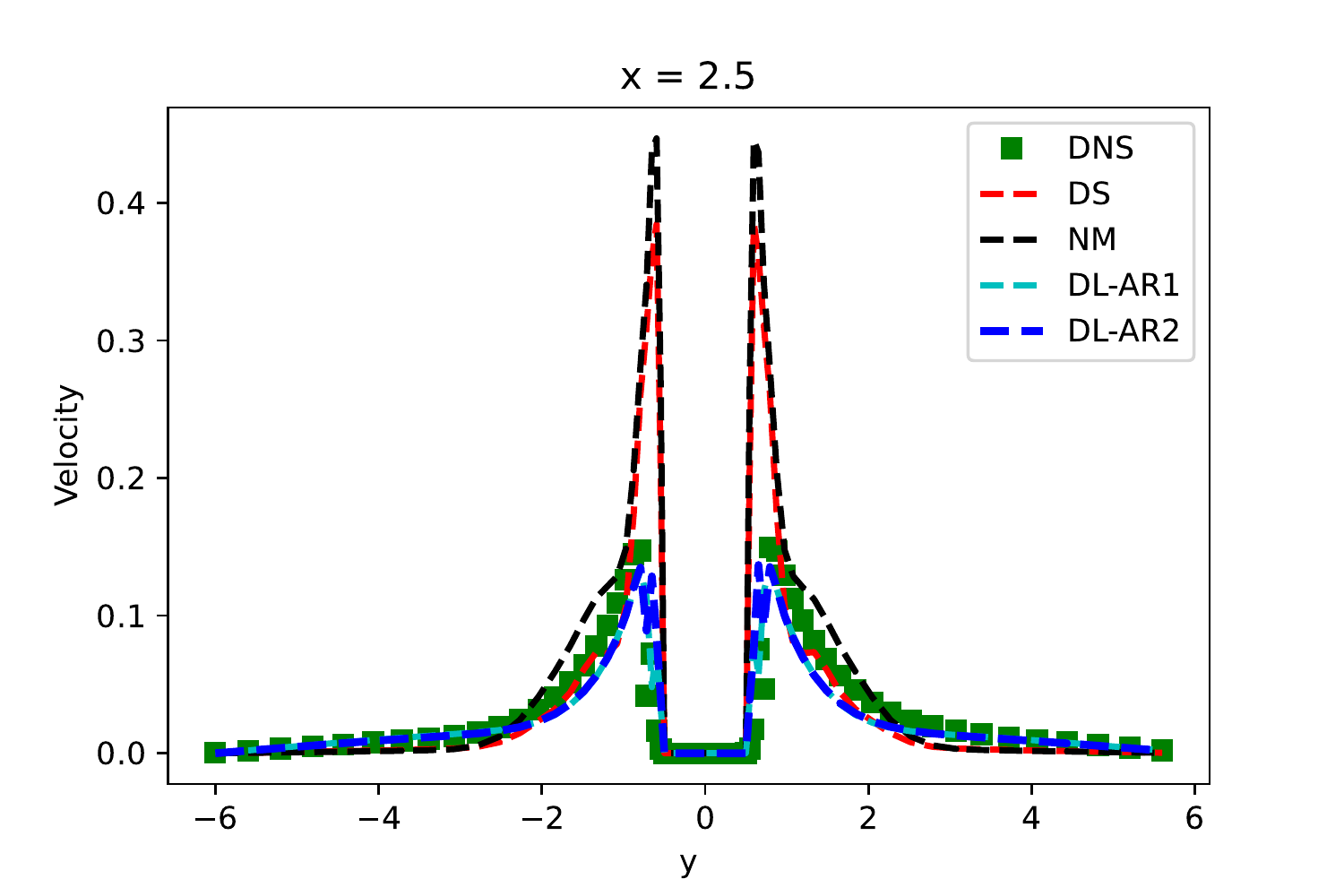}
\includegraphics[width=5cm]{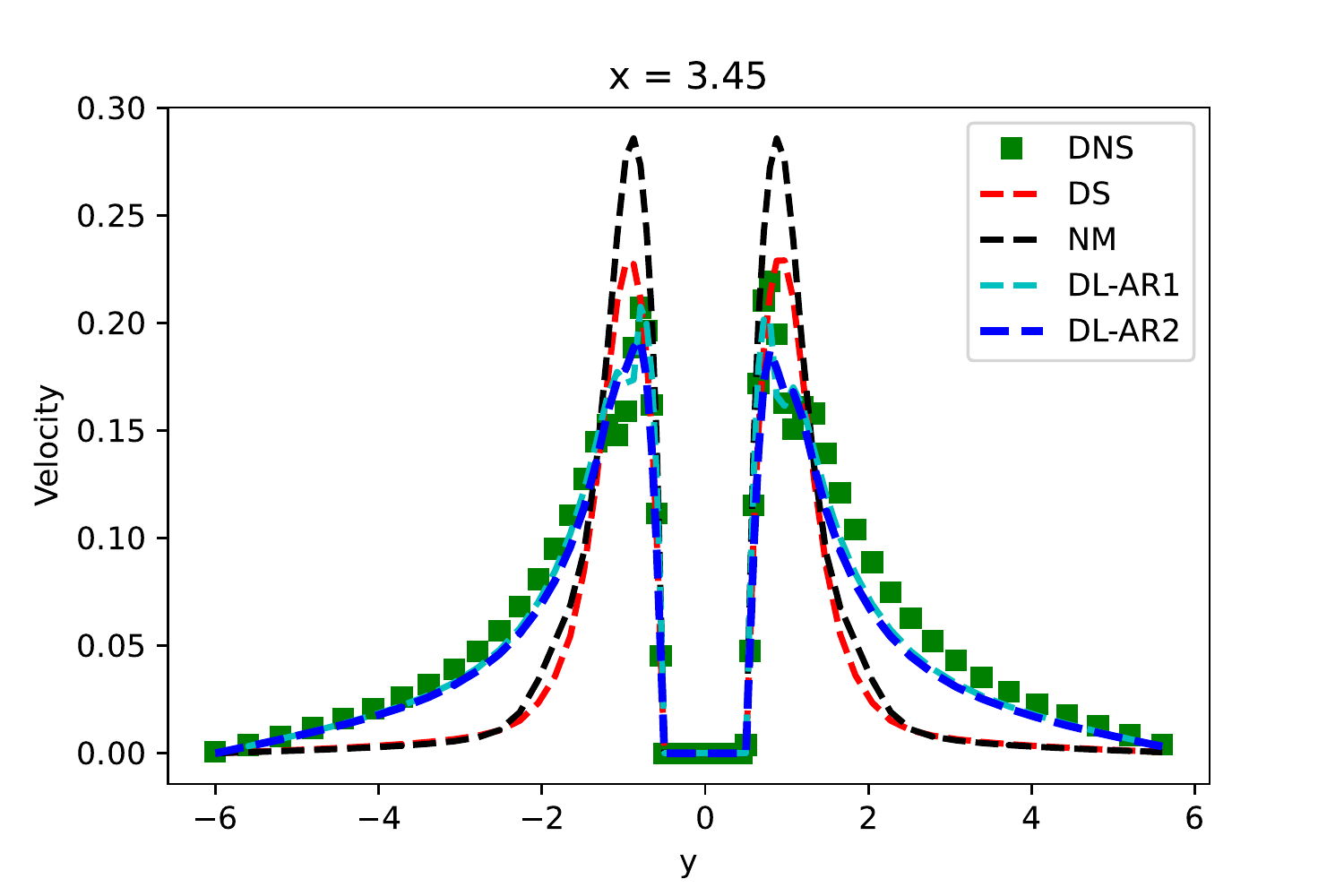}
\includegraphics[width=5cm]{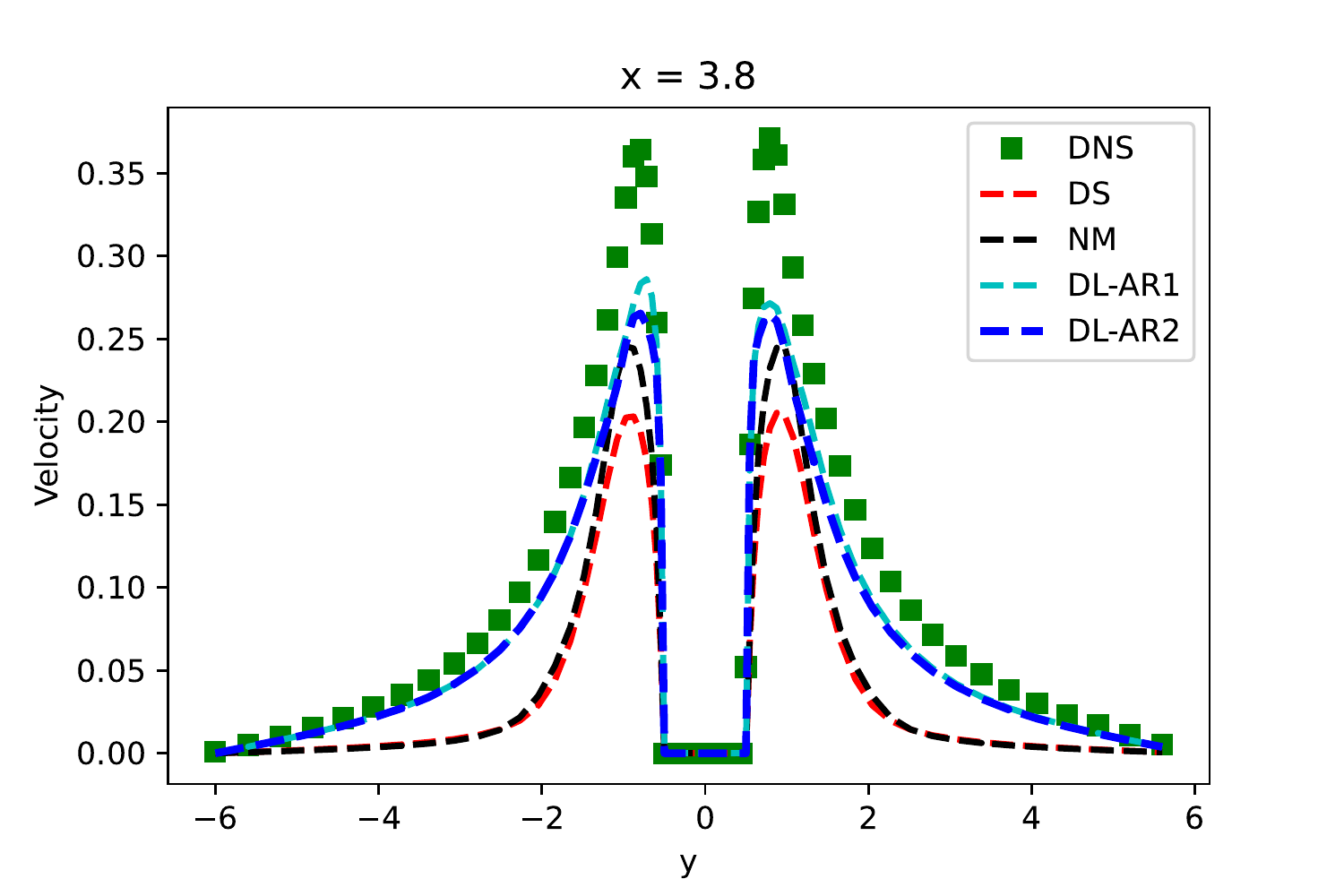}
\includegraphics[width=5cm]{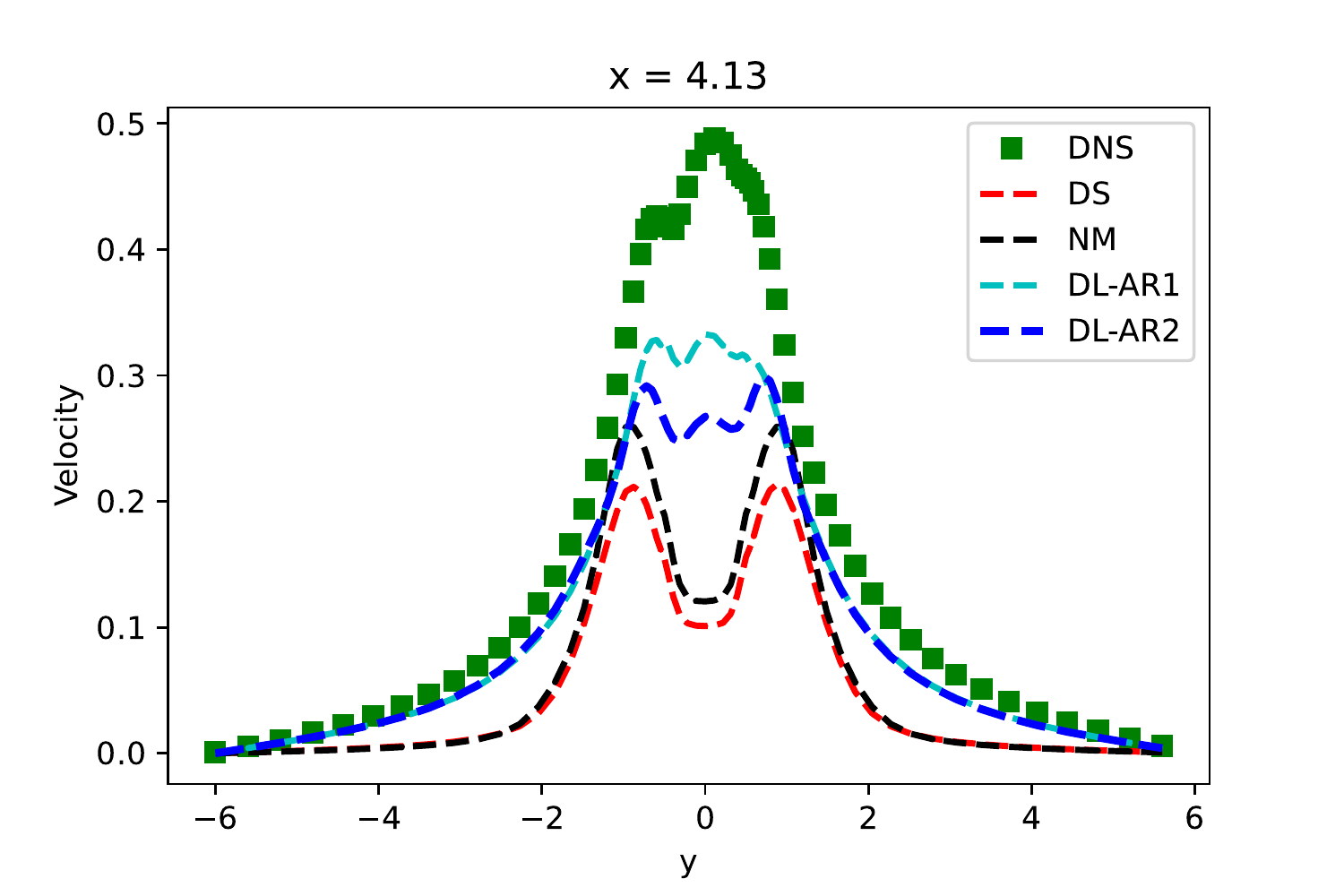}
\includegraphics[width=5cm]{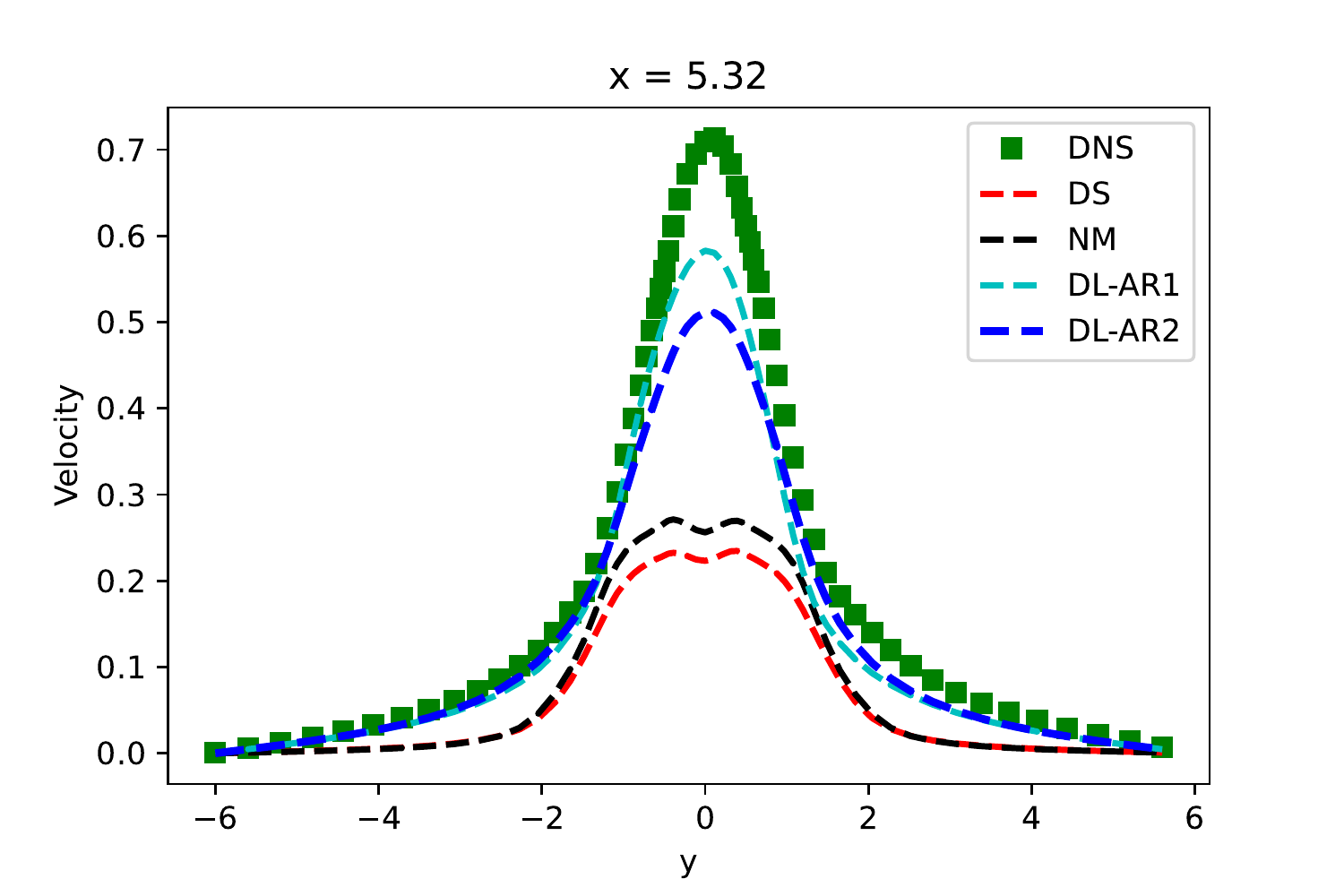}
\includegraphics[width=5cm]{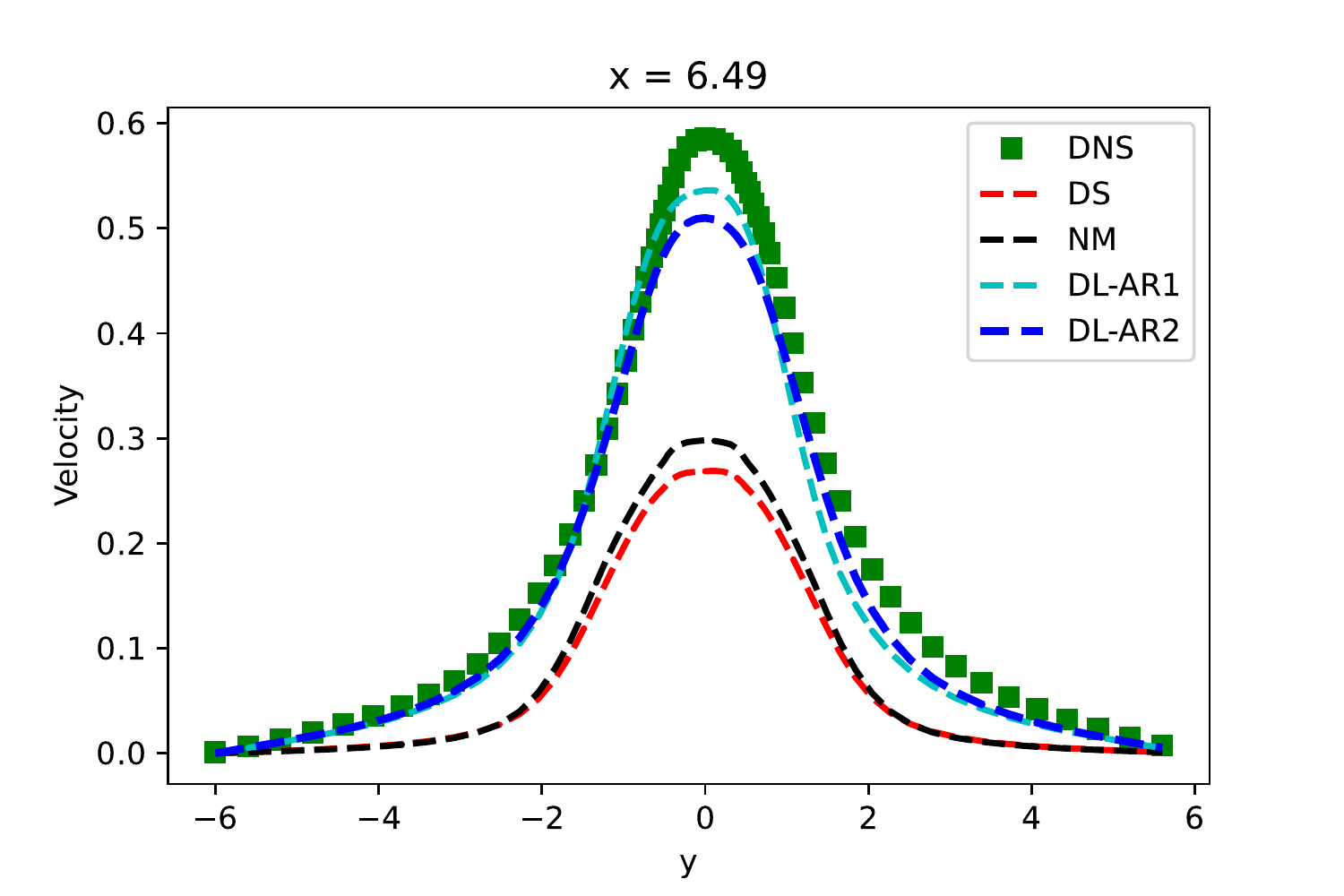}
\includegraphics[width=5cm]{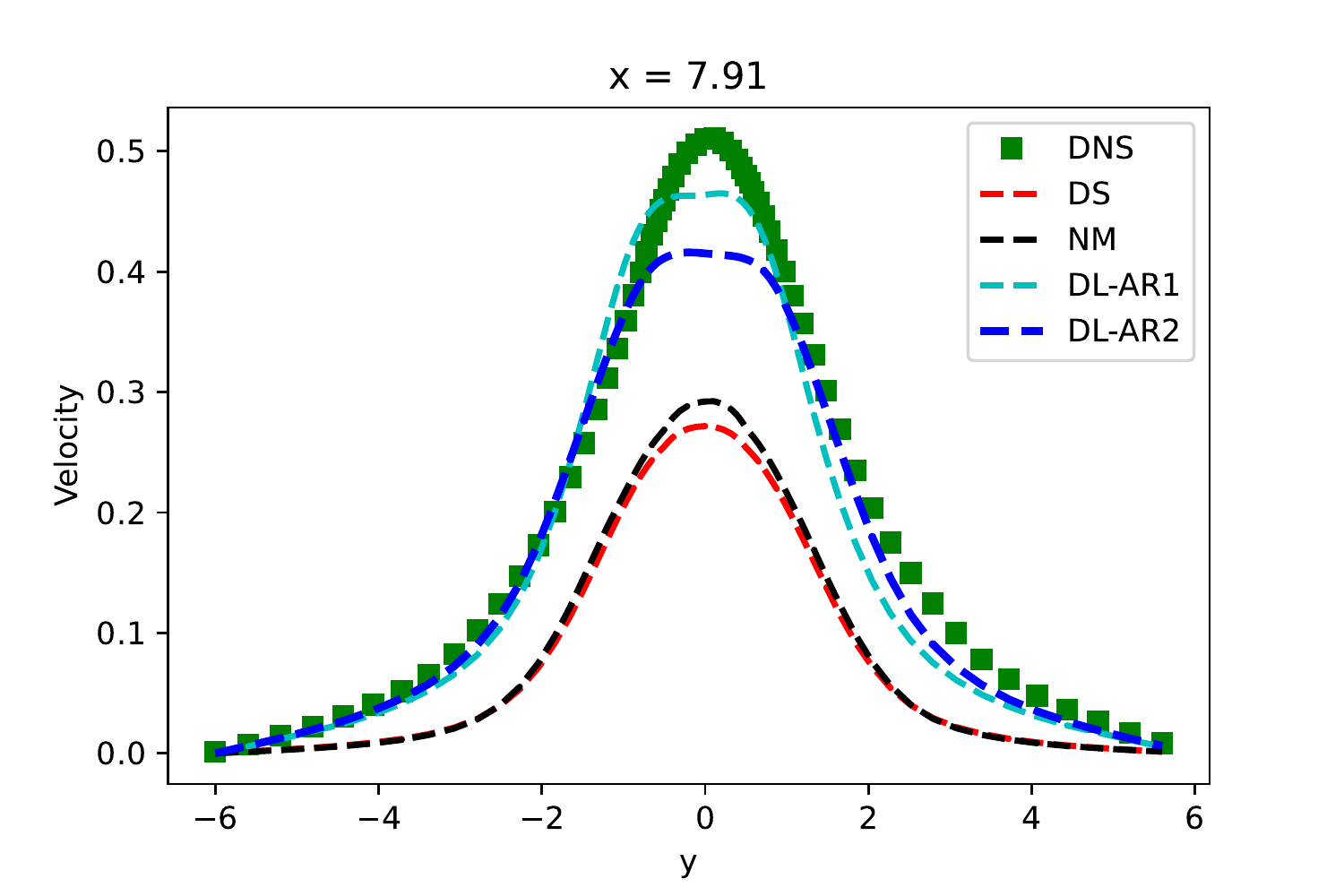}
\includegraphics[width=5cm]{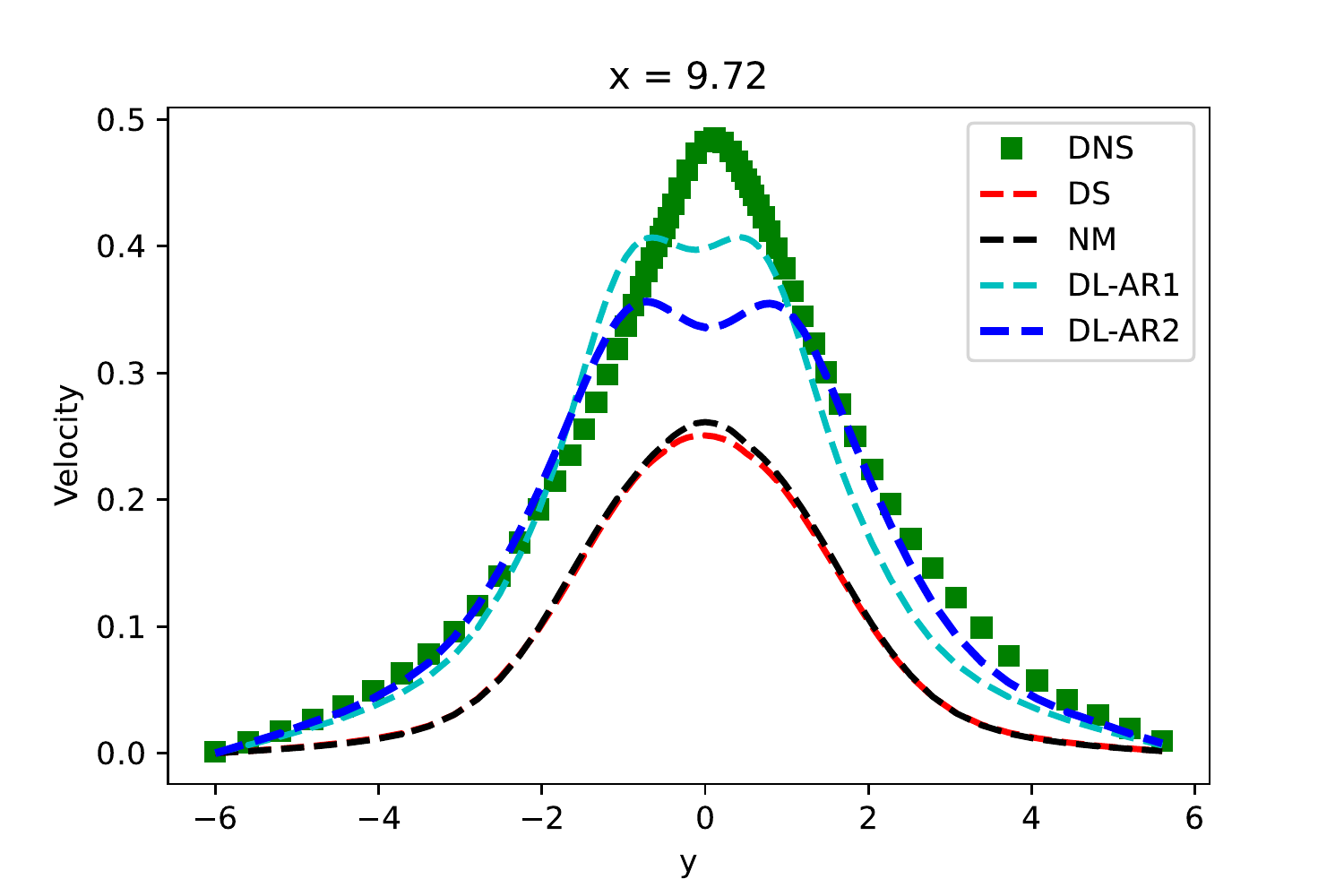}
\includegraphics[width=5cm]{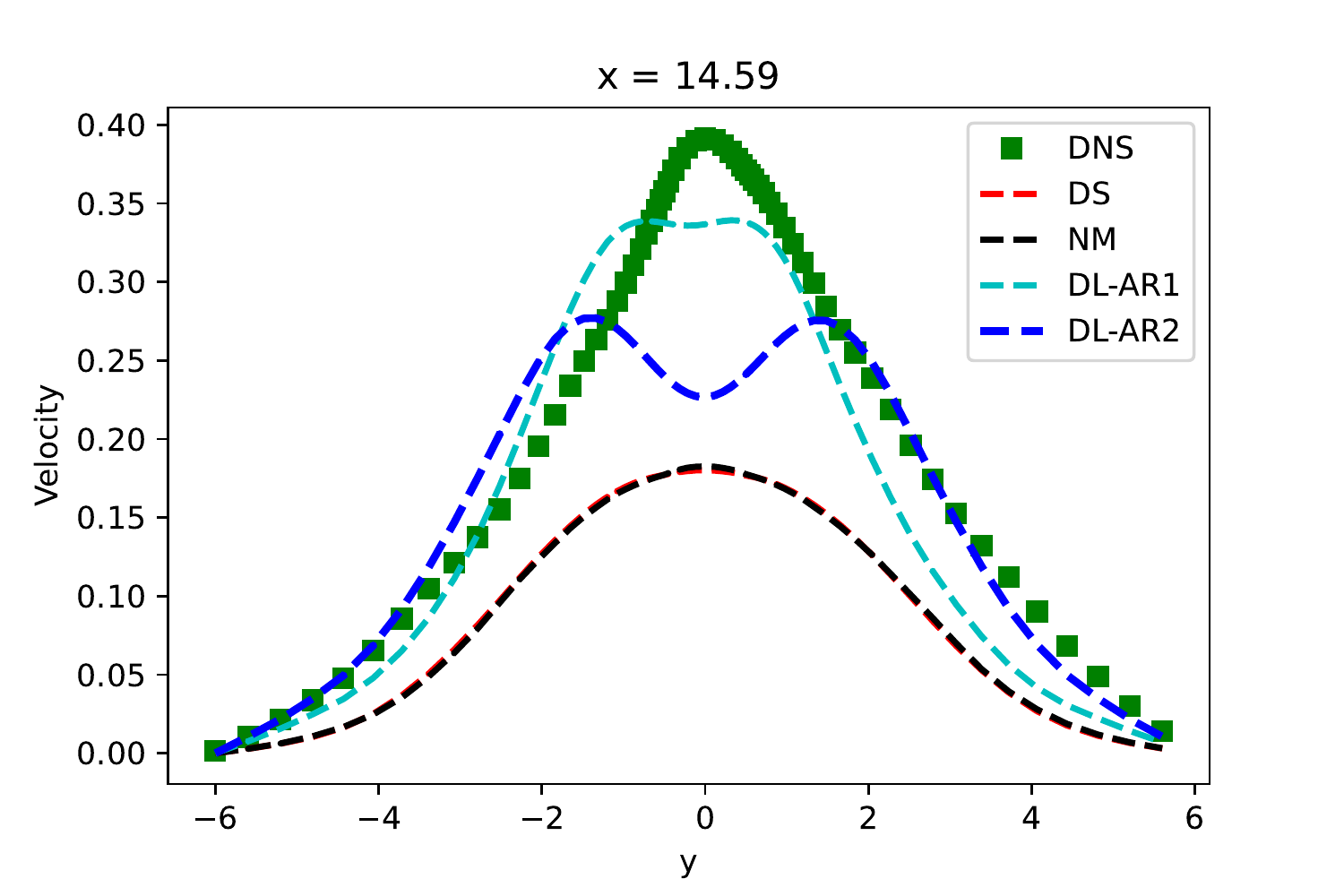}
\label{f1}
\caption{RMS profile for $u_2$ for AR2-Re$1,000$ configuration.}
\end{figure}

\begin{figure}[H]
\centering
\includegraphics[width=5cm]{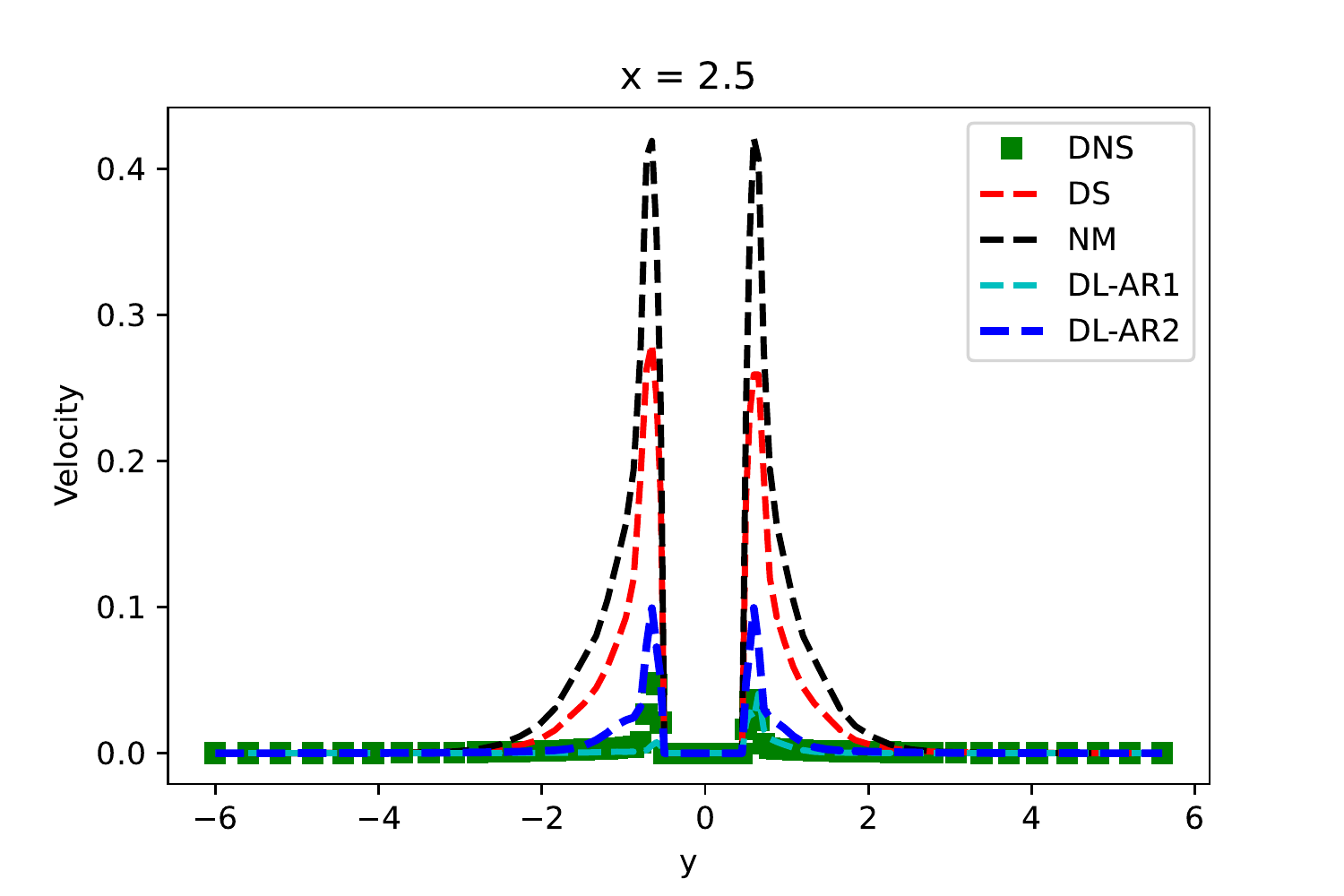}
\includegraphics[width=5cm]{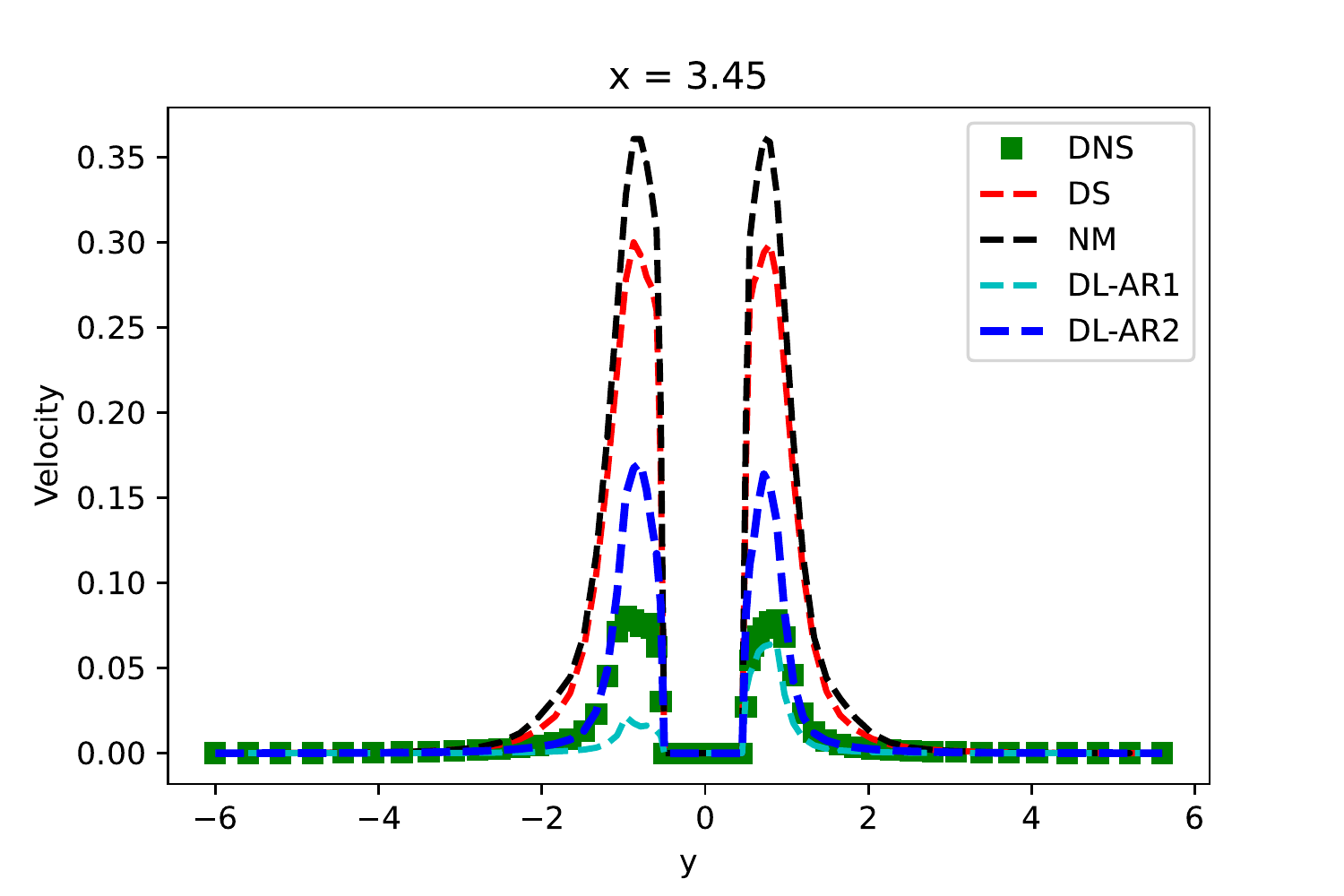}
\includegraphics[width=5cm]{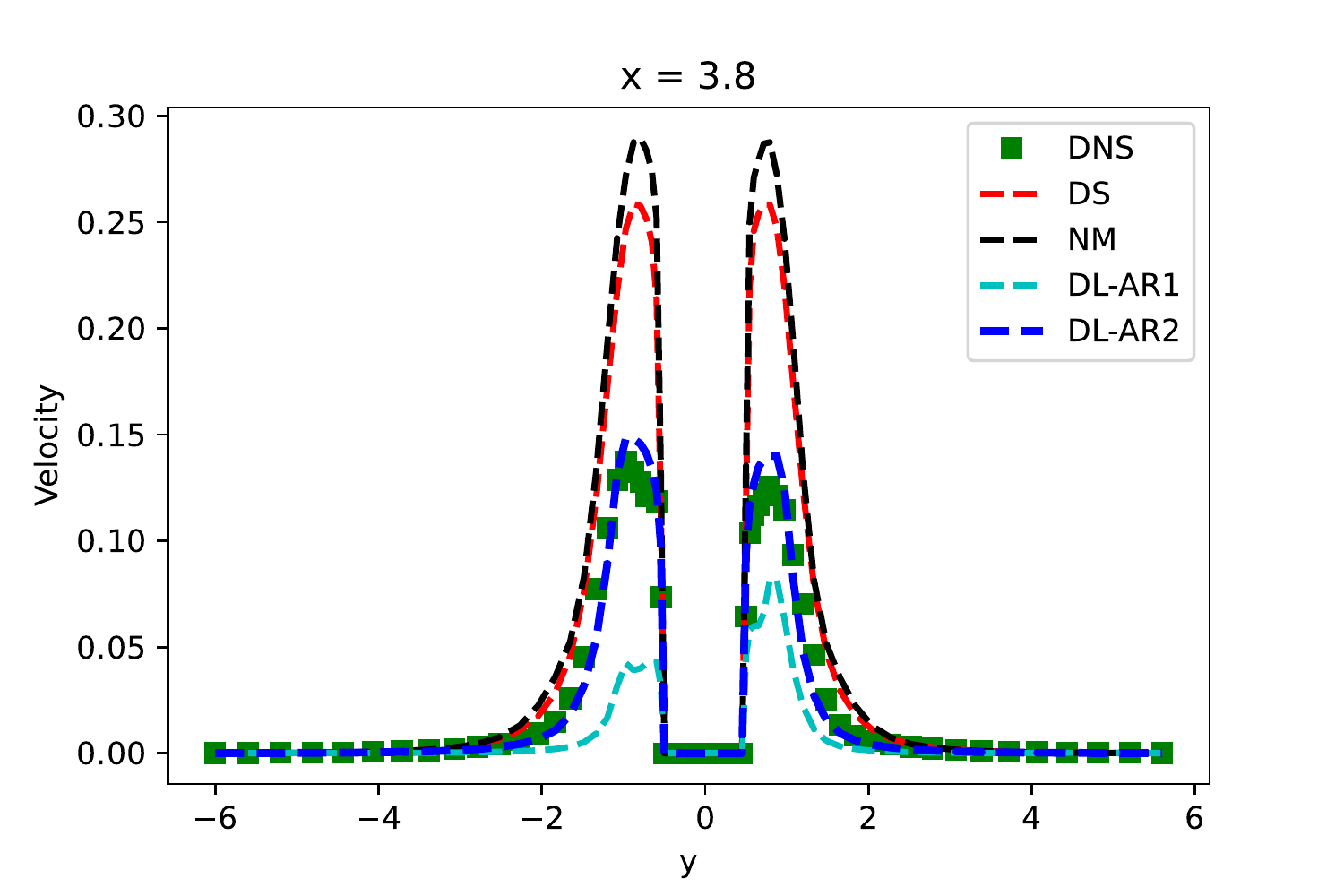}
\includegraphics[width=5cm]{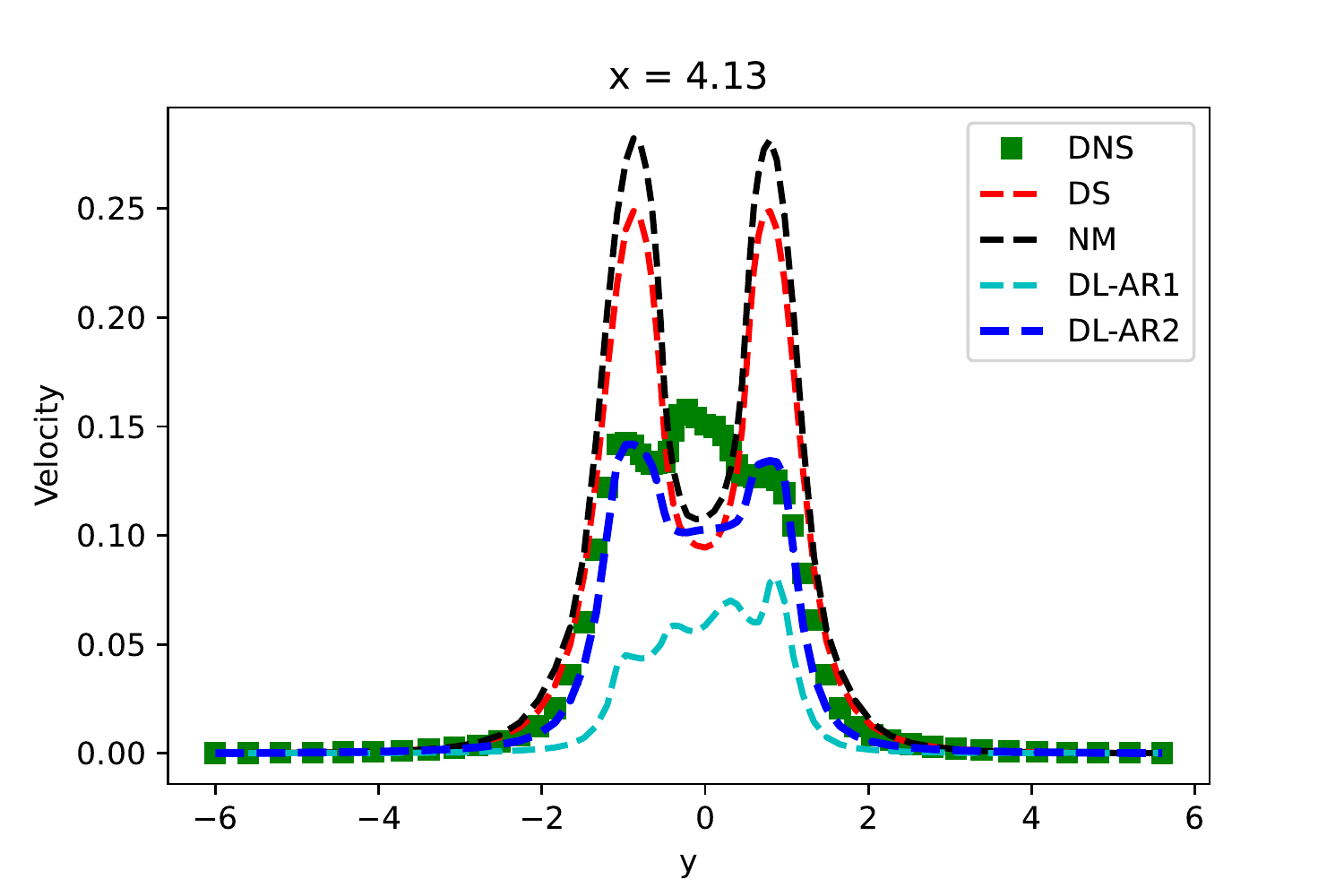}
\includegraphics[width=5cm]{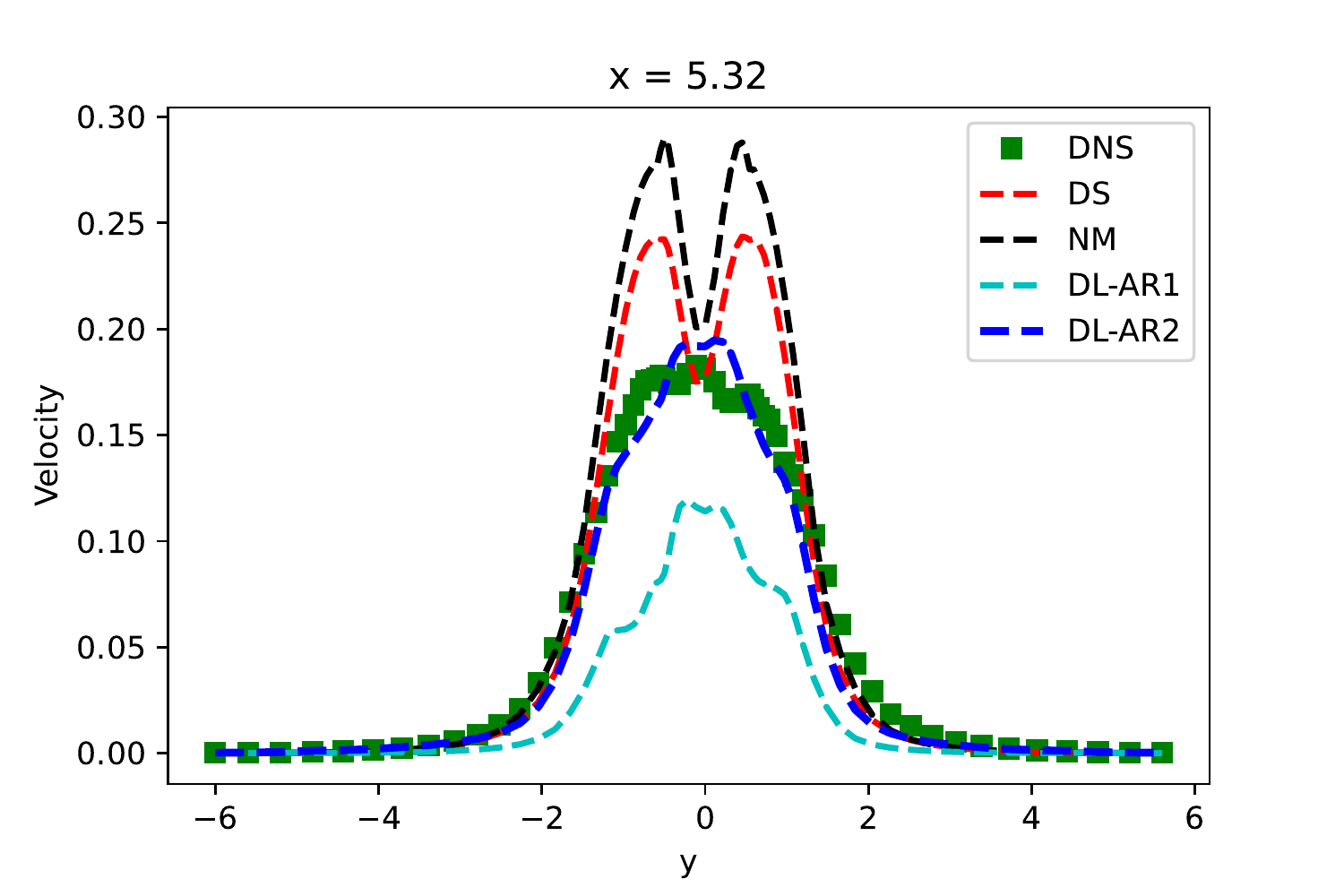}
\includegraphics[width=5cm]{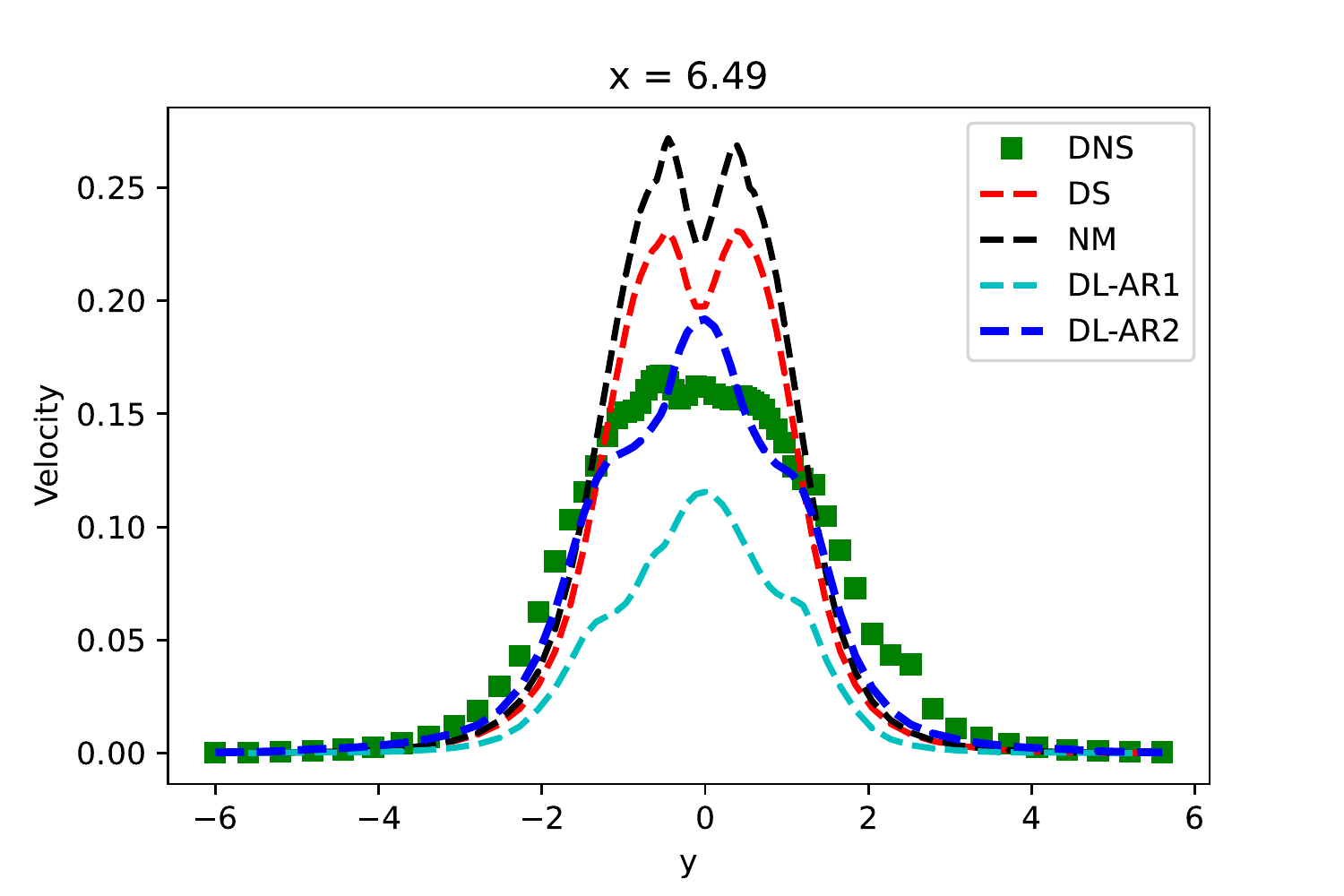}
\includegraphics[width=5cm]{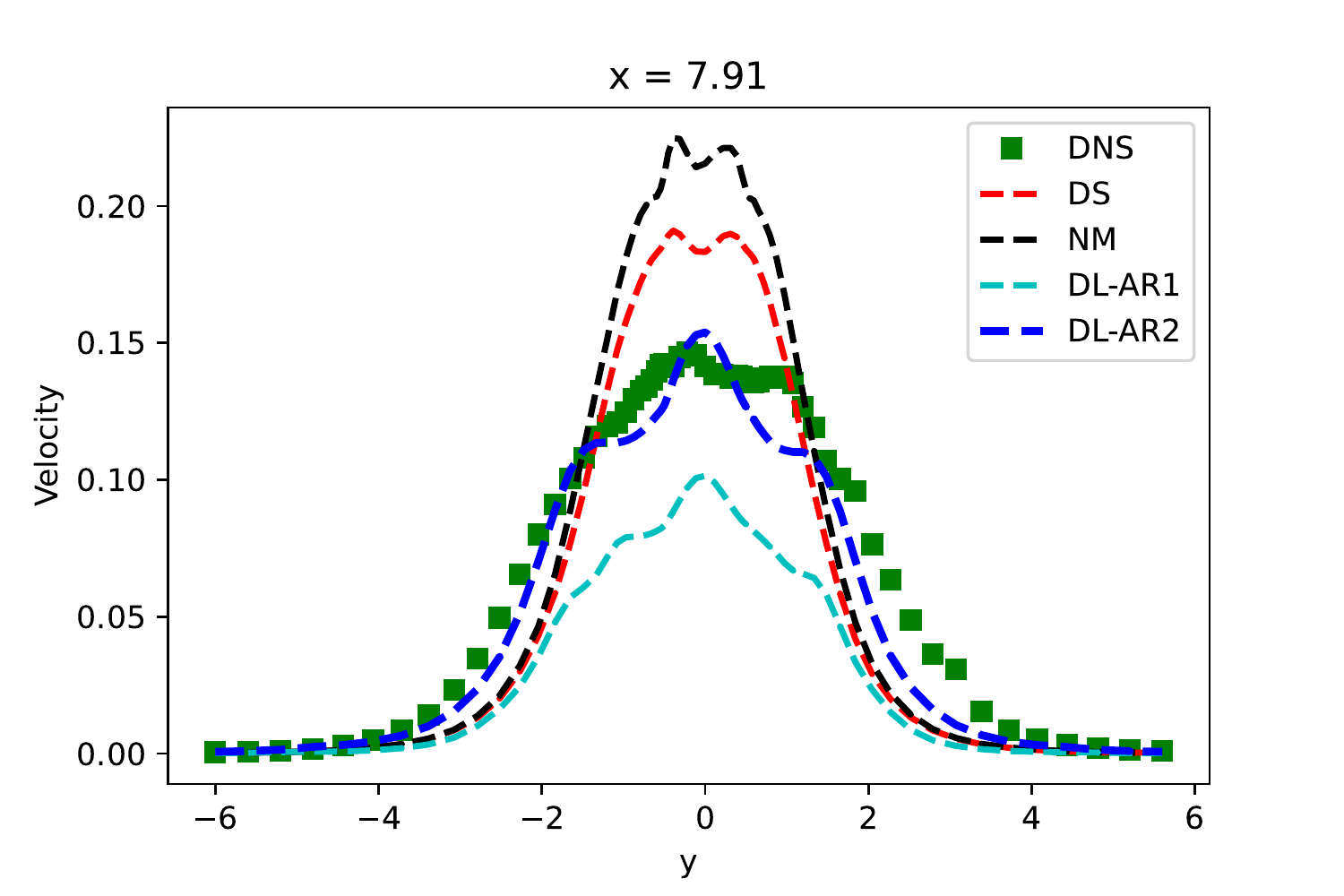}
\includegraphics[width=5cm]{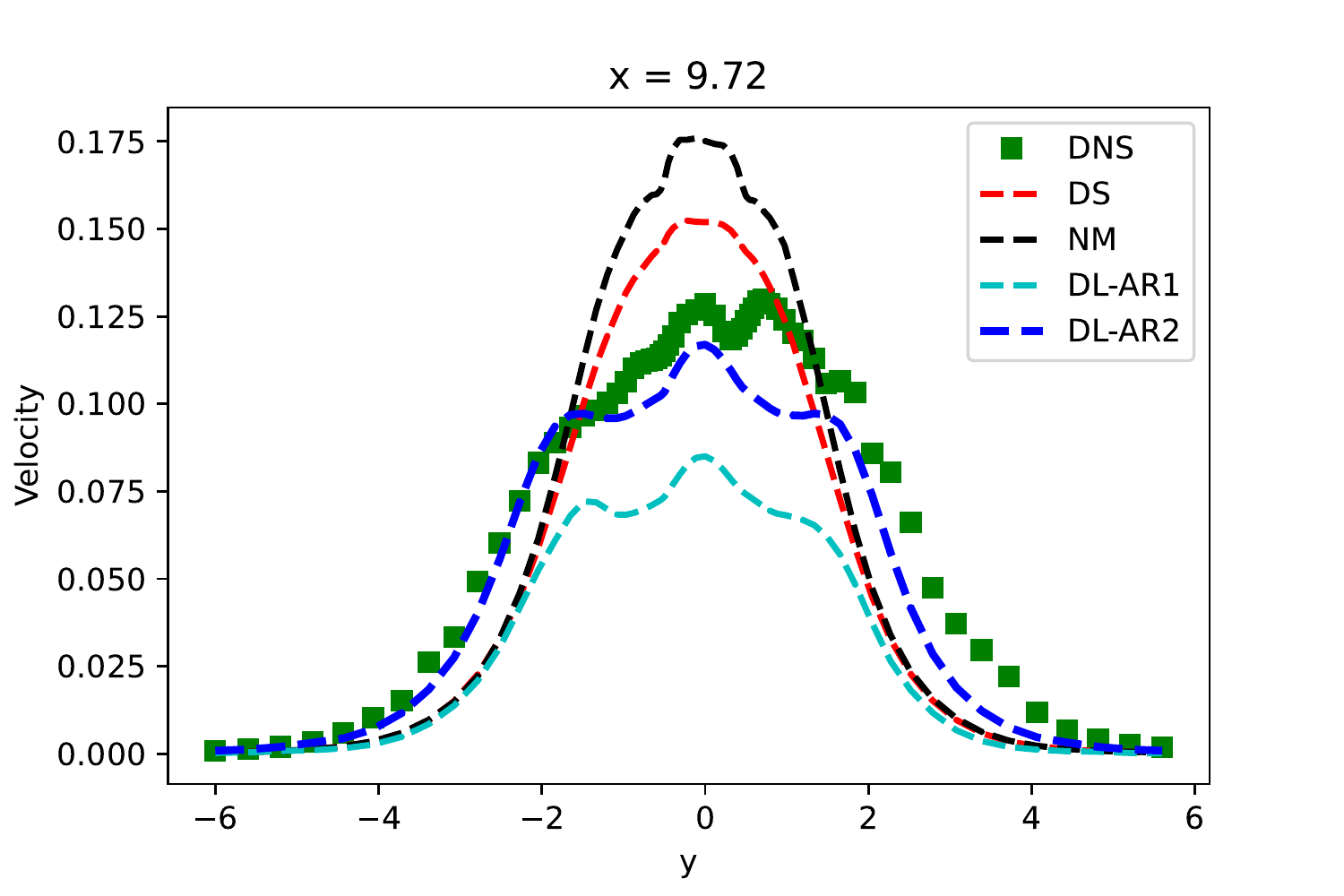}
\includegraphics[width=5cm]{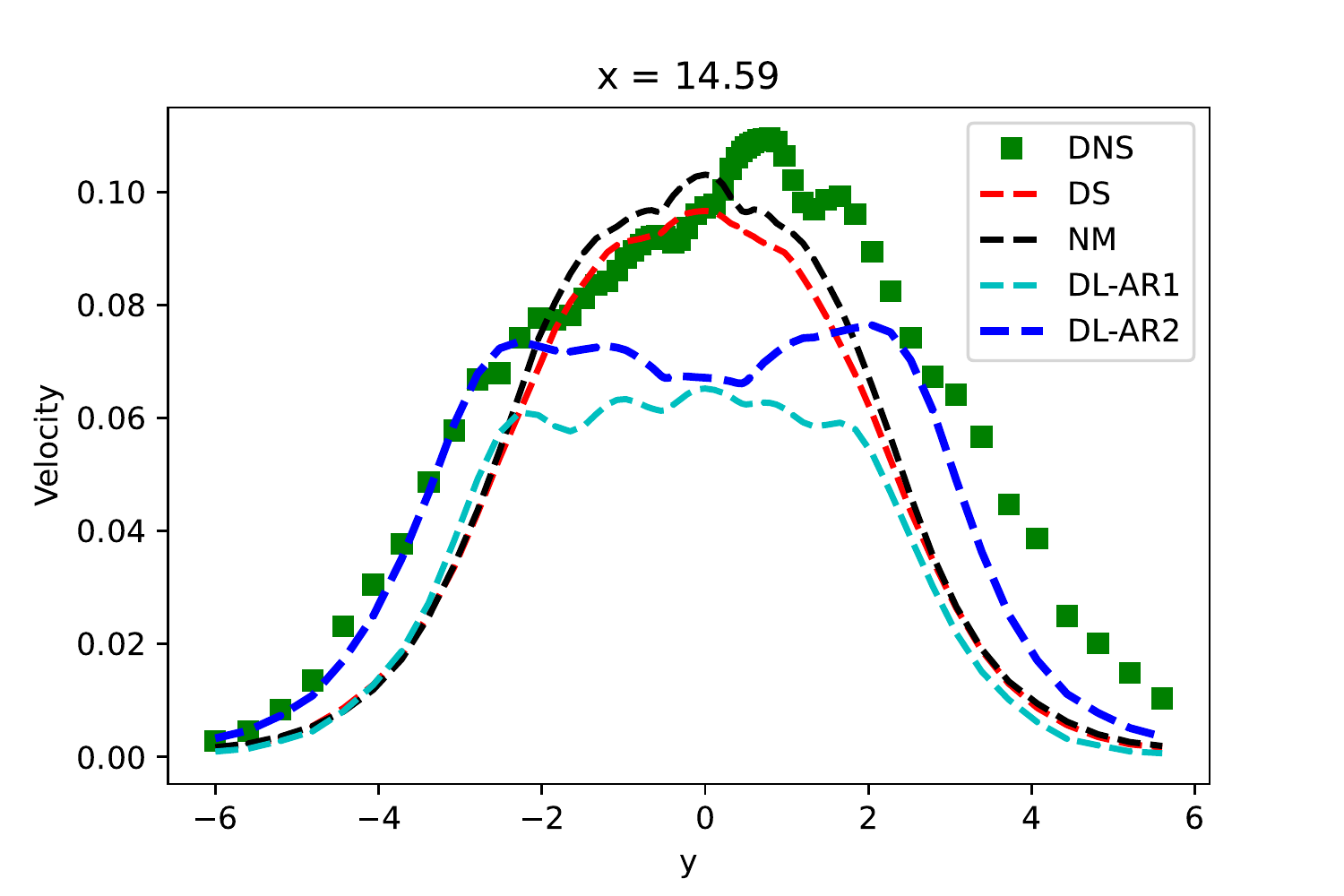}
\label{f1}
\caption{RMS profile for $u_3$ for AR2-Re$1,000$ configuration.}
\end{figure}

\begin{figure}[H]
\centering
\includegraphics[width=5cm]{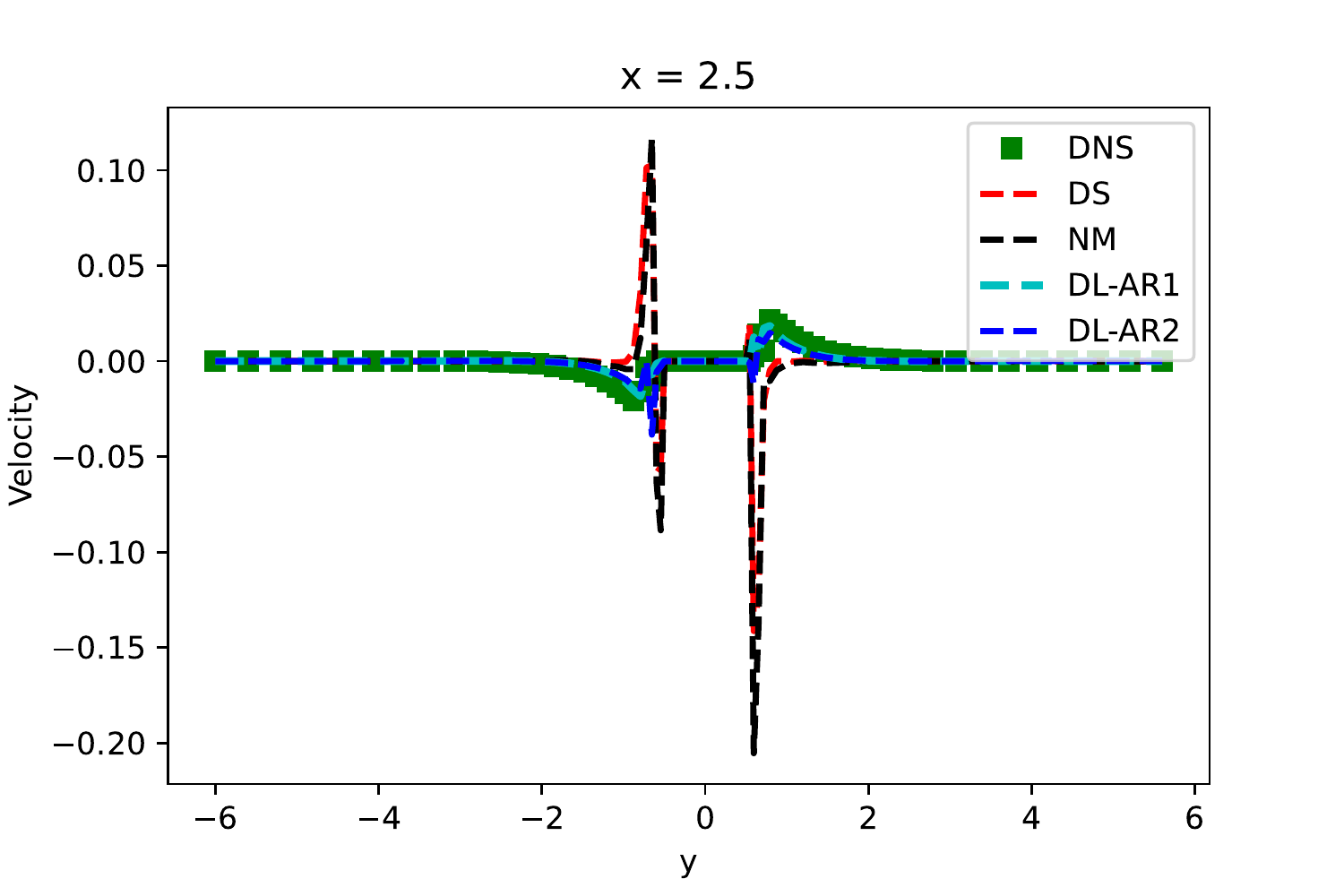}
\includegraphics[width=5cm]{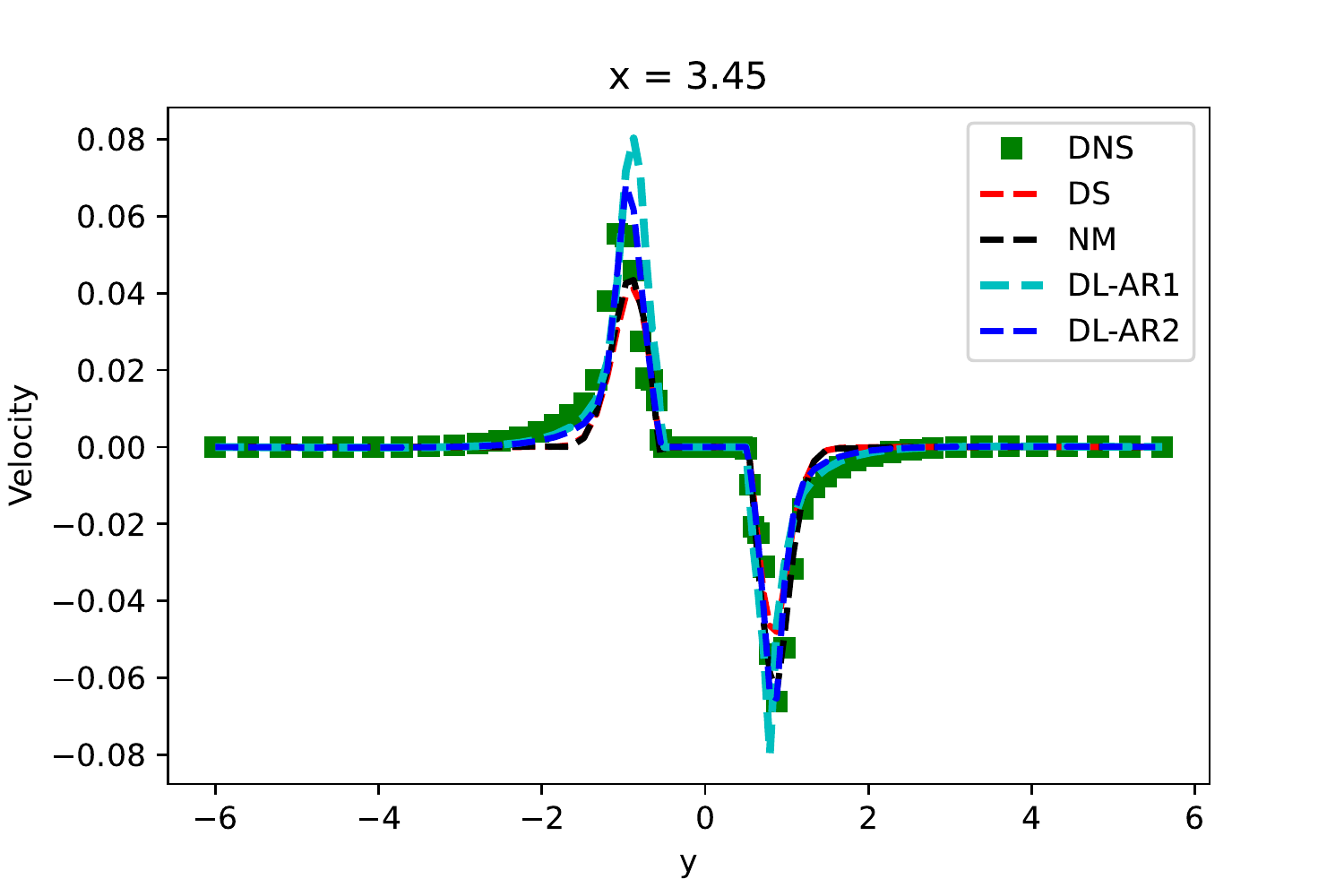}
\includegraphics[width=5cm]{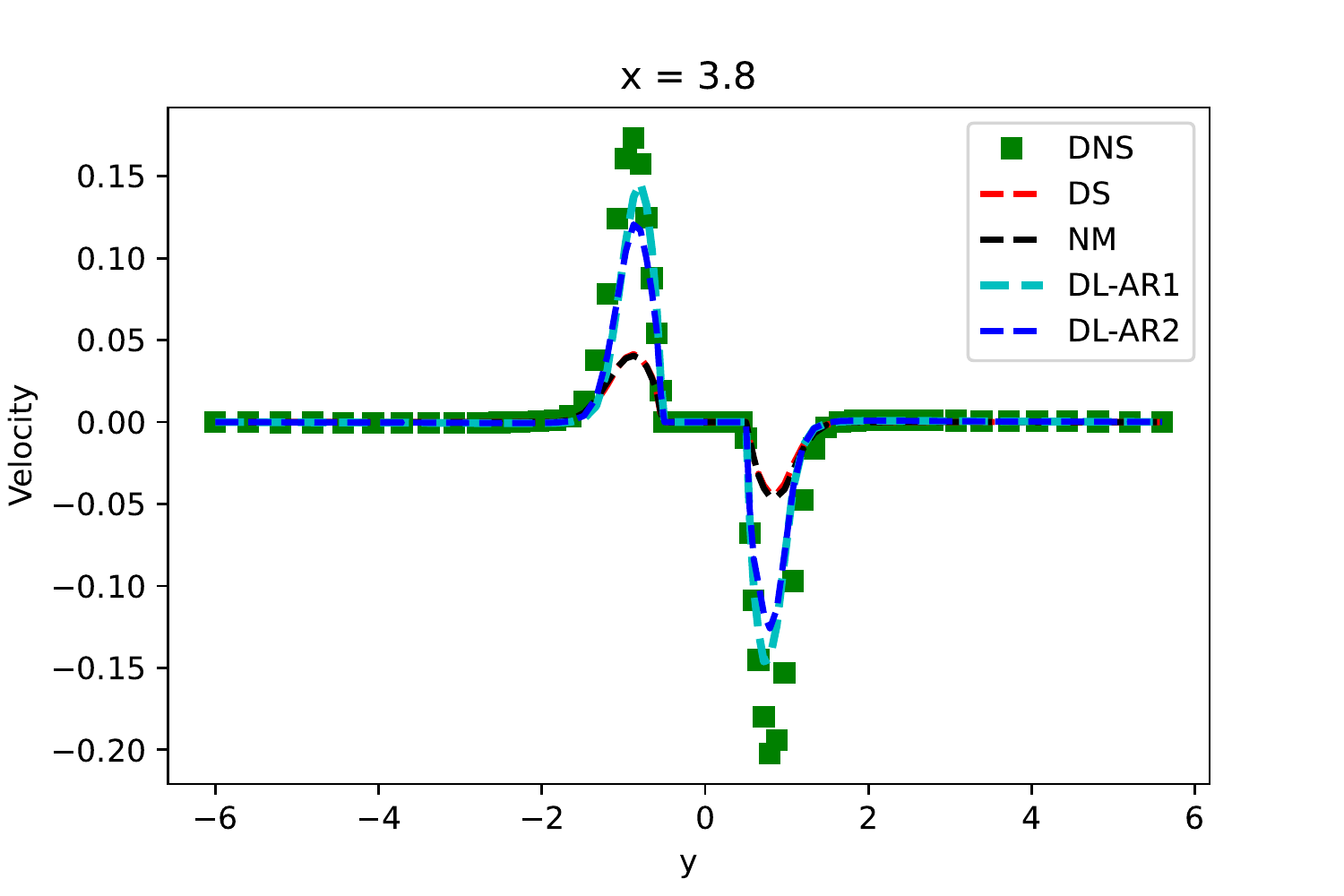}
\includegraphics[width=5cm]{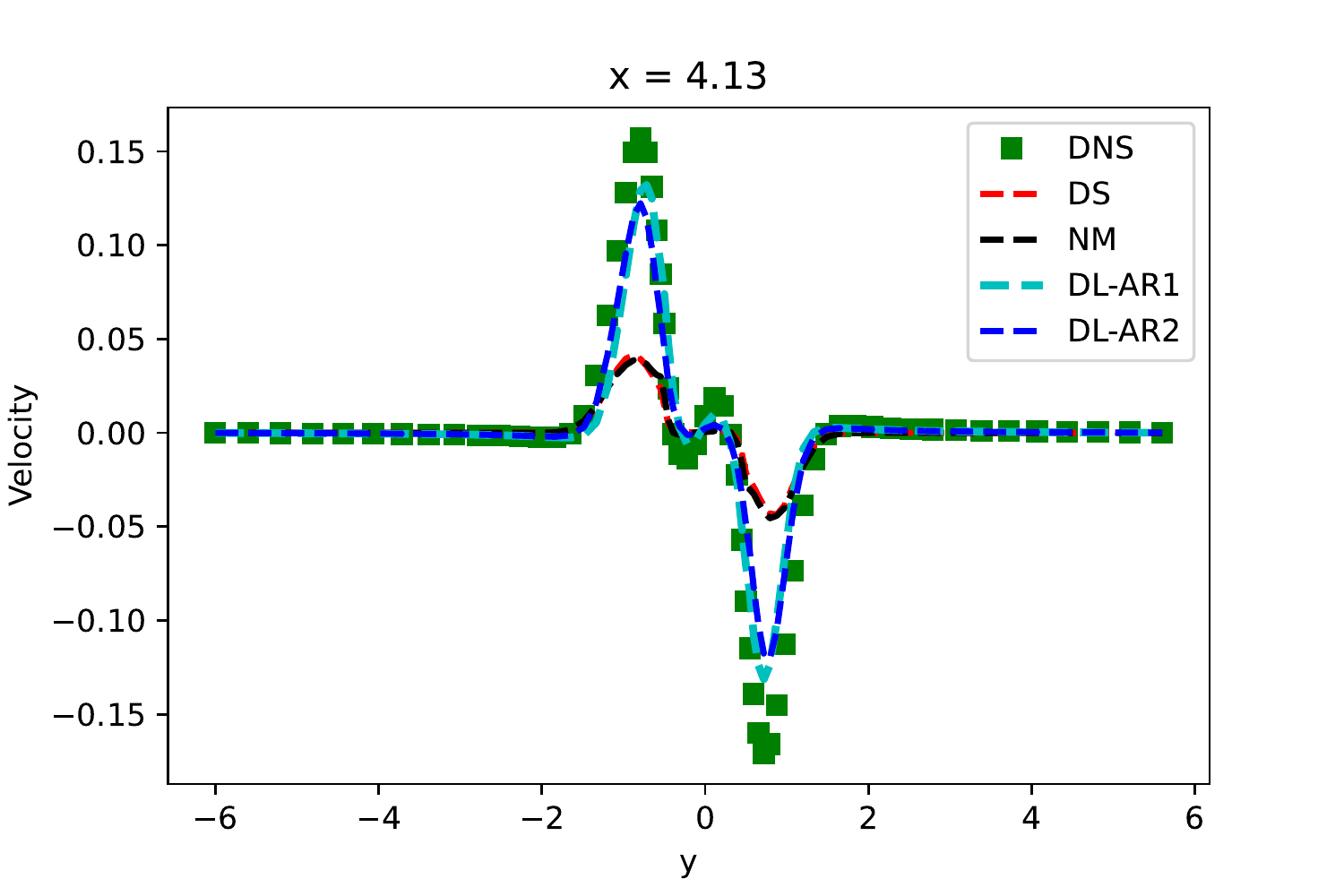}
\includegraphics[width=5cm]{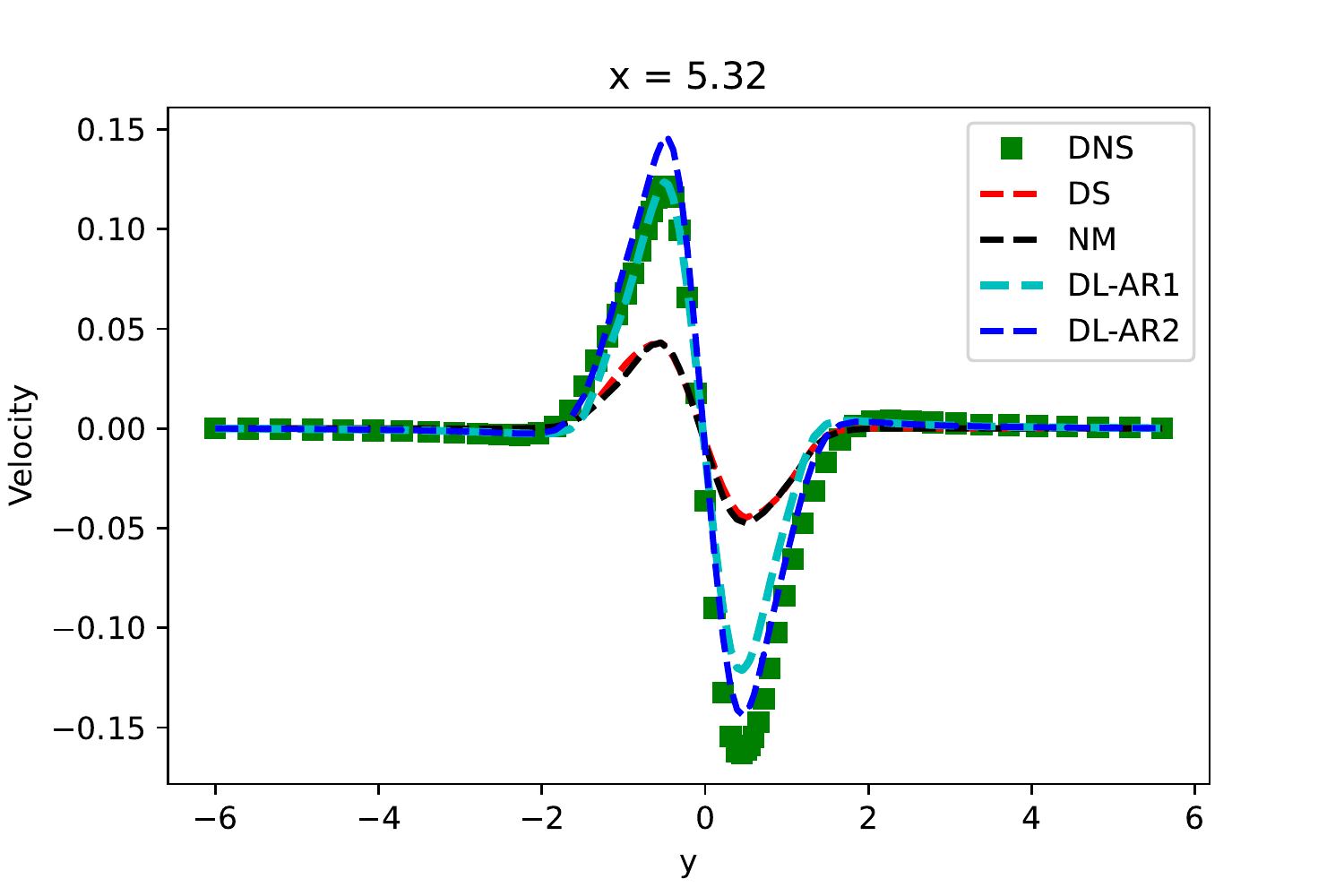}
\includegraphics[width=5cm]{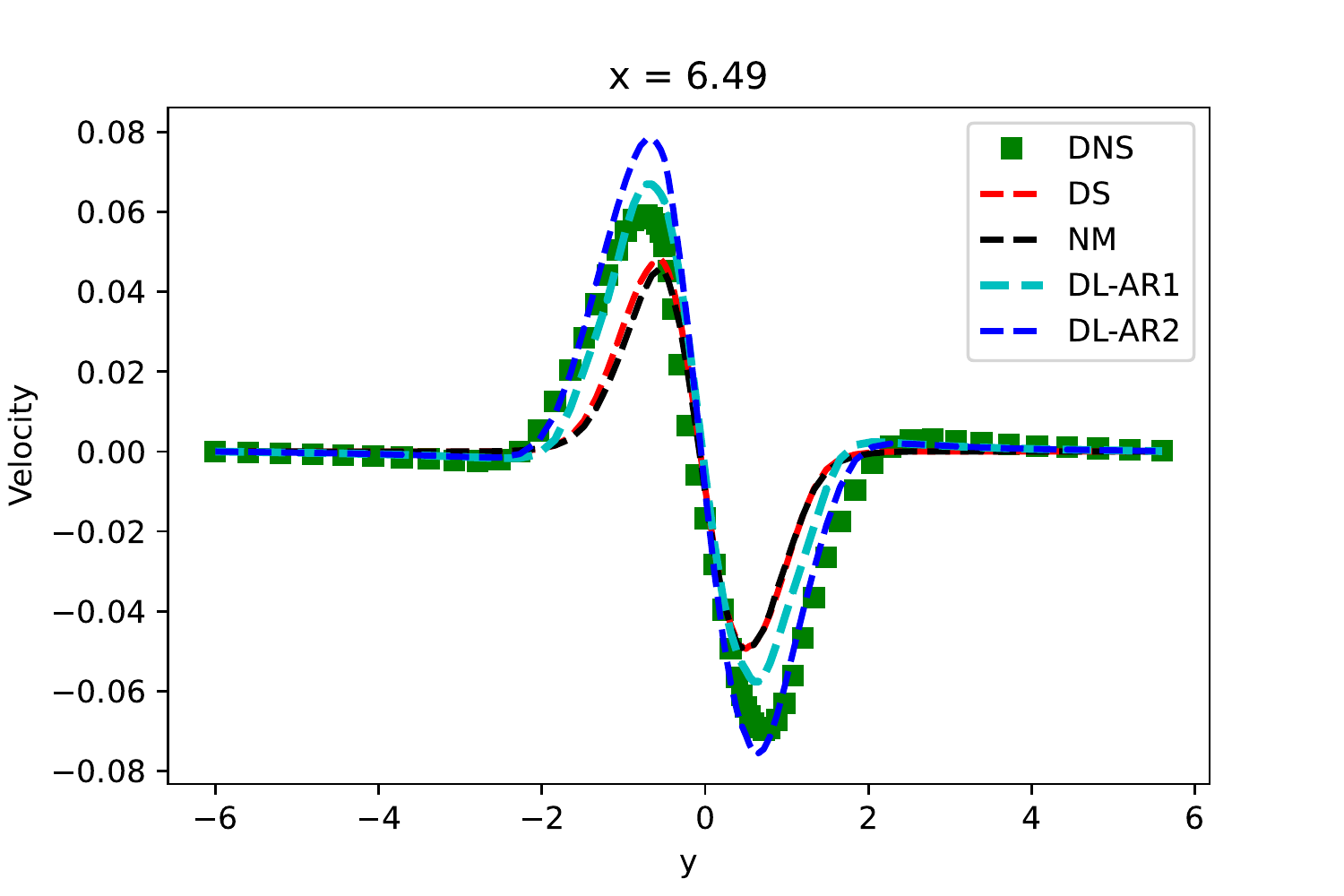}
\includegraphics[width=5cm]{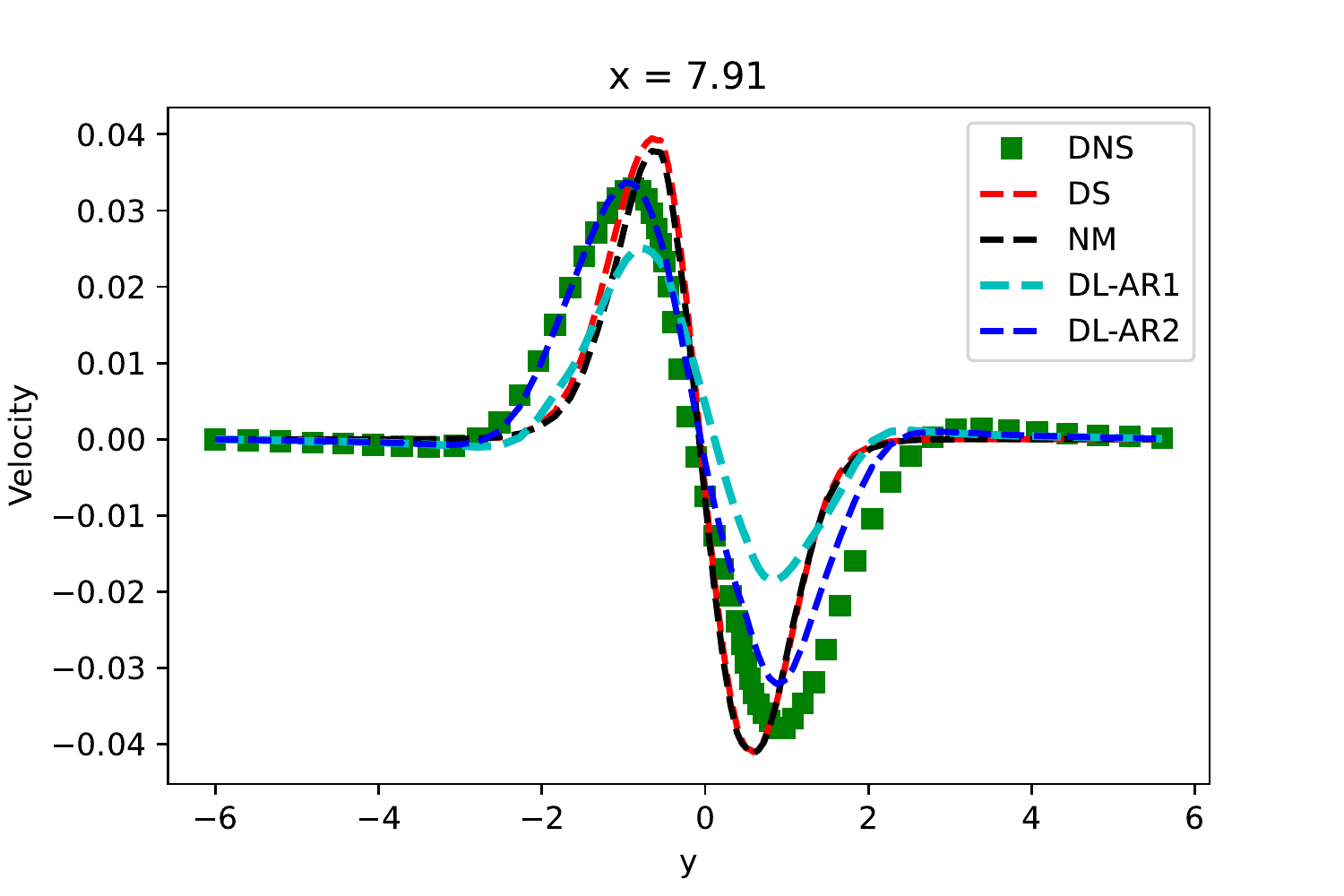}
\includegraphics[width=5cm]{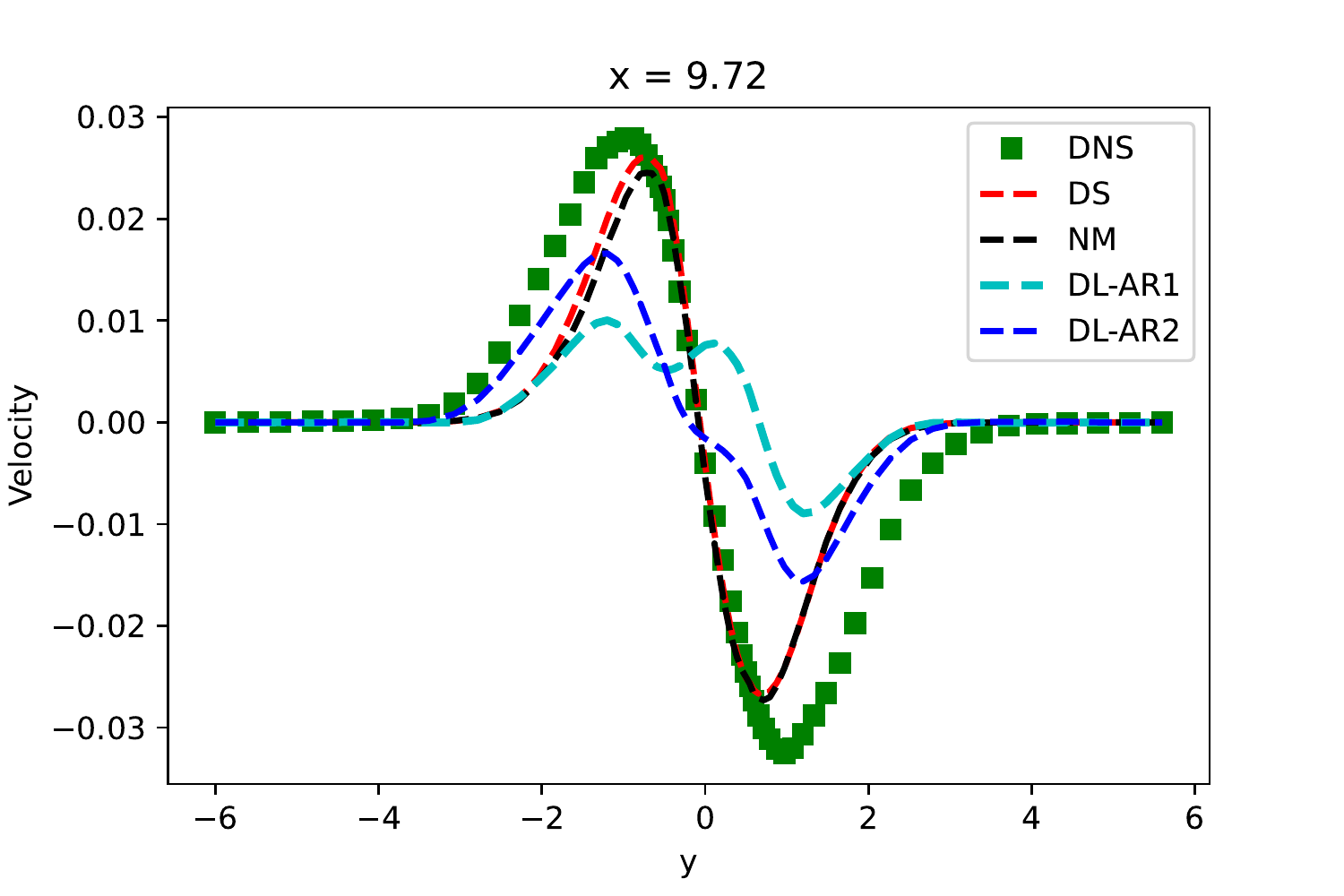}
\includegraphics[width=5cm]{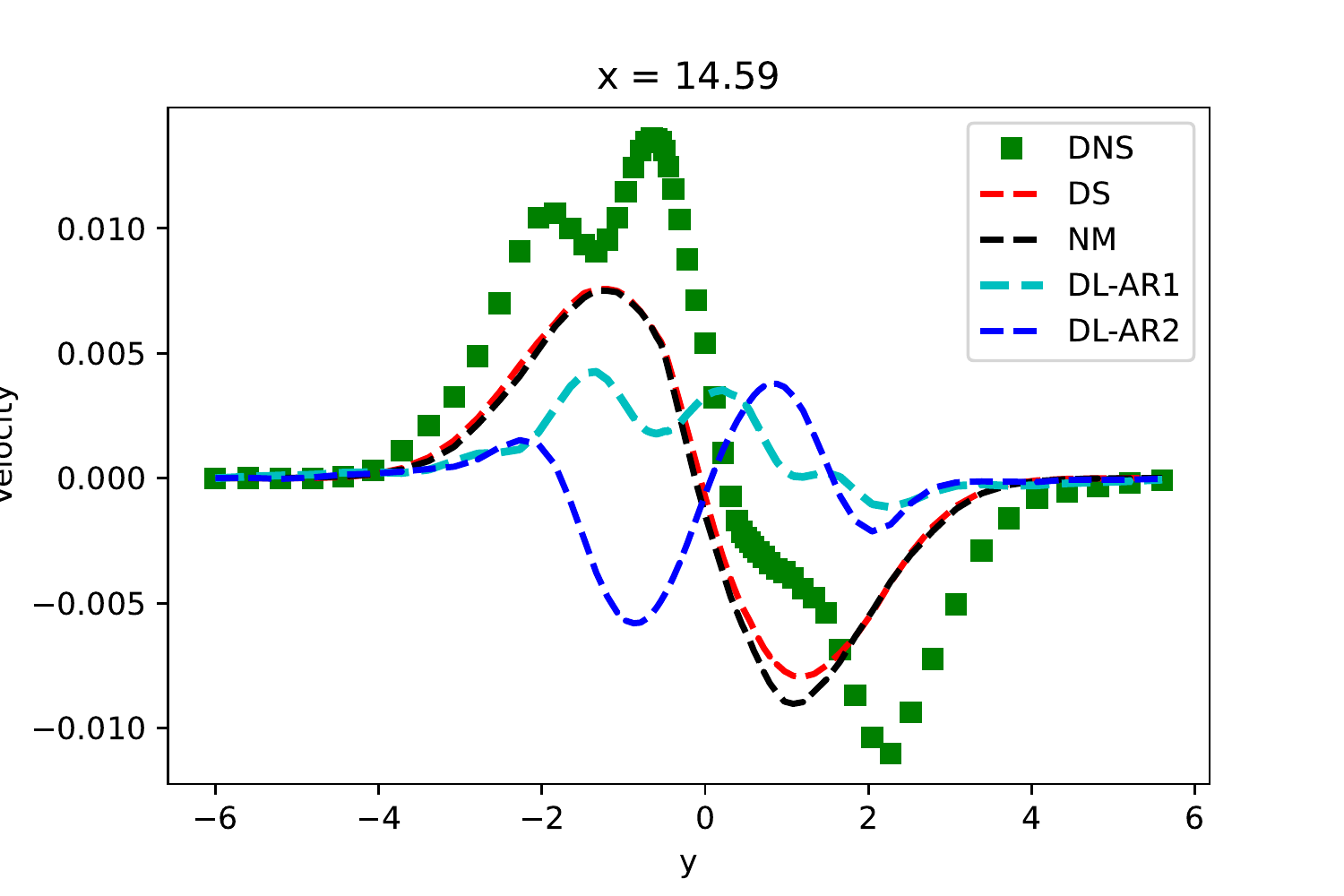}
\label{f1}
\caption{$\tau_{12}$ for AR2-Re$1,000$ configuration.}
\end{figure}

\subsubsection{AR2 $+$ Re$2,000$}

\begin{figure}[H]
\centering
\includegraphics[width=5cm]{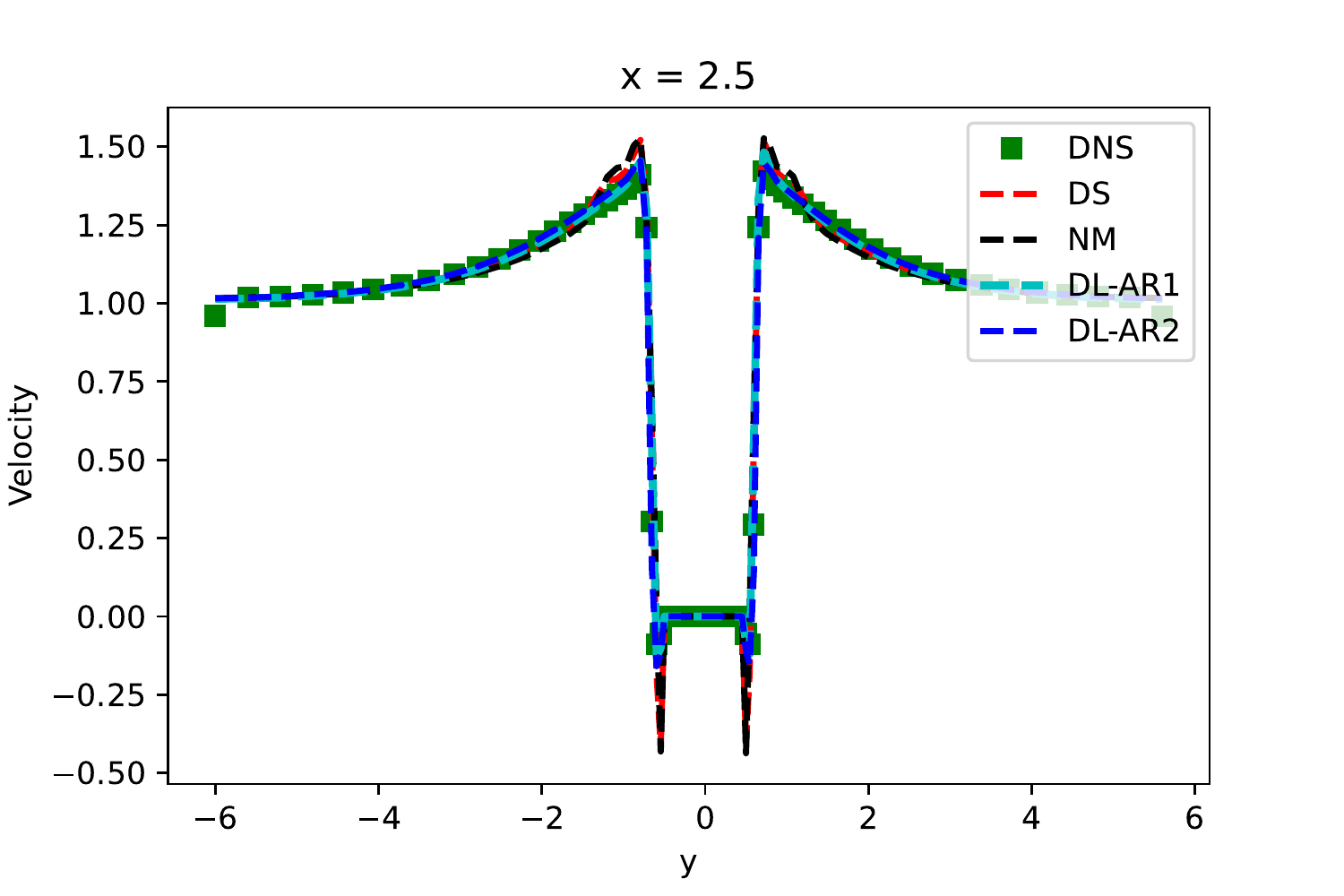}
\includegraphics[width=5cm]{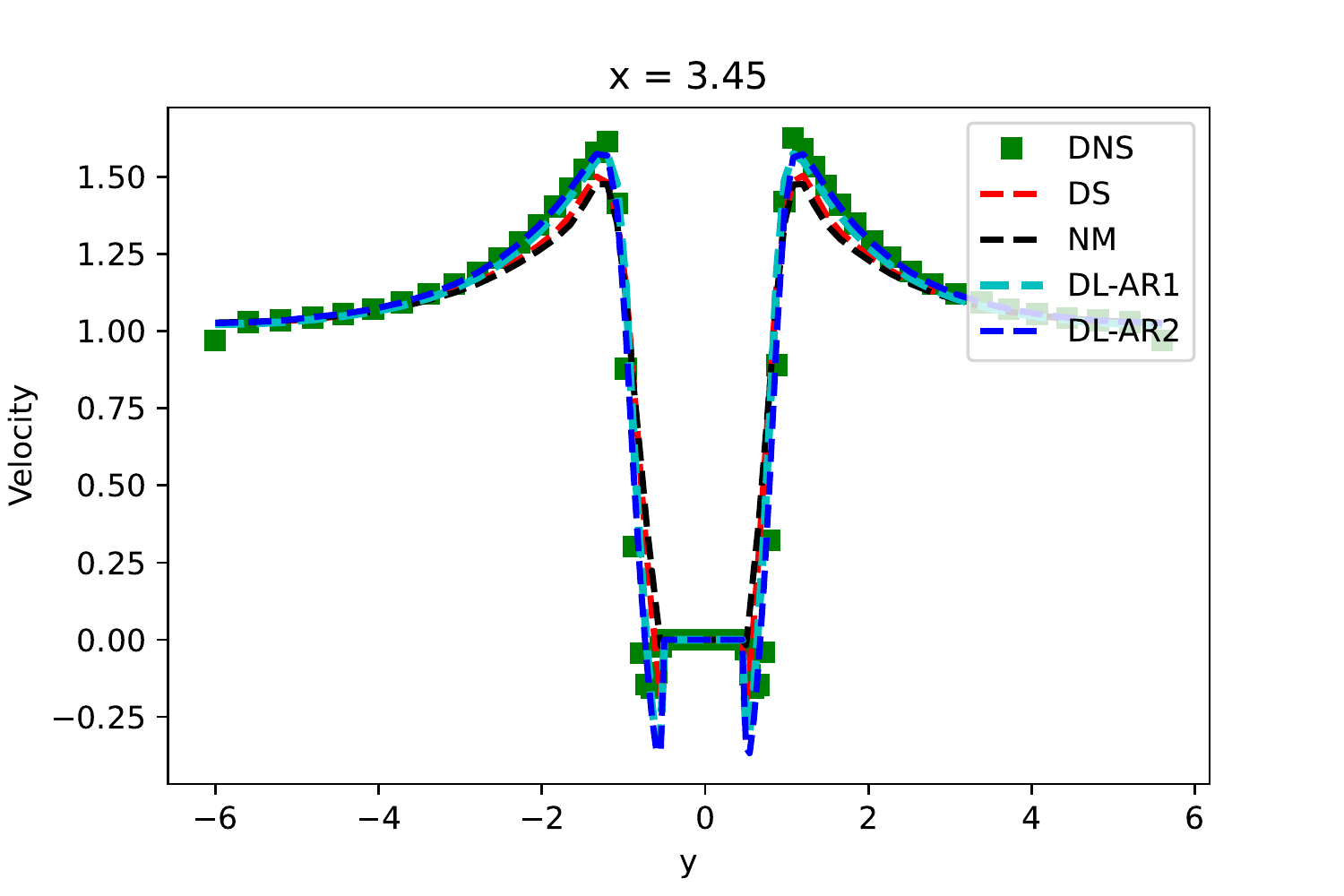}
\includegraphics[width=5cm]{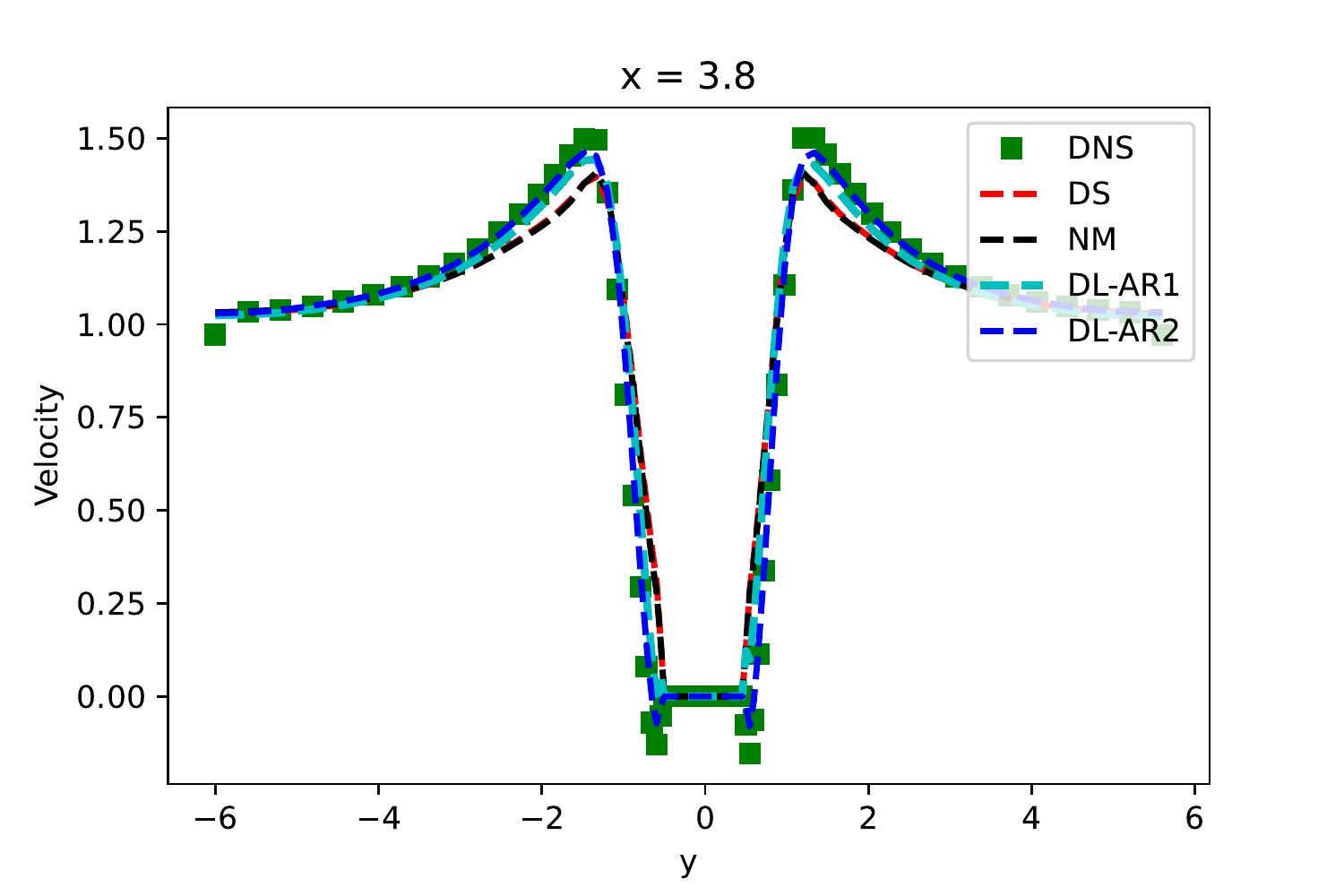}
\includegraphics[width=5cm]{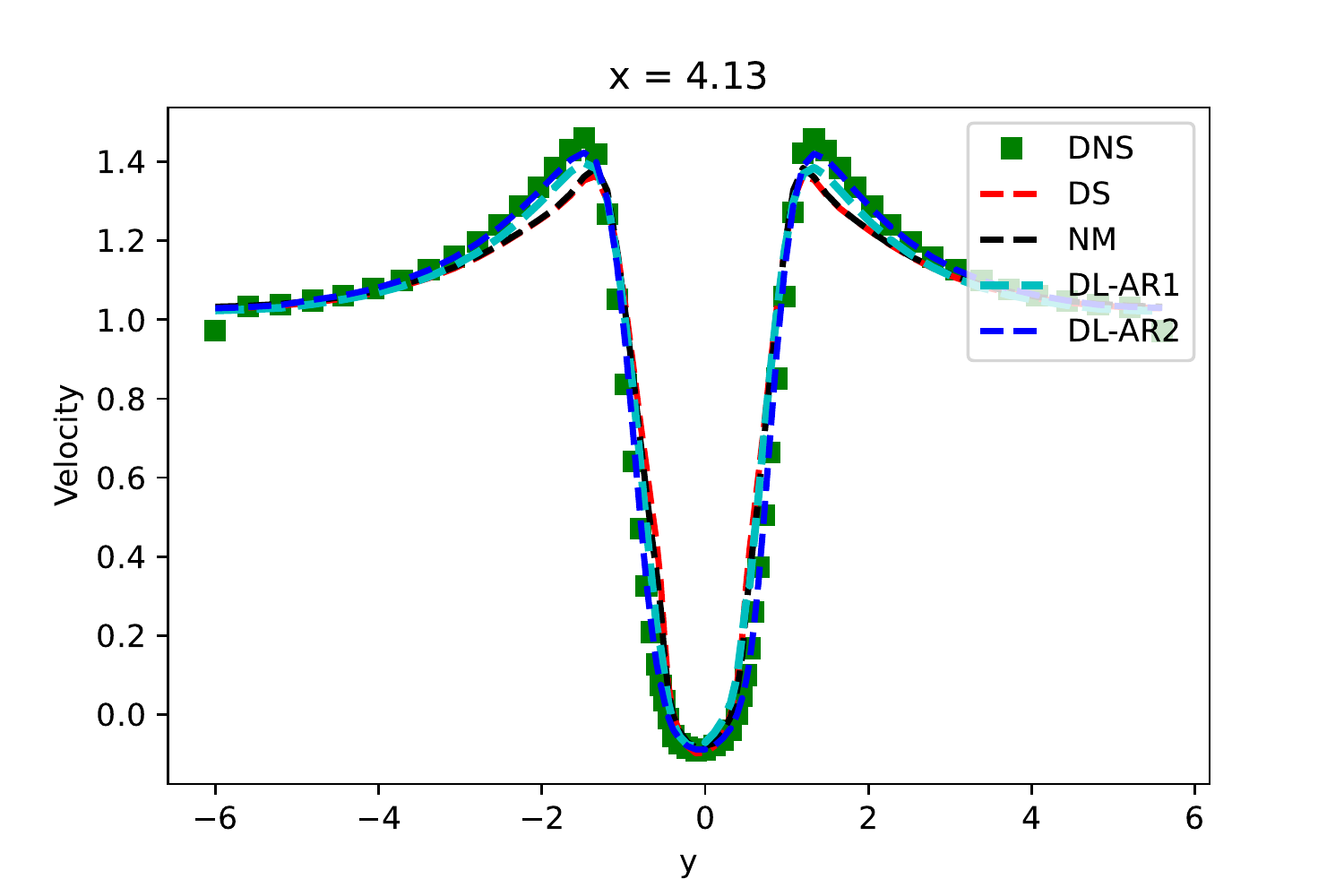}
\includegraphics[width=5cm]{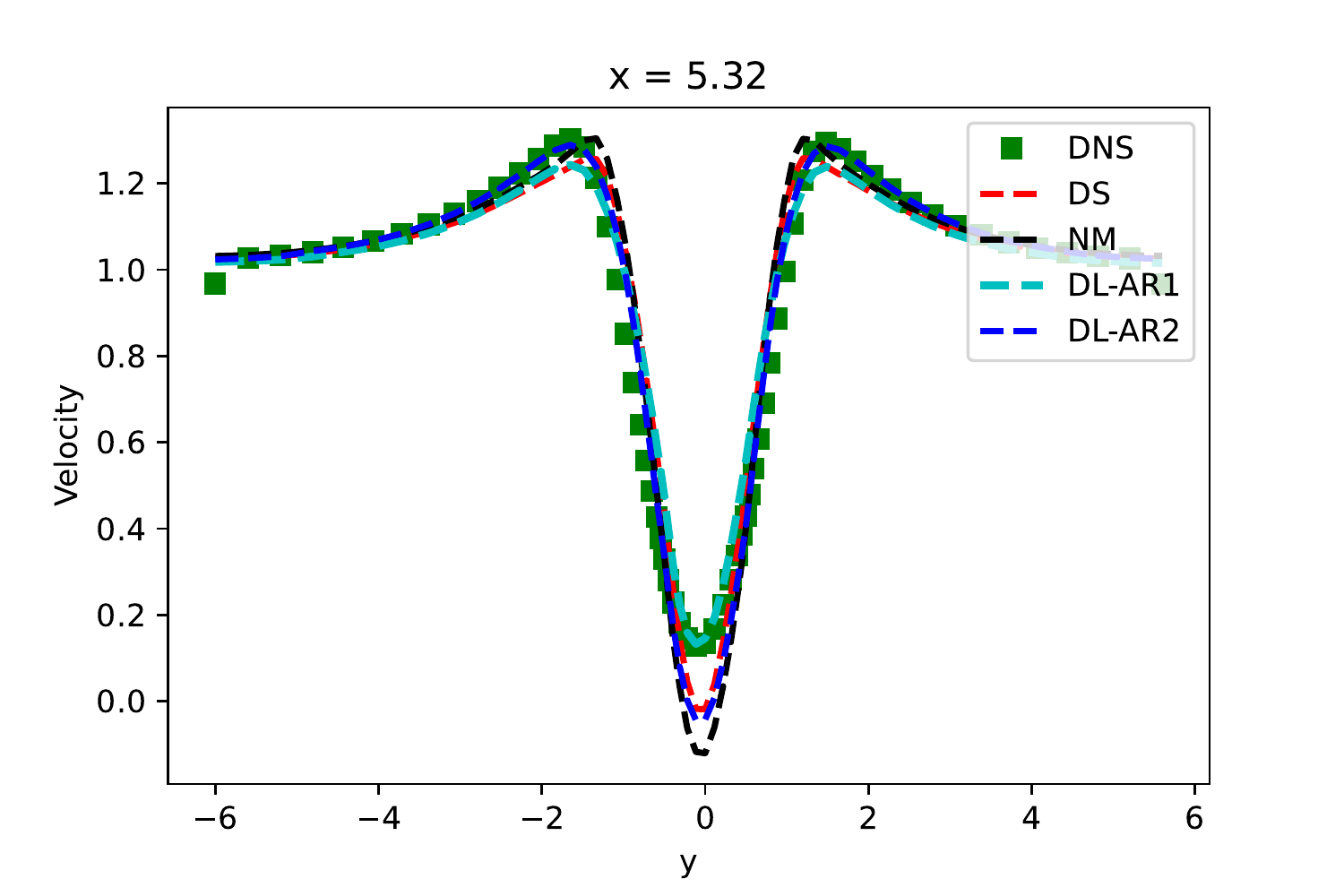}
\includegraphics[width=5cm]{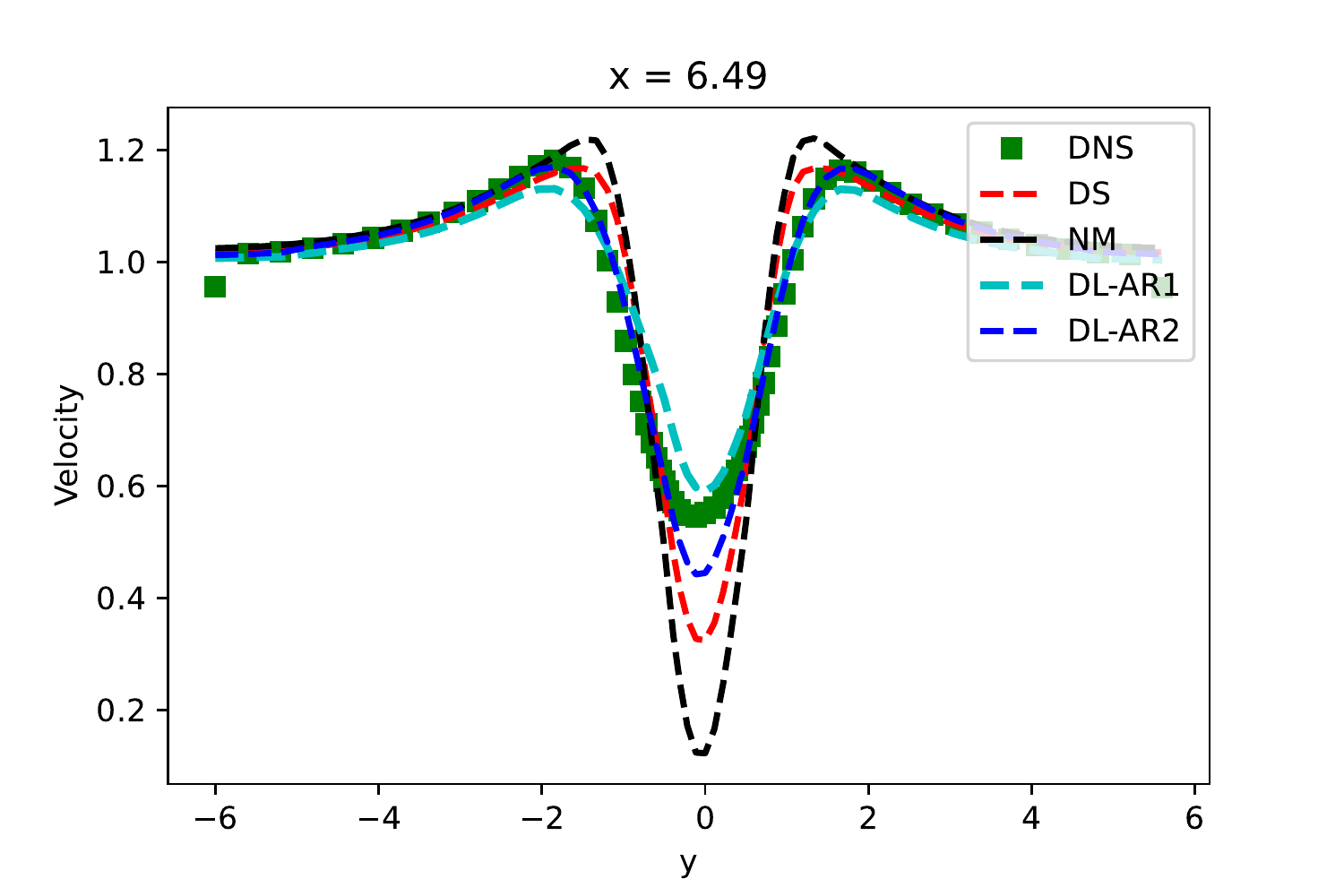}
\includegraphics[width=5cm]{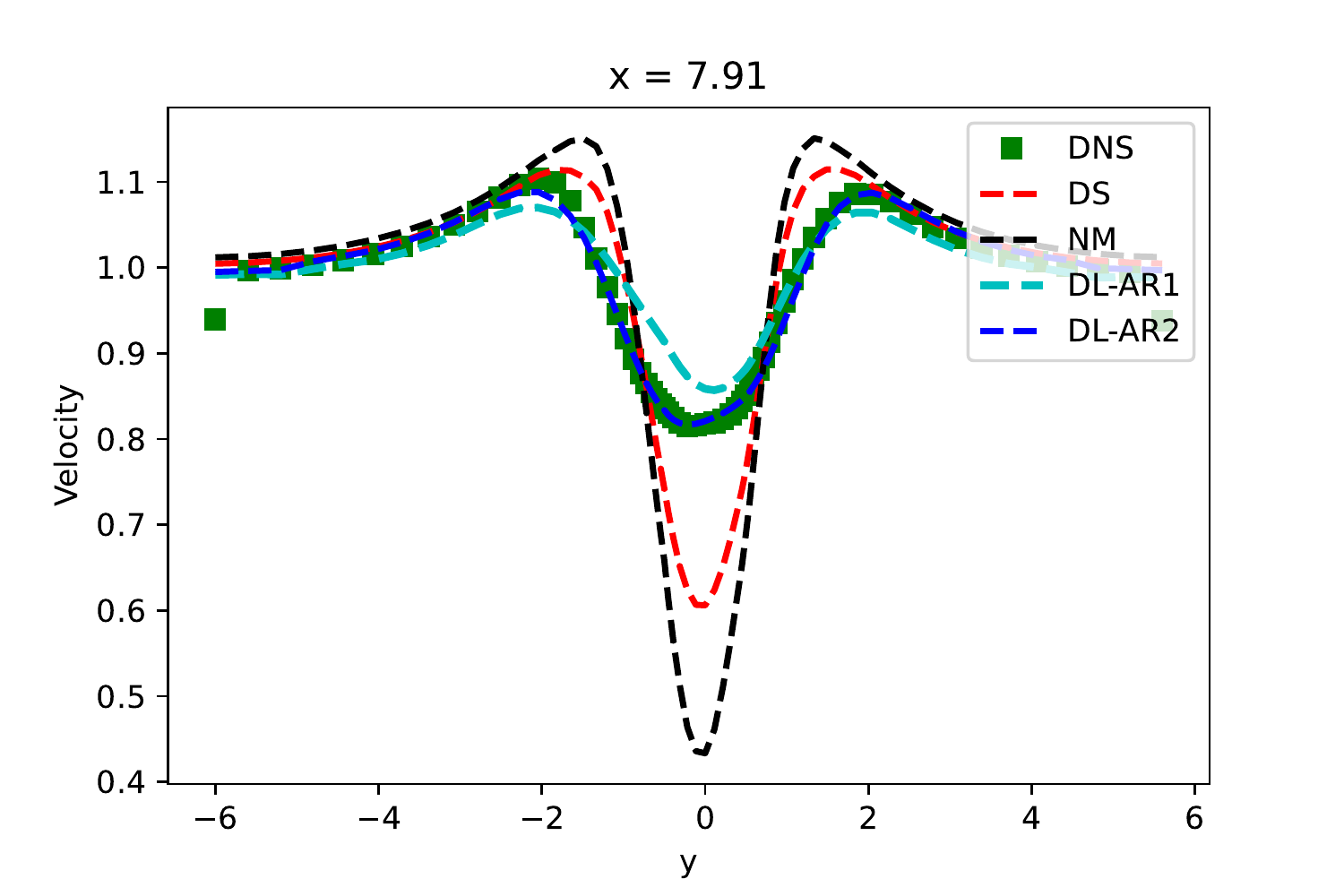}
\includegraphics[width=5cm]{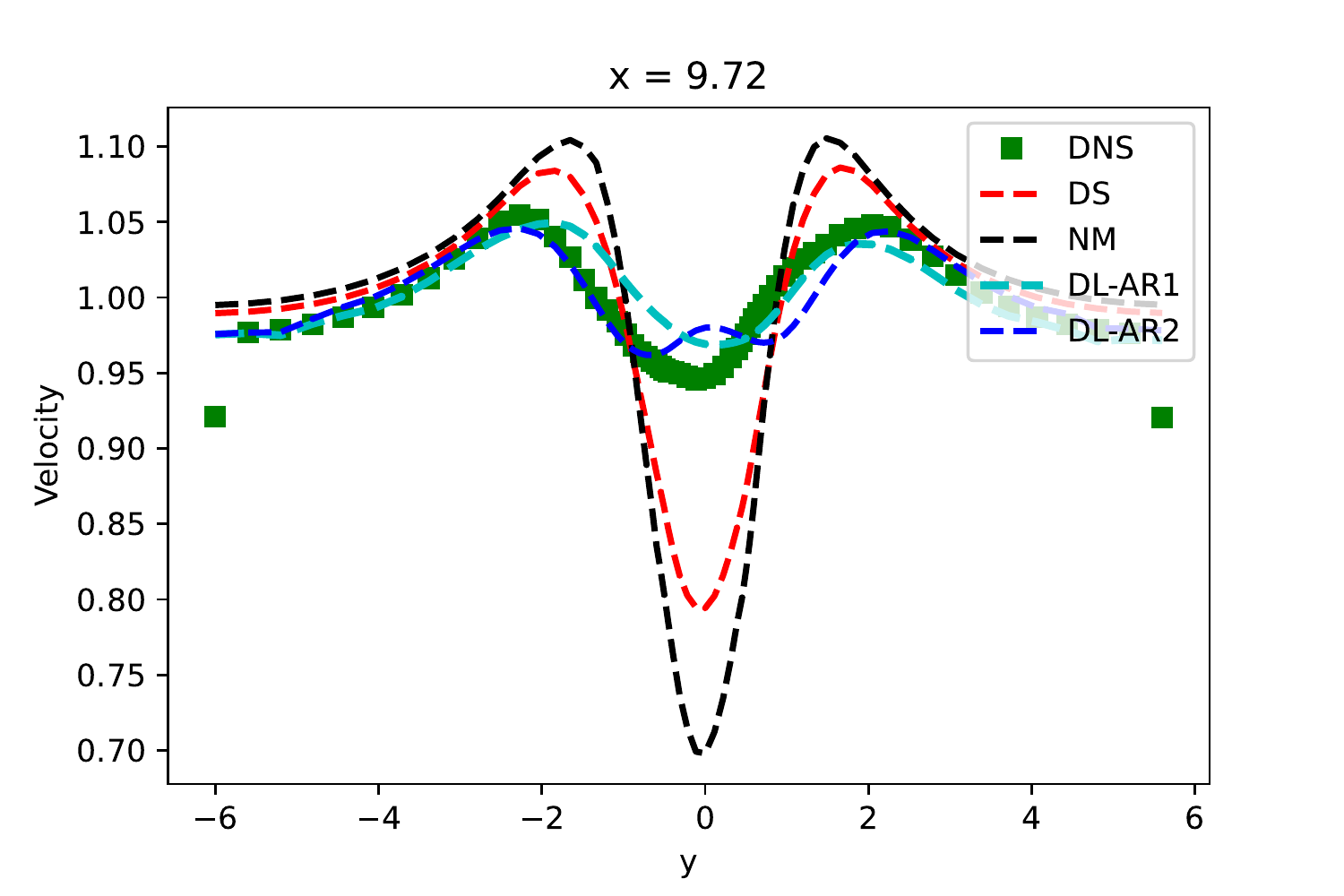}
\includegraphics[width=5cm]{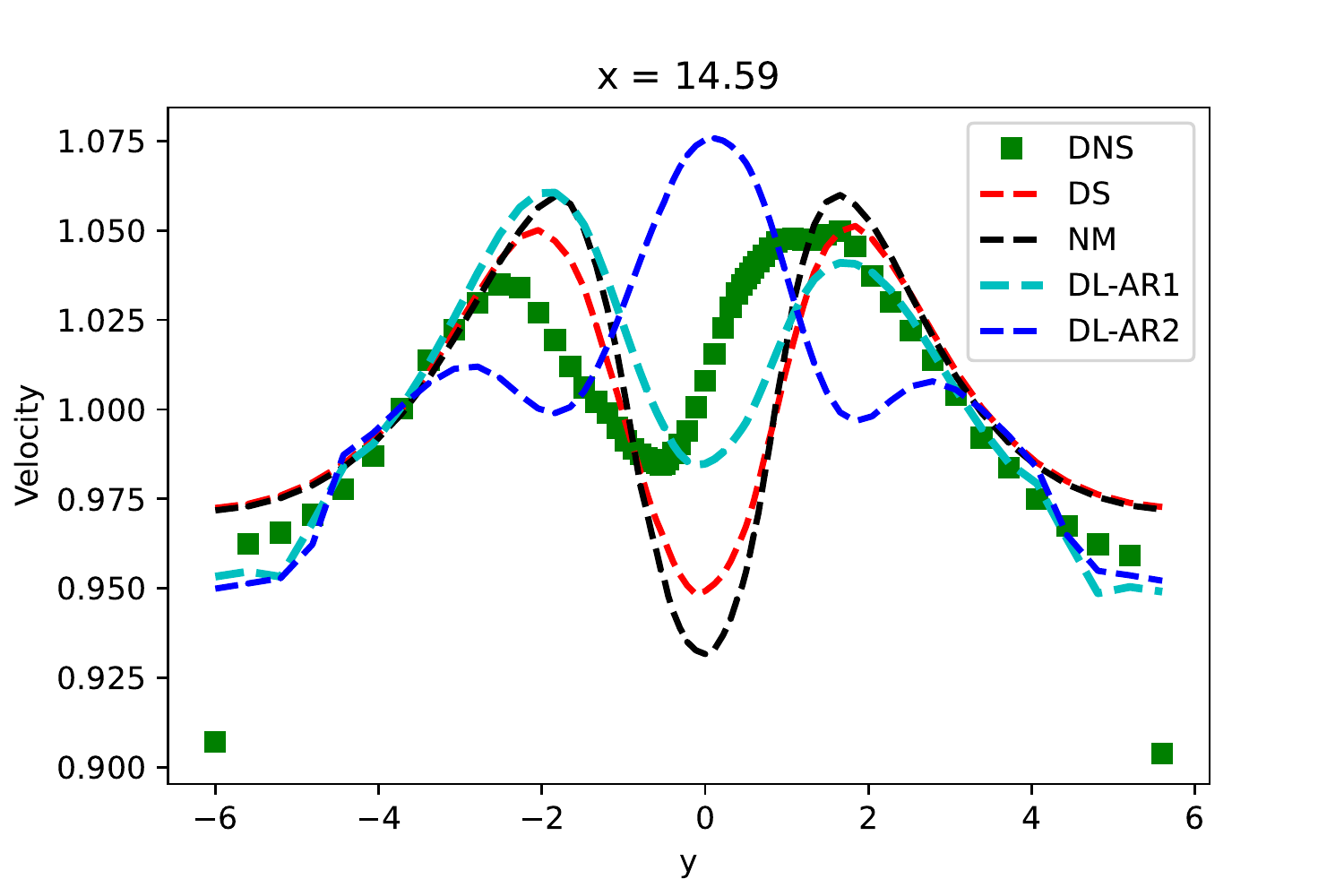}
\label{f1}
\caption{Mean profile for $u_1$ for AR2-Re$2,000$ configuration.}
\end{figure}

\begin{figure}[H]
\centering
\includegraphics[width=5cm]{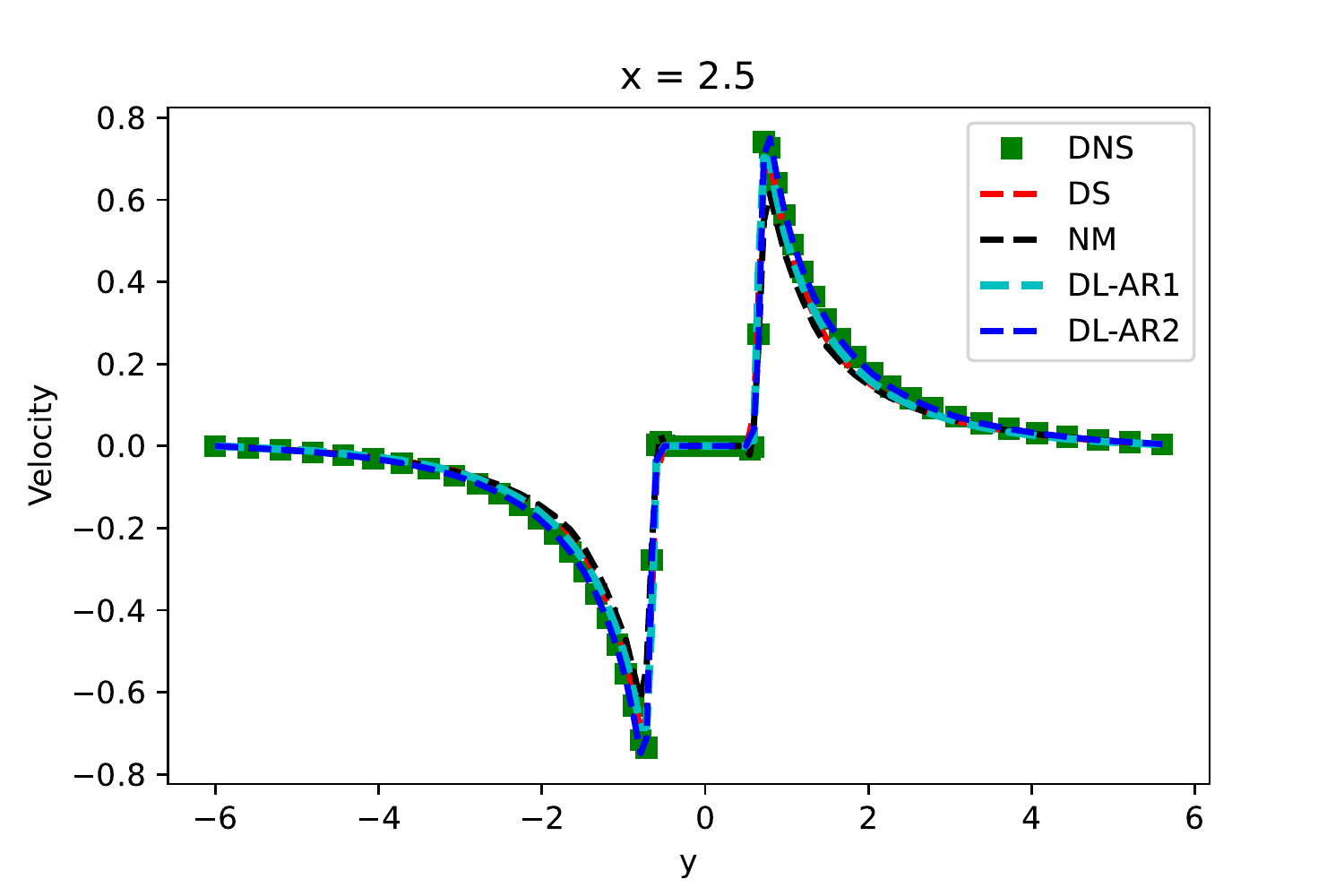}
\includegraphics[width=5cm]{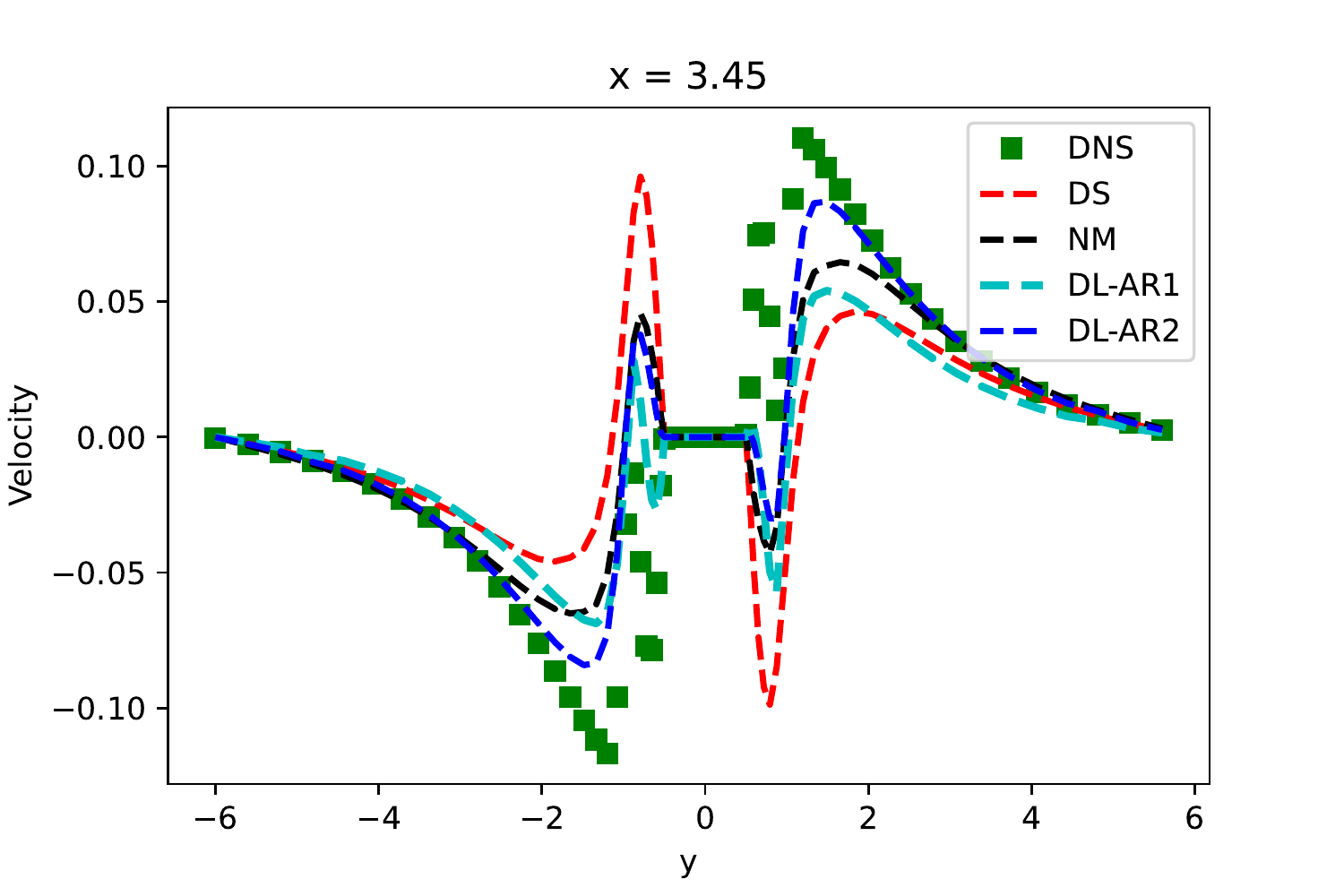}
\includegraphics[width=5cm]{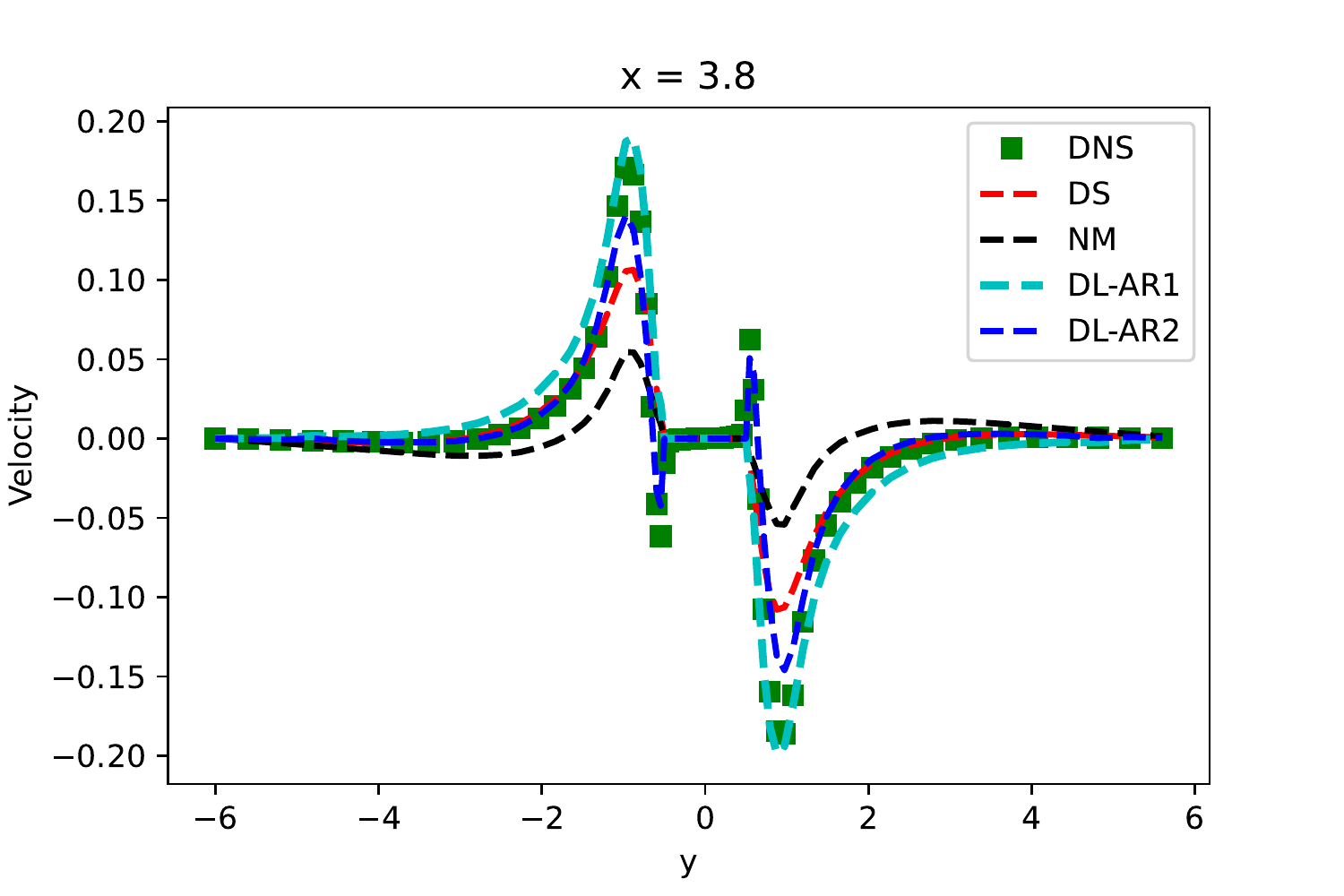}
\includegraphics[width=5cm]{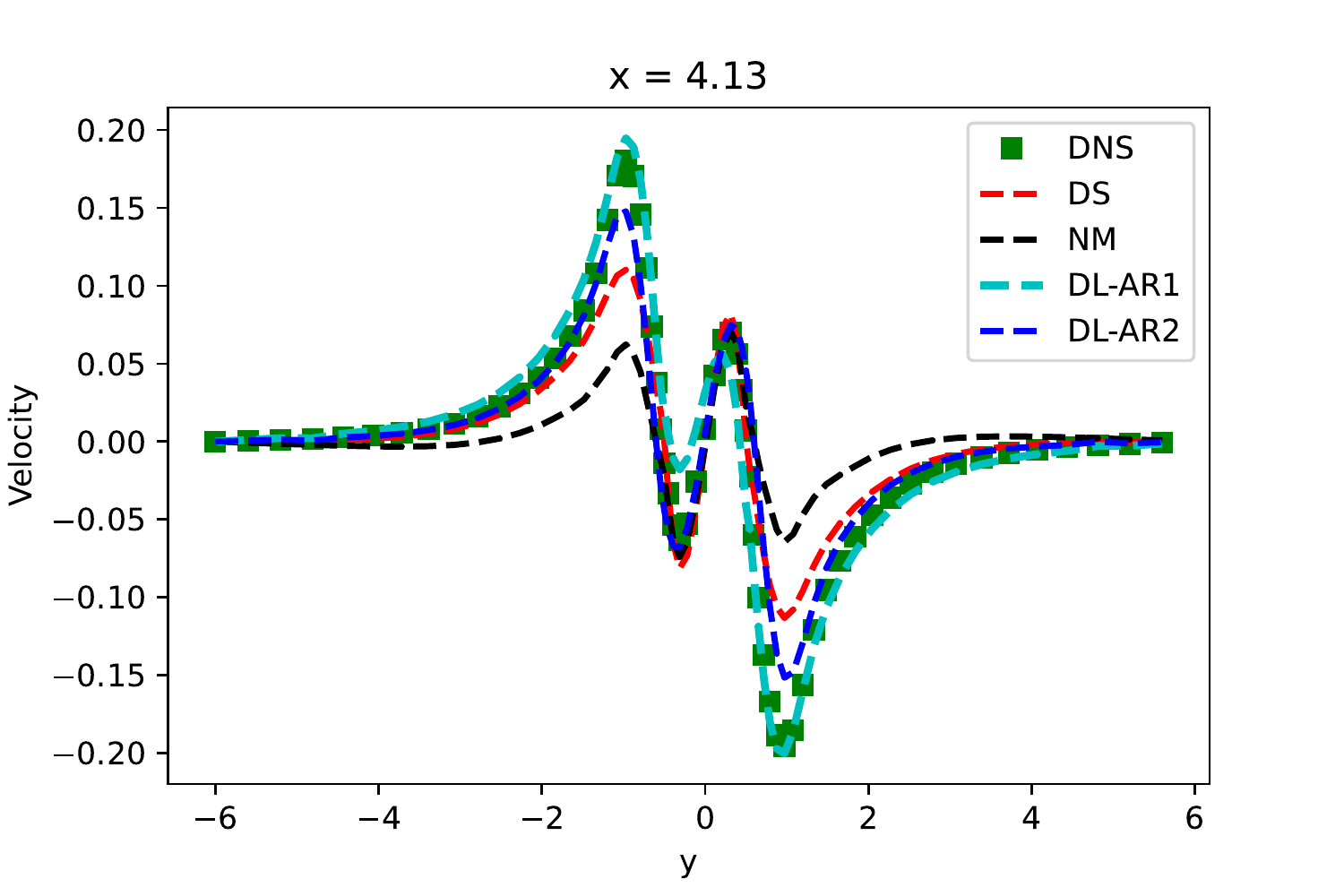}
\includegraphics[width=5cm]{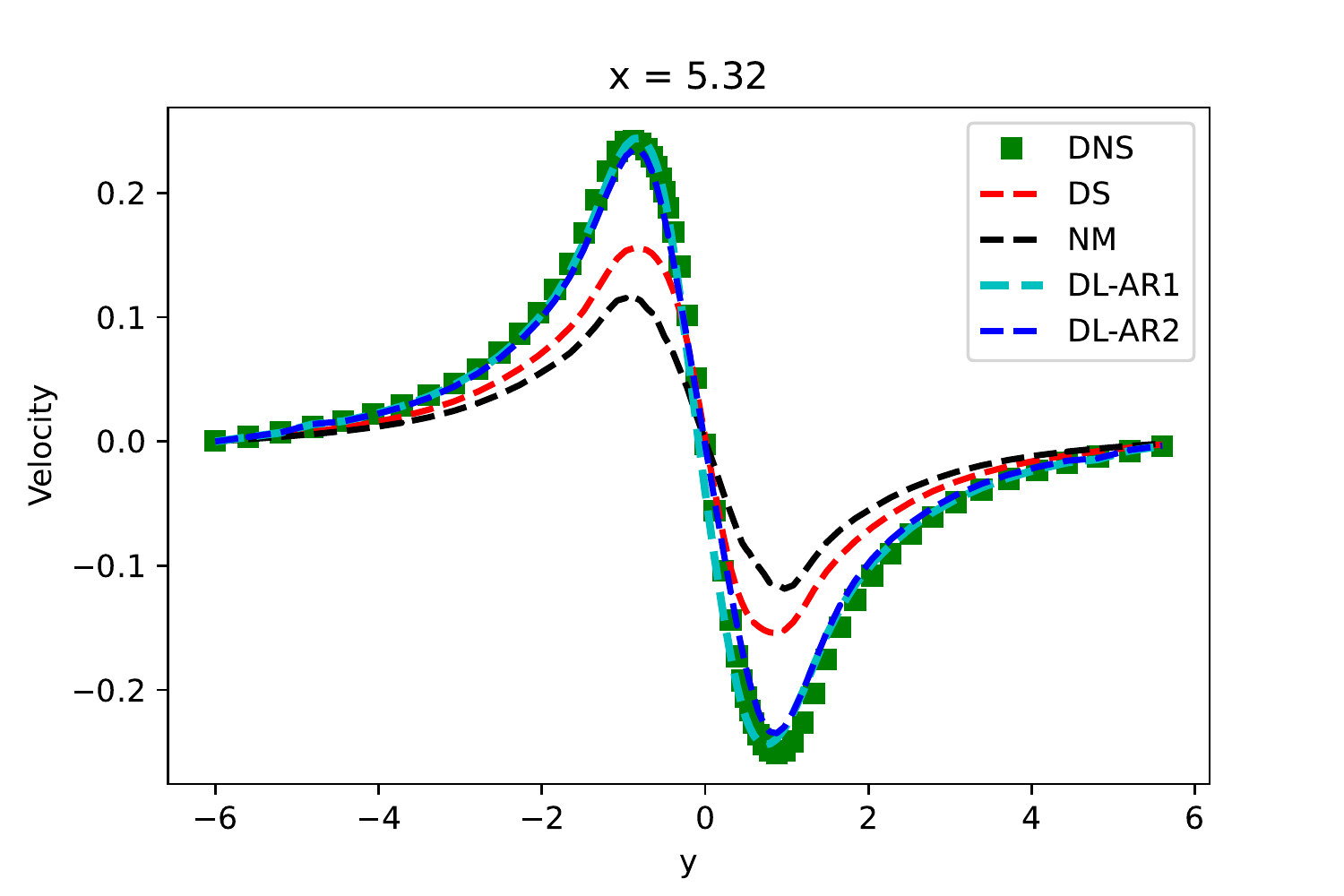}
\includegraphics[width=5cm]{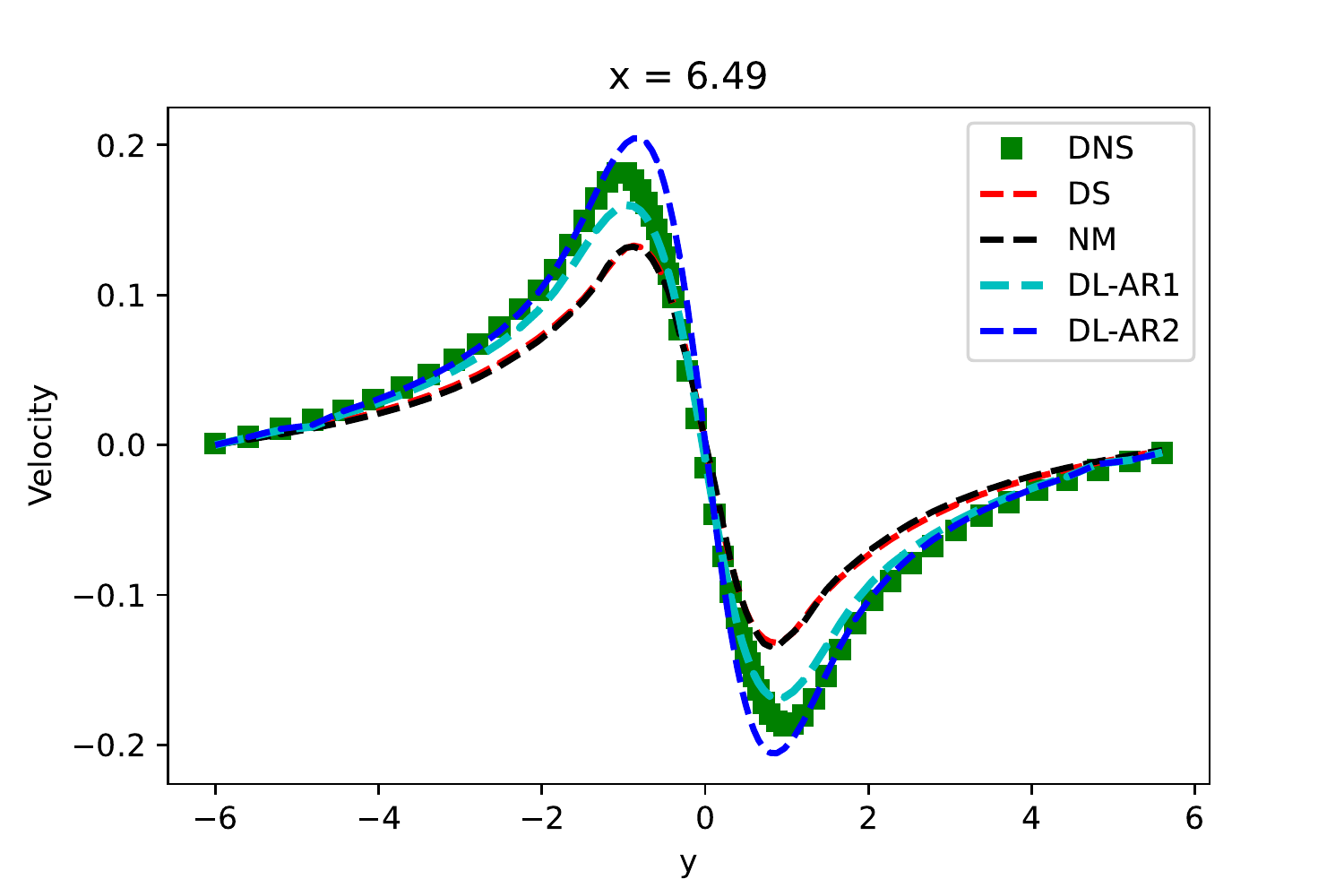}
\includegraphics[width=5cm]{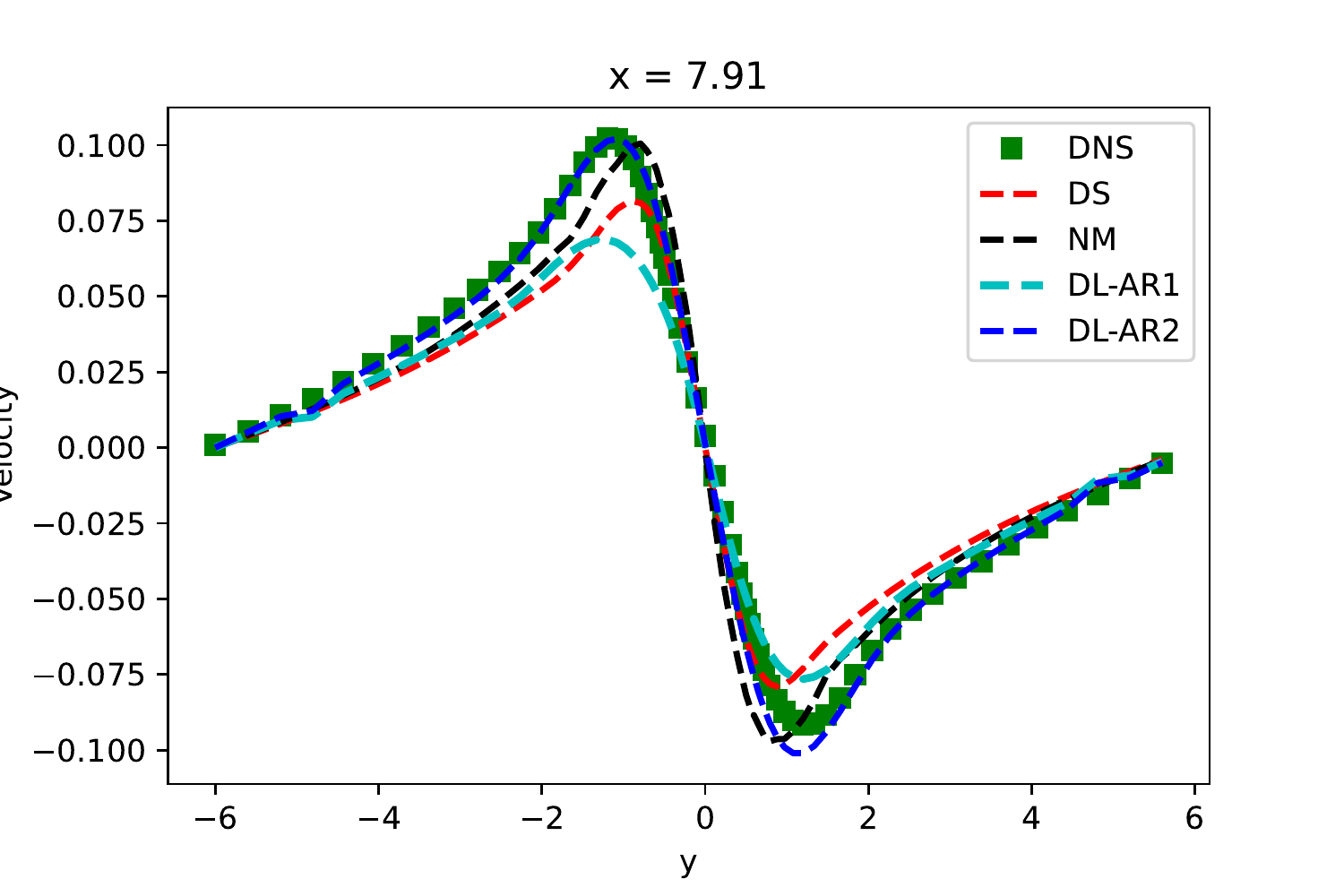}
\includegraphics[width=5cm]{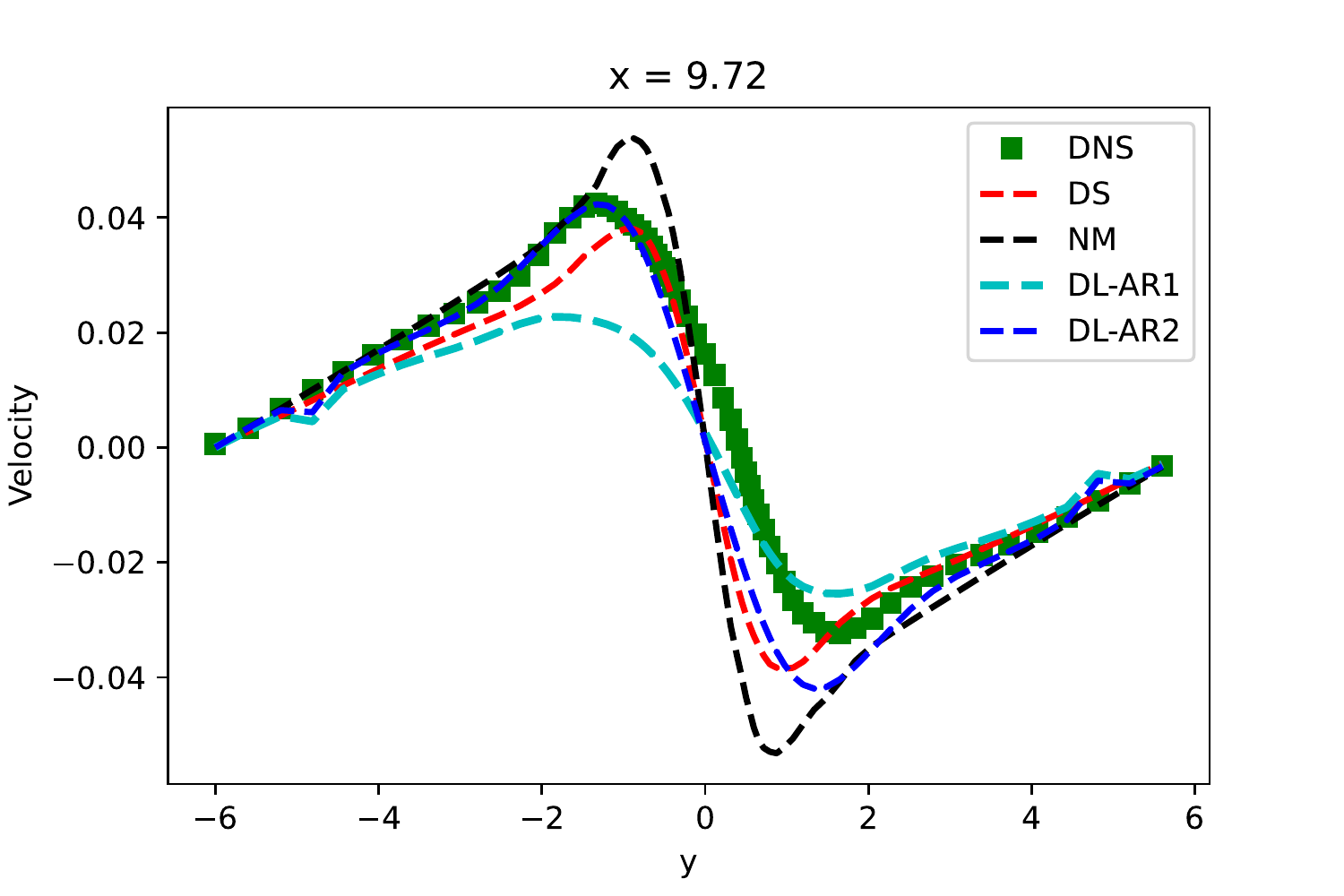}
\includegraphics[width=5cm]{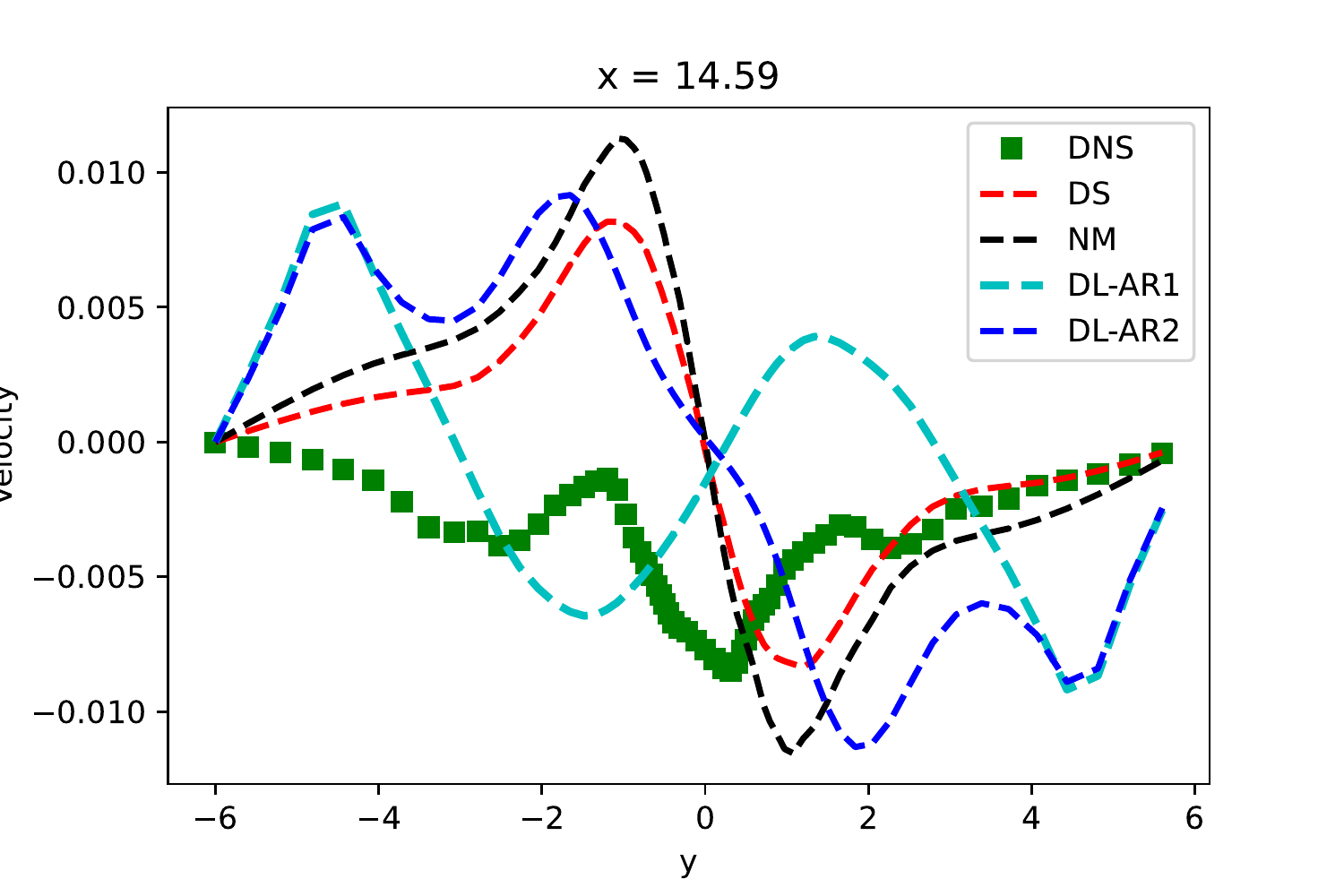}
\label{f1}
\caption{Mean profile for $u_2$ for AR2-Re$2,000$ configuration.}
\end{figure}

\begin{figure}[H]
\centering
\includegraphics[width=5cm]{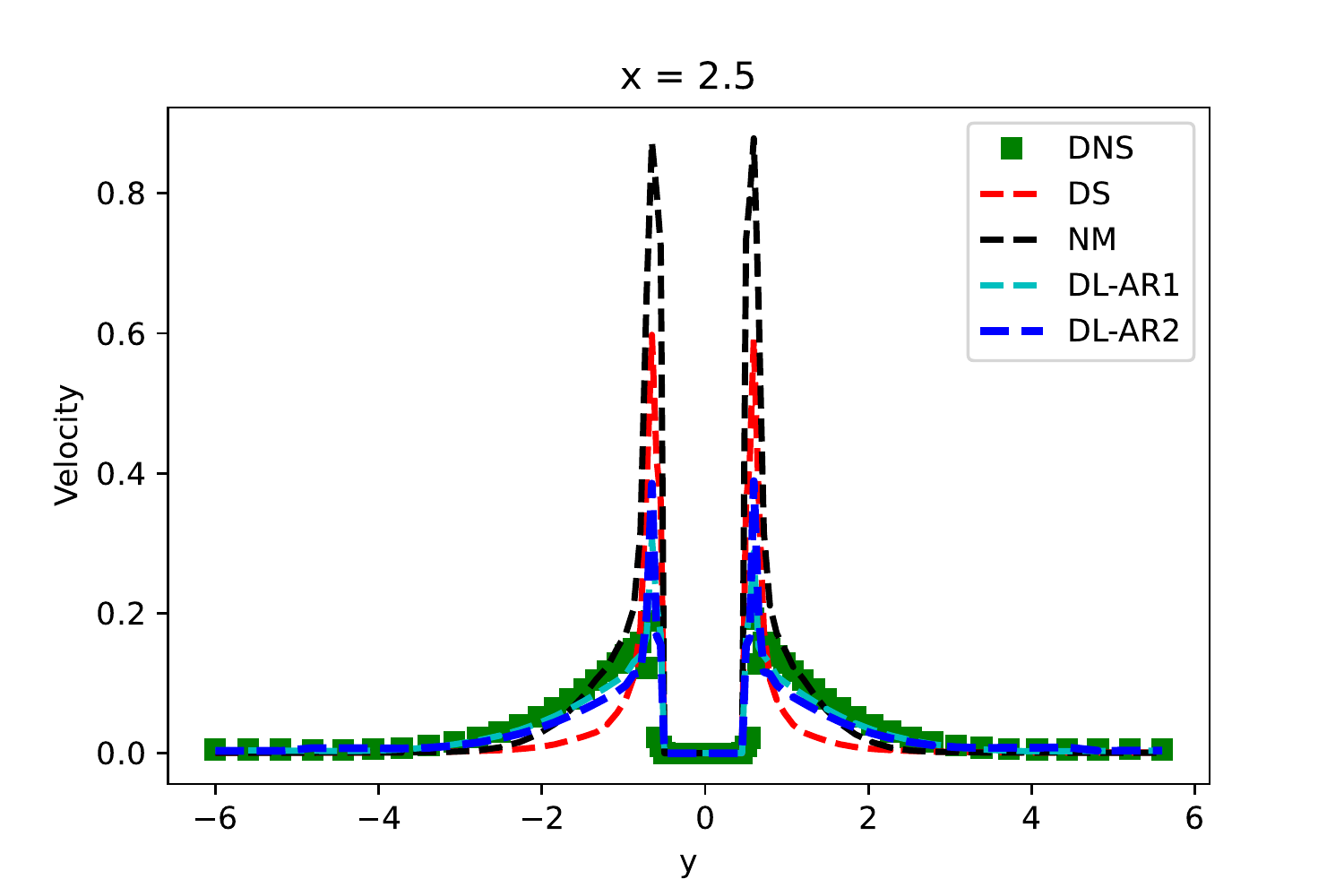}
\includegraphics[width=5cm]{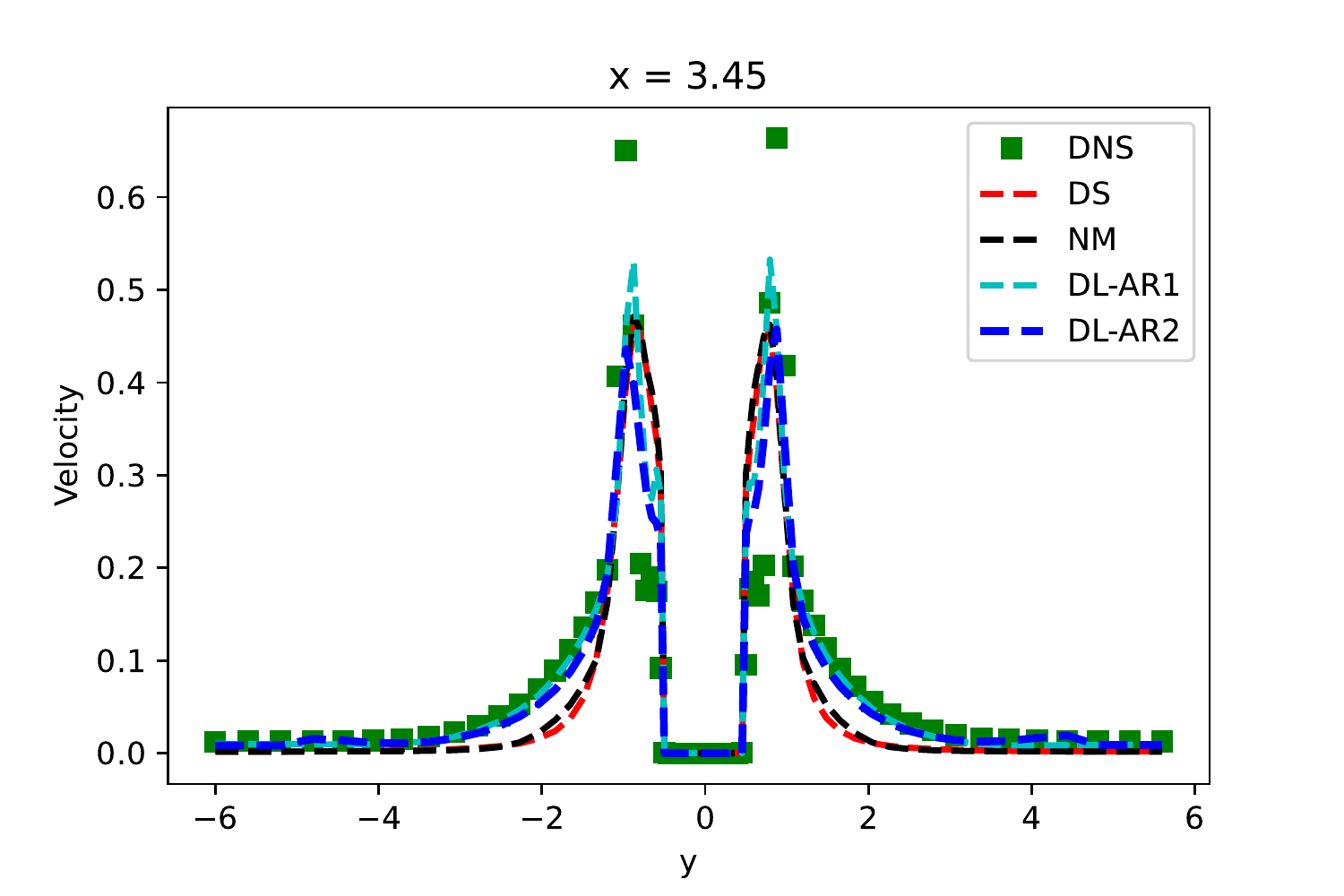}
\includegraphics[width=5cm]{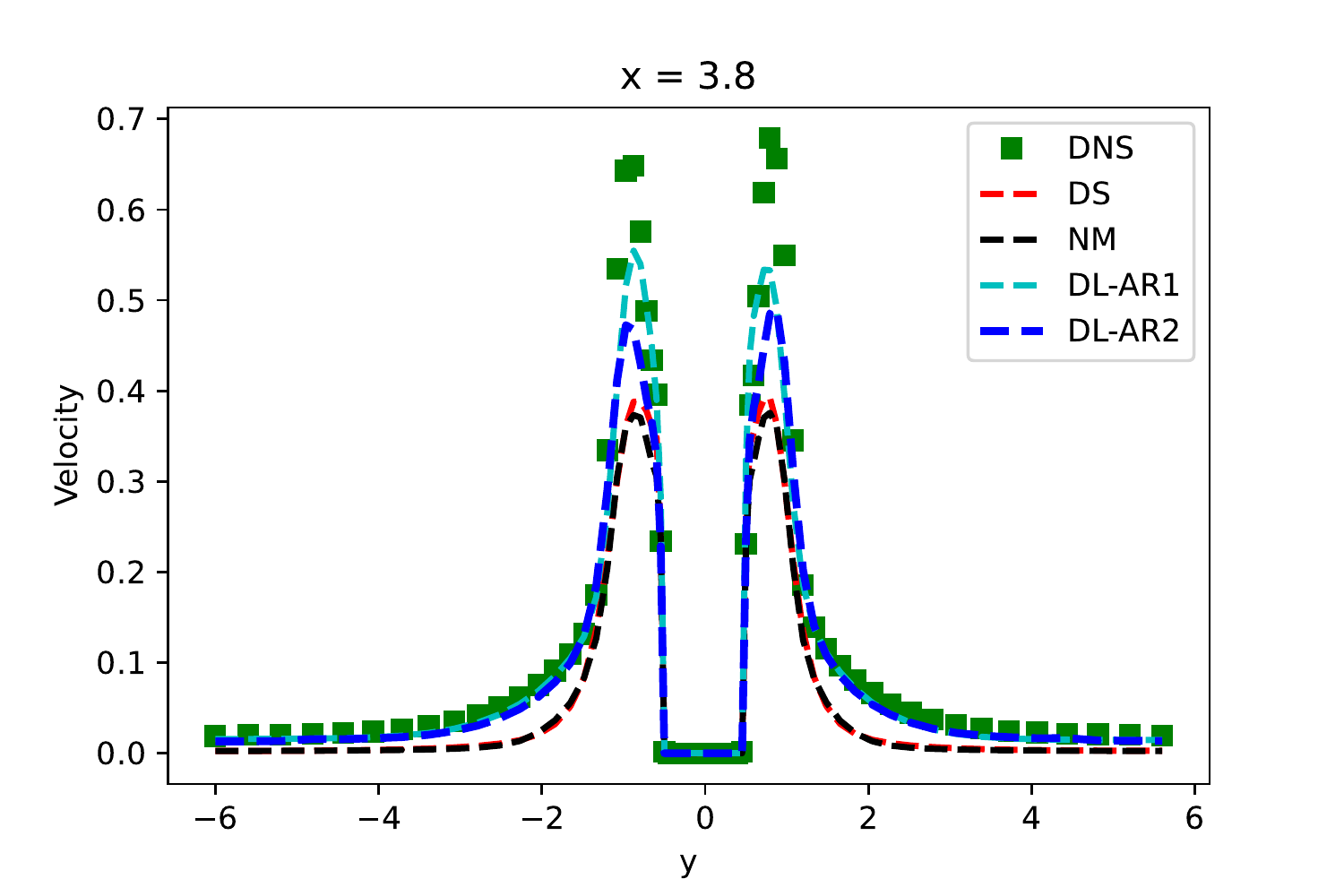}
\includegraphics[width=5cm]{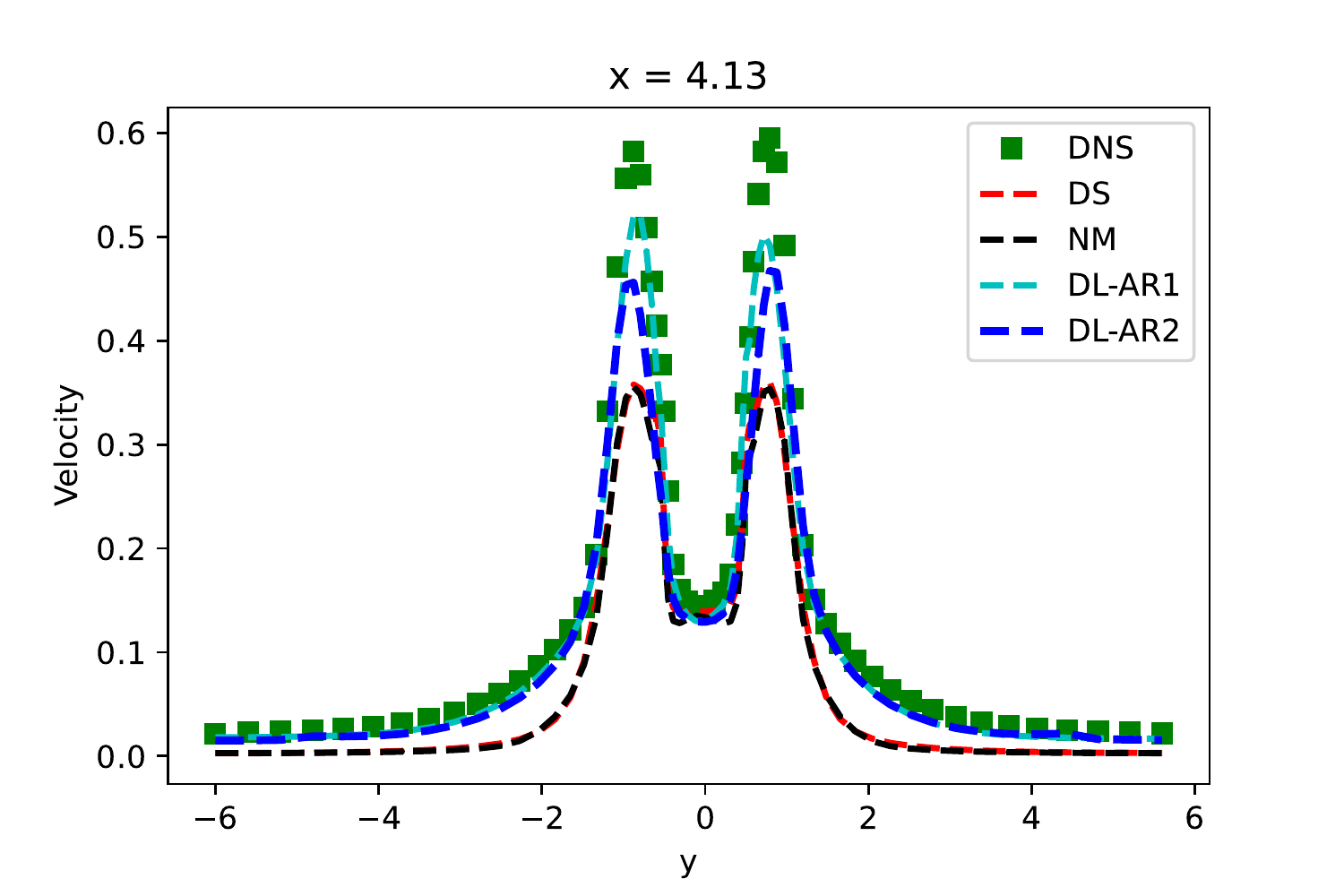}
\includegraphics[width=5cm]{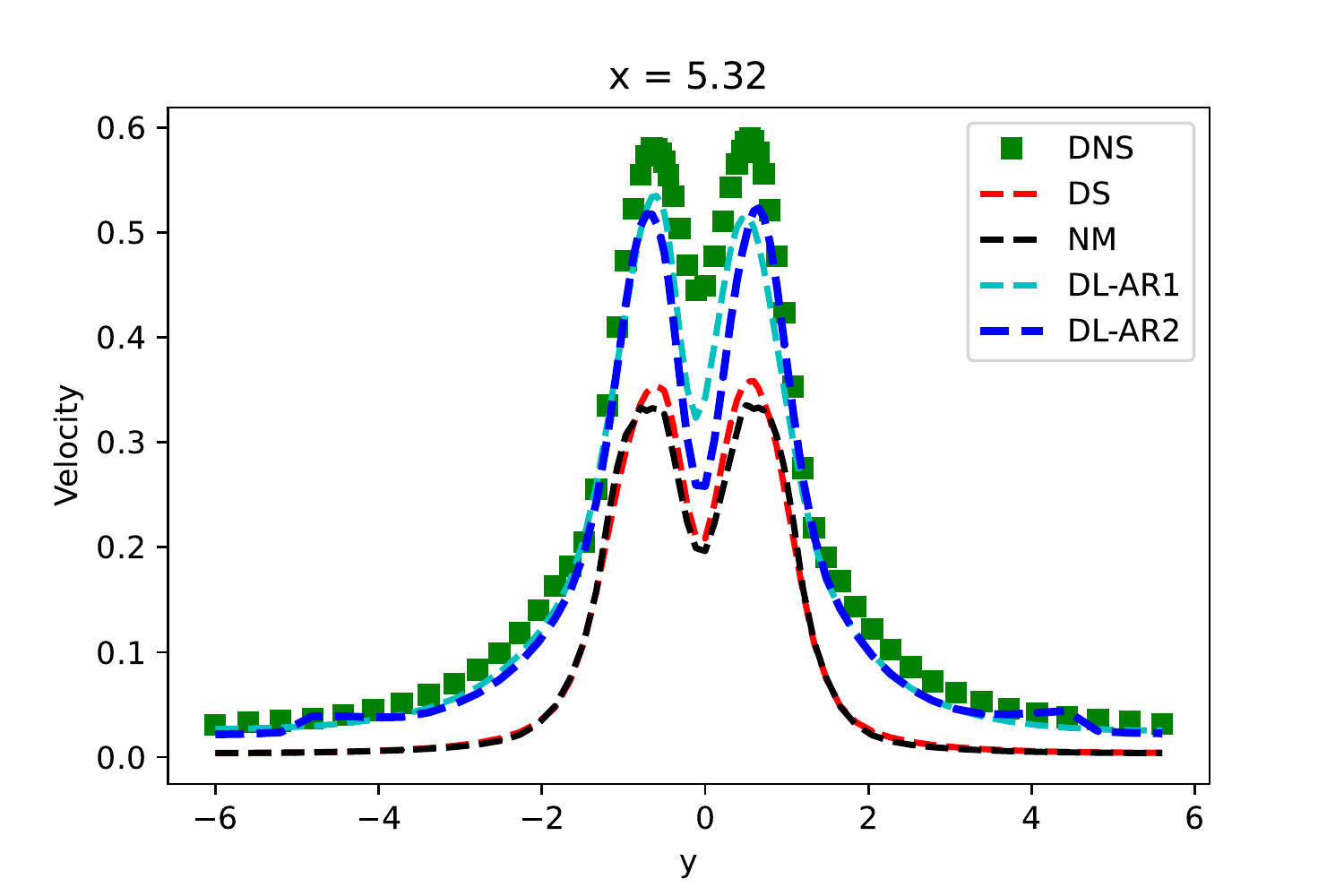}
\includegraphics[width=5cm]{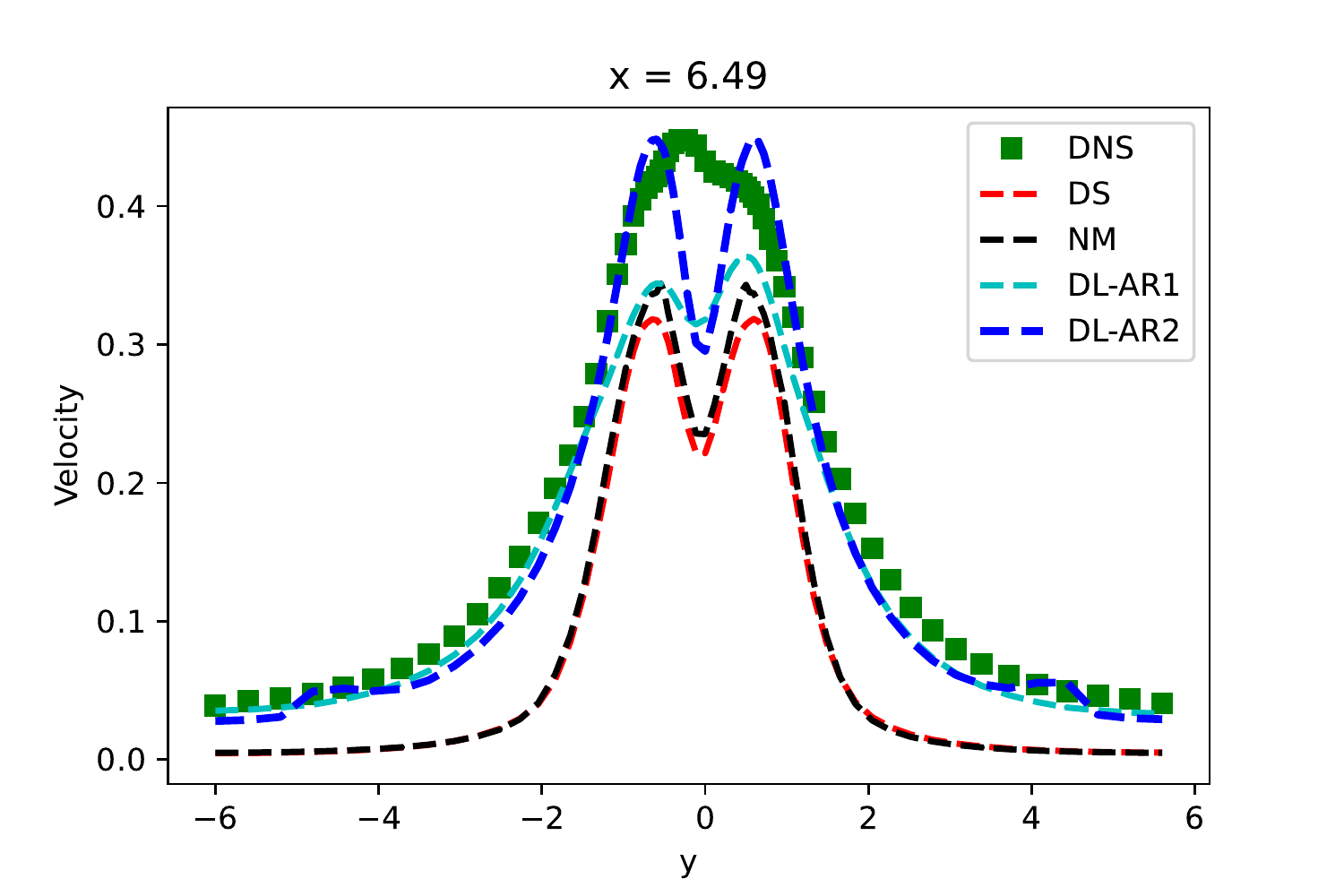}
\includegraphics[width=5cm]{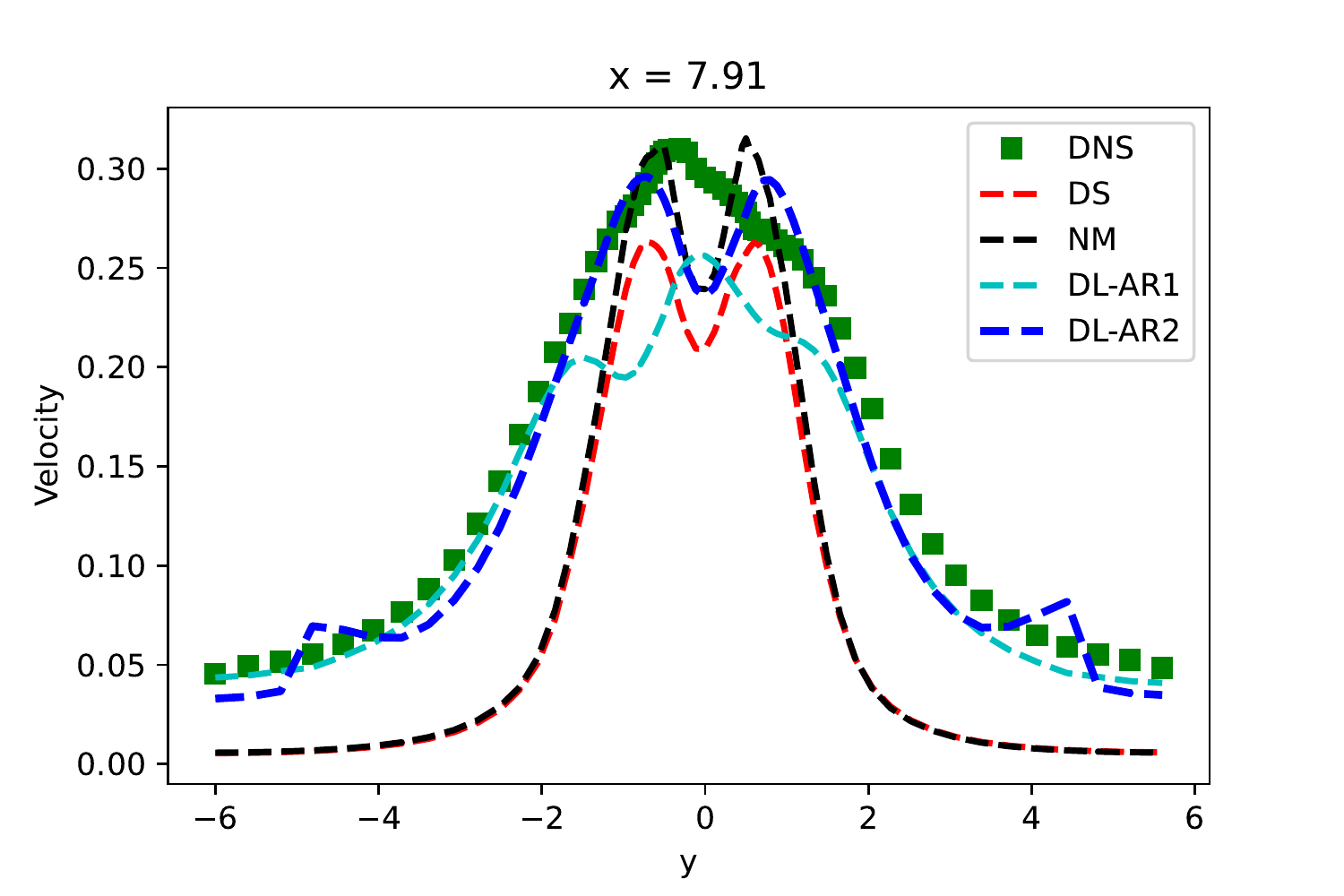}
\includegraphics[width=5cm]{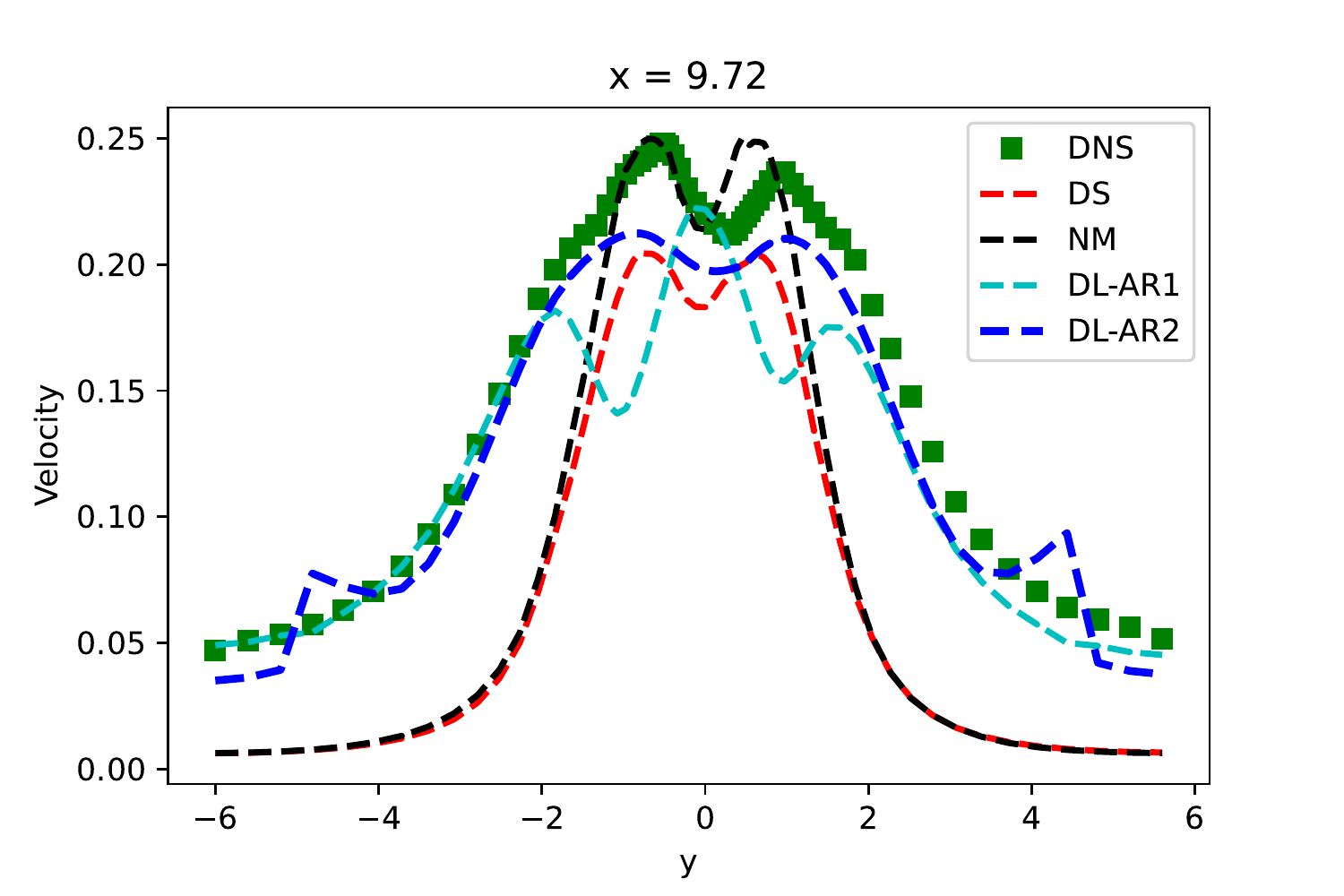}
\includegraphics[width=5cm]{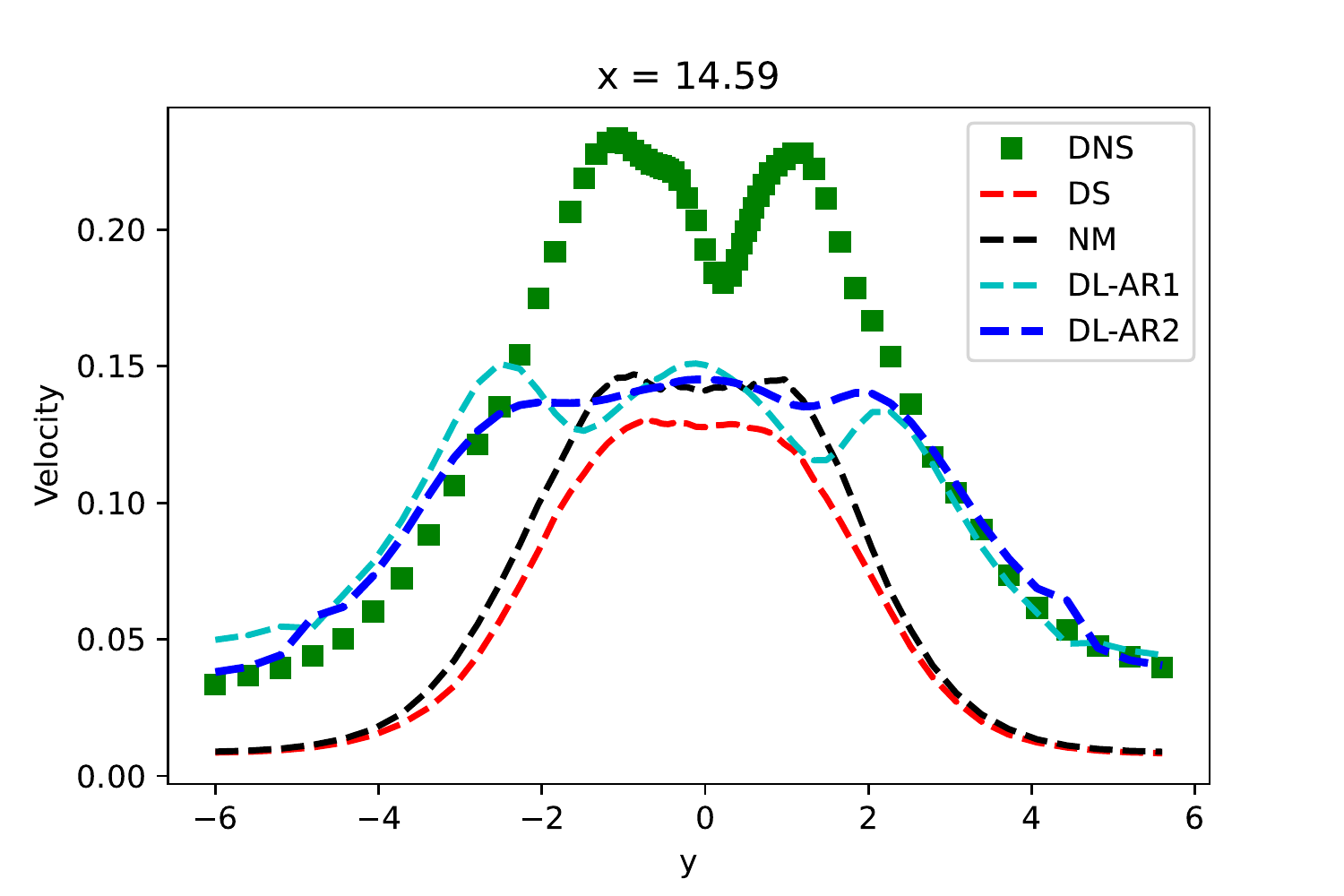}
\label{f1}
\caption{RMS profile for $u_1$ for AR2-Re$2,000$ configuration.}
\end{figure}

\begin{figure}[H]
\centering
\includegraphics[width=5cm]{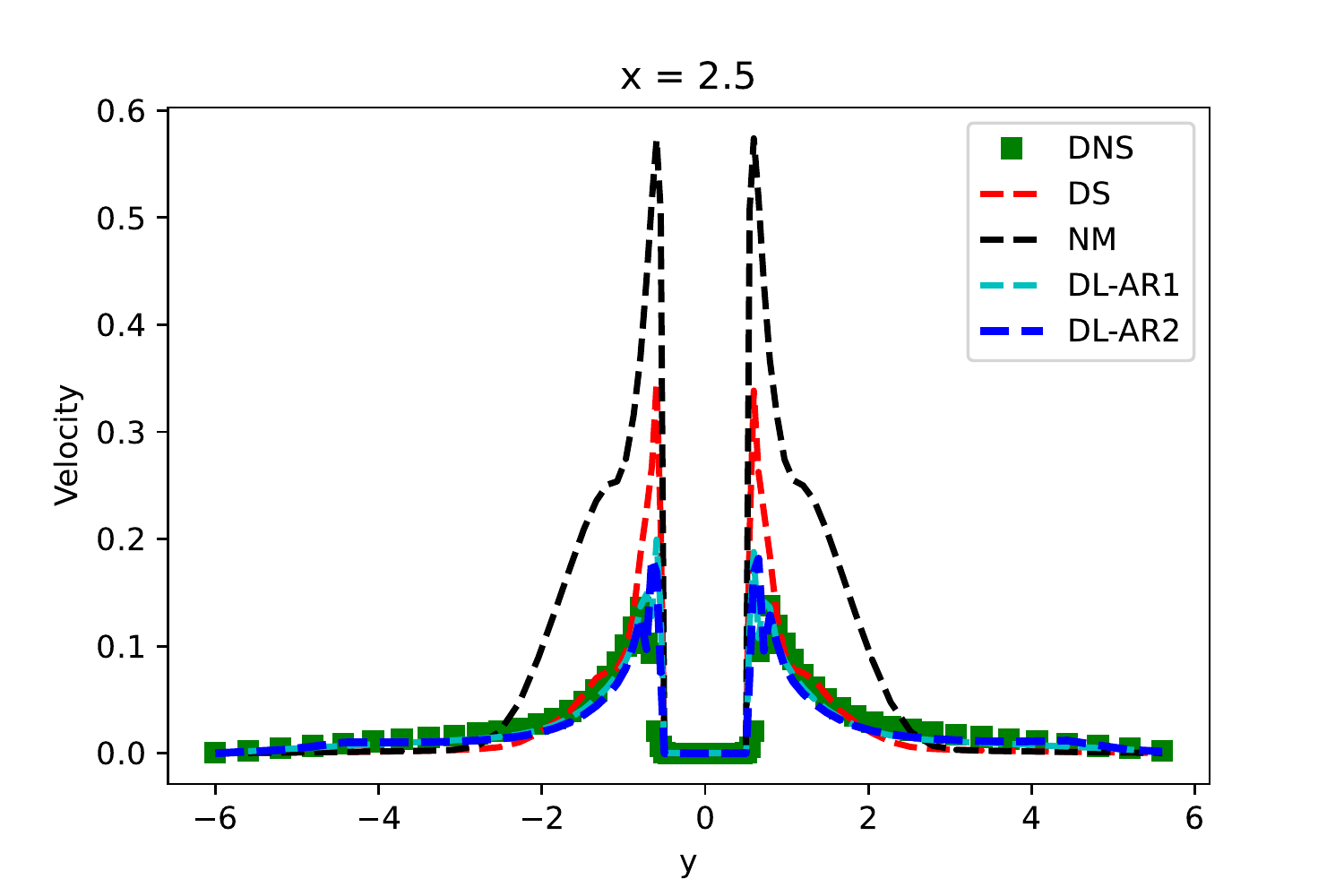}
\includegraphics[width=5cm]{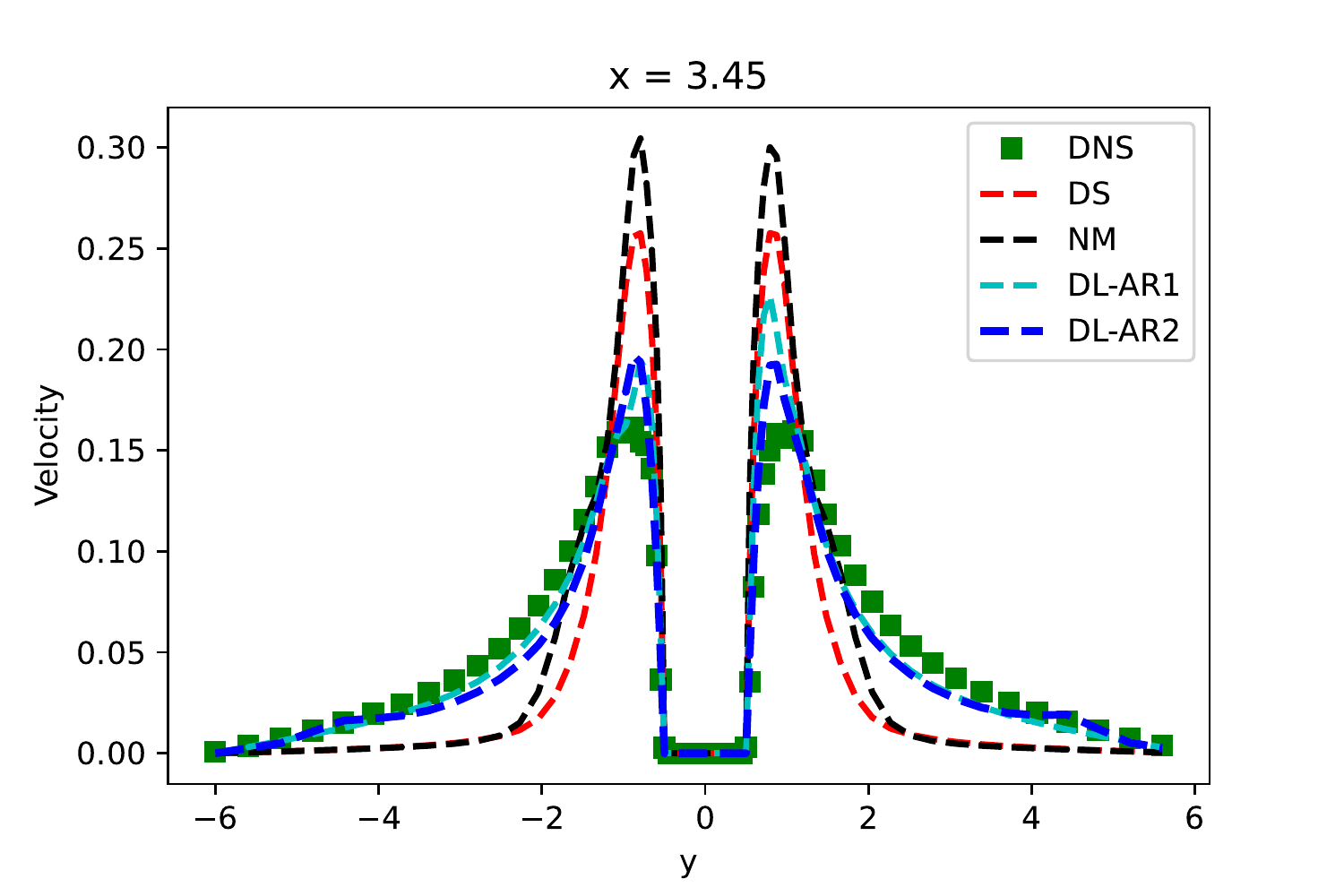}
\includegraphics[width=5cm]{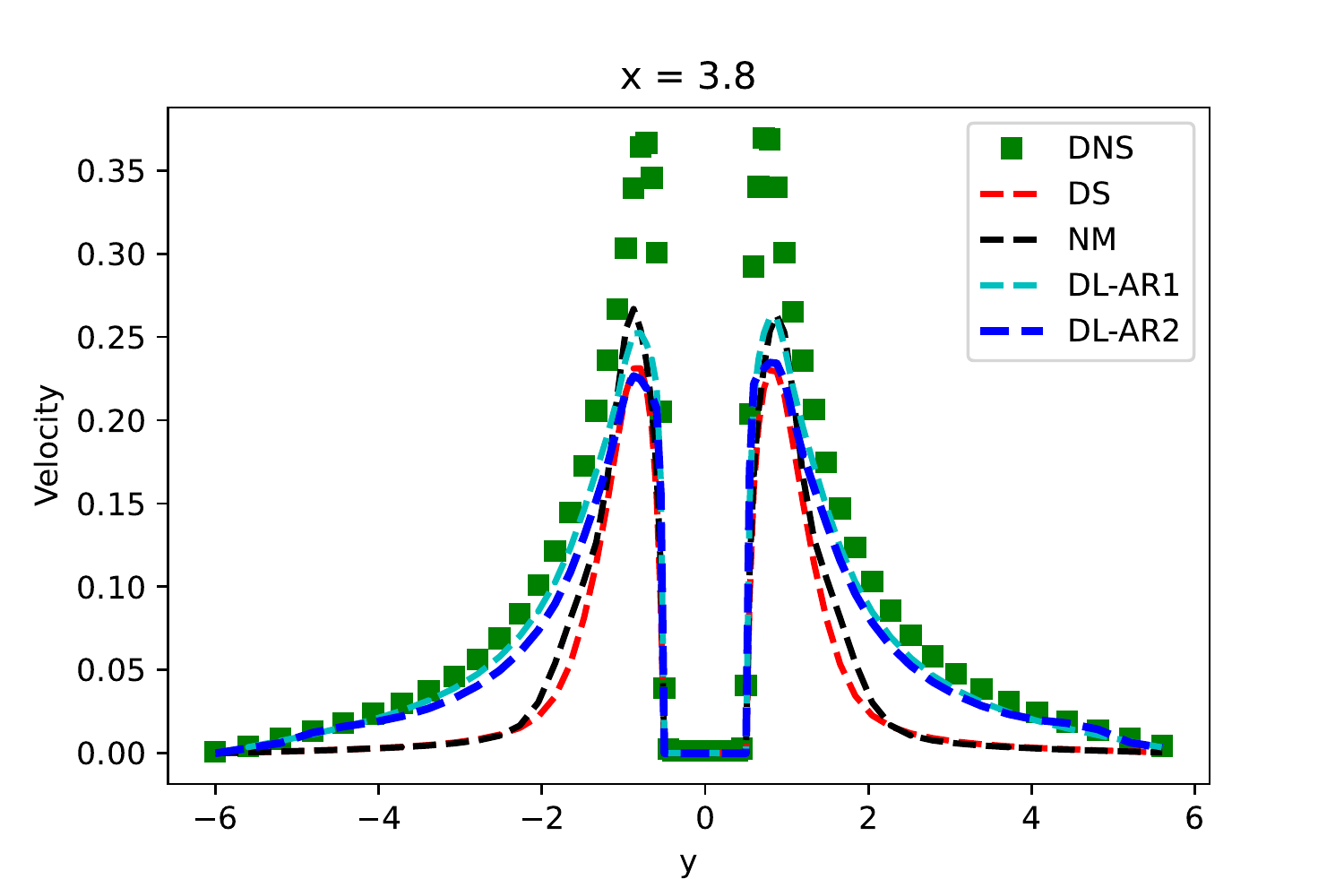}
\includegraphics[width=5cm]{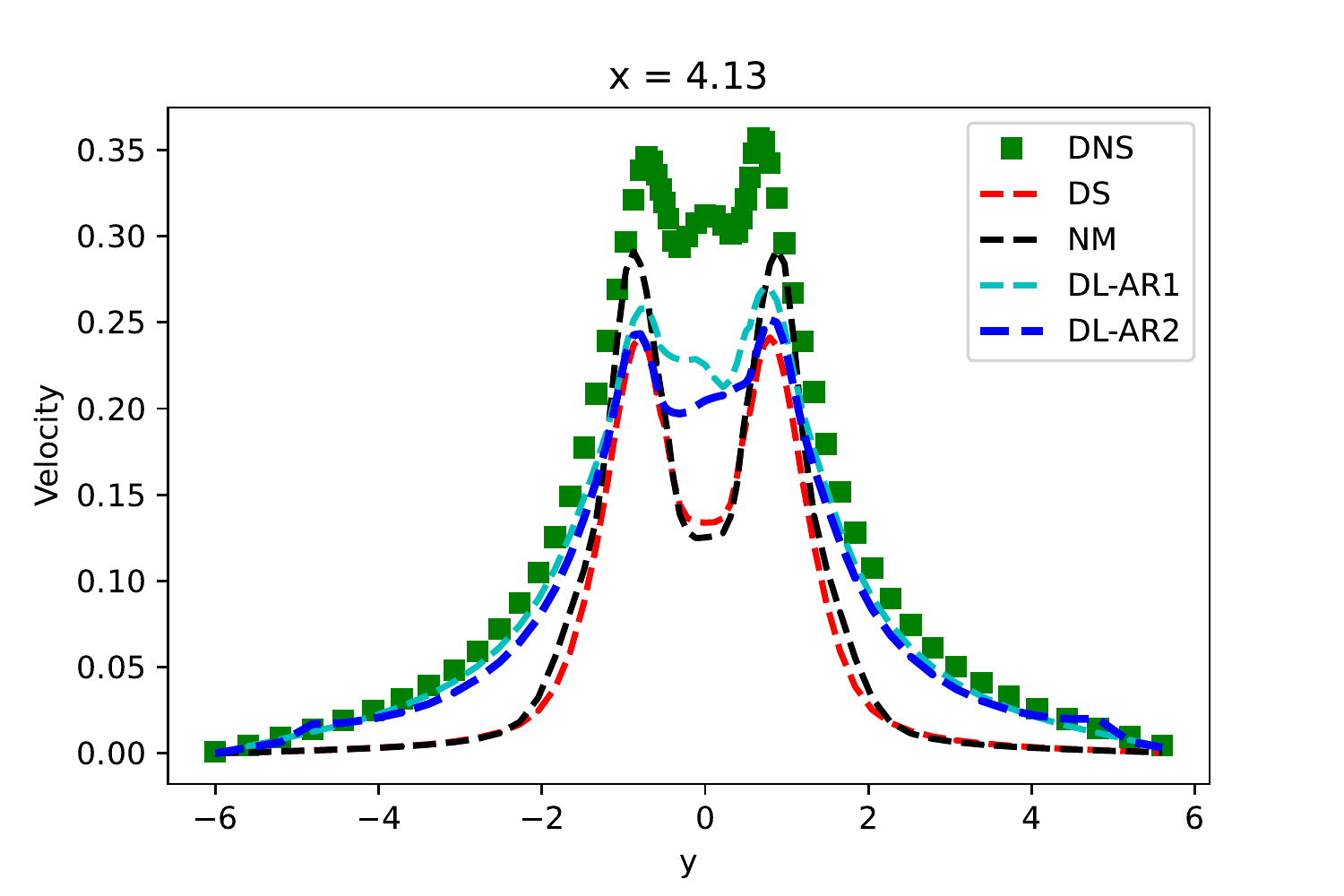}
\includegraphics[width=5cm]{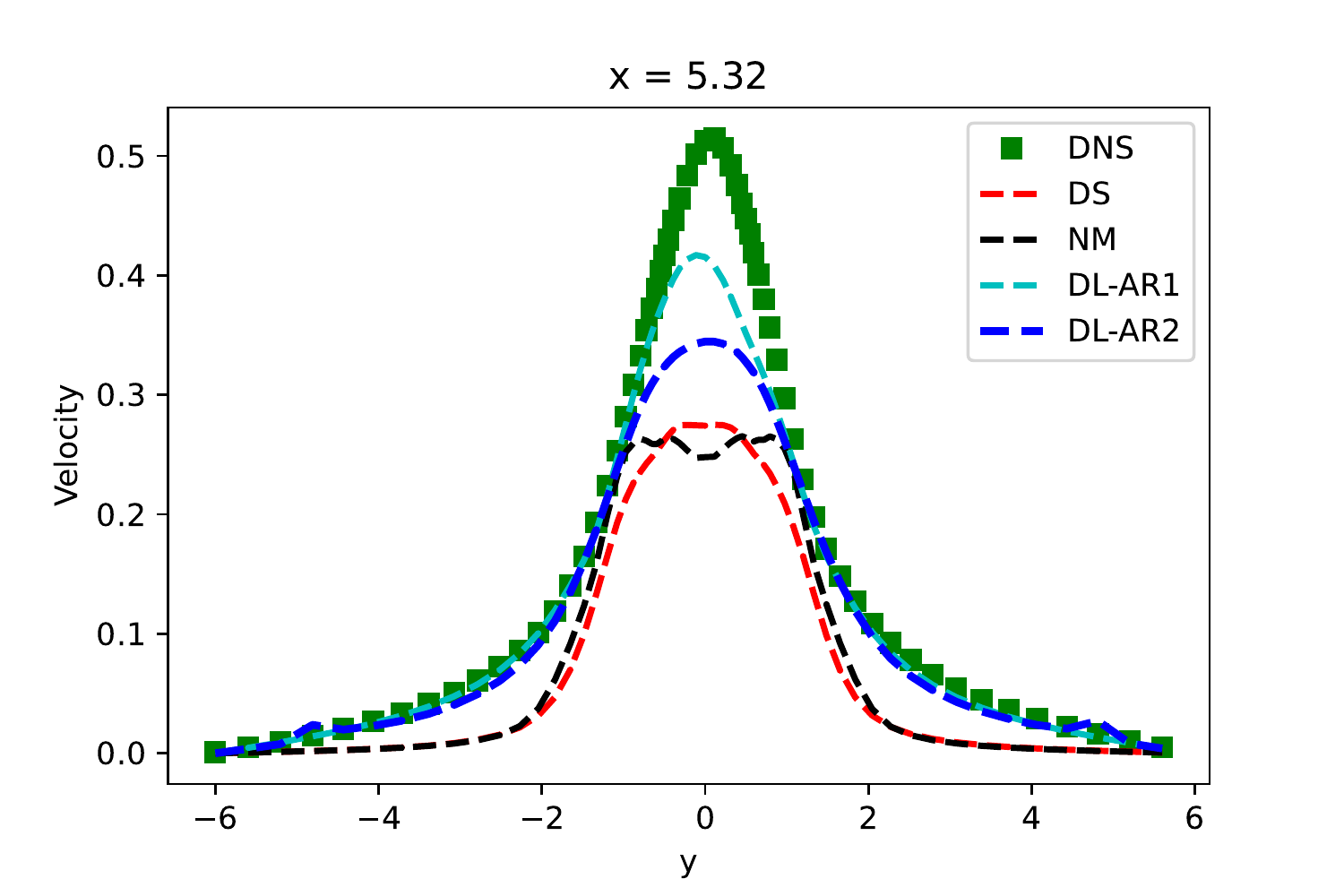}
\includegraphics[width=5cm]{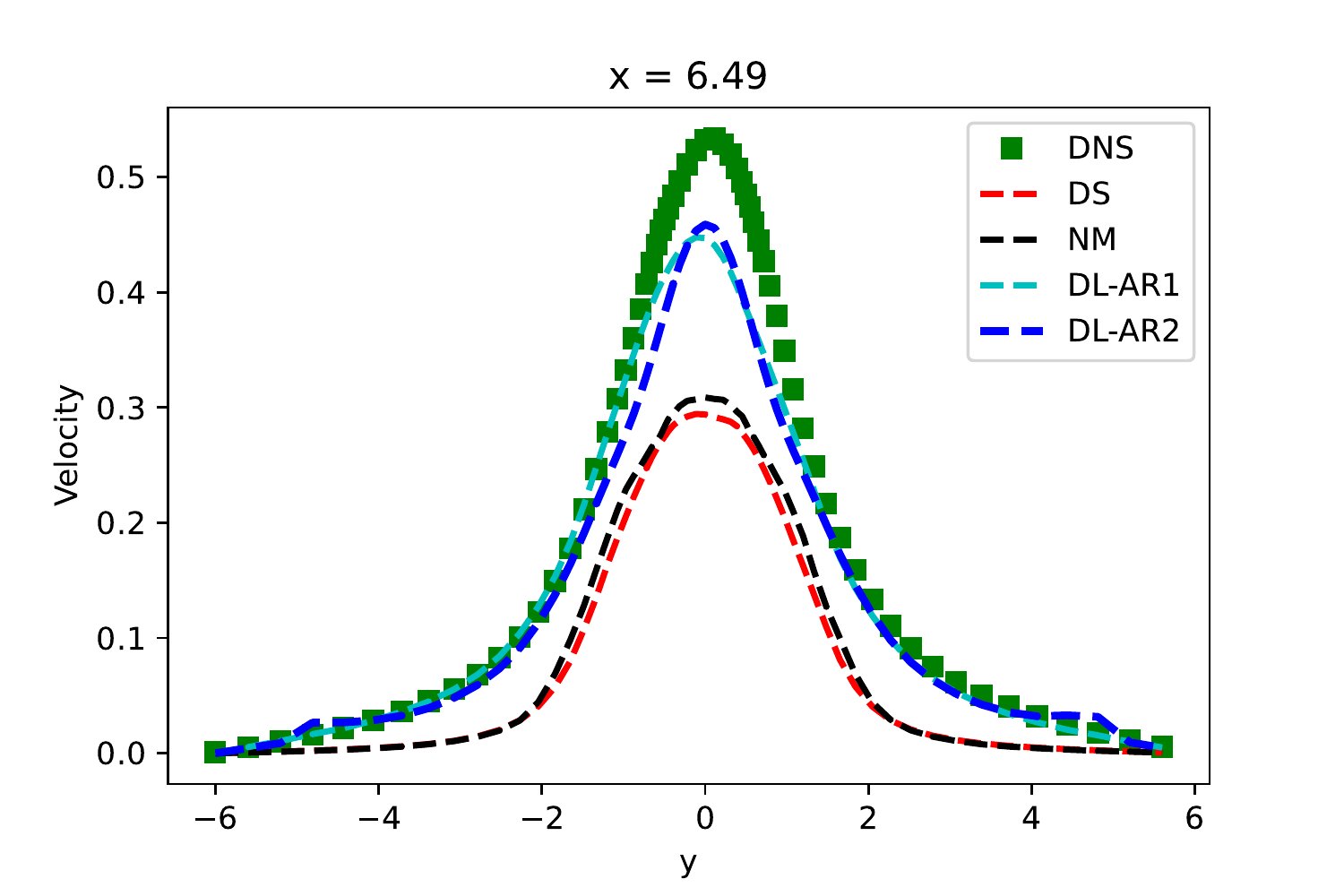}
\includegraphics[width=5cm]{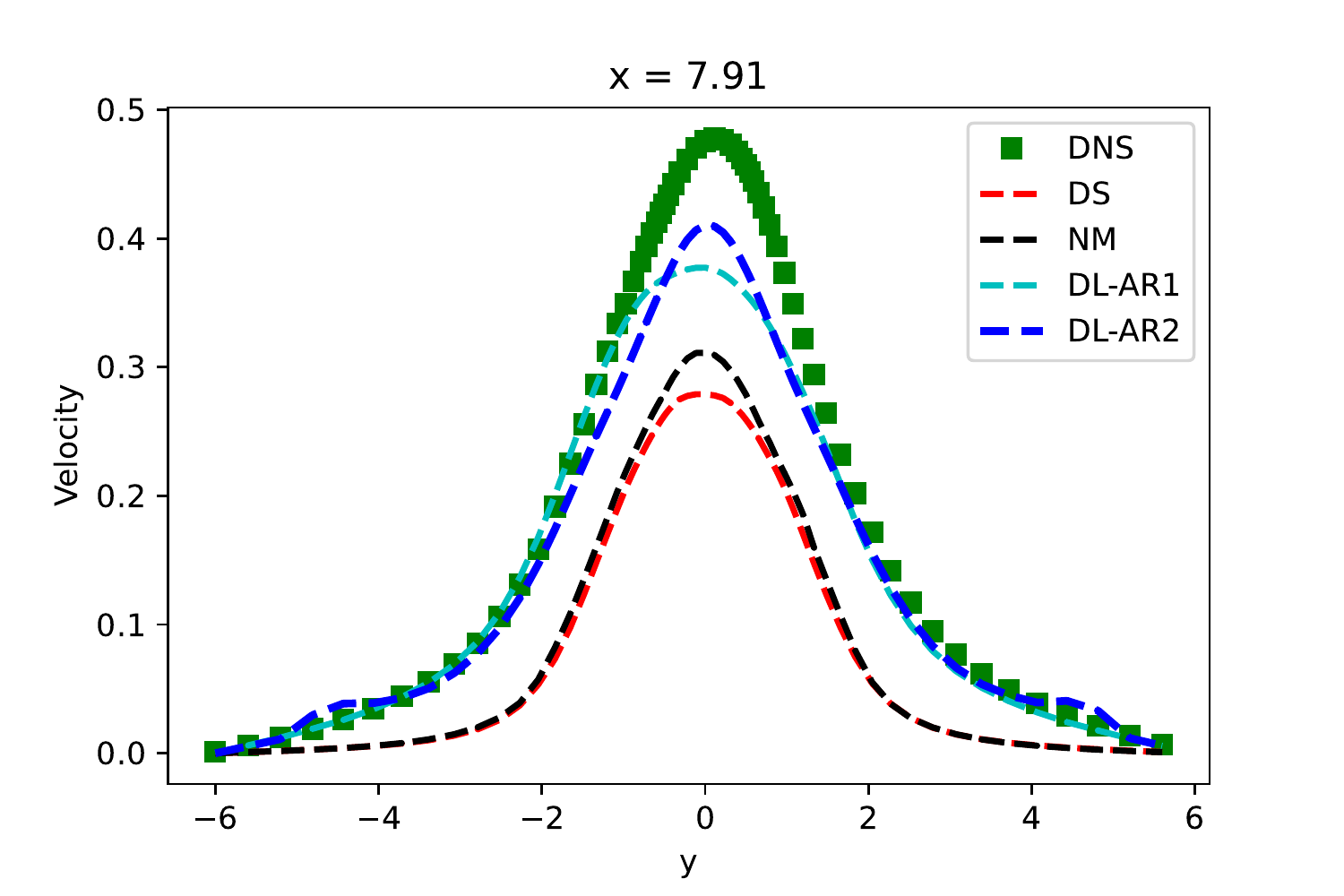}
\includegraphics[width=5cm]{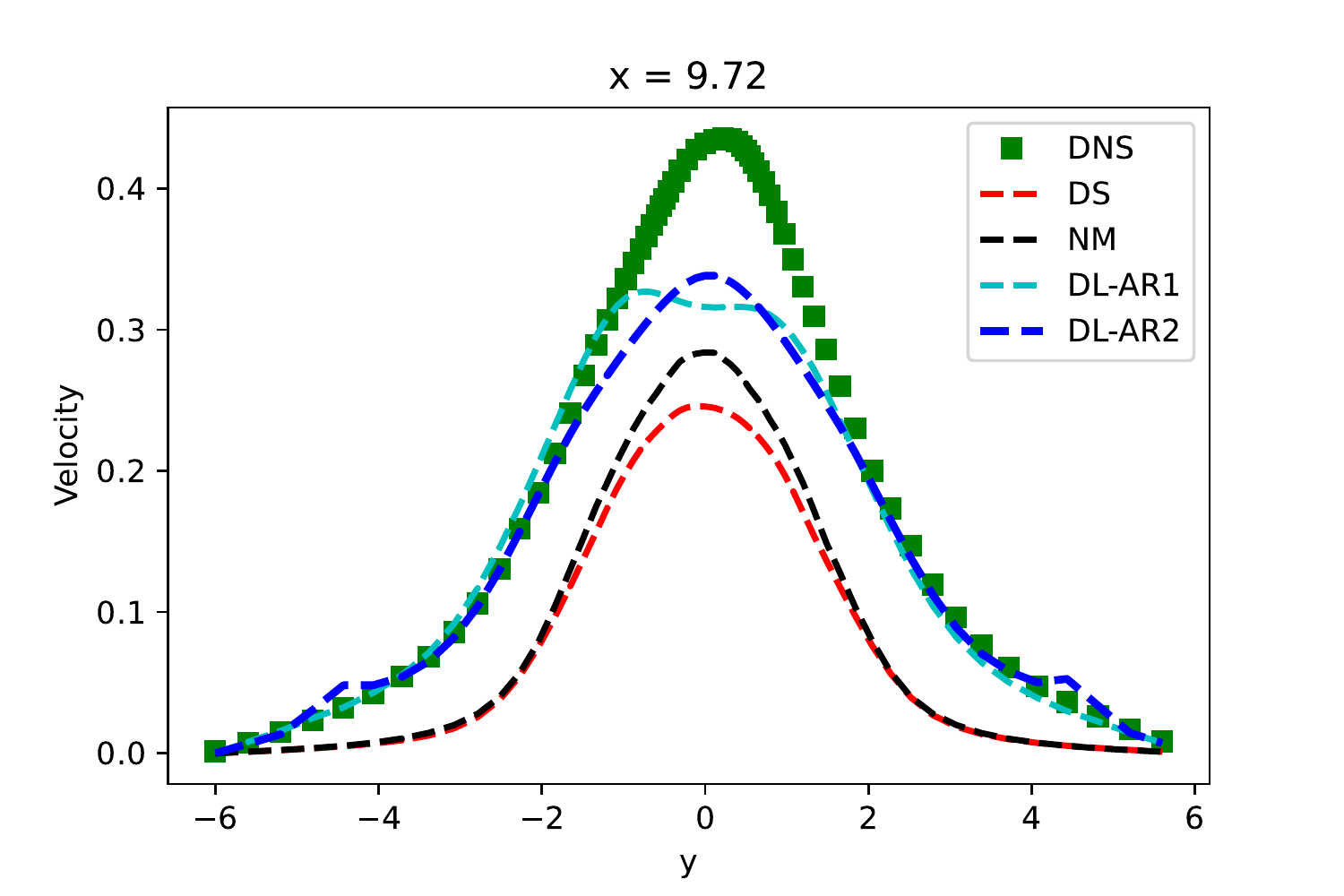}
\includegraphics[width=5cm]{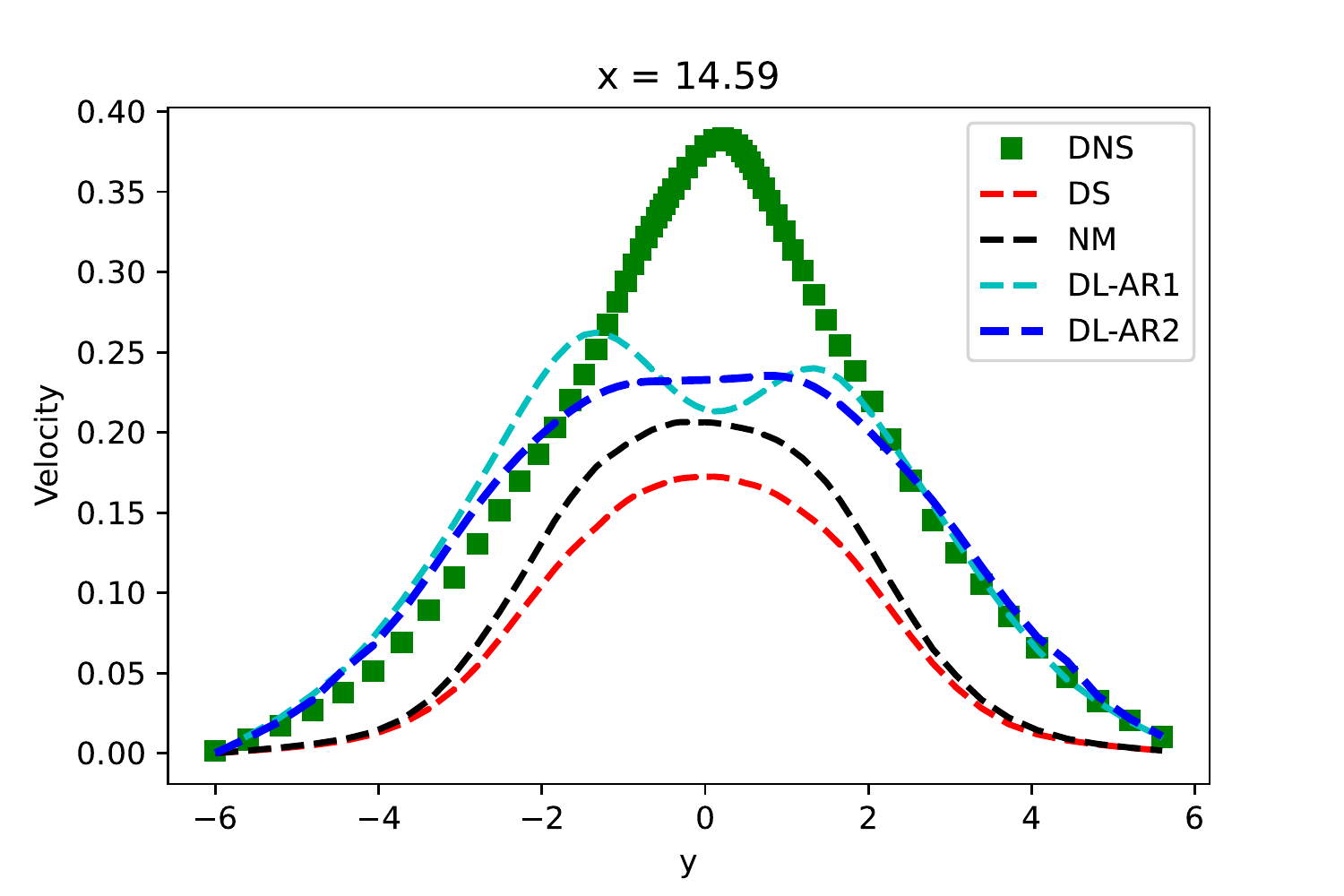}
\label{f1}
\caption{RMS profile for $u_2$ for AR2-Re$2,000$ configuration.}
\end{figure}

\begin{figure}[H]
\centering
\includegraphics[width=5cm]{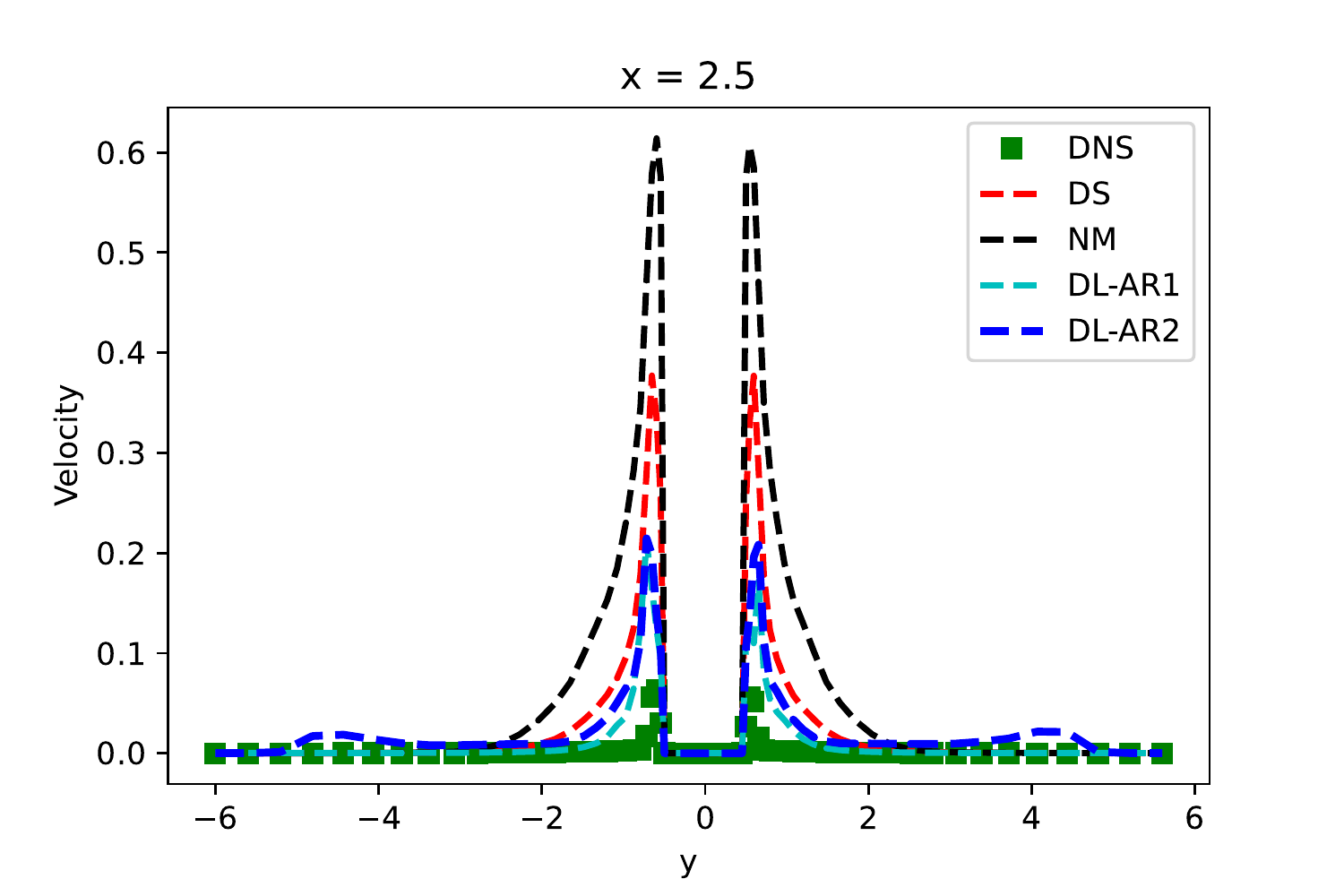}
\includegraphics[width=5cm]{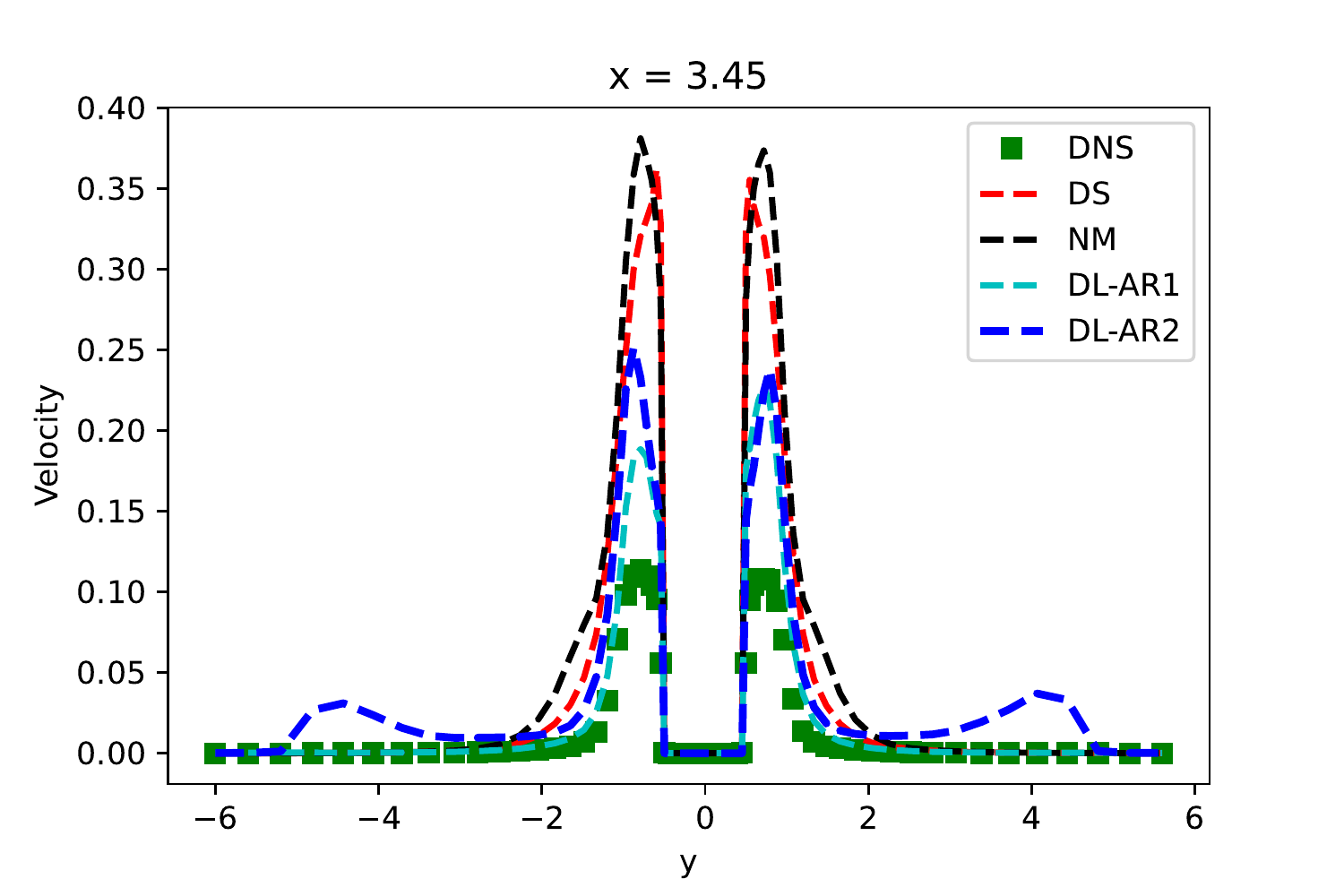}
\includegraphics[width=5cm]{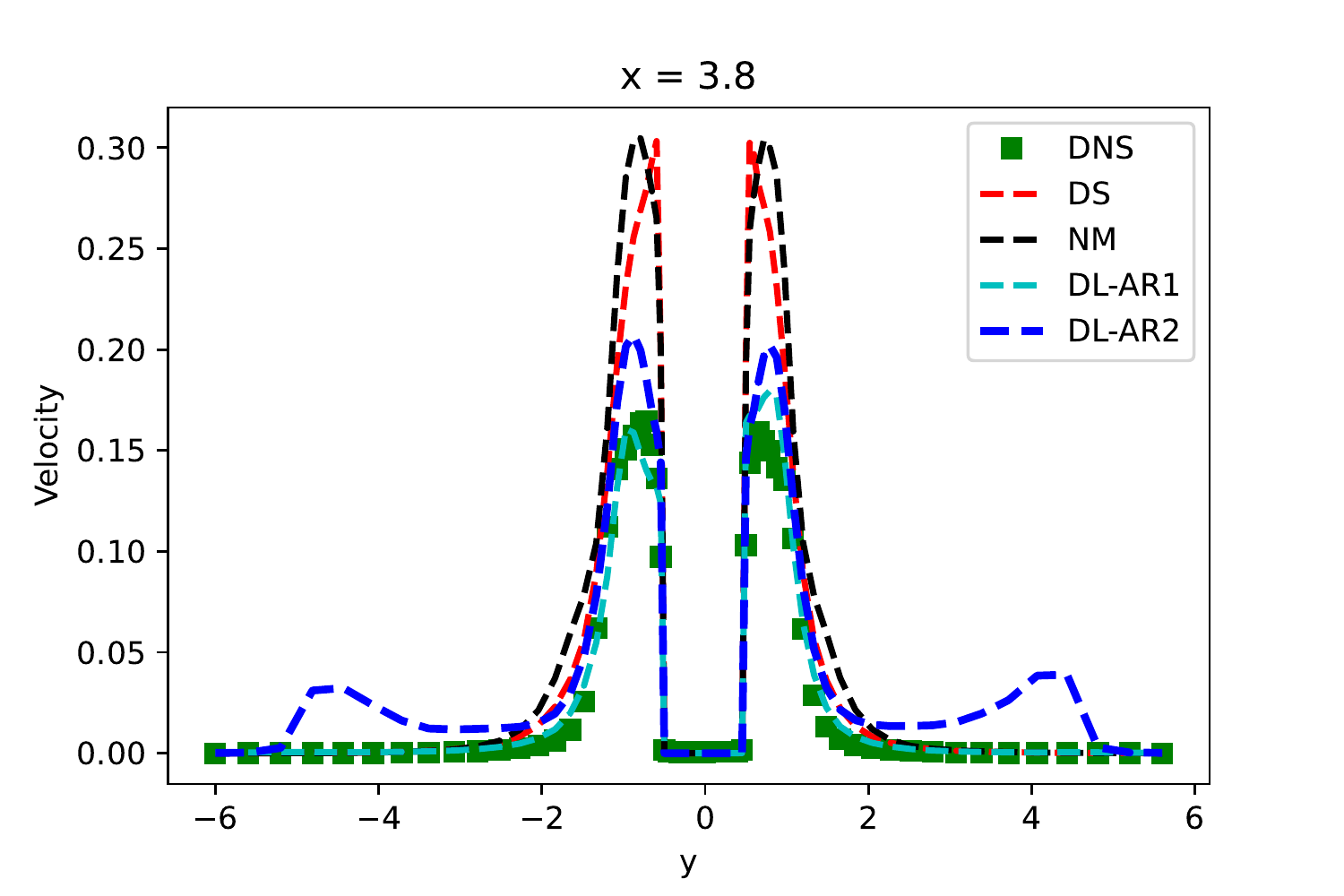}
\includegraphics[width=5cm]{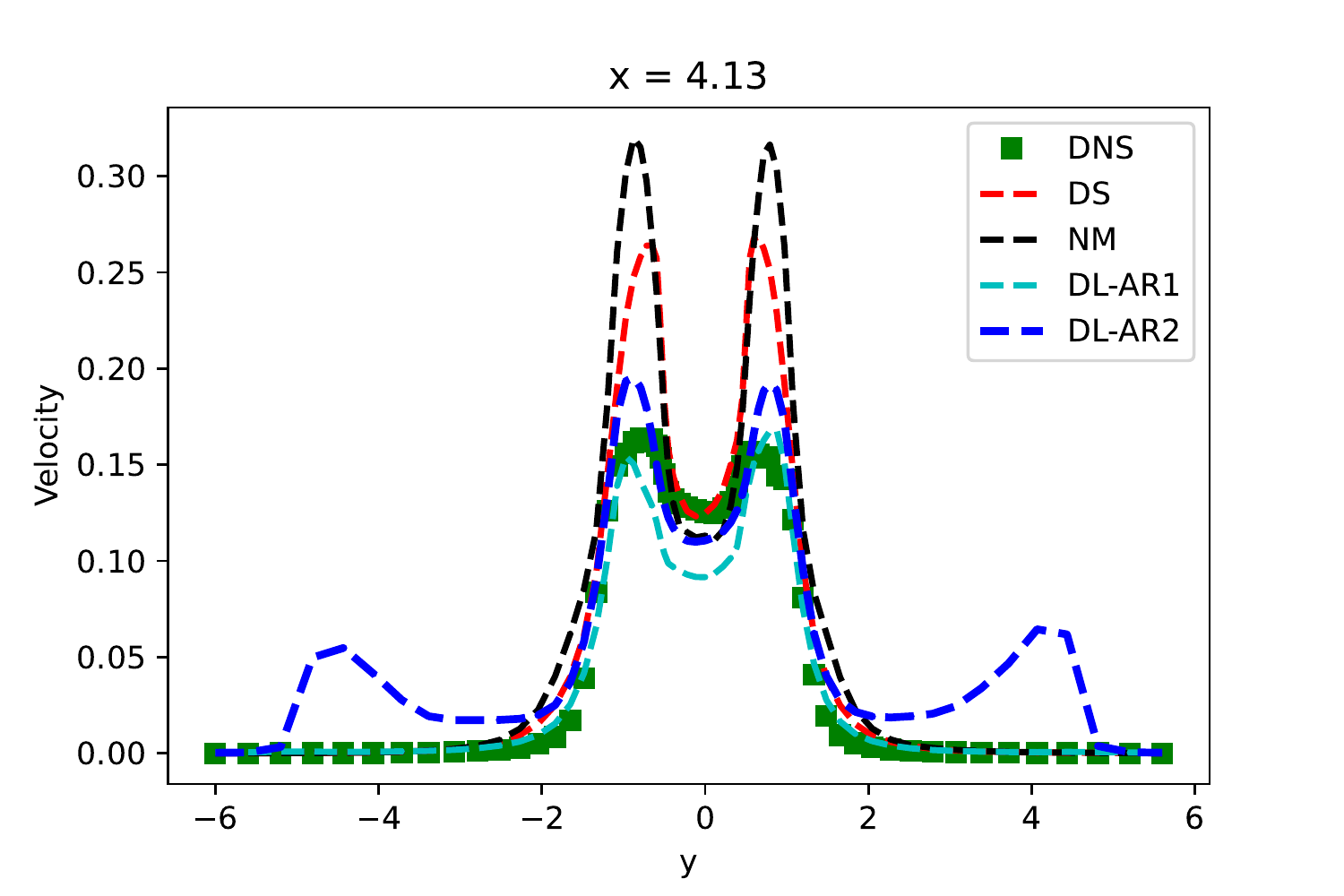}
\includegraphics[width=5cm]{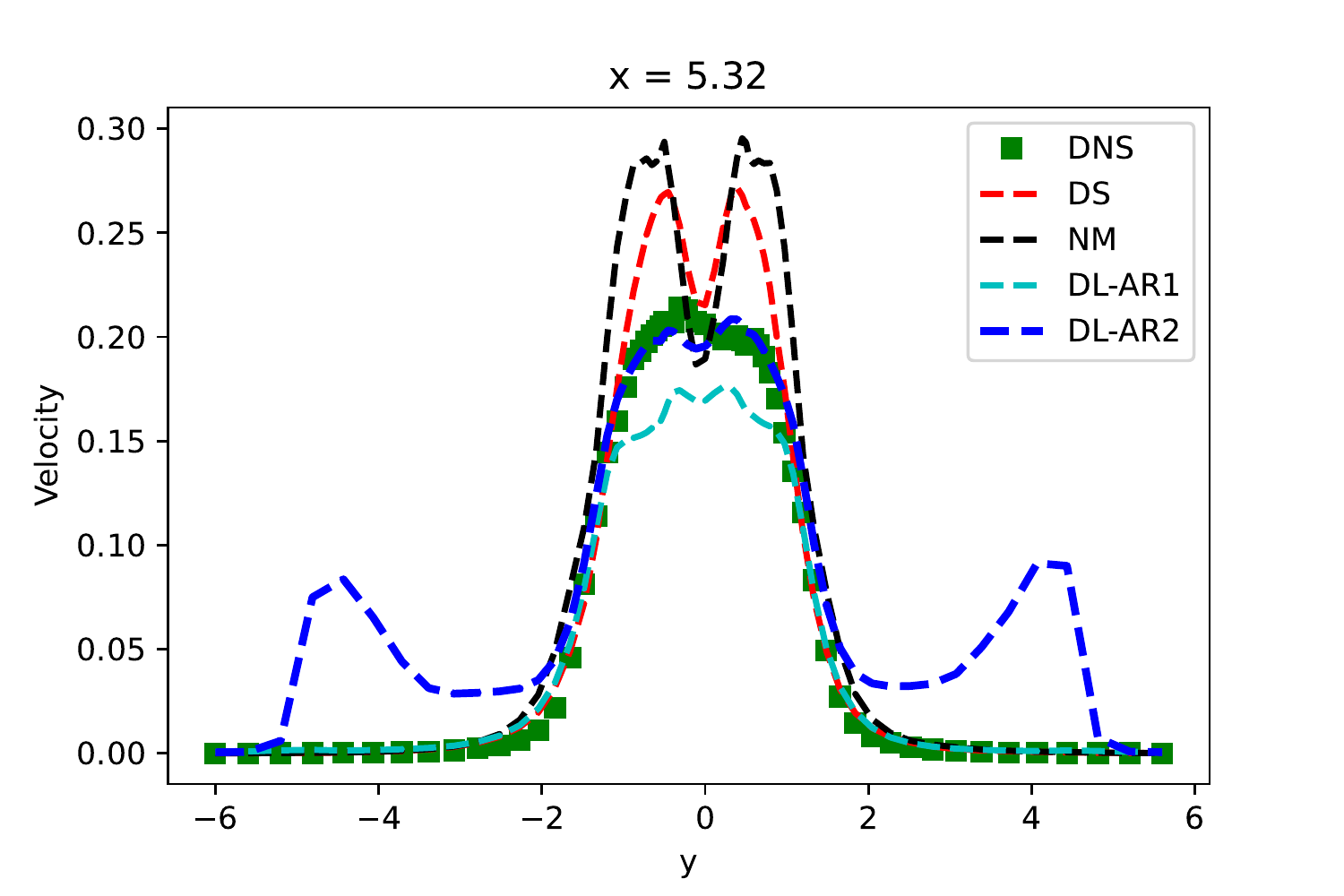}
\includegraphics[width=5cm]{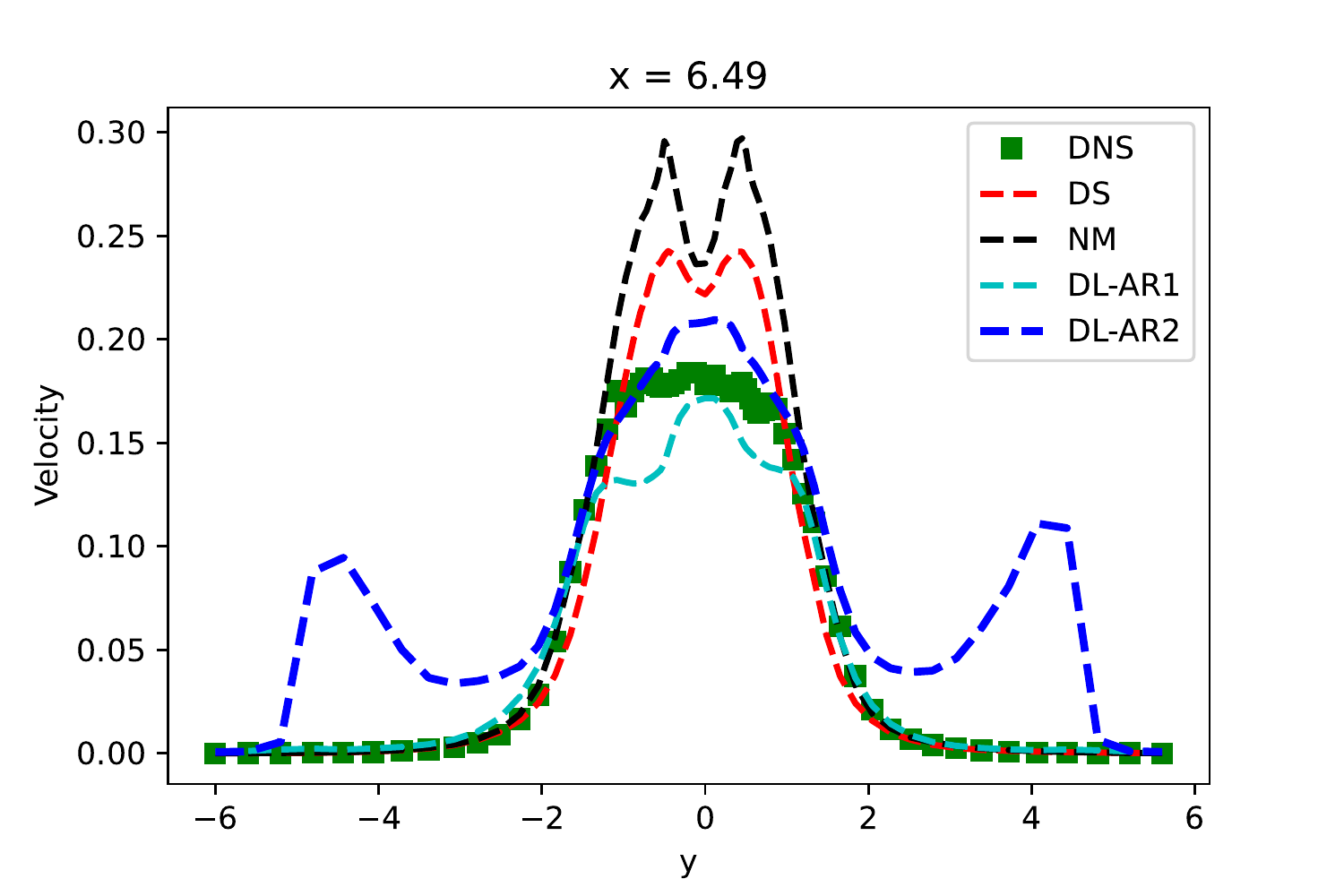}
\includegraphics[width=5cm]{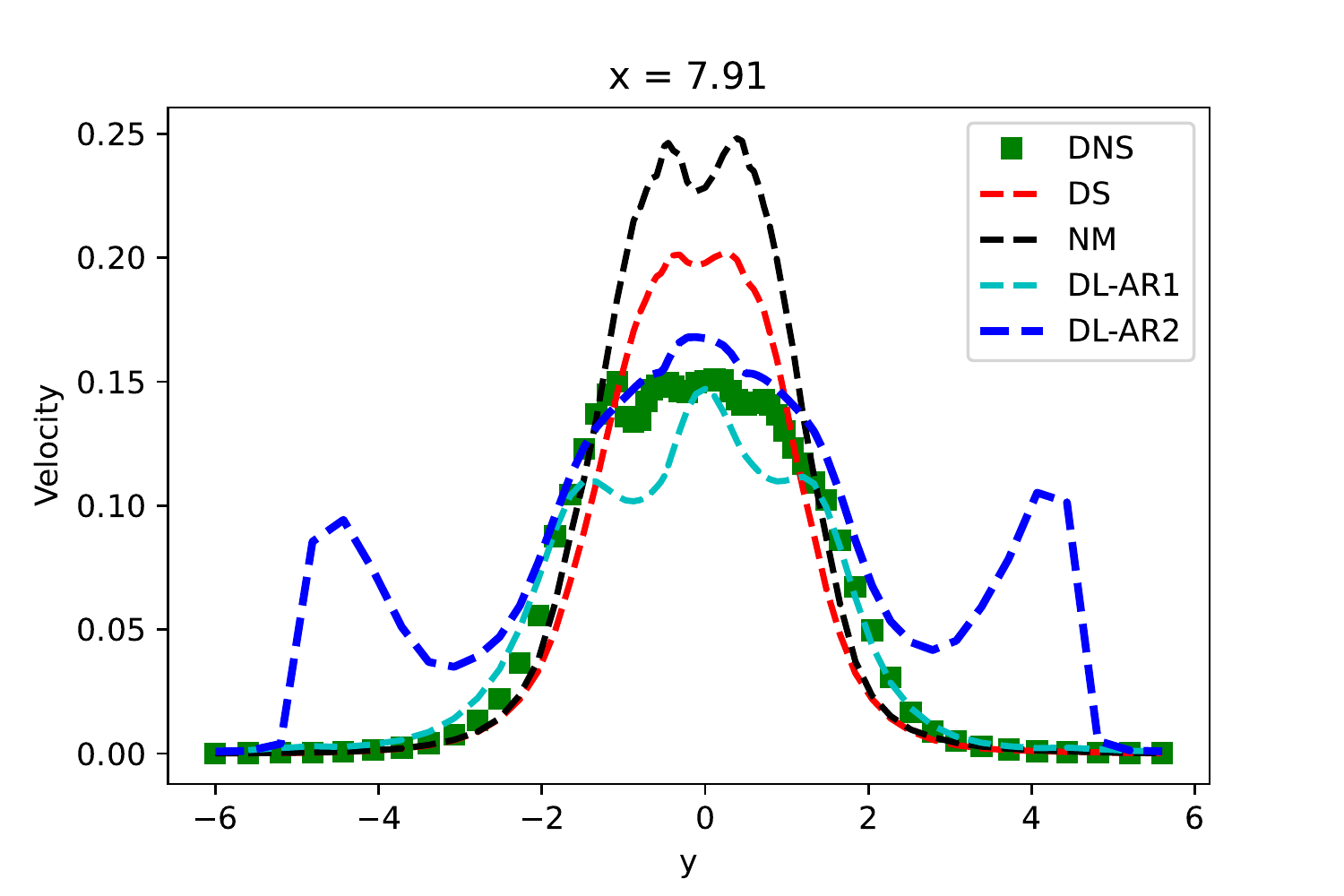}
\includegraphics[width=5cm]{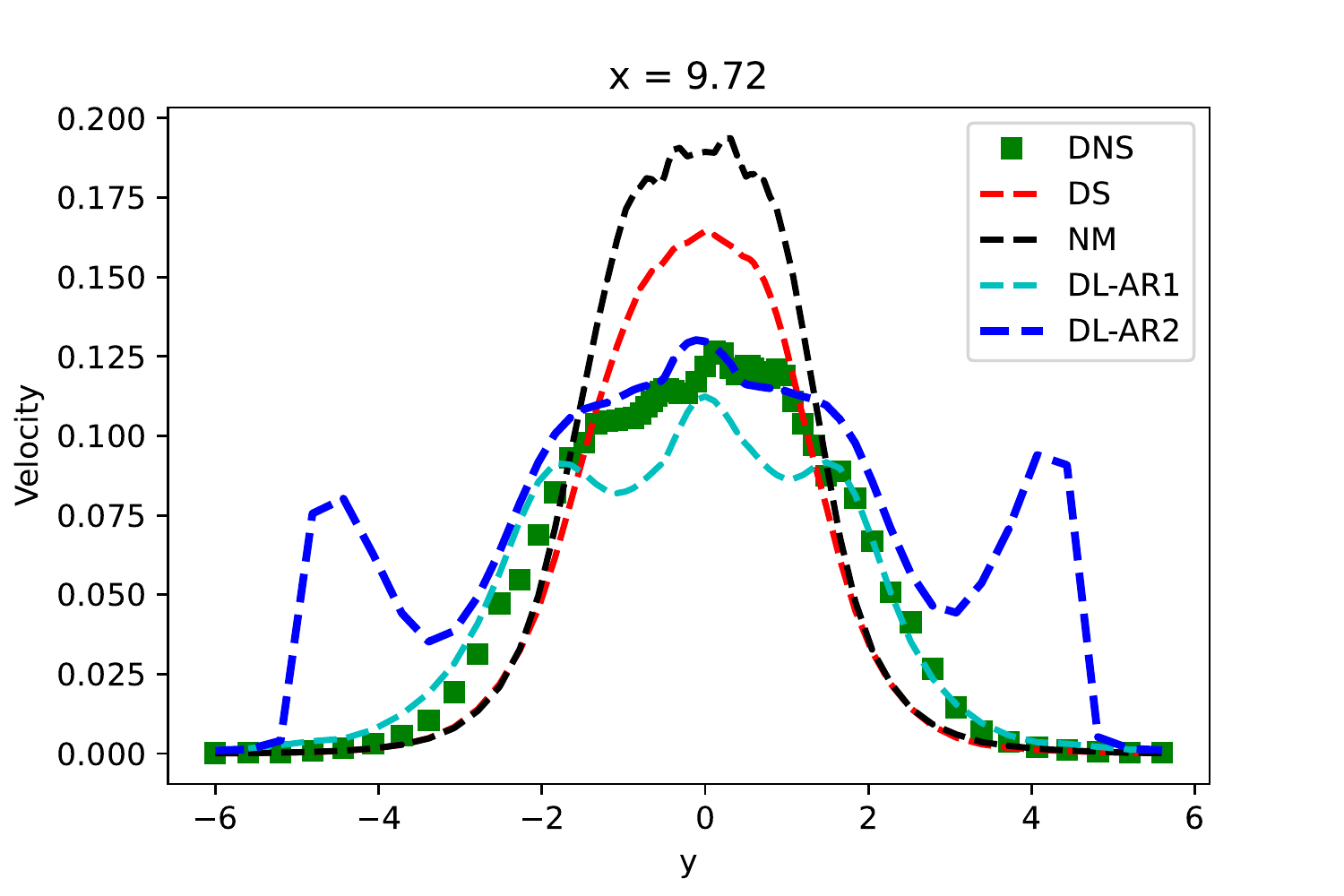}
\includegraphics[width=5cm]{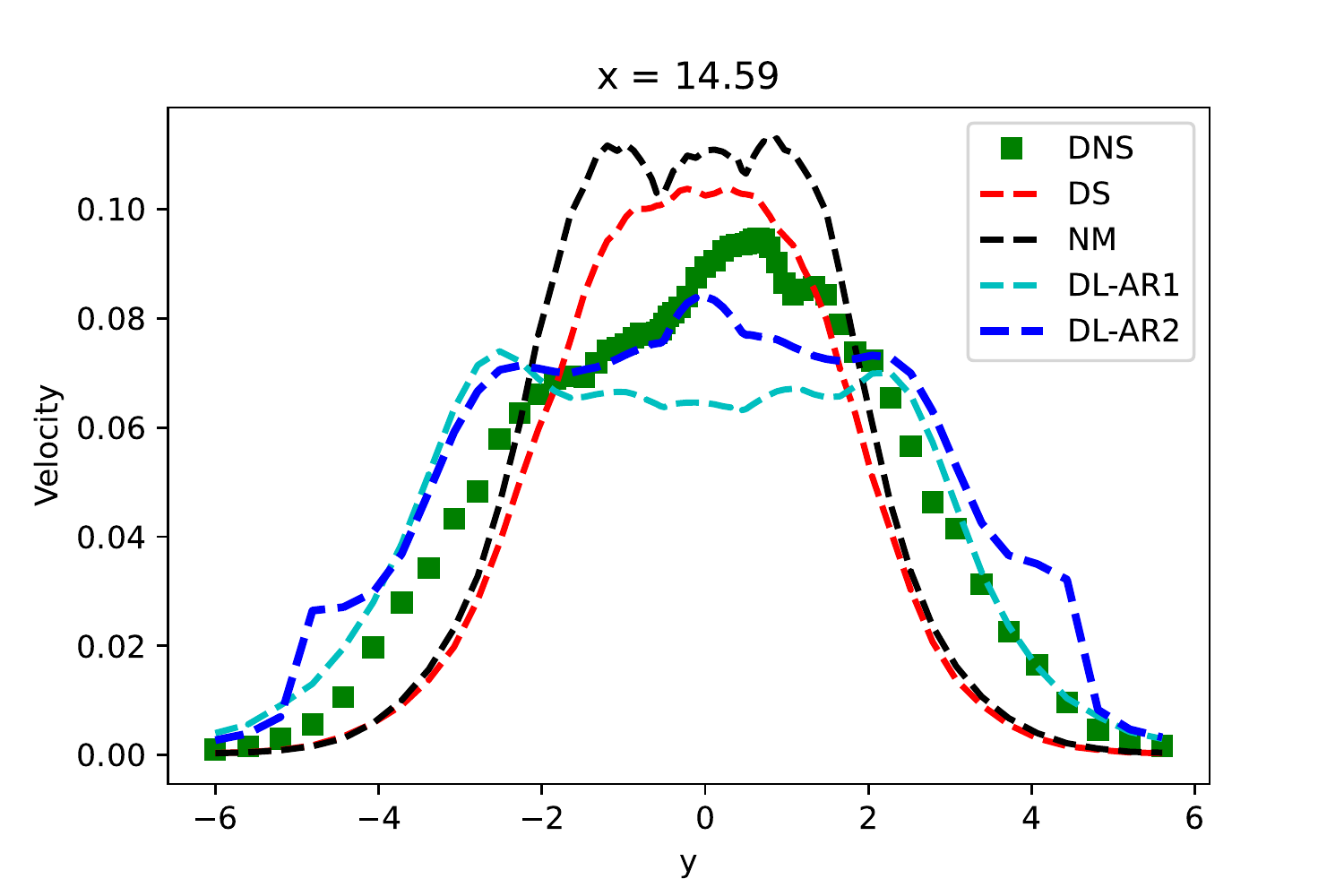}
\label{f1}
\caption{RMS profile for $u_3$ for AR2-Re$2,000$ configuration.}
\end{figure}

\begin{figure}[H]
\centering
\includegraphics[width=5cm]{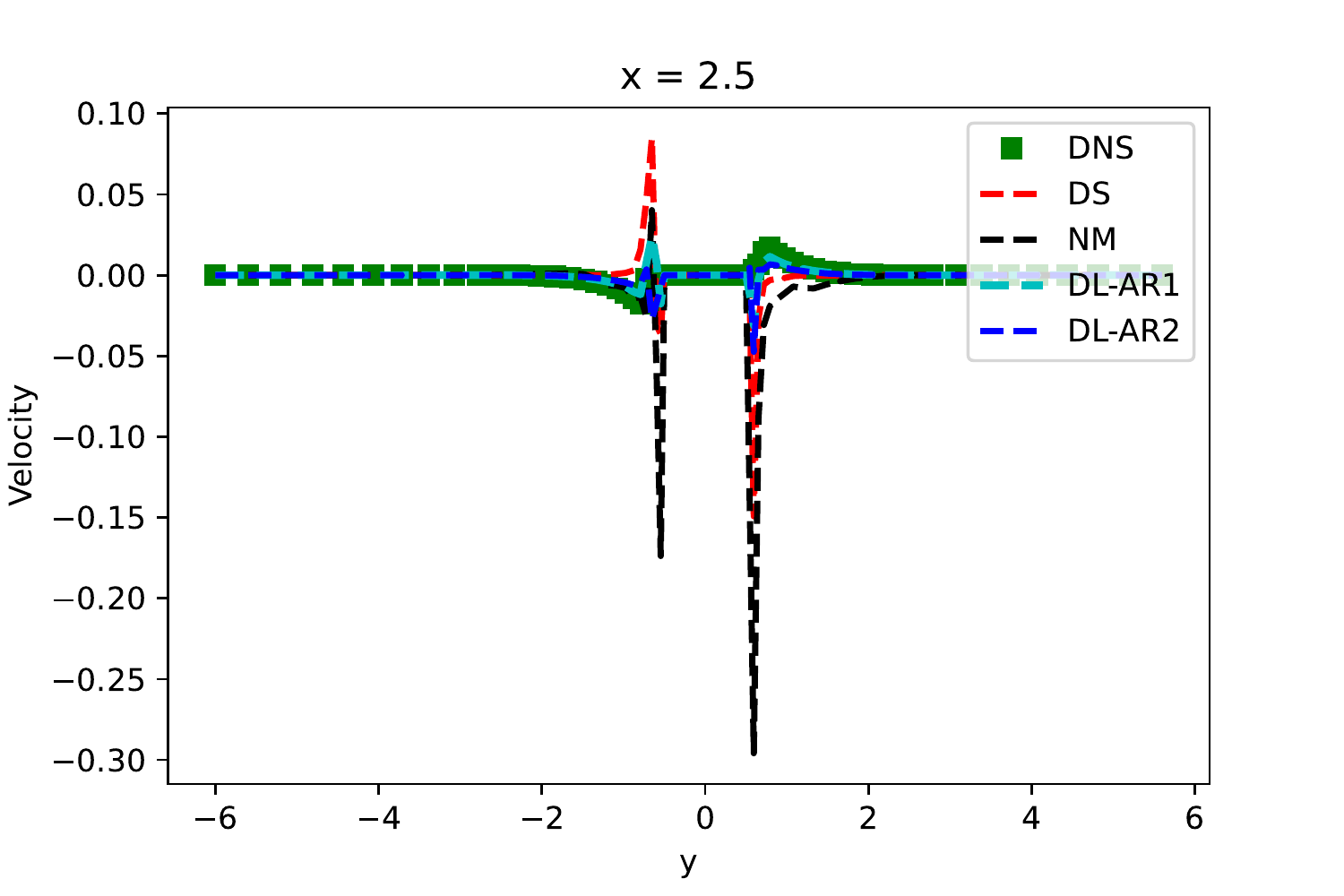}
\includegraphics[width=5cm]{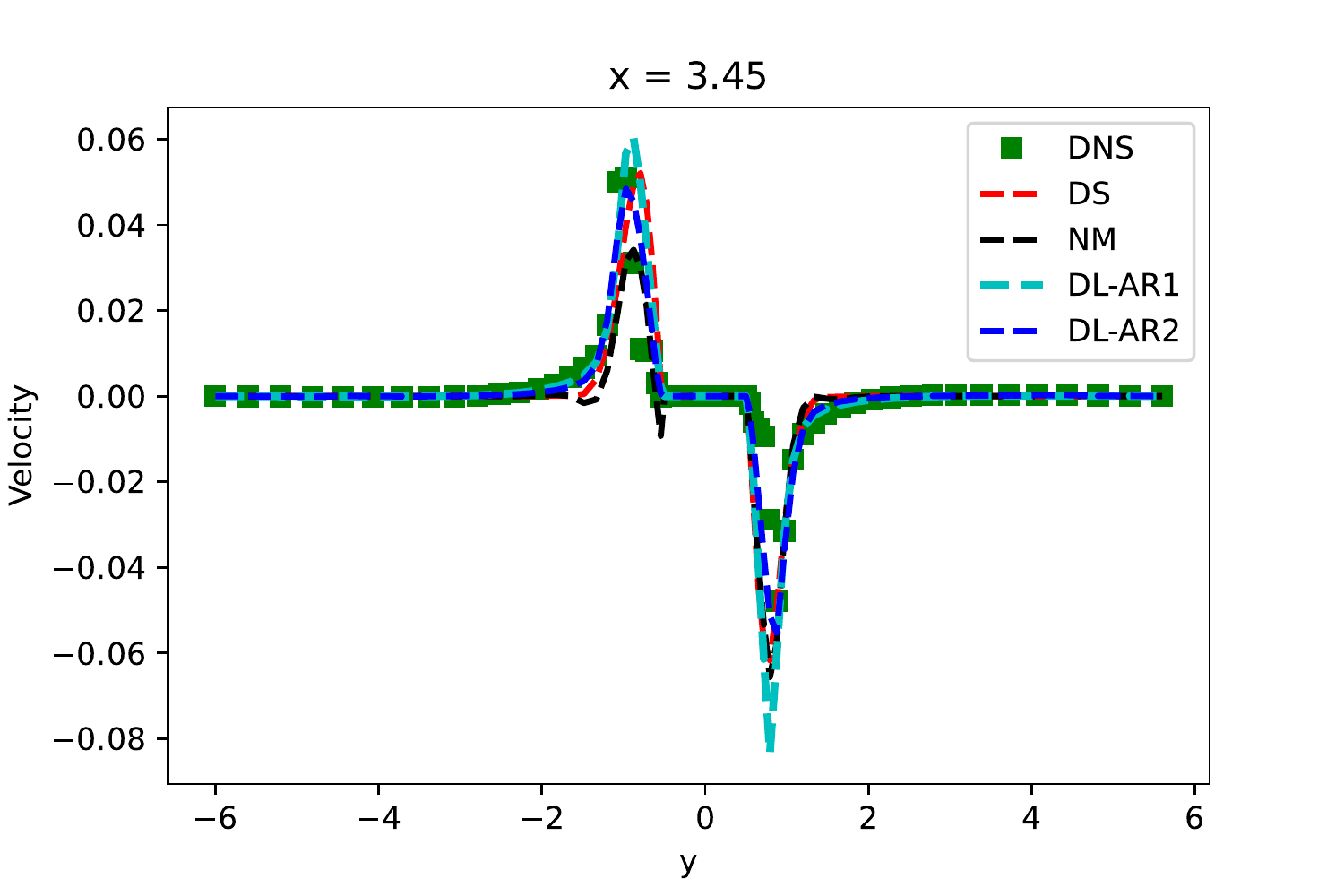}
\includegraphics[width=5cm]{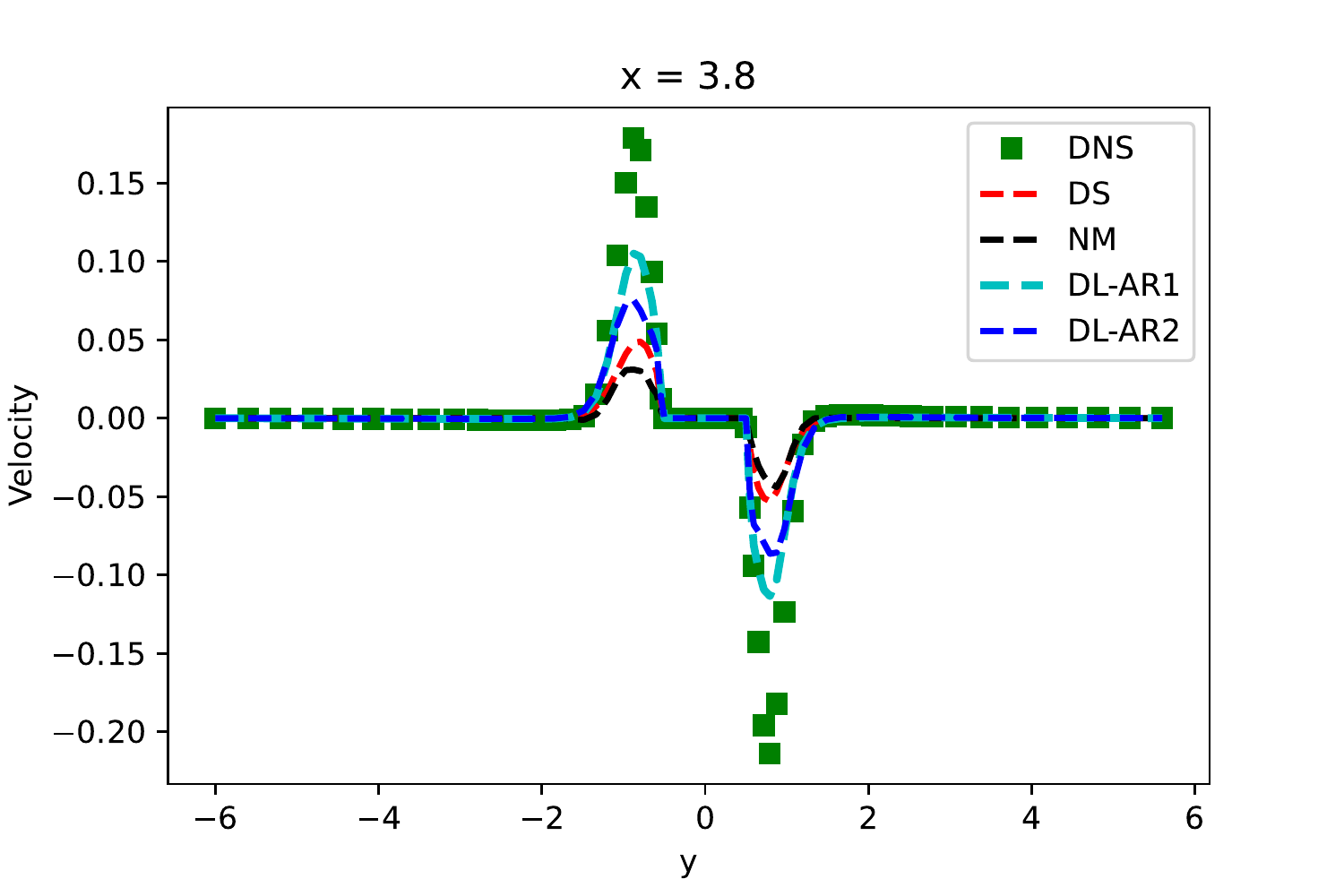}
\includegraphics[width=5cm]{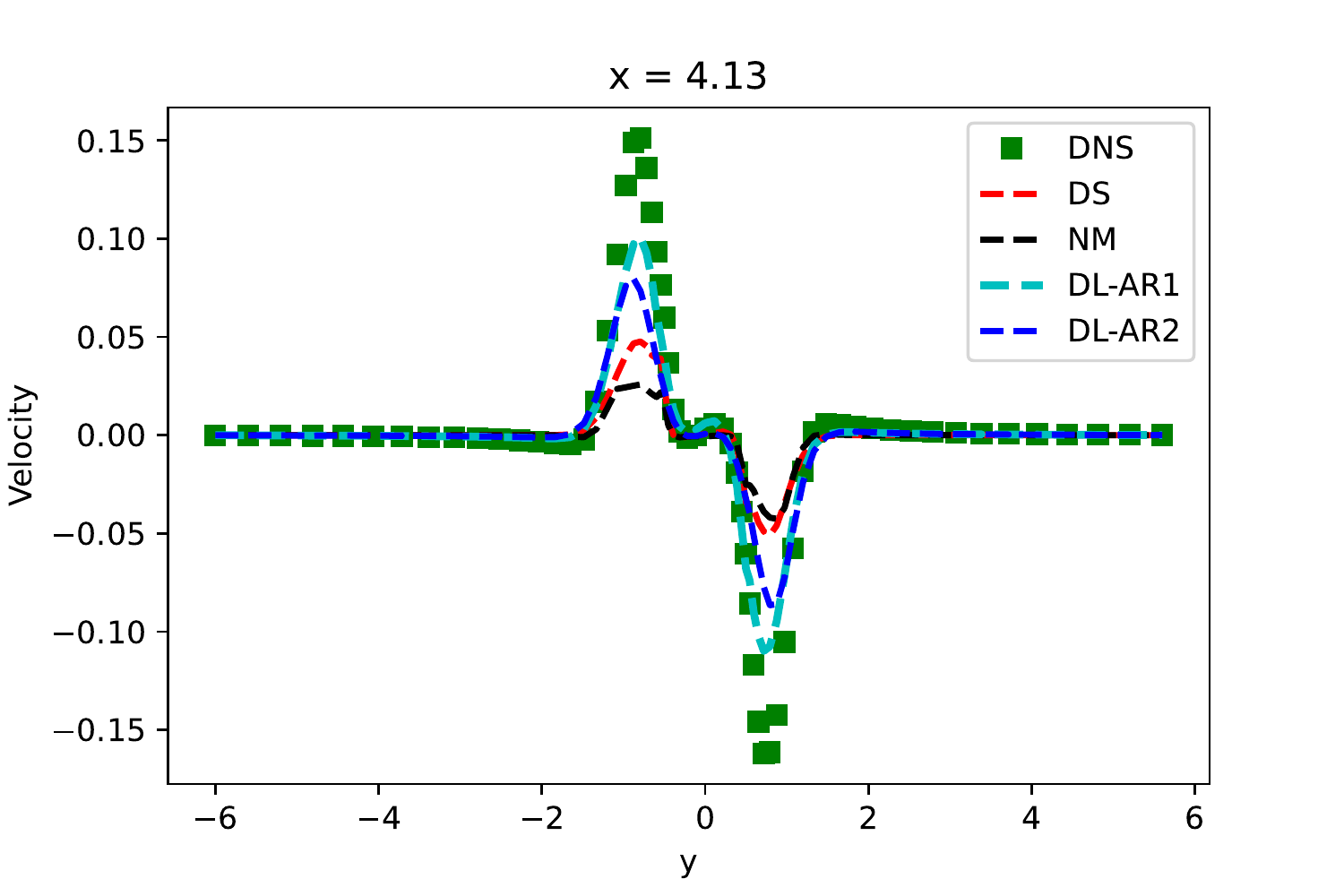}
\includegraphics[width=5cm]{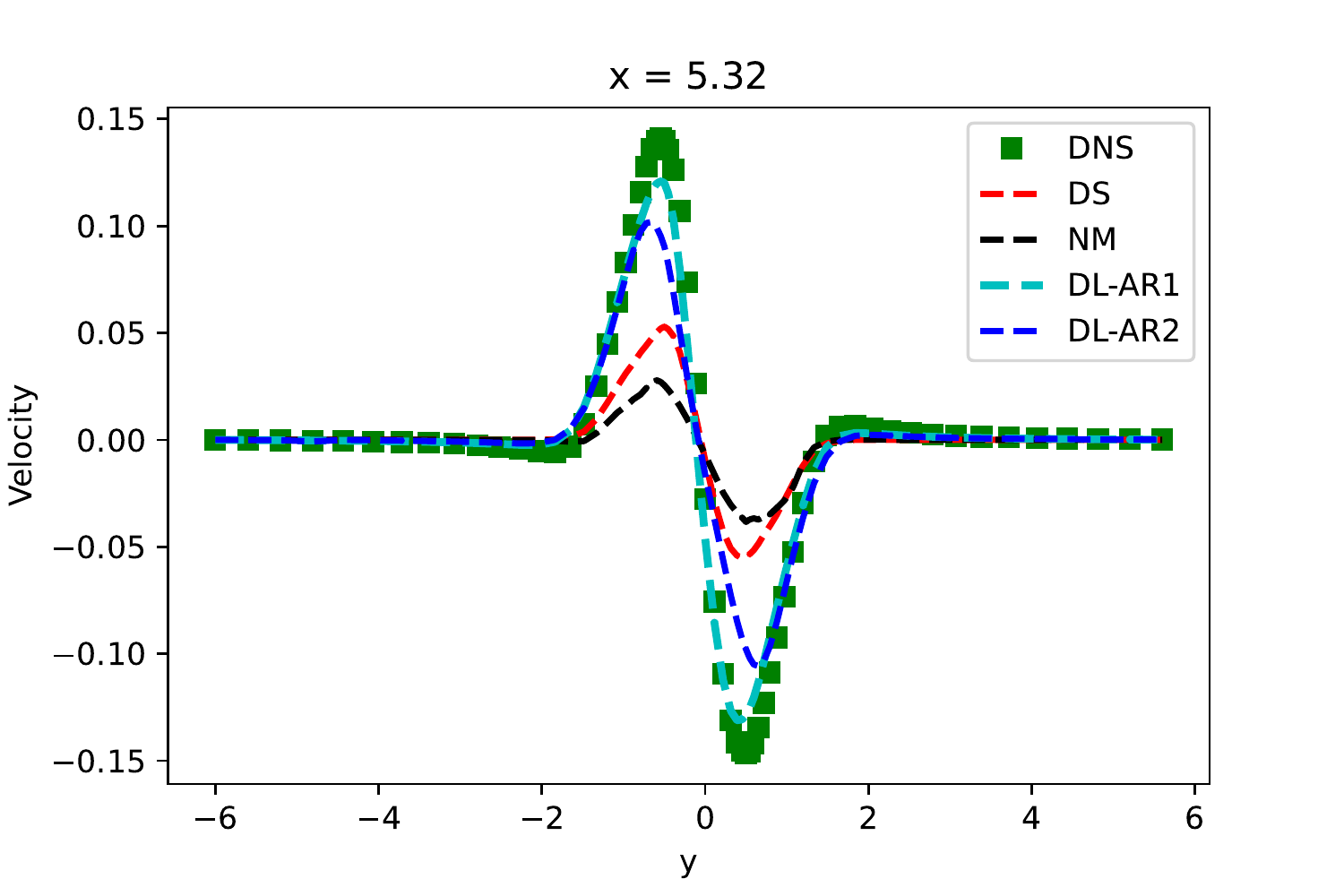}
\includegraphics[width=5cm]{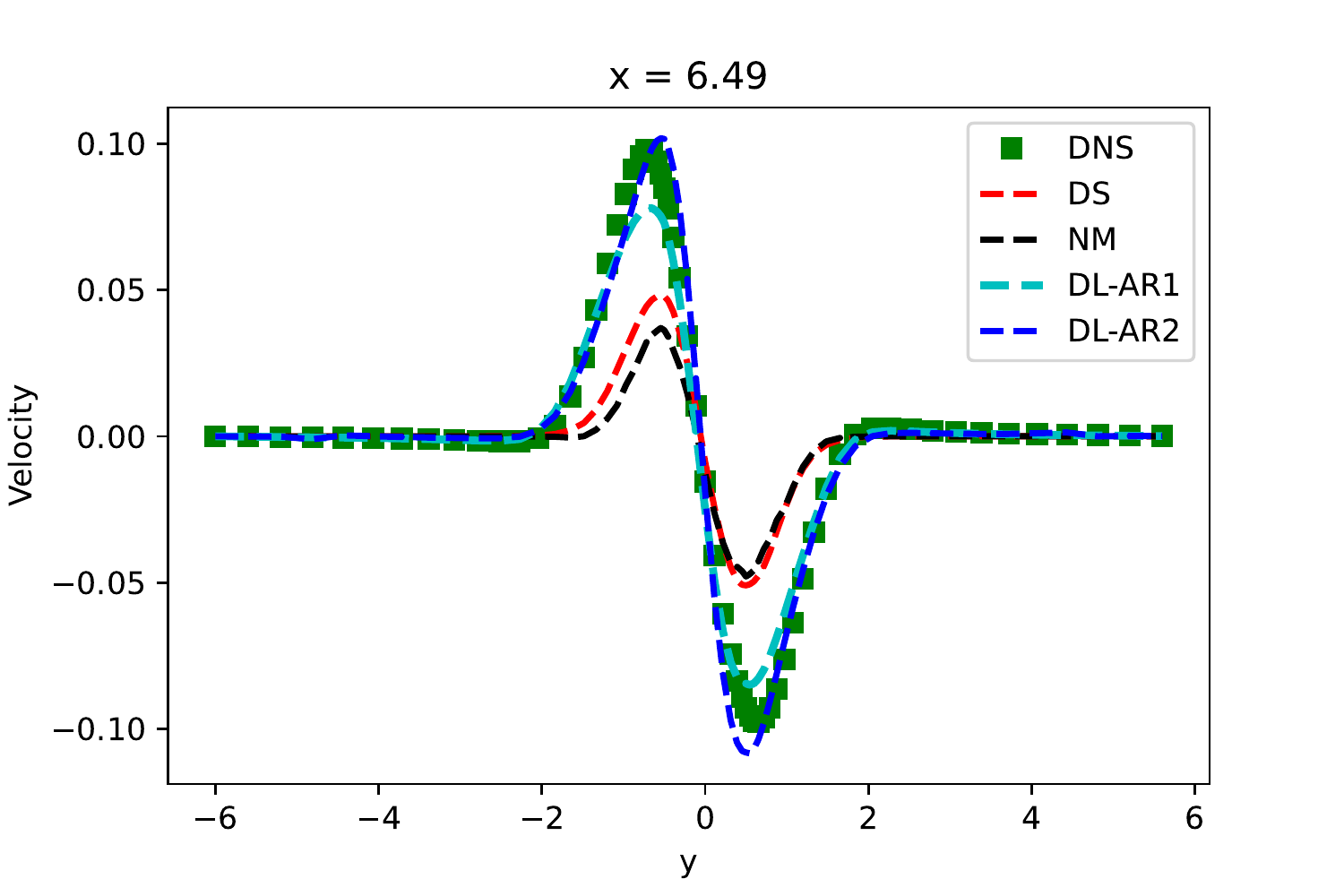}
\includegraphics[width=5cm]{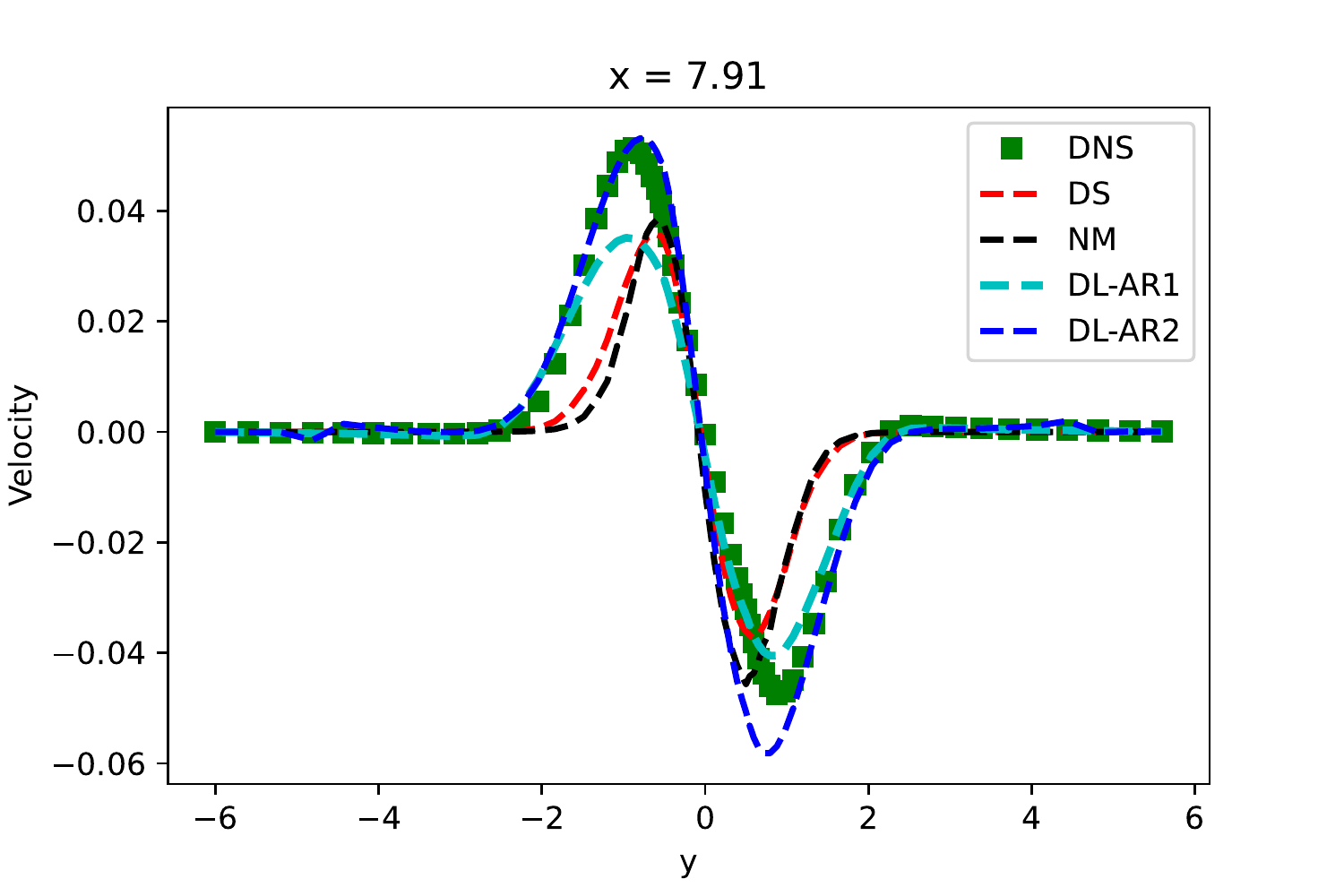}
\includegraphics[width=5cm]{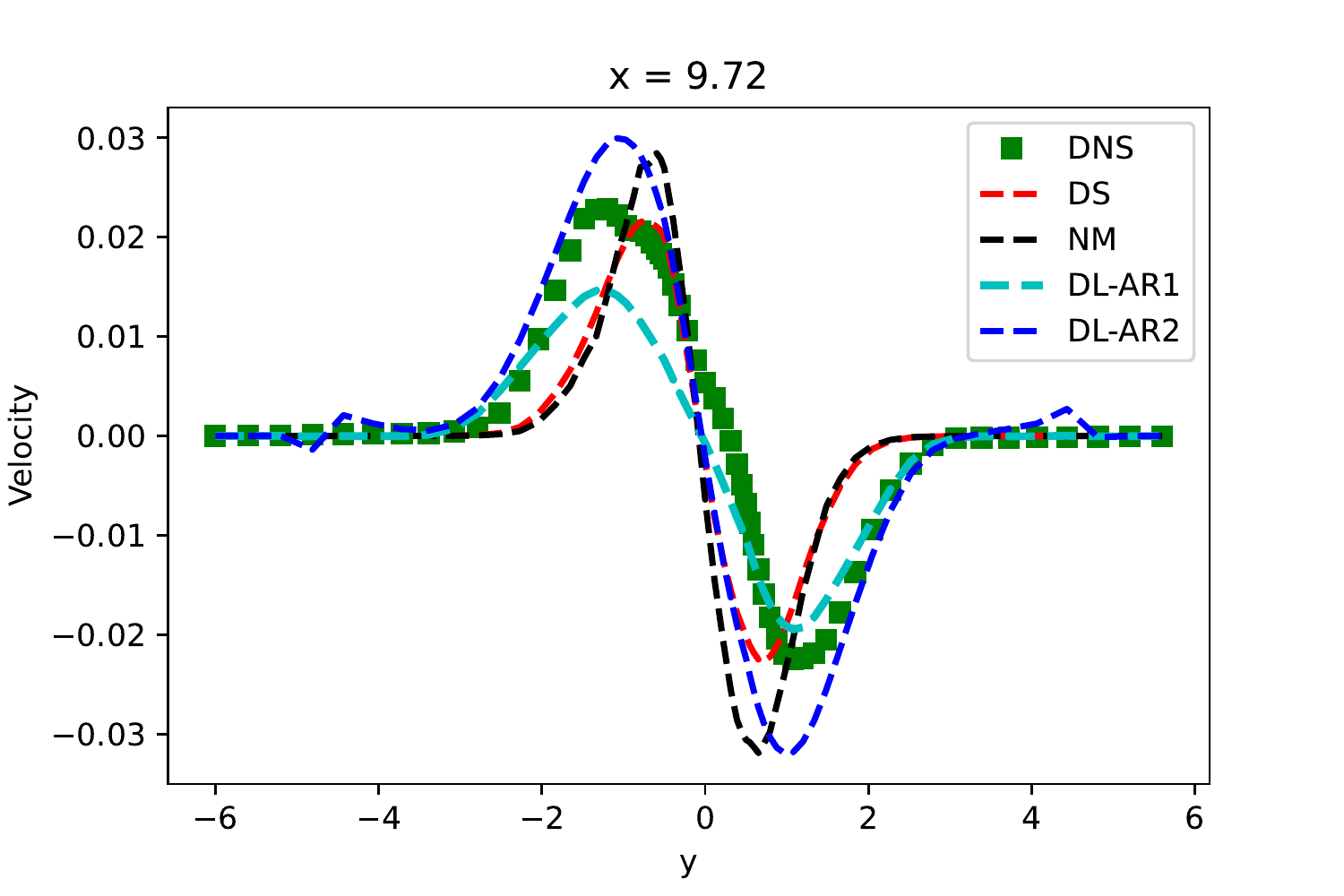}
\includegraphics[width=5cm]{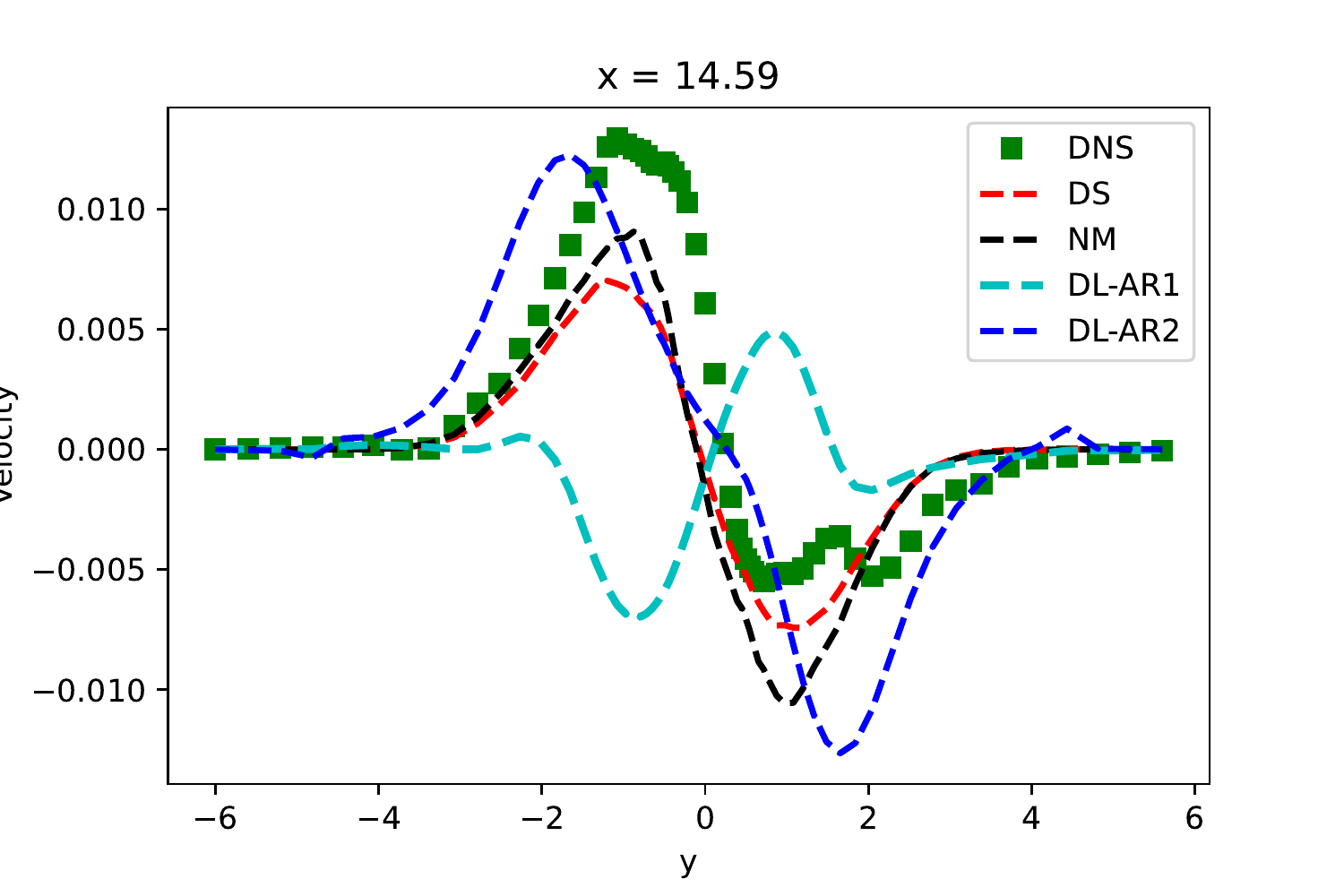}
\label{f1}
\caption{$\tau_{12}$ for AR2-Re$2,000$ configuration.}
\end{figure}

\subsubsection{AR4}

\begin{figure}[H]
\centering
\includegraphics[width=5cm]{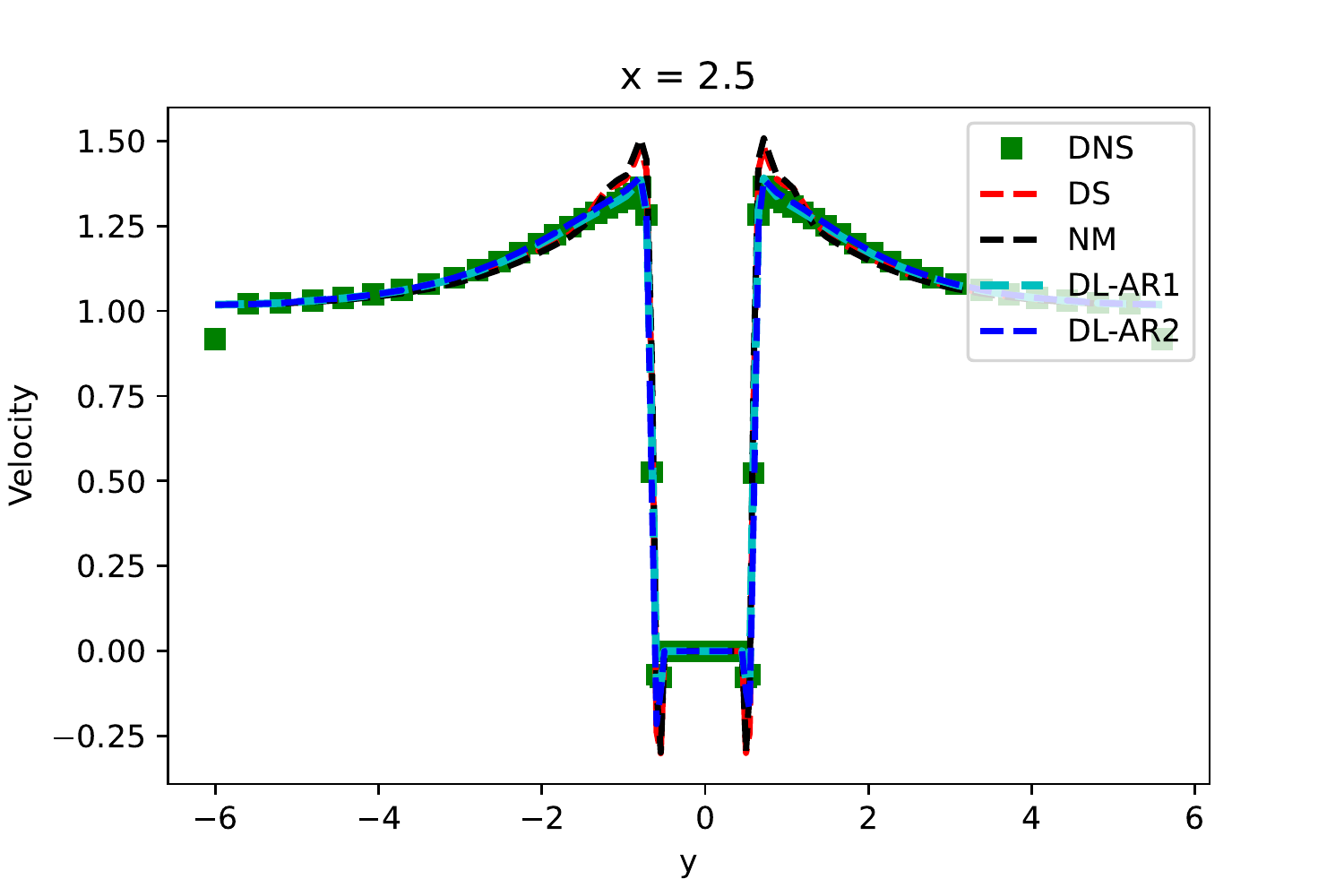}
\includegraphics[width=5cm]{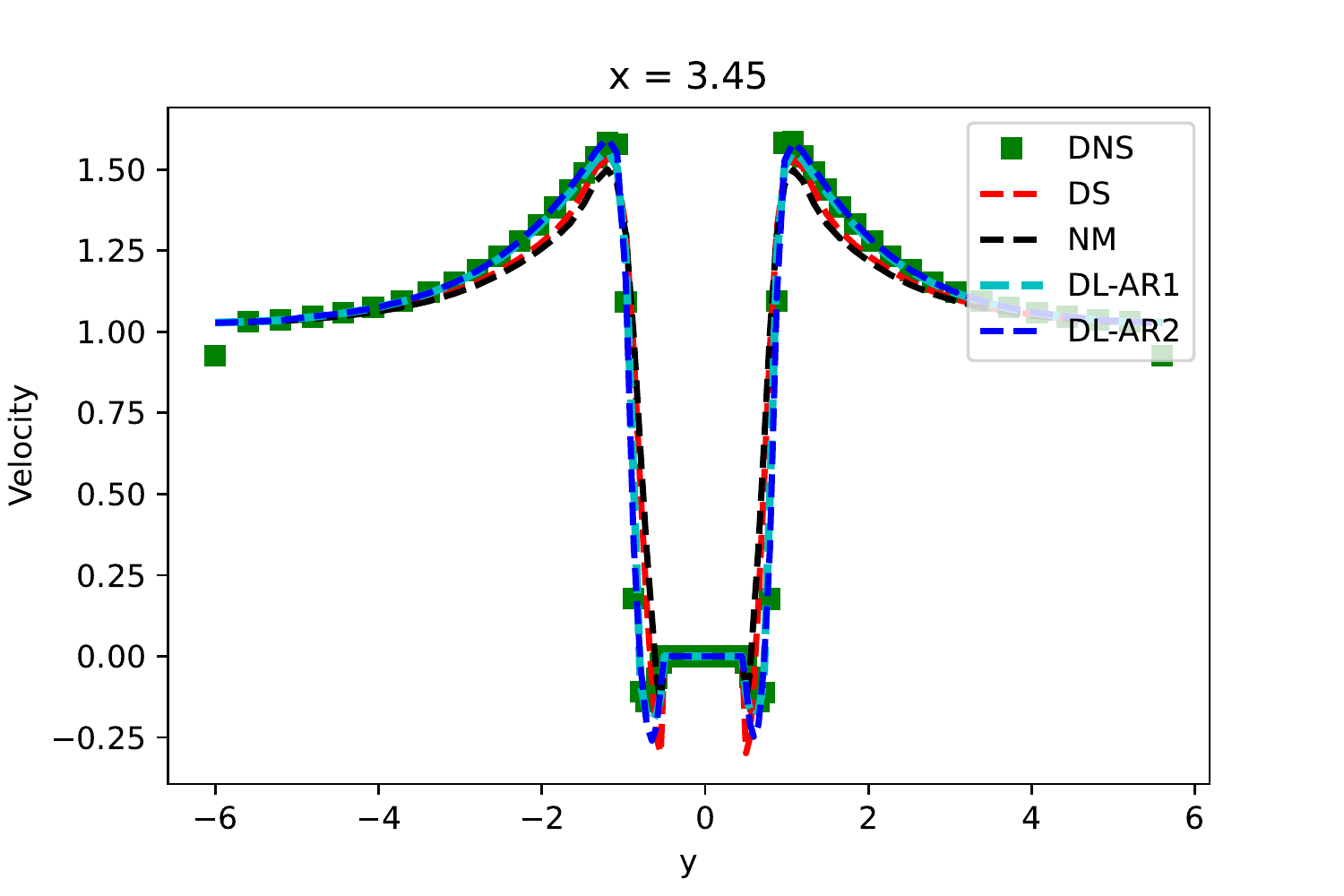}
\includegraphics[width=5cm]{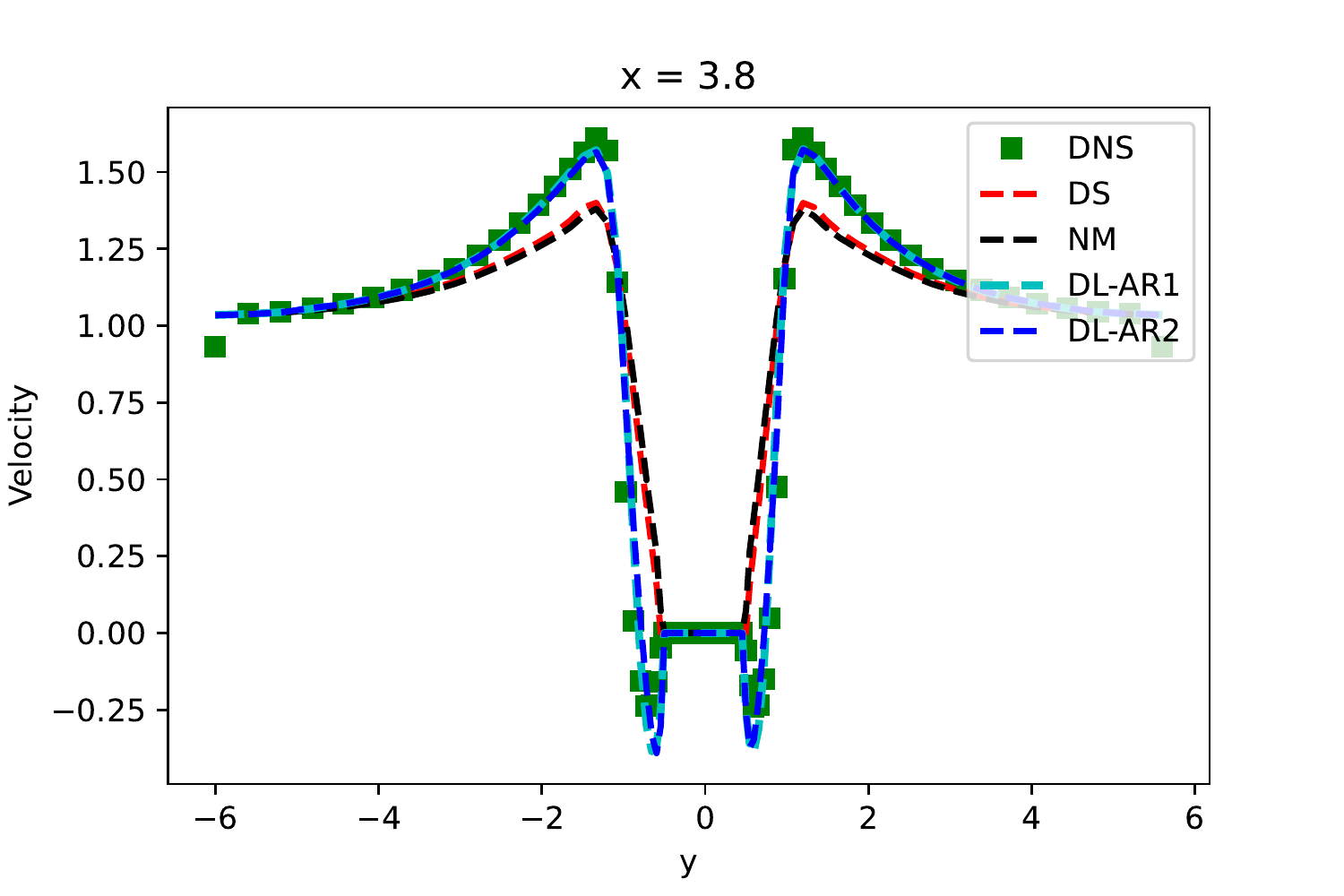}
\includegraphics[width=5cm]{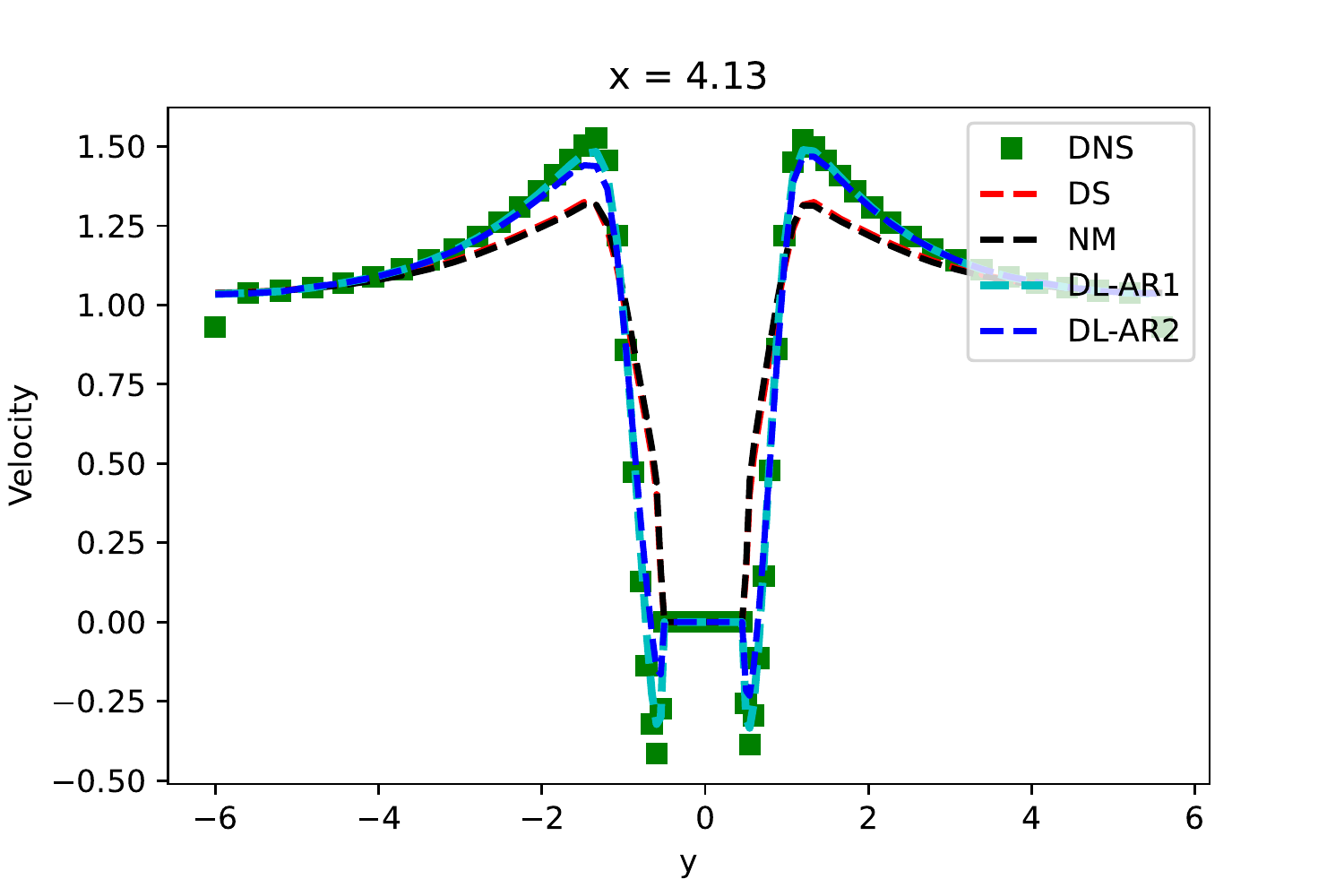}
\includegraphics[width=5cm]{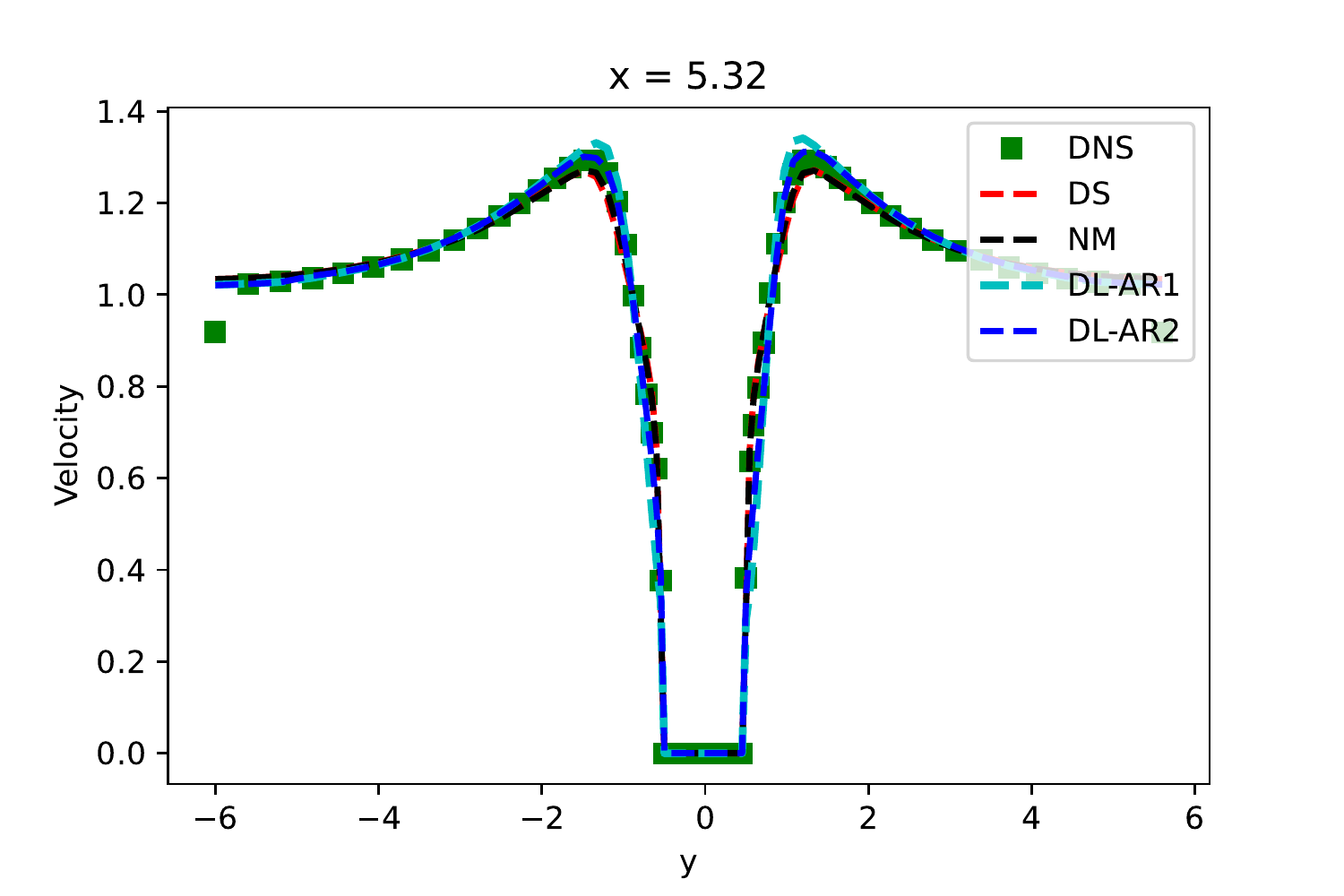}
\includegraphics[width=5cm]{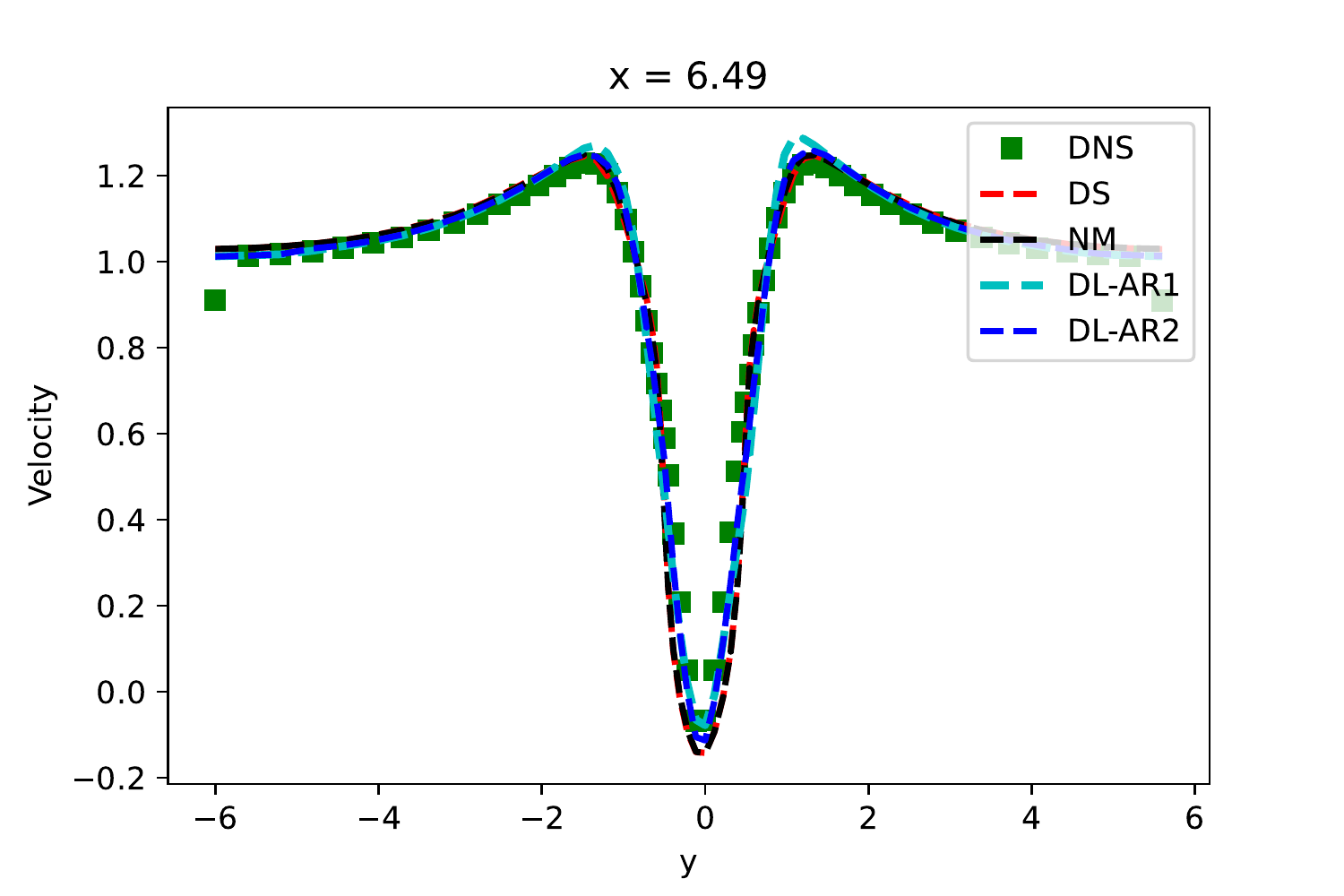}
\includegraphics[width=5cm]{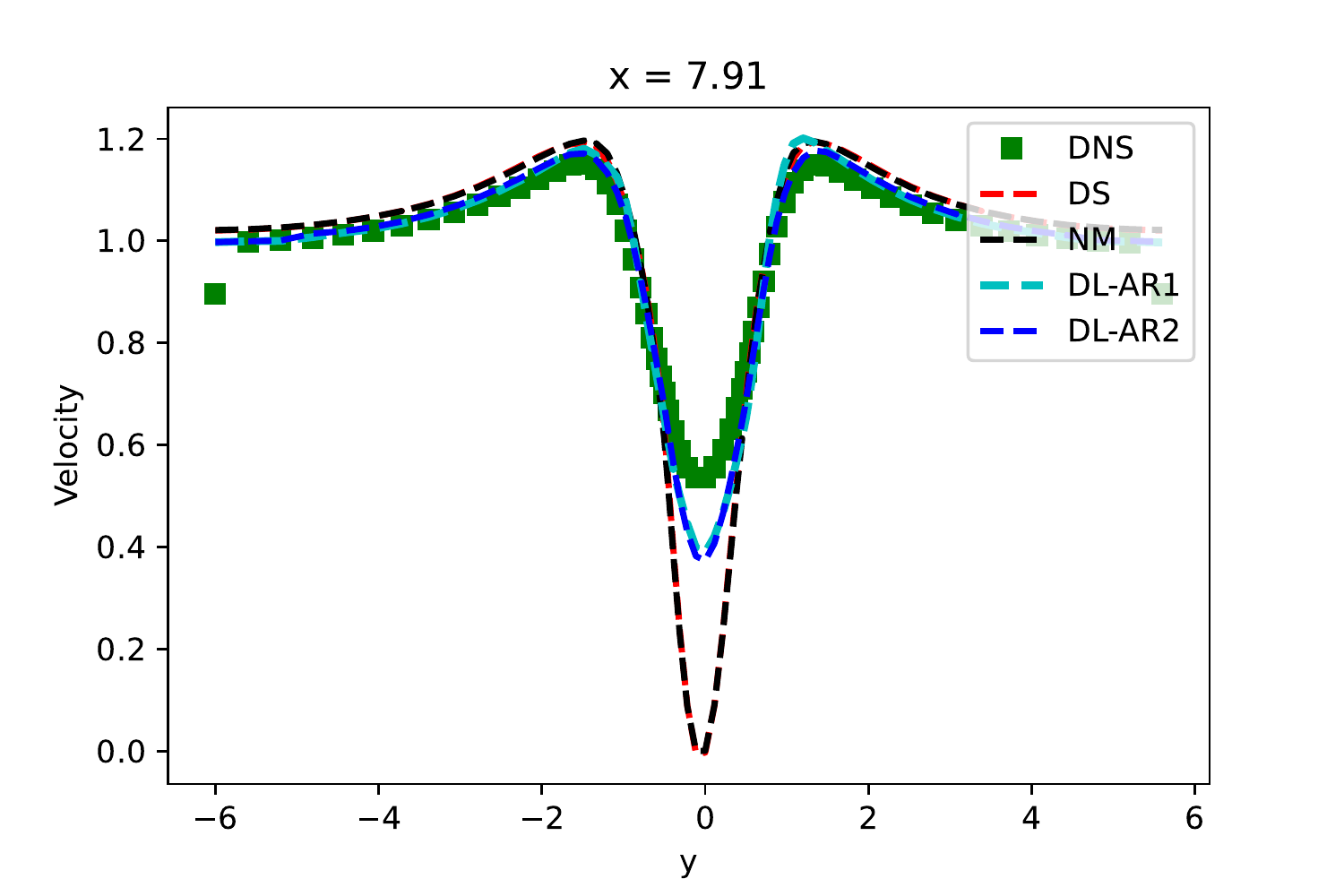}
\includegraphics[width=5cm]{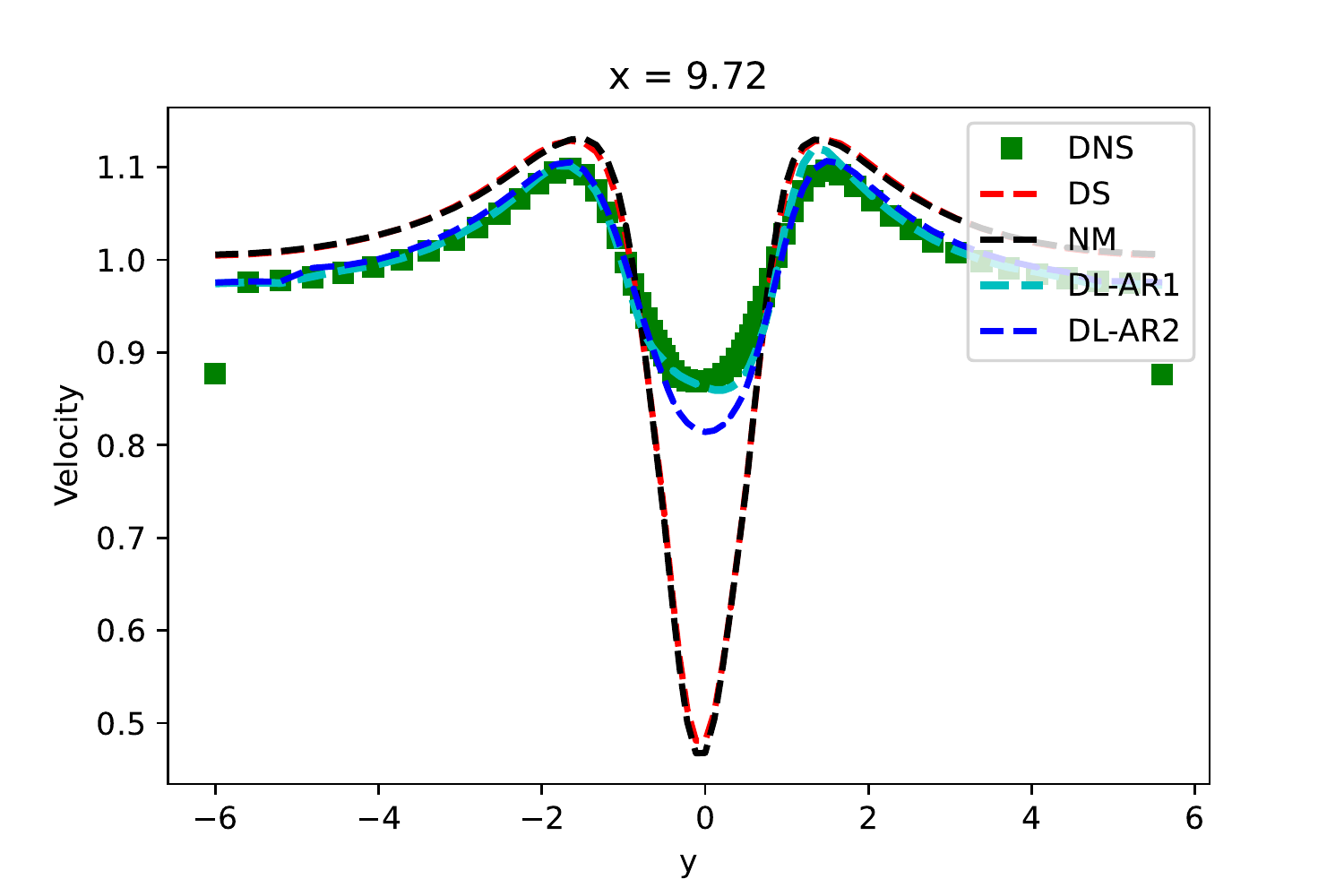}
\includegraphics[width=5cm]{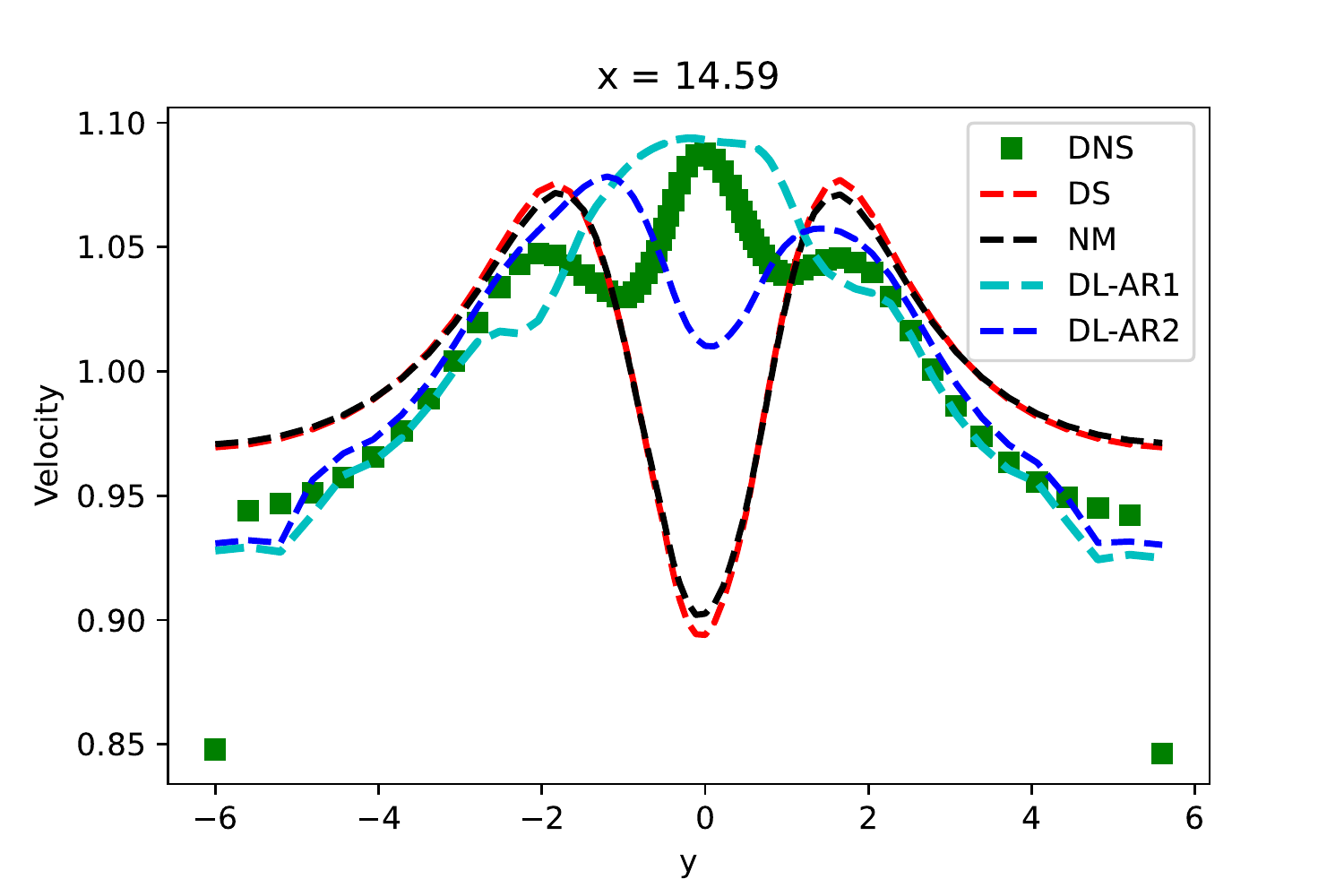}
\label{f1}
\caption{Mean profile for $u_1$ for AR4 configuration.}
\end{figure}

\begin{figure}[H]
\centering
\includegraphics[width=5cm]{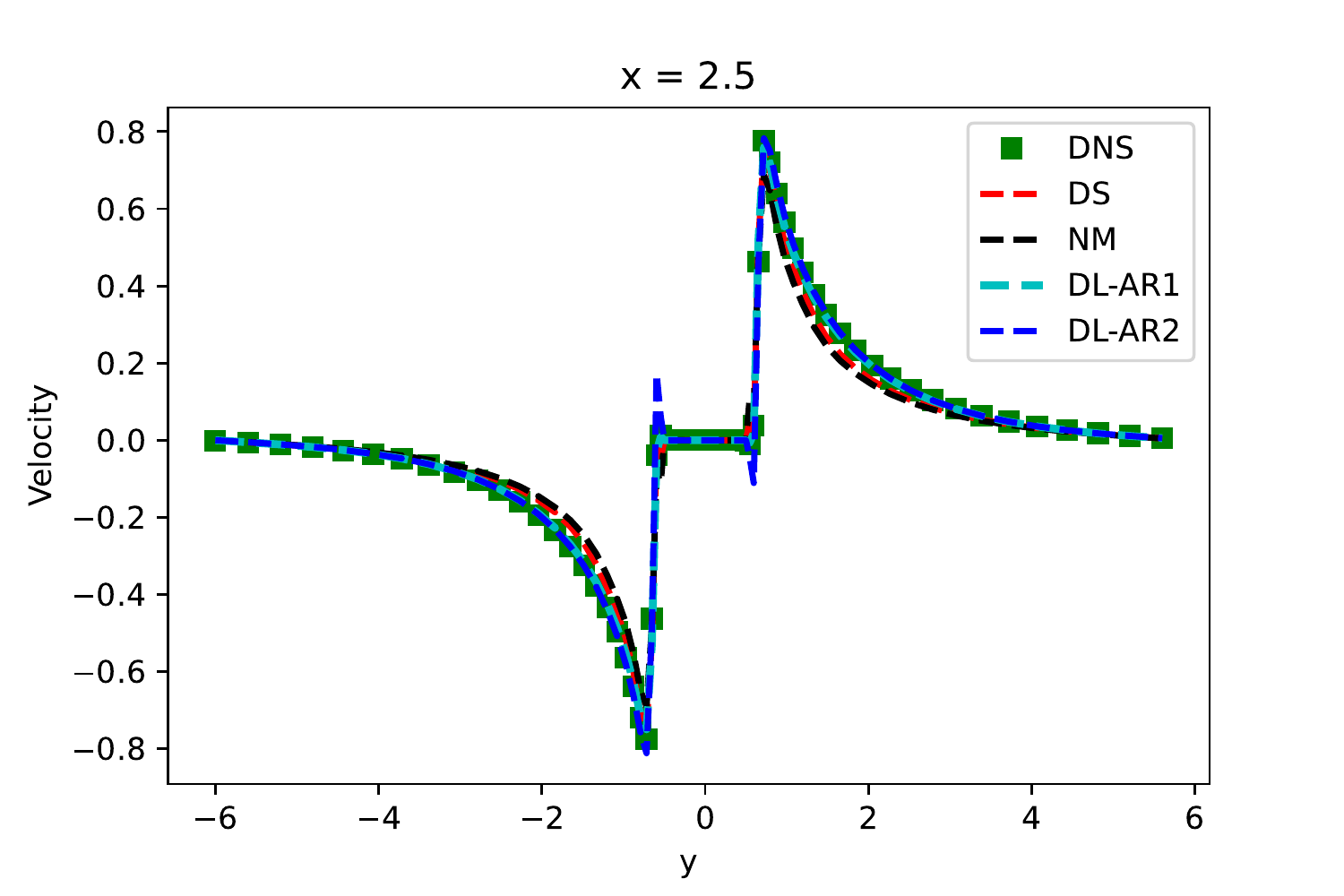}
\includegraphics[width=5cm]{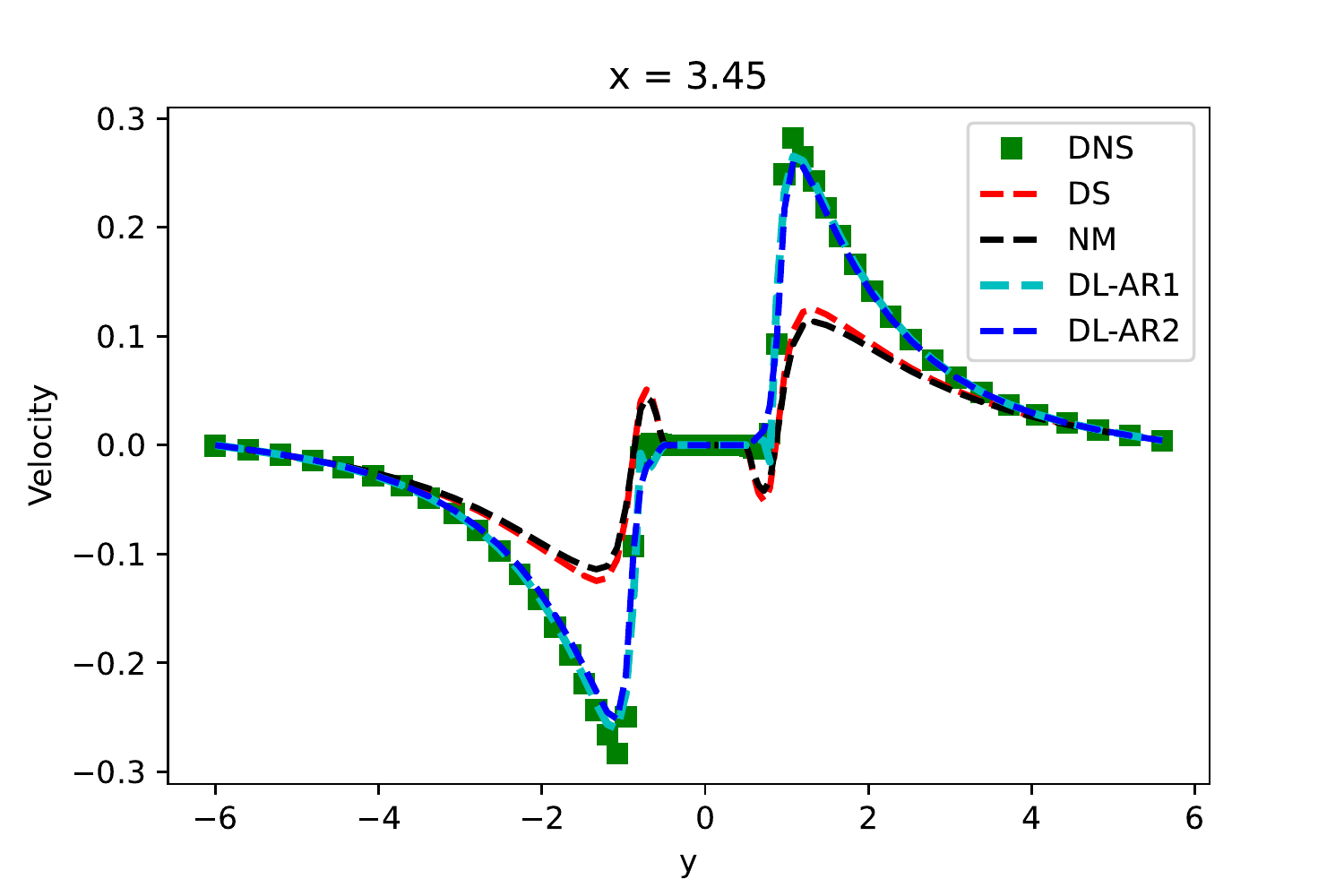}
\includegraphics[width=5cm]{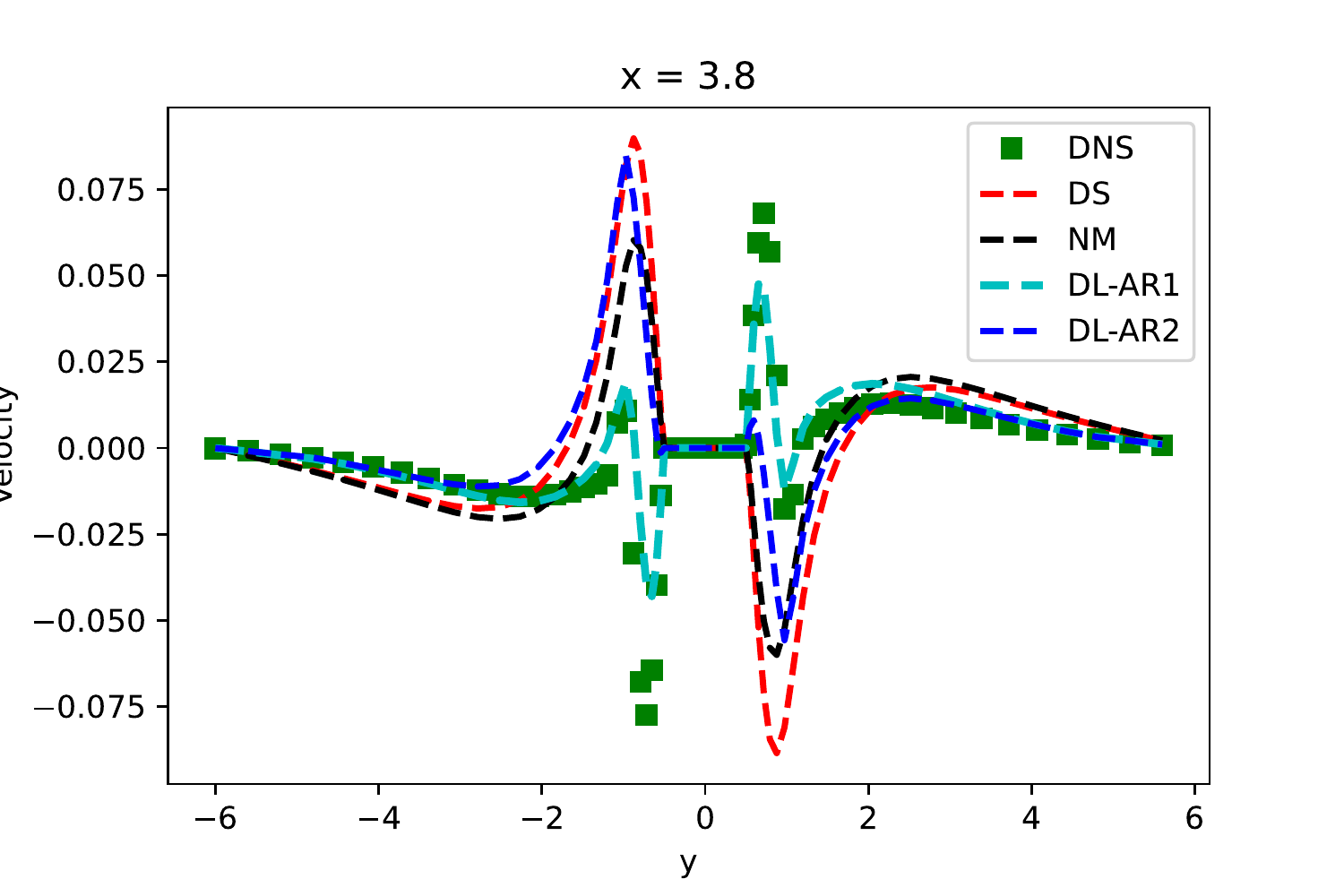}
\includegraphics[width=5cm]{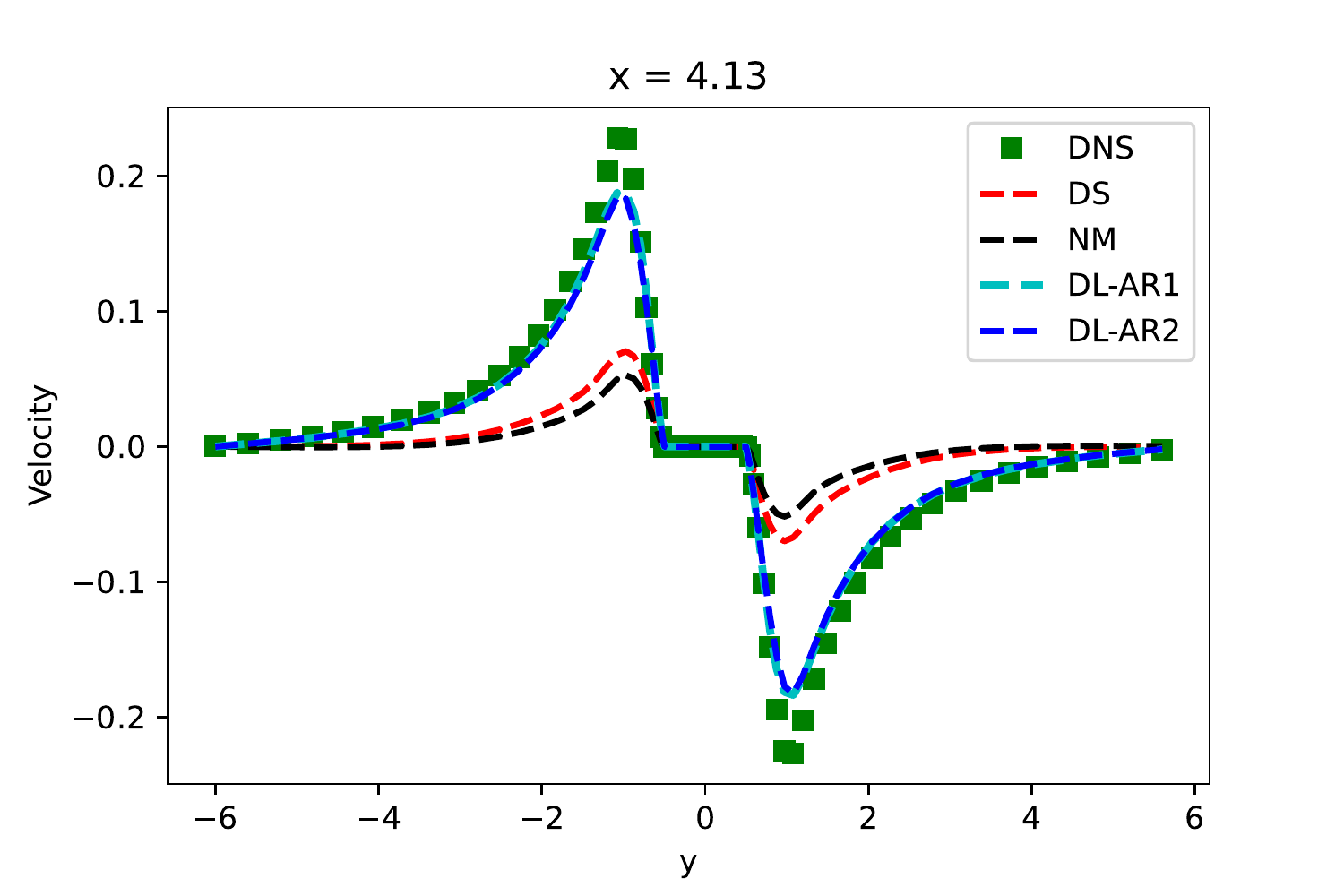}
\includegraphics[width=5cm]{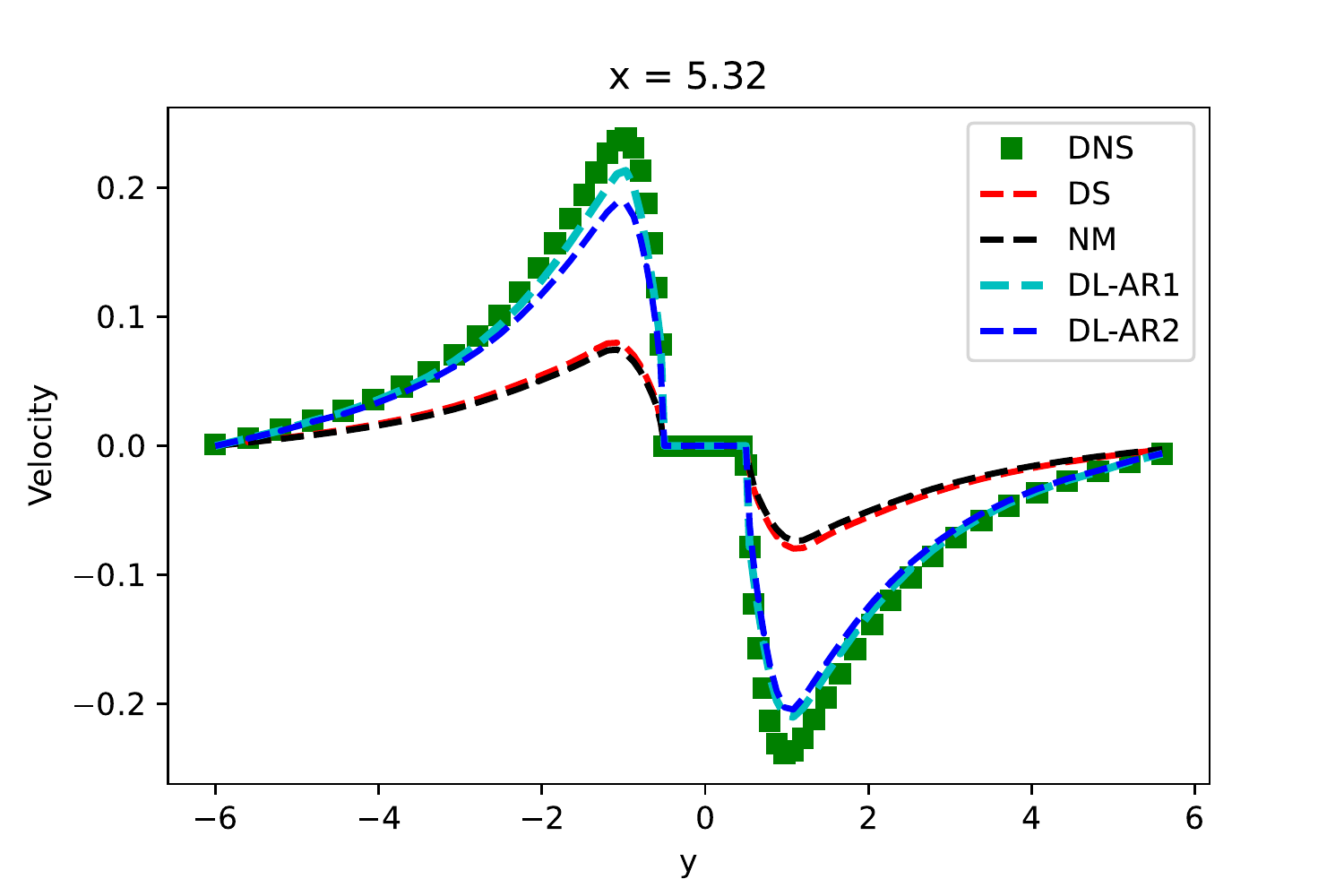}
\includegraphics[width=5cm]{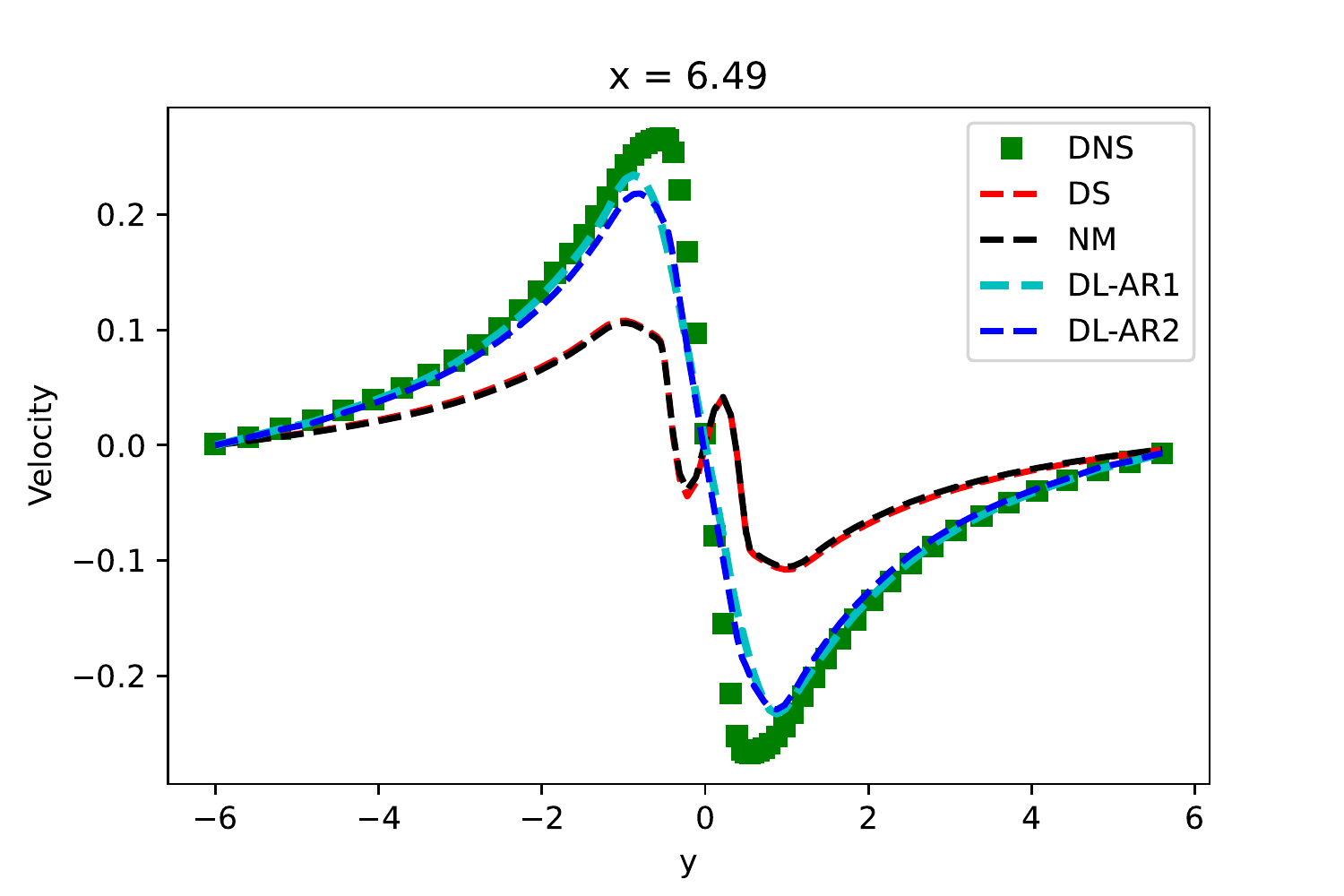}
\includegraphics[width=5cm]{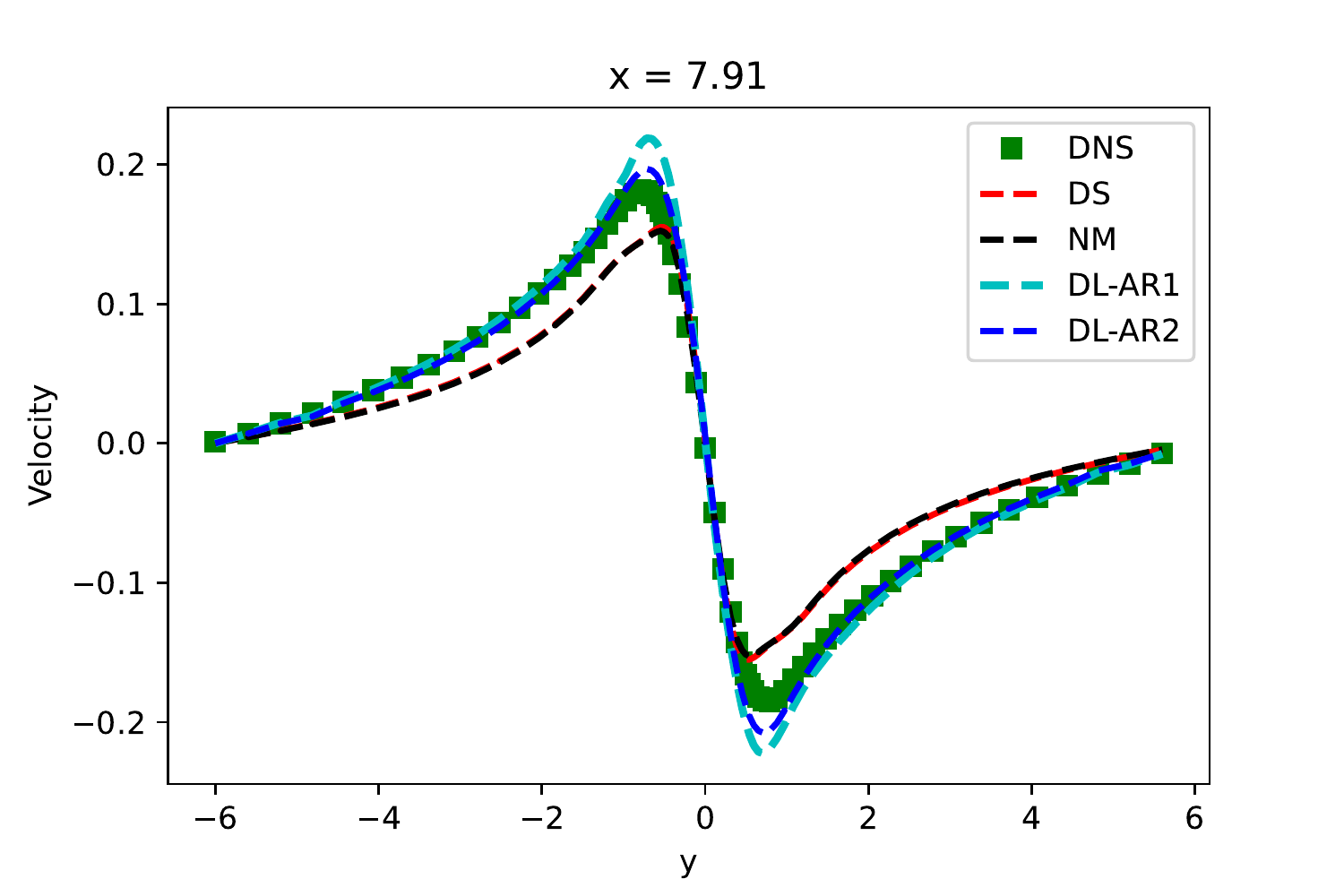}
\includegraphics[width=5cm]{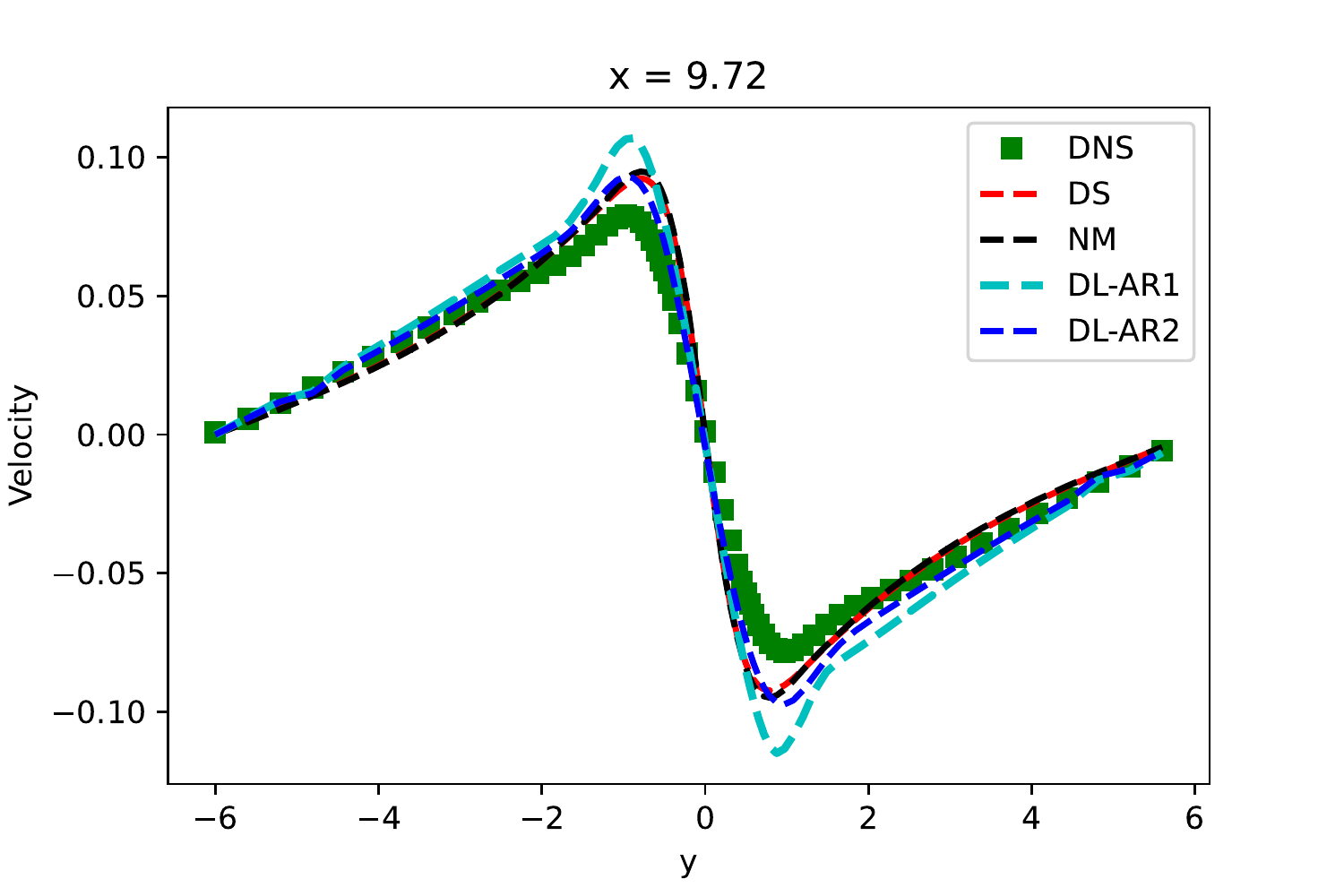}
\includegraphics[width=5cm]{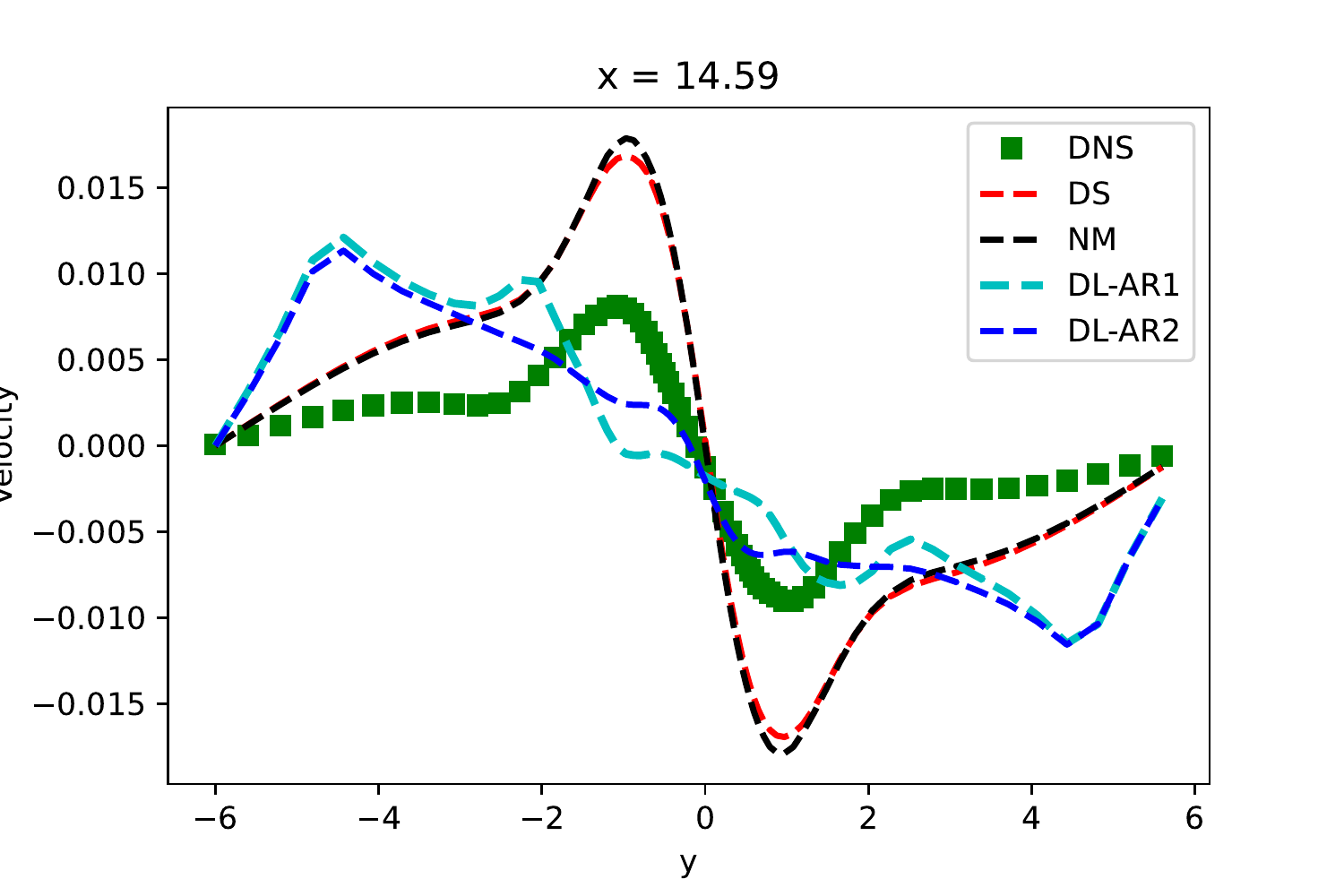}
\label{f1}
\caption{Mean profile for $u_2$ for AR4 configuration.}
\end{figure}

\begin{figure}[H]
\centering
\includegraphics[width=5cm]{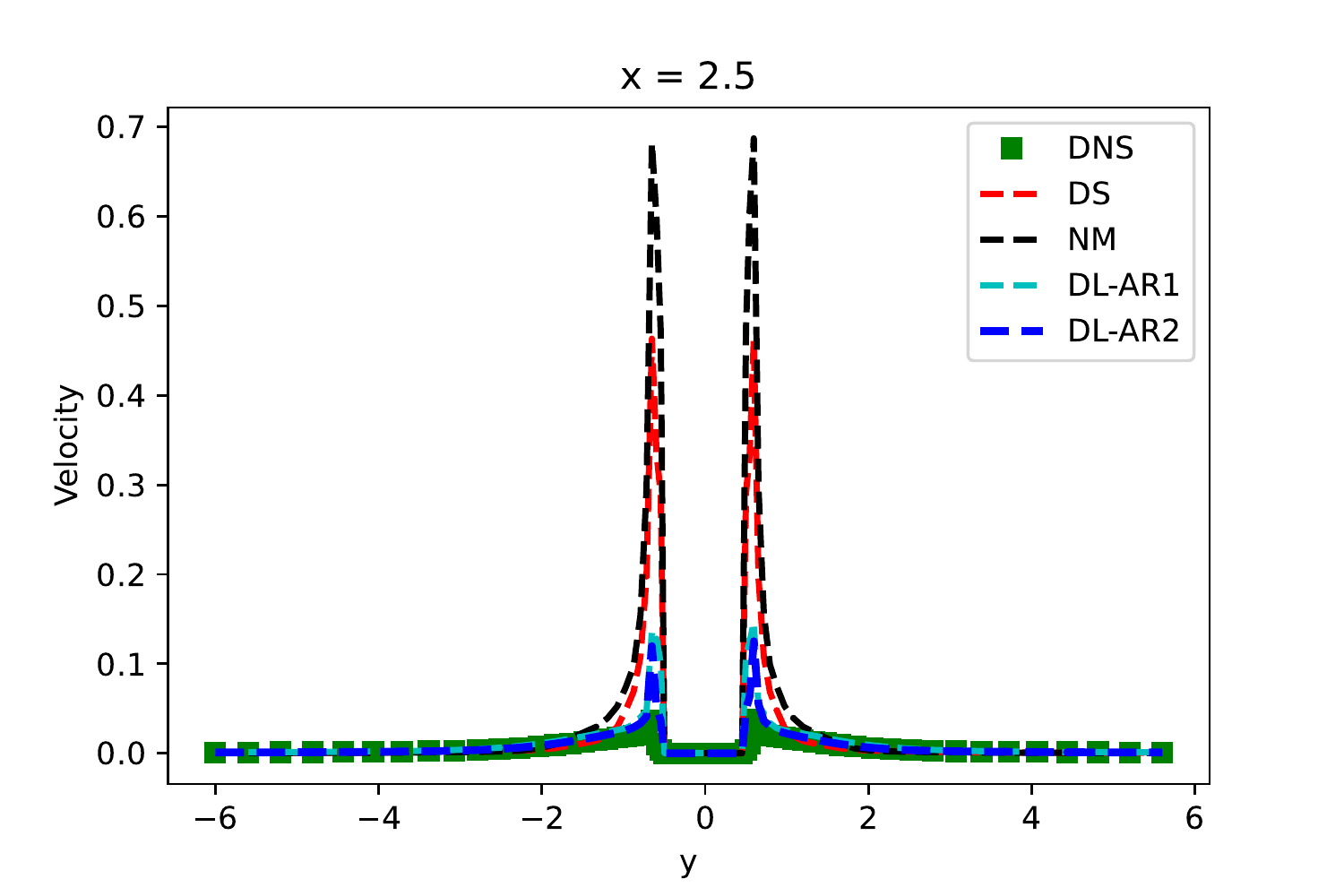}
\includegraphics[width=5cm]{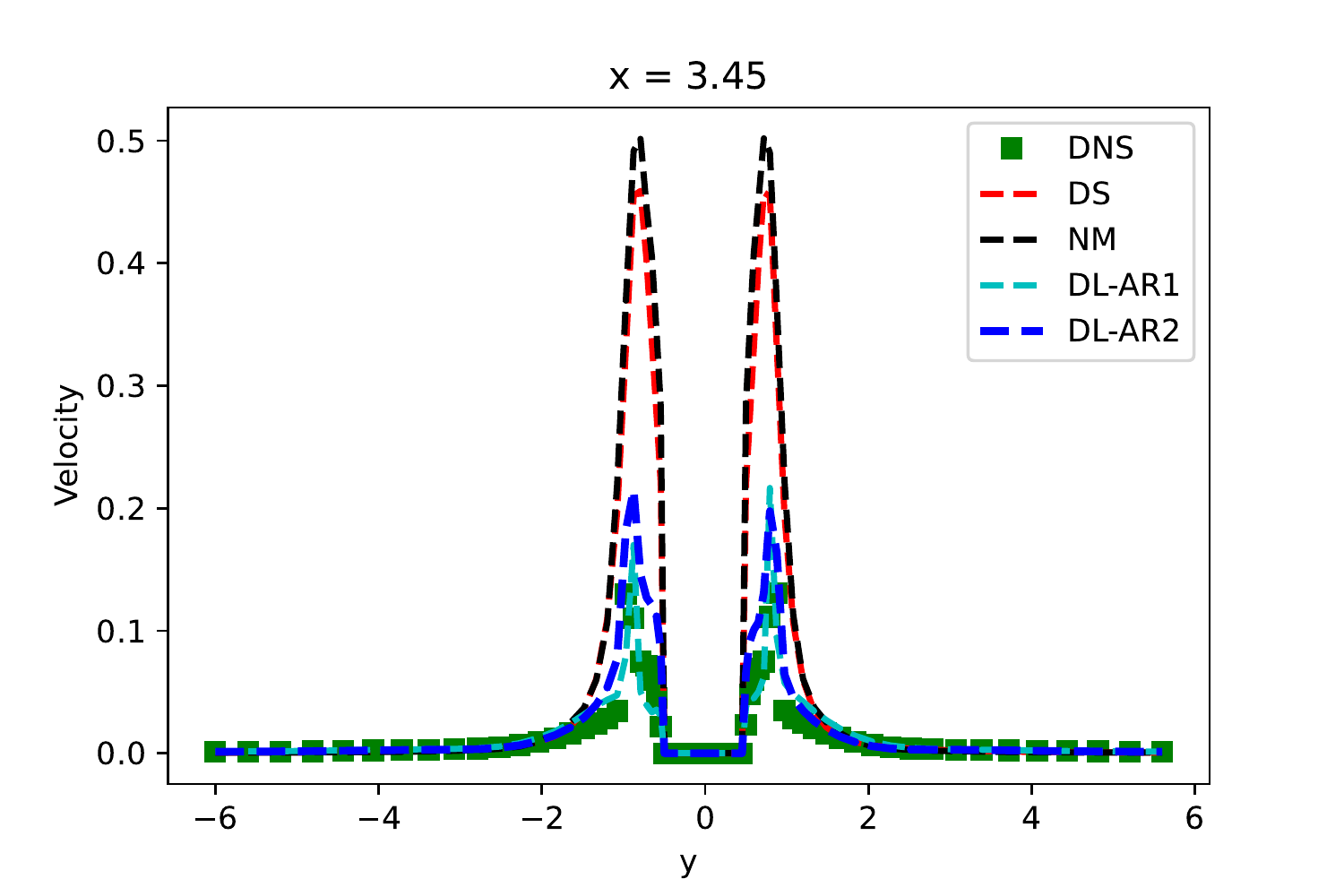}
\includegraphics[width=5cm]{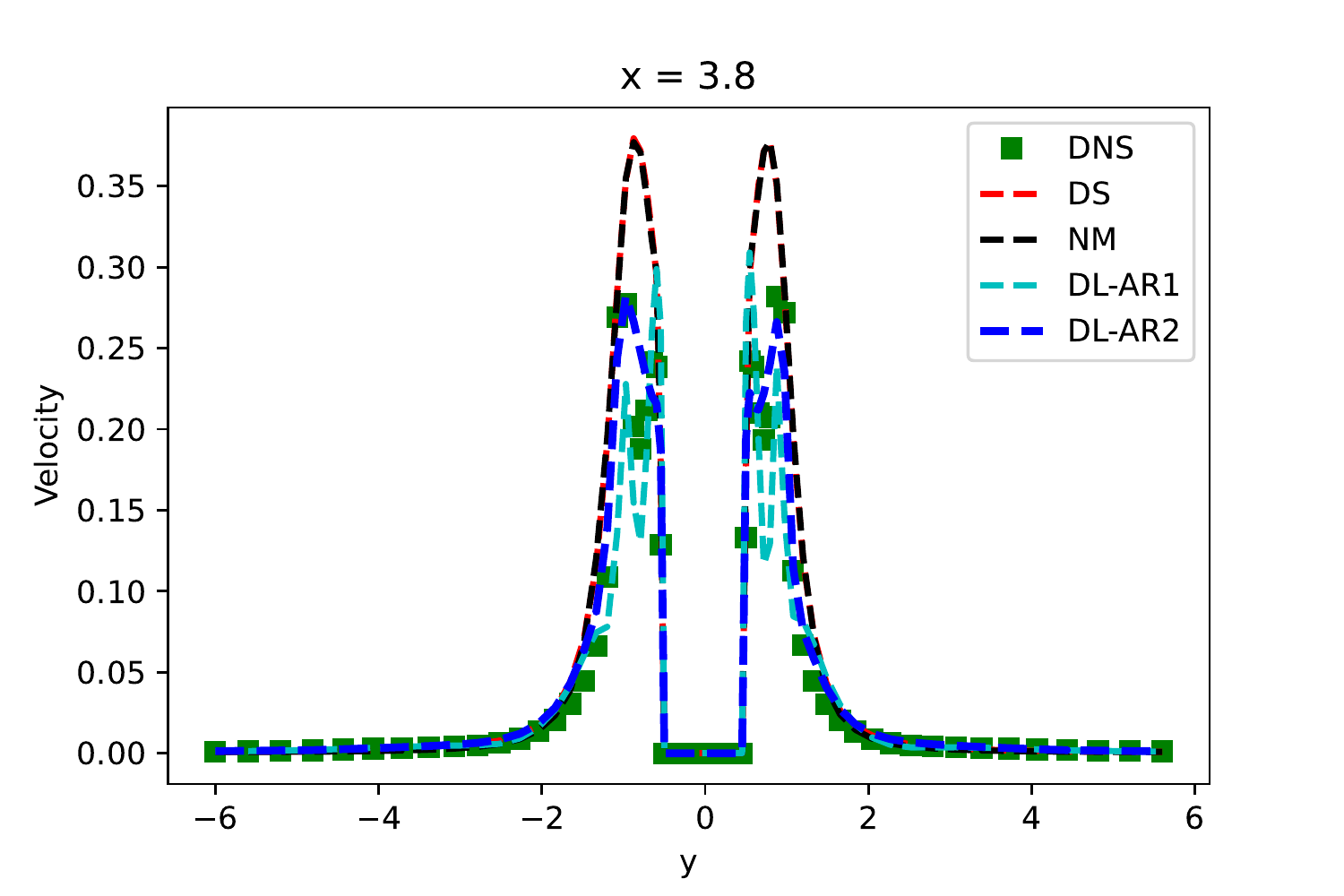}
\includegraphics[width=5cm]{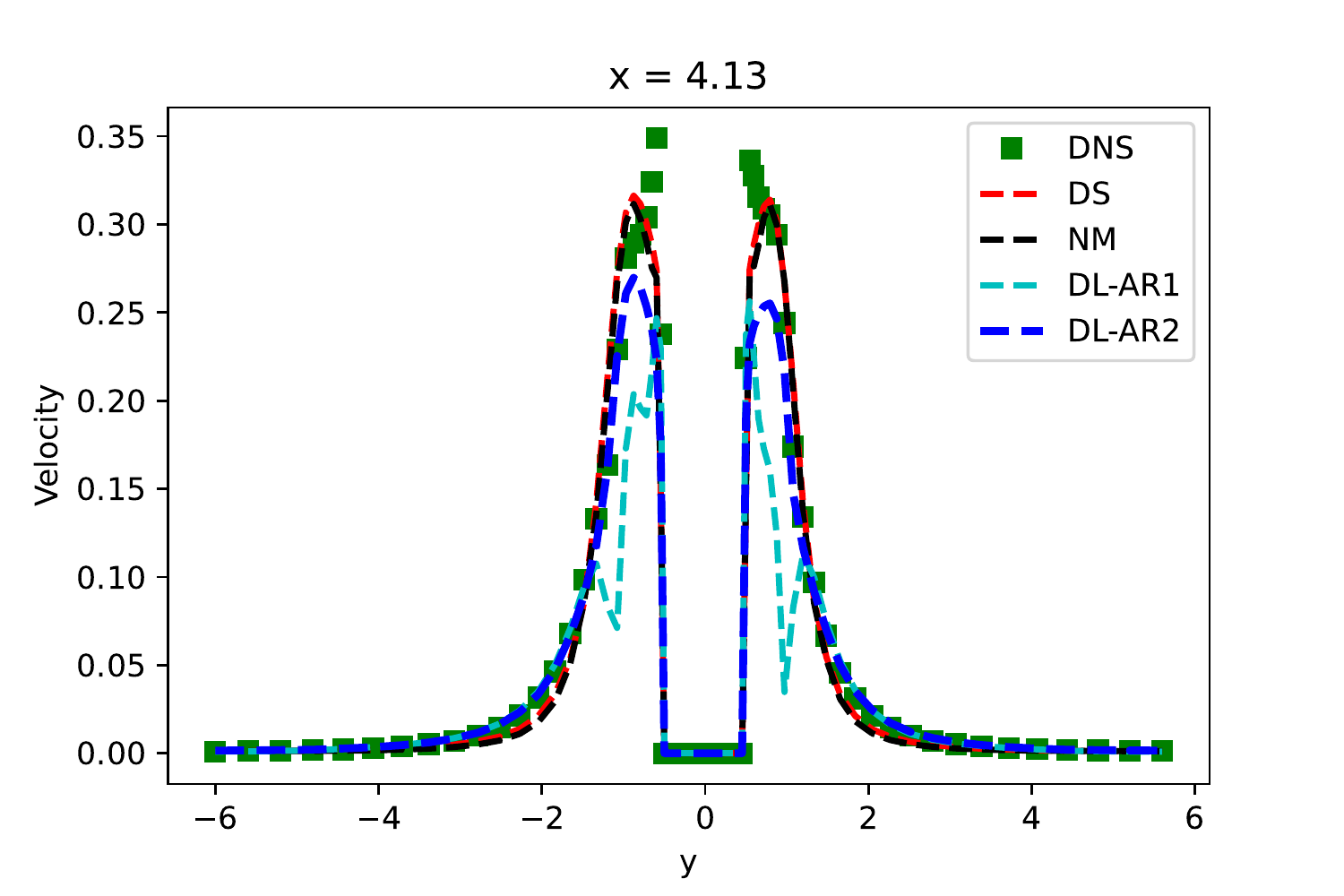}
\includegraphics[width=5cm]{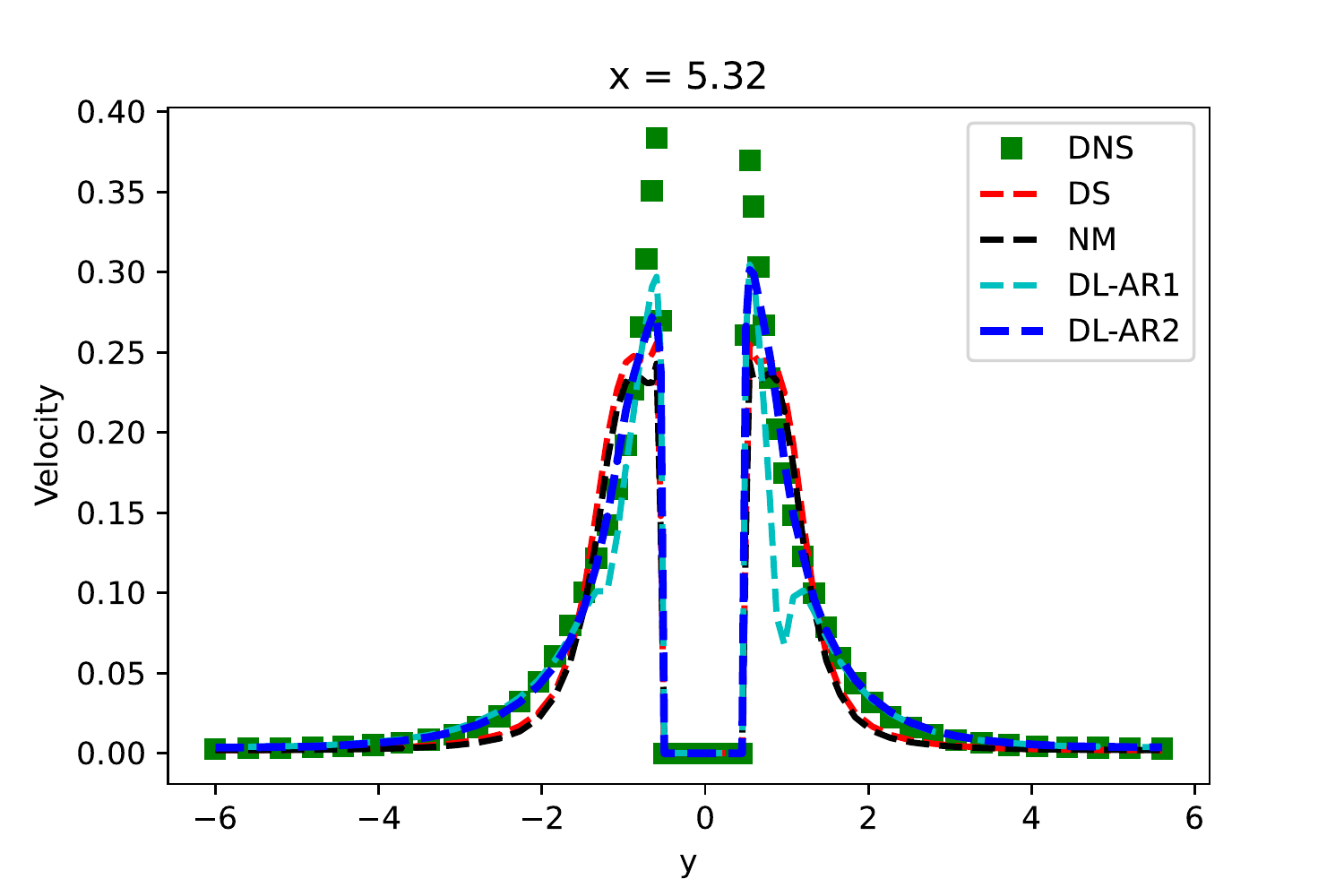}
\includegraphics[width=5cm]{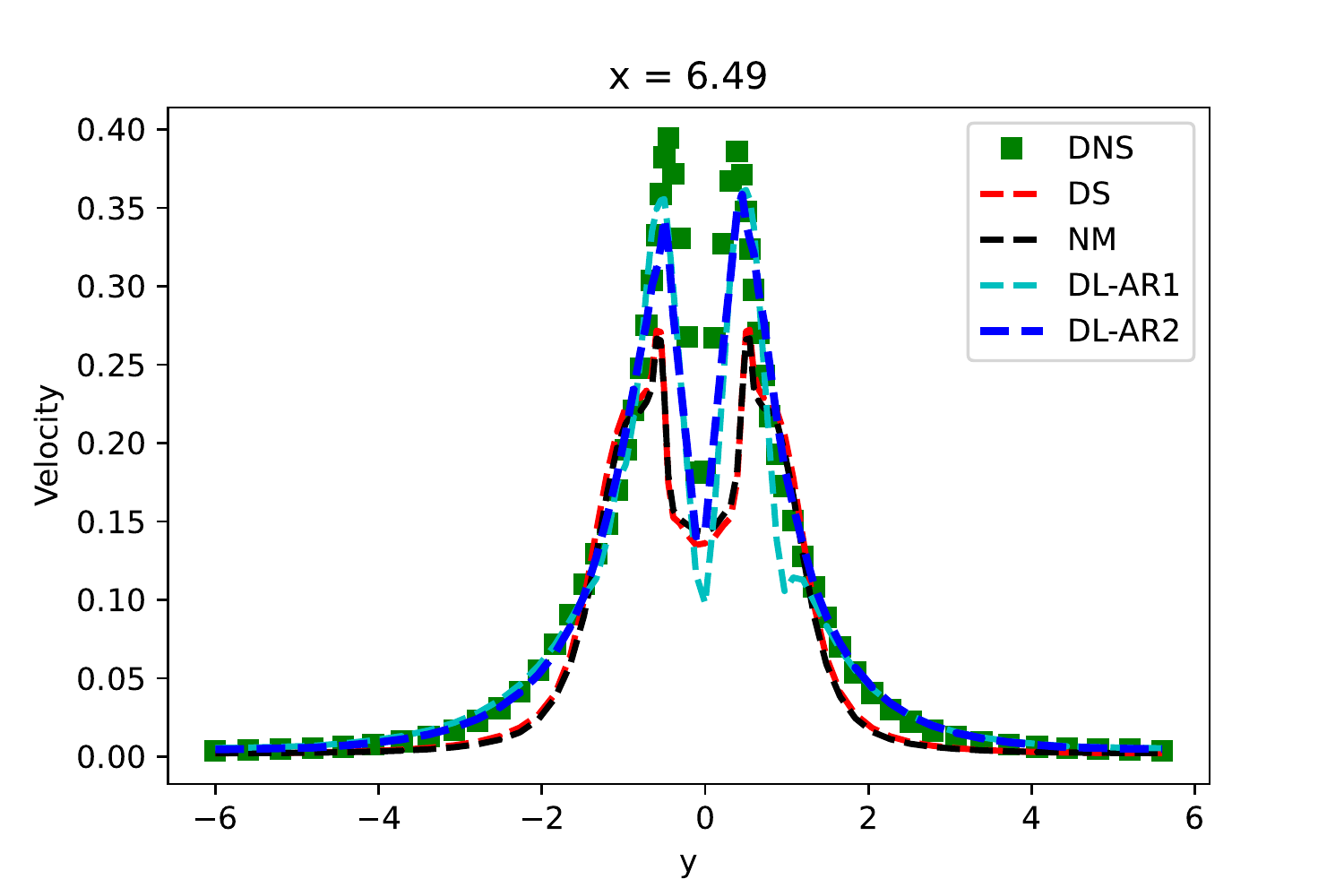}
\includegraphics[width=5cm]{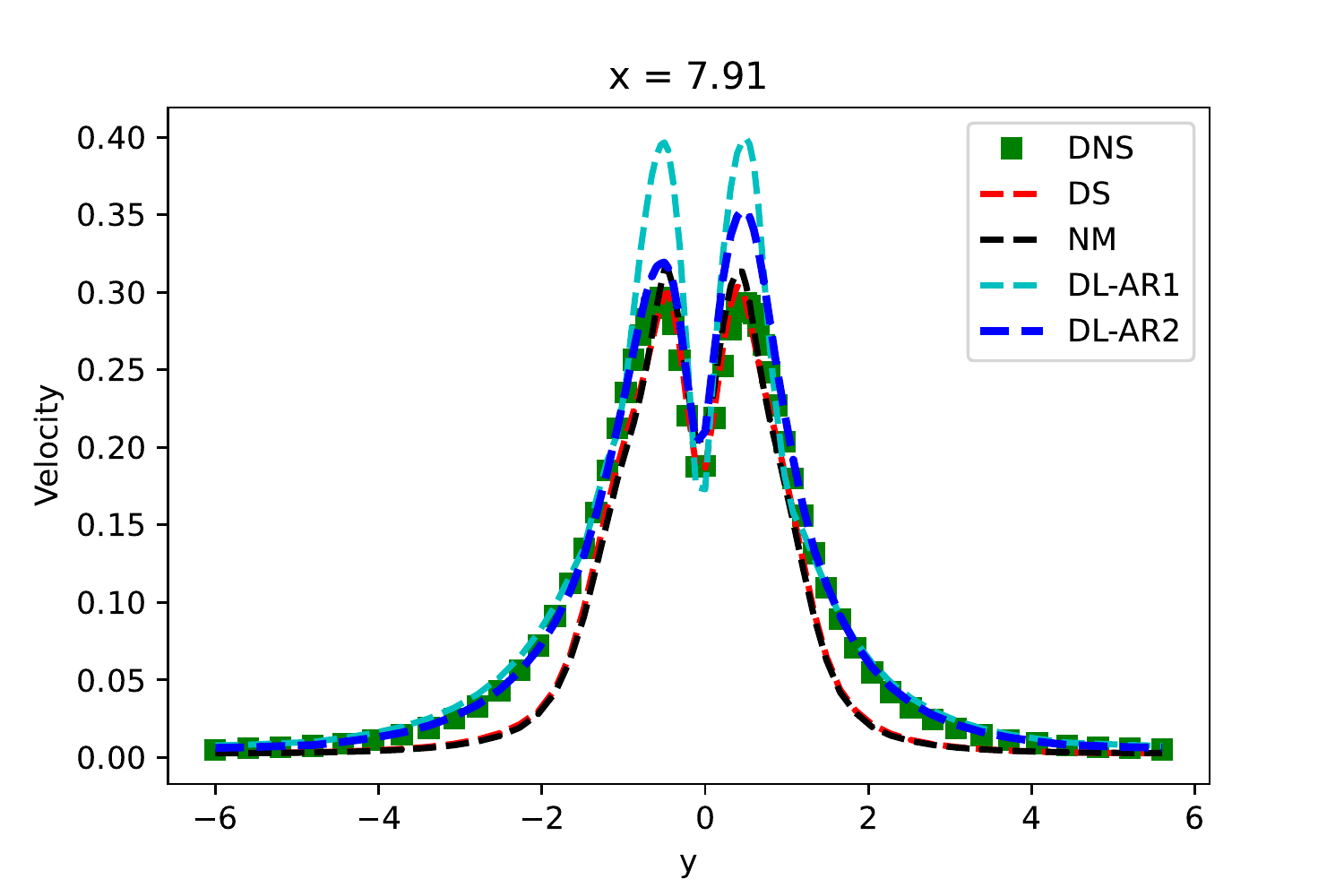}
\includegraphics[width=5cm]{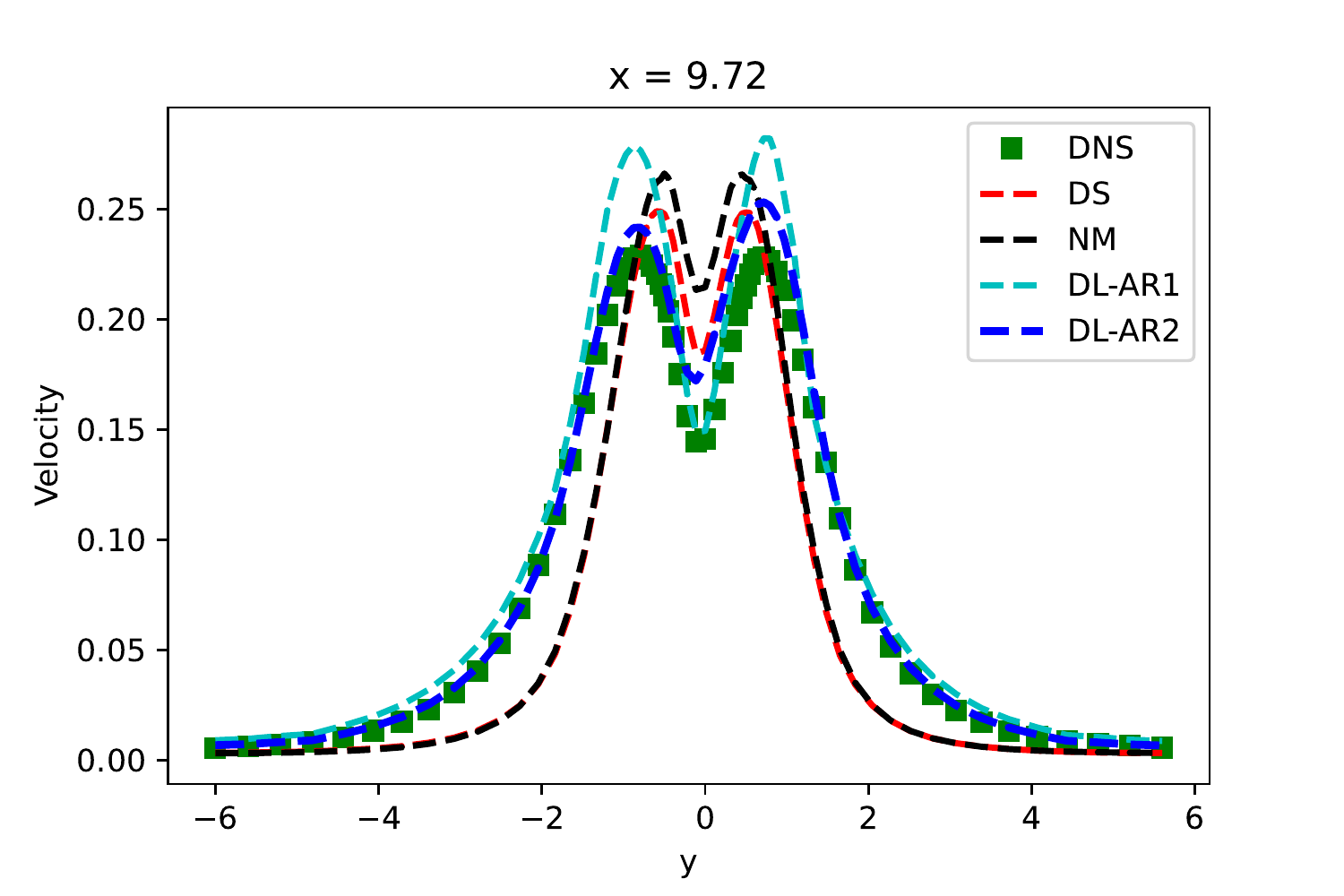}
\includegraphics[width=5cm]{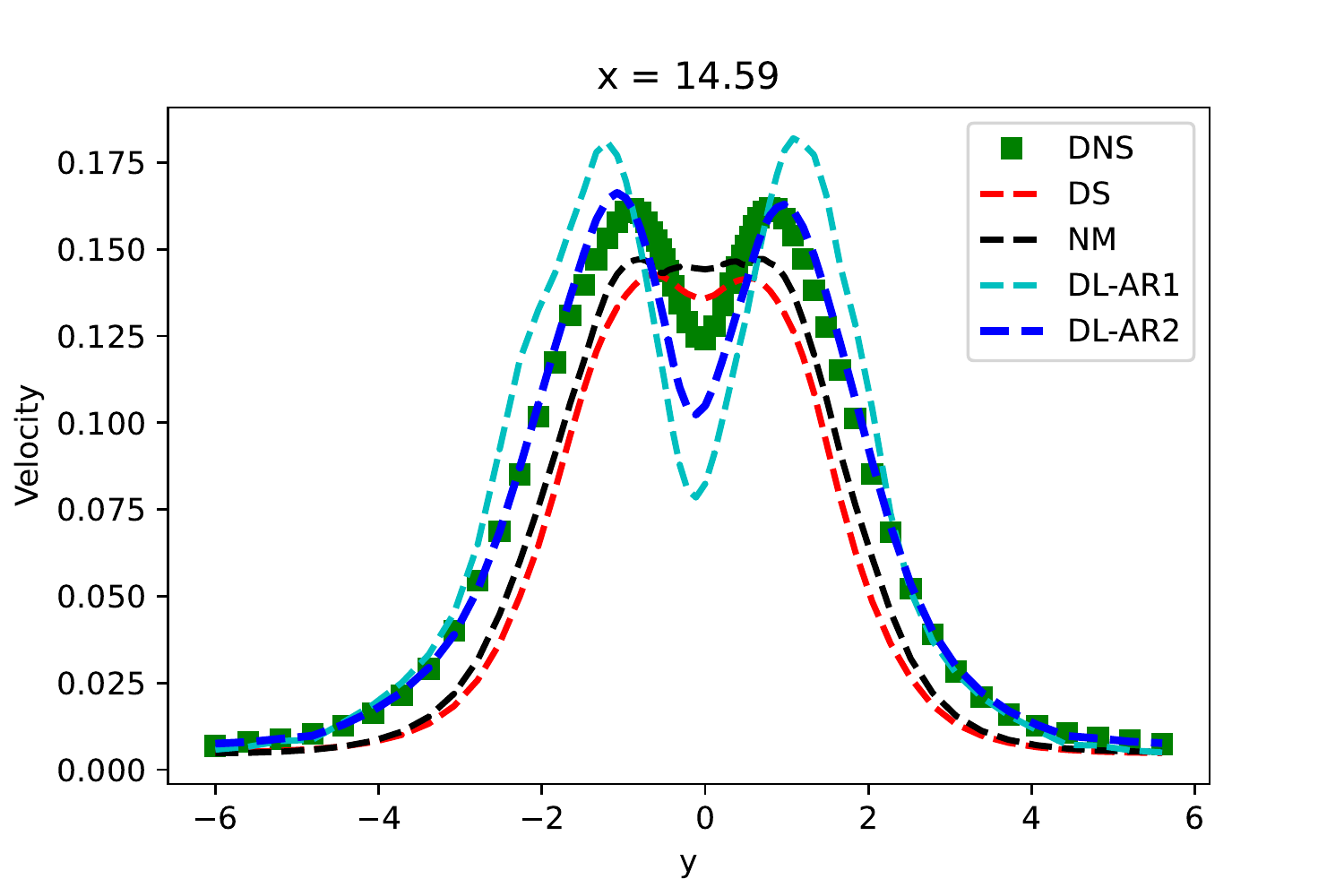}
\label{f1}
\caption{RMS profile for $u_1$ for AR4 configuration.}
\end{figure}

\begin{figure}[H]
\centering
\includegraphics[width=5cm]{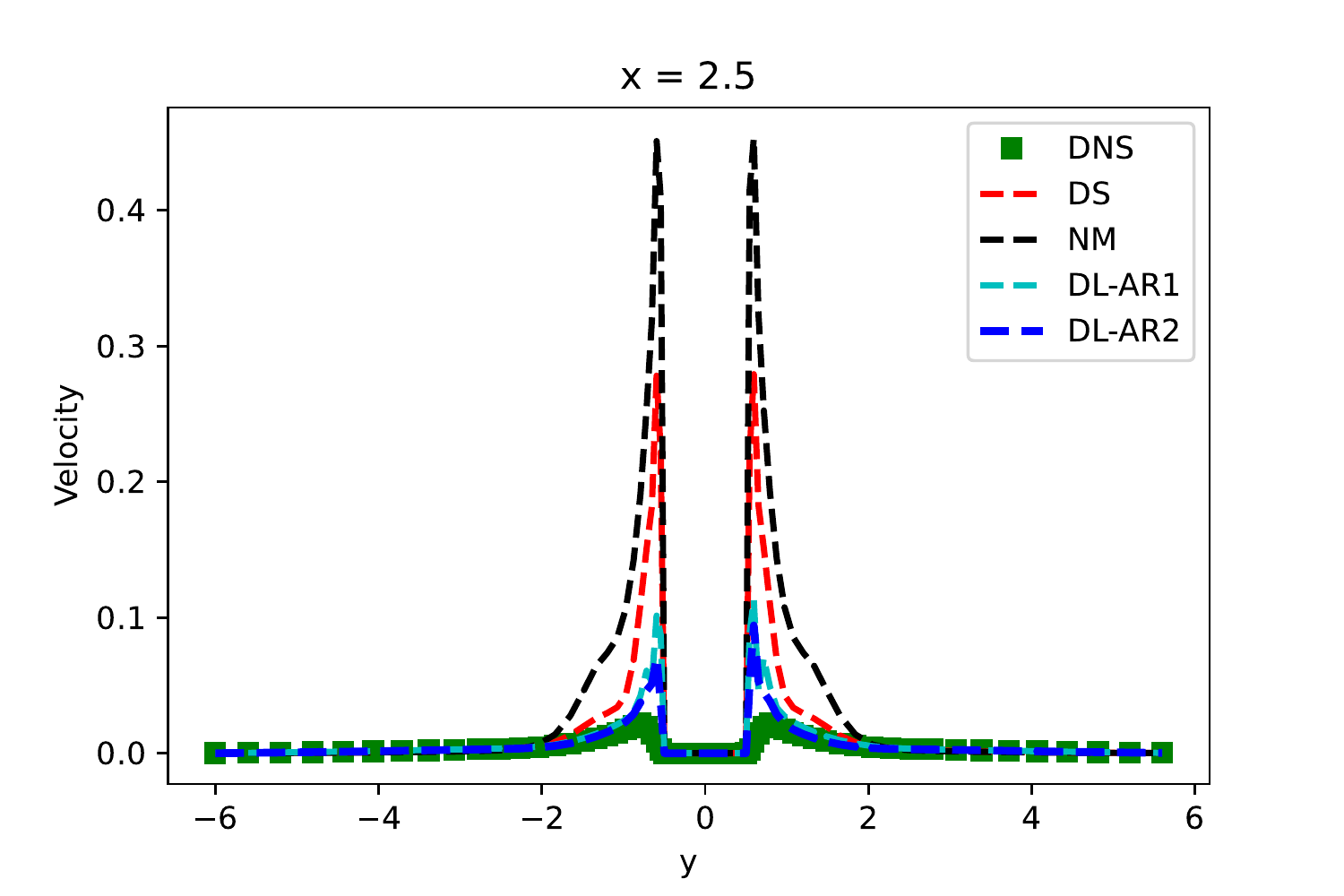}
\includegraphics[width=5cm]{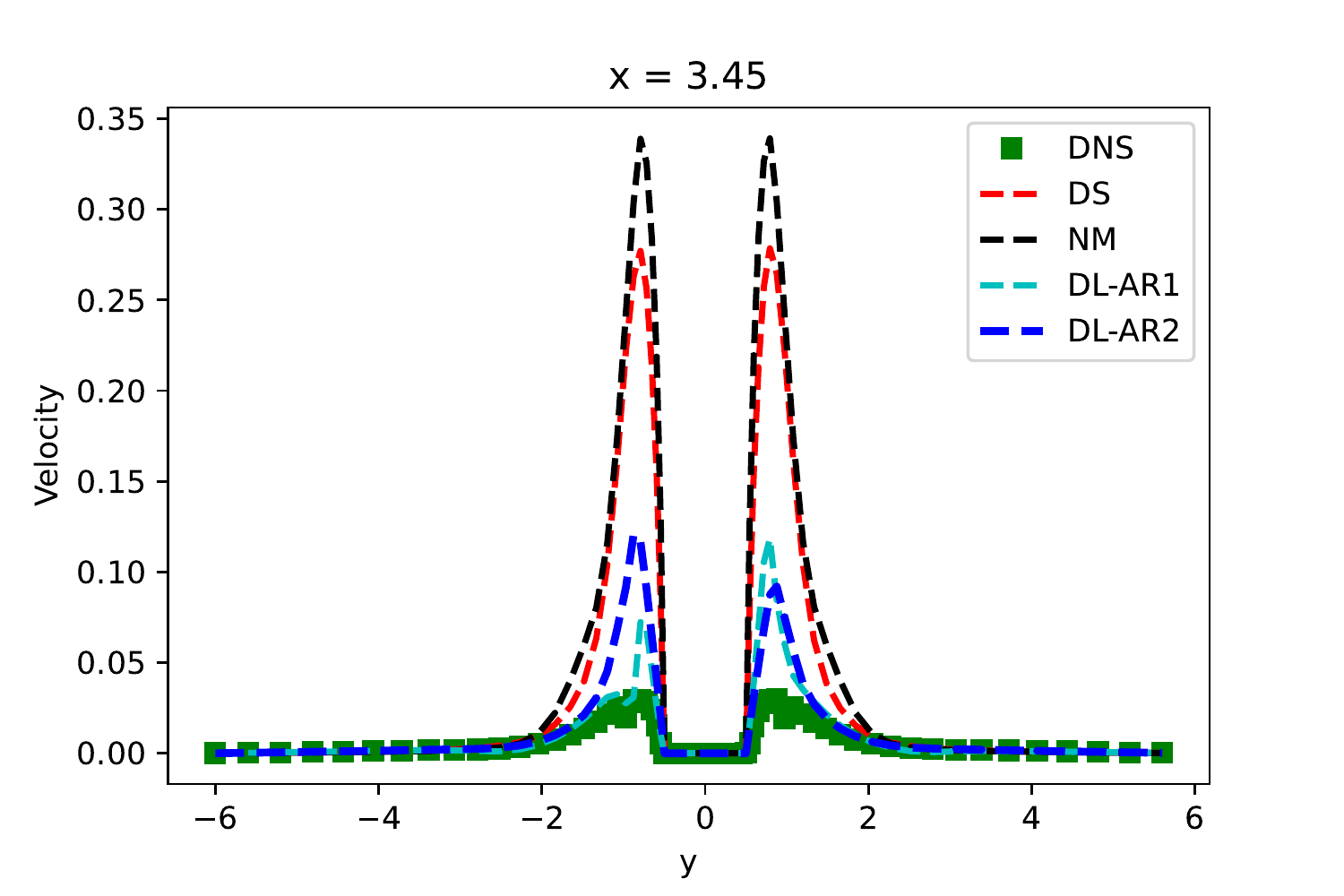}
\includegraphics[width=5cm]{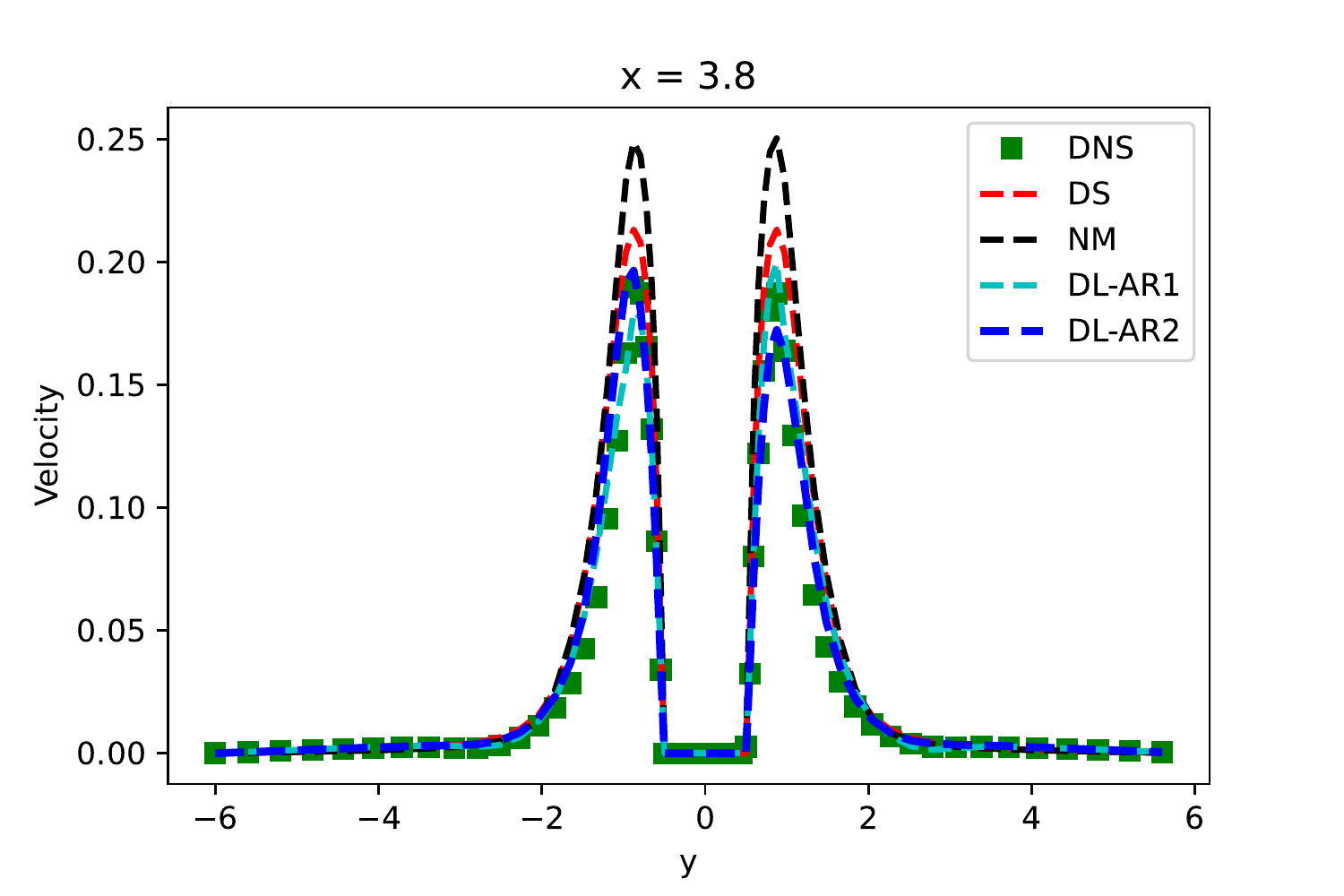}
\includegraphics[width=5cm]{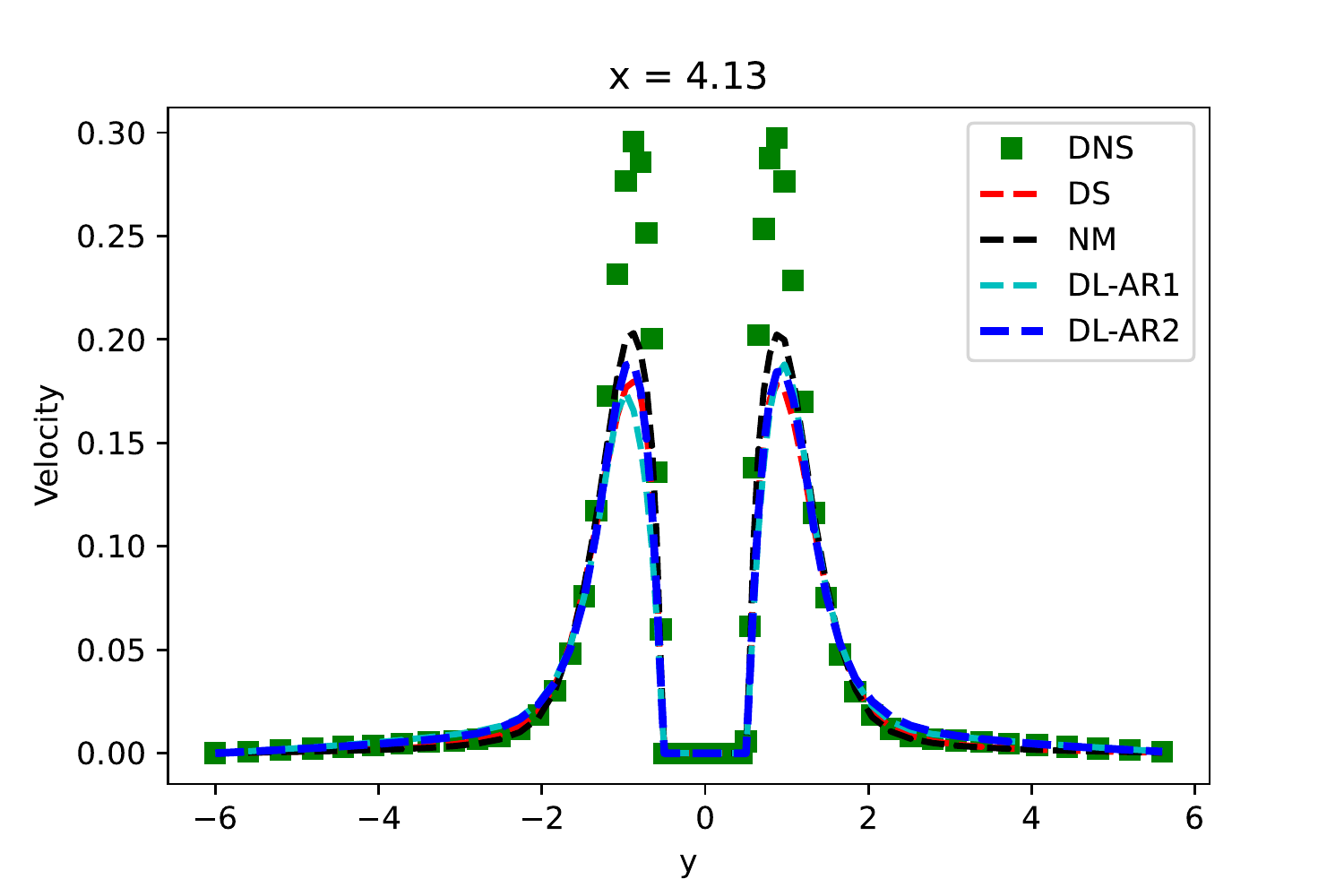}
\includegraphics[width=5cm]{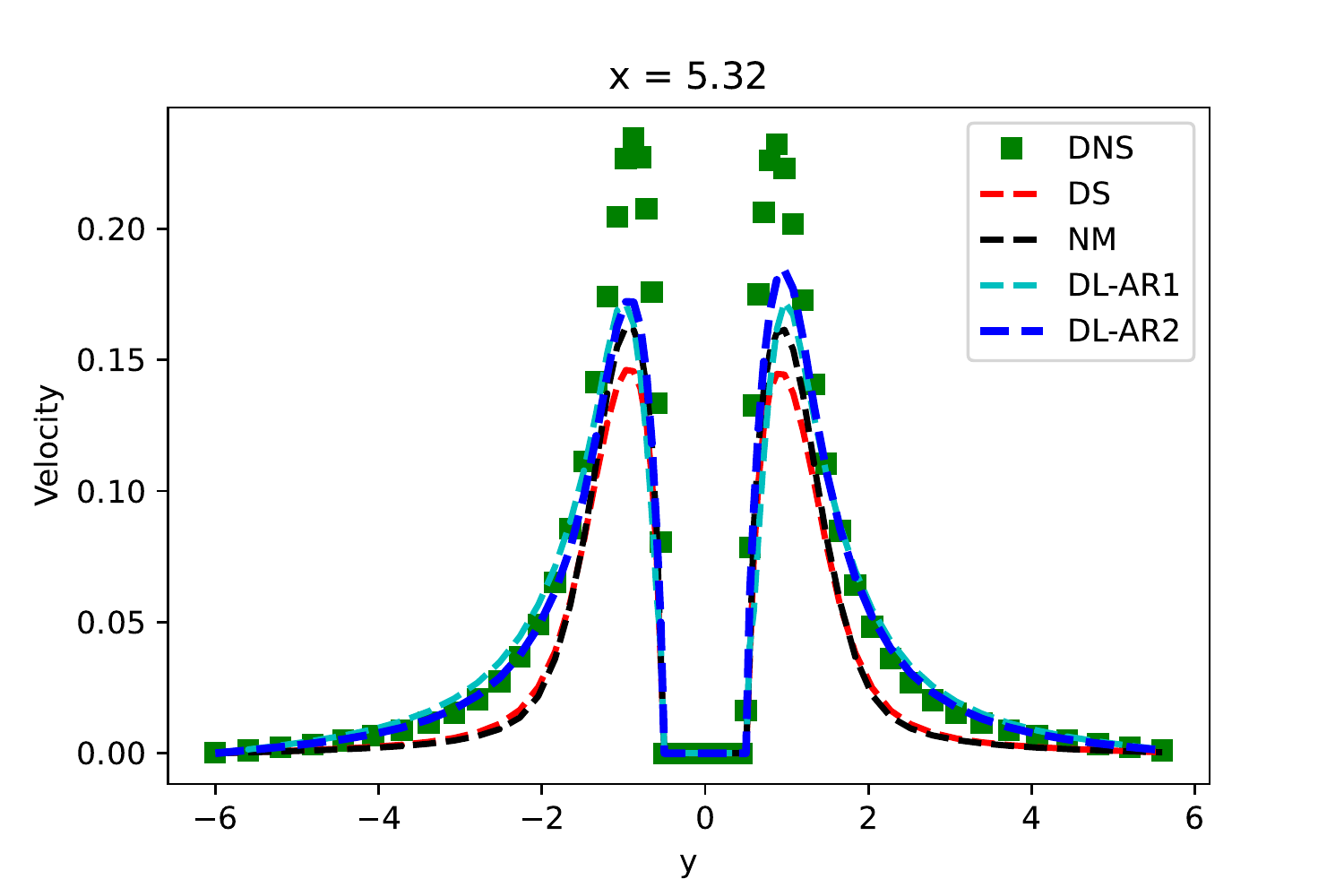}
\includegraphics[width=5cm]{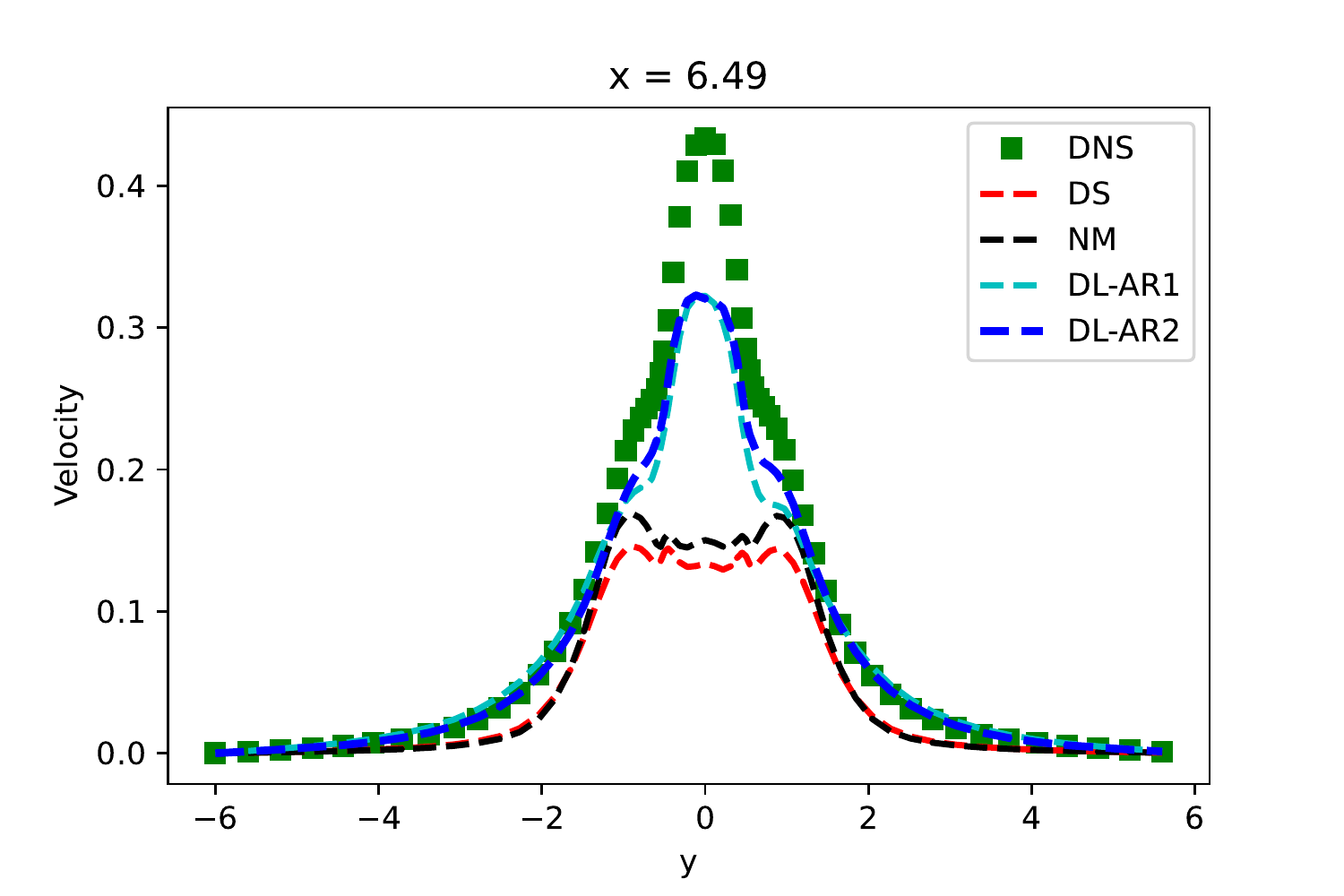}
\includegraphics[width=5cm]{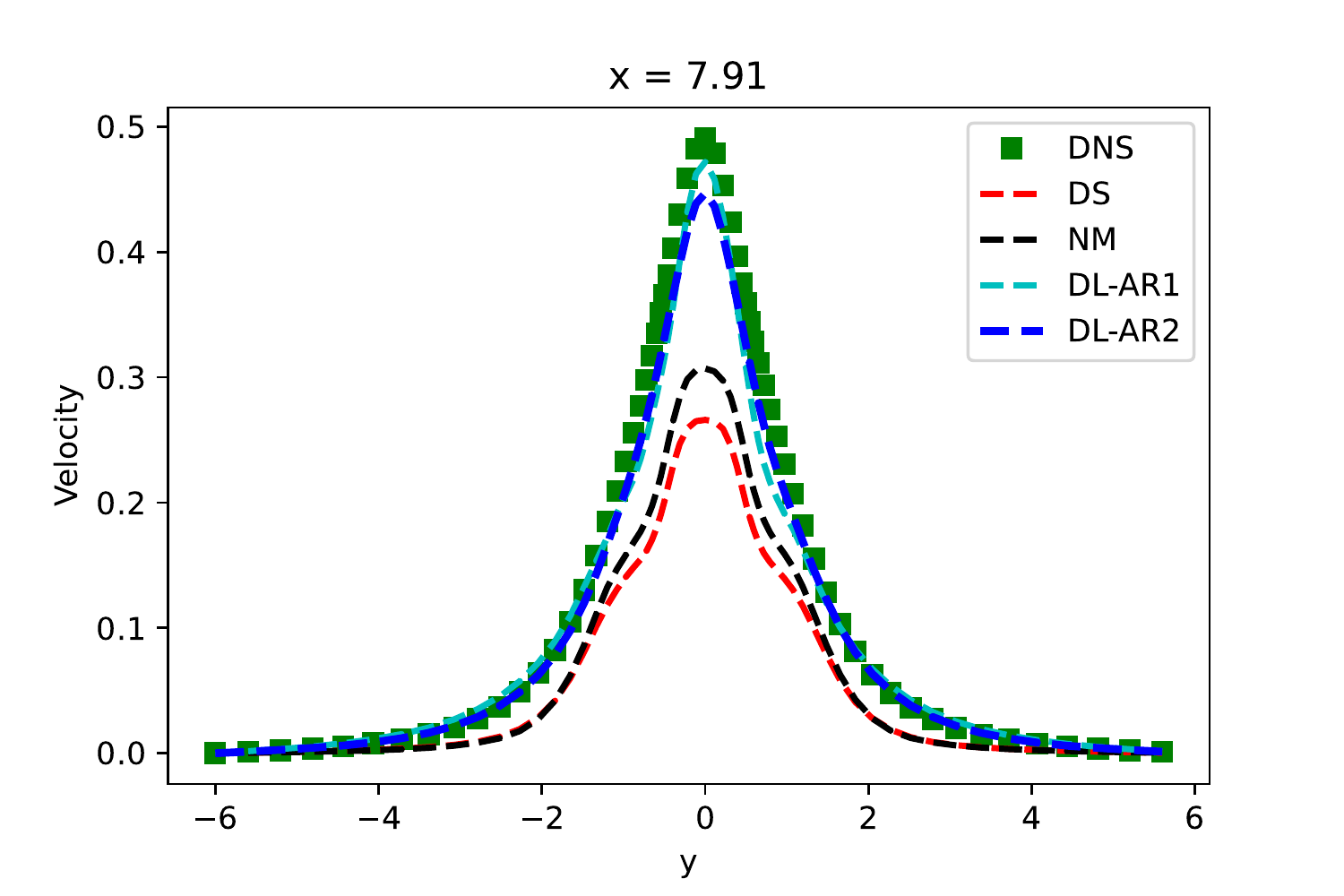}
\includegraphics[width=5cm]{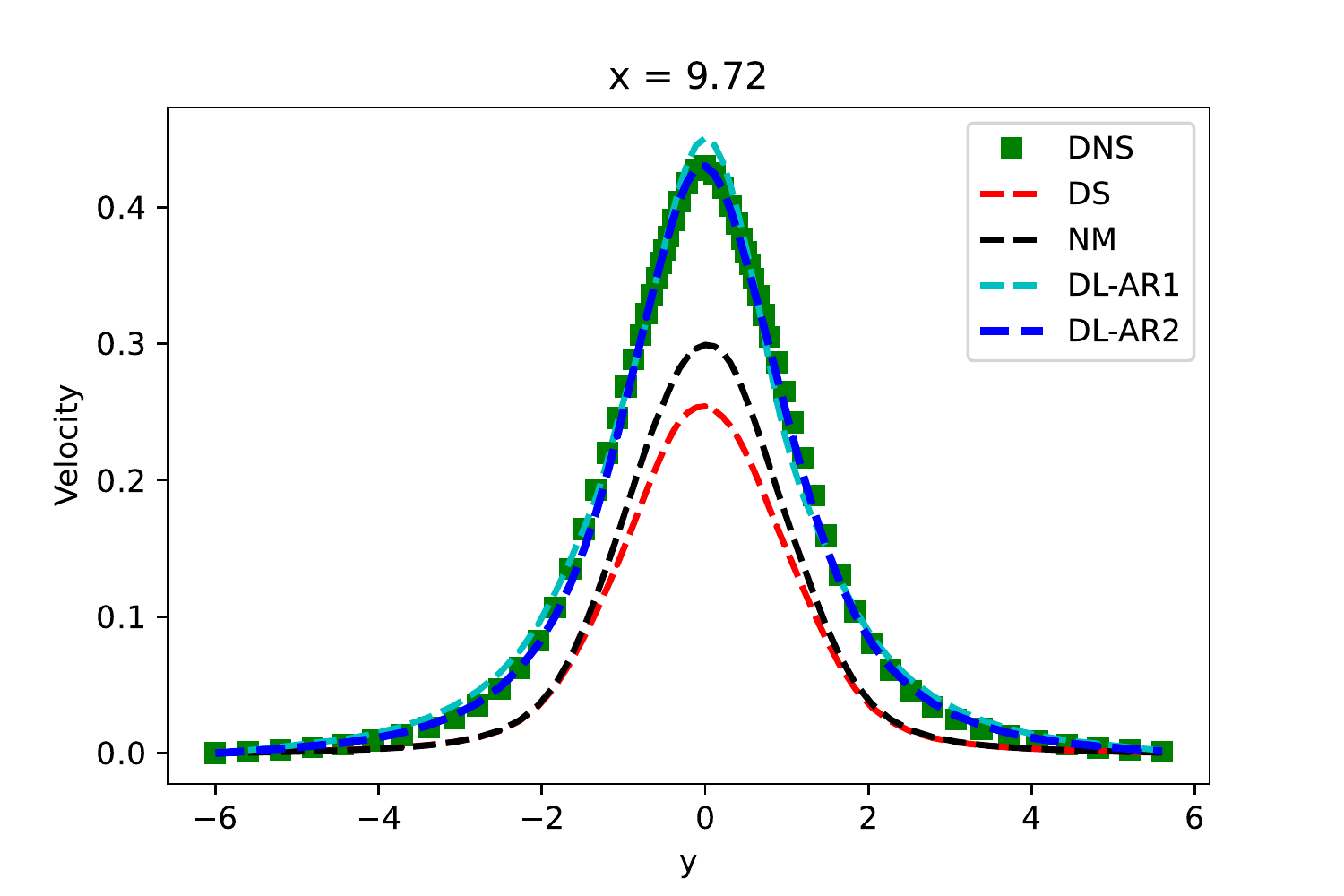}
\includegraphics[width=5cm]{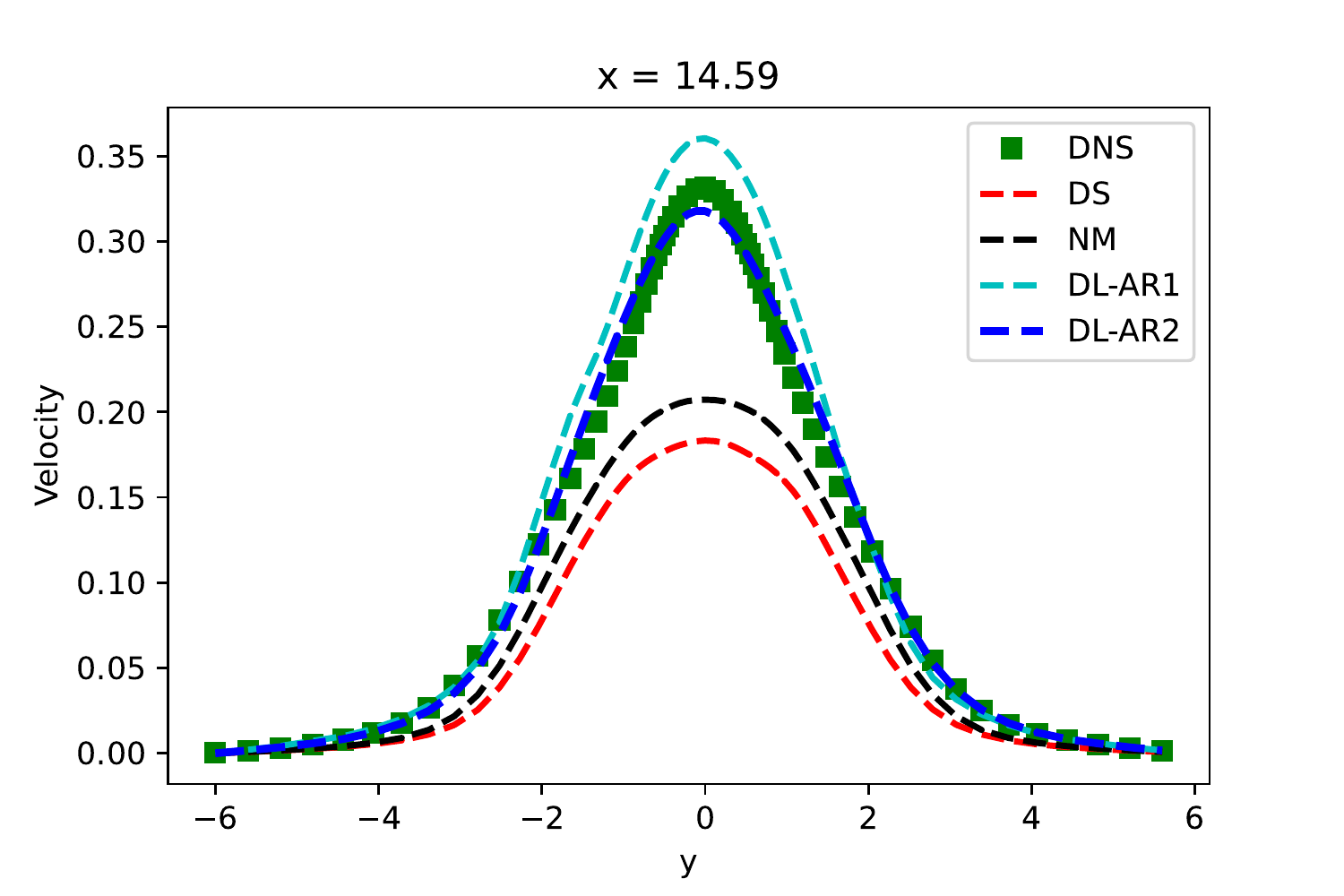}
\label{f1}
\caption{RMS profile for $u_2$ for AR4 configuration.}
\end{figure}

\begin{figure}[H]
\centering
\includegraphics[width=5cm]{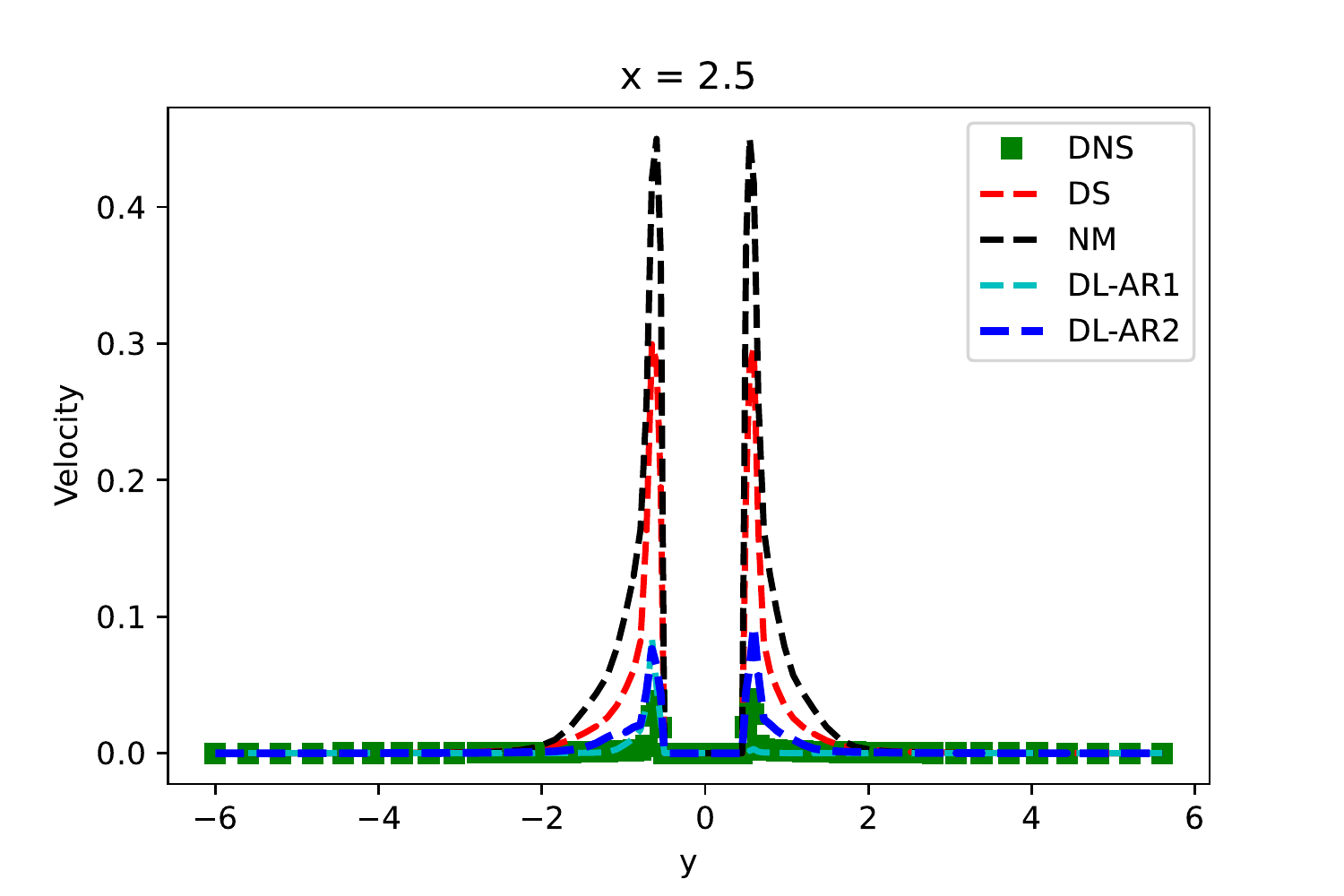}
\includegraphics[width=5cm]{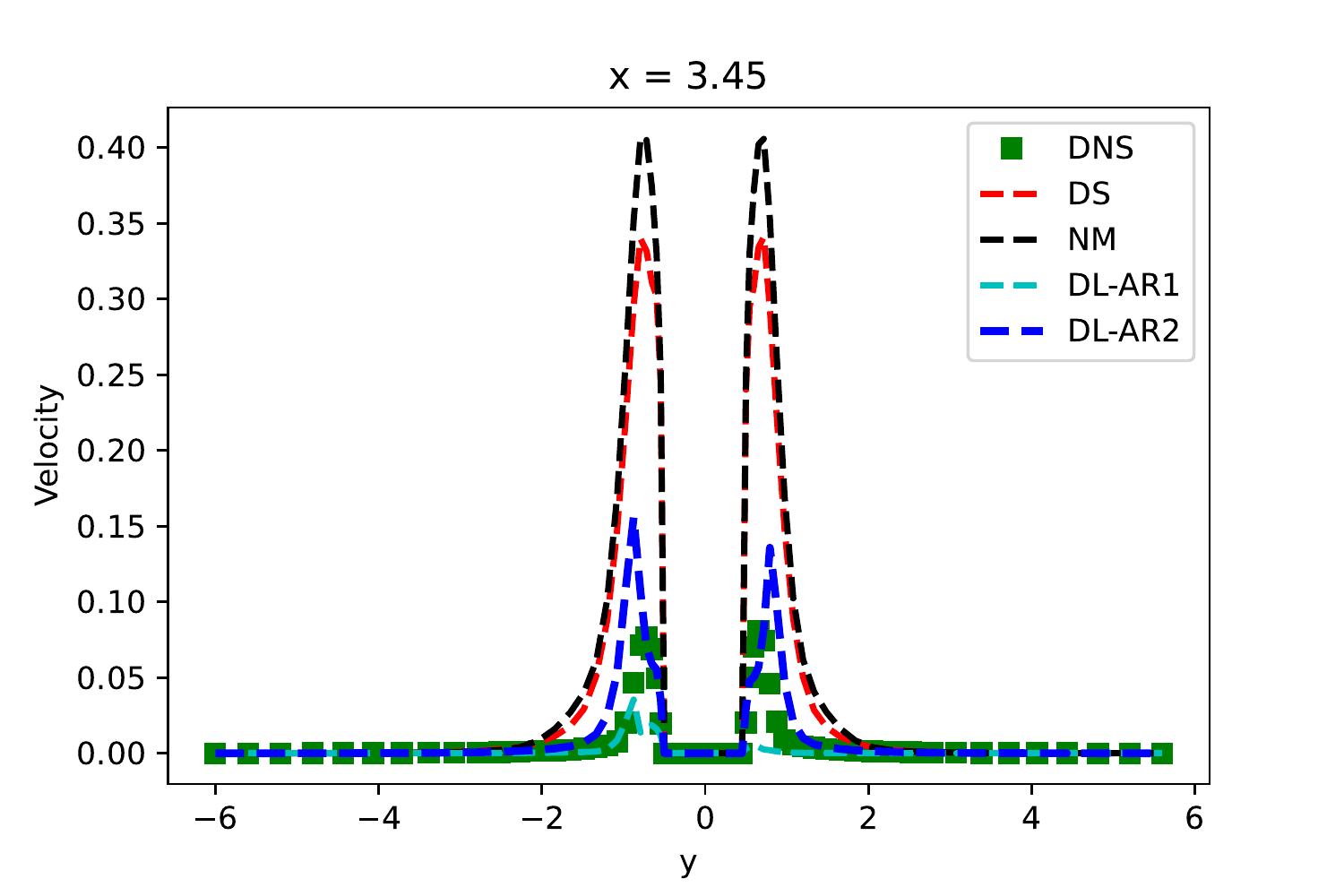}
\includegraphics[width=5cm]{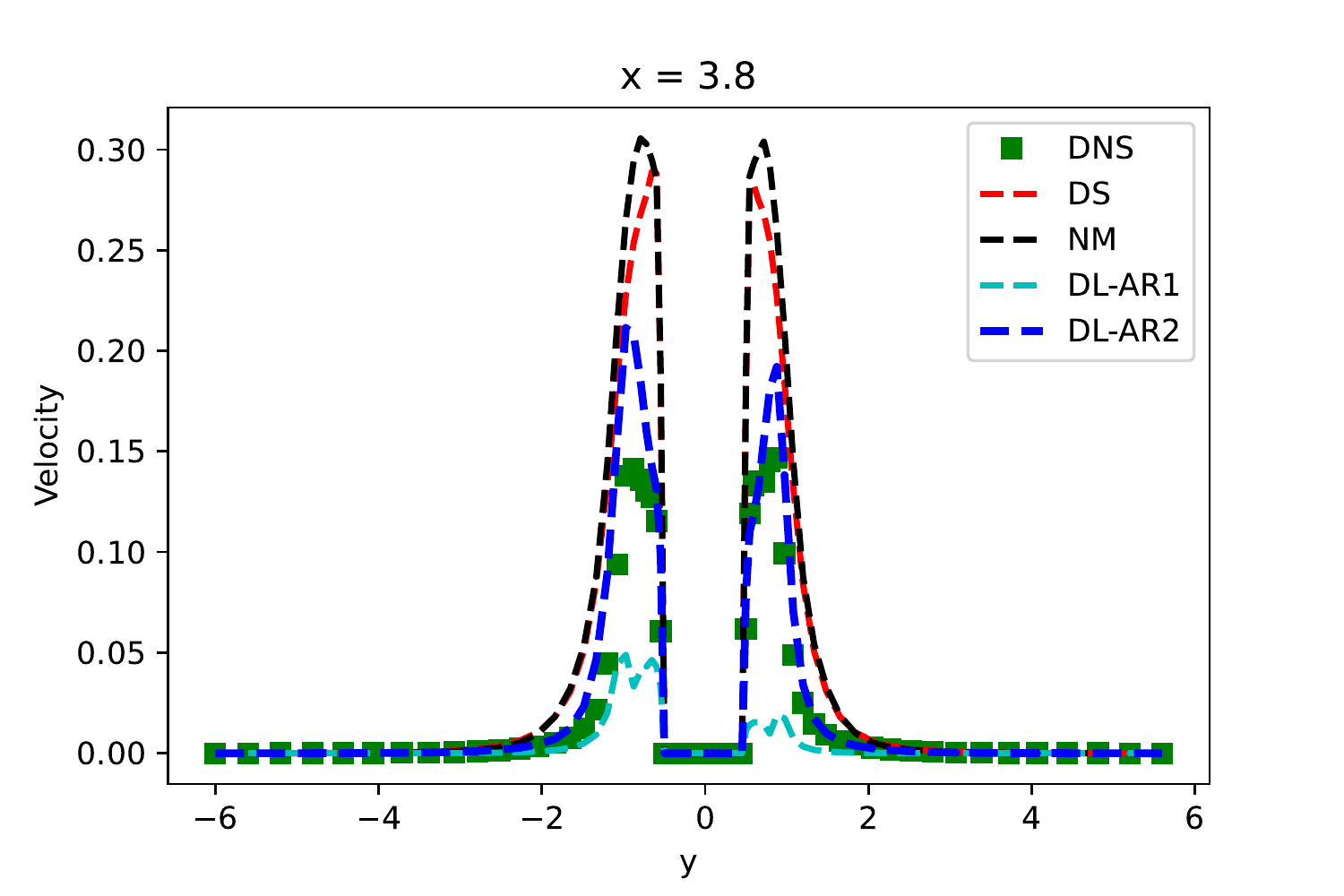}
\includegraphics[width=5cm]{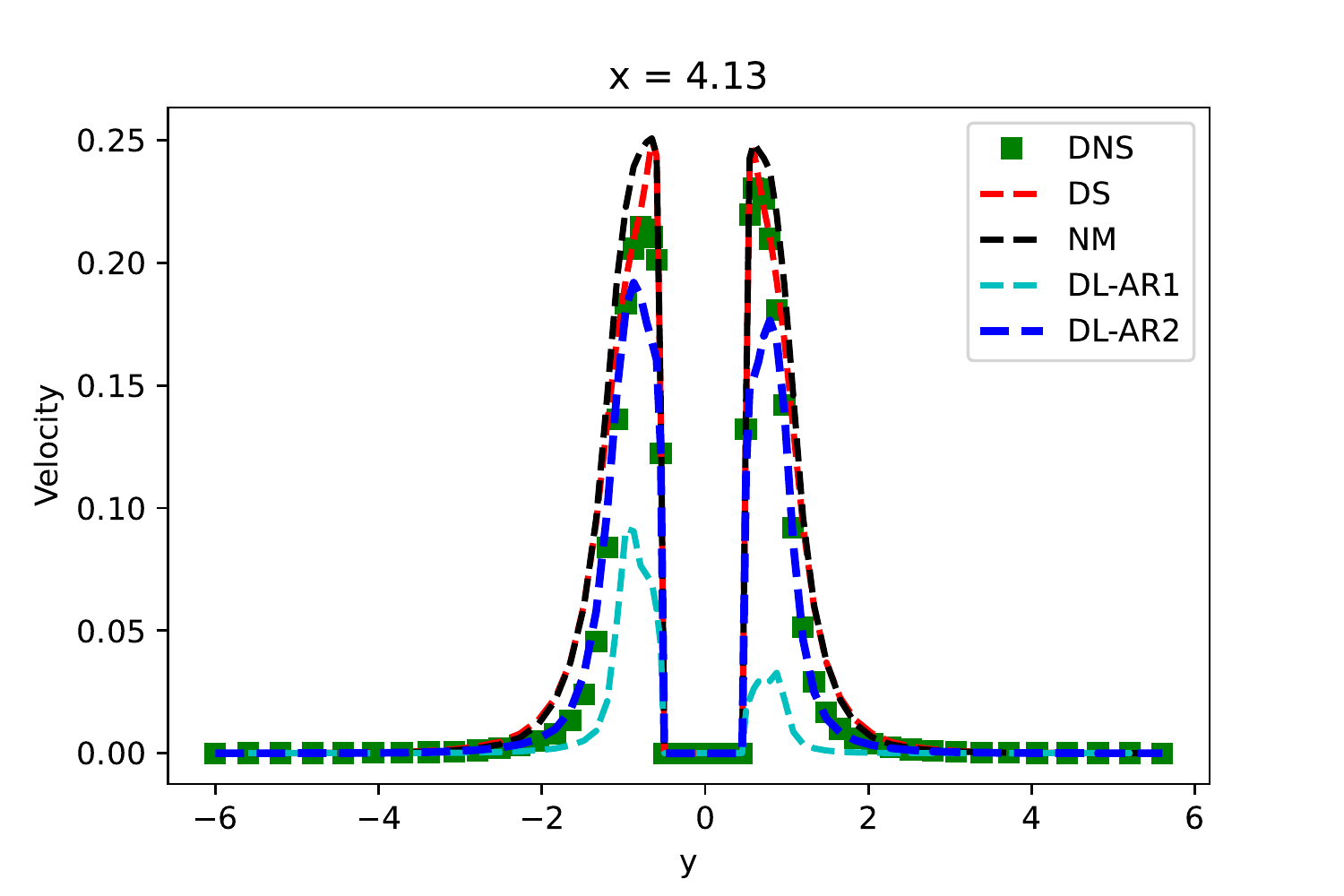}
\includegraphics[width=5cm]{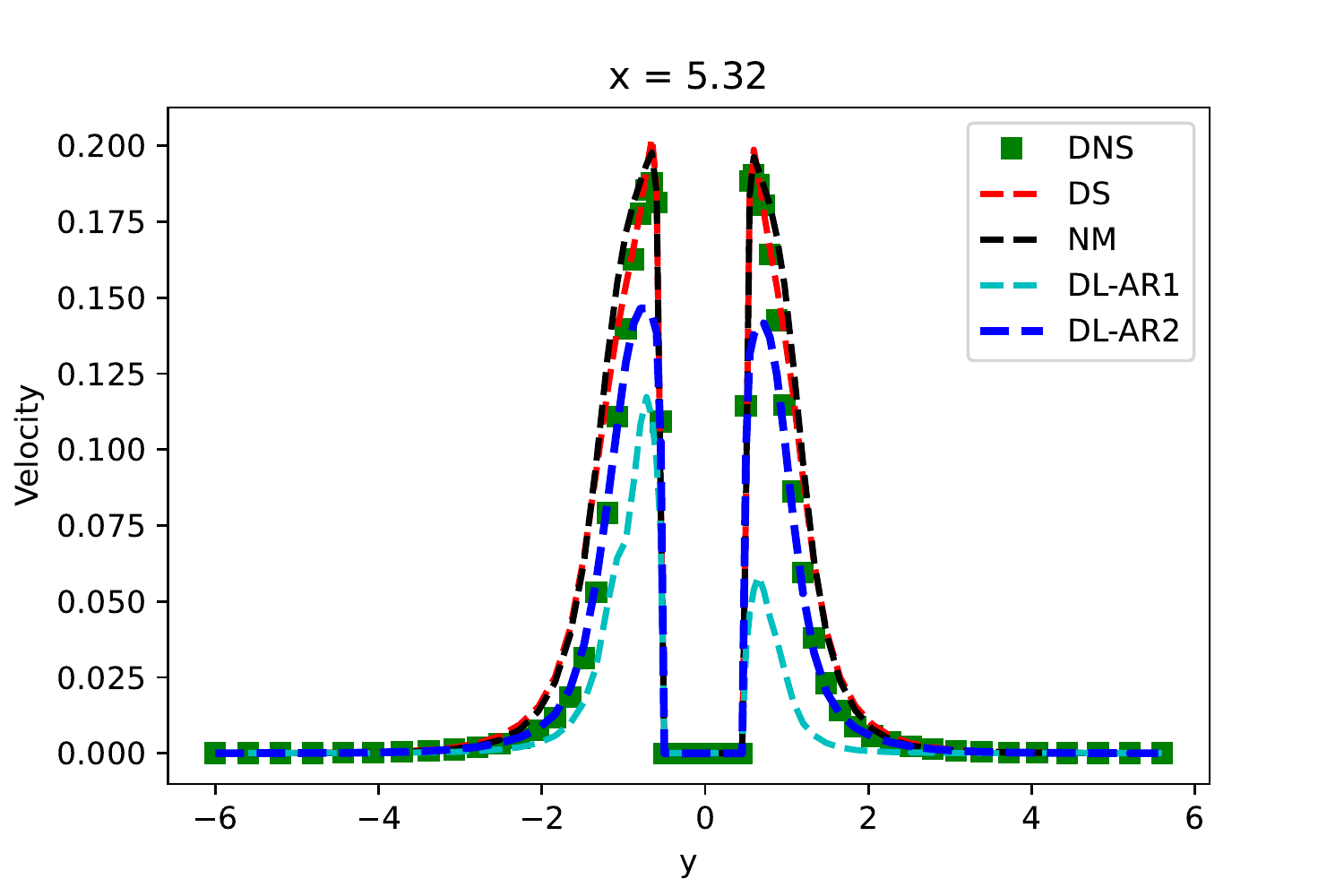}
\includegraphics[width=5cm]{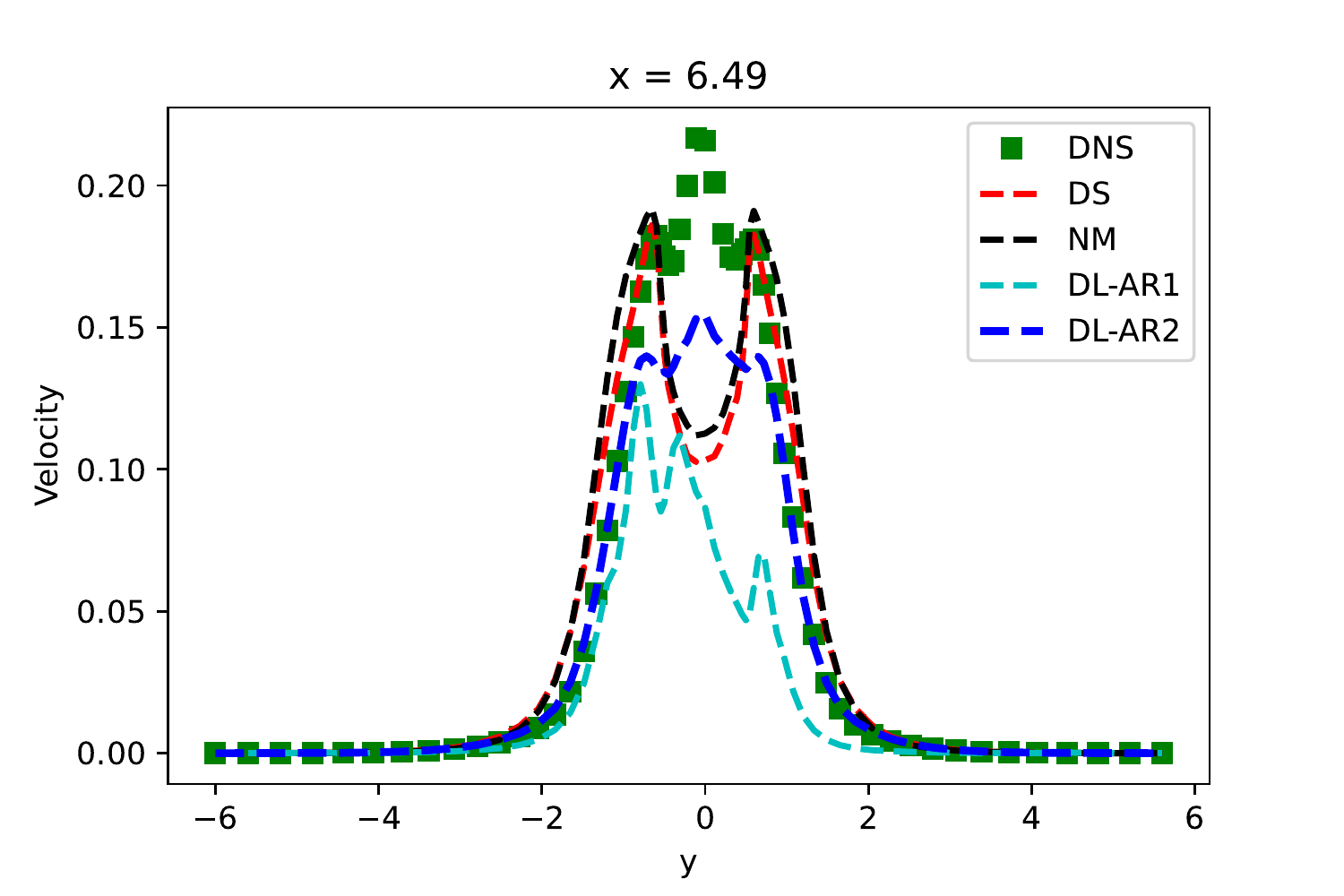}
\includegraphics[width=5cm]{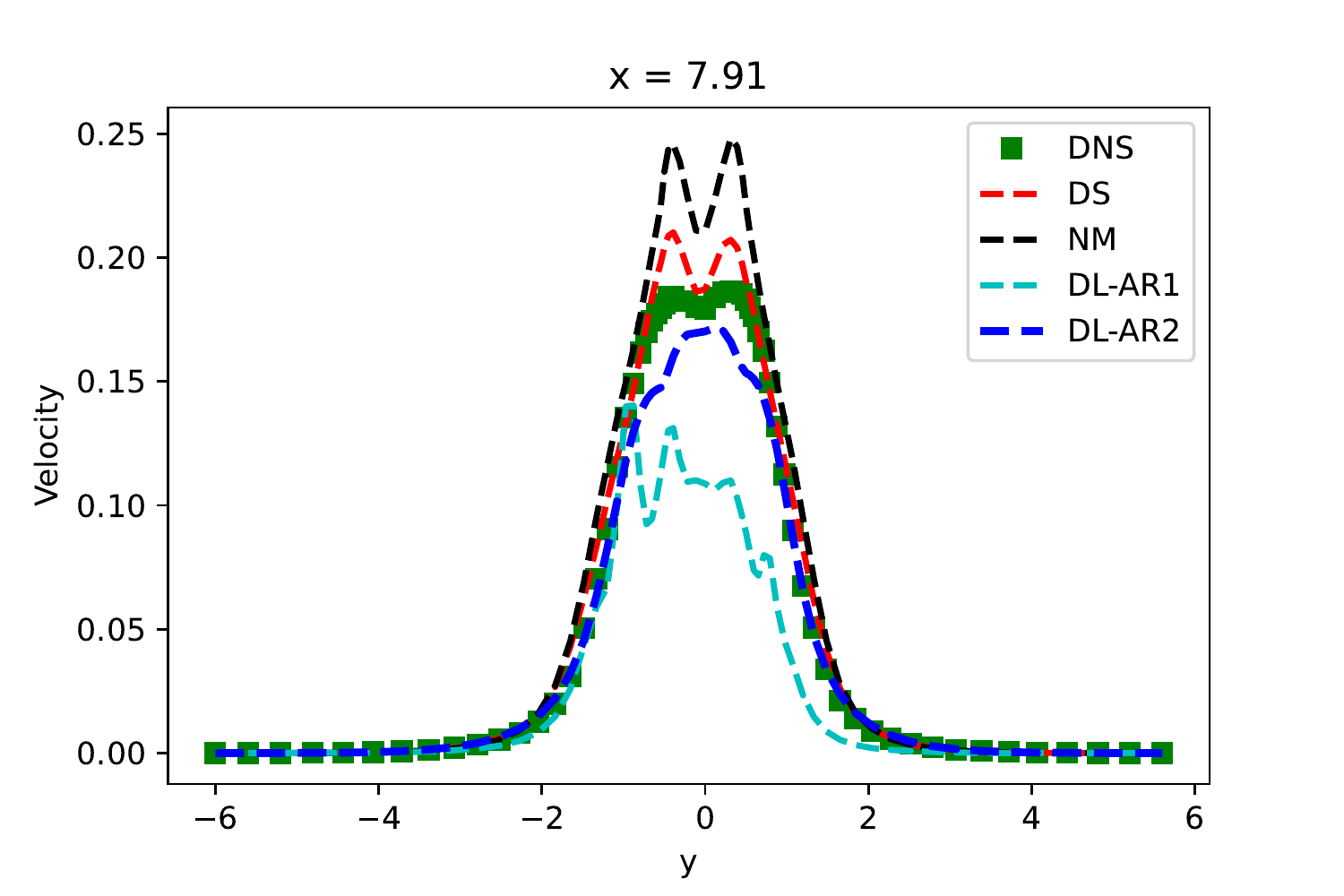}
\includegraphics[width=5cm]{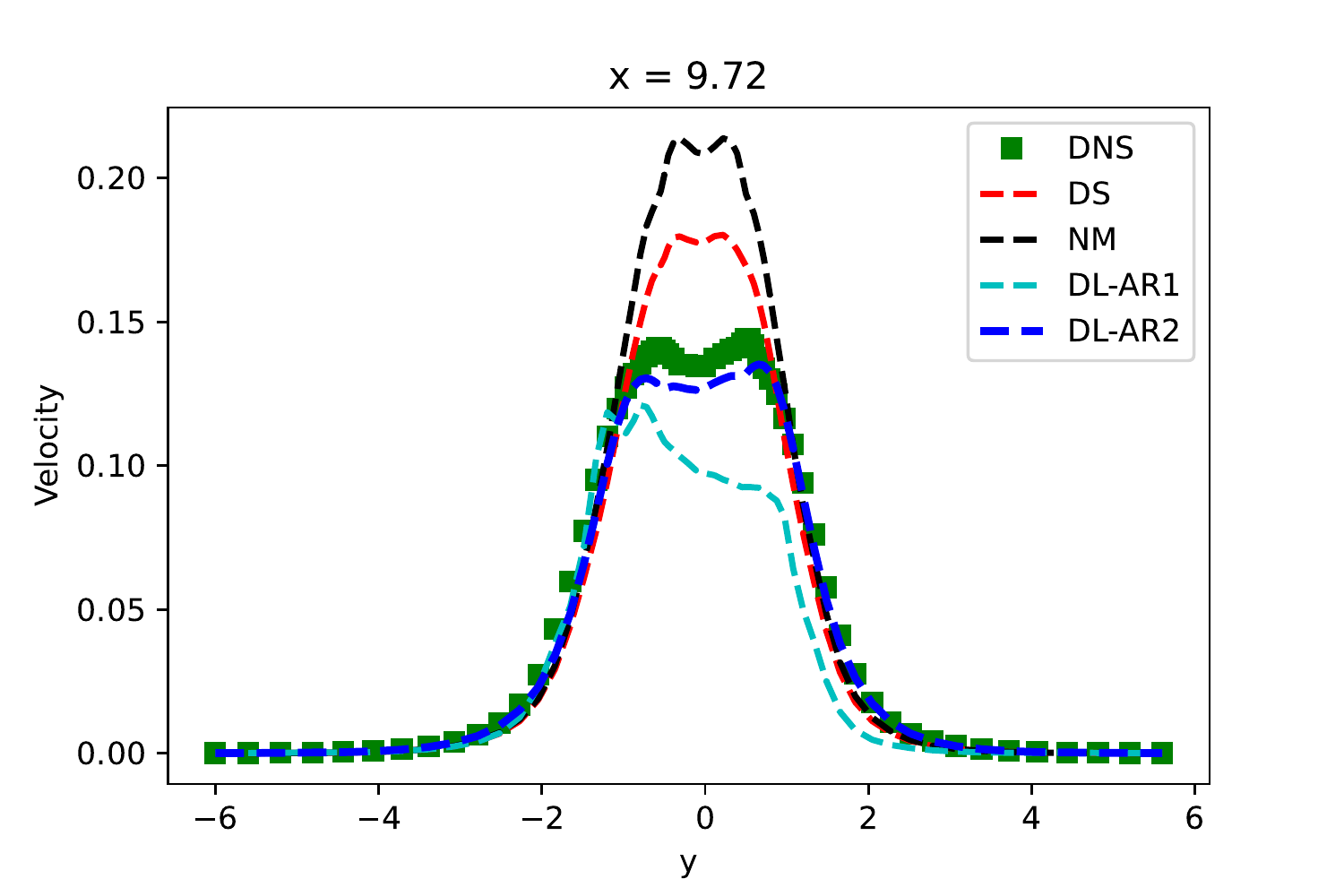}
\includegraphics[width=5cm]{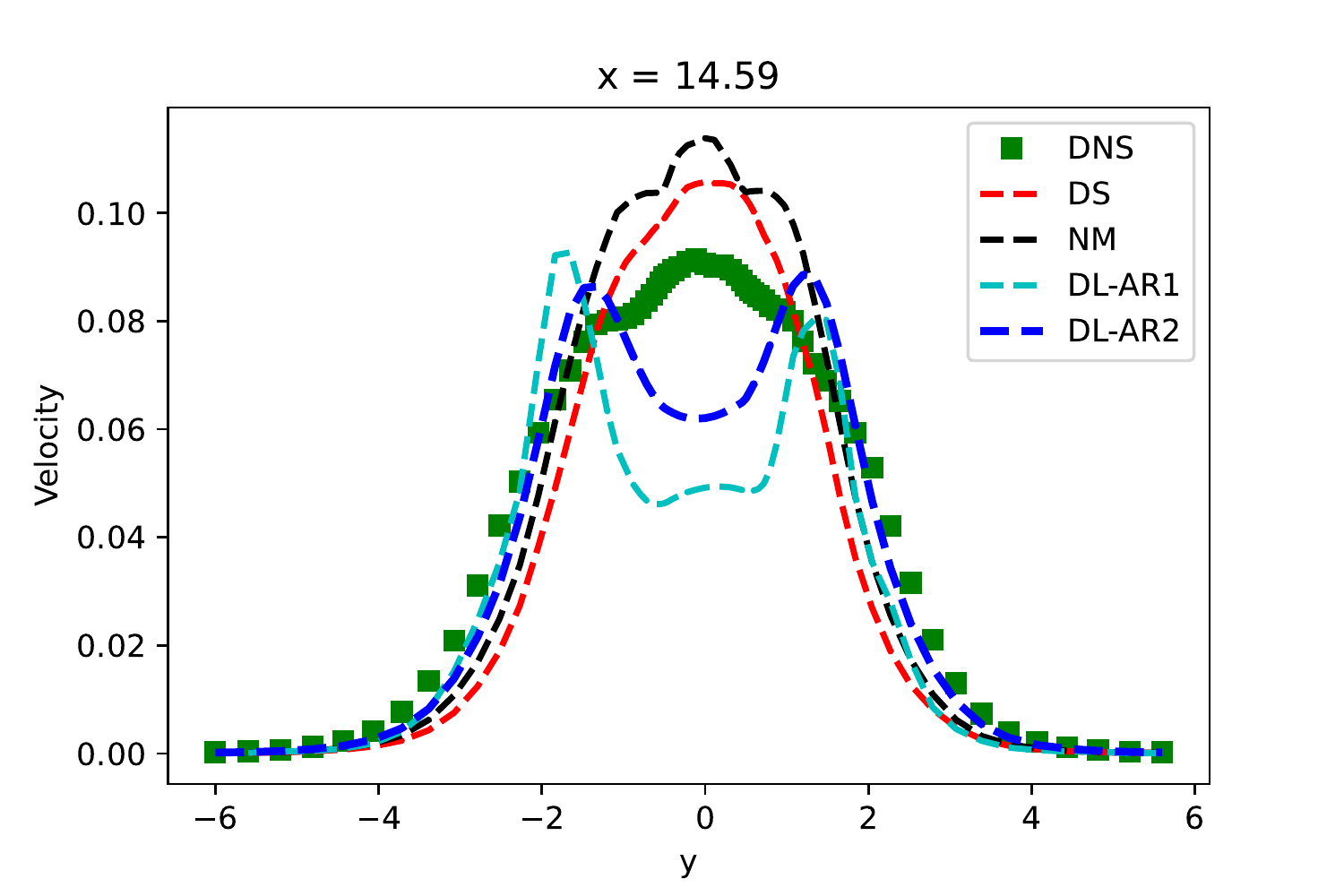}
\label{f1}
\caption{RMS profile for $u_3$ for AR4 configuration.}
\end{figure}

\begin{figure}[H]
\centering
\includegraphics[width=5cm]{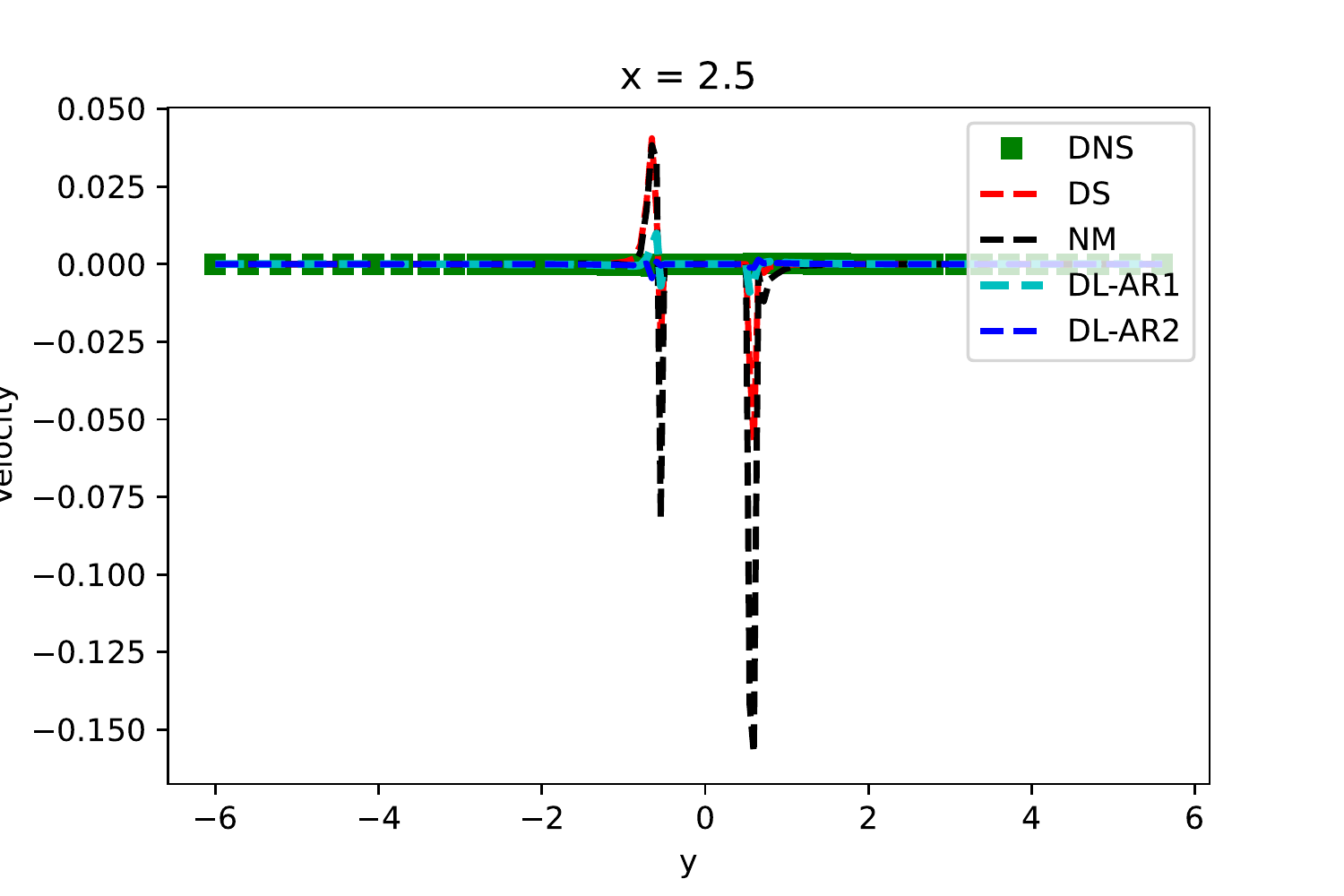}
\includegraphics[width=5cm]{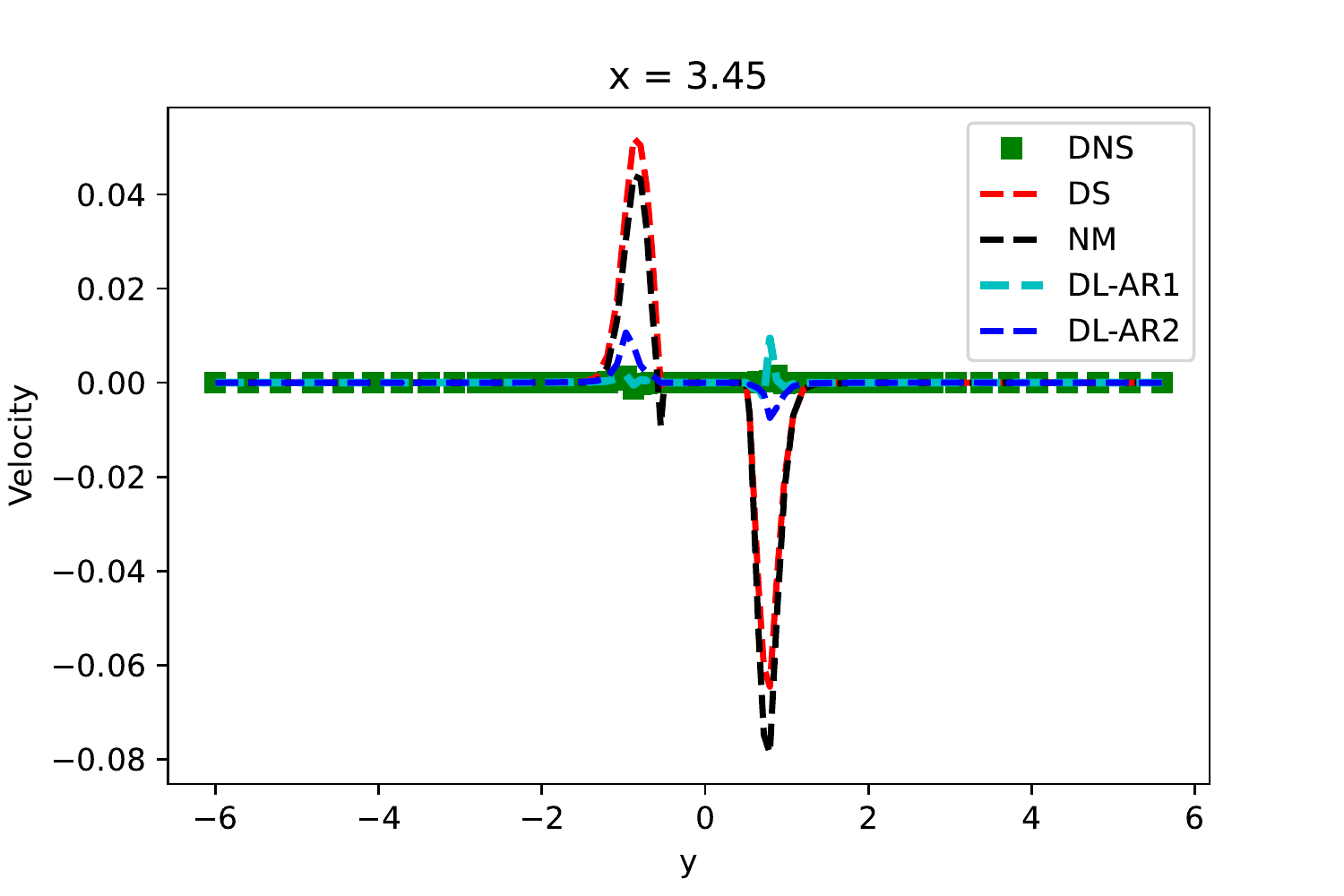}
\includegraphics[width=5cm]{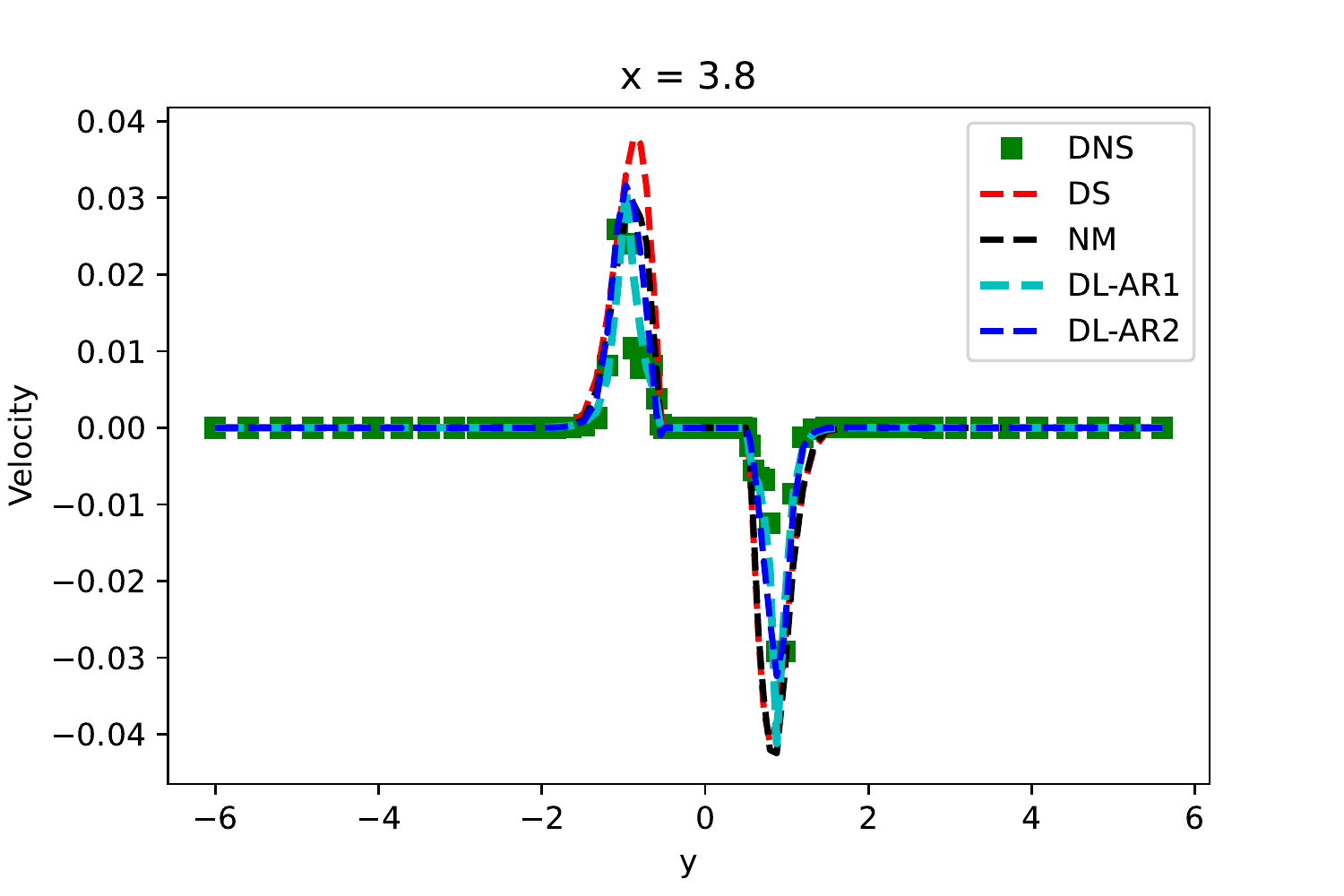}
\includegraphics[width=5cm]{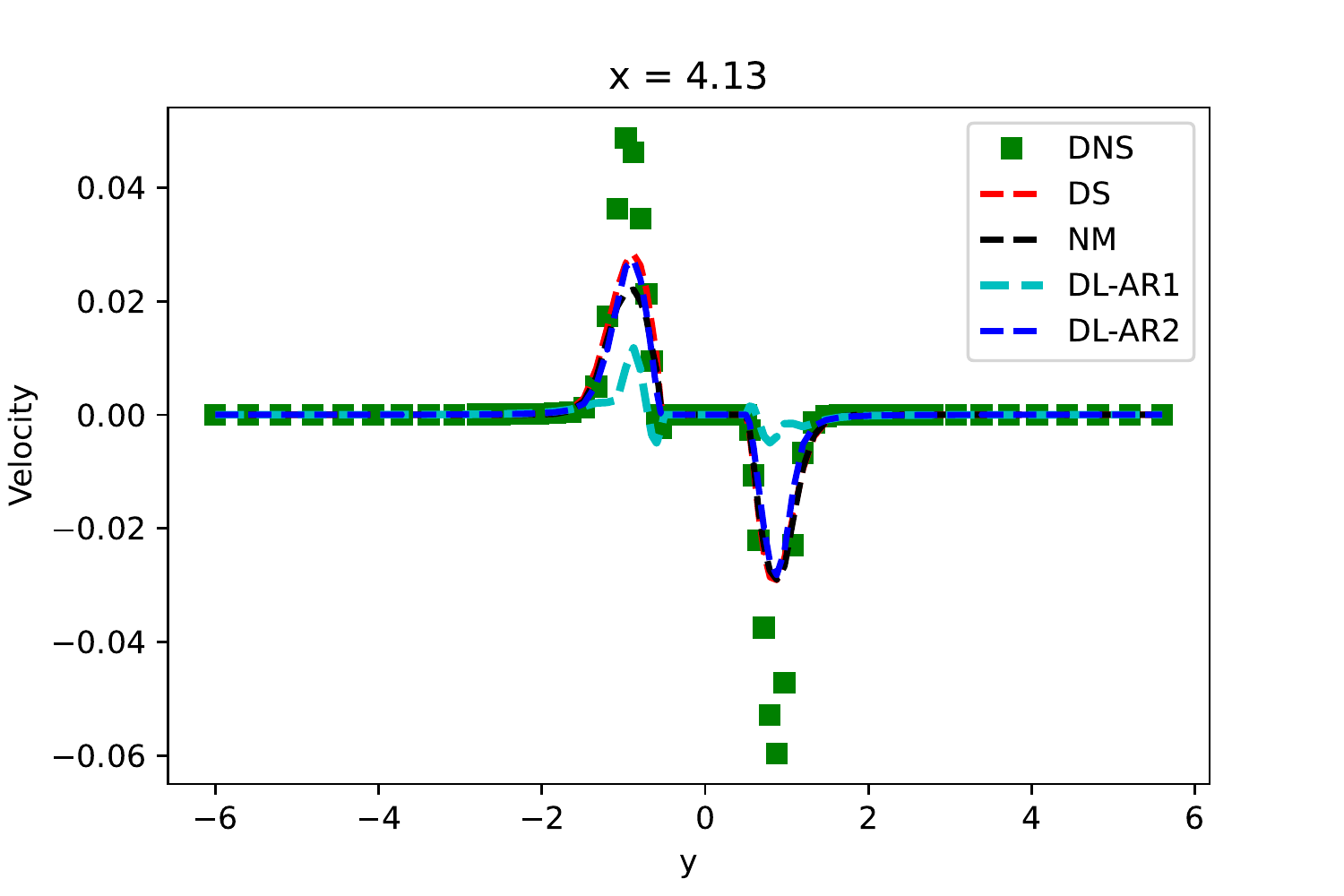}
\includegraphics[width=5cm]{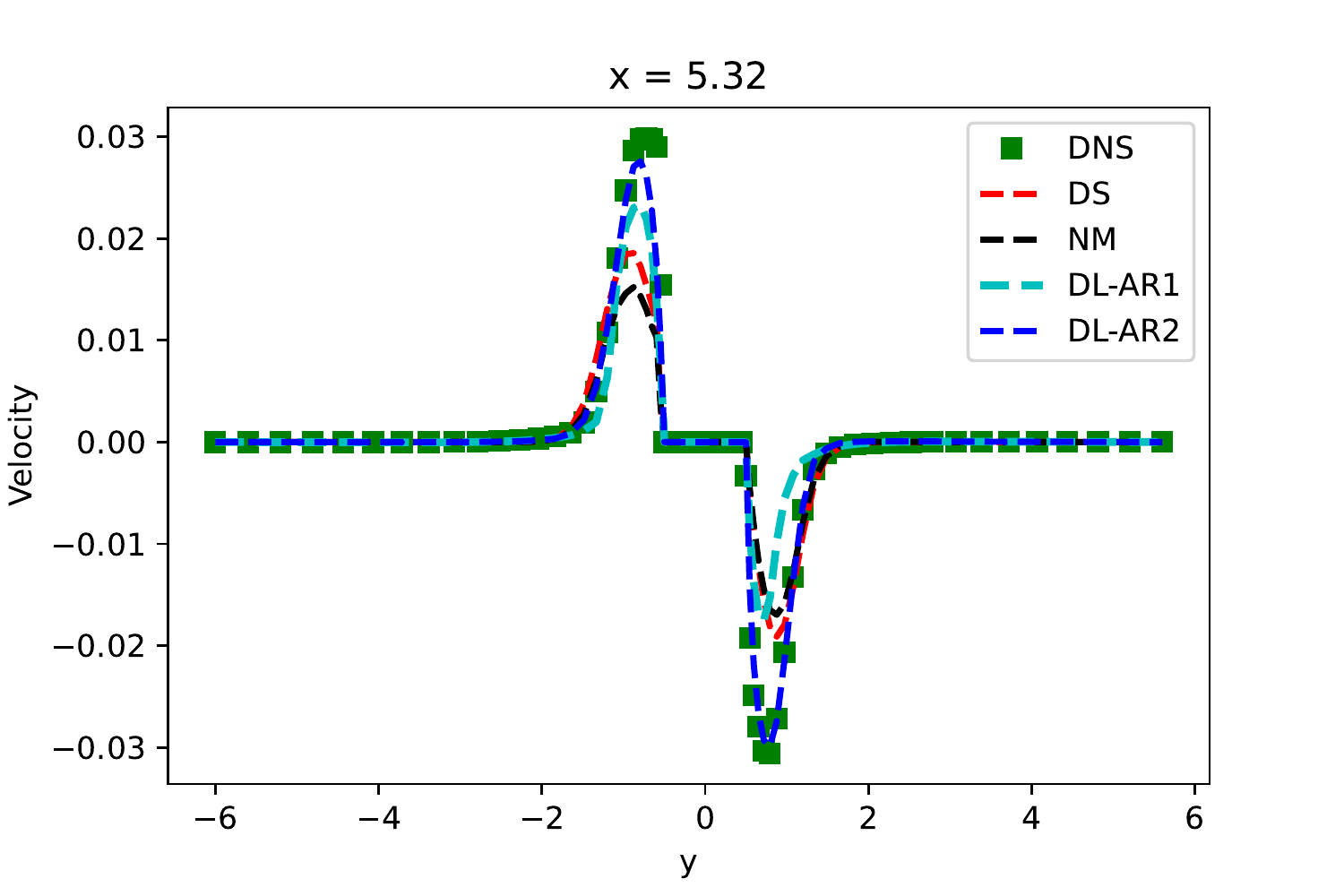}
\includegraphics[width=5cm]{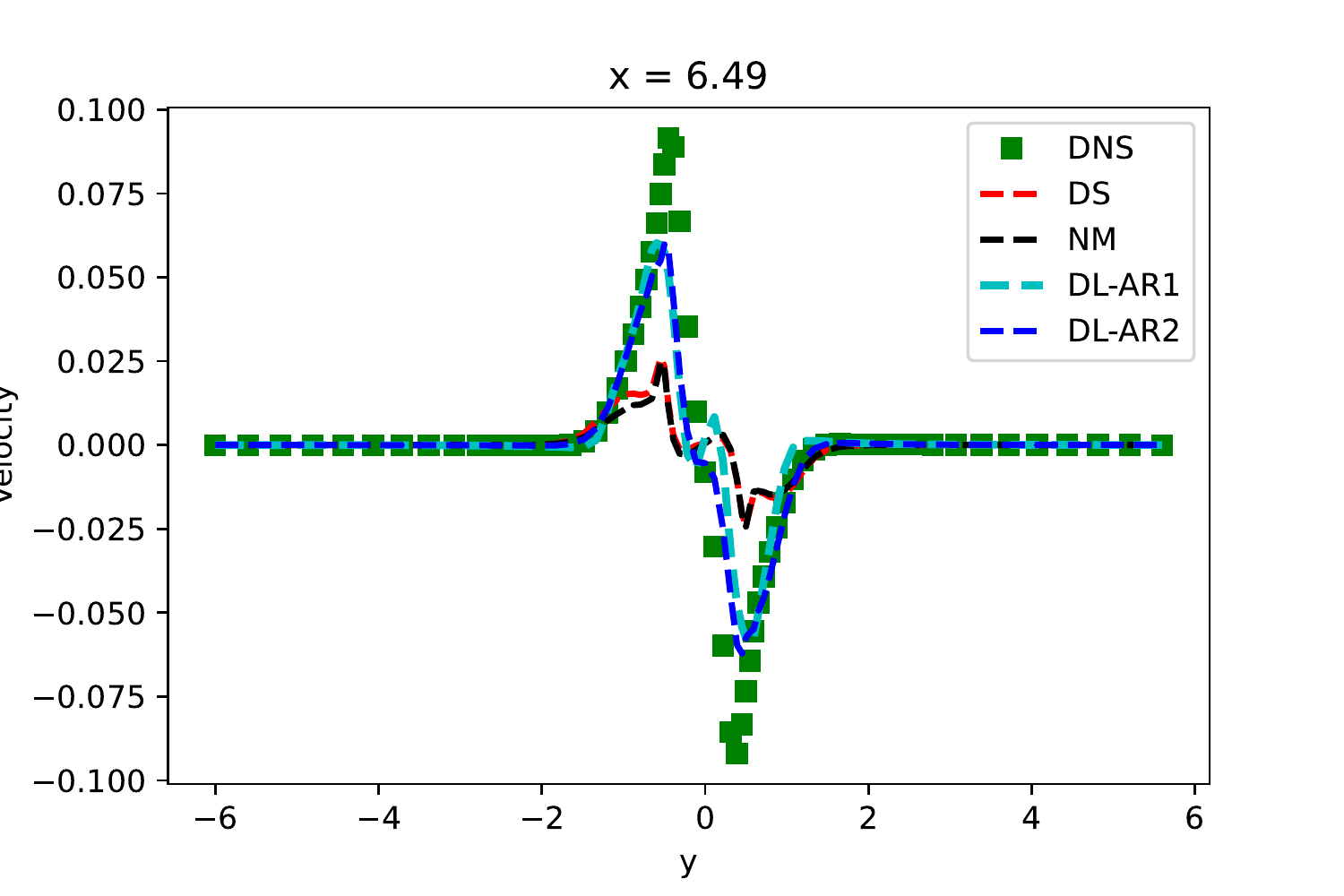}
\includegraphics[width=5cm]{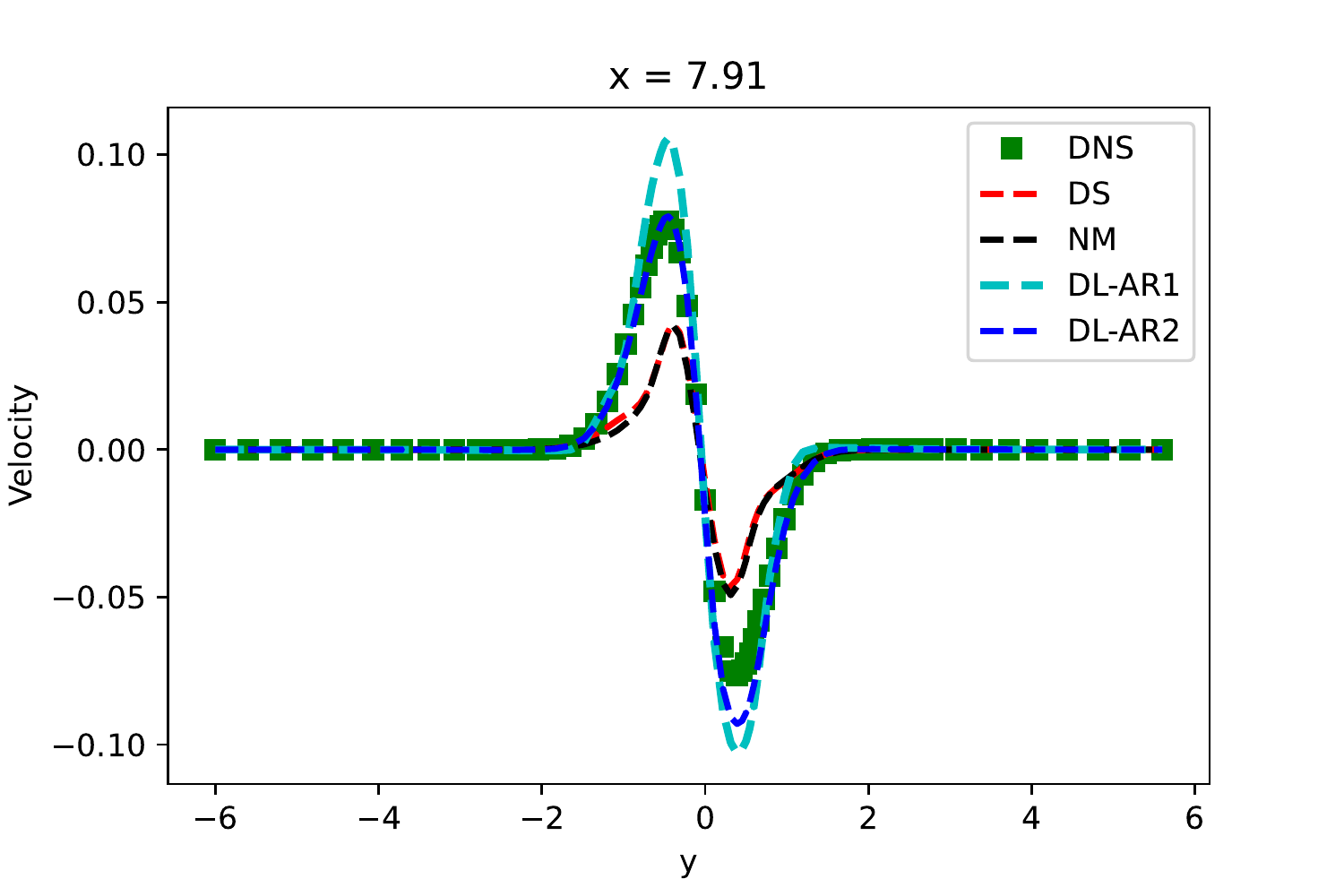}
\includegraphics[width=5cm]{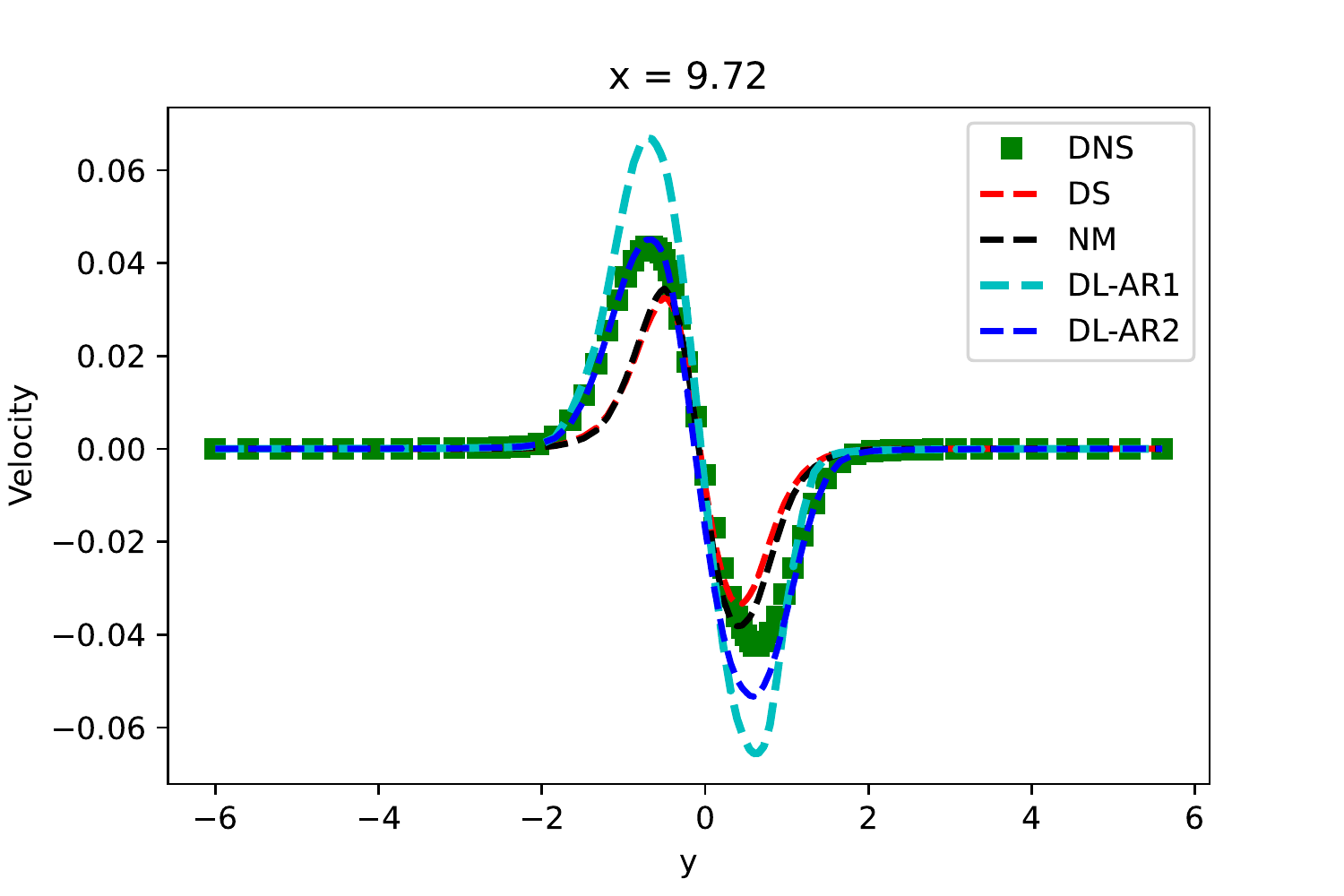}
\includegraphics[width=5cm]{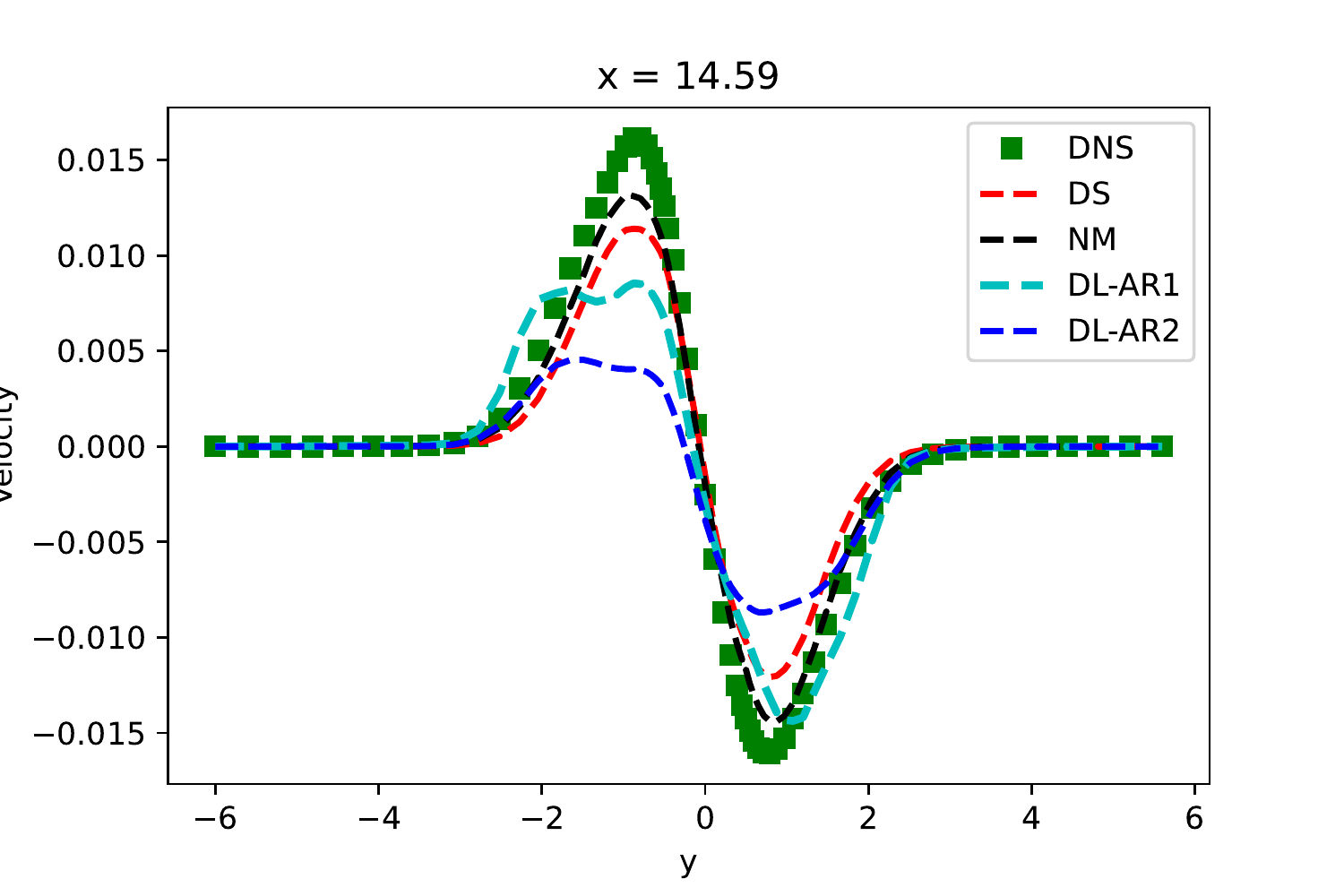}
\label{f1}
\caption{$\tau_{12}$ for AR4 configuration.}
\end{figure}

\bibliographystyle{my-elsarticle-num}
\bibliography{library}

\begin{thebibliography}{10}
\expandafter\ifx\csname url\endcsname\relax
  \def\url#1{\texttt{#1}}\fi
\expandafter\ifx\csname urlprefix\endcsname\relax\def\urlprefix{URL }\fi
\expandafter\ifx\csname href\endcsname\relax
  \def\href#1#2{#2} \def\path#1{#1}\fi

\bibitem{NASA2030}
D.~Mavriplis, E.~Lurie, W.~Gropp, D.~Darmofal, J.~Alonso, A.~Khodadoust,
  J.~Slotnick, {CFD vision 2030 study: A path to revolutionary computational
  aerosciences}, \emph{NASA Technical Report} (2014) 1--58.

\bibitem{Smagorinsky1963}
J.~Smagorinsky, {General circulation experiments with the primitive equations
  I. The basic experiment}, \emph{Monthly Weather Review} 91~(3) (1963)
  99--164.

\bibitem{Germano1991}
M.~Germano, U.~Piomelli, P.~Moin, W.~H. Cabot, {A dynamic subgrid-scale eddy
  viscosity model}, \emph{Physics of Fluids} 3 (1991) 1760--1765.

\bibitem{Lilly1992}
D.~K. Lilly, {A proposed modification of the Germano subgrid-scale closure
  method}, \emph{Physics of Fluids} 4 (1992) 633--635.

\bibitem{SpalartCFDinIndustry}
P.~Spalart, V.~Venkatakrishnan, {On the role and challenges of CFD in the
  aerospace industry}, \emph{The Aeronautical Journal} 120~(1223) (2016)
  209--232.

\bibitem{Bose2018a}
S.~T. Bose, G.~I. Park, {Wall-Modeled Large-Eddy Simulation for Complex
  Turbulent Flows}, \emph{Annual Review of Fluid Mechanics} 50 (2018) 535--561.

\bibitem{Kaltenbach1995}
H.-J. Kaltenbach, H.~Choi, {Large-eddy an airfoil simulation of flow around on
  a structured mesh}, in: Center for Turbulence Research Annual Research
  Briefs, 1995, pp. 51--60.

\bibitem{Moin2021}
K.~Goc, O.~Lehmkuhl, G.~Park, S.~Bose, P.~Moin, {Large eddy simulation of
  aircraft at affordable cost: a milestone in computational fluid mechanics},
  \emph{Flow} E14 (2021) 1.

\bibitem{Ling1}
J.~Ling, R.~Jones, J.~Templeton, {Machine learning strategies for systems with
  invariance properties}, \emph{Journal of Computational Physics} 318 (2016)
  22--35.

\bibitem{Ling2}
J.~Ling, A.~Kurzawski, J.~Templeton, {Reynolds averaged turbulence modelling
  using deep neural networks with embedded invariance}, \emph{Journal of Fluid
  Mechanics} 807 (2016) 155--166.

\bibitem{Sirignano2020}
J.~Sirignano, J.~F. MacArt, J.~B. Freund, {DPM: A deep learning PDE
  augmentation method with application to large-eddy simulation}, \emph{Journal
  of Computational Physics} 423 (2020) 109811.

\bibitem{MacArt2021}
J.~F. MacArt, J.~Sirignano, J.~B. Freund, {Embedded training of neural-network
  subgrid-scale turbulence models}, \emph{Physical Review Fluids} 6 (2021)
  050502.

\bibitem{Harlow1965}
F.~H. Harlow, J.~E. Welch, {Numerical calculation of time-dependent viscous
  incompressible flow of fluid with free surface}, \emph{Phys. Fluids} 8~(12)
  (1965) 2182--2189.

\bibitem{Kim1985}
J.~Kim, P.~Moin, {Application of a fractional-step method to incompressible
  Navier--Stokes equations}, \emph{J. Comp. Phys.} 59~(2) (1985) 308--323.
\newblock

\bibitem{Desjardins2008}
O.~Desjardins, G.~Blanquart, G.~Balarac, H.~Pitsch, {High order conservative
  finite difference scheme for variable density low Mach number turbulent
  flows}, \emph{J. Comp. Phys.} 227~(15) (2008) 7125--7159.

\bibitem{MacArt2016}
J.~F. MacArt, M.~E. Mueller, {Semi-implicit iterative methods for low Mach
  number turbulent reacting flows: Operator splitting versus approximate
  factorization}, \emph{J. Comp. Phys.} 326 (2016) 569--595.

\end{thebibliography}

\end{document}